\documentclass[sn-basic]{sn-jnl}% Basic Springer Nature Reference Style/Chemistry Reference Style
\usepackage{booktabs}  % possibly also \usepackage{longtable}
\usepackage{graphicx,mathrsfs}  % must come after epubtk
\usepackage{lscape}
\usepackage{subfigure}
\usepackage{dsfont}
\usepackage{cancel}
\usepackage{placeins}
\usepackage{mathtools}
\usepackage{enumerate}
\usepackage{amssymb,amsmath,amsthm,amsfonts,epsfig}
\usepackage{aas_macros}
\usepackage{epsf}
\usepackage{longtable}
\usepackage[usenames]{color}
\usepackage{epstopdf,pifont}
\usepackage{bm}
\usepackage{dcolumn}
\usepackage[latin1]{inputenc}
\usepackage{latexsym}
\usepackage{rotating}
\usepackage{bookmark}
\usepackage{longtable}
\usepackage{enumerate}
\usepackage{tensor}
\usepackage{mathtools}
\usepackage{nccmath}
\usepackage{multirow}
\usepackage{url}
\usepackage[normalem]{ulem} %for \sout
\usepackage[svgnames,dvipsnames,x11names]{xcolor}
\usepackage{varioref}
\usepackage{tikz}
\usepackage{feynmp-auto}
\usepackage{physics}
\usepackage[compat=1.1.0]{tikz-feynman}
\usepackage{bbm}
\usepackage{threeparttable}
\usepackage{lipsum}
\usetikzlibrary{shapes.geometric}
\usepackage{anyfontsize}
\usepackage{colortbl}
\definecolor{summersky}{cmyk}{0.71,0.33,0,0.5}
\definecolor{flamingo}{cmyk}{0,0.51,0.71,0.5}
\definecolor{rp}{cmyk}{0.2, 1, 0.6, 0}
\definecolor{pacificblue}{cmyk}{0.95,0.3,0, 0.5}
\definecolor{gray60}{cmyk}{0.4,0.4,0,0.8}

\definecolor{blue3}{RGB}{31, 119, 180}
\definecolor{red3}{RGB}{    214, 39, 40}
\definecolor{orange3}{RGB}{255, 127, 14}
\definecolor{green3}{RGB}{44, 160, 44}
\definecolor{repBlue}{RGB}{31, 119, 180}
\definecolor{repRed}{RGB}{  214, 39, 40}
\definecolor{repGreen}{RGB}{44, 160, 44}
\definecolor{MyRed}{RGB}{208,57,75}
\definecolor{MyBlue}{RGB}{68,123,178}
\definecolor{MyYellow}{RGB}{238,140,59}
\allowdisplaybreaks
\newcommand{\ben}{\begin{enumerate}}
\newcommand{\een}{\end{enumerate}}

\newcommand{\beq}{\begin{eqnarray}}
\newcommand{\eeq}{\end{eqnarray}} 
\newcommand{\ba}{\begin{align}}
\newcommand{\ea}{\end{align}}
\newcounter{mnotecount}[section]

\newcommand{\mnotex}[1]%{}
{\protect{\stepcounter{mnotecount}}$^{\mbox{\footnotesize
$%\!\!\!\!\!\!\,
\bullet$\themnotecount}}$ \marginpar{
\raggedright\tiny\em
$\!\!\!\!\!\!\,\bullet$\themnotecount: #1} }

\def\mbh{M_{\rm BH}}
\def\order{\mathcal{O}}

\def\be{\begin{equation}}
\def\ee{\end{equation}}
\def\bea{\begin{eqnarray}}
\def\eea{\end{eqnarray}}

\def\brho{\bar{\rho}}
\def\bp{\bar{p}}
\def\bpsi{\bar{\Psi}}
\renewcommand{\vec}[1]{\boldsymbol{#1}}
\renewcommand{\themnotecount}{\thesection.\arabic{mnotecount}}

\def\theequation{\thesection.\arabic{equation}}

\numberwithin{equation}{section}
\labelformat{section}{Sect.~#1} 
\labelformat{subsection}{Sect.~#1} 
\labelformat{subsubsection}{Sect.~#1}
\labelformat{subsubsubsection}{Sect.~#1}
\labelformat{paragraph}{Par.~#1}
\labelformat{equation}{Eq.~(#1)} 
\labelformat{figure}{Fig.~#1} 
\labelformat{subfigure}{Fig.~\thefigure#1} 
\labelformat{table}{Table~#1} 
\labelformat{appendix}{Appendix~#1}
\begin{document}
%%%%%%%%%%%%%%%%%%%%%%%%%%%%%%%%%%%%%%%%%%%%%%%%%%%%%%%%%%%%%%%%%%%%%%%%%%%%%%
%%%%%%%%%%%%%%%%%%%%%%%%%%%%%%%%%%%%%%%%%%%%%%%%%%%%%%%%%%%%%%%%%%%%%%%%%%%%%%
%%%%%%%%%%%%%%%%%%%%%%%%%%%%%%%%%%%%%%%%%%%%%%%%%%%%%%%%%%%%%%%%%%%%%%%%%%%%%%

\title[Tidal Response of Compact Objects]{Tidal Response of Compact Objects}

%%=============================================================%%
%% GivenName	-> \fnm{Joergen W.}
%% Particle	-> \spfx{van der} -> surname prefix
%% FamilyName	-> \sur{Ploeg}
%% Suffix	-> \sfx{IV}
%% \author*[1,2]{\fnm{Joergen W.} \spfx{van der} \sur{Ploeg} 
%%  \sfx{IV}}\email{iauthor@gmail.com}
%%=============================================================%%

\author*[1]{\fnm{Sumanta} \sur{Chakraborty}}\email{tpsc@iacs.res.in}

\author[2]{\fnm{Paolo} \sur{Pani}}\email{paolo.pani@uniroma1.it}
\equalcont{These authors contributed equally to this work.}

\affil*[1]{\orgdiv{School of Physical Sciences}, \orgname{Indian Association for the Cultivation of Science}, \orgaddress{\street{Street}, \city{Kolkata}, \postcode{700032}, \country{India}}}

\affil[2]{\orgdiv{Dipartimento di Fisica}, \orgname{Sapienza Universit\`a di Roma \& Sezione INFN Roma}, \orgaddress{\street{Piazzale Aldo Moro 5}, \city{Rome}, \postcode{00185}, \country{Italy}}}

%%%%%%%%%%%%%%%%%%%%%%%%%%%%%%%%%%%%%%%%%%%%%%%%%%%%%%%%%%%%%%%%%%%%%%%%%%%%%%
%%%%%%%%%%%%%%%%%%%%%%%%%%%%%%%%%%%%%%%%%%%%%%%%%%%%%%%%%%%%%%%%%%%%%%%%%%%%%%
%%%%%%%%%%%%%%%%%%%%%%%%%%%%%%%%%%%%%%%%%%%%%%%%%%%%%%%%%%%%%%%%%%%%%%%%%%%%%%
%%FS \date{\today}
%%FS \maketitle
%%%%%%%%%%%%%%%%%%%%%%%%%%%%%%%%%%%%%%%%%%%%%%%%%%%%%%%%%%%%%%%%%%%%%%%%%%%%%%
%%%%%%%%%%%%%%%%%%%%%%%%%%%%%%%%%%%%%%%%%%%%%%%%%%%%%%%%%%%%%%%%%%%%%%%%%%%%%%
%%%%%%%%%%%%%%%%%%%%%%%%%%%%%%%%%%%%%%%%%%%%%%%%%%%%%%%%%%%%%%%%%%%%%%%%%%%%%%
\abstract{
The tidal response of compact objects provides a powerful probe of their internal structure and of the surrounding gravitational field. We provide a comprehensive and unified overview of tidal effects in black holes, neutron stars, and exotic compact objects, with emphasis on both static and dynamical responses to external fields, encoded in Love numbers and dissipation numbers. We discuss the vanishing of static bosonic Love numbers for black holes in vacuum General Relativity, their modifications in alternative theories, in non-standard models of compact objects, and in the presence of matter, as well as their role in testing deviations from Einstein's theory and environmental effects. A fundamental distinction between bosonic and fermionic perturbations is highlighted, as the latter yield nonzero static Love numbers even for black holes in General Relativity. For neutron stars, we overview the dependence of tidal Love numbers on the equation of state, the emergence of quasi-universal relations, and the impact of rotation, nonlinearities, and dynamical effects. Exotic compact objects typically feature nonvanishing static tidal Love numbers --- a striking observational signature that differentiates them from black holes. We further review how tidal effects influence the gravitational-wave signals from binary inspirals, and explore their implications for gravitational-wave astronomy. In particular, we stress their significance for current and future detectors as tools to test General Relativity, constrain the nuclear equation of state, and probe the fundamental nature of compact objects and their environments.
}
%%%%%%%%%%%%%%%%%%%%%%%%%%%%%%%%%%%%%%%%%%%%%%%%%%%%%%%%%%%%%%%%%%%%%%%%%%%%%%
%%%%%%%%%%%%%%%%%%%%%%%%%%%%%%%%%%%%%%%%%%%%%%%%%%%%%%%%%%%%%%%%%%%%%%%%%%%%%%
%%%%%%%%%%%%%%%%%%%%%%%%%%%%%%%%%%%%%%%%%%%%%%%%%%%%%%%%%%%%%%%%%%%%%%%%%%%%%%
%%%%%%%%%%%%%%%%%%%%%%%%%%%%%%%%%%%%%%%%%%%%%%%%%%%%%%%%%%%%%%%%%%%%%%%%%%%%%%
%%%%%%%%%%%%%%%%%%%%%%%%%%%%%%%%%%%%%%%%%%%%%%%%%%%%%%%%%%%%%%%%%%%%%%%%%%%%%%
%%%%%%%%%%%%%%%%%%%%%%%%%%%%%%%%%%%%%%%%%%%%%%%%%%%%%%%%%%%%%%%%%%%%%%%%%%%%%%

\maketitle

\pagebreak
\tableofcontents
\newpage

%%%%%%%%%%%%%%%%%%%%%%%%%%%%%%%%%%%%%%%%%%%%%%%%%%%%%%%%%%%%%%%%%%%%%%%%%%%%%%
%%%%%%%%%%%%%%%%%%%%%%%%%%%%%%%%%%%%%%%%%%%%%%%%%%%%%%%%%%%%%%%%%%%%%%%%%%%%%%
%%%%%%%%%%%%%%%%%%%%%%%%%%%%%%%%%%%%%%%%%%%%%%%%%%%%%%%%%%%%%%%%%%%%%%%%%%%%%%
\section*{List of Acronyms}

\begin{tabular}{ll}
%%FS \hline
   BH  & black hole   \\
   CE  & Cosmic Explorer \\
   EFT  & effective field theory   \\
   EMRI  & extreme mass-ratio inspiral   \\
   EOB  & effective-one-body   \\
   EoS  & equation of state   \\
   ET  & Einstein Telescope   \\
   GR  & General Relativity   \\
   GW  & gravitational wave   \\
   LIGO  & laser interferometer gravitational-wave observatory  \\
   LISA  & laser interferometric space antenna   \\
   LN  & Love number   \\
   LVK   & LIGO-Virgo-KAGRA\\
   NS  & neutron star  \\
   PN  & post-Newtonian  \\
   QNM & quasinormal mode\\
%%FS   \hline
\end{tabular}

\bigskip

%%%%%%%%%%%%%%%%%%%%%%%
\section{Introduction} \label{sec:intro}
%%%%%%%%%%%%%%%%%%%%%%%
The study of tidal deformations of self-gravitating bodies dates back to 1909, when Augustus Edward Hough Love analyzed how the shape of the Earth responds to the external gravitational fields of the Moon and the Sun, introducing the quantities now known as \emph{tidal Love numbers}~(LNs)\footnote{Strictly speaking, LNs quantify the gravitational response of an object to external \emph{tidal} fields. Throughout this review we adopt a slightly broader terminology and use the term LNs also for scalar, electromagnetic, and fermionic susceptibilities of compact objects, even when these are not directly associated with gravitational tides. When needed, we will refer explicitly to \emph{tidal} LNs to emphasize the gravitational case.}. 
While today part of standard textbook material, tidal interactions
remain conceptually subtle and encode deep information about the structure of gravitational interactions. 
A simple example illustrates this point: although the Sun exerts a stronger gravitational force on the Earth than the Moon, terrestrial tides are dominated by the latter. 
The reason is that tidal effects arise not from the gravitational force itself, but from its spatial \emph{variation}. 

In Newtonian gravity, tides are governed by second derivatives of the gravitational potential. 
In General Relativity~(GR), this idea is sharpened by the equivalence principle: the gravitational field can always be locally eliminated, but its gradients --- encoded in the Riemann curvature tensor --- cannot. 
Tidal effects therefore provide the most direct and invariant characterization of gravitational interactions. 
Indeed, the strength of tidal forces measures the intensity of spacetime curvature and can diverge near spacetime singularities, such as those inside black holes~(BHs).

For nearly a century, the study of tidal deformations remained largely confined to planetary and stellar physics. 
Only relatively recently has the problem been developed systematically within relativistic gravity, and its importance dramatically elevated by the advent of gravitational-wave~(GW) astronomy. 
The inspiral phase of compact binaries encodes tidal effects in the GW signal, allowing the internal structure of the binary components to be probed through precision waveform measurements. 

For neutron stars~(NSs), tidal deformability provides direct information about the equation of state~(EoS) of ultra-dense matter, one of the central open problems in nuclear physics. 
For BHs, the situation is even more striking: in four-dimensional vacuum GR, static bosonic tidal LNs vanish exactly. 
This remarkable property reflects deep structural features of classical gravity and has important implications for effective field theory~(EFT), symmetry principles, and the nature of horizons. 
Deviations from this behavior can signal new physics, environmental effects, modified gravity, or the presence of exotic compact objects~(ECOs), thereby turning tidal interactions into probes of fundamental physics.

The first observational breakthrough came with the detection of GW170817 by the LIGO-Virgo-KAGRA~(LVK) Collaboration \citep{LIGOScientific:2017vwq}, which provided the first direct constraints on NS tidal LNs. 
Since then, theoretical progress has accelerated rapidly. 
Static and dynamical tidal responses have been analyzed within perturbation theory and EFT; dissipative effects such as tidal heating have been systematically incorporated; rotational and nonlinear corrections have been computed; and symmetry-based explanations for the vanishing of BH LNs have been uncovered. 
At the same time, the prospect of next-generation GW detectors promises percent-level measurements of tidal parameters, elevating tidal effects from largely theoretical concepts to precision diagnostics of strong gravity.

Over the past decade, tidal interactions have emerged as a unifying framework connecting disparate areas of gravitational physics, from dense-matter theory to scattering amplitudes and horizon symmetries.

The interplay between conservative and dissipative responses, the appearance of logarithmic running in dynamical tides, and the structural origin of vanishing BH LNs is reshaping our understanding of compact-object physics. 

The purpose of this review is to provide a comprehensive and self-contained account of these developments. 
We aim to bridge perturbation theory, worldline EFT, symmetry-based approaches, and GW phenomenology, presenting both conceptual foundations and key technical results.

The structure of this work is as follows. 
In the remainder of this introductory section, we present a historical overview of major milestones and a primer on the computation of tidal deformability in Newtonian gravity and GR, establishing notation and conventions used throughout the review.

\ref{sec:staticBHs} discusses static bosonic deformations of BHs in four-dimensional vacuum GR, while~\ref{sec:BHBGR} reviews extensions including surrounding matter, modified gravity, and extra dimensions. 
\ref{sec:fermionic} presents recent results on the \emph{fermionic} response of BHs in GR, highlighting that fermionic LNs are nonzero. 
\ref{sec:dynBHs} addresses the dynamical tidal response of BHs in vacuum GR, focusing on horizon absorption, dissipative effects, and outstanding issues related to conservative dynamical contributions.

\ref{sec:NSs} reviews tidal LNs of NSs in both static and dynamical regimes, including rotational and other subleading corrections. 
\ref{sec:ECOs} examines the tidal response of ECOs and BH mimickers, introducing a unified framework to describe tidal susceptibilities of BHs, NSs, and horizonless objects within extensions of the membrane paradigm. 
Finally,~\ref{sec:GW} discusses how tidal effects are modelled in GW signals and their observational implications for GW astronomy, including constraints on the NS EoS, tests of gravity and of the environment around compact objects, as well as prospects for distinguishing BHs from alternative compact objects.

We will work in units where $G = c = 1$, reinstating the appropriate factors of $G$ in selected sections whenever needed for clarity.

{\bf Note added:}
After completion of this work, we became aware of another comprehensive review on the same topic~\citep{OtherReview}. While there is partial overlap, the two works have distinct emphases and should be regarded as complementary.

%%%%%%%%%%%%%%%%%%%%%%%
\subsection{Milestones}
%%%%%%%%%%%%%%%%%%%%%%%
Despite being a century-old problem, the study of the tidal deformability of self-gravitating objects has recently flourished due to its impactful implications on GW astronomy.
While summarizing all major developments in the field is infeasible, we believe it is both useful and instructive to present a (necessarily incomplete and biased) timeline highlighting some of the most relevant milestones. A more comprehensive and up-to-date list of references can be found throughout this work.
%

%%FS \begin{itemize}[leftmargin=0.9cm, rightmargin=0cm]
\begin{itemize}
    \item[1909] \citet{Love1909} introduces Earth's susceptibilities, now known as the (tidal) LNs, which characterize the response of a planet to external tidal forces.
    \item [1933] \citet{Chandrasekhar1933} calculates the static tidal deformations of polytropic Newtonian stars.
    \item [1984] Damour presents the first computation of the static LNs of a Schwarzschild BH, in the proceedings of a Les Houches Summer School on Gravitational Radiation \citep{Deruelle:1984hq}.
    \item [2005] Goldberger and Rothstein develop a long wavelength effective point-particle description of a static compact object, systematically accounting for finite size (i.e., tidal) effects \citep{Goldberger:2004jt,Goldberger:2005cd,Porto:2016pyg}.
    \item [2005] In the same year, \citet{Porto:2005ac} includes spin effects in the worldline EFT. This was later specialized to the case of slowly-rotating objects \citep{Endlich:2015mke}. 
    \item [2007] In a seminal work, \citet{Hinderer:2007mb} computes the relativistic static LNs of a non-rotating NS. Based on this, \citet{Flanagan:2007ix} quantify the effect of the NS tidal deformability on the  gravitational inspiral waveform. 
    \item [2008] \citet{Taylor:2008xy} compute the post-Newtonian~(PN) metric of a tidally deformed binary system (see also \citealt{Poisson:2005pi}).
    \item [2009] In two independent papers, \citet{Binnington:2009bb}, and \citet{Damour:2009vw} (the latter building on the formalism in \citet{Damour:1991yw,Damour:1990pi,Damour:1992qi,Damour:1993zn}) develop the theory of the relativistic static LNs for static BHs and NSs in GR, also introducing the classification between electric (i.e., even-parity) and magnetic (i.e., odd-parity) tidal perturbations. First detailed proof of the vanishing of the static bosonic LNs for static BHs in GR, which also underscores a naturalness problem from the EFT perspective \citep{Porto:2016zng}.
    \item [2012] \citet{Kol:2011vg} compute the static electric LNs of a non-rotating and spherically symmetric BH in arbitrary dimensions. This settles some ambiguities in the definition of the static LNs and demonstrates that the BH LNs do not vanish in higher dimensions. 
    \item [2013] \citet{Yagi:2013bca,Yagi:2013awa} show that the moment of inertia, the (quadrupolar, electric) LN, and the spin-induced quadrupole moment of a NS satisfy approximately universal relations that are only mildly sensitive to the EoS. Analogous approximately universal relations were later found among LNs of higher multipolar order or different parity, and for binary systems (see \citealt{Yagi:2016bkt} for a review).
    \item [2014] \citet{PoissonWill} publish the textbook ``Gravity'', developing a modern framework to study the tidal deformabilities of compact objects and their impact on the GW signal from binary coalescence.
    \item [2015] The static LNs of rotating objects are computed perturbatively in the spin \citep{Poisson:2014gka,Pani:2015hfa,Pani:2015nua,Landry:2015zfa}, to linear and quadratic order, for NSs and BHs, respectively \citep{Pani:2015hfa,Pani:2015nua}. This introduces a new class of rotational LNs and suggests that the static LNs of a spinning BH are also zero.
    \item [2015] The static LNs of ECOs are computed first for gravastars \citep{Pani:2015tga}, and later for boson stars \citep{Cardoso:2017cfl,Sennett:2017etc} and other models of BH mimickers \citep{Cardoso:2017cfl}.
    \item [2015] \citet{Landry:2015cva} introduce the magnetic tidal LNs of an irrotational fluid. Few years later, the difference between magnetic tidal LNs of a static and irrotational fluid is clarified \citep{Pani:2018inf}.
    \item [2016] An effective action for dynamical tides in GR is developed \citep{Steinhoff:2016rfi}, extending earlier Newtonian results \citep{Flanagan:2007ix}.
    \item [2017] The LVK Collaboration measures the static LNs of a NS for the first time, from the signal of GW170817, the first  binary NS coalescence ever detected \citep{LIGOScientific:2017vwq}. This event provides an upper bound on the tidal deformability, ruling out some of the softest EoS \citep{Abbott:2018exr}.
    \item [2017] The static LNs of BHs beyond GR are computed \citep{Cardoso:2017cfl,Cardoso:2018ptl} and shown to be nonzero in certain theories.
    \item [2019] The effect of nonzero tidal deformability is estimated in an extreme mass-ratio inspiral, showing that the primary LN affect the waveform at leading order in the mass ratio \citep{Pani:2019cyc}.
    \item [2020] Working independently, \citet{Chia:2020yla} and \citet{LeTiec:2020spy} show that the static tidal LNs exactly vanishes for a Kerr BH of any angular momentum.
    \item [2020] The vanishing of the static bosonic LNs of a Kerr BH is explained in terms of special symmetries of the static perturbations of the Kerr metric \citep{Hui:2020xxx,Charalambous:2021mea,Charalambous:2021kcz,Hui:2021vcv,Gounis:2024hcm,Kehagias:2024rtz,Combaluzier-Szteinsznaider:2024sgb} and, more generally, of any static vacuum GR solutions even at the nonlinear level \citep{Parra-Martinez:2025bcu}.
    \item [2021] A series of studies starts investigating the dynamical (i.e., frequency-dependent) LNs of BHs and other objects \citep{Bonelli:2021uvf,Creci:2021rkz, Consoli:2022eey, Saketh:2023bul, Perry:2023wmm, Chakraborty:2023zed}.
    \item [2022] Tidal effects in a  binary are studied in terms of scattering amplitudes \citep{Mougiakakos:2022sic, Saketh:2023bul, Saketh:2024juq, Ivanov:2026icp}.
    \item [2023] The static LNs of a BH are shown to vanish also at the non-linear level \citep{DeLuca:2023mio,Riva:2023rcm}. Few years later, the quadratic corrections to the LNs of a NS are computed for the first time \citep{Pitre:2025qdf, Pani:2025qxs}, based on the formalism in~\citet{Poisson:2020vap}.
    \item [2025] BHs in GR are shown to have \emph{nonzero} static LNs and zero dissipation numbers for fermionic perturbations \citep{Chakraborty:2025zyb}, highlighting a fundamental distinction between bosonic and fermionic perturbations of a BH, which can be interpreted as a breaking of the hidden symmetries that underlie the vanishing of the LNs in the bosonic sector.
    The vanishing of the fermionic dissipation numbers is tied to the absence of superradiance for fermions \citep{Unruh:1973bda, Brito:2015oca}. 
\end{itemize}

%%\newpage

\subsection{Tidal Love numbers: A primer}
In Newtonian gravity, the gravitational force is given by the gradient of the gravitational potential, while tidal forces correspond to the spatial derivatives of the gravitational force --~that is, the second derivatives of the potential. In GR, the metric replaces the gravitational potential as the fundamental quantity, and tidal forces are instead encoded in the Riemann curvature tensor. An alternative interpretation arises from the geodesic deviation equation, which describes the relative acceleration of nearby geodesics in terms of the Riemann tensor. Since the latter vanishes only in flat spacetimes (unlike the metric, which can always be locally transformed to the Minkowski form in a local inertial frame), tidal forces offer the most direct way to characterize the strength of the gravitational field.  For instance, strong singularities --~such as the one inside a Schwarzschild BH~-- are associated with diverging tidal forces, whereas weak singularities --~like the mass inflation singularity at the Cauchy horizon of a Reissner--Nordstr\"om BH \citep{Poisson:1990eh,Ori:1991zz}~-- might exhibit finite tidal forces.

%%%%%%%%%%%%%%%%%%%%%%%%%%%%%%%%%%%%%%%%%%%%%%%%%%%%%%%%%%%%%
%%%%%%%%%%%%%%%%%%%%%%%%%%%%%%%%%%%%%%%%%%%%%%%%%%%%%%%%%%%%%
%%%%%%%%%%%%%%%%%%%%%%%%%%%%%%%%%%%%%%%%%%%%%%%%%%%%%%%%%%%%%
\subsubsection{Newtonian case: Electric tidal Love numbers}

The concept of tidal response was originally formulated within the framework of Newtonian gravity. In this context, one considers a static, spherically symmetric, and non-rotating body of mass $M$ immersed in an external tidal field. The total gravitational potential at a distance $r$ from the center of mass of the body is then given by the superposition of the potential generated by the body itself, denoted by $U_{\rm body}$, and the external tidal potential, denoted by $U_{\rm tidal}$.
The self-potential $U_{\rm body}$ can be further decomposed into two contributions: (a) the unperturbed potential arising from the body's spherically symmetric mass distribution, and (b) the perturbative response of the body to the external tidal field, which encodes its deformation and gives rise to higher multipole moments. As a result, the total gravitational potential takes the following form \citep{PoissonWill,LeTiec:2020bos}:
%%%%%%%%%%%%%%%%%%%%%%%%%%%%%%%%%%%%%%%%%%%%%%%%%%%%%%%%%%%%%%%%%%%%%%%%%%%%%%
\begin{equation}
U=\underbrace{\frac{M}{r}+\sum_{\ell=2}^{\infty}\sum_{m=-\ell}^{\ell}\frac{(2\ell-1)!!}{\ell!}\frac{M_{\ell m}}{r^{\ell+1}}Y_{\ell m}}_{U_{\text{body}}}-\underbrace{\sum_{\ell=2}^{\infty}\sum_{m=-\ell}^{\ell}\frac{(\ell-2)!}{\ell!}\mathcal{E}_{\ell m}r^\ell Y_{\ell m}}_{U_{\text{tidal}}}~.
\label{total_potential}
\end{equation}
%%%%%%%%%%%%%%%%%%%%%%%%%%%%%%%%%%%%%%%%%%%%%%%%%%%%%%%%%%%%%%%%%%%%%%%%%%%%%%
Here, $(M/r)$ is the potential of the unperturbed spherically symmetric mass distribution, $Y_{\ell m}(\theta,\phi)$ are the standard spherical harmonics, while $M_{\ell m}$ are the mass multipole moments induced by the tidal deformation encoded in the tidal moments $\mathcal{E}_{\ell m}$ \citep{1955MNRAS.115..101B, PoissonWill, Hinderer:2007mb, Binnington:2009bb, Damour:2009vw}. The moments $M_{\ell m}$ and $\mathcal{E}_{\ell m}$ must be related to each other, since the cause of tidal deformation is the presence of the tidal moments $\mathcal{E}_{\ell m}$. For small tidal fields, the response, i.e., the mass multipole moments $M_{\ell m}$, can be written as \citep{PoissonWill,LeTiec:2020bos}: 
%%%%%%%%%%%%%%%%%%%%%%%%%%%%%%%%%%%%%%%%%%%%%%%%%%%%%%%%%%%%%%%%%%%%%%%%%%%%%%
\begin{equation}
M_{\ell m}=-\frac{(\ell-2)!}{(2\ell-1)!!}\left[2k^{\rm E}_{\ell m}\mathcal{E}_{\ell m}-M\nu^{\rm E}_{\ell m}\dot{\mathcal{E}}_{\ell m}-2k^{\rm E, \omega^{2}}_{\ell m}M^{2}\ddot{\mathcal{E}}+\cdots\right]R^{2\ell +1}\,,
\label{I_def}
\end{equation}
%%%%%%%%%%%%%%%%%%%%%%%%%%%%%%%%%%%%%%%%%%%%%%%%%%%%%%%%%%%%%%%%%%%%%%%%%%%%%%
where $k^{\rm E}_{\ell m}$'s are the static electric-type LNs, ${\nu}^{\rm E}_{\ell m}$ are the tidal dissipation numbers\footnote{Sometimes $k^{\rm E}_{\ell m}$'s are called the \emph{conservative} LNs, since they are related to the tidal field $\mathcal{E}_{\ell m}$ and are invariant under time-reversal, while ${\nu}^{\rm E}_{\ell m}$'s are referred to as \emph{dissipative} LNs, since they are related to the time derivative of the tidal field and are not invariant under time reversal.}, and $k^{\rm E, \omega^{2}}$ is the dynamical electric-type LNs. Note that the dissipation number is proportional to $\tau_{\rm d}$, which is the viscosity induced time-delay \citep{PoissonWill}, while $R$ is a characteristic length scale related to the unperturbed size of the deformed body (typically, its radius). In the above formula, no summation is left implicit and the dots represent higher-order time derivatives of the tidal field.
Note that the pre-factor ${-(\ell-2)!/(2\ell-1)!!}$ in~\ref{I_def} is a matter of convention and may differ from that adopted in other works. However, this choice does not affect the definition of the LNs and is therefore irrelevant for our purposes. The LNs, as presented above, have also been defined differently in some of the literature, primarily due to variations in normalization conventions. For completeness, in~\ref{conventionLN} we shall provide the relevant conversion factors among the different definitions of the LNs used in the literature and the ones adopted in this review.

In the Newtonian limit, and for static perturbations, we will only have to consider the LNs $k_{\ell m}^{\rm E}$, which are also static. In this case, we can write $g_{tt}=-(1-2U)$, and hence the perturbation of the $g_{tt}$ component, as presented in~\ref{total_potential}, reads,
%%%%%%%%%%%%%%%%%%%%%%%%%%%%%%%%%%%%%%%%%%%%%%%%%%%%%%%%%%%%%%%%%%%%%%%%%%%%%%
\begin{align}
\delta g_{tt}|_{\rm static}&=\sum_{\ell m}\left[\frac{2}{r^{\ell+1}}\left\{\frac{(2\ell-1)!!M_{\ell m}}{\ell!}\right\}-\frac{2\mathcal{E}_{\ell m}r^{\ell}}{\ell(\ell-1)}\right]Y_{\ell m}
\nonumber
\\
&=-\sum_{\ell,m}\frac{2\mathcal{E}_{\ell m}r^{\ell}}{\ell(\ell-1)}\left[1+2k_{\ell m}^{\rm E}\left(\frac{R}{r}\right)^{2\ell+1}\right]Y_{\ell m}~.
\label{deltag}
\end{align}
%%%%%%%%%%%%%%%%%%%%%%%%%%%%%%%%%%%%%%%%%%%%%%%%%%%%%%%%%%%%%%%%%%%%%%%%%%%%%%
Thus, for static perturbations in Newtonian gravity, the LNs are simply the ratio of the induced mass multipole moments to the external tidal moments, with a factor $R^{-2\ell-1}$ included to render the LNs $k_{\ell m}^{\rm E}$, dimensionless.

In a dynamical context, one must use~\ref{I_def}. Since it is often convenient to work in the frequency domain, the relation between the multipole moments and the tidal moments can be obtained by replacing each time derivative of the tidal moments with $-i\omega\mathcal{E}_{\ell m}$. Therefore, the potential (and hence the perturbation of the $g_{tt}$ component of the metric) in the frequency domain becomes:
%%%%%%%%%%%%%%%%%%%%%%%%%%%%%%%%%%%%%%%%%%%%%%%%%%%%%%%%%%%%%%%%%%%%%%%%%%%%%%
\begin{equation}
\delta g_{tt}|_{\rm dynamic}=-\sum_{\ell m}\frac{2(\ell-2)!}{\ell!}\mathcal{E}_{\ell m}r^{\ell}\left[1+\mathcal{F}_{\ell m}(\omega)\left(\frac{R}{r}\right)^{2\ell+1}\right]Y_{\ell m}~,
\label{total_potential_Fourier_space}
\end{equation}
%%%%%%%%%%%%%%%%%%%%%%%%%%%%%%%%%%%%%%%%%%%%%%%%%%%%%%%%%%%%%%%%%%%%%%%%%%%%%%
where $\omega$ is the mode frequency and
%%%%%%%%%%%%%%%%%%%%%%%%%%%%%%%%%%%%%%%%%%%%%%%%%%%%%%%%%%%%%%%%%%%%%%%%%%%%%%
\begin{equation}
\mathcal{F}_{\ell m}(\omega)=2k^{\rm E}_{\ell m}+iM\omega \nu^{\rm E}_{\ell m}+2k^{\rm E, \omega^{2}}_{\ell m}M^{2}\omega^{2}+\mathcal{O}(M^{3}\omega^{3})
\label{def:F}
\end{equation}
%%%%%%%%%%%%%%%%%%%%%%%%%%%%%%%%%%%%%%%%%%%%%%%%%%%%%%%%%%%%%%%%%%%%%%%%%%%%%%
is the tidal response function. The dynamical LNs can be obtained by taking the real part of the response function, i.e., $k_{\ell m}^{\rm E}(\omega)=k^{\rm E}_{\ell m}+k^{\rm E, \omega^{2}}_{\ell m}M^{2}\omega^{2}+\mathcal{O}(M^{4}\omega^{4})=(1/2)\textrm{Re}\mathcal{F}_{\ell m}$. This formalism can also be applied to slowly rotating self-gravitating bodies with $\omega$ being replaced by $\bar{\omega}$, the mode frequency in the body's co-rotating frame of reference \citep{1970A&A.....4..452Z, PoissonWill, Charalambous:2021mea}. 

%%%%%%%%%%%%%%%%%%%%%%%%%%%%%%%%%%%%%%%%%%%%%%%%%%%%%%%%%%%%%
%%%%%%%%%%%%%%%%%%%%%%%%%%%%%%%%%%%%%%%%%%%%%%%%%%%%%%%%%%%%%
%%%%%%%%%%%%%%%%%%%%%%%%%%%%%%%%%%%%%%%%%%%%%%%%%%%%%%%%%%%%%
\subsubsection{Analogy between Newtonian tidal deformability and electromagnetic susceptibility}

An instructive analogy can be drawn between tidal deformability in Newtonian gravity and susceptibility in electromagnetism, both describing the linear response of a system to an external field.
In classical electromagnetism, a dielectric material placed in an external electric field develops an induced polarization, proportional to the applied field through the electric susceptibility.
Within the linear regime, the induced polarization $\mathbf{P}$ in response to an external electric field $\mathbf{E}$ reads
%%%%%%%%%%%%%%%%%%%%%%%%%%%%%%%%%%%%%%%%%%%%%%%%%%%%%%%%%%%%%%%%%%%%%%%%%%%%%%
\begin{equation}
\mathbf{P}=\chi_{\rm e}\,\mathbf{E}\,,
\end{equation}
%%%%%%%%%%%%%%%%%%%%%%%%%%%%%%%%%%%%%%%%%%%%%%%%%%%%%%%%%%%%%%%%%%%%%%%%%%%%%%
where $\chi_{\rm e}$ is the susceptibility.  
In both the gravitational and electromagnetic cases, the response function encodes intrinsic information about the internal structure and composition of the system.  
Thus, LNs can be interpreted as generalized gravitational ``susceptibilities,'' quantifying how the internal structure of an object resists or accommodates external gravitational gradients.  
Just as a highly polarizable material has a large susceptibility and responds strongly to electric fields, a highly deformable object has a large LN and develops substantial multipolar deformations under tidal forces.  

Being electromagnetism a relativistic theory, there also exists a susceptibility associated with the response of a material to an external magnetic field.  
In this case, an object acquires a magnetization $\mathbf{M}$ when placed in a magnetic field $\mathbf{B}$, such that
%%%%%%%%%%%%%%%%%%%%%%%%%%%%%%%%%%%%%%%%%%%%%%%%%%%%%%%%%%%%%%%%%%%%%%%%%%%%%%
\begin{equation}
\mathbf{M}=\chi_{\rm m}\,\mathbf{B}\,.
\end{equation}
%%%%%%%%%%%%%%%%%%%%%%%%%%%%%%%%%%%%%%%%%%%%%%%%%%%%%%%%%%%%%%%%%%%%%%%%%%%%%%
As we will show, this analogy extends naturally to gravity in the relativistic case, where both electric and magnetic LNs arise.  
Moreover, just as there are materials with $\chi_{\rm e/m}>0$, known as paraelectric/paramagnetic, and with $\chi_{\rm e/m}<0$, known as dielectric/diamagnetic, there exist compact objects whose LNs can be positive or negative.  
Finally, as we shall discuss, vacuum BHs in GR possess vanishing static LNs, analogous to perfect conductors in electromagnetism, for which internal fields precisely cancel the applied ones, leaving no net polarization.

%%%%%%%%%%%%%%%%%%%%%%%%%%%%%%%%%%%%%%%%%%%%%%%%%%%%%%%%%%%%%
%%%%%%%%%%%%%%%%%%%%%%%%%%%%%%%%%%%%%%%%%%%%%%%%%%%%%%%%%%%%%
%%%%%%%%%%%%%%%%%%%%%%%%%%%%%%%%%%%%%%%%%%%%%%%%%%%%%%%%%%%%%
\subsubsection{Relativistic case: Static bosonic Love numbers} \label{sec:relLNs}

In the relativistic case, one needs to study the perturbed Einstein's equations, and hence derive the LNs from the multipolar decomposition of the perturbing fields\footnote{Another concept is that of the \emph{surficial} LNs \citep{Landry:2014jka}, which quantify how the surface geometry of a self-gravitating body is deformed by an external tidal field (see~\ref{app:surficial}).}. We will discuss and determine the LNs associated with scalar, electromagnetic (EM), as well as gravitational perturbations. Even though, strictly speaking, only the LNs associated with gravitational perturbations are referred to as the LNs, we will refer to the ``susceptibilities'' associated with all possible perturbations as LNs, in order to simplify the notation and since confusions are not likely to arise. 
For clarity, we will first consider the LNs of non-rotating, spherical objects, in terms of metric perturbations. The case of spinning objects will be discussed later on.
In the static, spherically symmetric case the background spacetime is:
%%%%%%%%%%%%%%%%%%%%%%%%%%%%%%%%%%%%%%%%%%%%%%%%%%%%%%%%%%%%%%%%%%%%%%%%%%%%%%%%%    
\begin{align}\label{met_sph_stat}
d\bar{s}^{2}=-f(r)dt^{2}+\frac{dr^{2}}{g(r)}+r^{2}d\Omega^{2}~,
\end{align}
%%%%%%%%%%%%%%%%%%%%%%%%%%%%%%%%%%%%%%%%%%%%%%%%%%%%%%%%%%%%%%%%%%%%%%%%%%%%%%%%%
where $f$ and $g$ are arbitrary functions of the radial coordinate alone. If there exists some radius $r_{\rm h}$, for which $f(r_{\rm h})=0=g(r_{\rm h})$, then the above spacetime depicts a BH. As the above spacetime is perturbed, assuming that the background geometry has non-trivial scalar and EM contributions, we have\footnote{Note that we denote background (or, unperturbed) quantities with an `bar' over them.} 
%%%%%%%%%%%%%%%%%%%%%%%%%%%%%%%%%%%%%%%%%%%%%%%%%%%%%%%%%%%%%%%%%%%%%%%%%%%%%%%%%    
\begin{align}\label{pert_all}
g_{\mu \nu}=\bar{g}_{\mu \nu}+h_{\mu \nu}~;
\quad
A_{\mu}=\bar{A}_{\mu}+\delta A_{\mu}~;
\quad
\Phi=\bar{\Phi}+\delta \Phi~.
\end{align}
%%%%%%%%%%%%%%%%%%%%%%%%%%%%%%%%%%%%%%%%%%%%%%%%%%%%%%%%%%%%%%%%%%%%%%%%%%%%%%%%%
It is assumed that all the perturbation quantities are much smaller than the background quantities. Since the background spacetime is static and spherically symmetric, we can expand the perturbations using spherical harmonics. Further, the perturbations can be decomposed into axial and polar sectors, each sector transforming differently under parity. The scalar perturbation can be expanded as
%%%%%%%%%%%%%%%%%%%%%%%%%%%%%%%%%%%%%%%%%%%%%%%%%%%%%%%%%%%%%%%%%%%%%%%%%%%%%%%%%    
\begin{align}\label{scalarpertss}
\delta \Phi=\int d\omega \sum_{\ell m}e^{-i\omega t}\delta \phi_{\ell m}(r)Y_{\ell m}(\theta,\phi)~.
\end{align}
%%%%%%%%%%%%%%%%%%%%%%%%%%%%%%%%%%%%%%%%%%%%%%%%%%%%%%%%%%%%%%%%%%%%%%%%%%%%%%%%%
Note that scalar perturbations always contribute to the polar sector, while vector perturbations contribute to both the axial and polar sectors. Axial vector perturbations are
%%%%%%%%%%%%%%%%%%%%%%%%%%%%%%%%%%%%%%%%%%%%%%%%%%%%%%%%%%%%%%%%%%%%%%%%%%%%%%%%%   
\begin{align}
\delta A_{\mu}^{\rm axial}=\int d\omega \sum_{\ell m}e^{-i\omega t}\delta a^{\ell m}_{4}(r)\mathcal{A}^{\ell m}_{\mu}(\theta,\phi)~,
\end{align}
%%%%%%%%%%%%%%%%%%%%%%%%%%%%%%%%%%%%%%%%%%%%%%%%%%%%%%%%%%%%%%%%%%%%%%%%%%%%%%%%%
where only the angular components of vector $\mathcal{A}^{\ell m}_{\mu}(\theta,\phi)$ are non-zero and have the following expressions: $\mathcal{A}^{\ell m}_{\theta}=-(1/\sin \theta)(\partial_{\phi}Y_{\ell m})$, $\mathcal{A}^{\ell m}_{\phi}=\sin \theta \partial_{\theta}Y_{\ell m}$. Note that the axial sector is determined by the radial function $\delta a^{\ell m}_{4}(r)$ alone. The polar sector of the vector perturbation has the following expression: 
%%%%%%%%%%%%%%%%%%%%%%%%%%%%%%%%%%%%%%%%%%%%%%%%%%%%%%%%%%%%%%%%%%%%%%%%%%%%%%%%%    
\begin{align}
\delta A_{\mu}^{\rm polar}=\int d\omega \sum_{\ell m}e^{-i\omega t}\Big[\left\{\delta a^{\ell m}_{1}(r)\delta_{\mu}^{t}+\delta a^{\ell m}_{2}(r)\delta_{\mu}^{r}\right\}Y_{\ell m}+\delta a^{\ell m}_{3}(r)\mathcal{P}_{\mu}^{\ell m}(\theta,\phi)\Big]~.
\end{align}
%%%%%%%%%%%%%%%%%%%%%%%%%%%%%%%%%%%%%%%%%%%%%%%%%%%%%%%%%%%%%%%%%%%%%%%%%%%%%%%%%
Here we have three unknown radial functions, $\delta a^{\ell m}_{1}(r)$, $\delta a^{\ell m}_{2}(r)$ and $\delta a^{\ell m}_{3}(r)$, with the four-vector $\mathcal{P}_{\mu}^{\ell m}(\theta,\phi)$ having non-zero components only in the angular part: $\mathcal{P}^{\ell m}_{\theta}=\partial_{\theta}Y_{\ell m}$, and $\mathcal{P}^{\ell m}_{\phi}=\partial_{\phi}Y_{\ell m}$. For the metric perturbation $h_{\mu \nu}$, its decomposition into the axial and the polar sector, in the Regge--Wheeler gauge \citep{Regge:1957td}, yields
%%%%%%%%%%%%%%%%%%%%%%%%%%%%%%%%%%%%%%%%%%%%%%%%%%%%%%%%%%%%%%%%%%%%%%%%%%%%%%%%%   
\begin{align}\label{pert_grav}
h_{\mu \nu}=\int d\omega \sum_{\ell m}e^{-i\omega t}\left(\mathbb{H}^{\ell m~(\textrm{axial})}_{\mu \nu}+\mathbb{H}^{\ell m~(\textrm{polar})}_{\mu \nu}\right)~,
\end{align}
%%%%%%%%%%%%%%%%%%%%%%%%%%%%%%%%%%%%%%%%%%%%%%%%%%%%%%%%%%%%%%%%%%%%%%%%%%%%%%%%%
where the axial part of the perturbation, namely $\mathbb{H}^{\ell m~(\textrm{axial})}_{\mu \nu}$ can be expressed as,
%%%%%%%%%%%%%%%%%%%%%%%%%%%%%%%%%%%%%%%%%%%%%%%%%%%%%%%%%%%%%%%%%%%%%%%%%%%%%%%%%    
\begin{align}\label{pert_grav_ax}
\mathbb{H}^{\ell m~(\textrm{axial})}_{\mu\nu}=
\left(
\begin{array}{cccc}
0 & 0 & -\frac{h^{\ell m}_{0}(r)}{\sin \theta}\partial_{\phi}Y_{\ell m} & h^{\ell m}_{0}(r)\sin \theta \partial_{\theta}Y_{\ell m}\\
0 & 0 & -\frac{h^{\ell m}_{1}(r)}{\sin \theta}\partial_{\phi}Y_{\ell m} & h^{\ell m}_{1}(r)\sin \theta \partial_{\theta}Y_{\ell m} \\
\textrm{symm} & \textrm{symm} & 0& 0 \\
\textrm{symm} & \textrm{symm} & 0 & 0\\
\end{array}
\right)~.
\end{align}
%%%%%%%%%%%%%%%%%%%%%%%%%%%%%%%%%%%%%%%%%%%%%%%%%%%%%%%%%%%%%%%%%%%%%%%%%%%%%%%%%
Therefore, in the Regge--Wheeler gauge, the axial metric perturbations are determined by two unknown radial functions: $h^{\ell m}_{0}(r)$ and $h^{\ell m}_{1}(r)$. The components of the polar metric perturbation $\mathbb{H}^{\ell m~(\textrm{polar})}_{\mu \nu}$ read
%%%%%%%%%%%%%%%%%%%%%%%%%%%%%%%%%%%%%%%%%%%%%%%%%%%%%%%%%%%%%%%%%%%%%%%%%%%%%%%%%    
\begin{align}\label{pert_grav_pol}
\mathbb{H}^{\ell m~(\textrm{polar})}_{\mu \nu}=
\left(
\begin{array}{cccc}
f(r)H^{\ell m}_{0}(r)& H^{\ell m}_{1}(r)& 0 & 0\\
H^{\ell m}_{1}(r) & g^{-1}(r)H^{\ell m}_{2}(r)& 0 & 0\\
0 & 0 & r^{2}K_{\ell m}(r)& 0 \\
0 & 0 & 0 & r^{2} \sin^{2}\theta K_{\ell m}(r) \\
\end{array}
\right)Y_{\ell m}(\theta,\phi)~,
\end{align}
%%%%%%%%%%%%%%%%%%%%%%%%%%%%%%%%%%%%%%%%%%%%%%%%%%%%%%%%%%%%%%%%%%%%%%%%%%%%%%%%%
where we have four unknown functions: $H^{\ell m}_{0}(r)$, $H^{\ell m}_{1}(r)$, $H^{\ell m}_{2}(r)$ and $K_{\ell m}(r)$ determining the polar sector. 

Among the various perturbations of the background spacetime, our primary interest --~motivated by the Newtonian analysis presented above and relevant to the study of LNs~-- lies in the multipolar expansion of specific components of these perturbations. For instance, in the case of EM perturbations, the charge multipole moments are encoded in the multipolar expansion of $A_t$, while the current multipole moments are obtained from the expansion of $A_\phi$. Similarly, for gravitational perturbations, the mass multipole moments arise from the expansion of $g_{tt}$, which is related to the Newtonian potential in the weak field limit, whereas the spin multipole moments are extracted from the expansion of $g_{t\phi}$. 
These multipolar expansions are carried out in an asymptotically cartesian and mass-centered (ACMC) coordinate system \citep{Thorne:1980ru} and in the static limit of the perturbations ($\omega\to0$). Assuming that the spatial metric of the background spacetime is asymptotically flat, the leading order expansion in the \emph{intermediate} region $R\ll r\ll 1/\omega$, where $R$ is the radius of the object, reads \citep{Thorne:1980ru,Cardoso:2017cfl} (in the static limit, these become the asymptotic expansion\footnote{Note that we have slightly changed the normalization of the vector tidal part compared to~\citet{Cardoso:2017cfl} to remove a pole at $\ell=1$.}):
%%%%%%%%%%%%%%%%%%%%%%%%%%%%%%%%%%%%%%%%%%%%%%%%%%%%%%%%%%%%%%%%%%%%%%%%%%%%%%%%%   
%\resizebox{.9\linewidth}{!}{
%\begin{minipage}{\linewidth}
\begin{align}
\Phi&=\Phi_{0}+\sum_{\ell \geq 1}\left[\frac{1}{r^{\ell+1}}\left\{\Phi_{\ell}Y_{\ell 0}+\ell'<\ell~\textrm{pole}\right\}-r^{\ell}\left\{\mathcal{E}_{\ell}^{\rm s}Y_{\ell 0}+\ell'<\ell~\textrm{pole}\right\}\right]~, \label{expscalar}
\\
\medmath{A_{t}}&\medmath{=-\frac{Q}{r}+\sum_{\ell\geq 1}\left[\frac{2}{r^{\ell+1}}\left\{\sqrt{\frac{4\pi}{2\ell+1}}Q_{\ell}Y_{\ell 0}+\ell'<\ell~\textrm{pole}\right\}-\frac{2r^{\ell}}{\ell}\left\{\mathcal{E}_{\ell}^{\rm v}Y_{\ell 0}+\ell'<\ell~\textrm{pole}\right\}\right]\,,}
\\
\medmath{A_{\phi}}&\medmath{=\sin \theta\sum_{\ell\geq 1}\left[\frac{2}{r^{\ell}}\left\{\sqrt{\frac{4\pi}{2\ell+1}}\frac{J_{\ell}}{\ell}\partial_{\theta}Y_{\ell 0}+\ell'<\ell~\textrm{pole}\right\}+\frac{2r^{\ell+1}}{3\ell}\left\{\mathcal{B}_{\ell}^{\rm v}\partial_{\theta}Y_{\ell 0}+\ell'<\ell~\textrm{pole}\right\}\right]\,,}
\\
\medmath{g_{tt}}&\medmath{=-1+\frac{2M}{r}+\sum_{\ell \geq2}\left[\frac{2}{r^{\ell+1}}\left\{\sqrt{\frac{4\pi}{2\ell+1}}M_{\ell}Y_{\ell 0}+\ell'<\ell~\textrm{pole}\right\}-\frac{2r^{\ell}}{\ell(\ell-1)}\left\{\mathcal{E}_{\ell}Y_{\ell 0}+\ell'<\ell~\textrm{pole}\right\}\right]\,,}
\label{polar_gravity}
\\
\medmath{g_{t\phi}}&\medmath{=\frac{2J}{r}\sin^{2}\theta+\sin \theta \sum_{\ell\geq 2}\left[\frac{2}{r^{\ell}}\left\{\sqrt{\frac{4\pi}{2\ell+1}}\frac{S_{\ell}}{\ell}\partial_{\theta}Y_{\ell 0}+\ell'<\ell~\textrm{pole}\right\}+\frac{2r^{\ell+1}}{3\ell(\ell-1)}\left\{\mathcal{B}_{\ell}\partial_{\theta}Y_{\ell 0}+\ell'<\ell~\textrm{pole}\right\}\right]\,,}
\label{axial_gravity}
\end{align}
%\end{minipage}
%}
%%%%%%%%%%%%%%%%%%%%%%%%%%%%%%%%%%%%%%%%%%%%%%%%%%%%%%%%%%%%%%%%%%%%%%%%%%%%%%%%%
Here, $M_{\ell}$ are the mass multipole moments induced by the electric part of the tidal field $\mathcal{E}_{\ell}$, while $S_{\ell}$ are the spin multipole moments induced by the magnetic part of the tidal field $\mathcal{B}_{\ell}$. This is for gravitational tidal perturbations. For EM tidal perturbations, $Q_{\ell}$ are the charge moments induced by the electric tidal field $\mathcal{E}_{\ell}^{\rm v}$, while $J_{\ell}$ are the current moments induced by the magnetic tidal field $\mathcal{B}_{\ell}^{\rm v}$, respectively. For a scalar perturbation, only the polar (electric) parity matters, with $\Phi_{\ell}$ being the moment induced by the tidal field $\mathcal{E}_{\ell}^{\rm s}$. Note that, due to spherical symmetry, all the multipole moments are independent of the azimuthal number $m$, and hence for spherically symmetric configurations the LNs depend only on the angular number $\ell$. 
The normalization of the above expressions is arbitrary, but has been fixed in such a way that $M_\ell$ and $S_\ell$ coincide with the multipole moments defined by Geroch and Hansen for an asymptotically flat spacetime \citep{Geroch:1970cd,Hansen:1974zz}. In such a notation, $M=M_0$, $Q=Q_0$, and $\Phi_0$ are the monopoles of the gravitational, electric, and scalar field, respectively, whereas $J=S_0$ is the angular momentum of the spacetime (the current dipole of the gravitational field).

Given the above multipolar expansion, the LNs are defined as the ratio of the coefficient of the decaying mode, scaling as $r^{-\ell-1}$ for the polar sector, and as $r^{-\ell}$ for the axial sector, to the coefficient of the growing mode, scaling as $r^{\ell}$ for the polar sector, and as $r^{\ell+1}$ for the axial sector. For concreteness, let us focus on the most interesting case of gravitational perturbations. Given,~\ref{polar_gravity} and~\ref{axial_gravity}, the asymptotic limits for the metric perturbations $H_{0}$ and $h_{0}$ takes the following form, 
%%%%%%%%%%%%%%%%%%%%%%%%%%%%%%%%%%%%%%%%%%%%%%%%%%%%%%%%%%%%%%%%%%%%%%%%%%%%%%%%%    
\begin{align}
H_{0}(r)&=\frac{2M_{\ell}}{r^{\ell+1}}\sqrt{\frac{4\pi}{2\ell+1}}\left(1+\cdots\right)-\left(\frac{2}{\ell(\ell-1)}\right)\mathcal{E}_{\ell}r^{\ell}\left(1+\cdots\right)~,
\label{Hasymp}
\\
h_{0}(r)&=\frac{2S_{\ell}}{\ell r^{\ell}}\sqrt{\frac{4\pi}{2\ell+1}}\left(1+\cdots\right)+\left(\frac{2}{3\ell(\ell-1)}\right)\mathcal{B}_{\ell}r^{\ell+1}\left(1+...\right)~.
\label{h0asymp}
\end{align}
%%%%%%%%%%%%%%%%%%%%%%%%%%%%%%%%%%%%%%%%%%%%%%%%%%%%%%%%%%%%%%%%%%%%%%%%%%%%%%%%%
where the dots represent lower multipoles having $(\ell'<\ell)$. In the Newtonian limit, the metric perturbation $h_{0}$ is non-existent, and hence there are no magnetic LNs, while the metric perturbation $H_{0}$ will be identical to the perturbation of the Newtonian potential, presented in~\ref{I_def}, in the static limit. Here, we are interested in the leading order behavior. Given the above asymptotic expansion, the static LNs, both electric types and magnetic types are defined as,
%%%%%%%%%%%%%%%%%%%%%%%%%%%%%%%%%%%%%%%%%%%%%%%%%%%%%%%%%%%%%%%%%%%%%%%%%%%%%%%%%    
\begin{align}
k_{\ell}^{\rm E}&\equiv-\frac{\ell(\ell-1)}{2R^{2\ell+1}}\sqrt{\frac{4\pi}{2\ell+1}}\frac{M_{\ell}}{\mathcal{E}_{\ell}}~,
\label{tlne}
\\
k_{\ell}^{\rm B}&\equiv-\frac{3\ell(\ell-1)}{2(\ell+1)R^{2\ell+1}}\sqrt{\frac{4\pi}{2\ell+1}}\frac{S_{\ell}}{\mathcal{B}_{\ell}}~,
\label{tlnb}
\end{align}
%%%%%%%%%%%%%%%%%%%%%%%%%%%%%%%%%%%%%%%%%%%%%%%%%%%%%%%%%%%%%%%%%%%%%%%%%%%%%%%%%
where $R$ is the size of the unperturbed compact object whose deformation we are measuring. Due to spherical symmetry of the background spacetime, it follows that the LNs are independent of the magnetic number $m$. 

Similarly, for the EM perturbations, in the electric sector, we will obtain an expansion for $\delta a^{\ell m}_{1}(r)$ as a linear combination of $Q_{\ell}r^{-\ell-1}$ and $\mathcal{E}_{\ell}^{\rm v}r^{\ell}$. In the magnetic sector, the perturbation of the vector potential will be given by the linear combination of $J_{\ell}r^{-\ell}$ and $\mathcal{B}_{\ell}^{\rm v}r^{\ell+1}$. Same consideration applies to the scalar perturbation as well. Therefore, the LNs for the electric and magnetic sectors, associated with the EM perturbations, and the LNs for the scalar perturbation will be given by,
%%%%%%%%%%%%%%%%%%%%%%%%%%%%%%%%%%%%%%%%%%%%%%%%%%%%%%%%%%%%%%%%%%%%%%%%%%%%%%%%%    
\begin{align}
k^{\rm E}_{\ell~(\textrm{EM})}&=-\frac{\ell(\ell-1)}{2R^{2\ell+1}}\sqrt{\frac{4\pi}{2\ell+1}}\frac{Q_{\ell}}{\mathcal{E}^{\rm v}_{\ell}}~,
\\
k^{\rm B}_{\ell~(\textrm{EM})}&=-\frac{3\ell(\ell-1)}{2(\ell+1)R^{2\ell+1}}\sqrt{\frac{4\pi}{2\ell+1}}\frac{J_{\ell}}{\mathcal{B}^{\rm v}_{\ell}}~,
\\
k_{\ell~(\textrm{scalar})}&=-\frac{1}{2}\frac{\Phi_{\ell}}{\mathcal{E}_{\ell}^{\rm s}}~.
\end{align}
%%%%%%%%%%%%%%%%%%%%%%%%%%%%%%%%%%%%%%%%%%%%%%%%%%%%%%%%%%%%%%%%%%%%%%%%%%%%%%%%%
Roughly speaking, irrespective of the nature of the perturbation, the LNs are effectively given by the following quantities:
%%%%%%%%%%%%%%%%%%%%%%%%%%%%%%%%%%%%%%%%%%%%%%%%%%%%%%%%%%%%%%%%%%%%%%%%%%%%%%%%%    
\begin{align}
k_{\ell}^{\rm E}&=\frac{1}{2R^{2\ell+1}}\left(\frac{\textrm{coefficient~of}~r^{-\ell-1}}{\textrm{coefficient~of}~r^{\ell}}\right)~,
\\
k_{\ell}^{\rm B}&=-\left(\frac{\ell}{2(\ell+1)R^{2\ell+1}}\right)\left(\frac{\textrm{coefficient~of}~r^{-\ell}}{\textrm{coefficient~of}~r^{\ell+1}}\right)~,
\end{align}
Thus, their determination boils down to computing the ratio between the corresponding growing and decaying component of the field at large distance.
This steps requires solving the perturbation equations for a given background and matter content. Regularity condition at the inner boundary (either the center of the matter distribution or the horizon, in case of a BH), uniquely fixes the ratio among the tidal coefficients and the object's response coefficient at large distances, hence determining the LNs. We will discuss specific examples in the next sections.

%%%%%%%%%%%%%%%%%%%%%%%%%%%%%%%%%%%%%%%%%%%%%%%%%%%%%%%%%%%%%%%%%%%%%%%%%%%%%%%%%

%%%%%%%%%%%%%%%%%%%%%%%%%%%%%%%%%%%%%%%%%%%%%%%%%%%%%%%%%%%%%
%%%%%%%%%%%%%%%%%%%%%%%%%%%%%%%%%%%%%%%%%%%%%%%%%%%%%%%%%%%%%
%%%%%%%%%%%%%%%%%%%%%%%%%%%%%%%%%%%%%%%%%%%%%%%%%%%%%%%%%%%%%
\subsubsection{Connecting different conventions for Love numbers}
\label{conventionLN}

Unfortunately, there exist several different definitions of the LNs adopted in the literature. We find it useful here to provide the mapping between different definitions. We will mainly provide this mapping among the LNs by (a) \citet{Damour:2009vw}, (b) \citet{Binnington:2009bb} and (c) \citet{Yagi:2013bca}. We first relate the LNs in the electric sector, defined in~\ref{tlne}, with this corresponding ones in \citet{Binnington:2009bb} and \citet{Damour:2009vw}, denoted by $k_{\ell}^{\rm BP}$ and $k_{\ell}^{\rm DN}$, respectively, which will subsequently be related to the ones defined in \citet{Yagi:2013bca}, denoted by $\bar{\lambda}_{\ell}^{\rm YY}$. Such a relation, for generic $\ell$ mode, yields
%%%%%%%%%%%%%%%%%%%%%%%%%%%%%%%%%%%%%%%%%%%%%%%%%%%%%%%%%%%%%%%%%%%%%%%%%%%%%%
\begin{equation}
k_{\ell}^{\rm E}=k_{\ell}^{\rm BP}=k_{\ell}^{\rm DN}=\frac{(2\ell-1)!!}{2}\left(\frac{M}{R}\right)^{2\ell+1}\bar{\lambda}_{\ell}^{\rm YY}~.
\label{eq:conventionElectric}
\end{equation}
%%%%%%%%%%%%%%%%%%%%%%%%%%%%%%%%%%%%%%%%%%%%%%%%%%%%%%%%%%%%%%%%%%%%%%%%%%%%%%
Therefore, the electric LNs defined here are identical to the ones adopted in \citet{Binnington:2009bb, Damour:2009vw}. 
%
% Further, here and also in the subsequent sections, we will keep the explicit dependence of the LNs on the $(M/R)$ ratio, and shall not use the compactness, as different papers uses different definitions for compactness, leading to potential confusion. Therefore, showing the $(M/R)$ dependence explicitly can get rid of such confusions.

The mapping among the magnetic LNs is slightly more complex and it is hard to provide a relation valid for generic $\ell$'s. Since, as we shall discuss, only the first multipole moments can possibly be phenomenologically relevant, we will focus on $\ell=2,3$. In contrast to the approach taken in this review, \citet{Damour:2009vw,Yagi:2013bca} determine the magnetic LNs by the asymptotic expansion of the following quantity $(rh_{0}'-2h_{0})$, and not simply $h_{0}$. This yields the following relations between the respective LNs: 
%%%%%%%%%%%%%%%%%%%%%%%%%%%%%%%%%%%%%%%%%%%%%%%%%%%%%%%%%%%%%%%%%%%%%%%%%%%%%%
\begin{equation}\label{eq:conventionMagnetic}
k_{2}^{\rm B}=\frac{1}{12}j_{2}^{\rm DN}=4\left(\frac{M}{R}\right)^{5}\bar{\sigma}_{2}^{\rm YY}~,
\end{equation}
%%%%%%%%%%%%%%%%%%%%%%%%%%%%%%%%%%%%%%%%%%%%%%%%%%%%%%%%%%%%%%%%%%%%%%%%%%%%%%
where $j_{\ell}^{\rm DN}$ and $\bar{\sigma}_{\ell}^{\rm YY}$ are the magnetic LNs defined in \citet{Damour:2009vw} and in \citet{Yagi:2013bca}, respectively. For $\ell=3$, on the other hand, we obtain,
%%%%%%%%%%%%%%%%%%%%%%%%%%%%%%%%%%%%%%%%%%%%%%%%%%%%%%%%%%%%%%%%%%%%%%%%%%%%%%
\begin{equation}\label{conn_3}
k_{3}^{\rm B}=\frac{3}{20}j^{\rm DN}_{3}=\frac{45}{2}\left(\frac{M}{R}\right)^{7}\bar{\sigma}_{3}^{\rm YY}~.
\end{equation} 
%%%%%%%%%%%%%%%%%%%%%%%%%%%%%%%%%%%%%%%%%%%%%%%%%%%%%%%%%%%%%%%%%%%%%%%%%%%%%%
It is possible to compute the mapping for any given $\ell$. On the other hand, \citet{Landry:2015cva} defines another magnetic LN, associated with the tidal deformation of a compact object, which consists of an irrotational fluid, as 
%%%%%%%%%%%%%%%%%%%%%%%%%%%%%%%%%%%%%%%%%%%%%%%%%%%%%%%%%%%%%%%%%%%%%%%%%%%%%%
\begin{equation}
k_{\ell}^{\rm B}=\widetilde{k}_{\ell}^{\rm LP}\left(\frac{2M}{R}\right)=\frac{\ell(\ell-1)}{2(\ell+2)(\ell+1)}j_{\ell}^{\rm DN}\,,
\end{equation} 
%%%%%%%%%%%%%%%%%%%%%%%%%%%%%%%%%%%%%%%%%%%%%%%%%%%%%%%%%%%%%%%%%%%%%%%%%%%%%%
where the last line follows from \citet{Pani:2018inf}. One can explicitly verify that for $\ell=2$ and $\ell=3$, we get back the identities in \ref{eq:conventionMagnetic} and \ref{conn_3}. We should emphasize that the magnetic LNs defined by \citet{Binnington:2009bb} are for static fluid, which are of limited astrophysical interest and hence will not be presented here, though we will have a short discussion about the choice of static versus irrotational fluid in a subsequent section.

%%%%%%%%%%%%%%%%%%%%%%%%%%%%%%%%%%%%%%%%%%%%%%%%%%%%%%%%%%%%%
%%%%%%%%%%%%%%%%%%%%%%%%%%%%%%%%%%%%%%%%%%%%%%%%%%%%%%%%%%%%%
%%%%%%%%%%%%%%%%%%%%%%%%%%%%%%%%%%%%%%%%%%%%%%%%%%%%%%%%%%%%%
\subsubsection{Relativistic dynamical Love numbers}

For static perturbations, the tidal fields are taken to be at infinity. Since, the source of the tidal field, if located at a finite distance, will interact gravitationally with the tidally deformed object, leading to relative motion, and hence the tidal field will no longer be static. Thus, in any practical situation, e.g., during the inspiral of any binary system emitting GWs, the tidal field is never static. Therefore, the static LNs should be thought of as $\omega \to 0$ limit of the dynamical LNs. 

The determination of the dynamical LNs requires defining an intermediate zone, since unlike the case of static tidal field, in the dynamical case, the tidal field is at a distance of $\mathcal{O}(1/\omega)$. Therefore, all the calculations must be done well within the tidal field. Thus, the asymptotic expansion of relevant perturbation variables, in the dynamical context, does not imply $r\to \infty$, rather to $r\to \omega^{-1}$. The dynamical LNs can be determined by solving the evolution equations for the metric perturbations in the case of static and spherically symmetric background spacetimes, or, by solving the Weyl scalars for rotating as well as non-rotating backgrounds. For generality, and keeping future prospects in mind, we discuss the case of Weyl scalar $\Psi_{4}\equiv W_{\mu \nu \alpha \beta}\bar{m}^{\mu}k^{\nu}\bar{m}^{\alpha}k^{\beta}$. Here, $W_{\mu \nu \alpha \beta}$ is the Weyl tensor, $k^{\alpha}$ is one of the principal null vector and $\bar{m}^{\mu}$ is the complex null vector on the 2-sphere. The expansion of $\Psi_{4}$ in the intermediate zone reads ($\Psi_{4}$ is related to the Newtonian potential $U$ in the Newtonian limit), 
%%%%%%%%%%%%%%%%%%%%%%%%%%%%%%%%%%%%%%%%%%%%%%%%%%%%%%%%%%%%%%%%%%%%%%%%%%%%%%
\begin{equation}\label{psi_4_intermediate}
\medmath{\Psi_{4}^{\text{(int)}}=\frac{1}{4}\sum_{\ell m}\sqrt{\frac{(\ell+2)(\ell+1)}{\ell(\ell-1)}} \,\mathcal{E}_{\ell m}(\omega)\,r^{\ell-2}\left[(1+\cdots)+\mathcal{F}_{\ell m}(\omega)\left(\frac{R}{r}\right)^{2\ell+1}(1+\cdots)\right]\,_{-2}Y_{\ell m}~,}
\end{equation}
%%%%%%%%%%%%%%%%%%%%%%%%%%%%%%%%%%%%%%%%%%%%%%%%%%%%%%%%%%%%%%%%%%%%%%%%%%%%%%
where $\,_{-2}Y_{\ell m}$ are spin-2 spheroidal harmonics \citep{Berti:2005gp}  and the terms denoted by `dots' are the lower multipole moments. Alike the Newtonian case, see~\ref{total_potential_Fourier_space}, here also $\mathcal{F}_{\ell m}(\omega)$ is the dynamical response function. The dynamical LNs and the dynamical dissipation numbers are defined as, 
%%%%%%%%%%%%%%%%%%%%%%%%%%%%%%%%%%%%%%%%%%%%%%%%%%%%%%%%%%%%%%%%%%%%%%%%%%%%%%
\begin{equation}\label{dynamicLNDN}
k_{\ell m}(\omega)=\frac{1}{2}\textrm{Re}\,\mathcal{F}_{\ell m}(\omega)\,;
\qquad
\nu_{\ell m}(\omega)=\textrm{Im}\,\mathcal{F}_{\ell m}(\omega)\,,
\end{equation}
%%%%%%%%%%%%%%%%%%%%%%%%%%%%%%%%%%%%%%%%%%%%%%%%%%%%%%%%%%%%%%%%%%%%%%%%%%%%%%
and each of them will have an expansion in powers of $\omega$. For static and spherically symmetric background, one has $k_{\ell m}(\omega)=k_{\ell m}^{(0)}+M^{2}\omega^{2}\,k_{\ell m}^{(2)}+\cdots$, while, the dissipation number can be expanded as $\nu_{\ell m}(\omega)=M\omega\,\nu_{\ell m}^{(1)}+M^{3}\omega^{3}\,\nu_{\ell m}^{(3)}+\cdots$. Note that this expansion is valid for any object, including BHs, viscous stars and dissipative ECOs. On the other hand, for axisymmetric and stationary background, there can be dynamical LNs and dissipation numbers at all orders in $M\omega$, e.g., in the $\omega \to 0$ limit $\nu_{\ell m}$ will be non-zero and there will be a $\mathcal{O}(M\omega)$ term in the small frequency expansion of the dynamical LNs. 

To summarize, static LNs and dissipation numbers are about determining two quantities $k_{\ell m}^{(0)}$ and $\nu_{\ell m}^{(0)}$, with $\nu_{\ell m}^{(0)}$ vanishing for static and spherically symmetric background metric. These can be determined, either by solving the equations for metric perturbations $H_{0}$ and $h_{0}$ in the static limit, or, by solving the Teukolsky equations and then taking the $\omega \to 0$ limit. It is expected that these two approaches will provide an identical result, however there can be surprises, which we will discuss later on.

%%%%%%%%%%%%%%%%%%%%%%%%%%%%%%%%%%%%%%%%%%%%%%%%%%%%%%%%%%%%%
%%%%%%%%%%%%%%%%%%%%%%%%%%%%%%%%%%%%%%%%%%%%%%%%%%%%%%%%%%%%%
%%%%%%%%%%%%%%%%%%%%%%%%%%%%%%%%%%%%%%%%%%%%%%%%%%%%%%%%%%%%%
\subsubsection{Ambiguities in defining the Love numbers and how to cure them} \label{sec:ambiguities}

Various ambiguities in the definition of the relativistic LNs have been discussed in the literature. 
First, the definition adopted in the previous section relies on metric perturbations formulated in the Regge--Wheeler gauge, which introduces potential gauge ambiguities in the results. 
To extract the LNs and characterize the tidal response in a gauge-invariant manner, it is useful --~as argued in \citet{LeTiec:2020bos, Chia:2020yla, Bhatt:2023zsy}~-- to compute the Newman-Penrose scalar $\Psi_4$ associated with the gravitational perturbations. 
Notably, $\Psi_4$ is directly related to the gravitational potential in the Newtonian limit \citep{LeTiec:2020bos}, thereby providing a physically transparent quantity for defining the tidal response. 
Furthermore, in the case of BHs, $\Psi_4$ can be computed directly from the Teukolsky equation \citep{Teukolsky:1972my}, even in the presence of spin, thus offering a robust and gauge-invariant framework for the study of tidal effects in rotating spacetimes.

The dynamical LNs are determined, in a gauge invariant approach, through the expansion of the Weyl scalar $\Psi_{4}$, in the intermediate regime ($r\gg M$, $r\omega\ll1$), which has been presented in~\ref{psi_4_intermediate}.
%$\Psi_4$ has the following behavior
%%%%%%%%%%%%%%%%%%%%%%%%%%%%%%%%%%%%%%%%%%%%%%%%%%%%%%%%%%%%%%%%%%%%%%%%%%%%%%
%\begin{equation}\label{psi_4_intermediate}
%\Psi_{4}^{\text{intermediate}}=\frac{1}{4}\sum_{\ell m}\sqrt{\frac{(\ell+2)(\ell+1)}{\ell(\ell-1)}} \,\mathcal{E}_{\ell m}\,r^{l-2}\left[(1+...)+\mathcal{F}_{\ell m}(\omega)\left(\frac{R}{r}\right)^{2\ell+1}(1+...)\right]\,_{-2}Y_{\ell m}~,
%\end{equation}
%%%%%%%%%%%%%%%%%%%%%%%%%%%%%%%%%%%%%%%%%%%%%%%%%%%%%%%%%%%%%%%%%%%%%%%%%%%%%%
%where ${-2}Y_{\ell m}$ are spin-2 spheroidal harmonics.
%
Here, the term involving $\mathcal{E}_{\ell m}$ is the tidal field, and the second term with $\mathcal{F}_{\ell m}$ corresponds to the response. Indeed, $\mathcal{F}_{\ell m}(\omega)$ is the \emph{tidal response function}: its real and imaginary parts are associated with the conservative and dissipative tidal response of the object, for a given $(\ell,m)$ mode. 

Since $\Psi_4$ is gauge invariant, the above definition is not associated with gauge issues. However, it still involves \emph{some} radial coordinate $r$ and hence ambiguities can arise from a coordinate transformation \citep{Pani:2015hfa,Pani:2015nua,Gralla:2017djj}.
For example, let us consider the following coordinate transformation at large distances ($r\ll \omega^{-1}$)
%%%%%%%%%%%%%%%%%%%%%%%%%%%%%%%%%%%%%%%%%%%%%%%%%%%%%%%%%%%%%%%%%%%%%%%%%%%%%%
\begin{equation}
r\to r\left[1+C\left(\frac{R}{r}\right)^{2\ell+1}\right]\,, \label{eq:coordtransf}
\end{equation}
%%%%%%%%%%%%%%%%%%%%%%%%%%%%%%%%%%%%%%%%%%%%%%%%%%%%%%%%%%%%%%%%%%%%%%%%%%%%%%
where $C={\cal O}(1)$ is some arbitrary constant and $R$ is a characteristic length scale associated with the object.
Under the above transformation the expansion of the Weyl scalar in the intermediate region becomes,
%%%%%%%%%%%%%%%%%%%%%%%%%%%%%%%%%%%%%%%%%%%%%%%%%%%%%%%%%%%%%%%%%%%%%%%%%%%%%%
\begin{equation}\label{psi_4_intermediateB}
\Psi_{4}^{\text{(int)}}\propto\mathcal{E}_{\ell m}(\omega)\,r^{\ell-2}\left[(1+\cdots)+\left\{C(\ell-2)+\mathcal{F}_{\ell m}(\omega)\right\}\left(\frac{R}{r}\right)^{2\ell+1}(1+\cdots)\right]\,_{-2}Y_{\ell m}~,
\end{equation}
%%%%%%%%%%%%%%%%%%%%%%%%%%%%%%%%%%%%%%%%%%%%%%%%%%%%%%%%%%%%%%%%%%%%%%%%%%%%%%
showing that the tidal response has been shifted by $C(\ell-2)$, despite the gauge invariance of the Weyl scalar $\Psi_4$.

There are two ways to resolve this ambiguity. The first uses an analytical continuation \citep{Kol:2011vg} of the angular momentum index $\ell$ (or, alternatively, of the number of spacetime dimensions $d$, the role of which will be discussed in~\ref{sec:higherD}). Indeed, if one is able to compute the perturbations analytically for generic $\ell$ as in~\ref{psi_4_intermediate}, then one can consider
$\ell\in\mathbb{C}$, to unambiguously identify the contributions from tidal deformation and dissipation \citep{LeTiec:2020bos}. Indeed, in such a case a coordinate transformation of the form given in~\ref{eq:coordtransf} would be forbidden, since coordinates must be real quantities. Unfortunately, it is not always possible to solve the problem for generic values of $\ell$, so this method can be used only in certain cases.

A more general resolution of the ambiguity involves matching the perturbation to some \emph{gauge-invariant coefficient}, i.e. a scalar quantity that does not depend on spacetime coordinates. 
For example, one could match with the binding energy of a binary system involving tidally deformed bodies \citep{Creci:2021rkz} or with the Wilson coefficients of an EFT, as discussed in~\ref{sec:EFT}.
Such matching might require specific coordinate transformations and might be complicated to perform in certain cases. However, once performed, it would unambiguously fix the tidal response.

%%%%%%%%%%%%%%%%%%%%%%%%%%%%%%%%%%%%%%%%%%%%%%%%%%%%%%%%%%%%%%%%%%%%%%%%%%%%%%
%%%%%%%%%%%%%%%%%%%%%%%%%%%%%%%%%%%%%%%%%%%%%%%%%%%%%%%%%%%%%%%%%%%%%%%%%%%%%%
%%%%%%%%%%%%%%%%%%%%%%%%%%%%%%%%%%%%%%%%%%%%%%%%%%%%%%%%%%%%%%%%%%%%%%%%%%%%%%
\section{Static bosonic Love numbers of black holes in four-dimensional vacuum General Relativity} \label{sec:staticBHs}

In this section, we focus on the static bosonic LNs of BHs in the framework of four-dimensional vacuum GR, where we will show that the static LNs identically vanish.
\ref{sec:BHBGR} will be instead devoted to BHs beyond vacuum GR, either due to the presence of matter, beyond-GR effects, or higher dimensions. 
The case of fermionic LNs will be discussed in~\ref{sec:fermionic}. 

%%%%%%%%%%%%%%%%%%%%%%%%%%%%%%%%%%%%%%%%%%%%%%%%%%%%%%%%%%%%%
%%%%%%%%%%%%%%%%%%%%%%%%%%%%%%%%%%%%%%%%%%%%%%%%%%%%%%%%%%%%%
%%%%%%%%%%%%%%%%%%%%%%%%%%%%%%%%%%%%%%%%%%%%%%%%%%%%%%%%%%%%%
% \subsection{Four dimensional black holes in vacuum General Relativity 
% %have vanishing static tidal Love number
% }\label{GRStatLN}

We will explicitly show that four-dimensional BHs in vacuum GR have vanishing static bosonic LNs. We will use present this result from different perspectives and using different approaches: BH perturbation theory,  symmetry arguments, EFT and scattering-amplitude calculations. We stress that, whenever we refer to LNs in this section, it will be for bosonic perturbations only.

%%%%%%%%%%%%%%%%%%%%%%%%%%%%%%%%%%%%
%%%%%%%%%%%%%%%%%%%%%%%%%%%%%%%%%%%%
%%%%%%%%%%%%%%%%%%%%%%%%%%%%%%%%%%%%
\subsection{Perturbation of a Schwarzschild black hole and Love numbers}
\label{vac_Sch_LN}

Let us start with the static perturbation equations associated with the scalar, electromagnetic, and gravitational (both axial and polar) perturbations of a static BH, described by the Schwarzschild metric. 
Static perturbations are obtained by taking the zero-frequency limit of the relevant equations in the dynamical context. The determination of the static LNs, requires solving the perturbation equations in the Schwarzschild spacetime due to some external static tidal field, and then the regularity at the BH horizon uniquely fixes the behavior at large distance of the perturbations. The background metric will be denoted as $g^{(0)}_{\mu\nu}$, and has the line element in~\ref{met_sph_stat} with $f(r)=1-2M/r=g(r)$, with $M$ being the ADM mass of the spacetime.
%%%%%%%%%%%%%%%%%%%%%%%%%%%%%%%%%%%%%%%%%%%%%%%%%%%%%%%%%%%%%%%%%%%%%%%%%%%%%%%%%

The simplest type of perturbation that can be considered in a BH spacetime is a scalar perturbation, governed by the massless Klein--Gordon equation $\square \Phi = 0$ on the background metric (this assumes that the scalar is minimally coupled). While test scalar perturbations are not directly relevant for observational purposes they serve as a useful toy model, capturing the essential features of LNs for vacuum BHs in GR.

The Klein--Gordon equation on the Schwarzschild background and in the static limit reduces to
%%%%
%%%%
\begin{equation}
\Psi''+\left(\frac{2}{r}+\frac{f'}{f}\right)\Psi'-\frac{\ell(\ell+1)}{fr^2}\Psi=0
\end{equation}
%%%
where we used the spherical harmonic decomposition in~\ref{scalarpertss} with $\delta \Phi=e^{-i\omega t}\Psi(r)Y_{\ell m}$, leaving the $\ell$-dependence in $\Psi$ implicit (the azimuthal number $m$ is degenerate due to the spherical symmetry of the background). The general solution is written in terms of Legendre polynomials
\begin{equation}
    \Psi= c_1 P_\ell(r/M-1)+c_2Q_\ell(r/M-1)
\end{equation}
where $P_\ell(r/M-1)\to r^\ell$ and $Q_\ell(r/M-1)\to r^{-1-\ell}$ at large distances. Comparing with the expansion in~\ref{expscalar}, $c_1$ and $c_2$ are proportional to the amplitude of the external scalar perturbation, $\mathcal{E}_\ell^{\rm s}$, and to the object's response, $\Phi_\ell$, respectively. However,
regularity at the horizon ($r=2M$) requires $c_2=0$ and hence the regular solution has no decaying mode. This shows that the static scalar LNs is zero for any $\ell$. 
In a similar vein, one can study LNs arising from test electromagnetic (vector) perturbations by solving Maxwell's equations on a Schwarzschild background. We postpone discussing this case to a later section, where we analyze perturbations of a spinning (Kerr) BH using the Teukolsky formalism, which offers a unified framework for scalar, electromagnetic, and gravitational perturbations relevant for GWs.\footnote{See also Appendix~B of \citet{Katagiri:2024fpn} for analytic expressions of the most general solutions describing static gravitational, electromagnetic, and scalar perturbations of a Schwarzschild BH in vacuum GR.}

For the time being, let us focus on the gravitational perturbations of a Schwarzschild BH, studying separately the axial and polar sectors, which decouple due to spherical symmetry (see~\ref{sec:relLNs}). The individual components of the axial and polar perturbations are given by~\ref{pert_grav_ax} and~\ref{pert_grav_pol}, respectively.  
The equations governing the static perturbations $\delta g_{\mu \nu}$ arises from the perturbation of the vacuum Einstein equations, i.e., from  $\delta R_{\mu\nu}=0$. In the axial sector, this yields three non-trivial Einstein's equations connecting $h_{0}$ and $h_{1}$. It turns out that any two of these equations can be combined appropriately to yield the third equation and hence, only two of them are independent. In particular, it follows that, the metric perturbation $h_{1}$ can be written in terms of $h_{0}$ and its derivative as
%%%%%%%%%%%%%%%%%%%%%%%%%%%%%%%%%%%%%%%%%%%%%%%%%%%%%%%%%%
\begin{align}\label{h1pfsch}
h_{1}=\frac{i\omega r^{2}\Big(rh_{0}'-2h_{0}\Big)}{r^{3}\omega^{2}-(\ell+2)(\ell-1)(r-2M)}~.
\end{align}
%%%%%%%%%%%%%%%%%%%%%%%%%%%%%%%%%%%%%%%%%%%%%%%%%%%%%%%%%%
We can then eliminate $h_{1}$ from the other perturbed Einstein's equations and obtain a single second order differential equation for $h_{0}$. Further, in the static limit, it follows that the metric perturbation $h_{1}$ vanishes identically and we obtain the following second order differential equation for $h_{0}$, 
%%%%%%%%%%%%%%%%%%%%%%%%%%%%%%%%%%%%%%%%%%%%%%%%%%%%%%%%%%
\begin{align}\label{h0stateq}
\frac{h_{0}''}{h_{0}}&=\frac{\ell(\ell+1)r-4M}{r^{2}(r-2M)}~.
\end{align}
%%%%%%%%%%%%%%%%%%%%%%%%%%%%%%%%%%%%%%%%%%%%%%%%%%%%%%%%%%
In the dynamical scenario, the above differential equation gets corrected at $\mathcal{O}(M^{2}\omega^{2})$. To linear order in the frequency,~\ref{h0stateq} can be solved explicitly in terms of hypergeometric and Meijer-G functions to compute the axial (odd-parity) static LNs. For concreteness, let us compute explicitly the $\ell=2$ case, which is also the most relevant one from the phenomenological point of view (see~\ref{sec:GW}). The differential equation for $h_{0}$ becomes,
%%%%%%%%%%%%%%%%%%%%%%%%%%%%%%%%%%%%%%%%%%%%%%%%%%%%%%%%%%
\begin{equation}
\dfrac{d^{2}h_{0}}{dx^{2}}=\frac{(6x-2)}{x^2(x-1)}h_{0}~,
\label{h0om0}
\end{equation}
%%%%%%%%%%%%%%%%%%%%%%%%%%%%%%%%%%%%%%%%%%%%%%%%%%%%%%%%%%
where we have introduced the redefined radial coordinate $x\equiv (r/2M)$. The above differential equation can be solved in terms of simple polynomials, and the general solution takes the following form, 
%%%%%%%%%%%%%%%%%%%%%%%%%%%%%%%%%%%%%%%%%%%%%%%%%%%%%%%%%%
\begin{align}
h^{\ell=2}_{0}&=\mathcal{A}_{1}x^{2}\left(x-1\right)+\mathcal{A}_{2}\frac{12x^{3}-6x^{2}-2x-1}{3x}+4\mathcal{A}_{2}x^{2}(x-1)\log\left(\frac{x-1}{x}\right)~.
\label{h0e}
\end{align}
%%%%%%%%%%%%%%%%%%%%%%%%%%%%%%%%%%%%%%%%%%%%%%%%%%%%%%%%%%
where $\mathcal{A}_1$ and $\mathcal{A}_2$ are the two dimensionless arbitrary constants, to be fixed later by appropriate boundary conditions at the event horizon. It is possible to determine the magnetic LN associated with the $\ell=2$ mode in terms of the ratio of these two arbitrary constants. For this purpose, we consider the limit $r\to \infty$\footnote{In the static case, the source of the tidal field is at infinity, since otherwise gravitational interaction would lead to a dynamical configuration. This is the reason behind determining the static LNs through asymptotic expansion. However, as we will discuss, in the dynamical situation, one can no longer perform a asymptotic expansion, rather one should consider the perturbations in an intermediate region $M\ll r\ll1/\omega$.} of~\ref{h0e}, such that the perturbation to the axial part of the metric has the following asymptotic behavior,
%%%%%%%%%%%%%%%%%%%%%%%%%%%%%%%%%%%%%%%%%%%%%%%%%%%%%%%%%%
\begin{equation}
h^{\ell=2}_{0}(r)\simeq \mathcal{A}_{1} x^{3}\left[1+\mathcal{O}(x^{-1})\right]+\frac{\mathcal{A}_{2}}{5x^{2}}\left[1+\mathcal{O}\left(x^{-1}\right)\right]~.
\label{sv}
\end{equation}
%%%%%%%%%%%%%%%%%%%%%%%%%%%%%%%%%%%%%%%%%%%%%%%%%%%%%%%%%%
A quick comparison of the above asymptotic expansion with~\ref{h0asymp}, yields, $S_{2}=(\mathcal{A}_{2}/5)(2M)^{2}$ and ${\cal B}_{2}=3\mathcal{A}_{1}\sqrt{(4\pi/5)}(2M)^{-3}$. Substituting these results in~\ref{tlnb}, we obtain the $\ell=2$ magnetic LN,
%%%%%%%%%%%%%%%%%%%%%%%%%%%%%%%%%%%%%%%%%%%%%%%%%%%%%%%%%%%%%%%%%%%%%%%%%%%%%%%%%
\begin{equation}
k^{\rm M}_{2}=-\frac{1}{15}\left(\frac{2M}{R}\right)^{5}\left(\frac{\mathcal{A}_{2}}{\mathcal{A}_{1}}\right)~.
\label{defkm}
\end{equation}
%%%%%%%%%%%%%%%%%%%%%%%%%%%%%%%%%%%%%%%%%%%%%%%%%%%%%%%%%%%%%%%%%%%%%%%%%%%%%%%%%
% It is worthwhile to mention that, instead of the redefined radial coordinate $x$, if we had worked with $\kappa x$, where $\kappa$ is any constant, the above LN would scale by $(\kappa/2)$. This ambiguity, however, does not affect the final results, since the ratio $(\mathcal{A}_{2}/\mathcal{A}_{1})$ needs to be evaluated by matching with the axial perturbation in the interior, and will also scale by the inverse factor. 
%
Similarly, for the $\ell=3$ mode, the differential equation for $h_{0}$, as presented in~\ref{h0stateq}, can be solved analytically. This yields,
%%%%%%%%%%%%%%%%%%%%%%%%%%%%%%%%%%%%%%%%%%%%%%%%%%%%%%%%%%%%%
\begin{align}
h^{\ell=3}_{0}&=\mathcal{C}_{1}\left(\frac{x^{2}}{3}\right)\left(x-1\right)(3x-2)
+\mathcal{C}_{2}\left(\frac{1+5x+30x^{2}-210x^{3}+180x^{4}}{2x}\right)
\nonumber
\\
&\qquad+30\mathcal{C}_{2}x^{2}(x-1)(3x-2)\log\left(\frac{x-1}{x}\right)~.
\label{h0e3}
\end{align}
%%%%%%%%%%%%%%%%%%%%%%%%%%%%%%%%%%%%%%%%%%%%%%%%%%%%%%%%%%
with $\mathcal{C}_{1}$ and $\mathcal{C}_{2}$ being two arbitrary constants of integration. Matching of the asymptotic expansion with~\ref{h0asymp} yields, $S_{3}=(3/14)\mathcal{C}_{2}(2M)^{3}$, as well as the tidal coefficient $\mathcal{B}_{3}=9\mathcal{C}_{1}\sqrt{(4\pi/7)}(2M)^{-4}$. Therefore, the $\ell=3$ magnetic LN of a compact object can be expressed as,
%%%%%%%%%%%%%%%%%%%%%%%%%%%%%%%%%%%%%%%%%%%%%%%%%%%%%%%%%%%%%%%%%%%%%%%%%%%%%%%%%
\begin{equation}
k^{\rm M}_{3}=-\frac{3}{56}\left(\frac{2M}{R}\right)^{7}\left(\frac{\mathcal{C}_{2}}{\mathcal{C}_{1}}\right)~.
\label{defkm3}
\end{equation}
%%%%%%%%%%%%%%%%%%%%%%%%%%%%%%%%%%%%%%%%%%%%%%%%%%%%%%%%%%%%%%%%%%%%%%%%%%%%%%%%%
The above result holds for the geometry outside any non-rotating spherically symmetric compact object and can be straightforwardly extended to any $\ell$. 

The determination of the arbitrary constants $(\mathcal{A}_{1},\mathcal{A}_{2})$ for the $\ell=2$ mode and $(\mathcal{C}_{1},\mathcal{C}_{2})$ for the $\ell=3$ mode, is where the interior of the compact object and the boundary conditions at its surface play a crucial role. For example, in the case of a Schwarzschild BH spacetime, the perturbation and its derivative should be regular at the event horizon, located at $r=2M$. One can check that the metric perturbation $h_{0}$ is well behaved at the horizon, however, its radial derivative is ill-behaved due to the logarithmic term in~\ref{sv}, requiring $\mathcal{A}_{2}=0=\mathcal{C}_{2}$. This in turn implies that both the $\ell=2$ and the $\ell=3$ magnetic LNs of a Schwarzschild BH identically vanish (the same applies to any other choices of angular number $\ell$ \citep{Binnington:2009bb}).

Magnetic LNs can also be defined using the Regge--Wheeler master function $\Psi_{\rm RW}$, which is typically employed in the analysis of quasi-normal modes, by considering the static limit. 
Care must be taken, since in the standard formulation the Regge--Wheeler function depends only on $h_{1}$, which vanishes in the static case. 
The correct procedure is therefore to first assume $\omega\neq 0$, rescale $\Psi_{\rm RW}$ by $i\omega$, and only then take the limit $\omega\to 0$, so that $\Psi_{\rm RW}$ is expressed in terms of $h_{0}$. 
For a Schwarzschild BH, the relation takes the form
%%%%%%%%%%%%%%%%%%%%%%%%%%%%%%%%%%%%%%%%%%%%%%%%%%%%%%%%%%%%%%%%%%%%%%%%%%%%%%%%%
\begin{equation}\label{h0RW}
h_{0}\propto \left[\, r f(r)\,\frac{d\Psi_{\rm RW}}{dr} + f(r)\,\Psi_{\rm RW} \,\right]~,
\end{equation}
%%%%%%%%%%%%%%%%%%%%%%%%%%%%%%%%%%%%%%%%%%%%%%%%%%%%%%%%%%%%%%%%%%%%%%%%%%%%%%%%%
showing that the metric perturbation $h_{0}$ is determined by the Regge--Wheeler function and its derivative. 
Since, for a BH, $h_{0}$ admits only a growing mode, it follows from~\ref{h0RW} that $\Psi_{\rm RW}$ must also grow, thereby yielding vanishing static axial LNs for a Schwarzschild BH.

In the polar sector, on the other hand, there are three non-trivial metric perturbations, $K$, $H_{1}$, and $H$. Among these, $H$ is directly related to the perturbation of the $g_{tt}$ component, and hence is the one we should solve for. In the static limit the equation for $H$ takes the following form,
%%%%%%%%%%%%%%%%%%%%%%%%%%%%%%%%%%%%%%%%%%%%%%%%%%%%%%%%%%
\begin{align}\label{Hsole}
r^2f\left[r^2 fH''+2(r-M)H'\right]-\left[\ell(\ell+1)r^2f+4M^{2})\right]H=0,
\end{align}
%%%%%%%%%%%%%%%%%%%%%%%%%%%%%%%%%%%%%%%%%%%%%%%%%%%%%%%%%%
which is again valid up to $\mathcal{O}(M\omega)$ since dynamical corrections enter at $\mathcal{O}(M^{2}\omega^{2})$. It turns out that the perturbation variable $K$ can be also expressed in terms of $H$ and its derivative as,
%%%%%%%%%%%%%%%%%%%%%%%%%%%%%%%%%%%%%%%%%%%%%%%%%%%%%%%%%%
\begin{align}\label{KHHfd}
K=\left[\frac{(\ell+2)(\ell-1)r(r-2M)+4M(r-M)}{(\ell-1)(\ell+2)r(r-2M)}\right]H
+\frac{2MH'}{(\ell-1)(\ell+2)}~,
\end{align}
%%%%%%%%%%%%%%%%%%%%%%%%%%%%%%%%%%%%%%%%%%%%%%%%%%%%%%%%%%
with frequency dependent corrections appearing at $\mathcal{O}(M^{2}\omega^{2})$. The other perturbation variable $H_{1}$ takes the following form, 
%%%%%%%%%%%%%%%%%%%%%%%%%%%%%%%%%%%%%%%%%%%%%%%%%%%%%%%%%%
\begin{align}\label{H1}
H_{1}&=-\left[\frac{2(\ell+2)(\ell-1)r^2 f+8(r-M)(r-3M)}{(\ell-1)\ell(\ell+1)(\ell+2)(r-2M)^{2}}\right]i\omega M H
\nonumber
\\
&\qquad -\left[\frac{2(\ell+2)(\ell-1)r^{3}f+4Mr(r-3M)}{(\ell-1)\ell(\ell+1)(\ell+2)(r-2M)}\right]i\omega H'~,
\end{align}
%%%%%%%%%%%%%%%%%%%%%%%%%%%%%%%%%%%%%%%%%%%%%%%%%%%%%%%%%%
where we have used~\ref{KHHfd}, and terms of $\mathcal{O}(M^{2}\omega^{2})$ have been ignored. It is evident that $H_{1}$ vanishes in the static limit. The differential equation for $H$, presented in~\ref{Hsole}, can be solved exactly in terms of associated Legendre polynomials, yielding
%%%%%%%%%%%%%%%%%%%%%%%%%%%%%%%%%%%%%%%%%%%%%%%%%%%%%%%%%%
\begin{align}\label{Hext}
H=\mathcal{P}_{1}P_{\ell}^{2}\left(1+2z\right)+\mathcal{P}_{2}Q_{\ell}^{2}\left(1+2z\right)~;
\quad 
z\equiv \frac{r}{2M}-1~.
\end{align}
%%%%%%%%%%%%%%%%%%%%%%%%%%%%%%%%%%%%%%%%%%%%%%%%%%%%%%%%%%
Here, $\mathcal{P}_{1}$ and $\mathcal{P}_{2}$ are two arbitrary constants, to be determined by applying appropriate boundary conditions at large distance and at the event horizon. In order to obtain the asymptotic (large-$r$) behavior of the metric perturbation $H$, we can express the associated Legendre polynomials in terms of the hypergeometric functions and then obtain the following asymptotic expansion for the metric perturbation, 
%%%%%%%%%%%%%%%%%%%%%%%%%%%%%%%%%%%%%%%%%%%%%%%%%%%%%%%%%%
\begin{align}
H_{\rm asymp}&=\mathcal{P}_{1}\frac{\Gamma(1+2\ell)}{(2M)^{\ell}\Gamma(\ell-1)\Gamma(1+\ell)}r^{\ell}
+\mathcal{P}_{2}(2M)^{\ell+1}\frac{\sqrt{\pi}\Gamma(3+\ell)}{2^{2(\ell+1)}\Gamma(\ell+\frac{3}{2})}r^{-\ell-1}~.
\end{align}
%%%%%%%%%%%%%%%%%%%%%%%%%%%%%%%%%%%%%%%%%%%%%%%%%%%%%%%%%%
where we have used the result that $\Gamma(-\ell)$ diverges, owing to the fact that $\ell$ is an integer. Therefore, a comparison with~\ref{Hasymp}, immediately yields the mass moments and the tidal moments, which for a generic $\ell$ mode reads,
%%%%%%%%%%%%%%%%%%%%%%%%%%%%%%%%%%%%%%%%%%%%%%%%%%%%%%%%%%
\begin{align}
2M_{\ell}=\mathcal{P}_{2}(2M)^{\ell+1}&\frac{\sqrt{\pi}\Gamma(3+\ell)}{2^{2(\ell+1)}\Gamma(\ell+\frac{3}{2})}~;
\\
-\left(\frac{2}{\ell(\ell-1)}\right)\sqrt{\frac{2\ell+1}{4\pi}}\mathcal{E}_{\ell}&=\mathcal{P}_{1}\frac{\Gamma(1+2\ell)}{(2M)^{\ell}\Gamma(\ell-1)\Gamma(1+\ell)}~.
\end{align}
%%%%%%%%%%%%%%%%%%%%%%%%%%%%%%%%%%%%%%%%%%%%%%%%%%%%%%%%%%
Given the mass and electric tidal moments, the electric-type LNs can be determined from~\ref{tlne}, by simply taking the ratio of these moments with appropriate normalization factors, yielding,
%%%%%%%%%%%%%%%%%%%%%%%%%%%%%%%%%%%%%%%%%%%%%%%%%%%%%%%%%%
\begin{align}\label{electrinLNgenl}
k_{\ell}^{\rm E}&=\frac{\mathcal{P}_{2}}{2^{2\ell+3}\mathcal{P}_{1}}\frac{\sqrt{\pi}\Gamma(3+\ell)\Gamma(\ell-1)\Gamma(1+\ell)}{\Gamma(\ell+\frac{3}{2})\Gamma(1+2\ell)}~.
\end{align}
%%%%%%%%%%%%%%%%%%%%%%%%%%%%%%%%%%%%%%%%%%%%%%%%%%%%%%%%%%
In this case as well, the electric LNs as presented above holds for the geometry external to any static and spherically symmetric compact object, not just BHs. The boundary condition at the surface of the compact object, which is the event horizon for a BH, distinguishes between various compact objects. It turns out that the polar perturbation $H$ is not regular at the BH horizon due to the $Q_{\ell}^{2}$ term and hence we must demand $\mathcal{P}_{2}=0$. This in turn leads to vanishing electric LNs for Schwarzschild BH. 

Similar to the axial case, the polar LNs can be computed from the Zerilli function. 
This involves relating the metric perturbation $H_{0}$ to the Zerilli master function $\Psi_{\rm Z}$. 
Unlike the Regge--Wheeler function, however, the Zerilli function is well defined in the static limit and therefore requires no additional rescaling. 
For our purposes, the asymptotic expansion of the relation between $H_{0}$ and $\Psi_{\rm Z}$ is sufficient, and reads
%%%%%%%%%%%%%%%%%%%%%%%%%%%%%%%%%%%%%%%%%%%%%%%%%%%%%%%%%%
\begin{align}\label{HoZ}
H_{0} \to \Psi_{\rm Z}' + \left(\frac{(\ell-1)(\ell+2)}{2}+1\right)\frac{\Psi_{\rm Z}}{r}\,,\qquad r\to\infty
\end{align}
%%%%%%%%%%%%%%%%%%%%%%%%%%%%%%%%%%%%%%%%%%%%%%%%%%%%%%%%%%
The exact expression for generic choices of the radial coordinate can be found in Appendix~D of \citet{Cardoso:2017cfl}. 
Thus, the polar metric perturbation $H_{0}$ is expressed in terms of the Zerilli function and its derivative. 
Since for a Schwarzschild BH the perturbation $H_{0}$ admits only a growing mode, it follows that the Zerilli function must also grow, implying once again that the polar static LNs vanish.  
In summary, Schwarzschild BHs have vanishing static tidal LNs for any $\ell$ and for both axial and polar perturbations.

%%%%%%%%%%%%%%%%%%%%%%%%%%%%%%%%%%%%
%%%%%%%%%%%%%%%%%%%%%%%%%%%%%%%%%%%%
%%%%%%%%%%%%%%%%%%%%%%%%%%%%%%%%%%%%
\subsection{Static bosonic Love numbers of a spinning black hole}\label{stat_Love_Kerr}

The perturbation of a Kerr BH is best performed using the Weyl scalars --- (a) $\Phi$ for scalar perturbations ($s=0$), (b) $\Phi_{0}$ and $\rho^{2}\Phi_{2}$ for EM perturbations, with $s=\pm1$, respectively, and (c) $\Psi_{0}$ and $\rho^{4}\Psi_{4}$ for gravitational, with $s=\pm 2$, respectively. Here, $\rho=r-ia\cos \theta$ is a spin coefficient of the Newman-Penrose formalism, where $(r,\theta)$ are the Boyer-Lindquist coordinates, $a=(J/M)$ is the rotation parameter, with $M$ being the mass and $J$ the angular momentum of the Kerr BH. The fact that $\rho$ is complex is consistent with the fact that all the Weyl scalars are generically complex quantities. Evolution and variation of each of these scalars are governed by the Teukolsky equations \citep{Teukolsky:1972my, Teukolsky:1973ha, Press:1973zz, Teukolsky:1974yv}. The starting point to compute the tidal response is to solve the Teukolsky equation in the near-zone regime in a small frequency approximation. For our purpose, it will be best to express the Teukolsky equation in ingoing Kerr coordinates: $\{v,r,\theta,\widetilde{\phi}\}$, whose relations to the Boyer-Lindquist coordinates $\{t,r,\theta,\phi\}$ are given by \citep{Teukolsky:1974yv}
%%%%%%%%%%%%%%%%%%%%%%%%%%%%%%%%%%%%%%%%%%%%%%%%%%%%%%%%%%%%%%%%%%%%%%%%%%%%%%
\begin{equation}
\mathrm{d}v=\mathrm{d}t+\frac{r^2+a^2}{\Delta}\mathrm{d}r\,, 
\qquad 
\mathrm{d}\widetilde{\phi} = \mathrm{d}\phi +\frac{a}{\Delta}\mathrm{d}r\,,
\end{equation}
%%%%%%%%%%%%%%%%%%%%%%%%%%%%%%%%%%%%%%%%%%%%%%%%%%%%%%%%%%%%%%%%%%%%%%%%%%%%%%
where $\Delta \equiv r^{2}-2Mr+a^{2}=(r-r_+)(r-r_-)$, with $r_{\pm}=M\pm\sqrt{M^2-a^2}$ being the locations of the event and the Cauchy horizons. The relevant Newman-Penrose scalars can be decomposed into radial and angular parts as \citep{Teukolsky:1974yv} 
% 
%%%%%%%%%%%%%%%%%%%%%%%%%%%%%%%%%%%%%%%%%%%%%%%%%%%%%%%%%%%%%%%%%%%%%%%%%%%%%%
\begin{equation}\label{gen_eq_decomp}
\rho^{-s+|s|}\zeta_{-s+|s|}=\int\mathrm{d}\omega\,e^{-i\omega v}\sum_{\ell m} e^{-im\widetilde{\phi}}\,_{s}S_{\ell m}(\theta)R_{s}(r)~,
\end{equation}
%%%%%%%%%%%%%%%%%%%%%%%%%%%%%%%%%%%%%%%%%%%%%%%%%%%%%%%%%%%%%%%%%%%%%%%%%%%%%%
where $\zeta$ corresponds to --- (a) $\Phi$ for a scalar perturbation, (b) $\Phi_{0}$, and $\Phi_{2}$ for EM perturbations, and (c) $\Psi_{0}$ and $\Psi_{4}$ for gravitational perturbations. The angular part, $\,_{s}S_{\ell m}(\theta)$, satisfies the equation for spin-weighted spheroidal harmonics, while $R_{s}(r)$ satisfies the source-free radial Teukolsky equation, which in the static limit reads \citep{Teukolsky:1974yv}
%%%%%%%%%%%%%%%%%%%%%%%%%%%%%%%%%%%%%%%%%%%%%%%%%%%%%%%%%%%%%%%%%%%%%%%%%%%%%%
\begin{equation}
\dfrac{\mathrm{d}^{2}R_{s}}{\mathrm{d}r^2}+\frac{2}{\Delta}\Big[(s+1)(r-M)+iam\Big]\dfrac{\mathrm{d}R_{s}}{\mathrm{d}r}+\Bigg[am\left(\frac{4is(r-M)}{\Delta^{2}}\right)+\frac{2(2s-1)i\omega r-\lambda}{\Delta}\Bigg]R_{s}=0\,,
\label{TEq}
\end{equation}
%%%%%%%%%%%%%%%%%%%%%%%%%%%%%%%%%%%%%%%%%%%%%%%%%%%%%%%%%%%%%%%%%%%%%%%%%%%%%%
where $\lambda\equiv E_{\ell m}-s(s+1)$ and $E_{\ell m}=\ell(\ell+1)$ is the eigenvalue of the angular equation in the zero-frequency limit \citep{Press:1973zz, Fackerell:1977shn, Seidel:1988ue, Berti:2005gp}. The radial function is often denoted as $_{s}R_{\ell m}(r)$ to emphasize its angular and spin dependence, but for notational simplicity we will use the shorter form $R_{s}(r)$ throughout. It is possible to re-express the radial Teukolsky equation, showing explicitly the associated singular points, by expressing it in the following form \citep{Chia:2020yla}
%%%%%%%%%%%%%%%%%%%%%%%%%%%%%%%%%%%%%%%%%%%%%%%%%%%%%%%%%%%%%%%%%%%%%%%%%%%%%%
\begin{multline}\label{Teqr}
\dfrac{\mathrm{d}^{2}R_{s}}{\mathrm{d}r^2}+\left(\frac{2iP_{0}+(s+1)}{r-r_{+}}-\frac{2iP_{0}-(s+1)}{r-r_{-}}\right)\dfrac{\mathrm{d}R_{\rm s}}{\mathrm{d}r} 
\\ 
+\left[\frac{2isP_{0}}{(r-r_+)^2}-\frac{2isP_{0}}{(r-r_-)^2}+\frac{\ell(\ell+1)-s(s+1)}{(r-r_-)(r_+-r_-)}-\frac{\ell(\ell+1)-s(s+1)}{(r-r_+)(r_+-r_-)}\right]R_{\rm s}=0\,,
\end{multline}
%%%%%%%%%%%%%%%%%%%%%%%%%%%%%%%%%%%%%%%%%%%%%%%%%%%%%%%%%%%%%%%%%%%%%%%%%%%%%%
where
%%%%%%%%%%%%%%%%%%%%%%%%%%%%%%%%%%%%%%%%%%%%%%%%%%%%%%%%%%%%%%%%%%%%%%%%%%%%%%
\begin{equation}\label{def_Ppm}
P_{0}\equiv\frac{am}{r_+-r_-}\,.
\end{equation}
%%%%%%%%%%%%%%%%%%%%%%%%%%%%%%%%%%%%%%%%%%%%%%%%%%%%%%%%%%%%%%%%%%%%%%%%%%%%%%
% Up to this point the analysis is valid for both non-extremal, as well as extremal BH, since in the extremal limit, the above equation will be independent of $P_{0}$. At the next step, we introduce 
%%%%%%%%%%%%%%%%%%%%%%%%%%%%%%%
%%%%%%%%%%%%%%%%%%%%%%%%%%%%%%%
%%%%%%%%%%%%%%%%%%%%%%%%%%%%%%%
%%%%%%%%%%%%%%%%%%%%%%%%%%%%%%%
\begin{figure}
    \centering
    \includegraphics[width=\linewidth]{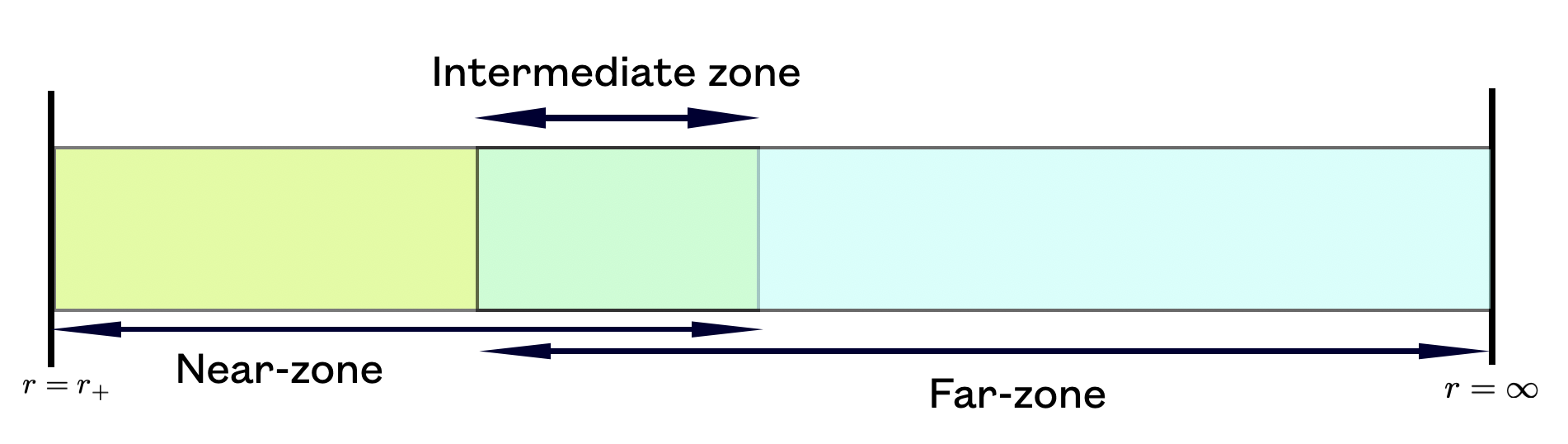}
    \caption{This figure schematically depicts different zones responsible for different behavior of the Weyl scalars. The near-zone is characterized by $\omega(r-r_{+})\ll 1$, and the intermediate zone corresponds to the region with $r\gg r_{+}$, is where  the source of the tidal field is located. While $\omega r\ll 1$, with $r\sim \omega^{-1}$ being the boundary of the near-zone. The far-zone corresponds to $r\gg r_+$, . See also \citet{Chia:2020yla, DeLuca:2024ufn}.
    }
    \label{fig:tidalzone}
\end{figure}
%%%%%%%%%%%%%%%%%%%%%%%%%%%%%%%
%%%%%%%%%%%%%%%%%%%%%%%%%%%%%%%
%%%%%%%%%%%%%%%%%%%%%%%%%%%%%%%
%%%%%%%%%%%%%%%%%%%%%%%%%%%%%%%
By introducing the re-scaled dimensionless radial coordinate $z\equiv (r-r_+)/(r_+-r_-)$,  the radial Teukolsky equation for non-extremal BH in the near zone\footnote{Note that the notion of near zone adopted here is frequency dependent and stems from the approach of \citet{page1976particle-25c, Starobinskij2, Teukolsky:1974yv, Cardoso:2008kj, Chia:2020yla}. 
An alternative splitting into near and far zones was proposed in \citet{Sasaki:2003xr}, where the near zone is defined by $r_+\leq r<\infty$ and the far zone by $r_+<r\leq \infty$. 
The overlap region $r_+<r<\infty$ allows for matching the two solutions. 
We will return to this approach later in the review.} ($M\omega z\ll 1$, see~\ref{fig:tidalzone}) takes the following form \citep{LeTiec:2020bos, Chakraborty:2023zed},
%%%%%%%%%%%%%%%%%%%%%%%%%%%%%%%%%%%%%%%%%%%%%%%%%%%%%%%%%%%%%%%%%%%%%%%%%%%%%%
\begin{align}\label{GMATE0}
\frac{\mathrm{d}^{2}R_{s}}{\mathrm{d}z^2}&+\left[\frac{2iP_{0}+(s+1)}{z}-\frac{2iP_{0}-(s+1)}{1+z}\right]\frac{\mathrm{d}R_{s}}{\mathrm{d}z}
\nonumber
\\
&+\left[\frac{2isP_{0}}{z^2}-\frac{2isP_{0}}{(z+1)^2}-\frac{\ell(\ell+1)-s(s+1)}{z(1+z)}\right]R_{s}=0\,,
\end{align}
%%%%%%%%%%%%%%%%%%%%%%%%%%%%%%%%%%%%%%%%%%%%%%%%%%%%%%%%%%%%%%%%%%%%%%%%%%%%%%
This is a second order differential equation with three regular singular points at $z=-1,\,0,\,\text{and}\,\infty$. Therefore, its solution can be written in terms of the Gauss hypergeometric function as,
%%%%%%%%%%%%%%%%%%%%%%%%%%%%%%%%%%%%%%%%%%%%%%%%%%%%%%%%%%%%%%%%%%%%%%%%%%%%%%
\begin{multline}\label{radialfngen0}
R_s(z)=(z+1)^{-s}\left[c_{1}z^{-s}\,_{2}F_{1}\left(1+\ell-s,-s-\ell;1-s+2iP_{0};-z\right)\right.
\\
\left.+c_2 z^{-2iP_{0}}\,_{2}F_{1}\left(\ell-2iP_{0}+1,-\ell-2iP_{0};1+s-2iP_{0};-z\right)\right]\,,
\end{multline}
%%%%%%%%%%%%%%%%%%%%%%%%%%%%%%%%%%%%%%%%%%%%%%%%%%%%%%%%%%%%%%%%%%%%%%%%%%%%%%
where $c_1$ and $c_2$ are the constants of integration, one of which can be fixed by considering the near horizon limit ($z\to 0$). Using properties of the hypergeometric function (see~\ref{app:identities}, in particular~\ref{hypzero}), the radial perturbation variable $R(z)$ in the near horizon limit can be approximately written as
%%%%%%%%%%%%%%%%%%%%%%%%%%%%%%%%%%%%%%%%%%%%%%%%%%%%%%%%%%%%%%%%%%%%%%%%%%%%%%
\begin{equation}
R_s^{\rm near}(z) \sim c_1 z^{-s}+c_2 z^{-2iP_{0}}\,.
\end{equation}
%%%%%%%%%%%%%%%%%%%%%%%%%%%%%%%%%%%%%%%%%%%%%%%%%%%%%%%%%%%%%%%%%%%%%%%%%%%%%%
The first and second terms correspond to the zero-frequency limit of an ingoing and outgoing wave at the future horizon, respectively \citep{Teukolsky:1974yv}. Imposing ingoing boundary conditions requires $c_2=0$, which can also be directly inferred by imposing regularity of the static solution in the Hartle-Hawking tetrad, i.e. $\Delta^s R_s$ must be $C^2$ (twice continuously differentiable) at the horizon \citep{LeTiec:2020bos}.  

Therefore, the regular radial Teukolsky function reduces to
%%%%%%%%%%%%%%%%%%%%%%%%%%%%%%%%%%%%%%%%%%%%%%%%%%%%%%%%%%%%%%%%%%%%%%%%%%%%%%
\begin{equation}\label{radialstatickerr}
R_s(z)=c_{1}z^{-s}(z+1)^{-s}\,_{2}F_{1}\left(1+\ell-s,-s-\ell;1-s+2iP_{0};-z\right)\,.
\end{equation}
%%%%%%%%%%%%%%%%%%%%%%%%%%%%%%%%%%%%%%%%%%%%%%%%%%%%%%%%%%%%%%%%%%%%%%%%%%%%%%
Given the above radial Teukolsky function, computing the tidal response requires computing the radial part of the appropriate Newman-Penrose scalar $\zeta_{-s+|s|}$ in the intermediate regime, obtained by taking the large-$r$ (or, equivalently large-$z$) limit of~\ref{radialstatickerr}, while keeping $r\ll \omega^{-1}$, i.e., in the intermediate regime (see~\ref{fig:tidalzone}). This yields
%%%%%%%%%%%%%%%%%%%%%%%%%%%%%%%%%%%%%%%%%%%%%%%%%%%%%%%%%%%%%%%%%%%%%%%%%%%%%%
\begin{multline}\label{psi4gen_stat}
\zeta_{-s+|s|}\propto \left(\frac{r}{r_{+}-r_{-}}\right)^{\ell-|s|}\left\{\frac{\Gamma\left(1-s+2iP_{0}\right)\Gamma\left(2\ell+1\right)}{\Gamma\left(\ell+2iP_{0}+1\right)\Gamma\left(1-s+\ell\right)}\right\}
\\
\times\left[1+\left(\frac{r}{r_{+}-r_{-}}\right)^{-2\ell-1}\left\{\frac{\Gamma\left(-2\ell-1\right)\Gamma\left(\ell+2iP_{0}+1\right)\Gamma\left(1-s+\ell\right)}{\Gamma\left(-s-\ell\right)\Gamma\left(-\ell+2iP_{0}\right)\Gamma\left(2\ell+1\right)}\right\}\right]\,.
\end{multline}
%%%%%%%%%%%%%%%%%%%%%%%%%%%%%%%%%%%%%%%%%%%%%%%%%%%%%%%%%%%%%%%%%%%%%%%%%%%%%%
We can compare the above expansion with the generalization of~\ref{psi_4_intermediate} for generic spin-$s$ perturbations, namely 
%%%
\begin{equation}\label{psi_4_intermediateB_genericspin}
\zeta_{-s+|s|}\sim \mathcal{N}r^{\ell-|s|}\left[1+\,_{s}\mathcal{F}_{\ell m}\left(\frac{R}{r}\right)^{1
+2\ell} \right]
\end{equation}
%%%
where, in the BH case, the characteristic radius becomes, $R=r_{+}$. Comparing~\ref{psi_4_intermediateB_genericspin} with~\ref{psi4gen_stat}, one can explicitly observe the existence of a non-zero static tidal response function for a Kerr BH under generic spin-$s$ perturbations,
%%%%%%%%%%%%%%%%%%%%%%%%%%%%%%%%%%%%%%%%%%%%%%%%%%%%%%%%%%%%%%%%%%%%%%%%%%%%%%
\begin{align}
\,_{s}\mathcal{F}^{\rm static}_{\ell m}&=\left(\frac{r_{+}-r_{-}}{r_{+}}\right)^{2\ell+1}\frac{\Gamma\left(-2\ell-1\right)\Gamma\left(\ell+2iP_{0}+1\right)\Gamma\left(1-s+\ell\right)}{\Gamma\left(-s-\ell\right)\Gamma\left(-\ell+2iP_{0}\right)\Gamma\left(2\ell+1\right)}\,,
\label{resp_func_arb_rotstatic}
\\
&=\left(-1\right)^{s+1}\left(\frac{iam}{r_{+}}\right)\frac{\left(\ell+s\right)!\left(\ell-s\right)!}{\left(2\ell+1\right)!\left(2\ell\right)!}\prod_{j=1}^{l}\left[j^2\left(1-\frac{r_{-}}{r_{+}} \right)^{2}+\left(\frac{2am}{r_{+}}\right)^2\right]\,,
\label{reducedstatickerrnonext}
\end{align}
%%%%%%%%%%%%%%%%%%%%%%%%%%%%%%%%%%%%%%%%%%%%%%%%%%%%%%%%%%%%%%%%%%%%%%%%%%%%%%
where in the second line, we have used the reflection formula for the $\Gamma$ functions (see~\ref{mirrorgamma} in~\ref{app:identities}), and the result is valid for integer values of $s$ (both positive and negative). (See~\ref{sec:fermionic} for the case of half-integer $s$.)
Note that the above response function is well-behaved in the extremal limit, i.e. $a\to M$, which also implies $r_{-}\to r_{+}$. The static response function of an extremal Kerr BH reads
%%%%%%%%%%%%%%%%%%%%%%%%%%%%%%%%%%%%%%%%%%%%%%%%%%%%%%%%%%%%%%%%%%%%%%%%%%%%%%
\begin{align}
\,_{s}\mathcal{F}^{\rm static}_{\ell m\textrm{(ext)}}&=i\left(-1\right)^{s+1}4^{\ell}m^{2\ell+1}\frac{\left(\ell+s\right)!\left(\ell-s\right)!}{\left(2\ell+1\right)!\left(2\ell\right)!}\,,
\end{align}
%%%%%%%%%%%%%%%%%%%%%%%%%%%%%%%%%%%%%%%%%%%%%%%%%%%%%%%%%%%%%%%%%%%%%%%%%%%%%%
Later on we will find the same result by studying directly the perturbations of an extremal Kerr BH. 

One can check that~\ref{reducedstatickerrnonext} correctly reproduces the results of \citet{LeTiec:2020spy, LeTiec:2020bos} for the spin value $s=-2$. 
Overall, the static response function is purely imaginary for generic spin-$s$ perturbations.
This shows that an arbitrarily rotating BH in vacuum GR has vanishing static bosonic LNs, but non vanishing dissipation numbers, 
%%%%%%%%%%%%%%%%%%%%%%%%%%%%%%%%%%%%%%%%%%%%%%%%%%%%%%%%%%%%%%%%%%%%%%%%%%%%%%
\begin{align}
\,_{s}k_{\ell m}^{\rm static}&=0\,,
\label{static_LNs}
\\
\,_{s}\nu^{\rm static}_{\ell m}&=\left(-1\right)^{s+1}\left(\frac{am}{r_{+}}\right)\frac{\left(\ell+s\right)!\left(\ell-s\right)!}{\left(2\ell+1\right)!\left(2\ell\right)!}\prod_{j=1}^{l}\left[j^2\left(1-\frac{r_{-}}{r_{+}} \right)^{2}+\left(\frac{2am}{r_{+}}\right)^2\right]\,,
\label{dissstatgen}
\\
\,_{s}k_{\ell m\textrm{(ext)}}^{\rm static}&=0\,,
\label{static_LNs_ext}
\\
\,_{s}\nu^{\rm static}_{\ell m\textrm{(ext)}}&=\left(-1\right)^{s+1}4^{\ell}m^{2\ell+1}\frac{\left(\ell+s\right)!\left(\ell-s\right)!}{\left(2\ell+1\right)!\left(2\ell\right)!}\,.
\label{dissstatgenext}
\end{align}
%%%%%%%%%%%%%%%%%%%%%%%%%%%%%%%%%%%%%%%%%%%%%%%%%%%%%%%%%%%%%%%%%%%%%%%%%%%%%%
Note that, for a Schwarzschild BH ($a=0$), as well as for axisymmetric tidal perturbations ($m=0$), the dissipation numbers vanish as well for any $\ell$. Moreover, for extremal BHs, the dissipation numbers grow as $m^{2\ell+1}$, and can in principle be very large for $\ell=m\gg 1$. 
% Given these interesting results, we now discuss four relevant aspects --- (a) the importance of analytic continuation; (b) the gauge invariant nature of the response function; (c) the extremal limit, done in a separate manner and finally (d) the large $\ell=m$ limit.

\paragraph{On the analytic continuation of $\ell$}~--- In the above analysis, the response function was determined by first treating $\ell$ as a complex number and then taking the integer limit at the end. 
To illustrate the necessity of this procedure, let us consider an example involving spin-$s$ perturbations of a Kerr BH. 
First, note that the asymptotic expansion of the Weyl scalar from~\ref{radialstatickerr}, including the subleading terms of the tidal field, is given by (see~\ref{hyp_z_2byz} and~\ref{hypzero} in~\ref{app:identities}), 
%%%%%%%%%%%%%%%%%%%%%%%%%%%%%%%%%%%%%%%%%%%%%%%%%%%%%%%%%%%%%%%%%%%%%%%%%%%%%%
\begin{align}\label{analytic_cont1}
\zeta_{-s+|s|}&\propto z^{-s+\ell}\Bigg[\underbrace{\frac{\Gamma(1-s+2iP_{0})\Gamma(-1-2\ell)}{\Gamma(-s-\ell)\Gamma(2iP_{0}-\ell)}z^{-1-2\ell}}_{\rm Response}+\frac{\Gamma(1-s+2iP_{0})\Gamma(1+2\ell)}{\Gamma(1+\ell-s)\Gamma(1+\ell+2iP_{0})}
\nonumber
\\
&\times\underbrace{\left(1+\cdots-\frac{\Gamma(-s-\ell+1+2n)\Gamma(-\ell-2iP_{0}+1+2n)\Gamma(-2\ell)}{\Gamma(-s-\ell)\Gamma(-\ell-2iP_{0})\Gamma(-2\ell+1+2n)\Gamma(2+2n)}z^{-1-2n}\right)}_{\rm Tidal}\Big]~,
\end{align}
%%%%%%%%%%%%%%%%%%%%%%%%%%%%%%%%%%%%%%%%%%%%%%%%%%%%%%%%%%%%%%%%%%%%%%%%%%%%%%
where $n$ is the nearest integer to $\ell$. In analytic continuation, $\ell\neq n$ and hence the response can be well separated from the sub-leading terms of the tidal field. 
However, if $\ell$ is considered to be an integer ($\ell=n$), then the leading power of $z$ in the response function coincides with the $(2\ell+1)$th sub-leading term in the tidal field, as evident from~\ref{analytic_cont1}. Therefore, the tidal field and the response function gets entangled with one another, and hence the coefficient of $z^{-1-2\ell}$ term becomes 
%%%%%%%%%%%%%%%%%%%%%%%%%%%%%%%%%%%%%%%%%%%%%%%%%%%%%%%%%%%%%%%%%%%%%%%%%%%%%%
\begin{align}\label{analytic_cont2}
&\textrm{coefficients~of}\,z^{-2\ell-1}|_{\textrm{integer}\,\ell}=\underbrace{\frac{\Gamma(1-s+2iP_{0})\Gamma(-1-2\ell)}{\Gamma(-s-\ell)\Gamma(2iP_{0}-\ell)}}_{\rm Response~(leading)}
\nonumber
\\
&\qquad -\underbrace{\frac{\Gamma(-s+\ell+1)\Gamma(\ell-2iP_{0}+1)\Gamma(-2\ell)}{\Gamma(-s-\ell)\Gamma(-\ell-2iP_{0})\Gamma(2+2\ell)}\times \frac{\Gamma(1-s+2iP_{0})\Gamma(1+2\ell)}{\Gamma(1+\ell-s)\Gamma(1+\ell+2iP_{0})}}_{\rm Tidal~(sub-leading)}\,.
\end{align}
%%%%%%%%%%%%%%%%%%%%%%%%%%%%%%%%%%%%%%%%%%%%%%%%%%%%%%%%%%%%%%%%%%%%%%%%%%%%%%
Using the reflection identity involving $\Gamma$ functions (see~\ref{mirrorgamma}), it follows that the coefficient of $z^{-2\ell-1}$ arising from the sub-leading contribution of the tidal field, as presented in~\ref{analytic_cont2}, can be expressed as
%%%%%%%%%%%%%%%%%%%%%%%%%%%%%%%%%%%%%%%%%%%%%%%%%%%%%%%%%%%%%%%%%%%%%%%%%%%%%%
\begin{align}
&\textrm{sub-leading~tidal~field}=
\frac{\Gamma(1-s+2iP_{0})\Gamma(1+2\ell)\Gamma(-2\ell)\Gamma(1+\ell-2iP_{0})}{\Gamma(-s-\ell)\Gamma(-\ell-2iP_{0})\Gamma(2+2\ell)\Gamma(1+\ell+2iP_{0})}
\nonumber
\\
&=\frac{\Gamma(1-s+2iP_{0})\Gamma(-1-2\ell)}{\Gamma(-s-\ell)\Gamma(-\ell+2iP_{0})}\frac{\Gamma(1+2\ell)\Gamma(-2\ell)}{\Gamma(2+2\ell)\Gamma(-1-2\ell)}\frac{\Gamma(1+\ell-2iP_{0})\Gamma(-\ell+2iP_{0})}{\Gamma(-\ell-2iP_{0})\Gamma(1+\ell+2iP_{0})}
\nonumber
\\
&=\frac{\Gamma(1-s+2iP_{0})\Gamma(-1-2\ell)}{\Gamma(-s-\ell)\Gamma(-\ell+2iP_{0})}
=\textrm{leading~response}\,,
\end{align}
%%%%%%%%%%%%%%%%%%%%%%%%%%%%%%%%%%%%%%%%%%%%%%%%%%%%%%%%%%%%%%%%%%%%%%%%%%%%%%
where we have used the fact that $\ell$ is an integer. Therefore, due to the relative minus sign (see~\ref{analytic_cont2}), the $z^{-1-2\ell}$ term arising from the leading-order response function cancels exactly against the subleading tidal-field contribution of the same order. Thus, had we worked directly with integer values of $\ell$, the mixing with the subleading tidal term would have obscured the fact that the response function is nonzero, making it appear as though the body does not respond to the external tidal field. To avoid such ambiguities, it is thus essential to adopt an analytic continuation in $\ell$.  

\paragraph{Gauge invariance}~---  Even though we are working with Weyl scalars which are by definition gauge invariant, it is illustrative to show the gauge invariance explicitly. For this purpose, we will derive the static response function in the Boyer-Lindquist coordinates and show that it is identical to~\ref{resp_func_arb_rotstatic}, which is the one derived from advanced null coordinates. First of all, the radial Teukolsky equation in Boyer-Lindquist coordinates, for generic spin-$s$ perturbations, using the rescaled radial coordinate $z$ and in the zero-frequency limit, takes the following form 
%%%%%%%%%%%%%%%%%%%%%%%%%%%%%%%%%%%%%%%%%%%%%%%%%%%%%%%%%%%%%%%%%%%%%%%%%%%%%%
\begin{equation}
z(1+z)\dfrac{d^{2}R_s}{dz^{2}}+(s+1)(1+2z)\dfrac{dR_s}{dz}+\left[\frac{P_{0}^{2}+is(1+2z)P_{0}}{z(1+z)}-\ell(\ell+1)+s(s+1)\right]R_s=0\,,
\end{equation}
%%%%%%%%%%%%%%%%%%%%%%%%%%%%%%%%%%%%%%%%%%%%%%%%%%%%%%%%%%%%%%%%%%%%%%%%%%%%%%
where $P_{0}$ is given by~\ref{def_Ppm}. The two solutions of the above differential equation has the following behavior in the near-horizon regime: $\sim c_{1}z^{iP_{0}-s}+c_{2}z^{-iP_{0}}$. In the Boyer-Lindquist coordinate the ingoing/regular mode behaves as $z^{-s}e^{iP_{0}\ln z}$, while the outgoing/irregular mode scales as, $e^{iP_{0}\ln z}$ \citep{Teukolsky:1974yv}.  Since we only have ingoing modes at the horizon of a BH, one must set $c_{2}=0$ and hence the solution of the radial perturbation equation in the Boyer Lindquist coordinate becomes
%%%%%%%%%%%%%%%%%%%%%%%%%%%%%%%%%%%%%%%%%%%%%%%%%%%%%%%%%%%%%%%%%%%%%%%%%%%%%%
\begin{equation}
R_s(z)=c_{1}z^{iP_{0}-s}(1+z)^{iP_{0}}\,_{2}F_{1}(-\ell+2iP_{0},1+\ell+2iP_{0};1-s+2iP_{0};-z)\,.
\end{equation}
%%%%%%%%%%%%%%%%%%%%%%%%%%%%%%%%%%%%%%%%%%%%%%%%%%%%%%%%%%%%%%%%%%%%%%%%%%%%%%
While the above solution is different from~\ref{radialstatickerr}, which was obtained in the advanced null coordinates, the asymptotic limit of the radial function in the Boyer-Lindquist coordinate reads
%%%%%%%%%%%%%%%%%%%%%%%%%%%%%%%%%%%%%%%%%%%%%%%%%%%%%%%%%%%%%%%%%%%%%%%%%%%%%%
\begin{equation}
R_s^{\infty}(z)\propto z^{\ell-s}\Bigg[1+\frac{\Gamma(-1-2\ell)\Gamma(1+\ell-s)\Gamma(1+\ell+2iP_{0})}{\Gamma(1+2\ell)\Gamma(-\ell-s)\Gamma(-\ell+2iP_{0})}z^{-1-2\ell}\Bigg]\,.
\end{equation}
%%%%%%%%%%%%%%%%%%%%%%%%%%%%%%%%%%%%%%%%%%%%%%%%%%%%%%%%%%%%%%%%%%%%%%%%%%%%%%
A comparison with~\ref{resp_func_arb_rotstatic} reveals that the response function in Boyer-Lindquist coordinates is identical to that in advanced null coordinates, as expected from a gauge invariant quantity.

\paragraph{Static LNs for extremal BHs}~--- 
For completeness, here we study the perturbations of the extremal Kerr BHs from the scratch, in order to match with our previous results regarding the extremal limit of the non-extremal BH. For that purpose, we first write down the radial Teukolsky equation for generic spin-$s$ perturbations of an extremal BH in the zero-frequency limit,
%%%%%%%%%%%%%%%%%%%%%%%%%%%%%%%%%%%%%%%%%%%%%%%%%%%%%%%%%%%%%%%%%%%%%%%%%%%%%%
\begin{equation}
\frac{\mathrm{d}^{2}R_{s}}{\mathrm{d}r^{2}}+\left[\frac{2(s+1)}{(r-M)}+\frac{2iMm}{(r-M)^2}\right]\frac{\mathrm{d}R_{s}}{\mathrm{d}r}+\left[-\frac{\ell(\ell+1)-s(s+1)}{(r-M)^{2}}+\frac{4imMs}{(r-M)^{3}}\right]R_s=0\,.
\end{equation}
%%%%%%%%%%%%%%%%%%%%%%%%%%%%%%%%%%%%%%%%%%%%%%%%%%%%%%%%%%%%%%%%%%%%%%%%%%%%%%
Introducing a dimensionless and re-scaled coordinate $\bar{z}=(r-M)/M$, the above equation becomes
%%%%%%%%%%%%%%%%%%%%%%%%%%%%%%%%%%%%%%%%%%%%%%%%%%%%%%%%%%%%%%%%%%%%%%%%%%%%%%
\begin{equation}
\frac{\mathrm{d}^{2}R_{s}}{\mathrm{d}\bar{z}^{2}}+\left[\frac{2(s+1)}{\bar{z}}+\frac{2im}{\bar{z}^{2}}\right]\frac{\mathrm{d}R_{s}}{\mathrm{d}\bar{z}}+\left[-\frac{\ell(\ell+1)-s(s+1)}{\bar{z}^{2}}+\frac{4ism}{\bar{z}^3}\right]R_{s}=0\,,
\end{equation}
%%%%%%%%%%%%%%%%%%%%%%%%%%%%%%%%%%%%%%%%%%%%%%%%%%%%%%%%%%%%%%%%%%%%%%%%%%%%%%
which has the following solution in terms of confluent hypergeometric functions,
%%%%%%%%%%%%%%%%%%%%%%%%%%%%%%%%%%%%%%%%%%%%%%%%%%%%%%%%%%%%%%%%%%%%%%%%%%%%%%
\begin{equation}
R_{s}(\bar{z})=c_{1}\bar{z}^{-s+\ell}\,\,_{1}F_{1}\left(-\ell-s;-2\ell;\frac{2im}{\bar{z}}\right)+c_{2}\bar{z}^{-\ell-s-1}\,\,_{1}F_{1}\left(\ell-s+1;2\ell+2;\frac{2im}{\bar{z}}\right)\,.
\end{equation}
%%%%%%%%%%%%%%%%%%%%%%%%%%%%%%%%%%%%%%%%%%%%%%%%%%%%%%%%%%%%%%%%%%%%%%%%%%%%%%
The arbitrary constants of integration $c_{1}$ and $c_{2}$ are determined by taking the $\bar{z}\to 0$ limit of the above equation, yielding
%%%%%%%%%%%%%%%%%%%%%%%%%%%%%%%%%%%%%%%%%%%%%%%%%%%%%%%%%%%%%%%%%%%%%%%%%%%%%%
\begin{align}\label{extremal_radial_stat}
R_{s}^{\rm near}&\sim e^{\frac{2im}{\bar{z}}}\left\{c_{2}\left(2im\right)^{-s-\ell-1} \frac{\Gamma(2\ell+2)}{\Gamma\left(-s+\ell+1\right)}+c_{1}\left(2im\right)^{-s+\ell}\frac{\Gamma(-2\ell)}{\Gamma\left(-s-\ell\right)}\right\}
\nonumber
\\
&+\bar{z}^{-2s}\left\{c_{2}(-2im)^{s-\ell-1}\frac{\Gamma(2\ell+2)}{\Gamma\left(s+\ell+1\right)}
+c_{1}(-2im)^{s+\ell} \frac{\Gamma(-2\ell)}{\Gamma\left(s-\ell\right)}\right\}\,,
\end{align}
%%%%%%%%%%%%%%%%%%%%%%%%%%%%%%%%%%%%%%%%%%%%%%%%%%%%%%%%%%%%%%%%%%%%%%%%%%%%%%
where the first part of the above solution, behaving as $\sim e^{1/\bar{z}}$ describes the outgoing mode at the horizon, while the second solution, scaling as $\sim \bar{z}^{-2s}$ is the ingoing mode at the horizon\footnote{The identification of the ingoing/irregular mode is straightforward, as for the radial Teukolsky function in the advanced null coordinates, the ingoing mode behaves as $\Delta^{-s}=(r-M)^{-2s}\sim \bar{z}^{-2s}$. The outgoing mode, on the other hand, scales as $e^{2i\bar{\omega}r_{*}}\sim e^{-2i(m/2M)r_{*}}$. For extremal BH, $r_{*}\sim (-2M/\bar{z})$, and hence, the outgoing mode for the Teukolsky function of an extremal BH scales as $e^{2i\bar{\omega}r_{*}}\sim e^{(2im/\bar{z})}$. These are obtained by taking the $\omega \to 0$ limit of the dynamical boundary conditions. These can also be motivated using the regularity of the radial Teukolsky function in Hartle-Hawking tetrad \citep{LeTiec:2020bos}.} Therefore, the condition that the perturbations must be purely ingoing at the horizon yields 
%%%%%%%%%%%%%%%%%%%%%%%%%%%%%%%%%%%%%%%%%%%%%%%%%%%%%%%%%%%%%%%%%%%%%%%%%%%%%%
\begin{equation}
\frac{c_{2}}{c_{1}}=-(2im)^{2\ell+1}\frac{\Gamma(-2\ell)\Gamma\left(-s+\ell+1\right)}{\Gamma\left(-s-\ell\right)\Gamma (2\ell+2)}\,.
\end{equation}
%%%%%%%%%%%%%%%%%%%%%%%%%%%%%%%%%%%%%%%%%%%%%%%%%%%%%%%%%%%%%%%%%%%%%%%%%%%%%%
To calculate the static tidal response function for an extremal BH, we need to determine the large $r$ limit of the radial Teukolsky function, which from~\ref{extremal_radial_stat} reads
%%%%%%%%%%%%%%%%%%%%%%%%%%%%%%%%%%%%%%%%%%%%%%%%%%%%%%%%%%%%%%%%%%%%%%%%%%%%%%
\begin{equation}\label{eq_bb}
R_{s}^{\rm int}(r)\sim \left[1+\frac{c_{2}}{c_{1}}\,\left(\frac{r}{r_{+}}\right)^{-2\ell-1}\right]\left(\frac{r}{r_{+}}\right)^{\ell-s}~,
\end{equation}
%%%%%%%%%%%%%%%%%%%%%%%%%%%%%%%%%%%%%%%%%%%%%%%%%%%%%%%%%%%%%%%%%%%%%%%%%%%%%%
and hence the ratio $(c_{2}/c_{1})$ describes the static tidal response function associated with the extremal Kerr BH, 
%%%%%%%%%%%%%%%%%%%%%%%%%%%%%%%%%%%%%%%%%%%%%%%%%%%%%%%%%%%%%%%%%%%%%%%%%%%%%%
\begin{align}
\,_{s}\mathcal{F}_{\ell m\,\textrm{(ext)}}^{\rm static}&=-(2im)^{2\ell+1}\frac{\Gamma(-2\ell)\Gamma(\ell-s+1)}{\Gamma(-\ell-s)\Gamma(2\ell+2)}
\nonumber
\\
&=-(2im)^{2\ell+1}\frac{(-1)^{s-\ell}}{2}\frac{\Gamma(\ell-s+1)\Gamma (1+\ell+s)}{\Gamma (2\ell+1)\Gamma(2\ell+2)}\,.
\label{responseext}
\end{align}
%%%%%%%%%%%%%%%%%%%%%%%%%%%%%%%%%%%%%%%%%%%%%%%%%%%%%%%%%%%%%%%%%%%%%%%%%%%%%%
In arriving at the second line, we have used the mirror formula for Gamma functions, as described in~\ref{mirrorgamma} of~\ref{app:identities}, as well as we have taken $(\ell,s)\in \mathbb{Z}$, such that $l\ge |s|$. Again, the static bosonic response function for an extremal BH is purely imaginary and hence the static bosonic LNs vanish, while the dissipation numbers are non-zero. In particular, one can show that the LNs and the dissipation numbers presented in~\ref{static_LNs_ext} and~\ref{dissstatgenext}, respectively, will match with the expression obtained from~\ref{responseext}. Thus, there exist a continuous limit from the non-extremal to the extremal BH. This suggests that there is no need to work out the response function for the extremal case separately, it should come from the $a\to M$ limit for the non-extremal BH. 

We have thus demonstrated the vanishing of static bosonic LNs from the perturbation theory; in the next sections we shall  arrive at the same result from various different perspectives, e.g., symmetry, EFT as well as using scattering amplitudes. 

\paragraph{The large $\ell=m$ limit}~---
As clear from~\ref{responseext}, the dissipation numbers \emph{grow exponentially} in the large $\ell=m$ limit, $\,_{s}\nu^{\rm static}_{\ell \ell\,\textrm{(ext)}}\sim \exp[2\ell (1-\ln 2)]$.
This exponential growth also persists for near-extremal BHs. In the large $\ell=m$ limit, and for $\chi\neq0$, we generically have
\begin{align}
  \,_{s}\nu^{\rm static}_{\ell \ell}\sim 
  2^{-2 \ell-1} \ell (-1)^{1+s} \left(2-\chi ^2+2 \sqrt{1-\chi ^2}\right)^{-\ell} e^{\frac{\ell \chi  \left(\pi -2 \tan ^{-1}\left(\sqrt{1-\chi
   ^2},\chi \right)\right)}{\sqrt{1-\chi ^2}}}
  \,,\qquad \ell=m\gg1
\end{align}
It is easy to see that the bosonic dissipation numbers grow exponentially whenever $\chi\gtrsim 0.94955$. To the best of our knowledge, this critical value was computed here for the first time.

%%%%%%%%%%%%%%%%%%%%%%%%%%%%%%%%%%%%
%%%%%%%%%%%%%%%%%%%%%%%%%%%%%%%%%%%%
%%%%%%%%%%%%%%%%%%%%%%%%%%%%%%%%%%%%
\subsection{Vanishing of static bosonic Love numbers from symmetry} \label{sec:ZeroLoveSymm}

The fact that asymptotically flat BHs in Einstein's theory have vanishing LNs can also be motivated from symmetry arguments. These include --- (a) certain symmetries of the linearized Einstein's equations, leading to ladder structure, conserved quantities and conformal Killing vectors, implying vanishing LNs \citep{Hui:2021vcv,Berens:2022ebl,BenAchour:2022uqo,Sharma:2024hlz}, (b) Symmetry of the near horizon geometry of BH spacetime, described by the $SL(2,\mathbb{R})$ group, the highest weight representation of which are the BHs, leading to vanishing LNs \citep{Charalambous:2021kcz}.  We will discuss the basic features, disparities and open issues associated with these symmetry based approaches. 

In the ladder symmetry approach, one relates the properties of generic $\ell$ modes with those of the $\ell=0$ mode. 
For $\ell=0$, static solutions contain both growing and decaying components at large distances. However, the ladder symmetry implies that the decaying component --- responsible for the tidal response --- must diverge at the horizon, thereby enforcing vanishing LNs for $\ell=0$ and, by symmetry, for all $\ell$. 
 Therefore, the vanishing of the bosonic LNs can be interpreted as arising from the symmetries of the linearized equation and the associated ladder structure. 
 The prohibition of the desired asymptotic fall-off behavior by the symmetry of the linearized Einstein's equations, is also related to the no-hair theorems \citep{Cardoso:2016ryw}, since any static bosonic hair of a four dimensional BH in GR will fall-off asymptotically just as the multipolar deformation. 

On the other hand, one can also argue for the symmetries of the linearized perturbation equation in the near zone region, which under certain approximation can be shown to be equivalent to the $SL(2,\mathbb{R})$ group. Then, one determines a finite dimensional representation of this group, which also has a ladder-like structure. Subsequently, it turns out that the symmetry of the highest weight representation, dubbed as the Love symmetry, forbids the existence of decaying solution for the solutions of linearized perturbation equation for generic $\ell$. In the absence of the decaying part the LNs vanish.  

The key differences between these approaches are as follows: (a) In the ladder symmetry approach the symmetry was for the linearized equation, without any approximations, while the Love symmetry approach works only in the near-zone region and the linearized equation need to be truncated in a somewhat arbitrary fashion. (b) The ladder symmetry approach connects higher $\ell$ modes with the $\ell=0$ mode, and all the analysis are performed for the $\ell=0$ mode alone. While in the Love symmetry approach, one works with non-zero values of $\ell$, starting from the highest weight representation of the Love symmetry group, one arrives at the zero weight representation, and then shows that the above exercise only allows growing mode, with $\ell=0$ playing no significant role here. Except for these two major differences, there are also smaller differences here and there, which we will highlight in the detailed discussion below. 

\paragraph{Ladder structure for scalar perturbations}~--- We start by discussing the ladder structure embedded in the scalar perturbation equation on the Schwarzschild background, using which properties of any $\ell$ mode can be associated with the $\ell=0$ mode. We start with a massless scalar field on the background of a Kerr BH, whose radial part $\phi_{\ell}(r)$, in the static limit and using the ingoing null coordinate system $(v,r,\theta,\widetilde{\phi})$, introduced earlier, satisfies~\ref{TEq} with $s=0$. Written explicitly, the function $\phi_{\ell}$ satisfies the following second order differential equation, 
% \sumantar{One interesting point: Even the ladder symmetry works in the null coordinate and not in `t' coordinate.}
%%%%%%%%%%%%%%%%%%%%%%%%%%%%%%%%%%%%%%%%%%%%%%%%%%%%%%%%%%%%%%%%%%%%%%%%%%%%%%
\begin{equation}
\dfrac{\mathrm{d}^{2}\phi_{\ell}}{\mathrm{d}r^2}+\left(\frac{\Delta'+2iam}{\Delta}\right)\dfrac{\mathrm{d}\phi_{\ell}}{\mathrm{d}r}-\Bigg[\frac{\ell(\ell+1)}{\Delta}\Bigg]\phi_{\ell}=0\,,
\label{TEqladder}
\end{equation}
%%%%%%%%%%%%%%%%%%%%%%%%%%%%%%%%%%%%%%%%%%%%%%%%%%%%%%%%%%%%%%%%%%%%%%%%%%%%%%
where $\Delta=(r-r_{+})(r-r_{-})$, with $r_{\pm}$ being the locations of the event and the Cauchy horizons, respectively, $a$ is the rotation parameter, $\ell$ and $m$ are the angular and azimuthal separation constants. To make the analysis consistent with our previous discussion, we introduce the dimensionless radial coordinate $z=\{(r-r_{+})/(r_{+}-r_{-})\}$, in terms of which $\Delta$ can be expressed as $\Delta=(r_{+}-r_{-})^{2}z(1+z)$. Note that $z=0$ denotes the location of the event horizon, and $z=-1$ corresponds to the Cauchy horizon. Multiplying~\ref{TEqladder} by $\Delta$, we can re-express the above equation as a Hamiltonian constraint: $H_{\ell}\phi_{\ell}=0$, where the Hamiltonian $H_{\ell}$ reads 
%%%%%%%%%%%%%%%%%%%%%%%%%%%%%%%%%%%%%%%%%%%%%%%%%%%%%%%%%%%%%%%%%%%%%%%%%%%%%%
\begin{align}
H_{\ell}&=-z(1+z)\partial_{z}\left[z(1+z)\partial_{z}\right]-\frac{2iam}{(r_{+}-r_{-})}\left[z(1+z) \partial_{z}\right]+\ell(\ell+1)z(1+z)~.
\end{align}
%%%%%%%%%%%%%%%%%%%%%%%%%%%%%%%%%%%%%%%%%%%%%%%%%%%%%%%%%%%%%%%%%%%%%%%%%%%%%%
Intriguingly, the above Hamiltonian can also be expressed in the following manner,
%%%%%%%%%%%%%%%%%%%%%%%%%%%%%%%%%%%%%%%%%%%%%%%%%%%%%%%%%%%%%%%%%%%%%%%%%%%%%%
\begin{align}
H_{\ell}&=D_{\ell+1}^{-}D_{\ell}^{+}-\frac{a^{2}m^{2}}{(r_{+}-r_{-})^{2}}-\frac{(\ell+1)^{2}}{4}~,
\\
&=D_{\ell-1}^{+}D_{\ell}^{-}-\frac{a^{2}m^{2}}{(r_{+}-r_{-})^{2}}-\frac{\ell^{2}}{4}~.
\end{align}
%%%%%%%%%%%%%%%%%%%%%%%%%%%%%%%%%%%%%%%%%%%%%%%%%%%%%%%%%%%%%%%%%%%%%%%%%%%%%%
The above structure is akin to that of harmonic oscillator in quantum mechanics, where the Hamiltonian can be expressed in terms of two sets of conjugate operators, the creation and annihilation operators, which increases or, decreases the energy eigenvalues, when acting on an energy eigenvector. Here, in the context of scalar perturbation, $D_{\ell}^{\pm}$ are playing the role of creation and annihilation operators, and the explicit expression of these two operators associated with the $\ell$ mode are given by, 
%%%%%%%%%%%%%%%%%%%%%%%%%%%%%%%%%%%%%%%%%%%%%%%%%%%%%%%%%%%%%%%%%%%%%%%%%%%%%%
\begin{align}
D_{\ell}^{+}&=-z(1+z)\partial_{z}-\frac{iam}{(r_{+}-r_{-})}-\frac{(\ell+1)}{2}(1+2z)\,,
\\
D_{\ell}^{-}&=z(1+z)\partial_{z}+\frac{iam}{(r_{+}-r_{-})}-\frac{\ell}{2}(1+2z)\,,
\end{align}
%%%%%%%%%%%%%%%%%%%%%%%%%%%%%%%%%%%%%%%%%%%%%%%%%%%%%%%%%%%%%%%%%%%%%%%%%%%%%%
Note that these two operators $D_{\ell}^{\pm}$, along with the Hamiltonain $H_{\ell}$ satisfy the following relations: 
%%%%%%%%%%%%%%%%%%%%%%%%%%%%%%%%%%%%%%%%%%%%%%%%%%%%%%%%%%%%%%%%%%%%%%%%%%%%%%
\begin{align}
H_{\ell+1}D_{\ell}^{+}=D_{\ell}^{+}H_{\ell}~;
\quad
H_{\ell-1}D_{\ell}^{-}=D_{\ell}^{-}H_{\ell}~;
\quad
D_{\ell+1}^{-}D_{\ell}^{+}-D_{\ell-1}^{+}D_{\ell}^{-}=\frac{2\ell+1}{4}~.
\end{align}
%%%%%%%%%%%%%%%%%%%%%%%%%%%%%%%%%%%%%%%%%%%%%%%%%%%%%%%%%%%%%%%%%%%%%%%%%%%%%%
Since $H_{\ell}\phi_{\ell}=0$, then it follows from the first of the above identities that $H_{\ell+1}(D_{\ell}^{+}\phi_{\ell})=0$, and the second identity guarantee that $H_{\ell-1}(D_{\ell}^{-}\phi_{\ell})=0$ as well. This implies, $D_{\ell}^{+}\phi_{\ell}\propto \phi_{\ell+1}$, as well as $D_{\ell}^{-}\phi_{\ell}\propto\phi_{\ell-1}$. This is why, the above structure is referred to as the ladder structure, given the solution $\phi_{\ell}$, we can determine $\phi_{\ell+1}$ by acting $D_{\ell}^{+}$ on it, and obtain $\phi_{\ell-1}$ by acting $D_{\ell}^{-}$. It is to be emphasized that despite a striking similarity with the harmonic oscillator, the ladder in harmonic oscillator transforms one state to another of the \emph{same} Hamiltonian, while here, the ladder operator transforms between states of different Hamiltonian. 

Therefore, given the $\ell=0$ solution for the scalar perturbation equation, namely $\phi_{0}$, one can generate the solution $\phi_{\ell}$, by repeatedly applying the raising operator, such that: $\phi_{\ell}=D_{\ell-1}^{+}D_{\ell-2}^{+}\cdots D_{1}^{+}D_{0}^{+}\phi_{0}$. Thus, if we can determine certain result associated with the $\phi_{0}$, then it can be applied to any $\ell$ modes by the repeated application of the raising operator $D_{\ell}^{+}$. The linearized equation, presented in~\ref{TEqladder}, for the $\ell=0$ mode has the following generic solution,
%%%%%%%%%%%%%%%%%%%%%%%%%%%%%%%%%%%%%%%%%%%%%%%%%%%%%%%%%%%%%%%%%%%%%%%%%%%%%%
\begin{align}\label{gen_zero_sol}
\phi_{0}=c_{1}+c_{2}\exp\left[-\frac{2iam}{r_{+}-r_{-}}\ln\left(\frac{z}{1+z}\right)\right]~,
\end{align}
%%%%%%%%%%%%%%%%%%%%%%%%%%%%%%%%%%%%%%%%%%%%%%%%%%%%%%%%%%%%%%%%%%%%%%%%%%%%%%
where $c_{1}$ and $c_{2}$ are the arbitrary constants of integration. In the case of Schwarzschild BH, the solution for the $\ell=0$ mode is simply, 
%%%%%%%%%%%%%%%%%%%%%%%%%%%%%%%%%%%%%%%%%%%%%%%%%%%%%%%%%%%%%%%%%%%%%%%%%%%%%%
\begin{align}
\phi_{0}^{\rm (Sch)}=c_{1}+c_{2}\ln\left(\frac{z}{1+z}\right)~.
\end{align}
%%%%%%%%%%%%%%%%%%%%%%%%%%%%%%%%%%%%%%%%%%%%%%%%%%%%%%%%%%%%%%%%%%%%%%%%%%%%%%
For both Schwarzschild and Kerr BH, the first solution is regular at the BH horizon, since it is merely a constant, while the second solution is not, due to the existence of the logarithmic behavior. Given the generic solution for the $\ell=0$ mode, and the structure of the ladder operator, at least for the case of Schwarzschild BH, it follows that upon application of $D_{0}^{+}$ on $\phi_{0}^{\rm (Sch)}$, one obtains,
%%%%%%%%%%%%%%%%%%%%%%%%%%%%%%%%%%%%%%%%%%%%%%%%%%%%%%%%%%%%%%%%%%%%%%%%%%%%%%
\begin{align}
\phi_{1}\propto D_{0}^{+}\phi_{0}^{\rm (Sch)}=-\frac{c_{1}(1+2z)}{2}-c_{2}\left[1-\frac{1+2z}{2}\ln\left(1+\frac{1}{z}\right) \right]\simeq -c_{1}r-\frac{c_{2}}{12r^{2}}~,
\end{align}
%%%%%%%%%%%%%%%%%%%%%%%%%%%%%%%%%%%%%%%%%%%%%%%%%%%%%%%%%%%%%%%%%%%%%%%%%%%%%%
which has a growing mode $\sim c_{1}r$ and a decaying mode $c_{2}/r^{2}$, asymptotically. Therefore, it follows that repeated application of $D^{+}$ on the zero mode solution $\phi_{0}=\textrm{constant}$ leads to, at leading order, $\phi_{\ell}\sim r^{\ell}$ at a large distance from the BH, signifying growing solution. While the other independent solution involving logarithmic terms, upon repeated application of $D^{+}$, at leading order, yields, $\phi_{\ell}\sim r^{-\ell-1}$ at a large distance from the BH, and hence corresponds to the decaying part. Thus, it is clear that the regularity at the horizon demands vanishing decaying part for the scalar perturbation at large distance from the BH, signaling vanishing static LNs. This derivation uses \emph{only} the ladder structure of the linearized scalar field equation around a Kerr background. 

However, it turns out that the above ladder structure is not unique \citep{BenAchour:2022uqo}, in particular, it is possible to rewrite~\ref{TEqladder}, the differential equation for $\phi_{\ell}$, as a Sturm-Liouville equation with an appropriate re-scaling: $\psi_{\ell}=\sqrt{z(1+z)}\phi_{\ell}=\{\sqrt{\Delta}/(r_{+}-r_{-})\}\phi_{\ell}$. Such that $\psi_{\ell}$ satisfies the following differential equation in the Boyer-Lindquist coordinate system \citep{BenAchour:2022uqo}, 
%%%%%%%%%%%%%%%%%%%%%%%%%%%%%%%%%%%%%%%%%%%%%%%%%%%%%%%%%%%%%%%%%%%%%%%%%%%%%%
\begin{align}
z(1+z)\left[\dfrac{d^{2}}{dz^{2}}+V_{\ell}(z)\right]\psi_{\ell}\equiv \widetilde{H}_{\ell}\psi_{\ell}=0~,
\quad
V_{\ell}=\frac{(r_{+}-r_{-})^{2}+4a^{2}m^{2}}{4(r_{+}-r_{-})^{2}z^{2}(1+z)^{2}}-\frac{\ell(\ell+1)}{z(1+z)}~.
\end{align}
%%%%%%%%%%%%%%%%%%%%%%%%%%%%%%%%%%%%%%%%%%%%%%%%%%%%%%%%%%%%%%%%%%%%%%%%%%%%%%
Note that only for $a=0$, one can compare the above Hamiltonian $\widetilde{H}_{\ell}$ in the Boyer-Lindquist coordinate with $H_{\ell}$, presented in the ingoing null coordinate. In that case, one obtains the following relations between the two: $\widetilde{H}_{\ell}=-H_{\ell}/\sqrt{z(1+z)}$. Moreover, alike the previous scenario, one can define two alternative ladder operators $\widetilde{D}^{+}_{\ell}$ and $\widetilde{D}^{-}_{\ell}$, associated with the above defined Hamiltonian operator, with the following structure,
%%%%%%%%%%%%%%%%%%%%%%%%%%%%%%%%%%%%%%%%%%%%%%%%%%%%%%%%%%%%%%%%%%%%%%%%%%%%%%
\begin{align}
\widetilde{D}_{\ell}^{+}&=z(1+z)\partial_{z}+\frac{(\ell-1)}{2}(1+2z)\,,
\\
\widetilde{D}_{\ell}^{-}&=z(1+z)\partial_{z}-\frac{(\ell+2)}{2}(1+2z)\,,
\end{align}
%%%%%%%%%%%%%%%%%%%%%%%%%%%%%%%%%%%%%%%%%%%%%%%%%%%%%%%%%%%%%%%%%%%%%%%%%%%%%%
which holds true in the rotating case as well. In the non-rotating case, one can compare these two operators with $D_{\ell}^{\pm}$, defined above, $\widetilde{D}_{\ell}^{\pm}=\mp D_{\ell}^{+}-(1+2z)$. Given the above operators, it immediately follows that, 
%%%%%%%%%%%%%%%%%%%%%%%%%%%%%%%%%%%%%%%%%%%%%%%%%%%%%%%%%%%%%%%%%%%%%%%%%%%%%%
\begin{align}
\widetilde{D}_{\ell}^{+}\widetilde{H}_{\ell-1}&-\widetilde{H}_{\ell}\widetilde{D}_{\ell}^{+}=-(1+2z)\widetilde{H}_{\ell-1}~,
\quad
\widetilde{D}_{\ell}^{-}\widetilde{H}_{\ell+1}-\widetilde{H}_{\ell}\widetilde{D}_{\ell}^{-}=-(1+2z)\widetilde{H}_{\ell+1}~.
\end{align}
%%%%%%%%%%%%%%%%%%%%%%%%%%%%%%%%%%%%%%%%%%%%%%%%%%%%%%%%%%%%%%%%%%%%%%%%%%%%%%
As a consequence, if we act the first identity on $\psi_{\ell-1}$, using $\widetilde{H}_{\ell-1}\psi_{\ell-1}=0$, it follows that: $\widetilde{H}_{\ell}(\widetilde{D}_{\ell}^{+}\psi_{\ell-1})=0$. Hence, $\widetilde{D}_{\ell}^{+}\psi_{\ell-1}\propto \psi_{\ell}$, as required by a raising operator. Similarly, acting the second operator identity on $\psi_{\ell+1}$, and using $\widetilde{H}_{\ell+1}\psi_{\ell+1}=0$, we obtain, $\widetilde{H}_{\ell}(\widetilde{D}_{\ell}^{-}\psi_{\ell+1})=0$, yielding $\widetilde{D}_{\ell}^{-}\psi_{\ell+1}\propto \psi_{\ell}$, typical of a lowering operator. In addition, we have the following two identities, which are useful in certain contexts, 
%%%%%%%%%%%%%%%%%%%%%%%%%%%%%%%%%%%%%%%%%%%%%%%%%%%%%%%%%%%%%%%%%%%%%%%%%%%%%%
\begin{align}
\widetilde{D}_{\ell+1}^{-}\widetilde{D}_{\ell}^{+}&-\widetilde{D}_{\ell-1}^{+}\widetilde{D}_{\ell}^{-}=-\frac{2\ell+1}{4}~,
\\
z(1+z)\widetilde{H}_{\ell}&=\widetilde{D}_{\ell}^{+}\widetilde{D}_{\ell-1}^{-}+\left(\frac{\ell^{2}-1}{4}\right)
+\left\{\frac{1}{4}+\left(\frac{am}{r_{+}-r_{-}}\right)^{2}\right\}
\end{align}
%%%%%%%%%%%%%%%%%%%%%%%%%%%%%%%%%%%%%%%%%%%%%%%%%%%%%%%%%%%%%%%%%%%%%%%%%%%%%%
The morale of the story being, there are ambiguities in the structure of the ladder operator, but the key idea remains the same, generic $\ell$ modes can be obtained by repeated application of the raising operator on the $\ell=0$ mode. In the rescaled ladder structure as well, we have the $\ell=0$ mode to be, $\psi_{0}=\sqrt{z(1+z)}\phi_{0}$, where $\phi_{0}$ has been presented in~\ref{gen_zero_sol}. Hence the logarithmic divergence at the horizon, forces the decaying part to be absent, which in turn implies vanishing tidal LN. Therefore, from purely mathematical ground, the logarithmic behavior is responsible for vanishing LN, however in what follows we will try to provide a more robust symmetry-based argument for vanishing of LNs.  

It is worth pointing out that the ladder symmetry, as described above, neither exists for generic static spherically symmetric spacetimes, nor for arbitrary stationary BH solutions in four dimensions. Considering scalar perturbations in a generic static and spherically symmetric spacetimes, the ladder symmetry between different $\ell$ modes of scalar perturbation will exist if --- (a) $-g_{tt}=g^{rr}$, and (b) $g_{tt}g_{\theta \theta}=-(r^{2}+ar+b)$, where $a$ and $b$ are $\ell$ independent constants \citep{Sharma:2024hlz}. In the stationary case, the situation is even more complicated, as the radial and angular parts of the Klein--Gordon equation for scalar perturbation is in general not separable. Thus, focusing on the Konoplya-Rezzolla-Zhidenko class of stationary solutions \citep{Konoplya:2016jvv, Konoplya:2018arm}~--- which allow for separability between radial and angular parts and describe the most general stationary, axisymmetric, asymptotically flat BH metric --- ladder symmetry between different $\ell$ modes will exist, provided the location of the horizon comes from a quadratic equation \citep{Sharma:2024hlz}. This once again highlights the fact that there is a close connection between the existence of a horizon and ladder symmetry. However, to comment on the vanishing of the LNs, existence of ladder symmetry is not enough. One must solve the $\ell=0$ mode and show that its ill-behaved part at the horizon is in one-to-one correspondence with the decaying mode, asymptotically. As we shall discuss in~\ref{sec:fermionic}, this does not occur for fermionic perturbations, which indeed yield nonzero LNs. The above analysis was for four dimensional BHs, while in higher dimensional case the situation changes, leading to non-zero LN even when ladder exists \citep{DeLuca:2024nih, Berens:2025jfs}.

\paragraph{Symmetries of linearized equations and conserved charges}~--- 
The above arguments, leading to the vanishing of the static LNs, associated with scalar perturbation of BHs, can be made more robust through the symmetry associated with the $\ell=0$ scalar perturbation equation. The existence of such a symmetry follows from the result that, if $\phi_{0}$ is a solution of the $H_{0}\phi_{0}=0$ equation, then 
%%%%%%%%%%%%%%%%%%%%%%%%%%%%%%%%%%%%%%%%%%%%%%%%%%%%%%%%%%%%%%%%%%%%%%%%%%%%%%
\begin{align}
\phi_{0}\to \phi_{0}+\delta_{0} \phi_{0}~;
\qquad 
\delta_{0}\equiv z(1+z)\partial_{z}~,
\end{align}
%%%%%%%%%%%%%%%%%%%%%%%%%%%%%%%%%%%%%%%%%%%%%%%%%%%%%%%%%%%%%%%%%%%%%%%%%%%%%%
is also a solution. This is what corresponds to the symmetry of the linearized equation associated with the $\ell=0$ mode of the scalar field --- under the action of $\delta_{0}$, one on-shell solution maps to another. This in turn implies that the operator $\delta_{0}$ must satisfy the following condition: $[\delta_{0},H_{0}]=0$. Taking a cue from the harmonic oscillator problem, one can determine such a symmetry operator associated with a generic $\ell$ mode, starting from $\delta_{0}$, as: $\delta_{\ell}=D^{+}_{\ell-1}\delta_{\ell-1}D^{-}_{\ell}$, satisfying $[\delta_{\ell},H_{\ell}]=0$. 
The existence of such symmetry operators and their role in constraining the space of regular solutions of the Teukolsky equation were clarified and systematized in \citet{Charalambous:2021kcz,Charalambous:2021mea}.
This symmetry implies that if $\phi_{\ell}$ is a solution of $H_{\ell}\phi_{\ell}=0$, then $\phi_{\ell}+\delta_{\ell}\phi_{\ell}$ is also a solution. Thus, among the solution space of the differential operator $H_{\ell}$, the symmetry $\delta_{\ell}$ maps one solution to another. Moreover, the above must also be a symmetry of the associated Lagrangian, and hence by the Noether's theorem, the Noether charge associated with this symmetry must be conserved. Given the Hamiltonian $H_{0}$ for the $\ell=0$ mode, the associated conserved Noether charge can be seen to be 
%%%%%%%%%%%%%%%%%%%%%%%%%%%%%%%%%%%%%%%%%%%%%%%%%%%%%%%%%%%%%%%%%%%%%%%%%%%%%%
\begin{equation}\label{charge1}
\mathcal{Q}_{0}=\left[z(1+z)\partial_{z}+\frac{2iam}{r_{+}-r_{-}}\right]\phi_{0}
\end{equation}
%%%%%%%%%%%%%%%%%%%%%%%%%%%%%%%%%%%%%%%%%%%%%%%%%%%%%%%%%%%%%%%%%%%%%%%%%%%%%%
Note that the equation $H_{0}\phi_{0}=0$ implies $\partial_{z}\mathcal{Q}_{0}=0$, so that $\mathcal{Q}_{0}$ must be conserved. 
However, this charge is not uniquely defined: one may freely add a constant term, or any function depending only on the angular coordinate, and the resulting expression would still be conserved. 
The charges associated with higher multipoles can be obtained by observing that $\phi_{0}$ can be generated from $\phi_{\ell}$ through successive applications of the lowering operator. 
Accordingly, for the growing branch of the solution ($\phi_{\ell}\propto r^{\ell}$), the charge $\mathcal{Q}_{\ell}$ is found to be proportional to the rotation parameter $a$ and to be purely imaginary:
%
%%%%%%%%%%%%%%%%%%%%%%%%%%%%%%%%%%%%%%%%%%%%%%%%%%%%%%%%%%%%%%%%%%%%%%%%%%%%%%
\begin{equation}\label{charge2}
\mathcal{Q}^{\rm growing}_{\ell}=\left(\frac{2iam}{r_{+}-r_{-}}\right)\prod_{j=1}^{\ell}\left[j^{2}+\left(\frac{2am}{r_{+}-r_{-}}\right)^{2}\right]~,
\end{equation}
%%%%%%%%%%%%%%%%%%%%%%%%%%%%%%%%%%%%%%%%%%%%%%%%%%%%%%%%%%%%%%%%%%%%%%%%%%%%%%
which vanishes for the Schwarzschild BH. The conserved charge associated with the decaying branch ($\phi_{\ell}\propto r^{-\ell-1}$), on the other hand, is non-zero for both Kerr and for Schwarzschild BH. 

At a first glance, it is clear that in the extremal limit ($a\to M$, leading to $r_{+}\to r_{-}$), both the conserved quantities in~\ref{charge1} and~\ref{charge2} diverge. This problem can be fixed by rescaling these conserved charges by $\{(r_{+}-r_{-})/r_{+}\}^{1+2\ell}$. Since this additional multiplicative factor is a constant, the rescaled charges remain conserved and are finite in the extremal limit.

For a Schwarzschild BH, the two branches of the solution $\phi_{\ell}$ are distinct. 
The growing branch, with asymptotic behavior $\phi_{\ell}\sim r^{\ell}$, carries a vanishing conserved charge. 
Since the charge is conserved (i.e., independent of the radial coordinate), it must connect to a solution with $\mathcal{Q}_{\ell}=0$ near the horizon as well. 
This corresponds to the constant solution, which is regular at the horizon. 
In contrast, the decaying branch, which behaves as $\phi_{\ell}\sim r^{-\ell-1}$ at large distances, carries a nonzero conserved charge and necessarily involves logarithmic terms, since the conserved charge must match onto the near-horizon solution behaving as $\ln\!\{z/(z+1)\}$. 
Thus, regularity at the horizon requires the conserved charge associated with the decaying branch to vanish, which in turn forces the decaying solution itself to vanish, leading to vanishing LNs.

For a Kerr BH, both the growing and the decaying modes yield nonzero conserved charges asymptotically. 
Near the horizon, the solution $\phi_{0}=\text{constant}$ has a nonzero conserved charge, while the other solution,
\[
\phi_{0}\propto\exp\!\left[-\frac{2iam}{r_{+}-r_{-}}\ln\!\left(\frac{z}{1+z}\right)\right],
\]
has a vanishing conserved charge. 
Therefore, the argument used for the Schwarzschild BH does not directly apply to Kerr, since both the growing and decaying modes contain a constant piece near the horizon, leaving the question of which mode carries the logarithmic contribution. 
To resolve this, one can employ another conserved quantity, namely the Wronskian of the two solutions. 
By construction of the linearized equations, the Wronskian is conserved across the radial direction and can thus be used for this purpose. 
Given the asymptotic structure of the solutions, the Wronskian is nonzero at infinity due to the decaying branch and nonzero near the horizon due to the logarithmic behavior \citep{Charalambous:2021mea}.
This links the two solutions, and the requirement of regularity of the Wronskian at the horizon again forces the decaying branch to vanish, leading to vanishing LNs.

We now provide further insight into the symmetry group responsible for the vanishing of the LNs in the Schwarzschild and Kerr BH cases. 
For this purpose, the structure of the Hamiltonian $\widetilde{H}_{\ell}$ proves most useful, since the techniques of Sturm-Liouville systems can be applied. 
As already noted, if $\psi_{1\ell}$ and $\psi_{2\ell}$ are two independent solutions of $\widetilde{H}_{\ell}\psi_{\ell}=0$, then the Wronskian, 
\[
W_{\ell}\equiv \psi_{1\ell}\,\partial_{z}\psi_{2\ell}-\psi_{2\ell}\,\partial_{z}\psi_{1\ell},
\]
is conserved. 
In addition, two further conserved charges can be defined, arising from a generic solution $\psi_{\ell}$, namely
%%%%%%%%%%%%%%%%%%%%%%%%%%%%%%%%%%%%%%%%%%%%%%%%%%%%%%%%%%%%%%%%%%%%%%%%%%%%%%
\begin{equation}
W_{1\ell}=\psi_{1\ell}\partial_{z}\psi_{\ell}-\psi_{\ell}\partial_{z}\psi_{1\ell}~,
\quad
W_{2\ell}=\psi_{2\ell}\partial_{z}\psi_{\ell}-\psi_{\ell}\partial_{z}\psi_{2\ell}~.
\end{equation}
%%%%%%%%%%%%%%%%%%%%%%%%%%%%%%%%%%%%%%%%%%%%%%%%%%%%%%%%%%%%%%%%%%%%%%%%%%%%%%
The above construction can be further extended to obtain an infinite class of conserved charges, e.g., any polynomial involving $W_{1\ell}^{n_{1}}W_{2\ell}^{n_{2}}$, for arbitrary $n_{1}$ and $n_{2}$ is also conserved. There exist a subclass of these infinite conserved charges that we will be interested in, these are, 
%%%%%%%%%%%%%%%%%%%%%%%%%%%%%%%%%%%%%%%%%%%%%%%%%%%%%%%%%%%%%%%%%%%%%%%%%%%%%%
\begin{align}
Y_{+\ell}&=W_{1\ell}~,\qquad Y_{-\ell}=W_{2\ell}~,
\\
Q_{+\ell}&=\frac{W_{1\ell}^{2}}{2}~,\qquad Q_{0\ell}=\frac{W_{1\ell}W_{2\ell}}{2}~,\qquad Q_{-\ell}=\frac{W_{2\ell}^{2}}{2}~,
\end{align}
%%%%%%%%%%%%%%%%%%%%%%%%%%%%%%%%%%%%%%%%%%%%%%%%%%%%%%%%%%%%%%%%%%%%%%%%%%%%%%
with the following non-zero Poisson bracket relations: 
%%%%%%%%%%%%%%%%%%%%%%%%%%%%%%%%%%%%%%%%%%%%%%%%%%%%%%%%%%%%%%%%%%%%%%%%%%%%%%
\begin{align}
\{Q_{+\ell},Q_{-\ell}\}&=2W_{\ell}Q_{0\ell}~,\quad \{Q_{0\ell},Q_{+\ell}\}=-W_{\ell}Q_{+\ell}~,\quad 
\{Q_{0\ell},Q_{-\ell}\}=W_{\ell}Q_{-\ell}~,
\label{Qchargepoisson}
\\
\{Q_{0\ell},Y_{\pm\ell}\}&=\mp\frac{W_{\ell}}{2}Y_{\pm\ell}~,\quad \{Q_{+\ell},Y_{-\ell}\}=W_{\ell}Y_{+\ell}~,
\quad \{Q_{-\ell},Y_{+\ell}\}=-W_{\ell}Y_{-\ell}~,
\\
\{Y_{+\ell},Y_{-\ell}\}&=W_{\ell}~.
\label{Ychargepoisson}
\end{align}
%%%%%%%%%%%%%%%%%%%%%%%%%%%%%%%%%%%%%%%%%%%%%%%%%%%%%%%%%%%%%%%%%%%%%%%%%%%%%%
As is evident, the above group generators are finite in number and close among themselves. 
Among the Poisson bracket relations, the one shown in~\ref{Qchargepoisson} corresponds to the algebra satisfied by the $SL(2,\mathbb{R})$ symmetry generators, while~\ref{Ychargepoisson} forms the Heisenberg algebra $\mathbb{H}$. 
Therefore, the full Poisson bracket structure defines the Schr\"{o}dinger algebra 
\[
\mathfrak{sh}(1) = SL(2,\mathbb{R}) \ltimes \mathbb{H},
\] 
with the Wronskian $W_{\ell}$ acting as the central charge. 
Consequently, the general symmetry of any Sturm-Liouville problem, including that of a scalar field on a Kerr BH background, is the finite-dimensional Schr\"{o}dinger group. 
The charges $Y_{\pm \ell}$ are generated by Galilean transformations:

%%%%%%%%%%%%%%%%%%%%%%%%%%%%%%%%%%%%%%%%%%%%%%%%%%%%%%%%%%%%%%%%%%%%%%%%%%%%%%
\begin{align}\label{Galilean}
z\rightarrow z~,\qquad \psi_{\ell}\to \psi_{\ell}+\eta_{+}\psi_{1\ell}+\eta_{-}\psi_{2\ell}~,
\end{align}
%%%%%%%%%%%%%%%%%%%%%%%%%%%%%%%%%%%%%%%%%%%%%%%%%%%%%%%%%%%%%%%%%%%%%%%%%%%%%%
while the charges $Q_{\pm \ell}$ and $Q_{0\ell}$ are generated by the following conformal transformation:
%%%%%%%%%%%%%%%%%%%%%%%%%%%%%%%%%%%%%%%%%%%%%%%%%%%%%%%%%%%%%%%%%%%%%%%%%%%%%%
\begin{align}\label{conformal}
z\rightarrow z+\alpha_{+}\psi_{1\ell}^{2}+\alpha_{-}\psi_{2\ell}^{2}+\alpha_{0}\psi_{1\ell}\psi_{2\ell}
\end{align}
%%%%%%%%%%%%%%%%%%%%%%%%%%%%%%%%%%%%%%%%%%%%%%%%%%%%%%%%%%%%%%%%%%%%%%%%%%%%%%
All of these results hold for any Sturm-Liouville problem, including that of a scalar field on a Schwarzschild or Kerr BH background. The interpretation of these symmetry structures in terms of near-horizon conformal dynamics and their connection to conserved charges and response functions has been further developed in \citet{Lupsasca:2025pnt}. 
Intriguingly, the $SL(2,\mathbb{R})$ symmetry can also be understood from the fact that the ladder operators correspond to generalized Darboux transformations --~similar to those used in the study of QNMs~-- which arise generically in separable, static, and spherically symmetric problems~\citep{DeLuca:2025zqr}.

To demonstrate the vanishing of the static LNs for Schwarzschild and Kerr BHs, additional input is required, in particular the behavior of the symmetry generators at the BH horizon. We present below the case of a Schwarzschild BH, which generalizes straightforwardly to Kerr \citep{BenAchour:2022uqo}. 
The two independent solutions of the Klein--Gordon equation for the $\ell=0$ mode in the Schwarzschild background are
\[
\psi_{1} = \sqrt{z(z+1)}, \qquad 
\psi_{2} = \sqrt{z(z+1)} \ln\!\left(\frac{z}{1+z}\right).
\]
Given these solutions, and noting that both $\psi_{2}$ and its derivative diverge at the horizon, it follows that the change in the action under the transformations in~\ref{Galilean} and~\ref{conformal} remains finite if and only if
\[
\alpha_{-}=0 = \alpha_{0} = \eta_{-}.
\]
Thus, regularity at the horizon requires that only the symmetry generators associated with $Q_{+}$ and $Y_{+}$ are nonzero, which preserve the asymptotically growing modes. 
Since the solutions must respect the Schr\"{o}dinger symmetry, the decaying modes are necessarily absent, leading to vanishing static LNs in the scalar sector.

For completeness, let us point out the notion of mass ladder, introduced in \citet{Cardoso:2017qmj}. This requires the existence of conformal Killing vector fields in the spacetime, which must be the eigenvectors of the Ricci tensor with constant eigenvalues. Intriguingly, this property is also connected to vanishing LNs of four dimensional BHs in GR \citep{Hui:2021vcv, DeLuca:2025zqr}.

\paragraph{Ladder symmetry for generic spin}~---
The ladder symmetry extends beyond the scalar case and persists for the static Teukolsky equation describing generic spin-$s$ bosonic perturbations of a Kerr BH. Starting from the general expression in~\ref{TEq}, the radial equation can be written as
%%%%%%%%%%%%%%%%%%%%%%%%%%%%%%%%%%%%%%%%%%%%%%%%%%%%%%%%%%%%%%%%%%%%%%%%%%%%%%
\begin{equation}
\dfrac{d}{dr}\left[\Delta^{s+1}\dfrac{\mathrm{d}R^{(s)}_{\ell m}}{\mathrm{d}r}\right]
+2iam \Delta^{s}\dfrac{\mathrm{d}R^{(s)}_{\ell m}}{\mathrm{d}r}
+\Bigg[am\left(\frac{2is\Delta'}{\Delta}\right)-\lambda\Bigg]\Delta^{s}R^{(s)}_{\ell m}=0\,.
\label{TEqnewspin1}
\end{equation}
%%%%%%%%%%%%%%%%%%%%%%%%%%%%%%%%%%%%%%%%%%%%%%%%%%%%%%%%%%%%%%%%%%%%%%%%%%%%%%
Introducing the rescaled field $\Psi^{(s)}_{\ell m}=\Delta^{s}R^{(s)}_{\ell m}$, one obtains
%%%%%%%%%%%%%%%%%%%%%%%%%%%%%%%%%%%%%%%%%%%%%%%%%%%%%%%%%%%%%%%%%%%%%%%%%%%%%%
\begin{equation}
\dfrac{d}{dr}\left[\Delta\dfrac{d\Psi^{(s)}_{\ell m}}{dr}\right]
+\left(2iam-s\Delta'\right)\dfrac{d\Psi^{(s)}_{\ell m}}{dr}
-(\ell+s)(\ell-s+1)\Psi^{(s)}_{\ell m}=0\,,
\label{TEqnewspin2}
\end{equation}
%%%%%%%%%%%%%%%%%%%%%%%%%%%%%%%%%%%%%%%%%%%%%%%%%%%%%%%%%%%%%%%%%%%%%%%%%%%%%%
where we used the Kerr identity
$-s\Delta''-\lambda=-(\ell+s)(\ell-s+1)$.
This equation can be written as a Hamiltonian constraint,
%%%%%%%%%%%%%%%%%%%%%%%%%%%%%%%%%%%%%%%%%%%%%%%%%%%%%%%%%%%%%%%%%%%%%%%%%%%%%%
\begin{equation}
H_{\ell}^{(s)}\Psi^{(s)}_{\ell m}=0~,
\qquad
H_{\ell}^{(s)}=
\dfrac{d}{dr}\!\left[\Delta\dfrac{d}{dr}\right]
+\left(2iam-s\Delta'\right)\dfrac{d}{dr}
-(\ell+s)(\ell-s+1)\,.
\label{TEqnewhamiltonspin}
\end{equation}
%%%%%%%%%%%%%%%%%%%%%%%%%%%%%%%%%%%%%%%%%%%%%%%%%%%%%%%%%%%%%%%%%%%%%%%%%%%%%%
In complete analogy with the scalar case ($s=0$), where the ladder operators
$D_\ell^\pm$ raise and lower the multipolar index $\ell$, the spin-$s$ Hamiltonian admits a factorized structure
%%%%%%%%%%%%%%%%%%%%%%%%%%%%%%%%%%%%%%%%%%%%%%%%%%%%%%%%%%%%%%%%%%%%%%%%%%%%%%
\begin{equation}
H_{\ell}^{(s)}
=E^{-}E_{s}^{+}-\ell(\ell+1)+s(s+1)
=E^{+}_{s-1}E^{-}-(\ell+s)(\ell-s+1)\,,
\end{equation}
%%%%%%%%%%%%%%%%%%%%%%%%%%%%%%%%%%%%%%%%%%%%%%%%%%%%%%%%%%%%%%%%%%%%%%%%%%%%%%
with
%%%%%%%%%%%%%%%%%%%%%%%%%%%%%%%%%%%%%%%%%%%%%%%%%%%%%%%%%%%%%%%%%%%%%%%%%%%%%%
\begin{equation}
E^{-}\equiv \partial_{r}~,
\qquad
E^{+}_{s}\equiv \Delta \partial_{r}+(2iam-s\Delta')\,.
\end{equation}
%%%%%%%%%%%%%%%%%%%%%%%%%%%%%%%%%%%%%%%%%%%%%%%%%%%%%%%%%%%%%%%%%%%%%%%%%%%%%%
These operators act as spin-lowering and spin-raising operators, respectively, and are closely related to generalized Teukolsky-Starobinsky identities and the hidden symmetry structure of Kerr perturbations \citep{Charalambous:2021kcz,Charalambous:2021mea}.

The operators satisfy
%%%%%%%%%%%%%%%%%%%%%%%%%%%%%%%%%%%%%%%%%%%%%%%%%%%%%%%%%%%%%%%%%%%%%%%%%%%%%%
\begin{equation}
E^{-}E^{+}_{s}-E^{+}_{s-1}E^{-}=-2s\,,
\end{equation}
%%%%%%%%%%%%%%%%%%%%%%%%%%%%%%%%%%%%%%%%%%%%%%%%%%%%%%%%%%%%%%%%%%%%%%%%%%%%%%
as well as the intertwining relations
%%%%%%%%%%%%%%%%%%%%%%%%%%%%%%%%%%%%%%%%%%%%%%%%%%%%%%%%%%%%%%%%%%%%%%%%%%%%%%
\begin{equation}
E^{+}_{s}H_{\ell}^{(s)}=H_{\ell}^{(s+1)}E^{+}_{s}~,
\qquad 
E^{-}H_{\ell}^{(s)}=H_{\ell}^{(s-1)}E^{-}\,.
\end{equation}
%%%%%%%%%%%%%%%%%%%%%%%%%%%%%%%%%%%%%%%%%%%%%%%%%%%%%%%%%%%%%%%%%%%%%%%%%%%%%%
Hence, if $\Psi^{(s)}_{\ell m}$ solves $H_{\ell}^{(s)}\Psi^{(s)}_{\ell m}=0$, then
$E^{+}_{s}\Psi^{(s)}_{\ell m}$ automatically solves the spin-$(s+1)$ equation. 
In this sense, $E^{+}_{s}$ and $E^{-}$ generate a ladder structure in spin space, reflecting the hidden symmetry of the Teukolsky system \citep{Charalambous:2021mea}.

As a consequence, the radial mode function for arbitrary $(\ell,s)$ can be generated from the scalar $\ell=0$ solution as
%%%%%%%%%%%%%%%%%%%%%%%%%%%%%%%%%%%%%%%%%%%%%%%%%%%%%%%%%%%%%%%%%%%%%%%%%%%%%%
\begin{equation}
\Psi^{(s)}_{\ell}
=\underbrace{E^{+}_{s-1}\cdots E^{+}_{0}}_{\rm spin\ raising}
\underbrace{D^{+}_{\ell-1}\cdots D^{+}_{0}}_{\rm angular\ raising}
\phi_{0}\,,
\end{equation}
%%%%%%%%%%%%%%%%%%%%%%%%%%%%%%%%%%%%%%%%%%%%%%%%%%%%%%%%%%%%%%%%%%%%%%%%%%%%%%
with the inverse relation obtained by successive lowering operations. 

This construction makes transparent the radial scaling of the two independent branches. 
Starting from the scalar solution~\ref{gen_zero_sol}, the constant piece generates a growing mode scaling as $r^{\ell+s}$, whereas the logarithmic piece generates a decaying mode scaling as $r^{-\ell+s-1}$. 
Importantly, the relative scaling between growing and decaying solutions remains $r^{2\ell+1}$, independently of the spin. 
For $\ell\geq |s|$, the logarithmic branch continues to correspond to the decaying mode. 

Therefore, imposing regularity at the horizon eliminates the logarithmic branch for any integer spin. The decaying mode is thus absent for all bosonic perturbations, implying the vanishing of static bosonic LNs independently of $s$.

More recently, these ladder symmetries and their associated conserved charges have been embedded into a broader near-horizon and conformal framework, providing a unified interpretation of static response, scattering amplitudes, and conserved fluxes \citep{Lupsasca:2025pnt}.

\paragraph{Near-zone symmetry of the Teukolsky equation}~---  
The vanishing of the bosonic LNs for BHs in GR has also been attributed to a hidden near-zone symmetry of the Teukolsky equation \citep{Charalambous:2021kcz}. We briefly outline this construction for Kerr BHs and then show how the static LNs of Schwarzschild BHs vanish identically, a result that extends straightforwardly to the Kerr case. We finally comment on some ambiguities inherent in this approach.

The starting point is the observation that, in the Kerr background, one can introduce the following vector fields \citep{Charalambous:2021kcz}
\begin{align}
L_{0}&=-\beta\partial_{t}~;\qquad 
\beta\equiv 2r_{+}\left(\frac{r_{+}+r_{-}}{r_{+}-r_{-}}\right)~,
\\
L_{\pm1}&=\exp\left(\pm t/\beta\right)\left[\mp \sqrt{\Delta}\partial_{r}
+\frac{\beta \Delta'}{2\sqrt{\Delta}}\partial_{t}
+\frac{a}{\sqrt{\Delta}}\partial_{\phi}\right]~,
\end{align}
whose commutators satisfy the $SL(2,\mathbb{R})$ algebra,
\begin{align}
[L_{m},L_{n}]=(m-n)L_{m+n}~,\qquad m,n=0,\pm1~.
\end{align}
These vector fields are regular at the horizon when expressed in appropriate coordinates: while $L_{\pm1}$ are not well defined at the horizon in Boyer--Lindquist coordinates, $L_{1}$ is regular in advanced coordinates $(v,r,\theta,\tilde\phi)$ and $L_{-1}$ in retarded coordinates $(u,r,\theta,\varphi)$. This near-zone symmetry is distinct from the conformal symmetry underlying the Kerr/CFT correspondence \citep{Guica:2008mu,Bredberg:2009pv}. Since the Kerr/CFT correspondence has no smooth Schwarzschild limit \citep{Charalambous:2021kcz}, but the above Love symmetry has a well defined Schwarzschild limit \citep{Bertini:2011ga}. 

The relevance of this algebra to BH perturbation theory follows from the quadratic Casimir operator,
\begin{align}
\mathcal{C}\equiv 
L_{0}^{2}-\frac{1}{2}\left(L_{-1}L_{1}+L_{1}L_{-1}\right)~,
\end{align}
which, upon substituting the above expressions for $L_{0}$ and $L_{\pm1}$, yields in a Kerr background
\begin{align}\label{CasimirSL2R}
\mathcal{C}=\partial_{r}\left(\Delta \partial_{r}\right)
+\frac{(2Mr_{+})^{2}}{\Delta}
\left[
\left(\omega-m\Omega_{\rm H}\right)^{2}
-4m\omega \Omega_{\rm H}
\left(\frac{r-r_{+}}{r_{+}-r_{-}}\right)
\right]~.
\end{align}

The usefulness of the Casimir operator lies in its connection to the perturbation equation of a test scalar field on the Kerr background. A scalar field decomposed as in~\ref{gen_eq_decomp} has a radial and an angular part, with the radial equation (the Teukolsky equation with $s=0$) in Boyer--Lindquist coordinates given by
\begin{align}\label{diffeqBL}
\dfrac{d}{dr}\left(\Delta \dfrac{dR_{0}}{dr}\right)
&+\frac{(2Mr_{+})^{2}}{\Delta}
\left[
\left(\omega-m\Omega_{\rm H}\right)^{2}
-4m\omega \Omega_{\rm H}
\left(\frac{r-r_{+}}{r_{+}-r_{-}}\right)
\right]R_{0}
\nonumber
\\
&+\epsilon 
\left[
\frac{2M(\beta am\omega+4M^{2}\omega^{2}r_{+})}{r_{+}(r-r_{-})}
+\omega^{2}\left(r^{2}+2Mr+4M^{2}\right)
\right]R_{0}
=E_{\ell}R_{0}~,
\end{align}
where $\Omega_{\rm H}=a/(2Mr_{+})$ and $E_{\ell}$ are the angular eigenvalues. For $a=0$, $E_{\ell}=\ell(\ell+1)$. The parameter $\epsilon$ is a bookkeeping device: $\epsilon=1$ for the full Kerr problem, while the near-zone regime ($M\omega\ll1$ and $\omega r\ll1$) is captured by formally taking $\epsilon\ll1$.

The crucial observation is that the first line of~\ref{diffeqBL} coincides with the Casimir operator of the $SL(2,\mathbb{R})$ algebra (more precisely, $SL(2,\mathbb{R})\times U(1)$, where $U(1)$ arises from axial symmetry). For static perturbations in the Schwarzschild background we have $\epsilon=0$, and the modes $\Phi_{\ell}=e^{-i\omega t}R_{0}(r)e^{im\phi}$ satisfy
\begin{align}
\mathcal{C}\Phi_{\ell}=\ell(\ell+1)\Phi_{\ell}~,
\qquad
L_{0}\Phi_{\ell}=i\beta\omega\,\Phi_{\ell}=h\Phi_{\ell}~,
\end{align}
realizing an infinite-dimensional representation of $SL(2,\mathbb{R})$ ($h$ is the eigenvalue of the $L_{0}$ operator with eigenvector $\Phi_{\ell}$).

One may instead consider finite-dimensional, non-unitary representations with bounded eigenvalues $|h|\leq \ell$, giving a $(2\ell+1)$-dimensional space. States can be labeled as $\Phi_{\ell_{1};\ell_{2}}$, satisfying
\begin{align}
\mathcal{C}\Phi_{\ell_{1};\ell_{2}}
&=\ell_{1}(\ell_{1}+1)\Phi_{\ell_{1};\ell_{2}}~,
\\
L_{0}\Phi_{\ell_{1};\ell_{2}}
&=\ell_{2}\Phi_{\ell_{1};\ell_{2}}~,
\qquad
-\ell_{1}\leq \ell_{2}\leq \ell_{1}~,
\end{align}
with $L_{\pm1}$ acting as lowering/raising operators. In particular,
\begin{align}
L_{1}\Phi_{\ell_{1};-\ell_{1}}=0~,
\qquad
L_{-1}\Phi_{\ell_{1};\ell_{1}}=0~.
\end{align}

To show the vanishing of static bosonic LNs for Schwarzschild BHs, consider the state $\Phi_{\ell_{1};0}$, which satisfies $L_{0}\Phi_{\ell_{1};0}=0$. Since $L_{0}$ generates time translations, this state is static and depends only on $r$ (setting $m=0$). Using $L_{1}^{\ell_{1}+1}\Phi_{\ell_{1};0}=0$, one finds at leading order $\partial_{r}^{\ell_{1}+1}\Phi_{\ell_{1};0}=0$, implying $\Phi_{\ell_{1};0}\sim r^{\ell_{1}}$. Hence the near-zone $SL(2,\mathbb{R})$ symmetry enforces a purely growing solution of the radial equation, with no decaying branch, which implies vanishing static bosonic LNs.

This argument extends to bosonic spin-$s$ perturbations by defining
\begin{align}
L_{0}^{(s)}&=L_{0}+s~,
\qquad
L_{\pm1}^{(s)}=L_{\pm1}
-s\,e^{\pm t/\beta}(1\pm1)\left(\partial_{r}\sqrt{\Delta}\right)~,
\end{align}
which satisfy the same $SL(2,\mathbb{R})$ algebra. The associated Casimir reads
\begin{align}
\mathcal{C}^{(s)}\Psi^{(s)}=
\Big[
\mathcal{C}
+s(\partial_{r}\Delta)\partial_{r}
+\frac{2Mr_{+}s(r_{+}-r_{-})}{\Delta}\partial_{t}
+\frac{2(r-M)as}{\Delta}\partial_{\phi}
+s^{2}+s
\Big]\Psi^{(s)}~,
\end{align}
which coincides with the Teukolsky operator in the near-zone limit. Therefore, the same symmetry argument implies that static LNs vanish identically for generic integer spin.

There are, however, ambiguities in this construction. The separation in~\ref{diffeqBL} through the parameter $\epsilon$ is not unique. Terms of ${\cal O}(a\omega)$ appear both in the $\epsilon$-independent and $\epsilon$-dependent pieces, and can in principle be reshuffled in infinitely many ways, only one of which reproduces the Casimir structure. Moreover, terms of ${\cal O}(M^{2}\omega^{2})$ are retained in the first line of~\ref{diffeqBL}, while terms of ${\cal O}(\epsilon a\omega)$ are neglected in the second, despite being parametrically comparable. This inconsistency clouds the robustness of the argument. A related symmetry structure, including a Heisenberg extension, arises in generic Sturm--Liouville systems and underlies the Love symmetry discussed previously. 

For recent developments unifying these approaches and clarifying their implications for the EFT description, see \citep{Berens:2025okm,Parra-Martinez:2025bcu}. 
In particular, \citep{Parra-Martinez:2025bcu} recently
found an accidental symmetry of the full vacuum Einstein's equations in the static case, which enforces the vanishing of the static LNs for spherically symmetric BHs at full nonlinear order in four-dimensional GR. 

%%%%%%%%%%%%%%%%%%%%%%%%%%%%%%%%%%%%
%%%%%%%%%%%%%%%%%%%%%%%%%%%%%%%%%%%%
%%%%%%%%%%%%%%%%%%%%%%%%%%%%%%%%%%%%
\subsection{Vanishing of static bosonic Love numbers from effective field theory} \label{sec:EFT}
Here we discuss the EFT approach to the vanishing of static bosonic LNs for BHs in GR, focusing first on a Schwarzschild BH and indicating the Kerr generalization at the end. 
The key point is that the conservative finite-size coefficients (the Wilson coefficients multiplying local tidal operators on the worldline) are not fixed by EFT alone, but must be \emph{matched} to a microscopic computation with the appropriate BH boundary conditions (regularity at the future horizon). 
For vacuum BHs in GR this matching yields vanishing static (time-reversal even) response \citep{Goldberger:2004jt,Goldberger:2005cd, Kol:2011vg, Hui:2012qt,Charalambous:2021kcz,Charalambous:2021mea}. 
We start with a scalar tidal field, where the bookkeeping is simplest, and then comment on the gravitational case and the associated linear response.

\paragraph{Macroscopic picture}~--- 
In the EFT approach there is a clean separation between a macroscopic (point-particle) description and a microscopic description that resolves the compact object. 
In the macroscopic picture, the object is represented by a worldline coupled to long-wavelength fields, and finite-size effects are encoded in an infinite tower of higher-derivative worldline operators \citep{Goldberger:2004jt,Goldberger:2005cd,Porto:2016pyg,Levi:2018nxp}. 
We first discuss the macroscopic picture before moving on to the microscopic one. 

To start, the action describing the EFT can be written schematically as
\[
\mathcal{A} = \mathcal{A}_{\Phi} + \mathcal{A}_{h} + \mathcal{A}_{\Phi h} + \mathcal{A}_{\rm pp} + \mathcal{A}_{\rm finite}\,,
\]
where (a) $\mathcal{A}_{\Phi}$ is the Klein--Gordon action for the external scalar tidal field, (b) $\mathcal{A}_{h}$ is the Einstein--Hilbert action expanded around flat space, (c) $\mathcal{A}_{\Phi h}$ describes the interaction between the gravitational and tidal fields, (d) $\mathcal{A}_{\rm pp}$ is the point-particle action for the worldline, and (e) $\mathcal{A}_{\rm finite}$ captures finite-size effects, including tidal deformations \citep{Goldberger:2004jt,Goldberger:2005cd,Porto:2016pyg}.

In the static configuration, $\partial_{t}\Phi = 0$, so that the Klein--Gordon Lagrangian for the scalar field is proportional to $\delta^{ij} \partial_{i}\Phi \partial_{j}\Phi$. 
To obtain the gravitational action, we write the metric as $g_{\mu\nu} = \eta_{\mu\nu} + h_{\mu\nu}$, where the non-zero components of the perturbation in a spherically symmetric coordinate system are $h_{tt} = -2H_{0}$ and $h_{rr} = 2H_{2}$. 
In the static case, $H_{0}=H_{0}(r)$ and $H_{2}=H_{2}(r)$, and to leading order in the weak-field expansion the Einstein-Hilbert Lagrangian density reduces (up to total derivatives) to
$\sqrt{-g}R=\sin\theta\,(2H_{2}^{2}-4rH_{2}H_{0}')$ in the present truncation \citep{Goldberger:2004jt}. 
The interaction term between the scalar and gravitational fields: $\sim \sqrt{-g}g^{\mu \nu}\partial_{\mu}\Phi\partial_{\nu}\Phi$, upon expressing $g_{\mu \nu}=\eta_{\mu \nu}+h_{\mu \nu}$ becomes proportional to $(H_{2} - H_{0}) (\partial_{r}\Phi)^{2}$, with $H_{2}$ and $H_{0}$ arising from $\sqrt{-g}$ and $g^{rr}$, respectively. 
Therefore, the total action describing the interaction of a static, spherically symmetric BH and its gravitational field with an external scalar field reads 
%%%%%%%%%%%%%%%%%%%%%%%%%%%%%%%%%%%%%%%%%%%%%%%%%%%%%%%%%%%%%%%%%%%%%%%%%%%%%%
\begin{align}\label{EFTactiontotal}
\mathcal{A}&=\underbrace{-\frac{1}{2}\int dtdrd\theta d\phi\, r^{2}\sin \theta\,\left(\delta^{ij}\partial_{i}\Phi\partial_{j}\Phi\right)}_{\mathcal{A}_{\Phi}}
\underbrace{+\frac{1}{16\pi G}\int dtdrd\theta d\phi\, \sin \theta\,\left(2H_{2}^{2}-4rH_{2}H_{0}'\right)}_{\mathcal{A}_{\rm h}}
\nonumber
\\
&\underbrace{+\frac{1}{2}\int dtdrd\theta d\phi\, r^{2}\sin \theta\,\left(H_{2}-H_{0}\right)\left(\partial_{r}\Phi\right)^{2}}_{\mathcal{A}_{\Phi h}}
\underbrace{-M\int d\tau \int d^{4}x\,\delta^{4}\left[x^{\mu}-X^{\mu}(\tau) \right]\left(1+H_{0}\right)}_{\mathcal{A}_{\rm pp}}
\nonumber
\\
&\underbrace{+\frac{1}{2}\int d\tau \int d^{4}x\,\delta^{4}\left[x^{\mu}-X^{\mu}(\tau) \right]M^{L}(\tau)\mathcal{E}_{L}(\tau)}_{A_{\rm finite}}~.
\end{align}
%%%%%%%%%%%%%%%%%%%%%%%%%%%%%%%%%%%%%%%%%%%%%%%%%%%%%%%%%%%%%%%%%%%%%%%%%%%%%%
Here, $X^{\mu}(\tau)$ denotes the worldline, and $M^{L}(\tau)$ are the induced mass multipole moments. 
Since the tidal field is generated by a scalar perturbation, spin multipoles and a magnetic tidal field are absent in this toy model. 
The electric tidal tensor is purely spatial, $\mathcal{E}_{L}=\mathcal{E}_{i_{1}\cdots i_{\ell}}$, and for a scalar tide it can be written as
$\mathcal{E}_{L}=\partial_{\langle i_{1}}\cdots \partial_{i_{\ell}\rangle}\Phi$, where $\langle\cdots\rangle$ denotes the STF projection \citep{Thorne:1980ru}. 
The perturbation $H_{0}$ appears in $\mathcal{A}_{\rm pp}$ through $\sqrt{-g_{\mu\nu}u^{\mu}u^{\nu}}$ and, in the static limit, only $u^{0}$ contributes at the order considered here \citep{Goldberger:2004jt,Goldberger:2005cd}.

The first step is to determine the gravitational field of the BH, in the absence of the scalar tidal field. Thus, only $\mathcal{A}_{h}$ and $\mathcal{A}_{\rm pp}$ are going to contribute. Variation of these two with respect to $H_{2}$ yields $H_{2}=rH_{0}'$, while variation with respect to $H_{0}$ yields 
%%%%%%%%%%%%%%%%%%%%%%%%%%%%%%%%%%%%%%%%%%%%%%%%%%%%%%%%%%%%%%%%%%%%%%%%%%%%%%
\begin{align}
\delta\left(\mathcal{A}_{h}+\mathcal{A}_{\rm pp}\right)&=\frac{1}{4\pi G}\int dtd^{3}x \delta H_{0}\left[\frac{1}{r}\dfrac{d}{dr}\left(r\dfrac{dH_{0}}{dr}\right)+r\dfrac{dH_{0}}{dr}
-4\pi GM\int d\tau \,\delta^{4}[x^{\mu}-X^{\mu}(\tau)]\right]~,
\end{align}
%%%%%%%%%%%%%%%%%%%%%%%%%%%%%%%%%%%%%%%%%%%%%%%%%%%%%%%%%%%%%%%%%%%%%%%%%%%%%%
where we have used the relation between $H_{2}$ and $H_{0}$. From the above variation of the gravitational action, we obtain the following equation of motion for $H_{0}$,
%%%%%%%%%%%%%%%%%%%%%%%%%%%%%%%%%%%%%%%%%%%%%%%%%%%%%%%%%%%%%%%%%%%%%%%%%%%%%%
\begin{align}
\nabla^{2}H_{0}=4\pi GM\int d\tau \,\delta^{4}[x^{\mu}-X^{\mu}(\tau)]\,,
\end{align}
%%%%%%%%%%%%%%%%%%%%%%%%%%%%%%%%%%%%%%%%%%%%%%%%%%%%%%%%%%%%%%%%%%%%%%%%%%%%%%
where $\nabla^{2}=(\partial^{2}/\partial r^{2})+(2/r)(\partial/\partial r)$ is the Laplacian for a purely radial function.
As we are interested in static situation, we have $X^{0}=\tau$ and $X^{i}=0$, such that the differential equation for $H_{0}$ boils down to $\nabla^{2}H_{0}=4\pi GM\delta^{3}(\vec{x})$. This is solved by
%%%%%%%%%%%%%%%%%%%%%%%%%%%%%%%%%%%%%%%%%%%%%%%%%%%%%%%%%%%%%%%%%%%%%%%%%%%%%%
\begin{align}
H_{0}=-\frac{GM}{r}=-\frac{r_{\rm s}}{2r}=-H_{2}~,
\end{align}
%%%%%%%%%%%%%%%%%%%%%%%%%%%%%%%%%%%%%%%%%%%%%%%%%%%%%%%%%%%%%%%%%%%%%%%%%%%%%%
where $r_{\rm s}=2M$ is the Schwarzschild radius. Thus, we have derived the Schwarzschild solution in the weak-field limit. Recently, the full Schwarzschild metric have been derived from the EFT action, see \citet{Mougiakakos:2024nku}.
The scalar tidal field interacts with the gravitational field through the $\mathcal{A}_{\Phi h}$ term, and also with the source itself due to its finite size, giving rise to $\mathcal{A}_{\rm finite}$. 
The finite-size action admits a natural split into a conservative (time-reversal even) sector and a dissipative (time-reversal odd) sector, the latter encoding absorption at the horizon in the BH case \citep{Goldberger:2005cd,Porto:2016pyg,Porto:2007qi,Goldberger:2020fot}.
Since the multipole moments are induced by the tidal field, for a static and non-rotating deformed object we write $M_{L}=\lambda_{\ell}\,\mathcal{E}_{L}$ \citep{Goldberger:2005cd,Porto:2016pyg,Levi:2018nxp}.
Therefore, the conservative part of $\mathcal{A}_{\rm finite}$, for the present case of a static gravitational source, reads
%%%%%%%%%%%%%%%%%%%%%%%%%%%%%%%%%%%%%%%%%%%%%%%%%%%%%%%%%%%%%%%%%%%%%%%%%%%%%%
\begin{align}
\mathcal{A}_{\rm finite}^{\rm conservative}&=\frac{\lambda_{\ell}}{2\ell!}\int d\tau \int d^{4}x\,\delta^{4}\left[x^{\mu}-X^{\mu}(\tau)\right]\mathcal{E}_{L}\mathcal{E}^{L}\,,
\nonumber
\\
&=\frac{\lambda_{\ell}}{2\ell!}\int d^{4}x\,\delta^{3}(\vec{x})
\partial_{\langle i_{1}}\partial_{i_{2}}\cdots \partial_{i_{\ell}\rangle}\Phi\partial^{\langle i_{1}}\partial^{i_{2}}\cdots \partial^{i_{\ell}\rangle}\Phi\,.
\end{align}
%%%%%%%%%%%%%%%%%%%%%%%%%%%%%%%%%%%%%%%%%%%%%%%%%%%%%%%%%%%%%%%%%%%%%%%%%%%%%%
Thus, the variation of $\mathcal{A}_{\Phi}$, $\mathcal{A}_{\Phi h}$, and $\mathcal{A}_{\rm finite}$ with respect to the scalar field $\Phi$ yields 
%%%%%%%%%%%%%%%%%%%%%%%%%%%%%%%%%%%%%%%%%%%%%%%%%%%%%%%%%%%%%%%%%%%%%%%%%%%%%%
\begin{align}
\nabla^{2}\Phi-\frac{2}{r^{2}}\partial_{r}\left(r^{2}H_{2}\partial_{r}\Phi\right)+(-1)^{\ell}\dfrac{\lambda_{\ell}}{\ell!}\partial_{\langle i_{1}}\partial_{i_{2}}\cdots \partial_{i_{\ell}\rangle}\left[\delta^{3}(\vec{x})\partial^{\langle i_{1}}\partial^{i_{2}}\cdots \partial^{i_{\ell}\rangle}\Phi\right]=0\,.
\end{align}
%%%%%%%%%%%%%%%%%%%%%%%%%%%%%%%%%%%%%%%%%%%%%%%%%%%%%%%%%%%%%%%%%%%%%%%%%%%%%%
We now decompose the scalar field as, $\Phi=\Phi_{\rm tidal}+\Phi_{h}+\Phi_{\rm Love}$, where $\Phi_{\rm tidal}$ is the tidal contribution, arising solely from $\mathcal{A}_{\Phi}$, $\Phi_{h}$ is the interaction between scalar and graviton, arising from $\mathcal{A}_{\Phi}+\mathcal{A}_{\Phi h}$, and finally $\Phi_{\rm Love}$ is the response, arising from $\mathcal{A}_{\Phi}+\mathcal{A}_{\rm finite}$. Thus, we obtain the following differential equations for each piece of the scalar field, 
%%%%%%%%%%%%%%%%%%%%%%%%%%%%%%%%%%%%%%%%%%%%%%%%%%%%%%%%%%%%%%%%%%%%%%%%%%%%%%
\begin{align}
\nabla^{2}\Phi_{\rm tidal}&=0\,,
\label{phitidal}
\\
\nabla^{2}\Phi_{h}&=\frac{1}{r^{2}}\partial_{r}\left(2H_{2}r^{2}\partial_{r}\Phi_{\rm tidal}\right)\,,
\label{phigrav}
\\
\nabla^{2}\Phi_{\rm Love}&+(-1)^{\ell}\dfrac{\lambda_{\ell}}{\ell!}\partial_{\langle i_{1}}\partial_{i_{2}}\cdots \partial_{i_{\ell}\rangle}\left[\delta^{3}(\vec{x})\partial^{\langle i_{1}}\partial^{i_{2}}\cdots \partial^{i_{\ell}\rangle}\Phi_{\rm tidal}\right]=0\,,
\label{phiresponse}
\end{align}
%%%%%%%%%%%%%%%%%%%%%%%%%%%%%%%%%%%%%%%%%%%%%%%%%%%%%%%%%%%%%%%%%%%%%%%%%%%%%%
In the static limit, the various contributions to the scalar field are decomposed into radial and angular parts, such that the tidal piece becomes $\Phi_{\rm tidal}=\sum_{\ell m}\mathcal{E}_{\ell m}(r)Y_{\ell m}(\theta,\phi)$. Therefore, the radial part, namely $\mathcal{E}_{\ell m}(r)$, satisfies the following differential equation,
%%%%%%%%%%%%%%%%%%%%%%%%%%%%%%%%%%%%%%%%%%%%%%%%%%%%%%%%%%%%%%%%%%%%%%%%%%%%%%
\begin{align}
\left[\dfrac{\partial^{2}}{\partial r^{2}}+\frac{2}{r}\dfrac{\partial}{\partial r}-\frac{\ell(\ell+1)}{r^{2}}\right]\mathcal{E}_{\ell m}(r)=0~,
\end{align}
%%%%%%%%%%%%%%%%%%%%%%%%%%%%%%%%%%%%%%%%%%%%%%%%%%%%%%%%%%%%%%%%%%%%%%%%%%%%%%
which has two independent solutions, a growing mode $\sim r^{\ell}$ and a decaying mode $\sim r^{-\ell-1}$. Since the scalar field is producing tidal field, it must grow with distance, and hence we have the following solution, $\mathcal{E}_{\ell m}\sim \mathcal{E}^{(0)}_{\ell m}r^{\ell}$. The interaction of the tidal field with graviton is described by $\Phi_{h}$, which is expressed as, $\Phi_{h}=\sum_{\ell m}\Phi_{h\,\ell m}(r)Y_{\ell m}(\theta,\phi)$, where the radial part, namely $\Phi_{h\,\ell m}(r)$, satisfies the following differential equation
%%%%%%%%%%%%%%%%%%%%%%%%%%%%%%%%%%%%%%%%%%%%%%%%%%%%%%%%%%%%%%%%%%%%%%%%%%%%%%
\begin{align}
\left[\dfrac{\partial^{2}}{\partial r^{2}}+\frac{2}{r}\dfrac{\partial}{\partial r}-\frac{\ell(\ell+1)}{r^{2}}\right]\Phi_{h\,\ell m}(r)=\frac{1}{r^{2}}\partial_{r}\left[2\left(\frac{r_{\rm s}}{2r}\right)r^{2}\partial_{r}\left(\mathcal{E}_{\ell m}^{(0)}r^{\ell}\right)\right]
=\mathcal{E}_{\ell m}^{(0)}r_{\rm s}\ell^{2}r^{\ell-3}~,
\end{align}
%%%%%%%%%%%%%%%%%%%%%%%%%%%%%%%%%%%%%%%%%%%%%%%%%%%%%%%%%%%%%%%%%%%%%%%%%%%%%%
which we obtain from~\ref{phigrav}. The above differential equation can be solved analytically, and one obtains the following expression for the radial function, $\Phi_{h\,\ell m}(r)=\mathcal{E}_{\ell m}^{(0)}(-r_{\rm s}\ell/2)r^{\ell-1}$. Finally, the static response of the body to the external scalar tidal field is to be determined from~\ref{phiresponse}. For this purpose, note that the tidal field $\Phi_{\rm tidal}$ can also be expressed as $\Phi_{\rm tidal}=\sum_{\ell}\mathcal{E}^{(0)}_{i_{1}\cdots i_{\ell}}x^{i_{1}}\cdots x^{i_{\ell}}$, such that $\partial_{\langle i_{1}}\partial_{i_{2}}\cdots \partial_{i_{\ell}\rangle}\Phi_{\rm tidal}=\mathcal{E}^{(0)}_{\langle i_{1}\cdots i_{\ell}\rangle}=\mathcal{E}^{(0)}_{L}$, and hence the differential equation satisfied by $\Phi_{\rm Love}$ becomes
%%%%%%%%%%%%%%%%%%%%%%%%%%%%%%%%%%%%%%%%%%%%%%%%%%%%%%%%%%%%%%%%%%%%%%%%%%%%%%
\begin{align}
\nabla^{2}\Phi_{\rm Love}+(-1)^{\ell}\dfrac{\lambda_{\ell}}{\ell!}\mathcal{E}^{(0)}_{i_{1}\cdots i_{\ell}}\partial^{\langle i_{1}}\partial^{i_{2}}\cdots \partial^{i_{\ell}\rangle}\delta^{3}(\vec{x})=0\,.
\end{align}
%%%%%%%%%%%%%%%%%%%%%%%%%%%%%%%%%%%%%%%%%%%%%%%%%%%%%%%%%%%%%%%%%%%%%%%%%%%%%%
This equation can be solved for in Fourier space, finally obtaining
%%%%%%%%%%%%%%%%%%%%%%%%%%%%%%%%%%%%%%%%%%%%%%%%%%%%%%%%%%%%%%%%%%%%%%%%%%%%%%
\begin{align}
\Phi_{\rm Love}=\frac{2^{\ell-2}}{\pi^{3/2}}\Gamma\left(\ell+\frac{1}{2}\right)\lambda_{\ell}\left(\mathcal{E}^{(0)}_{i_{1}\cdots i_{\ell}}x^{i_{1}}\cdots x^{i_{\ell}}\right)r^{-2\ell-1}\,.
\end{align}
%%%%%%%%%%%%%%%%%%%%%%%%%%%%%%%%%%%%%%%%%%%%%%%%%%%%%%%%%%%%%%%%%%%%%%%%%%%%%%
Thus, collecting the tidal piece, the leading graviton-induced correction, and the finite-size response, we obtain \citep{Hui:2012qt,Kol:2011vg,Charalambous:2021kcz}
%%%%%%%%%%%%%%%%%%%%%%%%%%%%%%%%%%%%%%%%%%%%%%%%%%%%%%%%%%%%%%%%%%%%%%%%%%%%%%
\begin{align}\label{EFTScalar}
\Phi=\sum_{\ell m}\mathcal{E}^{(0)}_{\ell m}Y_{\ell m}(\theta,\phi)r^{\ell}\left(\underbrace{1}_{\substack{\text{tidal}\\\text{field}}}-\underbrace{\frac{\ell}{2}\frac{r_{\rm s}}{r}}_{\substack{\text{graviton}\\\text{interaction}}}+\underbrace{\frac{2^{\ell-2}}{\pi^{3/2}}\Gamma\left(\ell+\frac{1}{2}\right)\frac{\lambda_{\ell}}{r^{1+2\ell}}}_{\rm induced~multipole}\right)~.
\end{align}
%%%%%%%%%%%%%%%%%%%%%%%%%%%%%%%%%%%%%%%%%%%%%%%%%%%%%%%%%%%%%%%%%%%%%%%%%%%%%%
Note that there is a clear distinction between the graviton-induced corrections to the tidal field and the induced multipole moments, since their coefficients are different. 
If one computes higher-order corrections to the tidal field due to the interaction of the scalar with gravitational perturbations, such corrections can be of the same order as the induced multipoles, but the coefficients will be entirely different.

In principle, since all possible corrections to the tidal field from its scattering with gravitons can be computed, one can always distinguish graviton-induced tidal effects from induced multipole moments. 
In practice, isolating the finite-size term from purely gravitational ``dressing'' effects requires a prescription; one convenient choice is the analytic-continuation method discussed in~\ref{sec:ambiguities}, which is closely related to the matching viewpoint in \citet{Kol:2011vg,Hui:2012qt,Charalambous:2021kcz}. 

At this stage it is useful to incorporate rotation. 
In the macroscopic EFT, a rotating BH is described by a worldline endowed with angular momentum, and rotation corrects the long-distance scattering of the tidal field already in the conservative sector; the genuinely BH-specific input still enters through the microscopic matching \citep{Porto:2016pyg,Charalambous:2021mea,Ivanov:2022hlo}.

In the slow-rotation approximation and for a scalar tide, the leading-order effect of rotation appears through the term $\mathcal{A}_{\Phi h}$, which describes the interaction between the scalar field and the gravitational perturbation around flat spacetime. 
In addition to the perturbation due to the mass of the BH, there is now a perturbation due to its rotation, so that the corresponding interaction action schematically reads \citep{Endlich:2015mke,Ivanov:2022hlo, Charalambous:2021mea}, 
%%%%%%%%%%%%%%%%%%%%%%%%%%%%%%%%%%%%%%%%%%%%%%%%%%%%%%%%%%%%%%%%%%%%%%%%%%%%%%
\begin{align}
\mathcal{A}_{\Phi h}=\int dtdrd\theta d\phi\,r^{2}\sin \theta \left[H_{2}\left(\partial_{r}\Phi\right)^{2}+\left(\frac{a^{2}}{2r^{4}}\right)\left(\partial_{\phi}\Phi\right)^{2}\right]\,.
\end{align}
%%%%%%%%%%%%%%%%%%%%%%%%%%%%%%%%%%%%%%%%%%%%%%%%%%%%%%%%%%%%%%%%%%%%%%%%%%%%%%
Note that, as we are dealing with scalar tidal perturbations, there are no spin-tide couplings at linear order in the BH spin. The latter are present for generic spin-$s$ perturbation, as we will discuss in the dynamical context and for slowly-spinning NSs. 

From the variation of $(\mathcal{A}_{\Phi}+\mathcal{A}_{\Phi h})$, it follows that the scalar field $\Phi$ satisfies the following equation, in the static limit, 
%%%%%%%%%%%%%%%%%%%%%%%%%%%%%%%%%%%%%%%%%%%%%%%%%%%%%%%%%%%%%%%%%%%%%%%%%%%%%%
\begin{align}
\left(\dfrac{\partial^{2}}{\partial r^{2}}+\frac{2}{r}\dfrac{\partial}{\partial r}-\frac{\ell(\ell+1)}{r^{2}} \right)\Phi=\dfrac{\partial}{\partial r}\left[\frac{r_{\rm s}}{r}\dfrac{\partial \Phi}{\partial r} \right]
+\dfrac{\partial}{\partial \phi}\left[\frac{a^{2}}{r^{4}}\dfrac{\partial \Phi}{\partial \phi} \right]\,,
\end{align}
%%%%%%%%%%%%%%%%%%%%%%%%%%%%%%%%%%%%%%%%%%%%%%%%%%%%%%%%%%%%%%%%%%%%%%%%%%%%%%
where we have used the result that $H_{2}=(r_{\rm s}/2r)$. We have ignored any finite size effects in this computation. Decomposition of the scalar as $\Phi=\Phi_{\rm tidal}+\Phi_{\rm h}+\Phi_{\rm rot}$, it follows that both $\Phi_{\rm tidal}$ and $\Phi_{\rm h}$ have the same structure as above, while the contribution coming from rotation of BH satisfies the following equation, 
%%%%%%%%%%%%%%%%%%%%%%%%%%%%%%%%%%%%%%%%%%%%%%%%%%%%%%%%%%%%%%%%%%%%%%%%%%%%%%
\begin{align}
\left(\dfrac{\partial^{2}}{\partial r^{2}}+\frac{2}{r}\dfrac{\partial}{\partial r}-\frac{\ell(\ell+1)}{r^{2}} \right)\Phi_{\rm rot}=\dfrac{\partial}{\partial \phi}\left[\frac{a^{2}}{r^{4}}\dfrac{\partial \Phi_{\rm tidal}}{\partial \phi} \right]\,,
\end{align}
%%%%%%%%%%%%%%%%%%%%%%%%%%%%%%%%%%%%%%%%%%%%%%%%%%%%%%%%%%%%%%%%%%%%%%%%%%%%%%
with the following solution, $\Phi_{\rm rot}=\sum_{\ell m}\mathcal{E}^{(0)}_{\ell m} \left(\frac{m^{2}a^{2}}{4\ell-2}\right)Y_{\ell m}r^{\ell-2}$, where we used $\Phi_{\rm tidal}=\sum_{\ell m}\mathcal{E}^{(0)}_{\ell m} Y_{\ell m}r^{\ell}$ from~\ref{EFTScalar} . Thus, scattering of the tidal field from rotating BH produces a relative sub-leading term of $\mathcal{O}(M^2/r^{2})$, one order lower than the scattering from mass.

In the presence of rotation the static response is described by a (generically anisotropic) Love tensor rather than a single number, and the conservative finite-size action  becomes \citep{LeTiec:2020bos,Ivanov:2022qqt,Saketh:2023bul}
%%%%%%%%%%%%%%%%%%%%%%%%%%%%%%%%%%%%%%%%%%%%%%%%%%%%%%%%%%%%%%%%%%%%%%%%%%%%%%
\begin{align}
\mathcal{A}_{\rm finite\,(rot)}^{\rm conservative}&=\frac{\lambda^{i_{1}i_{2}\cdots i_{\ell}}_{j_{1}j_{2}\cdots j_{\ell}}}{2\ell!}\int d^{4}x\,\delta^{3}(\vec{x})
\partial_{\langle i_{1}}\partial_{i_{2}}\cdots \partial_{i_{\ell}\rangle}\Phi\partial^{\langle j_{1}}\partial^{j_{2}}\cdots \partial^{j_{\ell}\rangle}\Phi~.
\end{align}
%%%%%%%%%%%%%%%%%%%%%%%%%%%%%%%%%%%%%%%%%%%%%%%%%%%%%%%%%%%%%%%%%%%%%%%%%%%%%%
Here, $\lambda^{i_{1}i_{2}\cdots i_{\ell}}_{j_{1}j_{2}\cdots j_{\ell}}$ is the Love tensor, which is symmetric and trace free in the upper and lower indices \citep{LeTiec:2020bos, Ivanov:2022qqt, Saketh:2023bul}. In the absence of rotation, the Love tensor can be expressed in terms of the LN $\lambda_{\ell}$, along with a collection of Kronecker deltas, as $\lambda^{i_{1}i_{2}\cdots i_{\ell}}_{j_{1}j_{2}\cdots j_{\ell}}=\lambda_{\ell}\delta^{\langle i_{1}}_{\langle j_{1}}\cdots \delta^{i_{\ell}\rangle}_{j_{\ell}\rangle}$, where $\langle \rangle$ denotes symmetrization and traceless part.

One can follow an identical procedure as before and hence obtain the following differential equation for $\Phi^{\rm rot}_{\rm Love}$,
%%%%%%%%%%%%%%%%%%%%%%%%%%%%%%%%%%%%%%%%%%%%%%%%%%%%%%%%%%%%%%%%%%%%%%%%%%%%%%
\begin{align}
\nabla^{2}\Phi^{\rm rot}_{\rm Love}+(-1)^{\ell}\dfrac{\lambda^{i_{1}i_{2}\cdots i_{\ell}}_{j_{1}j_{2}\cdots j_{\ell}}}{\ell!}\mathcal{E}^{(0)}_{i_{1}\cdots i_{\ell}}\partial^{\langle j_{1}}\partial^{j_{2}}\cdots \partial^{j_{\ell}\rangle}\delta^{3}(\vec{x})=0\,.
\end{align}
%%%%%%%%%%%%%%%%%%%%%%%%%%%%%%%%%%%%%%%%%%%%%%%%%%%%%%%%%%%%%%%%%%%%%%%%%%%%%%
The rest of the analysis remains identical: one moves to the Fourier domain and uses spherical polar coordinates in the Fourier space, leading to $\ell$ number of derivatives acting on $(1/r)$. Thus, the total solution including rotation effects reads
%%%%%%%%%%%%%%%%%%%%%%%%%%%%%%%%%%%%%%%%%%%%%%%%%%%%%%%%%%%%%%%%%%%%%%%%%%%%%%
\begin{align}\label{phi_eft_stat_rot}
\Phi&=\Phi_{\rm tidal}+\Phi_{h}+\Phi_{\rm rot}+\Phi_{\rm Love}^{\rm rot}=\sum_{\ell m}\mathcal{E}^{(0)}_{\ell' m'}Y_{\ell m}(\theta,\phi)r^{\ell}
\nonumber
\\
&\times\left(\underbrace{\delta_{\ell \ell'}\delta_{mm'}}_{\substack{\text{tidal}\\\text{field}}}-\underbrace{\left(\frac{\ell}{2}\frac{r_{\rm s}}{r}-\frac{m^{2}}{4\ell-2}\frac{a^{2}}{r^{2}}\right)}_{\substack{\text{graviton}\\\text{interaction}}}\delta_{\ell \ell'}\delta_{mm'}+\underbrace{\frac{2^{\ell-2}}{\pi^{3/2}}\Gamma\left(\ell+\frac{1}{2}\right)\frac{\lambda^{\ell' m'}_{\ell m}}{r^{1+2\ell}}}_{\rm induced~multipole}\right)~.
\end{align}
%%%%%%%%%%%%%%%%%%%%%%%%%%%%%%%%%%%%%%%%%%%%%%%%%%%%%%%%%%%%%%%%%%%%%%%%%%%%%%
Again, we have only depicted the leading order contributions from the graviton interactions. In principle, there are higher order terms in the graviton interaction sector, but they can be exactly determined using the EFT computation, and hence are distinguishable from the finite size effects. We now move on to the BH perturbation theory, which gives the microscopic picture. 

\paragraph{Microscopic picture}~--- 
In the microscopic picture one resolves the horizon and computes the tidal perturbation using BH perturbation theory; the BH nature enters through the boundary condition of regularity at the future horizon \citep{Teukolsky:1973ha,Press:1973zz,Chia:2020yla,Charalambous:2021mea}. 
We have already worked out the static solution in~\ref{stat_Love_Kerr} (using advanced null coordinates). The radial Teukolsky function regular at the future horizon takes the form (see~\ref{radialstatickerr} with $s=0$)
%%%%%%%%%%%%%%%%%%%%%%%%%%%%%%%%%%%%%%%%%%%%%%%%%%%%%%%%%%%%%%%%%%%%%%%%%%%%%%
\begin{equation}
R(z)=\mathcal{A}\,_{2}F_{1}\left(1+\ell,-\ell;1+2iP_{0};-z\right)\,,
\end{equation}
%%%%%%%%%%%%%%%%%%%%%%%%%%%%%%%%%%%%%%%%%%%%%%%%%%%%%%%%%%%%%%%%%%%%%%%%%%%%%%
where $\mathcal{A}$ is an integration constant and $P_{0}$ is defined in~\ref{def_Ppm}. In the asymptotic limit, keeping only the leading order pieces of the tidal field, and the finite size response,  we obtain from~\ref{psi4gen_stat},
%%%%%%%%%%%%%%%%%%%%%%%%%%%%%%%%%%%%%%%%%%%%%%%%%%%%%%%%%%%%%%%%%%%%%%%%%%%%%%
\begin{align}
R_{\infty}(r)&\propto \left(\frac{r}{r_{+}-r_{-}}\right)^{\ell}\times\left\{\frac{\Gamma\left(1+2iP_{+}\right)\Gamma\left(2\ell+1\right)}{\Gamma\left(\ell+2iP_{+}+1\right)\Gamma\left(1+\ell\right)}\right\}
\nonumber
\\
&\qquad \times \left[\left(1+\cdots\right)+\,_{0}\mathcal{F}^{\rm static}_{\ell m}\left(\frac{r}{r_{+}}\right)^{-2\ell-1}\left(1+\cdots \right)\right]\,,
\label{Weyl_Scalar}
\\
\,_{0}\mathcal{F}^{\rm static}_{\ell m}&=-\left(\frac{iam}{r_{+}}\right)\frac{\left(\ell!\right)^{2}}{\left(2\ell+1\right)!\left(2\ell\right)!}\prod_{j=1}^{\ell}\left[j^2\left(1-\frac{r_{-}}{r_{+}}\right)^{2}+\left(\frac{2am}{r_{+}}\right)^2\right]\,,
\label{response_stat_scalar}
\end{align}
%%%%%%%%%%%%%%%%%%%%%%%%%%%%%%%%%%%%%%%%%%%%%%%%%%%%%%%%%%%%%%%%%%%%%%%%%%%%%%
where the `dots' in the radial function denotes the contribution due to the interaction of tidal field with gravitational perturbations and are of order $(M/r)^{n}$, where $n$ is an integer. Also the definition of response function has a characteristic scale $R$, see~\ref{psi_4_intermediate}, which we have set to the horizon scale, i.e., $R=r_{+}$. 
Since we treat $\ell\in\mathbb{C}$, (a) no subleading graviton-interaction term can scale as $r^{-\ell-1}$ when $\ell$ is non-integer, and (b) real coordinate transformations cannot mimic the finite-size response, because they cannot generate the complex-$\ell$ branch structure. 
This analytic-continuation prescription provides a practical way to separate ``dressing'' terms from finite-size effects \citep{Kol:2011vg,Hui:2012qt,Charalambous:2021kcz}.
The appearance of ambiguity for integer values of $\ell$ as well as the gauge invariance of this formalism has been discussed in detail in~\ref{stat_Love_Kerr}. It is now time to match the above microscopic result arising from BH perturbation theory, with the macroscopic EFT picture.

\paragraph{Matching}~--- 
The EFT with finite-size operators encodes the response of a compact object to an external tidal field in the coefficient of the $r^{-2\ell-1}$ term in~\ref{phi_eft_stat_rot}, i.e.\ in the Wilson coefficients $\lambda$ determined by matching \citep{Goldberger:2005cd,Porto:2016pyg,Hui:2012qt}. 
Note that this applies to any compact object, not only BHs. 
In contrast, the Weyl scalar in the microscopic picture depends on the boundary conditions of the object (i.e., regularity at the horizon for a BH or regularity at the center for a star), and therefore yields different coefficients for the decaying term $r^{-2\ell-1}$. 
For vacuum BHs in GR, the Weyl scalar is given by~\ref{Weyl_Scalar}, with the coefficient of the $r^{-2\ell-1}$ term provided in~\ref{response_stat_scalar}. 
By equating these two coefficients between the microscopic and macroscopic pictures, we obtain 
%%%%%%%%%%%%%%%%%%%%%%%%%%%%%%%%%%%%%%%%%%%%%%%%%%%%%%%%%%%%%%%%%%%%%%%%%%%%%%
\begin{align}\label{EFT_matching}
\frac{2^{\ell-2}}{\pi^{3/2}}\Gamma\left(\ell+\frac{1}{2}\right)\lambda^{i_{1}\cdots i_{\ell'}}_{j_{1}\cdots j_{\ell}}\mathcal{E}^{(0)}_{i_{1}\cdots i_{\ell'}}n^{\langle j_{1}\cdots j_{\ell}\rangle}
=(r_{+}-r_{-})^{2\ell+1}\sum_{m=-\ell}^{\ell}\mathcal{E}_{\ell m}^{(0)}Y_{\ell m}\,_{0}\mathcal{F}_{\ell m}^{\rm static}
\end{align}
%%%%%%%%%%%%%%%%%%%%%%%%%%%%%%%%%%%%%%%%%%%%%%%%%%%%%%%%%%%%%%%%%%%%%%%%%%%%%%
We can express the spherical harmonic components in terms of the Cartesian symmetric and tracefree tensors, using the following identities:
%%%%%%%%%%%%%%%%%%%%%%%%%%%%%%%%%%%%%%%%%%%%%%%%%%%%%%%%%%%%%%%%%%%%%%%%%%%%%%
\begin{align}
Y_{\ell m}=\mathcal{Y}^{*\,i_{1}\cdots i_{\ell}}_{\ell m}n_{\langle i_{1}\cdots i_{\ell}\rangle}\,;
\qquad
\mathcal{E}_{\ell m}^{(0)}=\frac{4\pi \ell!}{(2\ell+1)!!}\mathcal{Y}^{i_{1}\cdots i_{\ell}}_{\ell m}\mathcal{E}_{i_{1}\cdots i_{\ell}}^{(0)}\,,
\end{align}
%%%%%%%%%%%%%%%%%%%%%%%%%%%%%%%%%%%%%%%%%%%%%%%%%%%%%%%%%%%%%%%%%%%%%%%%%%%%%%
where $\mathcal{Y}^{i_{1}\cdots i_{\ell}}_{\ell m}$ are symmetric and tracefree tensors, and are the basis for $(2\ell+1)$ dimensional vector space of the symmetric and tracefree tensors on the two sphere \citep{PoissonWill,Thorne:1980ru}. Therefore, by using the above identities in~\ref{EFT_matching}, one can read off the Love tensor as, 
%%%%%%%%%%%%%%%%%%%%%%%%%%%%%%%%%%%%%%%%%%%%%%%%%%%%%%%%%%%%%%%%%%%%%%%%%%%%%%
\begin{align}\label{Love_tensor}
\lambda^{i_{1}\cdots i_{\ell'}}_{j_{1}\cdots j_{\ell}}=\frac{(r_{+}-r_{-})^{2\ell+1}\pi^{3/2}}{2^{\ell-2}\Gamma\left(\ell+\frac{1}{2}\right)}\left(\frac{4\pi \ell!}{(2\ell+1)!!}\right)\sum_{m=-\ell}^{\ell}\,_{0}\mathcal{F}_{\ell m}^{\rm static}\mathcal{Y}^{*\,i_{1}\cdots i_{\ell'}}_{\ell m}\mathcal{Y}_{j_{1}\cdots j_{\ell}}^{\ell m}\,.
\end{align}
%%%%%%%%%%%%%%%%%%%%%%%%%%%%%%%%%%%%%%%%%%%%%%%%%%%%%%%%%%%%%%%%%%%%%%%%%%%%%%
In case of spherical symmetry, the response function $\,_{0}\mathcal{F}_{\ell m}^{\rm static}$ is independent of the magnetic number $m$, and hence one can take it outside the summation and, using the normalization condition for the spherical tensors $\mathcal{Y}_{j_{1}\cdots j_{\ell}}^{\ell m}$, one obtains
%%%%%%%%%%%%%%%%%%%%%%%%%%%%%%%%%%%%%%%%%%%%%%%%%%%%%%%%%%%%%%%%%%%%%%%%%%%%%%
\begin{align}\label{Love_tensorsphsymm}
\lambda^{i_{1}\cdots i_{\ell'}\,\textrm{(spherical symm)}}_{j_{1}\cdots j_{\ell}}=\underbrace{\frac{(r_{+}-r_{-})^{2\ell+1}\pi^{3/2}}{2^{\ell-2}\Gamma\left(\ell+\frac{1}{2}\right)}\,_{0}\mathcal{F}_{\ell}^{\rm static}}_{k_{\ell}}\delta^{\langle\,i_{1}}_{\langle j_{1}}\cdots \delta^{i_{\ell'}\rangle}_{j_{\ell}\rangle}\delta_{\ell \ell'}\,.
\end{align}
%%%%%%%%%%%%%%%%%%%%%%%%%%%%%%%%%%%%%%%%%%%%%%%%%%%%%%%%%%%%%%%%%%%%%%%%%%%%%%
From~\ref{response_stat_scalar}, it is clear that in the limit of $a\to 0$, i.e., for Schwarzschild BH, the response function $\,_{0}\mathcal{F}_{\ell}^{\rm static}$ vanishes identically (and hence is trivially independent of $m$), yielding vanishing scalar LNs: $k_{\ell}=0$. 

In the rotating case, i.e., for Kerr BH, using the explicit form of the response function, given in~\ref{response_stat_scalar}, we can write down the Love tensor for the $\ell=2$ mode as, 
%%%%%%%%%%%%%%%%%%%%%%%%%%%%%%%%%%%%%%%%%%%%%%%%%%%%%%%%%%%%%%%%%%%%%%%%%%%%%%
\begin{align}\label{Love_tensorfin}
\lambda^{i_{1}i_{2}}_{j_{1}j_{2}}=-\frac{4\pi \chi M^{5}}{135}\left[4(1-\chi^{2})^{2}\,_{1}I^{i_{1}i_{2}}_{j_{1}j_{2}}
+5\chi^{2}(1-\chi^{2})\,_{3}I^{i_{1}i_{2}}_{j_{1}j_{2}}+\chi^{4}\,_{5}I^{i_{1}i_{2}}_{j_{1}j_{2}}\right]\,
\end{align}
%%%%%%%%%%%%%%%%%%%%%%%%%%%%%%%%%%%%%%%%%%%%%%%%%%%%%%%%%%%%%%%%%%%%%%%%%%%%%%
where
%%%%%%%%%%%%%%%%%%%%%%%%%%%%%%%%%%%%%%%%%%%%%%%%%%%%%%%%%%%%%%%%%%%%%%%%%%%%%%
\begin{align}
\,_{n}I^{i_{1}\cdots i_{\ell'}}_{j_{1}\cdots j_{\ell}}=\frac{8\pi \ell!}{(2\ell+1)!!}\sum_{m=-\ell}^{\ell}\left(\frac{im^{2n-1}}{2}\right)\mathcal{Y}^{*\,i_{1}\cdots i_{\ell'}}_{\ell m}\mathcal{Y}_{j_{1}\cdots j_{\ell}}^{\ell m}\,.
\end{align}
%%%%%%%%%%%%%%%%%%%%%%%%%%%%%%%%%%%%%%%%%%%%%%%%%%%%%%%%%%%%%%%%%%%%%%%%%%%%%%
As evident, for $\ell=2$, owing to the symmetry of the tensors $I^{i_{1}i_{2}}_{j_{1}j_{2}}$ ($\{i_{1},i_{2}\}\to \{j_{1},j_{2}\}$), it follows that the Love tensor is antisymmetric under the interchange of upper and lower indices, which is associated to the dissipative nature of the Love tensor \citep{Charalambous:2021mea}. Hence the LNs, which arise from the conservative part of the response function and are symmetric under exchange of indices, vanish identically,
%%%%%%%%%%%%%%%%%%%%%%%%%%%%%%%%%%%%%%%%%%%%%%%%%%%%%%%%%%%%%%%%%%%%%%%%%%%%%%
\begin{align}
k^{\rm static}_{2}\sim \textrm{Symm.}\left(\lambda^{i_{1}i_{2}}_{j_{1}j_{2}}\right)=0\,.
\end{align}
%%%%%%%%%%%%%%%%%%%%%%%%%%%%%%%%%%%%%%%%%%%%%%%%%%%%%%%%%%%%%%%%%%%%%%%%%%%%%%
The antisymmetric part is related to the dissipation numbers and yields the same result as in~\ref{dissstatgen}. 

To summarize, matching the microscopic BH perturbation theory to the macroscopic EFT yields vanishing static (conservative) LNs for scalar perturbations of Kerr BHs in GR, while the antisymmetric part of the Love tensor captures dissipation and agrees with~\ref{dissstatgen} \citep{Charalambous:2021mea,Ivanov:2022hlo}. 
Generalizations to EM and gravitational perturbations lead to vanishing static bosonic LNs in those sectors as well \citep{Charalambous:2021mea,Ivanov:2022hlo}. 
We will now describe how the same conclusion can be obtained from scattering-amplitude computations (see also \citealt{Hui:2012qt} for the EFT/matching viewpoint).

%%%%%%%%%%%%%%%%%%%%%%%%%%%%%%%%%%%%
%%%%%%%%%%%%%%%%%%%%%%%%%%%%%%%%%%%%
%%%%%%%%%%%%%%%%%%%%%%%%%%%%%%%%%%%%
\subsection{Vanishing of static bosonic Love numbers from scattering} \label{sec:scattering}
The EFT action for a compact object, and how the vanishing of the static bosonic LNs for a vacuum GR BH follows from matching the EFT with BH perturbation theory, was discussed in the previous section. 
Here, we will use the same EFT action, but follow a different procedure, namely the analysis of a scattering process, to arrive at the same conclusion. 
The finite-size action will be treated as the interaction term between the compact object and the tidal field, leading to scattering. 
The idea is as follows: the quadrupolar deformation is generated by the absorption of a graviton, arising from the tidal field, by the compact object, and the response function is obtained from the subsequent emission of a graviton from the compact object. 
In short, one studies the scattering between a BH and a graviton.

For a generic spin perturbation, the above scattering process will be described by the scattering matrix (or, S-matrix)
%%%%%%%%%%%%%%%%%%%%%%%%%%%%%%%%%%%%%%%%%%%%%%%%%%%%%%%%%%%%%%%%%%%%%%%%%%%%%%
\begin{align}
\,_{\rm out}\langle \vec{k}',h'|\vec{k},h\rangle_{\rm in}\equiv \langle \vec{k}',h'|S|\vec{k},h\rangle\,,
\end{align}
%%%%%%%%%%%%%%%%%%%%%%%%%%%%%%%%%%%%%%%%%%%%%%%%%%%%%%%%%%%%%%%%%%%%%%%%%%%%%%
where we have assumed the perturbations to be small, so that the worldline of the BH is not modified. Here $h$ and $h'$ are the helicities of the perturbing field before and after the scattering, while $\vec{k}$ and $\vec{k}'$ are the three momenta of the incoming waves before and after the scattering, respectively. Due to energy conservation, on-shell we have the following relation: $\omega=|\vec{k}|=|\vec{k}'|=\omega'$ and the normalization condition for the states becomes $\langle \vec{k}',h'|\vec{k},h\rangle=2|\vec{k}|\times (2\pi)^{3}\delta^{3}(\vec{k}-\vec{k}')\delta_{hh'}$. Expressing the S-matrix as $S=1+iT$, we obtain the scattering amplitude $\mathcal{M}(\omega,\vec{k}\to \vec{k}',h\to h')$ in the Cartesian coordinate system as, 
%%%%%%%%%%%%%%%%%%%%%%%%%%%%%%%%%%%%%%%%%%%%%%%%%%%%%%%%%%%%%%%%%%%%%%%%%%%%%%
\begin{align}
\langle \vec{k}',h'|iT|\vec{k},h\rangle=2\pi \delta(\omega-\omega')\times i\mathcal{M}(\omega,\vec{k}\to \vec{k}',h\to h')\,.
\end{align}
%%%%%%%%%%%%%%%%%%%%%%%%%%%%%%%%%%%%%%%%%%%%%%%%%%%%%%%%%%%%%%%%%%%%%%%%%%%%%%
Keeping in mind the axisymmetry of the background spacetime, it is more convenient to express the above scattering amplitude in a spherical basis, which is characterized by $(\omega, \ell, m,h)$ and normalized as, $\langle \omega',\ell',m',h'|\omega,\ell,m,h\rangle=2\pi\delta(\omega-\omega')\delta_{\ell\ell'}\delta_{mm'}\delta_{hh'}$. The scattering amplitude in the spherical basis reduces to
%%%%%%%%%%%%%%%%%%%%%%%%%%%%%%%%%%%%%%%%%%%%%%%%%%%%%%%%%%%%%%%%%%%%%%%%%%%%%%
\begin{align}
\langle \omega',\ell',m',h'|iT|\omega,\ell,m,h\rangle=2\pi\delta(\omega-\omega')\times i\mathcal{A}(\omega,\ell\to\ell',m\to m',h\to h')\,.
\end{align}
%%%%%%%%%%%%%%%%%%%%%%%%%%%%%%%%%%%%%%%%%%%%%%%%%%%%%%%%%%%%%%%%%%%%%%%%%%%%%%
The mapping between the scattering amplitude $\mathcal{M}$ in the Cartesian basis and the scattering amplitude $\mathcal{A}$ in the spherical basis reads
%%%%%%%%%%%%%%%%%%%%%%%%%%%%%%%%%%%%%%%%%%%%%%%%%%%%%%%%%%%%%%%%%%%%%%%%%%%%%%
\begin{align}\label{spherical_scat_transform}
\mathcal{A}(\omega,\ell\to\ell',m\to m',h\to h')&=\int \frac{d^{3}\vec{k}_{1}}{(2\pi)^{3}2|\vec{k}_{1}|}\int \frac{d^{3}\vec{k}_{2}}{(2\pi)^{3}2|\vec{k}_{2}|}
\sum_{h_{1},h_{2}}\langle \omega',\ell',m',h'|\vec{k}_{2},h_{2}\rangle 
\nonumber
\\
&\times \mathcal{M}(\omega,\vec{k}_{1}\to \vec{k}_{2},h_{1}\to h_{2})
\langle \vec{k}_{1},h_{1}|\omega,\ell,m,h\rangle\,.
\end{align}
%%%%%%%%%%%%%%%%%%%%%%%%%%%%%%%%%%%%%%%%%%%%%%%%%%%%%%%%%%%%%%%%%%%%%%%%%%%%%%
The transformation matrices $\langle \omega,\ell,m,h|\vec{k},h\rangle$ can be related to Wigner matrices \citep{Ivanov:2022qqt, Saketh:2023bul}. The complex amplitude $\mathcal{A}$ can be expressed as
%%%%%%%%%%%%%%%%%%%%%%%%%%%%%%%%%%%%%%%%%%%%%%%%%%%%%%%%%%%%%%%%%%%%%%%%%%%%%%
\begin{align}\label{scatteringmatch}
i\mathcal{A}=1-r_{\ell m}e^{2i\delta_{\ell m}}\,,
\end{align}
%%%%%%%%%%%%%%%%%%%%%%%%%%%%%%%%%%%%%%%%%%%%%%%%%%%%%%%%%%%%%%%%%%%%%%%%%%%%%%
where $\delta_{\ell m}$ is the phase shift due to elastic scattering and $1-r_{\ell m}^{2}$ is the absorption probability, with $r_{\ell m}$ and $\delta_{\ell m}$ being real. Therefore, the computation of the static LNs follows naturally from the above description. 
On one hand, the scattering amplitude $\mathcal{A}$ can be computed using Feynman's diagrams, based on the interactions in the finite-size action $\mathcal{A}_{\rm finite}$. 
On the other hand, one computes the two real numbers $r_{\ell m}$ and $\delta_{\ell m}$ from BH perturbation theory. 
By equating the two through~\ref{scatteringmatch}, one obtains an expression for the static LNs in the $\omega \to 0$ limit, assuming that the scattering amplitude is analytic in $\omega$. 
This method also applies to the dynamical scenario, which will be discussed in~\ref{sec:dynBHs}.

From the BH perturbation theory side, the scattering of a generic spin-$s$ particle by the BH is an well-studied subject, see e.g. \citet{Starobinskij2} and \citet{Brito:2015oca} for a review. In the case of a spinning BH, this starts from solving the Teukolsky equation. Since regular asymptotic behavior is the one of interest here, we consider the Teukolsky master variable defined in~\ref{gen_eq_decomp}, for the $-s$ modes. The radial function associated with these modes has the following asymptotic behavior, 
%%%%%%%%%%%%%%%%%%%%%%%%%%%%%%%%%%%%%%%%%%%%%%%%%%%%%%%%%%%%%%%%%%%%%%%%%%%%%%
\begin{align}
\,_{-s}R_{\ell m}^{\infty}=\,_{-s}B_{\ell m}^{\rm in}\frac{e^{-i\omega r}}{r}+\,_{-s}B_{\ell m}^{\rm out}\frac{e^{i\omega r}}{r^{-2s+1}}\,
\end{align}
%%%%%%%%%%%%%%%%%%%%%%%%%%%%%%%%%%%%%%%%%%%%%%%%%%%%%%%%%%%%%%%%%%%%%%%%%%%%%%
where $\,_{-s}B_{\ell m}^{\rm in}$ is the incoming amplitude, while $\,_{-s}B_{\ell m}^{\rm out}$ is the outgoing/reflected amplitude of the perturbation. Thus, the scattering amplitude must be related to the ratio $(\,_{-s}B_{\ell m}^{\rm out}/\,_{-s}B_{\ell m}^{\rm in})$, which has dimension $[\textrm{Length}]^{-2s}$. Further, there will also be contributions coming from the normalization of the angular sector. Combining all of these, we obtain, the following expression for the scattering amplitude (modulo a factor of order unity \citep{Sasaki:2003xr, Teukolsky:1974yv, Starobinskij2}),
%%%%%%%%%%%%%%%%%%%%%%%%%%%%%%%%%%%%%%%%%%%%%%%%%%%%%%%%%%%%%%%%%%%%%%%%%%%%%%
\begin{align}\label{scattAmpBHPT}
\,_{s}r_{\ell m}e^{2i\,_{s}\delta_{\ell m}}=(-1)^{\ell+1}\frac{\textrm{Re}\left[\,_{s}C_{\ell m}(a\omega)\right]}{(2\omega)^{2s}}\left(\frac{\,_{-s}B_{\ell m}^{\rm out}}{\,_{-s}B_{\ell m}^{\rm in}}\right)\,.
\end{align}
%%%%%%%%%%%%%%%%%%%%%%%%%%%%%%%%%%%%%%%%%%%%%%%%%%%%%%%%%%%%%%%%%%%%%%%%%%%%%%
Here, the factor $\,_{s}C_{\ell m}(a\omega)$ is the Starobinsky--Churilov constant \citep{Starobinskij2,Teukolsky:1974yv} and $(2\omega)^{2s}$ makes the above quantity dimensionless. Note that a similar expression arises also in the context of superradiance, as there also one is interested in the scattering of an incident wave by a rotating BH \citep{Brito:2015oca}. It also follows that the expression for scattering amplitude, as described by~\ref{scattAmpBHPT} also arises in the helicity reversing part of the scattering cross-section of a particle of spin $s$ by a rotating BH \citep{Saketh:2023bul}. 

Since we are working in the EFT picture, it is important to confine ourselves in the infrared regime, which translates to low frequency approximation. Intriguingly, Teukolsky equation can be solved exactly in the low frequency regime using the Mano-Suzuki-Takasugi~(MST) method\footnote{In the MST method, one solves the Teukolsky equation in the low-frequency, near-zone approximation, obtaining a solution expressed as an infinite series of hypergeometric functions. In the same low-frequency regime, the asymptotic solution can instead be written as an infinite series of Coulomb functions. Matching the near-zone and asymptotic solutions then determines the ratio between the outgoing and ingoing amplitudes.} \citep{Mano:1996vt, Mano:1996gn, Mano:1996mf, Sasaki:2003xr}, and hence the ratio $(\,_{-s}B_{\ell m}^{\rm out}/\,_{-s}B_{\ell m}^{\rm in})$ can be found, which reads %(for a derivation see~\ref{App:MST})
%%%%%%%%%%%%%%%%%%%%%%%%%%%%%%%%%%%%%%%%%%%%%%%%%%%%%%%%%%%%%%%%%%%%%%%%%%%%%%
\begin{align}\label{outbyinscatt}
\frac{\,_{-s}B_{\ell m}^{\rm out}}{\,_{-s}B_{\ell m}^{\rm in}}=\omega^{2s}\underbrace{\frac{1+ie^{i\pi \hat{\ell}}\left(\frac{K_{-\hat{\ell}-1;-s}}{K_{\hat{\ell};-s}}\right)}{1-ie^{-i\pi \hat{\ell}}\frac{\sin[\pi(\hat{\ell}+s+2iM\omega)]}{\sin[\pi(\hat{\ell}-s-2iM\omega)]}\left(\frac{K_{-\hat{\ell}-1;-s}}{K_{\hat{\ell};-s}}\right)}}_{\rm Near\,Zone}\underbrace{\frac{A^{\hat{\ell}}_{-;-s}}{A^{\hat{\ell}}_{+;-s}}e^{2iM\omega[2\log(2M\omega)-1+\sqrt{1-\chi^{2}}]}}_{\rm Far\,Zone}\,.
\end{align}
%%%%%%%%%%%%%%%%%%%%%%%%%%%%%%%%%%%%%%%%%%%%%%%%%%%%%%%%%%%%%%%%%%%%%%%%%%%%%%
The above expression for the ratio of outgoing versus ingoing amplitude at infinity gets separated into two factors, one arises from the near zone, while the other arises from the far zone. This distinction stems from the origin of the terms.  
The contributions involving $K_{\alpha;-s}$ arise from the solution of the Teukolsky equation in the near-zone regime and are therefore referred to as \emph{near-zone} terms.  
In contrast, the contributions involving $A^{\hat{\ell}}_{\pm;-s}$ originate from the solution of the Teukolsky equation in the asymptotic region and are thus associated with the \emph{far-zone} \citep{Ivanov:2022qqt, Mougiakakos:2022sic}. The accompanying exponential factor accounts for tail effects \citep{Poisson:1994yf, DeLuca:2024ufn, Saketh:2023bul,Saketh:2024juq,Ivanov:2026icp}.

In~\ref{outbyinscatt}, $\hat{\ell}=\ell+\mathcal{O}(M^{2}\omega^{2})$ is the frequency-dependent angular number (also called renormalized angular momentum) and we have the following expression for the ratio between outgoing and ingoing amplitude in the static limit (see~\ref{App:NearZoneRatioMST}), 
%%%%%%%%%%%%%%%%%%%%%%%%%%%%%%%%%%%%%%%%%%%%%%%%%%%%%%%%%%%%%%%%%%%%%%%%%%%%%%
\begin{align}\label{Bratio}
\frac{\,_{-s}B_{\ell m}^{\rm out}}{\,_{-s}B_{\ell m}^{\rm in}}=\omega^{2s}\left(\frac{A_{-}^{\nu}}{A_{+}^{\nu}}\right)\frac{1+\frac{1}{2}i(-1)^{s}\left[2\omega(r_{+}-r_{-})\right]^{2\ell+1}\frac{\left(\ell+s\right)!\left(\ell-s\right)!}{\left(2\ell+1\right)!\left(2\ell\right)!}\mathcal{I}_{-s\ell m}}{1-\frac{1}{2}i(-1)^{s}\left[2\omega(r_{+}-r_{-})\right]^{2\ell+1}\frac{\left(\ell+s\right)!\left(\ell-s\right)!}{\left(2\ell+1\right)!\left(2\ell\right)!}\mathcal{I}_{-s\ell m}}\,,
\end{align}
%%%%%%%%%%%%%%%%%%%%%%%%%%%%%%%%%%%%%%%%%%%%%%%%%%%%%%%%%%%%%%%%%%%%%%%%%%%%%%
where we have introduced the quantity $\mathcal{I}_{s\ell m}$, having the following definition,
%%%%%%%%%%%%%%%%%%%%%%%%%%%%%%%%%%%%%%%%%%
\begin{align}
\mathcal{I}_{s\ell m}\equiv i(-1)^{s+1}\frac{\left(\ell+s\right)!\left(\ell-s\right)!}{\left(2\ell+1\right)!\left(2\ell\right)!}
\left(\frac{-\tau}{2}\right)\prod_{k=1}^{\ell}\left(k^{2}+\tau^{2}\right)\,.
\end{align}
%%%%%%%%%%%%%%%%%%%%%%%%%%%%%%%%%%%%%%%%%%
where $\tau=\{2M/\sqrt{1-\chi^{2}}\}\{\omega-(am/2M^{2})\}$. Note that $\tau$  is not proportional to $\bar{\omega}$, but still depends linearly on $\omega$. Thus, in the static limit, i.e., for $\epsilon \to 0$, we have, $-\tau=(ma/M\sqrt{1-\chi^{2}})=\{2ma/(r_{+}-r_{-})\}=2P_{0}$, where $P_{0}$ is defined in~\ref{def_Ppm}. Using this connection between $\tau$ and $P_{0}$, we can present the above expression as
%%%%%%%%%%%%%%%%%%%%%%%%%%%%%%%%%%%%%%%%%%
\begin{align}
\mathcal{I}_{s\ell m}=i(-1)^{s+1}P_{0}\frac{\left(\ell+s\right)!\left(\ell-s\right)!}{\left(2\ell+1\right)!\left(2\ell\right)!}
\prod_{k=1}^{\ell}\left(k^{2}+4P_{0}^{2}\right)\,.
\end{align}
%%%%%%%%%%%%%%%%%%%%%%%%%%%%%%%%%%%%%%%%%%
Note that $\mathcal{I}_{s\ell m}=\mathcal{I}_{-s\ell m}$. Substituting the ratio of outgoing and ingoing amplitude from~\ref{Bratio}, and using the small frequency assumption, we obtain the scattering amplitude and phase:
%%%%%%%%%%%%%%%%%%%%%%%%%%%%%%%%%%%%%%%%%%
\begin{align}\label{BHPTscattering}
&\,_{s}r_{\ell m}e^{2i\,_{s}\delta_{\ell m}}
\nonumber
\\
&\approx \frac{(-1)^{\ell+1}}{2^{2s}}\left(\textrm{Re}\left[\,_{s}C_{\ell m}(a\omega)\right]\times \frac{A_{-}^{\nu}}{A_{+}^{\nu}}\right)\frac{1+\frac{1}{2}i(-1)^{s}\left[2\omega(r_{+}-r_{-})\right]^{2\ell+1}\frac{\left(\ell+s\right)!\left(\ell-s\right)!}{\left(2\ell+1\right)!\left(2\ell\right)!}\mathcal{I}_{-s\ell m}}{1-\frac{1}{2}i(-1)^{s}\left[2\omega(r_{+}-r_{-})\right]^{2\ell+1}\frac{\left(\ell+s\right)!\left(\ell-s\right)!}{\left(2\ell+1\right)!\left(2\ell\right)!}\mathcal{I}_{-s\ell m}}
\nonumber
\\
&\simeq 1+P_{0}\left[2\omega(r_{+}-r_{-})\right]^{2\ell+1}\left(\frac{\left(\ell+s\right)!\left(\ell-s\right)!}{\left(2\ell+1\right)!\left(2\ell\right)!}\right)^{2}\prod_{k=1}^{\ell}\left(k^{2}+4P_{0}^{2}\right)\,.
\end{align}
%%%%%%%%%%%%%%%%%%%%%%%%%%%%%%%%%%%%%%%%%%
Here, we have used the result that $(-1)^{\ell+1}2^{-2s}(\textrm{Re}[\,_{s}C_{\ell m}(a\omega)]\times \{A_{-}^{\nu}/A_{+}^{\nu}\})=1$ \citep{Sasaki:2003xr}. It follows from the above expression that $\,_{s}r_{\ell m}e^{2i\,_{s}\delta_{\ell m}}$ is real, showing that the scattering phase $\,_{s}\delta_{\ell m}=0$, while the amplitude $\,_{s}r_{\ell m}$ is non-zero. This in turn provides the following intriguing relations between the scattering and absorption cross-section of GWs by BHs (with $m=s$), with the amplitude $\,_{s}r_{\ell s}$ and phase $\,_{s}\delta_{\ell s}$, such that \citep{Ivanov:2022qqt, dolan2008scattering-43d},
%%%%%%%%%%%%%%%%%%%%%%%%%%%%%%%%%%%%%%%%%%
\begin{align}
\sigma_{\rm elastic}&=\frac{4\pi}{\omega^{2}}\sum_{\ell=|s|}^{\infty}\left(2\ell+1\right)\sin^{2}\,_{s}\delta_{\ell s}=0\,,
\\
\sigma_{\rm abs}&=\frac{\pi}{\omega^{2}}\sum_{\ell=|s|}^{\infty}\left(2\ell+1\right)\left(1-\,_{s}r_{\ell s}^{2}\right)
\nonumber
\\
&=-\frac{2\pi P_{0}}{\omega^{2}}\sum_{\ell=|s|}^{\infty}\left(2\ell+1\right)\left[2\omega(r_{+}-r_{-})\right]^{2\ell+1}\left(\frac{\left(\ell+s\right)!\left(\ell-s\right)!}{\left(2\ell+1\right)!\left(2\ell\right)!}\right)^{2}\prod_{k=1}^{\ell}\left(k^{2}+4P_{0}^{2}\right)\,.
\end{align}
%%%%%%%%%%%%%%%%%%%%%%%%%%%%%%%%%%%%%%%%%%
Thus, there is non-trivial absorption, but zero elastic scattering. As we will show later, this relates the scattering cross sections to the LNs and dissipation numbers of BHs.   

Indeed, the scattering phase and the scattering amplitude can be related to the LN, as well as to the dissipation number, through computations of various scattering diagrams, see~\ref{scatteringmatch}. The leading order effect, associated with static gravitational perturbations, corresponds to the following process: BH+graviton $\to$ BH+graviton. At the leading order, the scattering amplitude for this process, in the language of Feynman diagrams becomes,
%%%%%%%%%%%%%%%%%%%%%%%%%%%%%%%%%%%%%%%%%%
\begin{equation}\label{static_scat}
  i \mathcal{M}(\boldsymbol{k}_{\rm in}\rightarrow \boldsymbol{k}_{\rm out},h \rightarrow h) = 
  \vcenter{\hbox{\begin{tikzpicture}[scale=0.7]
        \begin{feynman}
            \vertex (i) at (0,0);
            \vertex (e) at (0,3);
            %\node[circle, draw=Red, fill = Red, scale=0.5] (w1) at (0, 1.0);
            \node[circle, draw=Orange, fill = Orange, scale=0.5] (w2) at (0, 1.5);
            \vertex (f1) at (1.5,2.5) {};
            \vertex (fs) at (1.5,0.5) {};
            
            \diagram*{
                (i) -- [double, double, thick] (w2),
                (w1) -- [edge label = $\lambda^{E/B}$] (w1),
                %(w2) -- edge label = $\lambda^{E/B}$,
                (w2) -- [double, double, thick] (e),
                (f1) -- [boson, MyYellow, ultra thick] (w2),
                (fs) -- [boson, MyYellow, ultra thick] (w2)
            };
        \end{feynman}
    \end{tikzpicture}}}
    = - i \frac{\omega^4}{4 M_{\rm pl}^2}\Lambda^{\rm E/B}_{ij,kl}\epsilon_h^{kl}(\boldsymbol{k}_{\rm in}){\epsilon}_{h}^{*ij}(\boldsymbol{k}_{\rm out}) M (GM)^4 + {\rm magnetic} ~,
\end{equation}
%%%%%%%%%%%%%%%%%%%%%%%%%%%%%%%%%%%%%%%%%%
where the vertex corresponds to $-i\Lambda^{\rm E/B}_{ij,kl}M(GM)^{4}$, with $\Lambda^{\rm E/B}_{ij,kl}$ being the dimensionless Love tensor associated with the electric/magnetic sector. Note that this Love tensor is normalized with respect to the mass of the system and hence is related to the Love tensor defined here by a power of $(M/r_{+})^{2\ell+1}$. Each on-shell graviton external leg (depicted by yellow wavy lines) contributes the following term: $(\omega^{2}/2M_{\rm pl})\epsilon^{ij}_{h}$, where $\epsilon^{ij}_{h}$ is the polarization tensor, with $h$ being the helicity of the state. This is because, the tidal term depends on the second time derivative of the gravitational field and, for plane-wave states, space derivatives can be interchanged for time derivatives. The above amplitude can be transformed to the spherical amplitude $\mathcal{A}$ using~\ref{spherical_scat_transform}, yielding \citep{Saketh:2023bul},
%%%%%%%%%%%%%%%%%%%%%%%%%%%%%%%%%%%%%%%%%%
\begin{align}\label{scatteringampleading}
i\mathcal{A}(\omega;\ell=2,m,h \to \ell=2,m,h)=-\frac{iM(GM)^{4}\omega^{5}}{40\pi M_{\rm pl}^{2}}\left[\frac{2}{3}\left(\frac{r_{+}}{M}\right)^{5}k^{\rm E/B}_{2m}+i\frac{1}{3}\left(\frac{r_{+}}{M}\right)^{5}\nu^{\rm E/B}_{2m}\right]\,,
\end{align}
%%%%%%%%%%%%%%%%%%%%%%%%%%%%%%%%%%%%%%%%%%
where $k^{\rm E/B}_{2m}$ contains all the static coefficients associated with LNs in the Love tensor, while $\nu^{\rm E/B}_{2m}$ contains all the static coefficients associated with the dissipation numbers in the Love tensor. Thus, using~\ref{scatteringmatch} to match the scattering amplitude and phase computed from BH perturbation theory in~\ref{BHPTscattering} for $\ell=2$, with~\ref{scatteringampleading} and $M_{\rm pl}^{-2}=64\pi G$, we finally obtain:
%%%%%%%%%%%%%%%%%%%%%%%%%%%%%%%%%%%%%%%%%%
\begin{align}
k^{\rm E/B}_{\ell=2}&=0\,,
\label{love_static}
\\
\nu^{\rm E/B}_{\ell=2}&=-\frac{1}{15}m\chi \left(\frac{M}{r_{+}}\right)^{5}\left[1+(m^{2}-1)\chi^{2}\right]\left[4+(m^{2}-4)\chi^{2}\right]\,.
\label{diss_static}
\end{align}
%%%%%%%%%%%%%%%%%%%%%%%%%%%%%%%%%%%%%%%%%%
Thus, the static LN for the $\ell=2$ mode vanishes identically for Kerr BHs, as we had derived earlier from various different approaches. This holds for generic $\ell$ as well. The dissipation number, on the other hand, has a structure identical to~\ref{dissstatgen}, except for an overall scaling by a factor of two. To summarize, the static bosonic LNs of the Schwarzschild and Kerr BH vanish identically, irrespective of the adopted method to compute them.

%%%%%%%%%%%%%%%%%%%%%%%%%%%%%%%%%%%%
%%%%%%%%%%%%%%%%%%%%%%%%%%%%%%%%%%%%
%%%%%%%%%%%%%%%%%%%%%%%%%%%%%%%%%%%%
\subsection{Vanishing of non-linear static bosonic Love numbers} \label{sec:non-linear}

So far, we have only discussed the static bosonic LNs for BHs in GR, within linear perturbation theory and have shown them to be zero. The natural question, which we will review here, is whether the bosonic LNs remains zero even when non-linear effects are taken into account. We will address this problem using two different approaches. The first approach corresponds to the use of a suitable coordinate system in order to obtain the LNs in a non-perturbative manner, while the second one will be using a more conventional coordinate system and the LNs will be obtained by a perturbative approach, order by order in the perturbation.  

In the first approach, one starts from the Weyl form for the static and spherically symmetric spacetime, which expresses the metric in the cylindrical coordinate system $(t,\rho,\phi,z)$ and reads
%%%%%%%%%%%%%%%%%%%%%%%%%%%%%%%%%%%%%%%%%%
\begin{align}
ds^{2}=-f(\rho,z)dt^{2}+\frac{1}{f(\rho,z)}\left[e^{2\gamma(\rho,z)}\left(d\rho^{2}+dz^{2}\right)+\rho^{2}d\phi^{2}\right]\,,
\end{align}
%%%%%%%%%%%%%%%%%%%%%%%%%%%%%%%%%%%%%%%%%%
which has two unknowns $f(\rho,z)$ and  $\gamma(\rho,z)$, and can contrasted with~\ref{met_sph_stat}. When substituted in the vacuum Einstein's equations, $R_{\mu \nu}=0$, the above ansatz yields three independent equations for these two unknown functions:
%%%%%%%%%%%%%%%%%%%%%%%%%%%%%%%%%%%%%%%%%%
\begin{align}
f\nabla^{2}f&=\nabla f\cdot \nabla f\,,
\label{maineqnonlin}
\\
\dfrac{\partial \gamma}{\partial z}&=\frac{\rho}{2f^{2}}\left(\frac{\partial f}{\partial \rho}\right)\left(\frac{\partial f}{\partial z}\right)\,,
\\
\dfrac{\partial \gamma}{\partial \rho}&=\frac{\rho}{4f^{2}}\left[\left(\frac{\partial f}{\partial \rho}\right)^{2}-\left(\frac{\partial f}{\partial z}\right)^{2}\right]\,.
\end{align}
%%%%%%%%%%%%%%%%%%%%%%%%%%%%%%%%%%%%%%%%%%
Here $\nabla$ is the three dimensional gradient and $\nabla^{2}=\rho^{-1}(\partial/\partial\rho)\{\rho(\partial/\partial \rho)\}+\rho^{-2}(\partial^{2}/\partial \phi^{2})+(\partial^{2}/\partial z^{2})$ is the three dimensional Laplacian in cylindrical coordinates. One can show that the function $f(\rho,z)$ is actually the Ernest potential associated with this problem leading to mass multipole moments \citep{Ernst:1967wx,Ernst:1967by}. Further, expressing $f=e^{2U}$,~\ref{maineqnonlin} reduces to a very simple equation,
%%%%%%%%%%%%%%%%%%%%%%%%%%%%%%%%%%%%%%%%%%
\begin{equation}
    \nabla^{2}U=0\,,
\end{equation}
%%%%%%%%%%%%%%%%%%%%%%%%%%%%%%%%%%%%%%%%%%
which is a linear equation emerging out of the non-linear Einstein's equations. 
This property ultimately stems from the integrability of Einstein's equations in static and spherically symmetric configurations, which implies the existence of Killing vectors and higher-rank Killing tensors. The latter are closely related to the presence of Lax pairs in integrable systems~\citep{Rosquist:1994yd, Rosquist:1997fn}.

Subsequently, one transforms from cylindrical coordinates to prolate spheroidal coordinates  $(t,x,y,\phi)$, with $|x|\geq1$ and $|y|\leq 1$, such that $\rho=\rho_{0}\sqrt{x^{2}-1}\sqrt{1-y^{2}}$ and $z=\rho_{0}xy$. This will simplify the solution for $U$, by reducing the differential equation $\nabla^{2}U=0$ to the following form,
%%%%%%%%%%%%%%%%%%%%%%%%%%%%%%%%%%%%%%%%%%
\begin{align}
\bigg[\dfrac{\partial}{\partial x}\left\{(x^{2}-1)\dfrac{\partial}{\partial x}\right\}+\dfrac{\partial}{\partial y}\left\{(1-y^{2})\dfrac{\partial}{\partial y}\right\}\bigg]U(x,y)=0\,,
\end{align}
%%%%%%%%%%%%%%%%%%%%%%%%%%%%%%%%%%%%%%%%%%
with the following solution:
%%%%%%%%%%%%%%%%%%%%%%%%%%%%%%%%%%%%%%%%%%
\begin{align}\label{gensolnonlin}
U(x,y)=\sum_{\ell}\left\{c_{1}^{(\ell)}P_{\ell}(x)+c_{2}^{(\ell)}Q_{\ell}(x)\right\}P_{\ell}(y)\,,
\end{align}
%%%%%%%%%%%%%%%%%%%%%%%%%%%%%%%%%%%%%%%%%%
where $P_{\ell}$ is the Legendre function and $Q_{\ell}$ is the associated Legendre function. If we keep only the $\ell=0$ mode, i.e., for the monopole mode alone, we obtain, $U\sim \textrm{constant}+\ln \sqrt{\{(x-1)/(x+1)\}}$. This leads to the following $g_{tt}$ component, $f\sim \{(x-1)/(x+1)\}$, which results into the usual Schwarzschild spacetime. The quadrupolar contribution to the metric function $U$ can be obtained from the $\ell=2$ mode, which has the structure 
\begin{equation}
  U=c_{1}^{(2)}(\textrm{polynomial~in~}x)+c_{2}^{(2)}\{\textrm{polynomial~in~}x+\ln (x-1)\}  \,.
\end{equation}
Since the presence of the $\ln (x-1)$ term leads to a singular Kretschmann scalar at the Schwarzschild horizon (note that horizon with the monopole term may not coincide with the horizon with monopole+quadrupole term), one must set the arbitrary constant $c_{2}^{(2)}$ to zero. This implies that, except for the monopole term, all the higher order terms are growing radial functions, and hence the corresponding LNs will vanish for all orders of $\ell$ \citep{Kehagias:2024rtz,Gounis:2024hcm,Combaluzier-Szteinsznaider:2024sgb}. Note that the solution for $U$, presented in~\ref{gensolnonlin} is obtained from the full non-linear Einstein's equations and hence the result derived from its behavior is also a non-linear/non-perturbative phenomenon, suggesting that the vanishing of static bosonic LNs is not just a consequence of linear perturbation theory, but holds for full non-linear general relativity. This result has also been generalized to Kerr BHs \citep{Gounis:2024hcm} and it was shown that the vanishing of non-linear static LNs follow from the existence of ladder symmetry, as detailed in~\ref{sec:ZeroLoveSymm}.

Two comments regarding the above derivation are in order: (a) this result holds only for the electric LNs, and not for the axial ones, since the unperturbed metric for a static and spherically symmetric spacetime is always in the polar form, upon using the Regge--Wheeler gauge, and (b) the above approach uses BH perturbation theory alone, however, does not match this result with gauge invariant quantities, e.g., the couplings in the EFT. 

The second approach \citep{Riva:2023rcm,Iteanu:2024dvx}, on the other hand, is perturbative and hence is not as general as above, but it provides LNs for both electric and magnetic sectors, and one can match these results with appropriate Feynman diagrams in the scattering approach (see also \citealt{DeLuca:2023mio} for a similar computation, but for the scalar case). The details of the first order perturbation for a static and spherically symmetric background spacetime can be obtained from~\ref{vac_Sch_LN}, which shows that $H_{0}=-(\delta g_{tt}/f)$ (ignoring angular factors) is the key quantity for determination of the polar LNs. Similarly, $h_{0}=-\delta g_{t\phi}$ (again, ignoring angular factors) is the key quantity for the determination of the axial LNs. Following \citet{Riva:2023rcm,Iteanu:2024dvx}, we focus on the even sector, however, an identical procedure can be adopted for the axial sector as well. 

In the even sector, after fixing all the gauges appropriately, and then taking the static limit, we obtain, three non-zero perturbations $H_{0}$, $H_{2}$ and $K$. At linear order, one can show that $H_{0}=H_{2}$, $K$ is determined in terms of $H_{0}$ and its derivatives, with $H_{0}$ yielding a single master equation to solve (see~\ref{vac_Sch_LN} for details). At the quadratic order in the perturbation, a similar result follows, where the quadratic perturbation satisfies an identical equation, but now with a source term, which is quadratic in the first order perturbations:
%%%%%%%%%%%%%%%%%%%%%%%%%%%%%%%%%%%%%%%%%%%%%%%%%%%%%%%%%%%%%%%%%%%%%%%%%%%%%%
\begin{equation}
H^{(2)''}_{0}+\frac{2(r-M)}{r(r-2M)}H^{(2)'}_{0}-\frac{\ell(\ell+1)r(r-2M)+4M^{2}}{r^{2}(r-2M)^{2}}H^{(2)}_{0}=S(H_{0}^{2},H^{'2}_{0})\,.
\end{equation}
%%%%%%%%%%%%%%%%%%%%%%%%%%%%%%%%%%%%%%%%%%%%%%%%%%%%%%%%%%%%%%%%%%%%%%%%%%%%%%
Thus, one must solve for the first order perturbation, whose details are provided in~\ref{vac_Sch_LN}, and then substitute the same in the above equation and solve for the quadratic perturbation $H_{0}^{(2)}$. The relevant equations for $K$ and $H_{2}$ at quadratic order can be found in \citet{Riva:2023rcm,Iteanu:2024dvx}. The homogeneous solution for the quadratic perturbation is given by~\ref{Hext}, in terms of associated Legendre polynomials, which can also be expressed in terms of hypergeometric functions using~\ref{legPintHyp} and~\ref{legQHyp}, such that \citep{Katagiri:2024wbg, Chakraborty:2025wvs}, 
%%%%%%%%%%%%%%%%%%%%%%%%%%%%%%%%%%%%%%%%%%%%%%%%%%%%%%%%%%%%%%%%%%%%%%%%%%%%%%
\begin{align}\label{staticSchhyp}
H^{(2)}_{0}|_{\rm homog.}&=\mathcal{A}\underbrace{f\left(\frac{r}{2M}\right)^{\ell}\,_{2}F_{1}\left(2-\ell,-\ell;-2\ell;\frac{2M}{r}\right)}_{\textrm{ tidal\,part}=H_{\rm T}}
\nonumber
\\
&\qquad+\mathcal{B}\underbrace{f\left(\frac{2M}{r}\right)^{1+\ell}\,_{2}F_{1}\left(1+\ell,3+\ell;2+2\ell;\frac{2M}{r}\right)}_{\textrm{ response}=H_{\rm R}}\,.
\end{align}
%%%%%%%%%%%%%%%%%%%%%%%%%%%%%%%%%%%%%%%%%%%%%%%%%%%%%%%%%%%%%%%%%%%%%%%%%%%%%%
At the linear level there are no source terms in vacuum GR and hence the above is the full solution for the perturbations, where regularity at the horizon demands $\mathcal{B}$ to vanish and hence the LNs also vanish. However, at the quadratic order, there is a source term and hence it is not clear, a priori, if the quadratic LNs will vanish as well. Further, using~\ref{hypdiffarg}, it is possible to rewrite the hypergeometric function associated with $r^{-1-\ell}$, and using~\ref{hyp_z_2byz}, we can rewrite the $r^{\ell}$ term, such that,
%%%%%%%%%%%%%%%%%%%%%%%%%%%%%%%%%%%%%%%%%%%%%%%%%%%%%%%%%%%%%%%%%%%%%%%%%%%%%%
\begin{align}
H^{(2)}_{0}|_{\rm homog.}&\sim\frac{(-1)^{\ell+1}\mathcal{A}}{f}\,_{2}F_{1}\left(-\ell,1+\ell;3;\frac{r}{2M}\right)
\nonumber
\\
&\qquad+\frac{\mathcal{B}}{f}\left(\frac{2M}{r}\right)^{1+\ell}\,_{2}F_{1}\left(1+\ell,\ell-1;2+2\ell;\frac{2M}{r}\right)\,,
\end{align}
%%%%%%%%%%%%%%%%%%%%%%%%%%%%%%%%%%%%%%%%%%%%%%%%%%%%%%%%%%%%%%%%%%%%%%%%%%%%%%
which matches with \citep{Riva:2023rcm}. The inhomogeneous solution can be obtained using the Green's function technique and is given by, 
%%%%%%%%%%%%%%%%%%%%%%%%%%%%%%%%%%%%%%%%%%%%%%%%%%%%%%%%%%%%%%%%%%%%%%%%%%%%%%
\begin{align}
H_{0}^{(2)}=\mathcal{A}H_{\rm T}+\mathcal{B}H_{\rm R}+\int_{2M}^{\infty}dr'S(r')G(r,r')\,,
\end{align}
%%%%%%%%%%%%%%%%%%%%%%%%%%%%%%%%%%%%%%%%%%%%%%%%%%%%%%%%%%%%%%%%%%%%%%%%%%%%%%
where $H_{\rm T}$ denotes the tidal part, while $H_{\rm R}$ is the response part of the homogeneous solution. The inhomogeneous part involves the Green's function, which upon imposing regularity at the horizon, becomes
%%%%%%%%%%%%%%%%%%%%%%%%%%%%%%%%%%%%%%%%%%%%%%%%%%%%%%%%%%%%%%%%%%%%%%%%%%%%%%
\begin{align}
G(r,r')=\frac{r'(r'-2M)}{W_{0}(2M)^{2}}\left[H_{\rm R}(r)H_{\rm T}(r')\theta(r-r')+H_{\rm R}(r')H_{\rm T}(r)\theta(r'-r)\right]\,,
\end{align}
%%%%%%%%%%%%%%%%%%%%%%%%%%%%%%%%%%%%%%%%%%%%%%%%%%%%%%%%%%%%%%%%%%%%%%%%%%%%%%
since for $r\simeq 2M<r'$, only $H_{\rm T}$ matters, which is regular. For generic choices of $\ell$, it is difficult to determine the full solution for $\delta g_{tt}$, including the quadratic perturbation. Note that a given $\ell$ mode at the linear level can give rise to $0$, $\ell$, $2\ell$ modes at quadratic level, due to angular momentum coupling. In the case of quadrupolar perturbation ($\ell=2$ for both linear and quadratic perturbation), it follows that the perturbation, regular at the horizon, is purely growing \citep{Riva:2023rcm}, leading to vanishing quadratic polar LNs. Identically, both $K$ and $H_{2}$ also shows a growing behavior, as in the case of linear perturbation.

This result can be further verified using EFT/scattering approaches. In the previous EFT section, we have shown that there are typically five terms, the point particle, interaction between gravitational perturbation around flat spacetime, interaction between gravitational perturbation and the tidal field, dynamics of tidal field and finally finite size interactions. In the case of non-linear coupling involving the quadrupolar modes only, the finite size term can be expressed as, 
%%%%%%%%%%%%%%%%%%%%%%%%%%%%%%%%%%%%%%%%%%%%%%%%%%%%%%%%%%%%%%%%%%%%%%%%%%%%%%
\begin{align}
\mathcal{A}_{\rm finite}=\int d\tau \left[\lambda_{2}^{\rm E}E_{ij}E^{ij}+\lambda_{2}^{\rm B}B_{ij}B^{ij}+\lambda_{222}^{\textrm{E}^{2}}E_{ij}E^{i}_{k}E^{jk}+\cdots \right]\,,
\end{align}
%%%%%%%%%%%%%%%%%%%%%%%%%%%%%%%%%%%%%%%%%%%%%%%%%%%%%%%%%%%%%%%%%%%%%%%%%%%%%%
where, the ``$\cdots$" denotes purely magnetic as well as mixed terms involving both magnetic and electric tidal fields. Further generalizing the results in~\ref{EFTScalar}, \citet{Riva:2023rcm} computed the corrections to the tidal field at $\mathcal{O}(G^{2})$, and studied the quadratic response function. This will involve diagrams like
%%%%%%%%%%%%%%%%%%%%%%%%%%%%%%%%%%%%%%%%%%
\begin{equation}
  \vcenter{\hbox{\begin{tikzpicture}[scale=0.7]
        \begin{feynman}
            %\vertex (i) at (0,0);
            %\vertex (e) at (0,3);
            %\node[circle, draw=Red, fill = Red, scale=0.5] (w1) at (0, 1.0);
            \node[circle, draw=Orange, fill = Orange, scale=1] (w2) at (0, 0);
            %\vertex (f1) at (1.5,2.5) {};
            \vertex (fs) at (2.5,0) {};
            \vertex (f2) at (-1.5,-2.5) {};
            \vertex (fq) at (1.5,-2.5) {};
            
            \diagram*{
                %(i) -- [double, double, thick] (w2),
                %(w2) -- [edge label = $\lambda^{E/B}$] (w2),
                %(w2) -- edge label = $\lambda^{E/B}$,
                %(w2) -- [double, double, thick] (e),
                %(f1) -- [boson, MyYellow, ultra thick] (w2),
                (fs) -- [boson, MyYellow, ultra thick] (w2),
                (f2) -- [fermion, MyYellow, ultra thick] (w2),
                (fq) -- [fermion, MyYellow, ultra thick] (w2)
            };
        \end{feynman}
    \end{tikzpicture}}}
\end{equation}
%%%%%%%%%%%%%%%%%%%%%%%%%%%%%%%%%%%%%%%%%%
which denotes a one-point function in which two tidal legs interact with the point particle, leading to GW emission. Note that the number of legs, excluding the wavy line, indicates the order in the tidal field; for instance, two legs correspond to a quadratic tidal interaction. Evaluation of this diagram following \citet{Riva:2023rcm}, yields vanishing quadratic LNs, i.e., $\lambda_{222}=0$. Thus, the vanishing of LNs follow from quadratic terms in the finite size action within the EFT formalism.

%%%%%%%%%%%%%%%%%%%%%%%%%%%%%%%%%%%%%%%%%%%%%%%%%%%%%%%%%%%%%
%%%%%%%%%%%%%%%%%%%%%%%%%%%%%%%%%%%%%%%%%%%%%%%%%%%%%%%%%%%%%
%%%%%%%%%%%%%%%%%%%%%%%%%%%%%%%%%%%%%%%%%%%%%%%%%%%%%%%%%%%%%
\section{Static bosonic Love numbers of black holes beyond four-dimensional vacuum General Relativity} \label{sec:BHBGR}

We now turn to the discussion of LNs for BHs in various theories of gravity, as well as for BHs in GR with additional matter degrees of freedom. 
The key question is whether the static bosonic\footnote{The investigation of the fermionic LNs is very recent and so far done only for GR BHs in vacuum \citep{Chakraborty:2025zyb} (see~\ref{sec:fermionic}). In the remaining of this section we will restrict to bosonic LNs.} LNs for BHs beyond vacuum GR are nonzero. 
If so, they could provide tantalizing evidence for BH spacetimes that differ from the Schwarzschild and Kerr solutions, potentially detectable through GW signals. 
Broadly, there are three main directions in the study of BHs beyond vacuum GR: 
(a) four-dimensional BHs in GR, but with non-vacuum spacetimes; 
(b) BHs that are solutions to the vacuum Einstein equations in higher-dimensional spacetimes; and 
(c) four-dimensional vacuum BHs that are not solutions of Einstein's equations. 
We will consider each of these cases, and their sub-classes, in the following discussion.

%%%%%%%%%%%%%%%%%%%%%%%%%%%%%%%%%%%%
%%%%%%%%%%%%%%%%%%%%%%%%%%%%%%%%%%%%
%%%%%%%%%%%%%%%%%%%%%%%%%%%%%%%%%%%%
\subsection{Love numbers of black holes surrounded by a stress-energy tensor}

We will consider three representative cases:
\begin{enumerate}
    \item A BH embedded in a spherical matter distribution;
    \item A BH with an electric field, described as an exact solution to Einstein-Maxwell theory; 
    \item A BH surrounded by an accretion disk.
\end{enumerate}

We will discuss all of these cases separately, along with the relevant equations governing the tidal deformability and of course the determination of the LNs. We will mostly focus on static and spherically symmetric background, since in most of these cases the determination of the LNs associated with rotating cases are unknown.

\subsubsection{Gravitational perturbations of GR with an anisotropic fluid} 
The first two cases of the above list, namely BHs surrounded by an electric field or by a spherical matter distribution, can be studied within a unified framework by considering GR coupled to an anisotropic perfect fluid with stress-energy tensor
\begin{equation}
    T_{\mu \nu}=\rho u_{\mu}u_{\nu}+p_{r}v_{\mu}v_{\nu}+p_{t}\Pi_{\mu \nu}\,,
\end{equation}
where $\Pi_{\mu \nu}=g_{\mu \nu}+u_{\mu}u_{\nu}-v_{\mu}v_{\nu}$, 
$\rho$ depicts the energy density, $p_{r}$ is the radial pressure, and $p_{t}$ depicts the tangential pressures along the transverse directions. For the background static and spherically symmetric spacetime, we consider the following ansatz, 
%%%%%%%%%%%%%%%%%%%%%%%%%%%%%%%%%%%%%%%%%%%%%%%%%%%%%%%%%%%%%%%%%%%%%%%%%%%%%%
\begin{equation}
ds^{2}=-f(r)dt^{2}+\frac{dr^{2}}{g(r)}+r^{2}d\Omega^{2}~;
\qquad
g(r)\equiv 1-\frac{2m(r)}{r}~, \label{ansatzanisotropic}
\end{equation}
%%%%%%%%%%%%%%%%%%%%%%%%%%%%%%%%%%%%%%%%%%%%%%%%%%%%%%%%%%%%%%%%%%%%%%%%%%%%%%
where $f(r)$ and $g(r)$ (or, equivalently $m(r)$, the mass function) are arbitrary functions of the radial coordinate $r$. In this case, it follows that $u_{\mu}=-\sqrt{f(r)}\delta_{\mu}^{t}$, as well as $v^{\mu}=\sqrt{g(r)}\delta^{\mu}_{r}$.

We now consider linearized Einstein's equations due to perturbations in the geometry sector ($g_{\mu \nu}\to g_{\mu \nu}+\delta g_{\mu \nu}$) as well as perturbations in the matter sector ($T_{\mu \nu}\to T_{\mu \nu}+\delta T_{\mu \nu}$). To the linear order in the perturbations, there are equations coming from the perturbed Einstein's equations, as well as the perturbed conservation equation for the energy-momentum tensor, 
%%%%%%%%%%%%%%%%%%%%%%%%%%%%%%%%%%%%%%%%%%%%%%%%%%%%%%%%%%%%%%%%%%%%%%%%%%%%%%
\begin{equation}
\delta G_{\mu \nu}=8\pi \delta T_{\mu \nu}~;
\qquad 
\delta \left(\nabla_{\nu}T^{\mu \nu}\right)=0~.
\end{equation}
%%%%%%%%%%%%%%%%%%%%%%%%%%%%%%%%%%%%%%%%%%%%%%%%%%%%%%%%%%%%%%%%%%%%%%%%%%%%%%
The perturbation of the metric can be decomposed into axial and polar sectors, depending on the transformation of the metric components under parity, see~\ref{sec:relLNs}. The explicit expressions for the axial metric perturbations can be obtained from~\ref{pert_grav_ax}, and consist of two radial functions: $\{h_{0},h_{1}\}$.
%%%%%%%%%%%%%%%%%%%%%%%%%%%%%%%%%%%%%%%%%%%%%%%%%%%%%%%%%%
%\begin{align}
%\delta g_{t\theta}=\delta g_{\theta t}=-\sum_{\ell,m}e^{-i\omega t}\frac{h_0(r)}{\sin\theta}\partial_\phi Y_{\ell m}~;
%\quad 
%\delta g_{t\phi}=\delta g_{\phi t}=\sum_{\ell,m}e^{-i\omega t}h_0(r)\sin\theta\partial_\theta Y_{\ell m}~, 
%\\
%\delta g_{r\theta}=\delta g_{\theta r}=-\sum_{\ell,m}e^{-i\omega t}\frac{h_1(r)}{\sin\theta}\partial_\phi Y_{\ell m}~;
%\quad 
%\delta g_{r\phi}=\delta g_{\phi r}=\sum_{\ell,m}e^{-i\omega t}h_1(r)\sin\theta\partial_\theta Y_{\ell m}~. 
%\end{align}
%%%%%%%%%%%%%%%%%%%%%%%%%%%%%%%%%%%%%%%%%%%%%%%%%%%%%%%%%%
Here, and also below, all the radial functions depend on the angular separation constants $(\ell,m)$, which we have skipped for notational convenience. The polar metric perturbations, on the other hand, we have four arbitrary radial functions: $\{H_{0},H_{2},K,H_{1}\}$. The connection of these function with metric perturbation has been presented in~\ref{pert_grav_pol}.
%%%%%%%%%%%%%%%%%%%%%%%%%%%%%%%%%%%%%%%%%%%%%%%%%%%%%%%%%%
%\begin{align}
%\delta g_{tt}&=\sum_{\ell,m}e^{-i\omega t} f(r) H_0(r)Y_{\ell m}~;
%\qquad
%\delta g_{tr}=\delta g_{rt}=\sum_{\ell,m}e^{-i\omega t} H_1(r)Y_{\ell m}~,
%\\
%\delta g_{rr}&=\sum_{\ell,m}e^{-i\omega t} \frac{H_2(r)}{g(r)}Y_{\ell m}~;
%\qquad
%\delta g_{\theta\theta}=\sum_{\ell,m}e^{-i\omega t} r^2 K(r)Y_{\ell m}=\frac{\delta g_{\phi\phi}}{\sin^2\theta}~. 
%\end{align}
%%%%%%%%%%%%%%%%%%%%%%%%%%%%%%%%%%%%%%%%%%%%%%%%%%%%%%%%%%

In the matter sector, there will be contributions from the perturbations of the energy density ($\rho\to \rho+\delta \rho$), radial and tangential pressure ($p_{r}\to p_{r}+\delta p_{r}$ and $p_{t}\to p_{t}+\delta p_{t}$), as well as perturbations of the fluid four-velocity ($u^{\mu}\to u^{\mu}+\delta u^{\mu}$) and radial velocity ($v^{\mu}\to v^{\mu}+\delta v^{\mu}$). The scalar perturbations, namely $\delta \rho$, $\delta p_{r}$ and $\delta p_{t}$, takes the following form, owing to the static and spherically symmetric nature of the background spacetime:
%%%%%%%%%%%%%%%%%%%%%%%%%%%%%%%%%%%%%%%%%%%%%%%%%%%%%%%%%%
\begin{align}
\delta X(t,r,\theta,\phi)&=\sum_{\ell,m}\delta X(r)Y_{\ell m}e^{-i\omega t}~, 
% \\
% \delta p_{\rm r}(t,r,\theta,\phi)&=\sum_{\ell,m}e^{-i\omega t}\delta p_{\rm r}(r)Y_{\ell m}~,\\
% \delta p_{\rm t}(t,r,\theta,\phi)&=\sum_{\ell,m}e^{-i\omega t}\delta p_{\rm t}(r)Y_{\ell m}~.
\end{align}
%%%%%%%%%%%%%%%%%%%%%%%%%%%%%%%%%%%%%%%%%%%%%%%%%%%%%%%%%%
where $\delta X$ collectively denotes $(\delta\rho, \delta p_{\rm r}, \delta p_{\rm t})$.
These perturbations contribute to the polar sector of the field equations, as these are scalar quantities. On the other hand, the perturbations of the fluid four velocity $\delta u^{\mu}$ and the radial vector $\delta v^{\mu}$ will contribute to both the axial and the polar sectors. In particular,  for the static and spherically symmetric background spacetime, the axial sector of the perturbed four-vectors $\delta u^{\mu}_{\rm axial}$ and $\delta v^{\mu}_{\rm axial}$ can be decomposed in vector spherical harmonics as
%%%%%%%%%%%%%%%%%%%%%%%%%%%%%%%%%%%%%%%%%%%%%%%%%%%%%%%%%%
\begin{align}
\delta u_{\rm axial}^\theta &=-\sum_{\ell,m}\frac{\sqrt{f}U^{\rm{up}}(r)e^{-i\omega t}}{4\pi(p_{\rm t}+\rho)r^2}\frac{\partial_\phi Y_{\ell m}}{\sin\theta}\,; 
\quad 
\delta u_{\rm axial}^\phi=\sum_{\ell,m}\frac{\sqrt{f}U^{\rm{up}}(r)e^{-i\omega t}}{4\pi(p_{\rm t}+\rho) r^2\sin^2\theta }\sin\theta\partial_\theta Y_{\ell m}\,,
\\
\delta v_{\rm axial}^\theta &=-\sum_{\ell,m}\frac{\sqrt{g}V^{\rm{up}}(r)e^{-i\omega t}}{4\pi(p_{\rm t}+\rho)r^2}\frac{\partial_\phi Y_{\ell m}}{\sin\theta}\,;
\quad 
\delta v_{\rm axial}^\phi=\sum_{\ell,m}\frac{\sqrt{g}V^{\rm{up}}(r)e^{-i\omega t}}{4\pi(p_{\rm t}+\rho)r^2\sin^2\theta }\sin\theta \partial_\theta Y_{\ell m}\,.
\end{align}
%%%%%%%%%%%%%%%%%%%%%%%%%%%%%%%%%%%%%%%%%%%%%%%%%%%%%%%%%%
The two unknown functions $U^{\rm up}$ and $V^{\rm up}$, originating from the perturbation of the fluid four velocity and the perturbation of the radial vector, acts as a source term of the perturbed Einstein's equations \citep{Chakraborty:2024gcr}. One may choose them to be zero (as in the standard treatment of Einstein-Maxwell theory), or determine them from some additional principle, depending on the actual model at hand \citep{Chakraborty:2024gcr}. 
Since in general there is no first principle derivation for these components of perturbed four-vector, we will either set (a) $\delta u_{\rm axial}^{\mu}=0=\delta v_{\rm axial}^{\mu}$ (referred to as the `up' choice), which corresponds to $U^{\rm up}=0=V^{\rm up}$, or, (b) $\delta u^{\rm axial}_{\mu}=0=\delta v^{\rm axial}_{\mu}$ (referred to as the `down' choice), leading to, $U^{\rm{up}}=-4\pi(\rho+p_{\rm t})f^{-1}h_0$ and $V^{\rm{up}}=-4\pi(\rho+p_{\rm t})h_{1}$ \citep{Chakraborty:2024gcr}. 
It turns out that, these two choices for the perturbed four vectors will lead to different perturbed Einstein's equations. Thus, in the axial sector, there are two unknowns: $(h_{0},h_{1})$. Interestingly, in the axial sector the perturbation of the conservation relation, i.e., $\delta \left(\nabla_{\nu}T^{\mu \nu}\right)=0$ does not contribute, the only contribution comes from the perturbed Einstein's equations. Since our interest is in the determination of static LNs, we consider the relevant equations in the $\omega\to 0$ limit. In this case, it follows that $h_{1}=0$ and $h_{0}$ satisfies the following differential equations, depending on the `up' or `down' conventions, 
%%%%%%%%%%%%%%%%%%%%%%%%%%%%%%%%%%%%%%%%%%%%%%%%%%%%%%%%%%
\begin{align}
&gh_{0}''-4\pi r\left(p_{\rm r}+\rho\right)h_{0}'-\left[\frac{(\ell+2)(\ell-1)}{r^2}+\frac{2g}{r^2}+8\pi \left(\rho-p_{\rm r}+2p_{\rm t}\right)\right]h_{0}=0\,,
\quad 
\textrm{(up)}
\label{eqaxialBHDMup}
\\
&gh_{0}''-4\pi r\left(p_{\rm r}+\rho\right)h_{0}'-\left[\frac{(\ell+2)(\ell-1)}{r^2}+\frac{2g}{r^2}-8\pi\left(\rho+p_{\rm r}\right)\right]h_{0}=0~.
\quad 
\textrm{(down)}
\label{eqaxialBHDMdown}
\end{align}
%%%%%%%%%%%%%%%%%%%%%%%%%%%%%%%%%%%%%%%%%%%%%%%%%%%%%%%%%%
These are the key equations in the context of static tidal perturbations of an anisotropic fluid living in a curved background in the axial sector. The applicability of these equations are very general, ranging from Schwarzschild BHs immersed in a spherical matter distribution to magnetic LNs of anisotropic neutron star. In what follows, we will apply these equations to determine the LNs for Reissner--Nordstr\"om BH, as well as BH living in a dark matter halo.  

In the polar sector, on the other hand, the perturbed components of the four velocities depend on the metric perturbations, as well as two unknown functions $C^{\rm up}$ and $D^{\rm up}$, such that the non-zero components of these perturbed four velocities can be decomposed as
%%%%%%%%%%%%%%%%%%%%%%%%%%%%%%%%%%%%%%%%%%%%%%%%%%%%%%%%%%
\begin{align}
\delta u^{t}_{\rm polar}&=\sum_{\ell,m}e^{-i\omega t}\frac{H_{0}(r)}{2\sqrt{f}}Y_{\ell m}~;
\qquad
\delta u^{r}_{\rm polar}=\sum_{\ell,m}e^{-i \omega t}\frac{g}{\sqrt{f}r^2}\frac{D^{\rm up}(r)Y_{\ell m}}{16\pi(\rho+p_{r})}~,
\\
\delta u^{\theta}_{\rm polar}&=\sum_{\ell,m}e^{-i \omega t}\frac{\sqrt{f}C^{\rm up}(r)\partial_{\theta}Y_{\ell m}}{4\pi r^{2}(\rho+p_{t})}~;
\qquad 
\delta u^{\phi}_{\rm polar}=\sum_{\ell,m}e^{-i \omega t}\frac{\sqrt{f}C^{\rm up}(r)\partial_{\phi}Y_{\ell m}}{4\pi r^{2}\sin^{2}\theta(\rho+p_{t})}~,
\\
\delta v^{t}_{\rm polar}&=\sum_{\ell,m}e^{-i \omega t}\frac{\sqrt{g}}{f}\left[\frac{D^{\rm up}}{16\pi r^2 (\rho+p_{r})}+H_{1}\right]Y_{\ell m}~;
\quad
\delta v^{r}_{\rm polar}=-\sum_{\ell,m}e^{-i \omega t}\frac{\sqrt{g}H_{2}}{2}Y_{\ell m}~.
\end{align}
%%%%%%%%%%%%%%%%%%%%%%%%%%%%%%%%%%%%%%%%%%%%%%%%%%%%%%%%%%
Given the above components of the perturbed four velocities along with the perturbed energy density and pressures, there are contributions from the perturbed Einstein's equations, as well as perturbed conservation equation. In the polar sector there are nine unknowns to be solved for: $(\delta \rho, \delta p_{r}, \delta p_{t})$, $(C^{\rm up},D^{\rm up})$ and $(H_{0},H_{1},H_{2},K)$. As emphasized earlier, our interest is in the static LNs, and hence we consider the $\omega\to 0$ limit of the above perturbed equations. In this limit, the perturbation of the conservation equation fixes $\delta p_{\rm t}$ and $\delta \rho$ through the following equations,
%%%%%%%%%%%%%%%%%%%%%%%%%%%%%%%%%%%%%%%%%%%%%%%%%%%%%%%%%%
\begin{align}
\delta p_t&=\left(\frac{\rho+p_r}{2}\right) H\,, 
\label{deltapt}
\\
\left(\frac{f'}{2f}\right)\delta \rho&=-\left(\frac{1+3g+8\pi r^{2}\bar{p}_{r}}{2rg}\right)\delta p_{r}+\left(\frac{\bar{p}_{r}+\bar{\rho}}{r}\right)H-\delta p_{r}'+\left(\frac{\bar{p}_{r}+\bar{\rho}}{2}\right)H'
\nonumber
\\
&+\left[\frac{r\bar{p}_{r}'}{2}+\frac{(\bar{\rho}+\bar{p}_{r})(1-g+8\pi r^{2}\bar{p}_{r})}{4g}\right]K'~,
\label{deltarho}
\end{align}
%%%%%%%%%%%%%%%%%%%%%%%%%%%%%%%%%%%%%%%%%%%%%%%%%%%%%%%%%%
Where we have used the fact that $H_{0}=H_{2}\equiv H(r)$, which follows from $\delta G^{\theta}_{\theta}=\delta G^{\phi}_{\phi}$. In addition, the perturbation of the conservation equation provides the following relation between matter perturbations $C^{\rm up}$ and $D^{\rm up}$, 
%%%%%%%%%%%%%%%%%%%%%%%%%%%%%%%%%%%%%%%%%%%%%%%%%%%%%%%%%%
\begin{align}\label{CupDup}
\ell(\ell+1) C^{\rm up} = \frac{g}{4f}D^{\rm up \prime}+\frac{1-g+4\pi r^2(\bar p_r -\bar \rho)}{4 r f}D^{\rm up}\,. 
\end{align}
%%%%%%%%%%%%%%%%%%%%%%%%%%%%%%%%%%%%%%%%%%%%%%%%%%%%%%%%%%
This completes the set of the equations obtained from the perturbation of matter conservation. The perturbed Einstein's equations, on the other hand, yield one algebraic equation for the metric perturbation $H_{1}$,
%%%%%%%%%%%%%%%%%%%%%%%%%%%%%%%%%%%%%%%%%%%%%%%%%%%%%%%%%%
\begin{align}\label{H1aniso}
H_{1}=-\frac{1}{\ell(\ell+1)}D^{\rm up}~,
\end{align}
%%%%%%%%%%%%%%%%%%%%%%%%%%%%%%%%%%%%%%%%%%%%%%%%%%%%%%%%%%
along with two first order equations for the metric perturbations $K$ and $H$. These two first order equations can be massaged to obtain the metric perturbation $K$ in terms of $H$, its derivative, and $\delta p_{r}$,
%%%%%%%%%%%%%%%%%%%%%%%%%%%%%%%%%%%%%%%%%%%%%%%%%%%%%%%%%%
\begin{align}\label{Kstatic}
K&=\left[\frac{1-g+8\pi r^{2}\bar{p}_{r}}{(\ell-1)(\ell+2)}\right]\left(rH'\right)-\frac{16 \pi  r^{2}}{(\ell-1)(\ell+2)}\delta p_{r}
\nonumber
\\
&\qquad +\left[\frac{(\ell-1)(\ell+2)g-16\pi g r^{2}\bar{p}_{r}+(1+8\pi r^{2}\bar{p}_{r})^{2}-g^{2}}{(\ell-1)(\ell+2)g} \right]H~.
\end{align}
%%%%%%%%%%%%%%%%%%%%%%%%%%%%%%%%%%%%%%%%%%%%%%%%%%%%%%%%%%
Using the above equation for $K$, the first order equations will give rise to the following second order equation for the metric perturbation $H(r)$, 
%%%%%%%%%%%%%%%%%%%%%%%%%%%%%%%%%%%%%%%%%%%%%%%%%%%%%%%%%%
\begin{align}\label{Hdiffstatic}
H''&+\left[\frac{1+g+4\pi r^2 (\bar p_r-\bar\rho)}{rg}\right]H'+H\Bigg[\frac{(1-g+8\pi r^2 \bar p_r)\{2g-4\pi r^2 (\bar{\rho}+\bar p_r)\}}{r^{2}g^{2}}
\nonumber
\\
&+\dfrac{d}{dr}\left(\frac{1-g+8\pi r^{2}\bar{p}_{r}}{rg}\right)-\frac{\ell(\ell+1)-8\pi r^{2}(\bar{\rho}+\bar p_r)}{r^{2}g}\Bigg]+\frac{8\pi (\delta \rho+\delta p_{r})}{g}=0~.
\end{align}
%%%%%%%%%%%%%%%%%%%%%%%%%%%%%%%%%%%%%%%%%%%%%%%%%%%%%%%%%%
One can show that the remaining perturbed Einstein's equations provide no new equations. There are two additional second order differential equations for $H$ and $K$, both of which can be shown to be identical to the above second order equation, upon the use of the algebraic equation for $K$, as well as the perturbation equations arising from the conservation of the matter sector. 

Note that, the perturbation equations in the polar sector gets decoupled into two pieces. One of which consists of three perturbations $(H_{1},C^{\rm up},D^{\rm up})$, and the other one consists of the rest of the perturbations. Note that given $D^{\rm up}$, both $H_{1}$ and $C^{\rm up}$ can be uniquely determined from~\ref{H1aniso} and~\ref{CupDup}, respectively. On the other hand, the matter perturbations $\delta p_{t}$ and $\delta \rho$, as well as the metric perturbation $K$, can be determined in terms of the metric perturbation $H$ and the matter perturbation $\delta p_{r}$. Thus, the polar sector requires input for $D^{\rm up}$ and $\delta p_{r}$, which can be used to determine all the other matter and metric perturbations. In the static limit, to determine the LNs, we simply need to solve for the metric perturbation $H(r)$, from~\ref{Hdiffstatic}. This only requires the knowledge about the EoS for the radial pressure, so that we can determine $\delta \rho +\delta p_{r}$ appearing in~\ref{Hdiffstatic}. 
Alternatively, knowledge of $D^{\rm up}$, as well as a relation between $\delta \rho$ and $\delta p_{r}$ will also close the system of equations. In which case one needs to solve the coupled differential equations, i.e.,~\ref{deltarho} and~\ref{Hdiffstatic}. 

One natural assumption in the context of dark matter halo, is to take $\delta p_{r}=0$, since the radial pressure for the background dark matter fluid vanishes. With this assumption, the metric perturbation $H(r)$, satisfies the following master equation, 
%%%%%%%%%%%%%%%%%%%%%%%%%%%%%%%%%%%%%%%%%%%%%%%%%%%%%%%%%%
\begin{align}\label{omega0even}
H''+\left(\frac{1+g}{r g}+\frac{8\pi r\bar{\rho}}{1-g}\right)H'+\left(-\frac{1+(\ell+2)(\ell-1)g+g^2}{r^2 g^2}+\frac{8\pi \bar{\rho}(1+g^{2})}{g^{2}(1-g)}\right)H=0~.
\end{align}
%%%%%%%%%%%%%%%%%%%%%%%%%%%%%%%%%%%%%%%%%%%%%%%%%%%%%%%%%%
For vacuum background spacetime, one can check that the above equation reduces to~\ref{Hsole}. As we will demonstrate, the solution of this equation also provides a growing mode and a decaying mode, leading to the determination of electric-type LNs for a BH surrounded by anisotropic matter distribution. The equation derived above holds also for isotropic distributions.

\subsubsection{Einstein--Maxwell theory}
In the case of the Reissner--Nordstr\"om BH, the electric field due to the charge $Q$ behaves as an anisotropic perfect fluid with the energy momentum tensor being given by $T^{\mu}_{\nu}=\textrm{diag.}(-\bar{\rho},\bar{p}_{r},\bar{p}_{t},\bar{p}_{t})$, where 
%%%%%%%%%%%%%%%%%%%%%%%%%%%%%%%%%%%%%%%%%%%%%%%%%%%%%%%%%%%%%%%%%%%%%%%%%%%%%%
\begin{equation}\label{emtRN}
\bar{\rho}=\frac{Q^{2}}{8\pi r^{4}}=-\bar{p}_{r}~;\qquad \bar{p}_{\rm t}=\frac{Q^{2}}{8\pi r^{4}}~,
\end{equation}
%%%%%%%%%%%%%%%%%%%%%%%%%%%%%%%%%%%%%%%%%%%%%%%%%%%%%%%%%%%%%%%%%%%%%%%%%%%%%%
such that the energy momentum tensor is traceless, which is due to conformal invariance of the electromagnetic action in four spacetime dimensions.  Substitution of the above energy-momentum tensor in the Einstein's equations yields the Reissner--Nordstr\"om BH solution, corresponding to~\ref{ansatzanisotropic} with $f(r)=g(r)=1-2M/r+Q^{2}/r^{2}$. Thus, all the ingredients of the previous section will be directly applicable in this case as well. 

Here also we have two possibilities for perturbations in the axial sector, namely the up and the down conventions, as detailed in~\ref{eqaxialBHDMup} and~\ref{eqaxialBHDMdown}. Given the explicit expressions for the unperturbed density, radial pressure and tangential pressure, we obtain the following differential equations for axial metric perturbation $h_{0}$, in the static limit, as, 
%%%%%%%%%%%%%%%%%%%%%%%%%%%%%%%%%%%%%%%%%%%%%%%%%%%%%%%%%%%%%%%%%%%%%%%%%%%%%%
\begin{align}
r^{2}\left(1-\frac{2M}{r}+\frac{Q^{2}}{r^{2}}\right)\dfrac{d^{2}h_{0}}{dr^{2}}&-\left[\ell(\ell+1)-\frac{4M}{r}+\frac{6Q^{2}}{r^{2}}\right]h_{0}=0\,, \qquad \textrm{(up)}
\label{up_RN_ax}
\\
r^{2}\left(1-\frac{2M}{r}+\frac{Q^{2}}{r^{2}}\right)\dfrac{d^{2}h_{0}}{dr^{2}}&-\left[\ell(\ell+1)-\frac{4M}{r}+\frac{2Q^{2}}{r^{2}}\right]h_{0}=0\,, \qquad \textrm{(down)}
\label{down_RN_ax}
\end{align}
%%%%%%%%%%%%%%%%%%%%%%%%%%%%%%%%%%%%%%%%%%%%%%%%%%%%%%%%%%%%%%%%%%%%%%%%%%%%%%
Note that, while studying the axial perturbation of Reissner--Nordstr\"om BH,~\ref{up_RN_ax} is the one that has been used previously in the literature \citep{Cardoso:2017cfl, Cardoso:2019upw}. Here we will extract LNs from both up and down approaches. 

For determining the LNs, we introduce a perturbative scale $\epsilon=(Q/M)^{2}$, such that to linear order in $\epsilon$, we obtain, $h_{0}=h_{0}^{(0)}+\epsilon h_{0}^{(1)}$, such that the zeroth order piece and the first order piece satisfies the following equations, with $\ell=2$, in up/down conventions:
%%%%%%%%%%%%%%%%%%%%%%%%%%%%%%%%%%%%%%%%%%%%%%%%%%%%%%%%%%%%%%%%%%%%%%%%%%%%%%
\begin{align}
r^{2}\left(1-\frac{2M}{r}\right)\dfrac{d^{2}h^{(0)}_{0}}{dr^{2}}&-\frac{2(3r-2M)}{r}h^{(0)}_{0}=0\,,
\label{axzero}
\\
r^{2}\left(1-\frac{2M}{r}\right)\dfrac{d^{2}h^{(1)}_{0}}{dr^{2}}&-\frac{2(3r-2M)}{r}h^{(1)}_{0}=-M^{2}\left(\dfrac{d^{2}h^{(0)}_{0}}{dr^{2}}-\alpha\frac{h^{(0)}_{0}}{r^{2}}\right)\,,
\label{axone}
\end{align}
%%%%%%%%%%%%%%%%%%%%%%%%%%%%%%%%%%%%%%%%%%%%%%%%%%%%%%%%%%%%%%%%%%%%%%%%%%%%%%
where $\alpha=6$ for up convention, and $\alpha=2$ for down convention. The solution to the zeroth order piece can be obtained by solving~\ref{axzero}, which reads, 
%%%%%%%%%%%%%%%%%%%%%%%%%%%%%%%%%%%%%%%%%%%%%%%%%%%%%%%%%%%%%%%%%%%%%%%%%%%%%%
\begin{align}
h_{0}^{(0)}=A\left(\frac{r}{2M}\right)^{2}\left(\frac{r}{2M}-1\right)
\end{align}
%%%%%%%%%%%%%%%%%%%%%%%%%%%%%%%%%%%%%%%%%%%%%%%%%%%%%%%%%%%%%%%%%%%%%%%%%%%%%%
Thus, the source term in~\ref{axone} reads, $-A$, for the up convention, while it becomes $-(Ar/2M)$, for the down convention. Thus, if we solve~\ref{axone} with the appropriate source term, then we will have the solution to consist of two arbitrary constants $(c_{1},c_{2})$ appearing from the homogeneous piece, and the constant $A$ coming from the particular solution. Subsequently we must demand regularity of the perturbation everywhere outside and on the horizon, which for this perturbed system requires regularity at $r=2M$\footnote{Actually, the horizon will also be shifted by $r\to 2M-(\epsilon M/2)$, i.e., will be located within the horizon of the Schwarzschild solution. Thus, the perturbation must be regular everywhere outside the horizon, including the horizon. Thus, the solution must be regular at $r=2M$ as well.}. This fixes, $c_{2}=-A$ (for up case), or, $c_{2}=-(A/3)$ (for down case). After fixing the arbitrary constant accordingly, if we perform an asymptotic expansion, it will be evident that there are no decaying modes in the expression for the axial perturbation, up to linear order in $\epsilon$, and hence the magnetic LNs vanish identically.

For the polar sector, we can use~\ref{Hdiffstatic} where, by substituting the result $\delta \rho+\delta p_{r}=0$ (see~\ref{emtRN} for validity of this result), we get back a source free differential equation for the metric perturbation $H$. Writing down the metric components explicitly, we obtain the following differential equation describing static polar perturbation of the Reissner--Nordstr\"om BH, 
%%%%%%%%%%%%%%%%%%%%%%%%%%%%%%%%%%%%%%%%%%%%%%%%%%%%%%%%%%%%%%%%%%%%%%%%%%%%%%
\begin{align}
&r^{2}\left(1-\frac{2M}{r}+\frac{Q^{2}}{r^{2}}\right)^{2}\dfrac{d^{2}H}{dr^{2}}+2r\left(1-\frac{2M}{r}+\frac{Q^{2}}{r^{2}}\right)\left(1-\frac{M}{r}\right)\dfrac{dH}{dr}
\nonumber
\\
&\qquad -\left[\ell(\ell+1)\left(1-\frac{2M}{r}+\frac{Q^{2}}{r^{2}}\right)+\frac{4M^{2}}{r^{2}}-2\left(1+\frac{2M}{r}\right)\frac{Q^{2}}{r^{2}}+\frac{2Q^{4}}{r^{4}}\right]H=0\,.
\end{align}
%%%%%%%%%%%%%%%%%%%%%%%%%%%%%%%%%%%%%%%%%%%%%%%%%%%%%%%%%%%%%%%%%%%%%%%%%%%%%%
This matches with previous results in the literature \citep{Cardoso:2017cfl, Cardoso:2019upw}. Again, we need to expand the metric perturbation $H$ as a series in $\epsilon=(Q^{2}/M^{2})$, such that, $H=H^{(0)}+\epsilon H^{(1)}$. From regularity at the horizon, alike the axial sector, one can show that both $H^{(0)}$ and $H^{(1)}$ has only growing pieces present in the asymptotic limit, leading to vanishing electric LNs. 
% $g_{\mu \nu}=\textrm{diag.}(-f,f^{-1},r^{2},r^{2}\sin^{2}\theta)$, where $f=1-(2M/r)+(Q^{2}/r^{2})$, which has an event horizon, located at $r=r_{+}=M+\sqrt{M^{2}-Q^{2}}$, and a Cauchy horizon at $r=r_{-}=M-\sqrt{M^{2}-Q^{2}}$. 
% In the determination of the LNs, the event horizon plays the dominant role, while the Cauchy horizon is often considered as an unstable one\footnote{The existence of Cauchy horizon leads to issues with the deterministic nature of gravity}.

Thus, our results are consistent with \citet{Cardoso:2017cfl}, which had also studied directly the perturbations of the Einstein--Maxwell system, and computed the axial and polar tidal LNs of a Reissner--Nordstr\"om BH, showing that the are all zero (see also \citealt{Pereniguez:2021xcj}). This result was also interpreted in terms of generalized ladder symmetries \citep{Rai:2024lho}. Here, we have taken a more general route and all the previous results follow from it.

\paragraph{Non-zero bosonic Love numbers for charged BHs}~---
Finally, in the context of Einstein--Maxwell theory, \citet{Ma:2024few} studied \emph{charged} scalar perturbations of a Kerr--Newman BH. They found a discontinuity between neutral and charged scalars: the LNs associated with neutral scalar perturbations vanish identically (even for non-rotating BHs), whereas those for charged perturbations are inversely proportional to the electric charge of the scalar in the small-charge limit. This apparent discontinuity can be resolved by invoking charge quantization, which makes the zero-charge limit ill defined. Interestingly, the dissipation number was found to be insensitive to the scalar field charge. The main mathematical result that has gone into this analysis, corresponds to the possibility of effective angular numbers taking half-integer values. 

\paragraph{Magnetically charged BHs and non-zero bosonic LNs}~---
Surprisingly, it turns out that if we consider again a massless \emph{electrically} charged scalar (with charge $q_{\rm e}$), perturbing a \emph{magnetically} charged rotating BH (with magnetic charge $Q_{\rm m}$) in GR, the corresponding static LNs are non-zero, while the dissipation numbers vanish identically \citep{Pereniguez:2025jxq}. This result relies on the Dirac quantization condition $2q_{\rm e}Q_{\rm m}=\textrm{integer}=N$. Thus, upon using a modified angular function for the scalar field \citep{Pereniguez:2024fkn}, alike \citep{Ma:2024few}, it follows that the angular number will be half-integers. This gives non-zero LNs, which vanish in the extremal limit and generically scales as $T_{\rm H}^{\sqrt{(1+2\ell)^{2}-N^{2}}}$, with $\ell\leq (|N|/2)$ and $T_{\rm H}$ being the BH temperature. Thus, in this case the LNs vanish only in the extremal limit,
whereas the dissipation numbers associated with charged scalar perturbation vanish identically for any temperature \citep{Pereniguez:2025jxq}.   

\paragraph{Quantum corrections}~---
Unlike classical vacuum BHs, BHs embedded in the vacuum of a quantum field theory are not completely isolated. 
Quantum vacuum polarization effects effectively dress the geometry with surrounding vacuum bubbles, which induce nonvanishing tidal responses \citep{Barbosa:2025uau,Barbosa:2026qcv}. 
In this framework, charged BHs are particularly interesting, as the induced corrections can be significantly larger than in the neutral case, especially close to extremality. 

Moreover, because charged BHs generically exhibit coupled responses to gravitational and electromagnetic tidal fields, their tidal deformability is most naturally described in terms of Love matrices, which encode the mixing between the two sectors \citep{Barbosa:2026qcv}.

%%%%%%%%%%%%%%%%%%%%%%%%%%%%%%%%%%%%%%%%%%%%%%%%%%%%%%
\subsubsection{Love numbers of regular black holes} \label{sec:regularBH}
%%%%%%%%%%%%%%%%%%%%%%%%%%%%%%%%%%%%%%%%%%%%%%%%%%%%%%
Regular models of BHs replace the central singularity with a nonsingular spacetime region \citep{Lan:2023cvz,Carballo-Rubio:2018pmi}. 
This typically requires exotic matter sources that violate some of the energy conditions, nonlinear electrodynamics, or beyond-GR corrections. 
The corresponding matter fields are not necessarily confined within the horizon and often have support in the exterior geometry. 
In this case they can affect observable properties, including the tidal deformability, thereby providing a possible avenue to distinguish these objects from vacuum GR BHs \citep{Carballo-Rubio:2018pmi}. 

The LNs of regular BHs have been recently investigated for various phenomenological models \citep{Coviello:2025pla,Wang:2025oek}. 
In all cases, the tidal response is controlled by the mechanism that regularizes the inner geometry. 
The dimensionless LNs scale as
\begin{equation}
    {}_2 k_\ell^{E,B} \propto \left(\frac{L}{M}\right)^p \,,
\end{equation}
where $L$ denotes the regularization scale and the exponent $p>0$ depends on the specific model and on the sector (electric or magnetic), typically lying in the range $p\in[2,4]$ \citep{Coviello:2025pla,Wang:2025oek}. 
If $L$ is of the order of the Planck length, the LNs are nonvanishing but phenomenologically negligible. 
However, regular solutions generally exist for $L\lesssim {\cal O}(M)$. 
When this upper bound is saturated, $L={\cal O}(M)$, the tidal response is no longer suppressed and can span the range ${}_2 k_\ell^{E,B}\sim 10^{-2}$--$10^{2}$, depending on the model and on the sector.

The LNs of regular BHs arising in the context of Loop Quantum Gravity have also been recently analyzed in \citet{Motaharfar:2025typ,Motaharfar:2025ihv,Liu:2025iby}.

Very recently, \citet{Bhattacharyya:2026itm} computed the dynamical tidal response of several regular BH models, showing that it exhibits rich frequency-dependent features and encodes information about the near-horizon and interior structure that is inaccessible in the static limit.

%%%%%%%%%%%%%%%%%%%%%%%%%%%%%%%%%%%%%%%%%%%%%%%%%%%%%%
\subsubsection{BHs surrounded by dark matter halos} \label{sec:BHhalos}
%%%%%%%%%%%%%%%%%%%%%%%%%%%%%%%%%%%%%%%%%%%%%%%%%%%%%%
In the case of a BH surrounded by a spherically symmetric dark matter halo, the latter is typically modeled by an energy-momentum tensor of the form $T^{\mu}_{\nu}=\textrm{diag}(-\rho,0,p_{t},p_{t})$, i.e., with vanishing radial pressure. This assumption reflects the picture in which dark matter particles move on circular orbits around the central BH, with an accretion rate that is negligible on the timescales of interest. 

It is important to emphasize the limitations of this simplified description. First, in spiral galaxies dark matter halos are not exactly spherically symmetric, so the above model does not apply in that context. Second, spiral galaxies contain not only dark matter but also baryonic matter, which can accrete onto the central BH with a highly non-spherical profile, and whose effects should also be taken into account. Since dark matter interactions are negligible and we are interested in linear perturbation theory, these two effects can be studied independently, which justifies discussing the impact of the dark matter halo and of baryonic accretion on BH LNs separately. In addition, low surface brightness galaxies are believed to be dark-matter dominated and approximately spherically symmetric~\citep{Blok:2002tr}, making this setup particularly relevant. 
  
For a spherically symmetric matter distribution with energy density $\rho(r)$ and negligible radial pressure, the functions $f(r)$, and $m(r)$ satisfies the following equations, 
%%%%%%%%%%%%%%%%%%%%%%%%%%%%%%%%%%%%%%%%%%%%%%%%%%%%%%%%%%%%%%%%%%%%%%%%%%%%%%
\begin{equation}
\frac{dm(r)}{dr}=4\pi r^{2}\rho(r)~;
\qquad
r\dfrac{d\ln f(r)}{dr}=\frac{2m(r)}{r-2m(r)}~;
\qquad 
2p_{\rm t}=\frac{m(r)\rho(r)}{r-2m(r)}~,
\end{equation}
%%%%%%%%%%%%%%%%%%%%%%%%%%%%%%%%%%%%%%%%%%%%%%%%%%%%%%%%%%%%%%%%%%%%%%%%%%%%%%
where the last equation determines the tangential pressure $p_{\rm t}$.  Thus, given any energy/mass distribution for dark matter, the above set of equations can be solved to uniquely determine the geometry of a BH immersed in its halo. Typical density profiles include: Hernquist \citep{Hernquist:1990be}, Navarro-Frank-White \citep{NFW}, Einasto \citep{Haud:1986yj}, among others \citep{Graham:2005xx, Prada:2005mx}. The dark matter particles in these halos are adiabatically accreted inside the BH and hence the distribution function for the dark matter gets modified. 
If we consider the distribution function to satisfy non-relativistic Boltzmann equation, it follows that the density distribution will have a spike around $r\approx8M_{\rm BH}$ and the density vanishes at $r\approx6M_{\rm BH}$ \citep{Gondolo:1999ef}, while for relativistic Boltzmann equation, the density profiles spikes at $r\approx6M_{\rm BH}$ and vanishes at $r\approx4M_{\rm BH}$, where $M_{\rm BH}$ is the mass of the central BH \citep{Sadeghian:2013laa}. For rotating generalization, see \citep{Ferrer:2017xwm, Mitra:2025tag}. 

The static master equation for the axial variable $h_{0}$, is given by~\ref{eqaxialBHDMup} for the `up' case, or~\ref{eqaxialBHDMdown}, for the `down' case. For the polar sector, on the other hand, the master equation for $H$ in the strictly static case is given by~\ref{omega0even}. In each of these we will discuss the LNs associated with BHs surrounded by both relativistic and non-relativistic dark matter profiles. 
%%%%%%%%%%%%%%%%%%%%%%%%%%
%%%%%%%%%%%%%%%%%%%%%%%%%%
%%%%%%%%%%%%%%%%%%%%%%%%%%
\begin{itemize}
\item \textbf{Axial LNs for the non-relativistic Hernquist profile}~--- We start with the determination of the `up' LNs associated with a Schwarzschild BH surrounded by a non-relativistic Hernquist profile. In this case, the metric perturbation $h_{0}$ satisfies~\ref{eqaxialBHDMup}, which upon inserting the background fluid quantities from \citet{Cardoso:2021wlq}, becomes complicated enough that any closed-form solution for $h_0$ is impossible. However, one can typically consider the matter distribution as a further small perturbation. Namely, since $\epsilon\equiv(M_{\rm DM}/r_s)\ll 1$ (where $r_{s}$ is the characteristic scale associated with the dark matter halo), we can expand $h_0$ as $h_0=h_0^{(0)}+\epsilon h_0^{(1)}+\order(\epsilon^2)$, where the zeroth order term, for the $\ell=2$ (say) case, is given by~\ref{h0e}, with $x=(r/2M_{\rm BH})$. 
In order for the zeroth order metric perturbation $h_0^{(0)}$, as well as its derivatives, to be regular at the horizon, we must set $\mathcal{A}_{2}=0$, i.e., the coefficient of the $\log(1-x)$ term must vanish. Hence, the zeroth order metric perturbation in the axial sector is simply $h_{0}^{(0)}\propto r^{2}(r-2\mbh)$. 

The first order metric perturbation $h_{0}^{(1)}$, on the other hand, satisfies the following differential equation \citep{Chakraborty:2024gcr}, 
%%%%%%%%%%%%%%%%%%%%%%%%%%%%%%%%%%%%%%%%%%%%%%%%%%%%%%%%%%
\begin{align}
h^{(1)\prime\prime}_{0}&-\left[\frac{\ell(\ell+1)r-4\mbh}{r^{2}(r-2\mbh)}\right]h^{(1)}_{0}=\left[\frac{2r_s(r_s+2\mbh)}{(r+r_s)^{3}}\right]h^{(0)\prime}_{0}
\nonumber
\\
&\qquad +\left[\frac{2r_s(\ell-1)(\ell+2)}{r(r+r_s)^{2}}+\frac{4r_s(r_s+2\mbh)(r-\mbh)}{r(r+r_s)^{3}(r-2\mbh)}\right]h^{(0)}_{0}~.
\end{align}
%%%%%%%%%%%%%%%%%%%%%%%%%%%%%%%%%%%%%%%%%%%%%%%%%%%%%%%%%%
As expected, the zeroth order metric perturbation acts as a source for the first order term. Substituting $h_{0}^{(0)}=\mathcal{A}_{1}r^{2}(r-2\mbh)$, for $\ell=2$, the above differential equation becomes
%%%%%%%%%%%%%%%%%%%%%%%%%%%%%%%%%%%%%%%%%%%%%%%%%%%%%%%%%%
\begin{align}\label{h0fup}
\medmath{h^{(1)\prime\prime}_{0}-\left[\frac{6r-4\mbh}{r^{2}(r-2\mbh)}\right]h^{(1)}_{0}=\frac{2r_s r\mathcal{A}_{1}}{(r+r_s)^{3}}\left[4r^{2}+r\left(9r_s+2\mbh\right)-2\mbh(7r_s+6\mbh)\right]~.}
\end{align}
%%%%%%%%%%%%%%%%%%%%%%%%%%%%%%%%%%%%%%%%%%%%%%%%%%%%%%%%%%
This differential equation can be solved analytically in terms of two arbitrary constants, although the explicit solution is cumbersome. The homogeneous part of the solution resembles~\ref{h0e}, with the arbitrary constant $D_{1}$ describing the growing mode and the other arbitrary constant $D_{2}$ describing the decaying mode. The particular solution depends on $\mathcal{A}_{1}$. As in the case of Schwarzschild BH, here also we impose regularity at the horizon and discard the redundant growing mode by setting $D_{1}=0$. This yields a unique solution for $D_{2}$, proportional to $\mathcal{A}_1$. As usual, the $\ell=2$ magnetic LN is determined from the ratio of the $\order(r^{-2})$ term and the $\order(r^3)$ terms in the asymptotic expansion of the complete metric perturbation $h_{0}$. To leading order in $\epsilon$, we obtain from~\ref{tlnb},
%%%%%%%%%%%%%%%%%%%%%%%%%%%%%%%%%%%%%%%%%%%%%%%%%%%%%%%%%%
\begin{equation}
k^{\rm B(up)}_{\ell=2}= \frac{M_{\rm DM} r^4_s \left[5+12 \log (r_s/R)\right]}{3 R^5}~,
\end{equation}
%%%%%%%%%%%%%%%%%%%%%%%%%%%%%%%%%%%%%%%%%%%%%%%%%%%%%%%%%%
where we have introduced an effective radial scale $R$, to be discussed later, which also makes the argument of the logarithmic term, as well as the LN, dimensionless \citep{Cardoso:2021wlq,Chakraborty:2024gcr}. 

For the `down' LN, one can follow the same path,  
%by expanding out the metric perturbation $h_{0}$ in powers of $\epsilon$. The zeroth order term can be solved to yield~\ref{h0e}, and the identical discussion involving regularity of the solution follows. On the other hand, 
with the first order term, for the $\ell=2$ mode, satisfying the following differential equation \citep{Chakraborty:2024gcr},
%%%%%%%%%%%%%%%%%%%%%%%%%%%%%%%%%%%%%%%%%%%%%%%%%%%%%%%%%%
\begin{equation}
h^{(1)\prime\prime}_{0}-\left[\frac{6r-4\mbh}{r^{2}(r-2\mbh)}\right]h^{(1)}_{0}=\mathcal{A}_{1}\frac{2r_s r}{(r+r_s)^{3}}\left[4r^{2}+r\left(5r_s-6\mbh\right)-8r_s\mbh\right]\,.
\end{equation}
%%%%%%%%%%%%%%%%%%%%%%%%%%%%%%%%%%%%%%%%%%%%%%%%%%%%%%%%%%
Note that the source term of this equation is different from the one for the `up' case, see~\ref{h0fup}. Repeating the procedure described for the `up' case, we obtain the `down' axial LN for the $\ell=2$ mode,
%%%%%%%%%%%%%%%%%%%%%%%%%%%%%%%%%%%%%%%%%%%%%%%%%%%%%%%%%%
\begin{equation}
k^{\rm B(down)}_{\ell=2}=\frac{M_{\rm DM}r_s^{4}\left[1-4 \log(r_s/R)\right]}{3R^{5}}~.
\end{equation}
%%%%%%%%%%%%%%%%%%%%%%%%%%%%%%%%%%%%%%%%%%%%%%%%%%%%%%%%%%
Note that this axial LN is negative for $R\ll r_s$, which is the case here. It is clear that the LNs obtained using the `up' and the `down' approaches are indeed different. In particular, the fractional change in the `up' and `down' axial LNs, associated with the $\ell=2$ mode, is
%%%%%%%%%%%%%%%%%%%%%%%%%%%%%%%%%%%%%%%%%%%%%%%%%%%%%%%%%%
\begin{equation}
\Delta k^{\rm B}_{\ell=2}\equiv \frac{k^{\rm B (up)}_{\ell=2}-k^{\rm B(down)}_{\ell=2}}{k^{\rm B (up)}_{\ell=2}}
=\frac{4}{5}~,
\end{equation}
%%%%%%%%%%%%%%%%%%%%%%%%%%%%%%%%%%%%%%%%%%%%%%%%%%%%%%%%%%
which is a $\mathcal{O}(1)$ quantity.
% , suggesting the intriguing result that the difference between the `up' and `down' axial LNs is significant.  

A comment regarding the characteristic length scale $R$ is in order. Unlike the case of vacuum BHs, where $R$ is uniquely fixed to be the horizon radius, in the present context several definitions are possible. For instance, choosing $R=(M_{\rm DM}+M_{\rm BH})$ reproduces the expression for the axial `up' LN obtained in \citet{Cardoso:2021wlq}. 
However, in \citet{Chakraborty:2024gcr}, anticipating that the LN scales as 
$k^{\rm B(up/down)}_{\ell=2}\sim \mathcal{O}(\epsilon)$, the choice $R=r_{s}$ was adopted. 
With this prescription the LN is indeed linear in $\epsilon$, but it decreases as $r_{s}$ increases. 
This behavior is somewhat counterintuitive: for fixed total mass, increasing the effective radius corresponds to a more dilute configuration, which one would expect to be more easily deformable, and hence to exhibit a larger LN. 
Choosing instead $R=M_{\rm DM}+M_{\rm BH}$ preserves the linear scaling with $\epsilon$ while yielding an LN that increases with $r_{s}$, in qualitative agreement with the expected behavior of extended configurations \citep{Cardoso:2019upw,Cardoso:2021wlq}.

\item \textbf{Polar LNs for the non-relativistic Hernquist profile}~--- Alike the axial sector, here also we adopt the perturbative scheme and expand the polar metric perturbation $H$ as $H=H^{(0)}+\epsilon H^{(1)}+\order(\epsilon^2)$, where $\epsilon=(M_{\rm DM}/r_s)$, is the smallness parameter. The zeroth order term satisfies~\ref{Hsole}, while the first order term satisfies the following differential equation (assuming generic $\ell$ for the moment)
%%%%%%%%%%%%%%%%%%%%%%%%%%%%%%%%%%%%%%%%%%%%%%%%%%%%%%%%%%
\begin{align}
H^{(1)\prime\prime}&+\frac{2(r-\mbh)}{r(r-2 \mbh)}H^{(1)\prime}-\frac{4\mbh^{2}-2\ell(\ell+1)r\,\mbh+\ell(\ell+1)r^2}{r^2(r-2 \mbh)^2}H^{(1)}
\nonumber
\\
&=\frac{2r_s\left[\mbh (r_s+4 \mbh)-r (r_s+3 \mbh)\right]}{\mbh (r_s+r)^3}H^{(0)\prime}
\nonumber
\\
&+\frac{2 r_s \left[r_s \ell(\ell+1) \mbh-2 r_s r+\mbh \left\{\left(\ell^2+\ell-4\right) r+4 \mbh\right\}\right]}{\mbh r(r_s+r)^3}H^{(0)}~.
\end{align}
%%%%%%%%%%%%%%%%%%%%%%%%%%%%%%%%%%%%%%%%%%%%%%%%%%%%%%%%%%
Thus, alike the axial sector, here also the zeroth order solution sources the first order perturbation. The solution for the zeroth order perturbation $H^{(0)}$ has been presented in~\ref{Hext}, whose regularity at the horizon enforces the coefficient of the logarithmic term, $\mathcal{P}_{2}$, to vanish. Thus, for $\ell=2$ the solution for the zeroth order term, which is regular at the horizon becomes, $H^{(0)}\propto 3r(r-2\mbh)/\mbh^2$. Substituting the above solution for $H^{(0)}$, requiring regularity at the horizon, along with the appropriate numerical factors introduced in~\ref{electrinLNgenl}, the LN, obtained as the ratio of the coefficient of the $r^{-3}$ term to the coefficient of the $r^{2}$ term, yields,
%%%%%%%%%%%%%%%%%%%%%%%%%%%%%%%%%%%%%%%%%%%%%%%%%%%%%%%%%%
\begin{equation}
k^{\rm E}_{\ell=2}=\frac{2M_{\rm DM}r_{s}^{4}\left[1+6\log(r_s/R)\right]}{R^{5}}\,.
\end{equation}
%%%%%%%%%%%%%%%%%%%%%%%%%%%%%%%%%%%%%%%%%%%%%%%%%%%%%%%%%%
For the choice of the characteristic scale $R=M_{\rm DM}+M_{\rm BH}$, see~\ref{tlne}, as motivated in the axial sector, it follows that the polar LN also scales as $\epsilon$ and grows as the size of the halo increases. This is consistent with our findings in the axial sector as well as with the perturbative scheme employed here \citep{Chakraborty:2024gcr}. 

\item \textbf{Axial LNs for the relativistic Hernquist profile}~--- 
% In the case of a relativistic Hernquist dark matter profile, the static equation for the axial perturbation $h_{0}$ is still given by~\ref{eqaxialBHDMup} for the `up' convention, and by~\ref{eqaxialBHDMdown} for the `down' convention. 
The only modification arising in the relativistic context concerns the density profile. The relativistic dark-matter distribution is obtained by evolving the initial Hernquist profile through the collisionless Boltzmann equation in GR \citep{Sadeghian:2013laa, Chakraborty:2024gcr}. The resulting density profile reads,
%%%%%%%%%%%%%%%%%%%%%%%%%%%%%%%%%%%%%%%%%%%%%%%%%%%%%%%%%%
\begin{equation}
\bar{\rho}=\lambda \,\left(1.85\times 10^{22}\,\textrm{GeV}/\textrm{cm}^{3}\right)\left(1-\frac{4M_{\rm BH}}{r}\right)^{2.45}\left(\frac{M_{\rm BH}}{r}\right)^{2.33}\,,
\end{equation}
%%%%%%%%%%%%%%%%%%%%%%%%%%%%%%%%%%%%%%%%%%%%%%%%%%%%%%%%%%
with 
%%%%%%%%%%%%%%%%%%%%%%%%%%%%%%%%%%%%%%%%%%%%%%%%%%%%%%%%%%
\begin{equation}
\lambda=\left(\frac{M_{\rm DM}}{10^{12}M_{\odot}}\right)^{0.33}\left(\frac{M_{\rm BH}}{10^{6}M_{\odot}}\right)^{-1.66}\left(\frac{r_{s}}{20\,\textrm{kpc}}\right)^{-0.67}\,.
\end{equation}
%%%%%%%%%%%%%%%%%%%%%%%%%%%%%%%%%%%%%%%%%%%%%%%%%%%%%%%%%%
For a typical astrophysical BH embedded in a dark-matter halo, $\lambda$ is of order unity. The associated mass function, obtained by integrating the above profile, involves hypergeometric functions. Owing to the complexity of the mass and density functions, closed-form analytical solutions of the axial and polar perturbation equations are not available.  

In this relativistic setup, the relevant dimensionless parameter reads \citep{Figueiredo:2023gas, Sadeghian:2013laa, Ferrer:2017xwm, Dyson:2025dlj, Speeney:2022ryg}
%%%%%%%%%%%%%%%%%%%%%%%%%%%%%%%%%%%%%%%%%%%%%%%%%%%%%%%%%%
\begin{equation}
\epsilon_{\rm R} \equiv \lambda \left(1.85\times 10^{22}\,\textrm{GeV}/\textrm{cm}^{3}\right) M_{\rm BH}^2 \sim 5 \times 10^{-8}\lambda~, 
\end{equation} 
%%%%%%%%%%%%%%%%%%%%%%%%%%%%%%%%%%%%%%%%%%%%%%%%%%%%%%%%%%
which sets the scale of the dark-matter profile. In the following, both axial and polar perturbations are expanded to linear order in $\epsilon_{\rm R}$, neglecting $\order(\epsilon_{\rm R}^2)$ terms. For instance, the axial perturbation $h_{0}$ is written as $h_0=h_0^{(0)}+\epsilon_{\rm R}h_0^{(1)}+\order(\epsilon_{\rm R}^2)$, where the zeroth-order solution, $h_{0}^{(0)}=\mathcal{A}_{1}\,r^{2}(r-2M_{\rm BH})$, is regular at the horizon. As in the non-relativistic case, the first-order term $h_{0}^{(1)}$ satisfies an inhomogeneous equation with $h_{0}^{(0)}$ as the source. We now have three regions --- (a) the region between $2M_{\rm BH}$ and $4M_{\rm BH}$, where we have Schwarzschild background and the perturbation is given by $h_{0}^{(0)}$; (b) the region close to $4M_{\rm BH}$, where there are DM effects, and hence the equation for $h_{0}^{(1)}$ is solved numerically by imposing the boundary conditions $h_{0}^{(1)}(4M_{\rm BH})=0$ and $\partial_{r}h_{0}^{(1)}(4M_{\rm BH})=0$, to match with their Schwarzschild counterparts for $r<4M_{\rm BH}$; and finally (c) the region $r>4M_{\rm BH}$, where one needs to match the numerical solution with the asymptotic behavior at the intermediate zone.
This yields the magnetic `up' LN associated with the $\ell=2$ mode \citep{Chakraborty:2024gcr},
%%%%%%%%%%%%%%%%%%%%%%%%%%%%%%%%%%%%%%%%%%%%%%%%%%%%%%%%%%
\begin{equation}
k^{\rm B(up)}_{\ell=2}\simeq -279.86\,\frac{r_{s}^5}{R^5}\,\epsilon_{\rm R}\,.
\end{equation}
%%%%%%%%%%%%%%%%%%%%%%%%%%%%%%%%%%%%%%%%%%%%%%%%%%%%%%%%%%
An analogous procedure applied to the `down' axial perturbation equation, by treating $\epsilon_{\rm R}$ as the expansion parameter. The zeroth-order solution is the same as the `up' case, while the first-order equation differs from the `up' case, but can be solved numerically using identical techniques, as discussed above. This yields \citep{Chakraborty:2024gcr},
%%%%%%%%%%%%%%%%%%%%%%%%%%%%%%%%%%%%%%%%%%%%%%%%%%%%%%%%%%
\begin{equation}
k^{\rm B(down)}_{\ell=2}\simeq -102.1\,\frac{r_{s}^5}{R^5}\,\epsilon_{R}\,.
\end{equation}
%%%%%%%%%%%%%%%%%%%%%%%%%%%%%%%%%%%%%%%%%%%%%%%%%%%%%%%%%%
Both LNs scale linearly with $\epsilon_{R}$, consistent with the perturbative scheme. In this case, the simplest consistent choice for the characteristic scale is $R=M_{\rm DM}+M_{\rm BH}$, leading to increasing LN with increasing $r_{s}$. Note that both the `up' and `down' magnetic LNs for the $\ell=2$ mode turns out to be negative, while in the non-relativistic case, only the `down' mode has negative LN for $\ell=2$. 

\item \textbf{Polar LNs for a relativistic Hernquist profile}~--- The computation of the electric LNs of a BH immersed in a relativistic Hernquist dark-matter profile proceeds analogously to the axial case. By expanding the metric perturbation $H(r)$ in the small parameter $\epsilon_{\rm R}$, we first obtain the zeroth-order perturbation, regular at horizon to be, $H^{(0)} \propto \{3r(r-2M_{\rm BH})/M_{\rm BH}^2\}$. The first-order correction, $H^{(1)}$, is obtained numerically, subject to the boundary conditions $H^{(1)}(4M_{\rm BH})=0$ and $\partial_{r}H^{(1)}(4M_{\rm BH})=0$, as in the axial case. Therefore, by matching between the near- and intermediate-zone solutions, we obtain the following LN \citep{Chakraborty:2024gcr}:
%%%%%%%%%%%%%%%%%%%%%%%%%%%%%%%%%%%%%%%%%%%%%%%%%%%%%%%%%%
\begin{equation}
k^{\rm E}_{\ell=2} = 645.9\,\frac{r_{s}^5}{R^5}\,\epsilon_R \,,
\end{equation}
%%%%%%%%%%%%%%%%%%%%%%%%%%%%%%%%%%%%%%%%%%%%%%%%%%%%%%%%%%
which are positive, in contrast to the axial ones, which are negative, and in agreement with the non-relativistic case. Moreover, the choice $R=M_{\rm DM}+M_{\rm BH}$ results into a desirded behaviour of the deformability of a DM cloud surrounding a BH.

\end{itemize}
%%%%%%%%%%%%%%%%%%%%%%%%%%
%%%%%%%%%%%%%%%%%%%%%%%%%%
%%%%%%%%%%%%%%%%%%%%%%%%%%

To summarize, BHs, surrounded by dark-matter has non-zero bosonic LNs. These LNs depend crucially on the nature of dark-matter distribution, i.e., if the dark-matter satisfies Hernquist profile or, relativistic profile with spikes. Thus, if these LNs can be observed in the future, one can not only comment about the existence of dark-matter environment, but also possibly on the density of dark-matter distribution. 

Note that the LNs increases as the characteristic length scale of the dark-matter distribution grows; the actual scaling depends on the models.

%%%%%%%%%%%%%%%%%%%%%%%%%%%%%%%%%%%%
%%%%%%%%%%%%%%%%%%%%%%%%%%%%%%%%%%%%
%%%%%%%%%%%%%%%%%%%%%%%%%%%%%%%%%%%%

\subsubsection{BHs surrounded by an accretion disk}
So far we have focused on spherically symmetric matter distributions.
Another relevant case is that of BHs surrounded by accretion disks. 
Computing the tidal response of these systems would involve solving a set of linear partial differential equations, owning to the absence of symmetries of the background.
A simplification occurs for \emph{thin} accretion disks, which can be modelled as matter distributions localized on the BH equatorial plane \citep{Kotlarik:2018nbd,Kotlarik:2022spo}. Assuming the disk has a small backreaction on the metric, one can still separate the radial and angular perturbation variables by introducing couplings with different harmonics \citep{Chen:2023akf,Pani:2013pma,Cannizzaro:2024fpz}.
Using such an effective geometry,
 \citet{Cannizzaro:2024fpz} computed the 
\emph{scalar} LNs of a BH surrounded by a thin accretion disk, which turns out to be non-zero and has the following form:
%%%%%%%%%%%%%%%%%%%%%%%%%%%%%%%%%%%%%%%%%%%%%%%%%%%%%%%%%%
\begin{equation}
\label{spin0TLNjt=0}
{}_0k_{22} = \frac{\epsilon}{11520} \frac{\left(25920 \tilde{b}^6+87360 \tilde{b}^5+20880 \tilde{b}^4-37824 \tilde{b}^3-14964 \tilde{b}^2+1532 \tilde{b}+863\right)}{(2 \tilde{b}+1)^2}
\,.
\end{equation}
%%%%%%%%%%%%%%%%%%%%%%%%%%%%%%%%%%%%%%%%%%%%%%%%%%%%%%%%%%
where $\epsilon=M_{\rm disk}/M_{\rm BH} \ll 1$ is the expansion parameter, and $\tilde b=b/(2M_{\rm BH})$, with $b$ approximately describing the location of maximum density of the accretion profile. This LN has been computed, by solving the inhomogeneous equation for the perturbed scalar field as linear order in $\epsilon$, numerically, and then imposing appropriate boundary condition asymptotically. As expected, the LN is linear in the disk perturbation parameter $\epsilon$, and it has a leading behavior $\sim \tilde{b}^4$ when $\tilde b\gg1$, independently of the model parameters.
It would be interesting to extend this computation to the gravitational case.

\subsubsection{Other environmental effects}
As the above examples illustrate, environmental effects can drastically alter the nature of tidal interactions, possibly leading to nonvanishing LNs for non-vacuum BHs. Further examples include:
\begin{itemize}
    \item BHs surrounded by clouds of ultralight bosonic fields sourced by accretion or superradiance \citep{Baumann:2018vus, DeLuca:2021ite, DeLuca:2022xlz, Arana:2024kaz}, for which the LN scales as
    \begin{equation} \label{eq:cloud}
        {}_s k_{22} \propto \left(\frac{\lambda_{\rm scalar}}{M_{\rm BH}}\right)^8\,,
    \end{equation}
    %%%
    where $\lambda_{\rm scalar}$ is the Compton wavelength of the ultralight field so that, for a given BH mass, ultralight bosons induce a larger tidal deformability (making also the cloud more prone to tidal disruption \citep{DeLuca:2021ite, DeLuca:2022xlz}, see~\ref{sec:GW}).
    \item accreting BHs modelled as Vaidya spacetimes \citep{Capuano:2024qhv}, where
    \begin{equation}
        {}_s k_{22} \propto \dot M\,,
    \end{equation}
    %%%
    and $\dot M\ll1$ is the mass accretion rate.
    \item BHs immersed in matter fluids \citep{Cardoso:2019upw, Cardoso:2021wlq}, or enclosed by thin shells of matter \citep{Katagiri:2023yzm, DeLuca:2024uju}\,.
\end{itemize}
In all these cases, the LNs are nonzero. While the details depend on the matter distribution around BHs, they  generically increase as the compactness decreases. This behavior reflects the general expectation that the less compact the environment around the BH is, the more easily it can be deformed (and hence also more easily it can be disrupted in a binary coalescence).

%%%%%%%%%%%%%%%%%%%%%%%%%%%%%%%%%%%%
%%%%%%%%%%%%%%%%%%%%%%%%%%%%%%%%%%%%
%%%%%%%%%%%%%%%%%%%%%%%%%%%%%%%%%%%%
\subsection{Love numbers of higher dimensional black holes}\label{sec:higherD}

We have demonstrated that BHs in four dimensional vacuum general relativity has vanishing static LNs from several different computations. The previous section demonstrates that if we include matter degrees of freedom in the problem, so that we are considering non-vacuum spacetimes in GR, the static LNs can be non-zero. Here we would like to to show that relaxing the assumption of four-dimensional spacetime can also lead to non-zero LNs. In this case we will use three possible cases --- (a) considering higher dimensional BHs, e.g., Schwarzschild--Tangherlini BHs, as well as Myers-Perry BHs in general relativity, (b) from the braneworld scenario, where one considers a four dimensional BH solution, which is induced from a higher dimensional spacetime, and (c) BHs in Lovelock theories, which requires higher dimensions for their existence. In what follows we will discuss each of these scenarios and the static LNs derived in these contexts. 

\subsubsection{Higher-dimensional black holes in vacuum General Relativity}
The tidal response of $d$-dimensional Schwarzschild-Tangherlini BHs has been studied with various techniques, 
including perturbative analyses~\citep{Kol:2011vg,Chakravarti:2018vlt,Hui:2020xxx} and modern scattering-amplitude methods~\citep{Ivanov:2022qqt,Akhtar:2025nmt,Ivanov:2026icp}. 

The Schwarzschild--Tangherlini BH, which describes the unique stationary BH solution to vacuum GR in $d$ spacetime dimensions, reads:
%%%%%%%%%%%%%%%%%%%%%%%%%%%%%%%%%%%%%%%%%%%%%%%%%%%%%%%%%%%%%%%%%%%%%%%%%%%%%%
\begin{align}\label{hd_sch_metric}
ds^{2}=-f(r)dt^{2}+\frac{dr^{2}}{f(r)}+r^{2}d\Omega_{d-2}^{2}~;
\qquad f(r)=1-\left(\frac{r_{+}}{r}\right)^{d-3}~,
\end{align}
%%%%%%%%%%%%%%%%%%%%%%%%%%%%%%%%%%%%%%%%%%%%%%%%%%%%%%%%%%%%%%%%%%%%%%%%%%%%%%
where $r_{+}$ is the location of the horizon, related to the ADM mass $M$ of the spacetime through the following relation, $16\pi \Gamma[(d-1)/2](GM/r_{+}^{d-3})=2(d-2)\pi^{(d-1)/2}$.

\paragraph{(i) Scalar LNs}~--- The perturbation of the above spacetime background by an external massive scalar field, can be decomposed into a radial part $r^{(2-d)/2}\,_{0}\Psi_{\ell m}(r)$, with the angular part of the scalar field given by the $(d-2)$ dimensional hyperspherical harmonics $Y_{\ell}^{m}$ \citep{Hui:2020xxx}, and the time dependence is through $e^{-i\omega t}$. This follows from the $\mathbb{R}\times SO(d-1)$ symmetry group of the background spacetime. The radial function $\Psi^{(0)}_{\ell m}(r)$ satisfies the usual Schr\"{o}dinger-like equation, 
%%%%%%%%%%%%%%%%%%%%%%%%%%%%%%%%%%%%%%%%%%%%%%%%%%%%%%%%%%%%%%%%%%%%%%%%%%%%%%
\begin{align}
\dfrac{d^{2}\,_{0}\Psi_{\ell m}}{dr_{*}^{2}}&+\left(\omega^{2}-V_{0}\right)\,_{0}\Psi_{\ell m}=0~,
\end{align}
%%%%%%%%%%%%%%%%%%%%%%%%%%%%%%%%%%%%%%%%%%%%%%%%%%%%%%%%%%%%%%%%%%%%%%%%%%%%%%
where the effective potential $V_{0}$ reads,
%%%%%%%%%%%%%%%%%%%%%%%%%%%%%%%%%%%%%%%%%%%%%%%%%%%%%%%%%%%%%%%%%%%%%%%%%%%%%%
\begin{align}
V_{0}&=f\left[\frac{\ell(\ell+d-3)}{r^{2}}+\frac{(d-2)f'}{2r}+\frac{f(d-2)(d-4)}{4r^{2}}\right]~.
\end{align}
%%%%%%%%%%%%%%%%%%%%%%%%%%%%%%%%%%%%%%%%%%%%%%%%%%%%%%%%%%%%%%%%%%%%%%%%%%%%%%
Note that the potential $V_{0}$ depends on the spacetime dimensions as well as the angular momentum $\ell$. To determine the LNs, we consider the static limit ($\omega \to 0$) of this equation, also focus on the massless case. In this situation, the above differential equation can be solved exactly for generic $\ell$, depending on the values of the ratio $\hat{\ell}=\ell/(d-3)$. For integer, half-integer and non-integer values of $\hat{\ell}$ the solutions to the above equation are different. We write below the solutions in each of these three cases, 
%%%%%%%%%%%%%%%%%%%%%%%%%%%%%%%%%%%%%%%%%%%%%%%%%%%%%%%%%%%%%%%%%%%%%%%%%%%%%%
\begin{align}
\,_{0}\Psi^{\textrm{(non-integer)}}_{\ell m}&=\mathcal{A}^{\rm nI}_{1}\left(\frac{r_{+}}{r}\right)^{\frac{d+2\ell-4}{2}}\,_{2}F_{1}\left(\hat{\ell}+1,\hat{\ell}+1;2\hat{\ell}+2; \left(\frac{r_{+}}{r}\right)^{d-3}\right)
\nonumber
\\
&\qquad +\mathcal{A}^{\rm nI}_{2}\left(\frac{r_{+}}{r}\right)^{\frac{-d-2\ell+2}{2}}\,_{2}F_{1}\left(-\hat{\ell},-\hat{\ell};-2\hat{\ell};\left(\frac{r_{+}}{r}\right)^{d-3}\right)~,
\label{schDscalarni}
\\
\,_{0}\Psi^{\textrm{(half-integer)}}_{\ell m}&=\mathcal{A}^{\rm hI}_{1}\left(\frac{r_{+}}{r}\right)^{\frac{d+2\ell-4}{2}}\,_{2}F_{1}\left(\hat{\ell}+1,\hat{\ell}+1;2\hat{\ell}+2; \left(\frac{r_{+}}{r}\right)^{d-3}\right)
\nonumber
\\
&\qquad +\mathcal{A}^{\rm hI}_{2}\left(\frac{r_{+}}{r}\right)^{\frac{d+2\ell-4}{2}}\,_{2}F_{1}\left(\hat{\ell}+1,\hat{\ell}+1;1;1-\left(\frac{r_{+}}{r}\right)^{d-3}\right)~,
\\
\,_{0}\Psi^{\textrm{(integer)}}_{\ell m}&=\mathcal{A}^{\rm I}_{1}\left(\frac{r_{+}}{r}\right)^{\frac{d+2\ell-4}{2}}\,_{2}F_{1}\left(\hat{\ell}+1,\hat{\ell}+1;2\hat{\ell}+2; \left(\frac{r_{+}}{r}\right)^{d-3}\right)
\nonumber
\\
&\qquad +\mathcal{A}^{\rm I}_{2}(-1)^{-\hat{\ell}-1}\left(\frac{r_{+}}{r}\right)^{\frac{2-d}{2}}\,_{2}F_{1}\left(-\hat{\ell},\hat{\ell}+1;1;\left(\frac{r}{r_{+}}\right)^{d-3}\right)~.
\end{align}
%%%%%%%%%%%%%%%%%%%%%%%%%%%%%%%%%%%%%%%%%%%%%%%%%%%%%%%%%%%%%%%%%%%%%%%%%%%%%%
The finiteness of the scalar perturbation at the event horizon yields different criteria for the arbitrary constants appearing above, for different choices of $\hat{\ell}$. For example, when $\hat{\ell}$ is non-integer, then use of~\ref{hyplog} and~\ref{hyplog2} yields a logarithmic divergence at the event horizon. In order to avoid this divergence, we must have $\mathcal{A}^{\rm nI}_{1}=\{\Gamma(-2\hat{\ell}-1)/\Gamma(-\hat{\ell})^{2}\}A$, as well as $\mathcal{A}^{\rm nI}_{2}=\{\Gamma(2\hat{\ell}+1)/\Gamma(\hat{\ell}+1)^{2}\}A$. Upon substitution of the above relation between the arbitrary constants in~\ref{schDscalarni}, and using~\ref{hypz1mz}, one arrives at the following expression for the radial scalar perturbation for non-integer values of $\hat{\ell}$
%%%%%%%%%%%%%%%%%%%%%%%%%%%%%%%%%%%%%%%%%%%%%%%%%%%%%%%%%%%%%%%%%%%%%%%%%%%%%%
\begin{align}\label{scalarnonintschD}
\,_{0}\Psi^{\textrm{(non-integer)}}_{\ell m}&\propto \left(\frac{r_{+}}{r}\right)^{\frac{d+2\ell-4}{2}}\,_{2}F_{1}\left(\hat{\ell}+1,\hat{\ell}+1;1;1-\left(\frac{r_{+}}{r}\right)^{d-3}\right)~.
\end{align}
%%%%%%%%%%%%%%%%%%%%%%%%%%%%%%%%%%%%%%%%%%%%%%%%%%%%%%%%%%%%%%%%%%%%%%%%%%%%%%
This shows how the regularity at the event horizon is manifested in the radial function. On the other hand, for half-integer values of $\hat{\ell}$, the solution with $\mathcal{A}^{\rm hI}_{1}$ has logarithmic divergence at the horizon, while the solution with $\mathcal{A}^{\rm hI}_{2}$ is finite there. Therefore, the regularity at the horizon demands $\mathcal{A}_{1}^{\rm hI}=0$. Similarly, for integer values of $\hat{\ell}$ also, it follows that the regularity at the horizon warrants $\mathcal{A}_{1}^{\rm I}=0$. These fix one of the two unknown coefficients appearing in the above solutions. The LN is then extracted by determining the expansion of the radial part of the scalar field $\phi$ at large $r$, which for each of these three cases yield,
%%%%%%%%%%%%%%%%%%%%%%%%%%%%%%%%%%%%%%%%%%%%%%%%%%%%%%%%%%%%%%%%%%%%%%%%%%%%%%
\begin{align}
&\,_{0}\Psi^{\textrm{(non-integer)}}_{\ell m}(r\to\infty)\simeq
\frac{\Gamma(2\hat{\ell}+1)}{\Gamma(\hat{\ell}+1)^{2}}\left(\frac{r}{r_{+}}\right)^{\ell}+\frac{\Gamma(-2\hat{\ell}-1)}{\Gamma(-\hat{\ell})^{2}}\left(\frac{r_{+}}{r}\right)^{\ell+d-3}~,
\\
&\,_{0}\Psi^{\textrm{(half-integer)}}_{\ell m}(r\to\infty)\simeq 
\left(\frac{r}{r_{+}}\right)^{\frac{d-2}{2}}\Big[(d-3)\left(\frac{r_{+}}{r}\right)^{d+\ell-3}\log\left(\frac{r_{+}}{r}\right)+\cdots
\nonumber
\\
&\qquad \qquad \qquad \qquad \qquad \qquad \quad +(-1)^{2\hat{\ell}}(2\hat{\ell})!(2\hat{\ell}+1)!\frac{\Gamma(-\hat{\ell})^{2}}{\Gamma(\hat{\ell}+1)^{2}}\left(\frac{r}{r_{+}}\right)^{\ell}\Big]~,
\\
&\,_{0}\Psi^{\textrm{(integer)}}_{\ell m}(r\to\infty)\simeq-\frac{\Gamma(1+2\hat{\ell})}{\Gamma(1+\hat{\ell})^{2}}\left(\frac{r}{r_{+}}\right)^{\ell}~.
\end{align}
%%%%%%%%%%%%%%%%%%%%%%%%%%%%%%%%%%%%%%%%%%%%%%%%%%%%%%%%%%%%%%%%%%%%%%%%%%%%%%
The asymptotic limit of the radial function for non-integer values of $\hat{\ell}$ follows from~\ref{hypzero}. For the case with half-integer $\hat{\ell}$, the asymptotic limit can be arrived at by the use of~\ref{hypcinteger} and~\ref{hypzero}, respectively. Note that, in this case, there will be several sub-leading terms, which we have not displayed here. Finally, for integer values of $\hat{\ell}$, using~\ref{hypnegativea}, the above asymptotic form for the radial function can be obtained. 

The LN for the scalar perturbation can be determined from the coefficient of the $r^{-2\ell-d+3}$ term, which is non-zero for both non-integer and half-integer values of $\hat{\ell}$. Though non-zero there is a striking difference between these two cases, in particular, for half-integer values of $\hat{\ell}$, the LN has a logarithmic behavior. While for integer values of $\hat{\ell}$, the scalar field only grows with the radial distance, leading to vanishing contribution to the LN. To summarize, we have the following expressions for the scalar LNs of a $d$-dimensional Schwarzschild BH \citep{Hui:2020xxx}:
%%%%%%%%%%%%%%%%%%%%%%%%%%%%%%%%%%%%%%%%%%%%%%%%%%%%%%%%%%%%%%%%%%%%%%%%%%%%%%
\begin{align}
&\,_{0}k_{\ell m}^{\textrm{(non-integer)}}=\left(\frac{2\hat{\ell}+1}{2\pi}\right)\frac{\Gamma(\hat{\ell}+1)^{4}}{\Gamma(2\hat{\ell}+2)^{2}}\tan(\pi \hat{\ell})~,
\label{HD_scalar_n}
\\
&\,_{0}k_{\ell m}^{\textrm{(half-integer)}}=\frac{(-1)^{2\hat{\ell}}(d-3)\Gamma(\hat{\ell}+1)^{2}}{(2\hat{\ell})!(2\hat{\ell}+1)!\Gamma(-\hat{\ell})^{2}}\log\left(\frac{r_{+}}{r}\right)~,
\label{HD_scalar_h}
\\
&\,_{0}k_{\ell m}^{\textrm{(integer)}}=0~.
\label{HD_scalar_I}
\end{align}
%%%%%%%%%%%%%%%%%%%%%%%%%%%%%%%%%%%%%%%%%%%%%%%%%%%%%%%%%%%%%%%%%%%%%%%%%%%%%%
Thus, higher dimensional BHs can have non-zero LNs under scalar perturbations. For example, if we consider $d=5$, then the $\ell=3$ mode will have non-zero LN, which will involve a logarithmic scaling. While for $d=6$, the $\ell=4$ mode will have non-zero LN, which will have no such radial dependence. 

\paragraph{(ii) Electromagentic LNs}~--- Having observed non-zero scalar LNs for higher dimensional Schwarzschild BH, we now move on to discuss the electromagnetic LNs, often referred to as electromagnetic polarizabilities. Since electromagnetic field is massless and described by a vector field, we will consider perturbations due to a massless vector field, which can be decomposed into parity even and parity odd sector, referred to as the scalar and the vector modes, respectively. In this case also the vector field can be decomposed into a radial part, along with vector hyperspherical harmonics. Given the static nature of the background, the time dependence of the perturbation remains as $e^{-i\omega t}$. The radial part of the massless vector perturbation satisfies the following Schr\"{o}dinger-like equations: 
%%%%%%%%%%%%%%%%%%%%%%%%%%%%%%%%%%%%%%%%%%%%%%%%%%%%%%%%%%%%%%%%%%%%%%%%%%%%%%
\begin{align}
&\dfrac{d^{2}\,_{1}\Psi_{\ell m\textrm{(s/v)}}}{dr_{*}^{2}}+\left(\omega^{2}-V_{1\textrm{(s/v)}}\right)\,_{1}\Psi_{\ell m\textrm{(s/v)}}=0~,
\end{align}
%%%%%%%%%%%%%%%%%%%%%%%%%%%%%%%%%%%%%%%%%%%%%%%%%%%%%%%%%%%%%%%%%%%%%%%%%%%%%%
where $\,_{1}\Psi_{\ell m\textrm{(s)}}$ is the scalar (even parity) and $\,_{1}\Psi_{\ell m\textrm{(v)}}$ is the vector (odd parity) mode associated with massless vector perturbation of the higher dimensional Schwarzschild BH. The effective potentials associated with each of these modes read:
%%%%%%%%%%%%%%%%%%%%%%%%%%%%%%%%%%%%%%%%%%%%%%%%%%%%%%%%%%%%%%%%%%%%%%%%%%%%%%
\begin{align}
&V_{1\textrm{(s)}}=f\left[\frac{\ell(\ell+d-3)}{r^{2}}+\frac{(d-4)\left\{(d-2)f-2rf'\right\}}{4r^{2}}\right]~,
\\
&V_{1\textrm{(v)}}=f\left[\frac{(\ell+1)(\ell+d-4)}{r^{2}}+\frac{(d-4)\left\{(d-6)f+2rf'\right\}}{4r^{2}}\right]~.
\end{align}
%%%%%%%%%%%%%%%%%%%%%%%%%%%%%%%%%%%%%%%%%%%%%%%%%%%%%%%%%%%%%%%%%%%%%%%%%%%%%%
Note that the function $f$ in the above expressions is simply the $-g_{tt}$ component of the higher dimensional Schwarzschild metric, described by~\ref{hd_sch_metric}. Also in $d=4$, both the potentials take identical values, but for $d>4$, the even and the odd sector exhibits different potentials. We first focus on the even sector and then discuss the LNs in the odd sector. In the even sector, alike the scalar case, there are three different scenarios to consider, (a) $\hat{\ell}=\ell/(d-3)$ is neither an integer, nor an half-integer, (b) $\hat{\ell}$ is an half-integer and finally, (c) $\hat{\ell}$ is an integer. In all of these cases, the solutions are in terms of various hypergeometric functions, which should satisfy regularity at the horizon, and asymptotically exhibits a growing mode ($r^{\ell}$) and a decaying mode ($r^{-\ell-d+3}$). The ratio of the coefficients among these modes provide us the LNs, which we summarize below: 
%%%%%%%%%%%%%%%%%%%%%%%%%%%%%%%%%%%%%%%%%%%%%%%%%%%%%%%%%%%%%%%%%%%%%%%%%%%%%%
\begin{align}
&\,_{1}k_{\ell m\textrm{(s)}}^{\textrm{(non-integer)}}=-\left(\frac{\hat{\ell}+1}{2^{2+4\hat{\ell}}\hat{\ell}}\right)\frac{\Gamma(\hat{\ell})\Gamma(\hat{\ell}+2)}{\Gamma(\hat{\ell}+\frac{3}{2})\Gamma(\hat{\ell}+\frac{1}{2})}\tan(\pi \hat{\ell})~,
\\
&\,_{1}k_{\ell m\textrm{(s)}}^{\textrm{(half-integer)}}=\frac{(-1)^{2\hat{\ell}}(d-3)\Gamma(\hat{\ell}+2)\Gamma(\hat{\ell})}{(2\hat{\ell}+1)!(2\hat{\ell})!\Gamma(1-\hat{\ell})\Gamma(-1-\hat{\ell})}\log\left(\frac{r_{+}}{r}\right)~,
\\
&\,_{1}k_{\ell m\textrm{(s)}}^{\textrm{(integer)}}=0~.
\end{align}
%%%%%%%%%%%%%%%%%%%%%%%%%%%%%%%%%%%%%%%%%%%%%%%%%%%%%%%%%%%%%%%%%%%%%%%%%%%%%%
Note that despite differences in the overall factors, the key features of the polar EM LNs are similar to the scalar LNs, namely, for non-integer and non-half-integer values of $\hat{L}$, the LN scales as $\tan(\pi \hat{\ell})$, for half-integer values of $\hat{\ell}$, the LN scales as $\log r$, while it vanishes for integer values of $\hat{\ell}$ \citep{Hui:2020xxx}. 

The axial/vector part of the EM perturbation, on the other hand, also satisfies a hypergeometric differential equation, however with the hypergeometric parameters dependent on both $\hat{\ell}$ and the spacetime dimension $d$ through the term $1/(d-3)$. This is unlike the previous two cases considered here, where the hypergeometric functions were dependent on $\hat{\ell}$ alone. Here we have four distinct cases --- (a) $\hat{\ell}$ is neither an integer, nor an half-integer; (b) it has two sub cases: (i) $\hat{\ell}$ is an half integer, along with $d>5$, and (ii) $\hat{\ell}$ is an integer, with $d\geq 5$; (c) it has two sub cases: (i) $d=4$ and $\ell$ is arbitrary, as well as, (ii) $d=5$ and $\hat{\ell}$ is an half integer; and finally (d) we have $\hat{\ell}\pm \{1/(d-3)\}$ to be an integer and $d>5$. In all of these cases, regularity at the horizon and the overall fall-off condition ($r^{-2\ell-d+3}$) fixes the LNs. Among these four choices, the options (c) and (d), as described above, does not have a decaying mode, and hence provides vanishing LNs for the axial vector perturbations. While they yield non-zero answers for the other two, namely options (a) and (b). In particular, for option (a), i.e., for non-integer and non-half-integer values of $\hat{\ell}$, we have, 
%%%%%%%%%%%%%%%%%%%%%%%%%%%%%%%%%%%%%%%%%%%%%%%%%%%%%%%%%%%%%%%%%%%%%%%%%%%%%%
\begin{align}\label{HDSchniEM}
\,_{1}k_{\ell m\textrm{(v)}}^{\textrm{(non-integer)}}&=\left(\frac{2\hat{\ell}+1}{\Gamma(2\hat{\ell}+2)^{2}}\right)\Gamma\left(\hat{\ell}+1-\frac{1}{d-3}\right)^{2}\Gamma\left(\hat{\ell}+1+\frac{1}{d-3}\right)^{2}
\nonumber
\\
&\qquad \qquad \times\frac{\sin[\pi(\hat{\ell}-\frac{1}{d-3})]\sin[\pi(\hat{\ell}+\frac{1}{d-3})]}{\pi\sin(2\pi \hat{\ell})}~.
\end{align}
%%%%%%%%%%%%%%%%%%%%%%%%%%%%%%%%%%%%%%%%%%%%%%%%%%%%%%%%%%%%%%%%%%%%%%%%%%%%%%
Finally, for half-integer values of $\hat{\ell}$ along with $d>5$, or, for integer values of $\hat{\ell}$ with $d\geq 5$, we will have the following non-zero LNs associated with the axial EM perturbations,
%%%%%%%%%%%%%%%%%%%%%%%%%%%%%%%%%%%%%%%%%%%%%%%%%%%%%%%%%%%%%%%%%%%%%%%%%%%%%%
\begin{align}\label{HDSchhiEM}
\,_{1}k_{\ell m\textrm{(v)}}^{\textrm{(half-integer/integer)}}=\frac{(-1)^{2\hat{\ell}}(d-3)\Gamma\left(\hat{\ell}+1-\frac{1}{d-3}\right)\Gamma\left(\hat{\ell}+1+\frac{1}{d-3}\right)}{(2\hat{\ell}+1)!(2\hat{\ell})!\Gamma\left(-\hat{\ell}-\frac{1}{d-3}\right)\Gamma\left(-\hat{\ell}+\frac{1}{d-3}\right)}\log\left(\frac{r_{+}}{r}\right)~.
\end{align}
%%%%%%%%%%%%%%%%%%%%%%%%%%%%%%%%%%%%%%%%%%%%%%%%%%%%%%%%%%%%%%%%%%%%%%%%%%%%%%
Therefore, unlike the four dimensional Schwarzschild BH, the higher dimensional Schwarzschild BH will have non-zero LNs under EM perturbation, for specific choices of the angular number $\ell$ and spacetime dimensions $d$ \citep{Hui:2020xxx}.

\paragraph{(iii) Gravitational LNs}~--- Finally, the gravitational perturbation can now be decomposed into three irreducible modes --- the scalar (parity even), the vector (parity odd) and the tensor (parity even). In four spacetime dimensions, the scalar mode is synonymous to the polar sector and the vector mode is identical to the axial sector, while the tensor mode has no analogue in four dimensions. 

Interestingly, the tensor mode of the gravitational perturbation associated with higher-dimensional Schwarzschild BH satisfies the same differential equation as that of the scalar perturbation. As a consequence, the LNs associated with the tensor modes are identical to those presented in~\ref{HD_scalar_n} -~\ref{HD_scalar_I}, and hence we will not discuss it any further. 

The vector modes of the gravitational perturbation for higher dimensional Schwarzschild BH can be combined to yield a Regge--Wheeler-like master function, which in the static limit, after appropriate radial re-scaling, will satisfy the Hypergeometric differential equation \citep{Hui:2021vcv}. It turns out that this equation has a structure similar to the one from axial EM perturbation, with $\{1/(d-3)\}\to 1+\{1/(d-3)\}$ and all the LNs presented in~\ref{HDSchniEM} and~\ref{HDSchhiEM} are modified accordingly. There is one point that needs discussion here, namely, the LNs so obtained are the Regge--Wheeler LNs, which are indeed connected with the LNs obtained from metric perturbations. In four dimensions, such a relation in the axial sector can be obtained by realizing that, if $\Psi_{\rm RW}\sim \mathcal{A}^{\rm RW}_{1}r^{\ell+1}+\mathcal{A}^{\rm RW}_{2}r^{-\ell}$, asymptotically, then the LN through Regge--Wheeler master function will be proportional to $(\mathcal{A}^{\rm RW}_{2}/\mathcal{A}^{\rm RW}_{1})$. Therefore, from~\ref{h0RW}, it follows that $h_{0}$ will have the following behavior in the asymptotic region, $h_{0}\sim \mathcal{A}^{\rm RW}_{1}(\ell+2)r^{\ell+1}+\mathcal{A}^{\rm RW}_{2}(-\ell+1)r^{-\ell}$, and hence the LN, as defined above, is related to the Regge--Wheeler LN through the following expression: 
%%%%%%%%%%%%%%%%%%%%%%%%%%%%%%%%%%%%%%%%%%%%%%%%%%%%%%%%%%%%%%%%%%%%%%%%%%%%%%%%%
\begin{equation}\label{h0RWLove4D}
\,_{2}k_{\ell}^{\rm B}=\left(\frac{1-\ell}{\ell+2}\right)k_{\ell}^{\rm RW}~.
\end{equation}
%%%%%%%%%%%%%%%%%%%%%%%%%%%%%%%%%%%%%%%%%%%%%%%%%%%%%%%%%%%%%%%%%%%%%%%%%%%%%%%%%
Here we will show the equivalent relation in higher spacetime dimensions. The first step in this direction is the generalization of~\ref{h0RW}, which takes the following form\footnote{Note that our relation differs from \citet{Hui:2020xxx} by an overall factor of $(1/\sqrt{2(\ell-1)(\ell+d-2)})$. However, this does not affect the LNs, since LNs come from the ratio between decaying and growing modes, it follows that they are independent of any overall normalization.}
%%%%%%%%%%%%%%%%%%%%%%%%%%%%%%%%%%%%%%%%%%%%%%%%%%%%%%%%%%%%%%%%%%%%%%%%%%%%%%
\begin{align}\label{h0RWD}
h_{0}=r^{\frac{6-d}{2}}f(r)\dfrac{d\Psi_{\rm RW}}{dr}+\left(\frac{d-2}{2}\right)r^{\frac{4-d}{2}}f(r)\Psi_{\rm RW}~.
\end{align}
%%%%%%%%%%%%%%%%%%%%%%%%%%%%%%%%%%%%%%%%%%%%%%%%%%%%%%%%%%%%%%%%%%%%%%%%%%%%%%
One can easily check that for $d=4$, the above reduces to~\ref{h0RW}. The Regge--Wheeler function, in $d$ spacetime dimensions, has the following asymptotic expansion, $\Psi_{\rm RW}\propto r^{\ell+(d-2/2)}(1+2k_{\ell}^{\textrm{RW}}r^{-2\ell-d+2})$, which when substituted in~\ref{h0RWD}, yields the following relation between the Regge--Wheeler and the metric LNs,
%%%%%%%%%%%%%%%%%%%%%%%%%%%%%%%%%%%%%%%%%%%%%%%%%%%%%%%%%%%%%%%%%%%%%%%%%%%%%%
\begin{align}\label{h0RWLoveD}
\,_{2}k_{\ell}^{\textrm{B}}=\left(\frac{1-\ell}{\ell+d-2}\right)k_{\ell}^{\textrm{RW}}~.
\end{align}
%%%%%%%%%%%%%%%%%%%%%%%%%%%%%%%%%%%%%%%%%%%%%%%%%%%%%%%%%%%%%%%%%%%%%%%%%%%%%%
As in four dimensions, here also the axial LNs and the Regge--Wheeler LNs have opposite signs. Thus, the axial LNs can be found from~\ref{HDSchniEM} and~\ref{HDSchhiEM} by --- (a) changing $1/(d-3)$ to $1+(1/d-3)$, and (b) rescaling the LNs by the multiplicative factor in~\ref{h0RWLoveD}. 

Finally, let us discuss the scalar part of the gravitational perturbation, which is equivalent to the polar perturbation in four dimensional Schwarzschild BH. For Schwarzschild-Tangherlini metric, the polar gravitational perturbation, at face value, satisfies the Heun equation. By an appropriate field redefinition, it can be transformed to a Zerilli-like variable, satisfying the hypergeometric differential equation. It turns out that at general spacetime dimension $d$, the Zerilli function has the following asymptotic behavior, $\Psi_{\rm Z}\sim r^{1+\ell+(4-d/2)}(1+k^{\textrm{Z}}_{\ell}r^{-2\ell-1-d+4})$, with the Zerilli LN $k^{\textrm{Z}}_{\ell}$ taking the following general form,
%%%%%%%%%%%%%%%%%%%%%%%%%%%%%%%%%%%%%%%%%%%%%%%%%%%%%%%%%%%%%%%%%%%%%%%%%%%%%%
\begin{align}
k^{\textrm{Z}}_{\ell}=-\frac{1}{4^{1+2\hat{\ell}}}\frac{(\ell+d-3)(\ell+d-2)^{2}}{\ell(\ell-1)^{2}}\frac{\Gamma(\hat{\ell})\Gamma(2+\hat{\ell})}{\Gamma(\hat{\ell}+\frac{1}{2})\Gamma(\hat{\ell}+\frac{3}{2})}\tan(\pi\hat{\ell})~.
\end{align}
%%%%%%%%%%%%%%%%%%%%%%%%%%%%%%%%%%%%%%%%%%%%%%%%%%%%%%%%%%%%%%%%%%%%%%%%%%%%%%
Alike the case of axial gravitational perturbation, the Zerilli LN can also be expressed in terms of the LN associated with the metric perturbation, namely $\,_{2}k_{\ell}^{\textrm{E}}$. Let us explore the connection between the two, first in four dimensions. 

Given~\ref{HoZ} and the fact that at large distance from the deformed object, the Zerilli function behaves as $\Psi_{\rm Z}=\mathcal{P}_{1}^{\rm Z}r^{\ell+1}+\mathcal{P}_{2}^{\rm Z}r^{-\ell}$, one can determine the metric perturbation $H_{0}$ as, $H_{0}=\mathcal{P}_{1}^{\rm Z}(\ell+2+\gamma_{\ell}/2)r^{\ell}+\mathcal{P}_{2}^{\rm Z}(-\ell+1+\gamma_{\ell}/2)r^{-\ell-1}$, where $\gamma_{\ell}\equiv (\ell-1)(\ell+2)$. Therefore, the electric-type LNs and the Zerilli LNs, for generic $\ell$, get related by, 
%%%%%%%%%%%%%%%%%%%%%%%%%%%%%%%%%%%%%%%%%%%%%%%%%%%%%%%%%%
\begin{align}\label{electric_Zerilli_four}
\,_{2}k_{\ell}^{\rm E}=\left(\frac{-\ell+1+\frac{\gamma_{\ell}}{2}}{\ell+2+\frac{\gamma_{\ell}}{2}}\right)k_{\ell}^{\rm Z}
=\left(\frac{\ell(\ell-1)}{(\ell+2)(\ell+1)}\right)k_{\ell}^{\rm Z}~.
\end{align}
%%%%%%%%%%%%%%%%%%%%%%%%%%%%%%%%%%%%%%%%%%%%%%%%%%%%%%%%%%
Intriguingly, unlike the Regge--Wheeler LNs, the Zerilli LNs have exactly the same sign as the electric-type LNs. In $d$ spacetime dimensions, the above relation between electric LNs and Zerilli LNs reduce to,
%%%%%%%%%%%%%%%%%%%%%%%%%%%%%%%%%%%%%%%%%%%%%%%%%%%%%%%%%%%%%%%%%%%%%%%%%%%%%%
\begin{align}
\,_{2}k_{\ell}^{\textrm{E}}=\frac{\ell(\ell-1)}{(\ell+d-3)(\ell+d-2)}k^{\textrm{Z}}_{\ell}~.
\end{align}
%%%%%%%%%%%%%%%%%%%%%%%%%%%%%%%%%%%%%%%%%%%%%%%%%%%%%%%%%%%%%%%%%%%%%%%%%%%%%%
Note that for $d=4$, the above matches with~\ref{electric_Zerilli_four}. Thus, in general, for arbitrary angular number $L$ and spacetime dimension $d$, the LN of a higher-dimensional Schwarzschild BH is non-zero. For some specific choices of both $\ell$ and $d$, in particular, if $\{\ell/(d-3)\}$ is an integer, then predominantly the LNs identically vanish (for details, see \citep{Hui:2020xxx}).

\paragraph{(iv) Scalar LNs for Myers-Perry BHs}~--- For completeness, let us briefly discuss the case of rotating higher dimensional BHs, in particular, the LNs of the Myers-Perry BH \citep{Charalambous:2023jgq,Myers:1986un}. As we have demonstrated, in the static limit, the Kerr BH (rotating BH in four dimensions) has zero tidal LN, but the higher dimensional rotating BH can have non-trivial LNs, alike the higher dimensional Schwarzschild BH. However, the equivalent of the Teukolsky equations do not exist for higher dimensional Myers-Perry BH, for electromagnetic and gravitational perturbations, thus we will concentrate on the scalar sector alone. Further, we will only consider the five-dimensional case, since the generic $d$ dimensional rotating Myers-Perry BH solution is complicated to solve for, even for scalar perturbation. The metric associated with five-dimensional Myers-Perry BH, with mass $M$ and angular momenta $J_{\phi}$ and $J_{\psi}$ along the two rotation axes, reads,
%%%%%%%%%%%%%%%%%%%%%%%%%%%%%%%%%%%%%%%%%%%%%%%%%%%%%%%%%%%%%%%%%%%%%%%%%%%%%%
\begin{align}
ds^{2}&=-dt^{2}+\frac{r_{\rm s}^{2}}{\Sigma}\left(dt-a\sin^{2}\theta d\phi-b\cos^{2}\theta d\psi\right)^{2}+\frac{r^{2}\Sigma}{\Delta}dr^{2}+\Sigma d\theta^{2}
\nonumber
\\
&\qquad +(r^{2}+a^{2})\sin^{2}\theta d\phi^{2}+(r^{2}+b^{2})\cos^{2}\theta d\psi^{2}~,
\end{align}
%%%%%%%%%%%%%%%%%%%%%%%%%%%%%%%%%%%%%%%%%%%%%%%%%%%%%%%%%%%%%%%%%%%%%%%%%%%%%%
where the mass and the angular momenta of the BH are described by the following parameters: $M=(3\pi/8)r_{\rm s}^{2}$, with, $J_{\phi}=(2/3)Ma$ and $J_{\psi}=(2/3)Mb$, denoting angular momentum along the $\phi$ and the $\psi$ directions, respectively. Further, the above metric uses the following two functions, $\Sigma=r^{2}+a^{2}\cos^{2}\theta+b^{2}\sin^{2}\theta$, and $\Delta=(r^{2}+a^{2})(r^{2}+b^{2})-r_{\rm s}^{2}r^{2}$, with $\Delta=0$ yielding the locations of the BH horizons. The above parameters, namely, $(r_{\rm s},a,b)$ depicts a BH geometry, provided they satisfy the following constraint: $|a|+|b|\leq r_{\rm s}$, which we will always assume to be true.
Given the fact that the background is stationary and axisymmetric, it follows that a scalar field living in this background will be separable into a temporal, axis-dependent parts and a $(r,\theta)$ dependent function. In addition, it follows that the radial and the angular parts also separates out,
%%%%%%%%%%%%%%%%%%%%%%%%%%%%%%%%%%%%%%%%%%%%%%%%%%%%%%%%%%%%%%%%%%%%%%%%%%%%%%
\begin{align}
\Phi=\sum_{\ell mj}\int d\omega e^{-i\omega t}e^{im\phi}e^{ij\psi}\Psi^{(0)}_{\ell mj}(\rho)S_{\ell mj}(\theta)~,
\end{align}
%%%%%%%%%%%%%%%%%%%%%%%%%%%%%%%%%%%%%%%%%%%%%%%%%%%%%%%%%%%%%%%%%%%%%%%%%%%%%%
where $\rho=r^{2}$, and the radial part satisfies the following differential equation,
%%%%%%%%%%%%%%%%%%%%%%%%%%%%%%%%%%%%%%%%%%%%%%%%%%%%%%%%%%%%%%%%%%%%%%%%%%%%%%
\begin{align}\label{genrad5dmp}
\Big[\partial_{\rho}\left(\Delta \partial_{\rho}\right)&+\frac{\omega^{2}}{4}\left(\rho+a^{2}+b^{2}\right)+\left(\frac{a^{2}-b^{2}}{4}\right)\left(\frac{m^{2}}{\rho+a^{2}}-\frac{j^{2}}{\rho+b^{2}}\right)
\nonumber
\\
&+\frac{(\rho+a^{2})(\rho+b^{2})r_{\rm s}^{2}}{4\Delta}\left(\omega-\frac{am}{\rho+a^{2}}-\frac{bj}{\rho+b^{2}}\right)^{2}\Big]\Psi^{(0)}_{\ell mj}=\hat{\ell}(\hat{\ell}+1)\Psi^{(0)}_{\ell mj}~,
\end{align}
%%%%%%%%%%%%%%%%%%%%%%%%%%%%%%%%%%%%%%%%%%%%%%%%%%%%%%%%%%%%%%%%%%%%%%%%%%%%%%
while the angular equation reduces to,
%%%%%%%%%%%%%%%%%%%%%%%%%%%%%%%%%%%%%%%%%%%%%%%%%%%%%%%%%%%%%%%%%%%%%%%%%%%%%%
\begin{align}
\Big[&-\frac{1}{4}\Delta^{(0)}_{\mathbb{S}^{3}}+\frac{\omega^{2}}{4}\left(a^{2}\sin^{2}\theta+b^{2}\cos^{2}\theta\right) \Big]S_{\ell mj}(\theta)=\hat{\ell}(\hat{\ell}+1)S_{\ell mj}~.
\end{align}
%%%%%%%%%%%%%%%%%%%%%%%%%%%%%%%%%%%%%%%%%%%%%%%%%%%%%%%%%%%%%%%%%%%%%%%%%%%%%%
Here, $\hat{\ell}=\ell/(d-3)$, and $\Delta^{(0)}_{\mathbb{S}^{3}}$ is the Laplace-Beltrami operator for scalars on the three-sphere, whose exact expression reads,
%%%%%%%%%%%%%%%%%%%%%%%%%%%%%%%%%%%%%%%%%%%%%%%%%%%%%%%%%%%%%%%%%%%%%%%%%%%%%%
\begin{align}
\Delta^{(0)}_{\mathbb{S}^{3}}&=\frac{1}{\sin \theta \cos \theta}\partial_{\theta}\left(\sin \theta \cos \theta\partial_{\theta}\right)-\frac{m^{2}}{\sin^{2}\theta}-\frac{j^{2}}{\cos^{2}\theta}~.
\end{align}
%%%%%%%%%%%%%%%%%%%%%%%%%%%%%%%%%%%%%%%%%%%%%%%%%%%%%%%%%%%%%%%%%%%%%%%%%%%%%%
The radial differential equation cannot be solved in closed form for generic frequencies, but can be determined under the near zone and small frequency approximation, which involves $\omega (r-r_{+})\ll 1$ and $\omega r_{+}\ll 1$. The solution so obtained, will be fixed by two boundary conditions --- (a) the radial function will be purely ingoing at the BH horizon, and (b) the behavior at large $r$ (here $r_{+}\ll r\ll \omega^{-1}$) will have a growing piece ($r^{\ell}\sim \rho^{\hat{\ell}}$), associated with the tidal field, and a decaying piece ($r^{-\ell-2}\sim \rho^{-\hat{\ell}-1}$), corresponding to the induced quadrupole moment. Here we will restrict ourselves to the static case alone, which is obtained by setting $\omega=0$. In this case, the angular function become an eigenfunction of the Laplace-Beltrami operator $\Delta^{(0)}_{\mathbb{S}^{3}}$, and the radial equation simplifies to, 
%%%%%%%%%%%%%%%%%%%%%%%%%%%%%%%%%%%%%%%%%%%%%%%%%%%%%%%%%%%%%%%%%%%%%%%%%%%%%%
\begin{align}\label{static5DMP}
\Bigg[\dfrac{d}{dx}\left\{x(1+x)\dfrac{d}{dx}\right\}+\frac{Z_{+}^{2}}{x}-\frac{Z_{+}^{2}}{1+x}-\hat{\ell}(\hat{\ell}+1)\Bigg]R_{\ell mj}=0~,
\end{align}
%%%%%%%%%%%%%%%%%%%%%%%%%%%%%%%%%%%%%%%%%%%%%%%%%%%%%%%%%%%%%%%%%%%%%%%%%%%%%%
where $x\equiv(\rho-\rho_{+})/(\rho_{+}-\rho_{-})$, is the new radial coordinate, $\rho_{\rm s}=r_{\rm s}^{2}$, and we have the following expressions for the quantities $\rho_{\pm}$ and $Z_{+}$,
%%%%%%%%%%%%%%%%%%%%%%%%%%%%%%%%%%%%%%%%%%%%%%%%%%%%%%%%%%%%%%%%%%%%%%%%%%%%%%
\begin{align}
\rho_{\pm}&=\frac{1}{2}\left[\left(\rho_{\rm s}-a^{2}-b^{2}\right)\pm\sqrt{\left(\rho_{\rm s}-a^{2}-b^{2}\right)^{2}-4a^{2}b^{2}} \right]
\\
Z_{+}&=\frac{r_{+}}{2}\frac{\rho_{s}}{\rho_{+}-\rho_{-}}\left(m\Omega_{\phi}+j\Omega_{\psi}\right)~;
\quad 
\Omega_{\phi}=\frac{a}{\rho_{+}+a^{2}}~;
\quad
\Omega_{\psi}=\frac{b}{\rho_{+}+b^{2}}~.
\end{align}
%%%%%%%%%%%%%%%%%%%%%%%%%%%%%%%%%%%%%%%%%%%%%%%%%%%%%%%%%%%%%%%%%%%%%%%%%%%%%%
We would like to emphasize that~\ref{static5DMP} uses near-horizon approximation. In particular, except for the $(1/\Delta)$ term and the derivative term in~\ref{genrad5dmp}, all the other terms have been ignored, and the coefficient of $(1/\Delta)$ has been approximated by replacing $\rho$ by $\rho_{+}$, in order to arrive at~\ref{static5DMP}. The above differential equation has three regular singular points, and hence can be expressed in terms of Hypergeometric functions, such that the general solution reads,
%%%%%%%%%%%%%%%%%%%%%%%%%%%%%%%%%%%%%%%%%%%%%%%%%%%%%%%%%%%%%%%%%%%%%%%%%%%%%%
\begin{align}
R_{\ell mj}&=c_{1}P_{\hat{\ell}}^{-2iZ_{+}}(1+2x)+c_{2}Q_{\hat{\ell}}^{-2iZ_{+}}(1+2x)
\nonumber
\\
&=\frac{c_{1}}{\Gamma(1+2iZ_{+})}\left(\frac{1+x}{x}\right)^{-iZ_{+}}\,_{2}F_{1}\left(1+\hat{\ell},-\hat{\ell};1+2iZ_{+};-x\right)
\nonumber
\\
&+\frac{c_{2}e^{2\pi Z_{+}}\Gamma(1+\hat{\ell})}{2\Gamma(2+2\hat{\ell})x^{1+\hat{\ell}}}\left(\frac{1+x}{x}\right)^{-iZ_{+}}
\Gamma(1+\hat{\ell}-2iZ_{+})\,_{2}F_{1}\left(1+\hat{\ell},1+\hat{\ell}+2iZ_{+};2+2\hat{\ell};-\frac{1}{x}\right)
\end{align}
%%%%%%%%%%%%%%%%%%%%%%%%%%%%%%%%%%%%%%%%%%%%%%%%%%%%%%%%%%%%%%%%%%%%%%%%%%%%%%
where, in arriving at the final expression, we have used~\ref{legPHyp} and~\ref{legQHyp}, respectively. In the limit $x\to 0$, the radial function behaves as, $R_{\ell mj}\sim c_{1}\exp(iZ_{+}r_{*})+c_{2}\exp(-iZ_{+}r_{*})$. For BHs, the one regular (ingoing) at the event horizon matters, and hence we must set $c_{2}=0$, such that the radial function becomes,
%%%%%%%%%%%%%%%%%%%%%%%%%%%%%%%%%%%%%%%%%%%%%%%%%%%%%%%%%%%%%%%%%%%%%%%%%%%%%%
\begin{align}
R_{\ell mj}=\frac{c_{1}}{\Gamma(1+2iZ_{+})}\left(\frac{1+x}{x}\right)^{-iZ_{+}}\,_{2}F_{1}\left(1+\hat{\ell},-\hat{\ell};1+2iZ_{+};-x\right)~.
\end{align}
%%%%%%%%%%%%%%%%%%%%%%%%%%%%%%%%%%%%%%%%%%%%%%%%%%%%%%%%%%%%%%%%%%%%%%%%%%%%%%
Note that, in the limit $a=0=b$, we have $Z_{+}=0$, in which case the use of~\ref{hypztozbyzm1} reduces the above radial function to~\ref{scalarnonintschD}. The response function is obtained by taking the asymptotic limit of the above equation, which can be derived by using~\ref{hyp_z_1byz} and~\ref{hypzero}, yielding the following scalar tidal response function,
%%%%%%%%%%%%%%%%%%%%%%%%%%%%%%%%%%%%%%%%%%%%%%%%%%%%%%%%%%%%%%%%%%%%%%%%%%%%%%
\begin{align}
F_{\ell mj}=-\frac{\Gamma(1+\hat{\ell})\Gamma(1+\hat{\ell}+2iZ_{+})\Gamma(-2\hat{\ell})}{\Gamma(1+2\hat{\ell})\Gamma(-\hat{\ell})\Gamma(-\hat{\ell}+2iZ_{+})}\left(\frac{\rho_{+}-\rho_{-}}{\rho_{\rm s}}\right)^{2\hat{\ell}+1}~.
\end{align}
%%%%%%%%%%%%%%%%%%%%%%%%%%%%%%%%%%%%%%%%%%%%%%%%%%%%%%%%%%%%%%%%%%%%%%%%%%%%%%
Here, we have introduced the factor $\rho_{\rm s}$, in order to make the response function dimensionless, and have assumed $\hat{\ell}$ to be a complex number, obtained by the analytic continuation. Use of~\ref{mirrorgamma}, reduces the above response function to the following form, 
%%%%%%%%%%%%%%%%%%%%%%%%%%%%%%%%%%%%%%%%%%%%%%%%%%%%%%%%%%%%%%%%%%%%%%%%%%%%%%
\begin{align}
F_{\ell mj}&=\frac{\Gamma(1+\hat{\ell})^{2}|\Gamma(1+\hat{\ell}+2iZ_{+})|^{2}}{2\pi\Gamma(1+2\hat{\ell})\Gamma(2+2\hat{\ell})}\left(\frac{\rho_{+}-\rho_{-}}{\rho_{\rm s}}\right)^{2\hat{\ell}+1}
\nonumber
\\
&\qquad \times \left[-i\sinh (2\pi Z_{+})+\tan(\pi \hat{\ell})\cosh(2\pi Z_{+}) \right]~.
\end{align}
%%%%%%%%%%%%%%%%%%%%%%%%%%%%%%%%%%%%%%%%%%%%%%%%%%%%%%%%%%%%%%%%%%%%%%%%%%%%%%
Here, we have used the results that $\sin(ix)=i\sinh(x)$ and $\cos(ix)=\cosh(x)$. Therefore, the dissipative part, which corresponds to $\textrm{Im}F_{\ell mj}$, turns out to be proportional to $\sinh(2\pi Z_{+})$, which vanishes in the limit of zero rotation. In other words, static tidal perturbation does not introduce dissipation for non-rotating BHs. The real part, on the other hand, provides the LNs, which read,
%%%%%%%%%%%%%%%%%%%%%%%%%%%%%%%%%%%%%%%%%%%%%%%%%%%%%%%%%%%%%%%%%%%%%%%%%%%%%%
\begin{align}
k^{(0)}_{\ell mj}=\frac{\Gamma(1+\hat{\ell})^{2}|\Gamma(1+\hat{\ell}+2iZ_{+})|^{2}}{4\pi\Gamma(1+2\hat{\ell})\Gamma(2+2\hat{\ell})}\left(\frac{\rho_{+}-\rho_{-}}{\rho_{\rm s}}\right)^{2\hat{\ell}+1}\tan(\pi \hat{\ell})\cosh(2\pi Z_{+})~.
\end{align}
%%%%%%%%%%%%%%%%%%%%%%%%%%%%%%%%%%%%%%%%%%%%%%%%%%%%%%%%%%%%%%%%%%%%%%%%%%%%%%
Note that in the limit of zero rotation, the above matches exactly with the scalar LNs of a higher dimensional Schwarzschild BH, presented in~\ref{schDscalarni}. As evident, for integer values of $\hat{\ell}$, i.e., for $\ell$ being an even integer, the LN vanishes identically. While, for half-integer values of $\hat{\ell}$, i.e., when $\ell$ is an odd integer, the LN diverges, suggesting running of the LN. This is consistent with the findings for higher dimensional Schwarzschild BH. 

In summary, the static LNs of higher dimensional BHs, both rotating and non-rotating, are generically non-zero, and depends on the choices for $\hat{\ell}\equiv \ell/(d-3)$, where $\ell$ is the angular number and $d$ is the spacetime dimensions. For integer values of $\hat{\ell}$, the static LNs vanish, irrespective of the nature of perturbation, while for non-integer choices of $\hat{\ell}$, the static LNs are non-zero, and typically scales as $\tan(\pi\hat{\ell})$. Finally, for half-integer values of $\hat{\ell}$, depending on the spin of the perturbation, the LNs are predominantly non-zero, and depict a running behavior, often through a $\ln (r/r_{+})$ term. The running of the LN is clearer from the EFT point of view \citep{Goldberger:2009qd}, which we have already discussed before.  

\subsubsection{Love numbers of lower dimensional BH}

The scalar LNs of a rotating (2+1) dimensional BTZ BH has been studied in \citet{DeLuca:2024ufn, Bhatt:2024mvr}. For lower dimensional BHs, the scalar perturbation can be exactly solved for generic frequencies, unlike the perturbations of four and higher dimensional BHs. Imposing ingoing boundary condition at the horizon and then determining the decaying ($\sim r^{-1}$) and growing ($\sim 1$) part one obtains a non-zero LN, which scales as $k_{\ell m}\sim L^{2}(\omega^{2}L^{2}-m^{2})(r_{+}^{2}-r_{-}^{2})^{-1}$ \citep{DeLuca:2024ufn, Bhatt:2024mvr}. There is also a logarithmic running present in the LN, which suggests an EFT orgin. Moreover, \citet{Bhatt:2024mvr} has also shown that charged BTZ BH also has non-zero LNs.

\subsubsection{Electrically-charged black holes}

Tensor and vector LNs of higher-dimensional Reissner--Nordstr\"om BHs were initially investigated in \citet{Pereniguez:2021xcj}. In the tensor sector, the LNs follow a power law in the BH temperature, $\sim T_H^{2\ell+1}$, and thus vanish only at extremality. Furthermore, in the charged case new modes of polarization in the vector sector are excited, due to the coupling between gravitational and EM perturbations.  

Recently, \citet{Xia:2025zfp} investigated the full spectrum of tidal LNs for Reissner--Nordstr\"om BHs in arbitrary spacetime dimensions. By deriving an effective two-dimensional quadratic action valid for tensor, vector, and scalar-type gravitational perturbations, their framework reproduces the results for tensor and vector LNs obtained in \citet{Pereniguez:2021xcj}. In addition, they found that the scalar-type LNs vanish when the effective multipolar index is an integer, while for half-integer values the LNs exhibit a characteristic logarithmic running \citep{Goldberger:2009qd}.

Finally, the emergence of Love symmetry and the associated ladder formalism has been extended to static and dynamical tidal response for charged, rotating BHs in five dimensions, uncovering new vanishing conditions in special limits such as BPS configurations \citep{Cvetic:2026wht}.

\subsubsection{Braneworld black holes} \label{sec:braneworld} 

In the braneworld scenario, our four dimensional spacetime, considered as the brane, is embedded into a higher-dimensional spacetime, known as the bulk \citep{Randall:1999ee, Csaki:2004ay}. The braneworld BHs are four-dimensional, but do not satisfy the four-dimensional Einstein's equations, rather the higher-dimensional bulk Einstein's equations projected on to the four-dimensional brane.  We will assume the bulk spacetime to be five dimensional, however, the analysis can be generalized for any higher dimensional bulk spacetime in a straightforward manner \citep{Dadhich:2000am, Shiromizu:1999wj, Chakraborty:2015bja, Seahra:2004fg, Maartens:2003tw}. The starting point is the five dimensional Einstein's equations on the bulk with a negative cosmological constant (we assume that except for gravity, no matter fields can probe higher dimensions), 
%%%%%%%%%%%%%%%%%%%%%%%%%%%%%%%%%%%%%%%%%%%%%%%%%%%%%%%%%%%%%%%%%%%%%%%%%%%%%%
\begin{align}
\,^{(5)}G_{AB}=\Lambda_{5} g_{AB}~.
\end{align}
%%%%%%%%%%%%%%%%%%%%%%%%%%%%%%%%%%%%%%%%%%%%%%%%%%%%%%%%%%%%%%%%%%%%%%%%%%%%%%
The corresponding four-dimensional equation can be derived by projecting the above equation onto the brane, which is a four-dimensional hypersurface, with the help of the Gauss--Codazzi equations. In general, the effective four-dimensional gravitational field equations on the brane involves geometric terms, e.g., the four dimensional Einstein tensor and a projected version of the bulk Weyl tensor, as well as matter energy momentum tensor and its quadratic combinations on the brane (further assuming that the bulk cosmological constant and the brane tensions cancel each other). Our interst in this section is on BHs on the brane, which are vacuum solutions on the brane. Therefore, the effective Einstein's equations on the brane reads, 
%%%%%%%%%%%%%%%%%%%%%%%%%%%%%%%%%%%%%%%%%%%%%%%%%%%%%%%%%%%%%%%%%%%%%%%%%%%%%%
\begin{align}
\,^{(4)}G_{\mu \nu}+E_{\mu \nu}=0~;
\qquad
E_{\mu \nu}=e^{A}_{\mu}n^{B}e^{C}_{\nu}n^{D}W_{ABCD}~,
\end{align}
%%%%%%%%%%%%%%%%%%%%%%%%%%%%%%%%%%%%%%%%%%%%%%%%%%%%%%%%%%%%%%%%%%%%%%%%%%%%%%
where $e^{A}_{\mu}$ are the projectors from the bulk to the brane, $n_{A}$ are the normal vectors to the brane hypersurface and $W_{ABCD}$ is the bulk Weyl tensor. Note that given the symmetries of the Weyl tensor, it follows that $E^{\mu}_{\mu}=0$, and the Bianchi identity further demands $\nabla_{\mu}E^{\mu}_{\nu}=0$. Both of these symmetry properties are identical to the stress-energy tensor of an EM field. Though $E_{\mu \nu}$ appears in the effective Einstein's equations on the brane with an opposite sign compared to the stress-energy tensor of an EM field. This suggests that the background solution of this vacuum gravitational field equations on the brane will be identical to that of the Reissner--Nordstr\"{o}m BH, but with $Q^{2}\to -M_{\rm BH}^{2}q$, yielding, 
%%%%%%%%%%%%%%%%%%%%%%%%%%%%%%%%%%%%%%%%%%%%%%%%%%%%%%%%%%%%%%%%%%%%%%%%%%%%%%
\begin{align}
ds^{2}=-\left(1-\frac{2M_{\rm BH}}{r}-\frac{M_{\rm BH}^{2}q}{r^{2}}\right)dt^{2}+\left(1-\frac{2M_{\rm BH}}{r}-\frac{M_{\rm BH}^{2}q}{r^{2}}\right)^{-1}dr^{2}+r^{2}d\Omega^{2}~,
\end{align}
%%%%%%%%%%%%%%%%%%%%%%%%%%%%%%%%%%%%%%%%%%%%%%%%%%%%%%%%%%%%%%%%%%%%%%%%%%%%%%
where $q$ can take either sign. For Reissner--Nordstr\"{o}m BH, $Q^{2}$ has the interpretation of an EM charge, while for braneworld BH, $q$ is connected with the size of the extra spatial dimension \citep{Chamblin:1999by, Dadhich:2000am}. Thus, any constraint on $q$ can be translated to a constraint on the size of the extra spatial dimension \citep{Deka:2024ecp, Krishnendu:2024jkj}.

Let us now consider the computation of the LNs associated with the braneworld BH. Since the background geometry of a braneworld BH is identical to that of the Reissner--Nordstr\"{o}m BH, it follows that the LNs associated with (neutral) scalar and electromagnetic perturbations will be identical for both of these BHs. Since these LNs vanish for Reissner--Nordstr\"{o}m BH, it follows that they will vanish for braneworld BH as well. However, the gravitational perturbation of the braneworld BH takes a different form than that of Reissner--Nordstr\"{o}m BH, as the field equations are structurally different, and hence may lead to non-zero LNs. To start with, we note that the background spacetime with a non-zero $E_{\mu \nu}$ is equivalent to an anisotropic fluid, with $\bar{\rho}=3U$, $\bar{p}_{r}=U+2P$ and $\bar{p}_{\rm t}=U-P$, ensuring traceless-ness of $E_{\mu \nu}$, such that
%%%%%%%%%%%%%%%%%%%%%%%%%%%%%%%%%%%%%%%%%%%%%%%%%%%%%%%%%%%%%%%%%%%%%%%%%%%%%%
\begin{align}
U=-\frac{2E_{\mu \nu}u^{\mu}u^{\nu}}{(8\pi G_{5})^{2}\lambda_{\rm b}}~;
\qquad
P=\frac{2E_{\mu \nu}h^{\mu \nu}}{3(8\pi G_{5})^{2}\lambda_{\rm b}}~,
\end{align}
%%%%%%%%%%%%%%%%%%%%%%%%%%%%%%%%%%%%%%%%%%%%%%%%%%%%%%%%%%%%%%%%%%%%%%%%%%%%%%
where $G_{5}$ is the five-dimensional gravitational constant, $\lambda_{\rm b}$ is the brane tension, $u^{\mu}$ depicts the four-velocity of static observers in the BH spacetime, and $h_{\mu \nu}=g_{\mu \nu}+u_{\mu}u_{\nu}$ is the induced metric orthogonal to these observers. The background solution, which has the same structure as that of the Reissner--Nordstr\"{o}m BH, is obtained by imposing the following EoS: $2U+P=0$, leading to the following radial behavior for $U$ and $P$, $U(r)=-(P_{0}/2r^{4})$ and $P(r)=(P_{0}/r^{4})$. The `charge' $q$ is determined in terms of $P_{0}$, as $q=12\pi P_{0}M_{\rm BH}^{-2}$, and the mass function reads, $m(r)=M_{\rm BH}+6\pi(P_{0}/r)$. We will assume that the perturbations to the energy density and the pressures are due to perturbations of $U$ and $P$ alone, and hence all the matter perturbations can be related to $\delta U$ and $\delta P$. 

With the above setup, it follows that, unlike the case of a BH immersed in a dark matter environment, or the Reissner--Nordstr\"{o}m BH, for braneworld BHs the perturbations in the matter sector, namely $\delta U$ and $\delta P$, are determined uniquely in terms of the metric perturbation $H(r)$, and the EoS $(\partial U/\partial P)$. This can be seen from~\ref{deltapt}, which yields, $\delta P=(P-U)\{(\partial U/\partial P)-1\}^{-1}H$. Therefore, both $\delta \rho$ and $\delta p_{r}$ gets related to the metric perturbation $H(r)$. In this case also, owing to the fact that $\delta G^{\theta}_{\theta}=\delta G^{\phi}_{\phi}$, for static perturbations, only $H$ and $K$ of the metric perturbations are relevant. The perturbation variable $K$ gets related to the metric perturbation $H$ and its derivative, following~\ref{Kstatic}, while $H$ satisfies the following second order differential equation: 
%%%%%%%%%%%%%%%%%%%%%%%%%%%%%%%%%%%%%%%%%%%%%%%%%%%%%%%%%%%%%%%%%%%%%%%%%%%%%%
\begin{align}
H''+\left[\frac{2}{r}\left(1+\frac{m(r)}{rg(r)}\right)+\frac{8\pi r}{g(r)}\left(P-U\right)\right]H'&+\Bigg[-\frac{\ell(\ell+1)}{r^{2}g(r)}-\left(\frac{f'(r)}{f(r)}\right)^{2}+\frac{48\pi}{g}\left(P+3U\right)
\nonumber
\\
&\quad +\frac{48\pi}{g}\left(\frac{2U+P}{\frac{\partial U}{\partial P}-1}\right)\Bigg]H=0\,.
\end{align}
%%%%%%%%%%%%%%%%%%%%%%%%%%%%%%%%%%%%%%%%%%%%%%%%%%%%%%%%%%%%%%%%%%%%%%%%%%%%%%
Here, $f(r)$ is the negative of the $g_{tt}$ component, $g(r)$ is the $g^{rr}$ component of the metric, and the mass function reads, $m(r)=M_{\rm BH}+12\pi\int^{r}dr\,r^{2}U(r)$. The `prime', as usual denotes derivative with respect to the radial coordinate $r$. The above differential equation can be derived from first principle, or, by substituting the relevant expressions for the background spacetime in~\ref{Hdiffstatic}. The final form of the differential equation, upon using the dimensionless coordinate $x\equiv(r/G_{4}M)-1$, is given by, 
%%%%%%%%%%%%%%%%%%%%%%%%%%%%%%%%%%%%%%%%%%%%%%%%%%%%%%%%%%%%%%%%%%%%%%%%%%%%%%
\begin{align}
\left(x^{2}-1-\beta\right)\partial_{x}^{2}H+2x\partial_{x}H&-\Bigg[\ell(\ell+1)+\frac{4(x+1+\beta)^{2}}{(1+x)^{2}(x^{2}-1-\beta)}-\frac{2\beta}{\left(1+x\right)^{2}}\Bigg]H=0\,,
\end{align}
%%%%%%%%%%%%%%%%%%%%%%%%%%%%%%%%%%%%%%%%%%%%%%%%%%%%%%%%%%%%%%%%%%%%%%%%%%%%%%
where, $\beta\equiv (12\pi P_{0}/G_{4}^{2}M^{2})$, is referred to as the tidal charge of the braneworld BH. Unfortunately, the above equation cannot be solved analytically, and hence we will determine the LNs of the braneworld BH by expanding the above equation in powers of $\beta$. up to linear order, expanding the metric perturbation $H$ as, $H=H^{(0)}+\beta H^{(1)}$, we obtain two differential equations, 
%%%%%%%%%%%%%%%%%%%%%%%%%%%%%%%%%%%%%%%%%%%%%%%%%%%%%%%%%%%%%%%%%%%%%%%%%%%%%%
\begin{align}
(x^{2}-1)d_{x}^{2}H^{(0)}&+2xd_{x}H^{(0)}+\left[-\ell(\ell+1)-\frac{4}{x^{2}-1}\right]H^{(0)}=0\,,
\\
(x^{2}-1)d_{x}^{2}H^{(1)}&+2xd_{x}H^{(1)}+\left[-\ell(\ell+1)-\frac{4}{x^{2}-1}\right]H^{(1)}
\nonumber
\\
&=d_{x}^{2}H^{(0)}+\left[\frac{8}{(x+1)(x^{2}-1)}+\frac{4}{(x^{2}-1)^{2}}+\frac{2}{(1+x)^{2}}\right]H^{(0)}\,.
\end{align}
%%%%%%%%%%%%%%%%%%%%%%%%%%%%%%%%%%%%%%%%%%%%%%%%%%%%%%%%%%%%%%%%%%%%%%%%%%%%%%
The solution for $H^{(0)}$ is in terms of associated Legendre polynomials and the regularity at the horizon reduces it to one of the associated Legendre polynomials, see~\ref{vac_Sch_LN}. The corresponding solution only has a growing mode leading to vanishing contribution to the LNs. The solution for $H^{(0)}$ can then be inserted in the second equation, which will now have a non-trivial source term. Again imposing regularity of $H^{(1)}$ at the horizon, it turns out it has both growing and decaying mode, leading to non-zero LNs. For the dominant $\ell=2$ mode the LN reads, $k_{2}=-(2\beta/3)$. Thus, as argued in \citet{Chakravarti:2018vlt} (see also \citealt{Tan:2020hog}),  though in an ad hoc manner, the conclusion that LNs of braneworld BHs are negative remains valid.

As we will show later, this is very similar to what happens for certain class of ECOs. This should not come as a surprise, since the braneworld BH solution has the interpretation as an ECO. This follows from the AdS/CFT duality, according to which the boundary theory of an AdS bulk is a CFT. In the braneworld scenario considered here, we have a similar setup, and as a consequence one obtains a four dimensional gravity theory coupled to CFT on the brane. In other words, the classical gravity on the bulk is the holographic dual of four dimensional gravity coupled to a cut-off CFT on the brane. Therefore, all the braneworld BHs are quantum corrected due to the back reaction on them, arising from the CFT coupling. Thus, the correct gravitational field equations on the brane correspond to $G_{\mu \nu}+E_{\mu \nu}=8\pi G_{4}\langle \hat{T}_{\mu \nu}\rangle$, where $\hat{T}_{\mu \nu}$ is the energy-momentum tensor operator in the CFT, and $\langle \hat{T}_{\mu \nu}\rangle$ is its expectation value in some quantum state of the CFT. As a consequence, the location of the horizon gets shifted by such quantum corrections, which reads \citep{Dey:2020lhq, Emparan:1999wa, Anderson:2004md, Fabbri:2007kr}, 
%%%%%%%%%%%%%%%%%%%%%%%%%%%%%%%%%%%%%%%%%%%%%%%%%%%%%%%%%%%%%%%%%%%%%%%%%%%%%%
\begin{align}
\Delta r_{+}=\frac{N^{2}L_{\rm pl}^{2}c^{2}}{GM_{\rm BH}}~.
\end{align}
%%%%%%%%%%%%%%%%%%%%%%%%%%%%%%%%%%%%%%%%%%%%%%%%%%%%%%%%%%%%%%%%%%%%%%%%%%%%%%
Here, $L_{\rm pl}=\sqrt{\hbar G/c^{3}}$ is the Planck length, $M_{\rm BH}$ is the BH mass and $N$ corresponds to the number of CFT degrees of freedom living on the brane, which has the following expression $N=(L/L_{\rm pl})\sim 10^{15}(L/1~\textrm{mm})$, where $L$ corresponds to the size of the bulk spacetime, or, equivalently the anti-de~Sitter radius. Expressing $\Delta r_{+}=\epsilon r_{+}$, for a sub-solar mass BH, we obtain $\epsilon \sim 10^{-40}$, for an extra dimension with anti-de~Sitter radius of 1~mm. The shift in the location of the horizon becomes smaller for smaller anti-de~Sitter radius of the bulk spacetime. This shows that braneworld BHs can be considered as an excellent BH mimicker, with its non-zero and negative static LNs further upholding the same conclusion.

\subsubsection{Black holes in Lovelock gravity} 
As GR provides second-order gravitational field equations, the field equations for Lovelock gravity are also of second order, even though Lovelock gravity consists of higher curvature terms and are relevant only in higher spacetime dimensions. This is due to a theorem by Lovelock \citep{Lovelock:1971yv, Padmanabhan:2013xyr}. Typically, by Lovelock gravity one takes the gravitational Lagrangian density to be: $R+\alpha (\textrm{curvature})^{2}+\beta (\textrm{curvature})^{3}+\cdots$. The sum truncates at Lovelock order $N$, such that $2N\leq(d-2)$, where $d$ is the spacetime dimension. For example, in $d=5$, only the $(\textrm{curvature})^{2}$ terms will remain in the action in addition to the Einstein-Hilbert term. As the complexity of the field equations grows with more and more Lovelock terms, we present here the result of Einstein-Gauss-Bonnet gravity in five spacetime dimensions, whose action involves the Einstein-Hilbert term $R$, in addition to the Gauss-Bonnet term $\alpha(R^{2}-4R_{\mu \nu}R^{\mu \nu}+R_{\mu \nu \alpha \beta}R^{\mu \nu \alpha \beta})$ \citep{Boulware:1985wk}. The exact BH solution in the Einstein-Gauss-Bonnet gravity is known as the Boulware-Deser solution \citep{Boulware:1985wk,Garraffo:2008hu} and one can study perturbations of it due to external bosonic fields \citep{Ishibashi:2003ap}. It turns out that the LNs of the Boulware-Deser BH are non-zero for generic choices of spacetime dimensions and for generic values of the angular number $\ell$, and they have the general structure, 
%%%%%%%%%%%%%%%%%%%%%%%%%%%%%%%%%%%%%%%%%%%%%%%%%%%%%%%%%%%%%%%%%%%%%%%%%%%%%%
\begin{align}
k_{\ell}^{\rm EGB}=k_{\ell}^{\rm E}+\alpha k_{\ell}^{\rm GB}~.
\end{align}
%%%%%%%%%%%%%%%%%%%%%%%%%%%%%%%%%%%%%%%%%%%%%%%%%%%%%%%%%%%%%%%%%%%%%%%%%%%%%%
where $k_{\ell}^{\rm E}$ are the LNs of higher dimensional Schwarzschild BH, as detailed earlier, and there are corrections of order $\alpha$, the Gauss-Bonnet coupling parameter. As previous discussed, there are situations where $k_{\ell}^{\rm E}$ vanishes, in which case the LNs are proportional to the coupling $\alpha$ \citep{Singha:2025xah}. 

One could also extend this for pure Lovelock theories of gravity \citep{Dadhich:2015nua, Gannouji:2013eka, Chakraborty:2020ifg}, where one considers a single term in the Lovelock polynomial. It turns out that there are non-trivial BH solutions also in these theories, for which one can study bosonic perturbations \citep{Ishibashi:2003ap, Singha:2025xah}. The LNs associated with these perturbations of pure Lovelock BHs are very similar to that of higher dimensional Schwarzschild BHs, they vanish for certain choices of spacetime dimensions $d$, Lovelock order $N$, and the angular number $\ell$, otherwise they are non-zero. 

To summarize, for Lovelock theories in higher spacetime dimensions the bosonic LNs are typically non-zero, while they vanish only in some special cases.

%%%%%%%%%%%%%%%%%%%%%%%%%%%%%%%%%%%%
%%%%%%%%%%%%%%%%%%%%%%%%%%%%%%%%%%%%
%%%%%%%%%%%%%%%%%%%%%%%%%%%%%%%%%%%%
\subsection{Love numbers of four dimensional black holes beyond General Relativity}  \label{sec:BH4DBGR}
%%%
The vanishing of the bosonic LNs for a BH is a prerogative of GR and is in fact very fragile \citep{Cardoso:2017cfl, Katagiri:2024fpn}: as we have seen, it does not hold for BHs surrounded by matter distributions or for BHs in higher dimensions. This property is also generically broken in four-dimensional extensions of GR.

The LNs of spherically symmetric BHs in a scalar-tensor (Brans-Dicke) theory (see, e.g, \citep{Berti:2015itd}) and in Chern--Simons gravity \citep{Alexander:2009tp} were computed in \citet{Cardoso:2017cfl}.
In both cases spherically symmetric and static BHs are described by the Schwarzschild metric, as in GR.
In Brans-Dicke theory the BH metric perturbations depend on scalar tides and one can in principle have a class of scalar tidal LNs. However, both scalar-led and gravitational-led tidal LNs were found to be zero.
For Chern-Simons gravity \citep{Alexander:2009tp}, polar perturbations are the same as in GR, therefore yielding vanishing polar tidal LNs. However, axial perturbations are different and one finds
\begin{equation}
    k^{\rm M}_\ell ={\cal O}(1) \frac{\alpha^2}{M^4}\,,
\end{equation}
where $\alpha$ is the dimensionful coupling constant of the theory. A similar result holds for \emph{both} polar and axial LNs in another theory with quadratic curvature corrections, namely scalar-Gauss--Bonnet gravity \citep{Kanti:1995vq}, where the background solution itself acquires corrections to the Schwarzschild metric. 

Indeed, nonzero BH LNs appear to be a generic feature of EFTs of gravity with higher-order operators, as explicitly shown in the case of quartic \citep{Cardoso:2018ptl} and cubic \citep{Cai:2019npx, DeLuca:2022tkm} curvature terms (see also \citealt{Cano:2025zyk} for a recent generalization) and in quadratic gravity \citep{Bhattacharyya:2025slf}.
Depending on the parity of the higher-order operators, a magnetic tidal field can excite an electric LNs and vice versa \citep{Cardoso:2018ptl,Cano:2025zyk}.

As an order of magnitude estimate, unless symmetry arguments set some sectors of the perturbations to coincide with GR, for a theory with a coupling $\alpha$ (with dimensions $[\alpha]={\rm mass}^{n}$) to extra fields and higher-curvature coupling $\beta$ (with dimensions $[\beta]={\rm mass}^{m}$), one expect that the tidal LNs in the small coupling limit scale as
%%%
%%
\begin{equation}
k_\ell ={\cal O}(1)\frac{\alpha^2}{M^{2n}}+\mathcal{O}(1)\frac{\beta}{M^{m}}\,, \label{kBGR}
\end{equation}
since the stress-energy tensor is typically quadratic in the coupling, while the higher curvature coupling coming into play in a linear fashion.
An example of the first class are the aforementioned Chern-Simons and Gauss-Bonnet theories, wherein a scalar field is coupled to quadratic curvature invariants, whereas an example of the second class are EFTs of gravity with high-order curvature operators, involving no extra fields.

Within EFTs, a key issue is the matching between the Wilson coefficients and the LNs extracted from the asymptotic expansion of the field, as this procedure can be theory dependent. This problem has been explored very recently in~\citet{Wang:2026qst}.

%%%%%%%%%%%%%%%%%%%%%%%%%%%%%%%%%%%%
%%%%%%%%%%%%%%%%%%%%%%%%%%%%%%%%%%%%
%%%%%%%%%%%%%%%%%%%%%%%%%%%%%%%%%%%%
\subsection{Love numbers of asymptotically non-flat black holes}

In all the previous sections we have assumed the background BH solution to be asymptotically flat. The reason is simple: in the perturbative computation of the LNs, the source of the tidal perturbation is localized within a distance of order $1/\omega$, so that only the behavior of the perturbation up to this scale matters, rather than its true asymptotic properties. It may therefore seem surprising that the computation of LNs, which extends only up to $r\ll \omega^{-1}$, can in fact probe the physics at asymptotic infinity. On the other hand, in the worldline EFT framework the asymptotic structure of spacetime plays a crucial role, since LNs arise from matching BH perturbation theory with the EFT description at infinity. Because the two approaches must yield identical results, it is important to clarify how the asymptotic structure of spacetime affects the LNs of BHs.

%%%%%%%%%%%%%
%%%%%%%%%%%%%
%%%%%%%%%%%%%
\subsubsection{Love numbers of asymptotically de Sitter black holes}

We start with the determination of static scalar LNs for asymptotically de Sitter BHs within Einstein's gravity, which are characterized by the mass $M$ and positive cosmological constant $\Lambda$, known as Schwarzschild-dS solution:
\begin{align}\label{SdSSol}
ds^{2}=-\left(1-\frac{2M}{r}-\frac{\Lambda}{3}r^{2}\right)dt^{2}+\left(1-\frac{2M}{r}-\frac{\Lambda}{3}r^{2}\right)^{-1}dr^{2}+r^{2}d\Omega^{2}\,
\end{align}
. Here we determine the LNs using a perturbative approach, while the discussion involving worldline EFT can be found in \citet{nair2024asymptotically-199}. We start with scalar perturbations.
Owing to the static and spherically symmetric nature of the background, the scalar field can be decomposed as in~\ref{scalarpertss}. From the Klein--Gordon equation, $\square \Phi=0$, it follows that the radial part of the scalar field is governed by 
%%%%%%%%%%%%%%%%%%%%%%%%%%%%%%%%%%%%%%%%%%%%%%%%%%%%%%%%%%%%%%%%%%%%%%%%%%%%%%
\begin{align}
\Delta \dfrac{d}{dr}\left[\Delta\dfrac{d\phi(r)}{dr}\right]+\left\{r^{4}\omega^{2}-\ell(\ell+1) \right\}\phi(r)=0\,;
\quad
\Delta=r^{2}f(r)=r^{2}\left(1-\frac{2M}{r}-\frac{\Lambda}{3}r^{2}\right)\,.
\end{align}
%%%%%%%%%%%%%%%%%%%%%%%%%%%%%%%%%%%%%%%%%%%%%%%%%%%%%%%%%%%%%%%%%%%%%%%%%%%%%%
The horizons of the Schwarzschild-dS spacetime can be determined from the zeroes of $\Delta$, yielding two positive and real solutions, $r=r_{+}$ and $r=r_{\rm c}$, where $r_{+}$ is the BH horizon while $r_{\rm c}$ is the cosmological horizon, respectively. In the near-zone regime, introducing a new dimensionless coordinate $z=(r-r_{+})/r_{+}$, we obtain $\Delta=(\Lambda r_{+}^{2}/3)(r_{\rm c}-r_{+})(r_{\rm c}+2r_{+})z(1+\alpha z)$, where we have ignored terms $\mathcal{O}(z^{3})$, and $\alpha=1-\{r_{+}/(r_{\rm c}-r_{+})\}+\{r_{+}/(r_{\rm c}+2r_{+})\}$. Therefore, the radial equation in the near zone becomes 
%%%%%%%%%%%%%%%%%%%%%%%%%%%%%%%%%%%%%%%%%%%%%%%%%%%%%%%%%%%%%%%%%%%%%%%%%%%%%%
\begin{align}
z(1+\alpha z)&\frac{d}{dz}\Big[z(1+\alpha z)\frac{d\phi^{\rm (near)}}{dz}\Big]
\nonumber
\\
&+\left(\frac{9\omega^2r_{+}^2}{\Lambda^{2}(r_\mathrm{c}-r_{+})^{2}(r_\mathrm{c}+2r_{+})^2}
-\frac{3\ell(\ell+1)z(1+\alpha z)}{\Lambda(r_\mathrm{c}-r_{+})(r_\mathrm{c}+2r_{+})}\right) \phi^{\rm (near)}=0\,.
\end{align}
%%%%%%%%%%%%%%%%%%%%%%%%%%%%%%%%%%%%%%%%%%%%%%%%%%%%%%%%%%%%%%%%%%%%%%%%%%%%%%
Keeping terms linear-in-frequency, as well as linear-in-cosmological constant, the above differential equation has the following solution, 
%%%%%%%%%%%%%%%%%%%%%%%%%%%%%%%%%%%%%%%%%%%%%%%%%%%%%%%%%%%%%%%%%%%%%%%%%%%%%%
\begin{align}\label{phinear}
\phi^{\rm (near)}(z)&=\mathcal{A}(1+\alpha z)^{2iM\omega}(\alpha z)^{-2 i M\omega}{}_{2}F_{1}[-\widetilde{\ell},\widetilde{\ell}+1,1-4iM\omega,-\alpha z]\,,
\\
\widetilde{\ell}&=\ell+\frac{8\ell(\ell+1)}{2\ell+1}\Lambda M^{2}\,,
\nonumber
\end{align}
%%%%%%%%%%%%%%%%%%%%%%%%%%%%%%%%%%%%%%%%%%%%%%%%%%%%%%%%%%%%%%%%%%%%%%%%%%%%%%
where we have imposed purely ingoing boundary boundary condition at the BH horizon. Note that, for $\Lambda\to 0$, we obtain $\widetilde{\ell}\to \ell$, as well as $\alpha \to 1$, and the solution reduces to the Schwarzschild result, i.e.~\ref{radialstatickerr} with $s=0=a$. 

Having obtained the near-zone solution, we consider it in the static limit and expand it to the intermediate region, which corresponds to $z\gg 1$, leading to,
%%%%%%%%%%%%%%%%%%%%%%%%%%%%%%%%%%%%%%%%%%%%%%%%%%%%%%%%%%%%%%%%%%%%%%%%%%%%%%
\begin{align}
\phi^{\rm (near)}_{\rm (int)}(z)&=\mathcal{A}
\frac{\Gamma(1+2\widetilde{\ell})}{\Gamma(1+\widetilde{\ell})^{2}}\left(\frac{\alpha r}{r_{+}}\right)^{\widetilde{\ell}}
\nonumber
\\
&\times\Bigg[1+\frac{\Gamma(1+\widetilde{\ell})^{2}\Gamma(-1-2\widetilde{\ell})}{\Gamma(1+2\widetilde{\ell})\Gamma(-\widetilde{\ell})^{2}}\left\{1+(\ell-\widetilde{\ell})\ln \left(\frac{\alpha r}{r_{+}}\right) \right\}\left(\frac{r}{r_{+}}\right)^{-2\ell-1}\Bigg]\,,
\end{align}
%%%%%%%%%%%%%%%%%%%%%%%%%%%%%%%%%%%%%%%%%%%%%%%%%%%%%%%%%%%%%%%%%%%%%%%%%%%%%%
Since $(\ell-\widetilde{\ell})\sim \Lambda M^{2}$, it follows that the term $\alpha \approx 1$ and $r_{+}\approx 2M$ within the logarithmic term, and hence the response function yields, 
%%%%%%%%%%%%%%%%%%%%%%%%%%%%%%%%%%%%%%%%%%%%%%%%%%%%%%%%%%%%%%%%%%%%%%%%%%%%%%
\begin{align}
F_{\ell\,\textrm{(dS)}}&=\frac{\Gamma(1+\widetilde{\ell})^{2}\Gamma(-1-2\widetilde{\ell})}{\Gamma(1+2\widetilde{\ell})\Gamma(-\widetilde{\ell})^{2}}\left\{1-\frac{8\ell(\ell+1)}{2\ell+1}\Lambda M^{2}\ln \left(\frac{r}{2M}\right)\right\}\,
\nonumber
\\
&\approx \frac{4\ell(\ell+1)}{2\ell+1}\Lambda M^{2}\frac{(\ell!)^{3}}{(2\ell)!(2\ell+1)!}+\mathcal{O}(\Lambda^{2}M^{4})\,.
\end{align}
%%%%%%%%%%%%%%%%%%%%%%%%%%%%%%%%%%%%%%%%%%%%%%%%%%%%%%%%%%%%%%%%%%%%%%%%%%%%%%
Hence, the static response function is purely real, and we obtain \citep{nair2024asymptotically-199} 
%%%%%%%%%%%%%%%%%%%%%%%%%%%%%%%%%%%%%%%%%%%%%%%%%%%%%%%%%%%%%%%%%%%%%%%%%%%%%%
\begin{align}
k_{\ell\,\textrm{(dS)}}^{\rm static}=\frac{2\ell(\ell+1)}{2\ell+1}\frac{(\ell!)^{3}}{(2\ell)!(2\ell+1)!}\Lambda M^{2}+\mathcal{O}(\Lambda^{2}M^{4})\,;
\qquad
\nu_{\ell\,\textrm{(dS)}}^{\rm static}=0+\mathcal{O}(\Lambda^{2}M^{4})\,.
\end{align}
%%%%%%%%%%%%%%%%%%%%%%%%%%%%%%%%%%%%%%%%%%%%%%%%%%%%%%%%%%%%%%%%%%%%%%%%%%%%%%
Thus, it follows that Schwarzschild-dS BH has non-zero static LNs, which are proportional to the cosmological constant $\Lambda$, while the static dissipation numbers vanish identically, modulo the fact that terms of $\mathcal{O}(\Lambda^{2}M^{4})$ have been neglected.

Let us briefly summarize the corresponding approach in the context of worldline EFT, where the asymptotic spacetime must satisfy the following conditions: (a) $[\nabla_{i},\square]=0$, (b) $\nabla_{i}=\partial_{i}$, and (c) $\partial_{i}\sqrt{-g}=0$. Intriguingly, due to the fact that the zero spatial curvature FRW spacetime is conformally flat, with the conformal factor depending on time alone, it follows that all the above conditions are identically satisfies. This allows one to solve for the scalar field equation in de~Sitter spacetime, which are again given by the Bessel functions. The determination of the LNs using EFT follows the following route --- (i) one first solves the scalar field equation in the near-zone region, with purely ingoing boundary condition at the horizon. The solution being given by~\ref{phinear}, (ii) subsequently, one solves the scalar field equation in the far zone, which involves two arbitrary constants $A^{\infty}_{\rm reg}$ and $A^{\infty}_{\rm irreg}$, whose ratio is given by the response function presented above, i.e., $(A^{\infty}_{\rm irreg}/A^{\infty}_{\rm reg})=F_{\ell\,\textrm{(dS)}}$, (iii) these two arbitrary constants are then related to the EFT and hence to the multipolar deformation as well as tidal multipoles, yielding the final response function. As shown in \citet{nair2024asymptotically-199}, the LNs obtained in this way is also non-zero and scales as $\Lambda M^{2}$, which is consistent with our previous findings.  

\subsubsection{Love Numbers of asymptotically anti-de Sitter black holes}

The metric of an Schwarzschild anti-de Sitter BH is identical to that of~\ref{SdSSol}, with $\Lambda<0$. In the Regge--Wheeler gauge, and in the static limit, one can reduce the metric perturbations to be described by the following two elements: $(H_{0},h_{0})$, with $H_{0}$ determining the polar perturbations and $h_{0}$ describing axial perturbations. The differential equation for $h_{0}$ takes the following form \citep{Franzin:2024cah}, 
%%%%%%%%%%%%%%%%%%%%%%%%%%%%%%%%%%%%%%%%%%%%%%%%%%%%%%%%%%%%%%%%%%%%%%%%%%%%%%
\begin{align}
h_{0}''-\frac{2\Lambda r^{3}-3\ell(\ell+1)r+12M}{r^{2}\left(\Lambda r^{3}-3r+6M\right)}h_{0}=0\,,
\end{align}
%%%%%%%%%%%%%%%%%%%%%%%%%%%%%%%%%%%%%%%%%%%%%%%%%%%%%%%%%%%%%%%%%%%%%%%%%%%%%%
while $H_{0}$, which also satisfies a second order differential equation, has a more complicated expression \citep{Franzin:2024cah}. The asymptotic behavior of the axial perturbation $h_{0}$ takes the following form, 
%%%%%%%%%%%%%%%%%%%%%%%%%%%%%%%%%%%%%%%%%%%%%%%%%%%%%%%%%%%%%%%%%%%%%%%%%%%%%%
\begin{align}\label{asymph0ads}
h_{0}=\mathcal{A}r^{2}\left[1+\frac{3(\ell-1)(\ell+2)}{2\Lambda r^{2}}+\cdots \right]
+\frac{\mathcal{B}}{r}\left[1-\frac{3(\ell-1)(\ell+2)}{10\Lambda r^{2}}+\cdots\right]\,.
\end{align}
%%%%%%%%%%%%%%%%%%%%%%%%%%%%%%%%%%%%%%%%%%%%%%%%%%%%%%%%%%%%%%%%%%%%%%%%%%%%%%
and hence one can easily identify $\mathcal{A}$ to be coefficient of the growing mode, and $\mathcal{B}$ to be the coefficient of the decaying mode, leading to the following expression for the axial LNs,
%%%%%%%%%%%%%%%%%%%%%%%%%%%%%%%%%%%%%%%%%%%%%%%%%%%%%%%%%%%%%%%%%%%%%%%%%%%%%%
\begin{align}
k^{\rm B}_{\ell}=L^{-3}\left(\frac{\mathcal{B}}{\mathcal{A}}\right)\,; 
\qquad L\equiv \sqrt{-\frac{3}{\Lambda}}\,.
\end{align}
%%%%%%%%%%%%%%%%%%%%%%%%%%%%%%%%%%%%%%%%%%%%%%%%%%%%%%%%%%%%%%%%%%%%%%%%%%%%%%
Note that unlike the definitions of LNs in the context of an asymptotically flat spacetime, here the scaling is different. Here the overall scaling is by $[\textrm{Length}]^{3}$, while in asymptotically flat case it is by $[\textrm{Length}]^{5}$. However, the length scale can also be chosen to be the horizon scale (or any combination of $M$ and $L$ with dimensions of $[\textrm{Length}]^{3}$), in which case the magnetic LNs would have a different numerical value. Imposing regularity at horizon, one solves the differential equation for $h_{0}$ numerically up to a large distance and then maps it to the asymptotic form in~\ref{asymph0ads} to determine the axial LNs. This leads to nonvanishing values \citep{Franzin:2024cah}. A similar technique can be applied to the polar sector, where the scaling of the LN is by $[\textrm{Length}]^{-1}$, unlike in asymptotically flat spacetimes, where it is given by $[\textrm{Length}]^{5}$. In this case also, choosing the length to be $L$, we get non-zero polar LNs \citep{Franzin:2024cah}. This result has also been generalized to rotating BHs in anti-de Sitter spacetime \citep{Yusmantoro:2025ylw}. Finally, in the context of anti-de Sitter BHs, \citet{Andrade:2019rpn} showed that tidal deformations of the horizon induce turbulent dynamics.

%%%%%%%%%
\subsection{Parametrized frameworks}
%%%%%%%%%%
As discussed in the previous sections, the tidal LNs of a BH are nonzero in theories beyond GR, in the presence of matter fields, in higher dimensions, and in the presence of a cosmological constant.
The variety of cases and models within each of these categories motivates developing model-independent parametrized approaches that can incorporate various cases. These approaches can be used to compute agnostic constraints on the tidal deformability, which can be eventually mapped to specific models/theories \emph{a posteriori}.

One of those approaches was developed in \citet{Katagiri:2023umb}, where one parametrizes effective potentials appearing in Regge--Wheeler, or Zerilli, equation as 
%%%%%%%%%%%%%%%%%%%%%%%%%%%%%%%%%%%%%%%%%%%%%%%%%%%%%%%%%%%%%%%%%%%%%%%%%%%%%%
\begin{align}
\dfrac{d^{2}\,_{s}\Psi_{\rm E/O}}{dr_{*}^{2}}+\left[\omega^{2}-f\left(\,_{s}V_{\ell}^{\rm E/O}+\delta V_{\ell;s}^{\rm E/O}\right)\right]=0\,,
\end{align}
%%%%%%%%%%%%%%%%%%%%%%%%%%%%%%%%%%%%%%%%%%%%%%%%%%%%%%%%%%%%%%%%%%%%%%%%%%%%%%
where $\,_{s}\Psi_{\rm E/O}$ is the master function for generic spin-s perturbations of the even (polar) and the odd (axial) sector, $\,_{s}V_{\ell}^{\rm E/O}$ is the effective potential for a Schwarzschild BH, and $\delta V_{\ell;s}^{\rm E/O}$ being the correction due to modified theories of gravity. One parametrizes this correction as, 
%%%%%%%%%%%%%%%%%%%%%%%%%%%%%%%%%%%%%%%%%%%%%%%%%%%%%%%%%%%%%%%%%%%%%%%%%%%%%%
\begin{align}\label{param_ln}
\delta V_{\ell;s}^{\rm E/O}=\frac{1}{r_{\rm H}^{2}}\sum_{j=3}\alpha_{j;s}^{\rm E/O}\left(\frac{r_{\rm H}}{r}\right)^{j}\,,
\end{align}
%%%%%%%%%%%%%%%%%%%%%%%%%%%%%%%%%%%%%%%%%%%%%%%%%%%%%%%%%%%%%%%%%%%%%%%%%%%%%%
where $j=0$ term would correspond to a mass term for the perturbation, the $j=1$ and $j=2$ terms will be comparable to the unperturbed potential, and that is why the above sum starts with $j=3$, and all the information about the theories of gravity are encoded in the parameter $\alpha_{j;s}^{\rm E/O}$. Thus, if one can determine LNs in terms of these parameters, given any theory, it will be straightforward to determine them. For example, following \citet{Katagiri:2023umb}, one can show that the following four cases can occur: 
\begin{itemize}

\item For $j\geq 2\ell+4$, for axial sector with generic integer spins and for $|s|=2$ for the polar sector, the LNs are nonzero and they do not inherit a logarithmic running. 

\item For $3\leq j \leq 2\ell+3$, the polar LNs are non-zero for $|s|=2$, but they have a logarithmic running. 

\item For $2|s|+3\leq j\leq 2\ell+3$, axial LNs are non-zero for generic $s$, and have a logarithmic running. 

\item For $3\leq j\leq 2|s|+2$, the axial LNs vanish identically for generic spin.

\end{itemize}

Thus, if one can determine the corrections to the Regge--Wheeler and Zerilli potentials (for gravitational perturbations) it is possible to comment on the nature of LNs, e.g., if it is zero or, non-zero, or, it has logarithmic running. To find exact values of LNs one must determine the coefficients $\alpha_{j;s}^{\rm E/O}$ by matching the perturbed potential in a given theory/model, with~\ref{param_ln}. 

Another important point to notice is the presence of a logarithmic term in the tidal response, highlighting a running behavior, associated with the existence of UV divergences in the corresponding graviton scattering processes \citep{Ivanov:2024sds, Saketh:2023bul,Saketh:2024juq,Ivanov:2026icp}. In fact, the coefficient of the logarithmic term typically arises from the beta function of the tidal coupling in the EFT \citep{Saketh:2023bul}, as we shall discuss in more details in~\ref{sec:dynBHs}. A possible connection between tail effects and the log term at ${\cal O}(\omega^2)$ is presented in \citet{Katagiri:2024wbg}.

Finally, \citet{Katagiri:2024fpn} provided a unified framework for quantifying tidal responses, in both conservative and dissipative contributions, for generic non-rotating objects beyond vacuum GR. In particular, it recovered and computed the LNs of BHs in several non-GR theories and, for the first time, the BH dissipative numbers beyond GR.

%%%%%%%%%%%%%%%%%%%%%%%%%%%%%%%%%%%%%%%%%%%%%%%%%%%%%%%%%%%%%%%%%%%%%%%%%%%%%%
%%%%%%%%%%%%%%%%%%%%%%%%%%%%%%%%%%%%%%%%%%%%%%%%%%%%%%%%%%%%%%%%%%%%%%%%%%%%%%
%%%%%%%%%%%%%%%%%%%%%%%%%%%%%%%%%%%%%%%%%%%%%%%%%%%%%%%%%%%%%%%%%%%%%%%%%%%%%%

\section{Static fermionic Love numbers of black holes}\label{sec:fermionic}
%%%%%%%%%%

In the previous section we focused on the \emph{bosonic} static response of a BH, considering scalar, electromagnetic, and gravitational perturbations. In this section we highlight that the \emph{fermionic} response is markedly different, as recently discovered \citep{Chakraborty:2025zyb}.

In the present context, the main difference between the two sectors relies in the conditions imposed by the regularity of the angular eigenfunctions of Teukolsky equations.
In the static ($\omega\to0$) limit, the angular equation reduces to the spin-weighted spherical harmonic equation \citep{Goldberg:1966uu,1977RSPSA.358...71B}, for which regularity at $\cos \vartheta=\pm 1$ requires $|m|\leq \ell$, $\ell\geq |s|$, and the quantization of $(\ell,m)$ in terms of $s$:
\begin{itemize}
    \item Bosonic perturbations: $(s,\ell,m)$ all integers;
    \item Fermionic perturbations: $(s,\ell,m)$ all half-integers.    
\end{itemize}
The corresponding regular angular waveforms are the spin-$s$ harmonics $S_s(\vartheta)e^{i m\varphi}={}_s Y_{\ell m}(\vartheta,\varphi)$ \citep{Goldberg:1966uu,1977RSPSA.358...71B}, in the static case.

Depending on whether $(s,\ell,m)$ are integer or half-integers, the behavior of~\ref{psi4gen_stat} is different. Simplifying\footnote{One can expand the sine factors and use standard identities involving Gamma functions with complex arguments, such as $\left| \Gamma(1+n+bi) \right|^2 
= \frac{\pi b}{\sinh(\pi b)} \prod_{k=1}^{n} \left(k^2 + b^2\right)$ and $\left| \Gamma\!\left(\tfrac{1}{2} + n + bi\right) \right|^2 
= \frac{\pi}{\cosh(\pi b)} \prod_{k=1}^{n} 
\left(\left(k-\tfrac{1}{2}\right)^2 + b^2\right)$.}~\ref{psi4gen_stat} in the fermionic sector, we get \citep{Chakraborty:2025zyb}
\begin{equation}
    {}_s\mathcal{F}_{\ell m} = \frac{(-1)^{\frac{1}{2} - s}}{2}\,
\frac{(\ell-s)!\,(\ell+s)!}{(2\ell+1)!\,(2\ell)!}\,
\prod_{k=1}^{\,\ell+\frac{1}{2}}
\left[ (k-\tfrac{1}{2})^2  \left(1-\frac{r_-}{r_+} \right)^2 + \left(\frac{2 a m}{r_+}\right)^2 \right]\,,
    \label{responseF}
\end{equation}
valid for any $(s,\ell,m)$ half-integers.
In fact, this result relies on the fact that a fermionic spin-$s$ perturbation on Kerr satisfy the Teukolsky equation with the corresponding value of $s$. This was explicitly proven for 
Dirac fields ($s=\pm \tfrac{1}{2}$, e.g. neutrinos) \citep{Brill:1957fx,ChandraBook,Chandrasekhar:1976ap,Unruh:1973bda,Lee:1977gk} and for Rarita-Schwinger fields ($s=\pm \tfrac{3}{2}$, e.g. gravitinos in supergravity) \citep{Gueven:1980be,TorresdelCastillo:1990aw}. Nevertheless, since elementary massless fermions with ${\rm spin} >3/2$ are not expected to exist in any consistent relativistic quantum field theory \citep{Weinberg:1980kq}, the above result encompasses all relevant cases. 

\subsection{Nonzero static fermionic Love numbers}
Remarkably, the fermionic response function is \emph{real} and nonvanishing for both Kerr \emph{and} Schwarzschild BHs, yielding a nonzero fermionic LN:
\begin{equation}
    {}_sk^{\rm static}_{\ell m} = \frac{(-1)^{\frac{1}{2} - s}}{4}\,
\frac{(\ell-s)!\,(\ell+s)!}{(2\ell+1)!\,(2\ell)!}\,
\prod_{k=1}^{\,\ell+\frac{1}{2}}
\left[ (k-\tfrac{1}{2})^2  \left(1-\frac{r_-}{r_+} \right)^2 + \left(\frac{2 a m}{r_+}\right)^2 \right]\,.
    \label{LoveF}
\end{equation}
In the Schwarzschild limit and for spin-$\frac{1}{2}$ fields, the fermionic LN simply reduces to 
\begin{align}
{}_{\pm\frac{1}{2}}k_{\ell m}^{\rm Schw}&=\pm 2^{-4 \ell-3}\,,
\end{align}
independent of the azimuthal number $m$, as expected. Furthermore, the LN for spin-up or spin-down perturbations is the same, modulo a sign, while in the bosonic case $\,_{s}\mathcal{F}_{\ell m}$ does not depend on the sign of $s$. This result was recently generalized to Reissner-N\"{o}rdstrom BHs and charged fermionic perturbations \citep{Pang:2025myy}.

Finally, the LN is finite for extremal BHs (obtained by taking $r_{-}\to r_{+}$ limit),
\begin{equation}
   {}_{\pm\frac{1}{2}}k_{\ell m}^{\,\rm extremal}= \pm 2^{2\ell-1} m^{2\ell+1}\frac{(\ell+1/2)! (\ell-1/2)!}{(2 \ell)! (2 \ell+1)!}\,.\label{responseExt}
\end{equation}
The above equation shows the same interesting feature noted for the dissipation numbers in the bosonic sector: in the extremal case the response grows as $\exp[2\ell (1-\ln 2)]$ in the large $\ell=m$ limit. 

As discussed for the bosonic sector, this exponential growth also persists for near-extremal BHs. In the large $\ell=m$ limit we generically have
\begin{equation}
  {}_{\pm\frac{1}{2}}k_{\ell \ell}^{\rm static}\sim \frac{(-1)^{\frac{1}{2}-s}}{4} \exp\left[{\frac{2 \ell \chi  \tan ^{-1}\left(\frac{\sqrt{1-\chi ^2}}{\chi }\right)}{\sqrt{1-\chi ^2}}}\right] \left(\frac{M}{2r_+}\right)^{2\ell+1} \,,\qquad \ell=m\gg1
\end{equation}
and again one can check that the fermionic LN grows exponentially whenever $\chi\gtrsim 0.94955$.

\paragraph{Breaking of hidden symmetries}~--- 
As discussed in the previous section, the vanishing of the BH static LNs in the bosonic sector can be interpreted in terms of hidden symmetries of the perturbations of the Kerr solution in the zero-frequency limit \citep{Hui:2020xxx, Charalambous:2021kcz, Hui:2021vcv, Berens:2022ebl, BenAchour:2022uqo, Charalambous:2022rre, Katagiri:2022vyz, Ivanov:2022qqt, DeLuca:2023mio, Charalambous:2021mea, Bonelli:2021uvf, Ivanov:2022hlo, Rai:2024lho, Bhatt:2023zsy, Sharma:2024hlz}.
In particular, can be understood in terms of ladder symmetries (see~\ref{sec:ZeroLoveSymm}), which imply that the decaying component of the $\ell=0$ mode must diverge at the horizon, thereby enforcing vanishing LNs for $\ell=0$ and, by the ladder symmetry, for all $\ell$.

Fermionic perturbations still enjoy the ladder symmetry, but their lowest multipole, $\ell=s\in \mathbb{Z}/2$, admits a regular decaying solution. This implies that the fermionic LNs are nonzero for any $\ell$.
In light of this,
an interesting problem is to revisit the $SL(2,\mathbb{R})$ ``Love'' symmetry of the near-horizon geometry\citep{Charalambous:2021kcz,Charalambous:2022rre} in the fermionic sector, as recently done in \citet{Parra-Martinez:2025bcu}.

\paragraph{Evading no-hair theorems with nonzero fermionic Love}~--- 
Finally, the vanishing of bosonic LNs for four dimensional BHs in GR is also tied to classical no-hair theorems, implying that a BH cannot sustain static bosonic hair decaying at infinity \citep{Bekenstein:1971hc,Bekenstein:1972ky,Hui:2021vcv}. 
In contrast, the fact that fermionic LNs are nonvanishing for four dimensional BHs in GR can be related to the possibility of evading the no-hair theorems with fermions, as illustrated by the recently discovered BH solutions with electroweak hair in Einstein--Weinberg--Salam theory \citep{Gervalle:2024yxj,Gervalle:2025awa}.

The nonzero fermionic response indicates that the lowest mode at $\ell = |s|$ corresponds to a static, normalizable mode (see~\ref{psi_4_intermediateB_genericspin}), i.e., a fermionic hair. 
In particular, for spin-$\tfrac{1}{2}$ the four-spinor is constant at infinity while remaining nontrivial at the horizon for $\ell = \tfrac{1}{2}$. 
The existence of such a mode could have been anticipated from supersymmetric extensions of gravity, since GR can be embedded into four-dimensional supergravity, where the fermionic superpartners of the bosonic (spin-$1$ and spin-$2$) fields defining the Kerr--Newman solution ensure the existence of spin-$\tfrac{1}{2}$ and spin-$\tfrac{3}{2}$ normalizable perturbations at the lowest value of $\ell$ in Kerr--Newman \citep{Chakraborty:2025zyb}. 
An open question is whether nontrivial fermionic hair can also be constructed for the full tower of $\ell$ values for which the response is nonzero, and whether such solutions persist at the nonlinear level.

\subsection{Zero static fermionic dissipation numbers}

Interestingly, the fact that the response in~\ref{responseF} is purely real implies that the fermionic dissipation numbers vanish identically for static perturbations,
\begin{equation}
    {}_{\pm\frac{1}{2}}\nu_{\ell m}^{\rm static}=0\,.
\end{equation}
%%%
This stands in sharp contrast with the bosonic sector, see~\ref{dissstatgen}. 

We can put the following result in a broader context by anticipating that, for the \emph{dynamical} response in the bosonic sector, one gets (see~\ref{sec:dynBHs})
%%%
\begin{equation}
    \Im({\cal F}_{s\ell m})\propto(\omega-m\Omega_H)\,,
\end{equation}
%%%
This frame-dragging factor \citep{Chia:2020yla,LeTiec:2020bos,LeTiec:2020spy} is responsible for a static bosonic dissipation number even in the static limit ($\omega\to0$), provided the BH is spinning. 

The absence of such an effect in the fermionic sector is directly tied to the absence of fermionic superradiance \citep{Unruh:1973bda,Brito:2015oca}. Indeed, the dissipation numbers are proportional to the energy absorbed at the horizon. For bosonic perturbations, the flux through the horizon scales as $\omega-m\Omega_H$, which enables superradiant energy extraction at low frequencies. In contrast, the stress-energy tensor---and its associated conserved current---for fermion fields ensures that the flux entering the horizon is always positive \citep{Unruh:1973bda}. In this case, the energy flux is simply proportional to $\omega$, even for rotating BHs, and therefore vanishes in the static limit.

\subsection{Summary: static response of a black hole under bosonic and fermionic perturbations}
%%%
We conclude this section by summarizing the properties of the static tidal response of a Kerr BH in GR. The latter is described by the following exact formula \citep{Chakraborty:2025zyb}
\begin{equation}
    {}_s\mathcal{F}_{\ell m} = \begin{dcases}
        &i\,(-1)^{1-s}\,\frac{a m}{r_+}\,
\frac{(\ell-s)!\,(\ell+s)!}{(2\ell+1)!\,(2\ell)!}\,
\prod_{k=1}^{\ell} \left[ k^2 \left(1-\frac{r_-}{r_+} \right)^2 + \left(\frac{2 a m}{r_+}\right)^2 \right]\quad\hspace{0.3cm}\text{for bosons}\\
&\frac{(-1)^{\frac{1}{2} - s}}{2}\,
\frac{(\ell-s)!\,(\ell+s)!}{(2\ell+1)!\,(2\ell)!}\,
\prod_{k=1}^{\,\ell+\frac{1}{2}}
\left[ (k-\tfrac{1}{2})^2  \left(1-\frac{r_-}{r_+} \right)^2 + \left(\frac{2 a m}{r_+}\right)^2 \right] \quad \text{for fermions},
    \end{dcases} \label{responseTOT}
\end{equation}
%%%%
where the first and second line is valid for $(s,\ell,m)$ integers and half-integers, respectively. The above formula
encapsulates the following properties:
\begin{itemize}
    \item For massless \emph{bosonic} fields (scalar, electromagnetic, gravitational for $s=0,
    \pm1,
    \pm2$, respectively), the static LNs are strictly zero, whereas the dissipation numbers are proportional to the BH angular momentum, vanishing in the Schwarzschild limit. The response is independent of the sign of $s$;
    \item For massless \emph{fermionic} fields (Dirac particles like neutrinos for $s=\pm1/2$, Rarita-Schwinger particles like gravitinos for $s=\pm3/2$), the static LNs are generically nonzero (including the non-rotating case), whereas the dissipation numbers are strictly zero, reflecting the absence of BH superradiance for fermions. The response is proportional of the sign of $s$;
    \item For both bosonic and fermionic perturbations, the response to large $\ell=m$ modes is exponential \citep{Chakraborty:2025zyb}, $\mathcal{F}_{s\ell \ell}\sim e^{2\ell x}$ where $x>0$ for $a>a_c \sim 0.94955M$, and $x<0$ otherwise.  This implies a sharp distinction in the response to large-$\ell$ perturbations between highly spinning BHs and those with lower spin.
    Possible phenomenological consequences of this exponential growth have not been studied yet.
\end{itemize}
%%%%

\begin{figure}[t!]
	\centering
 	\includegraphics[width=0.48\textwidth]{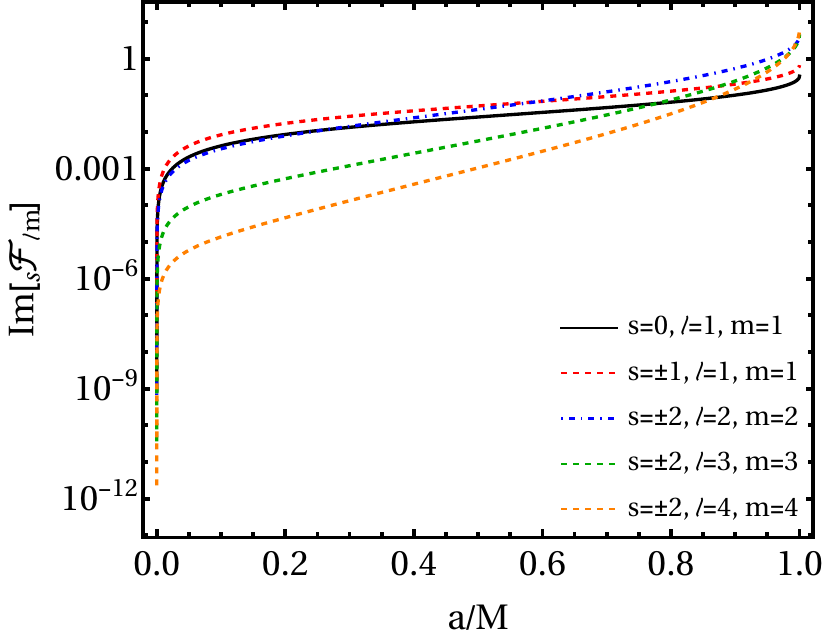}
    \includegraphics[width=0.48\textwidth]{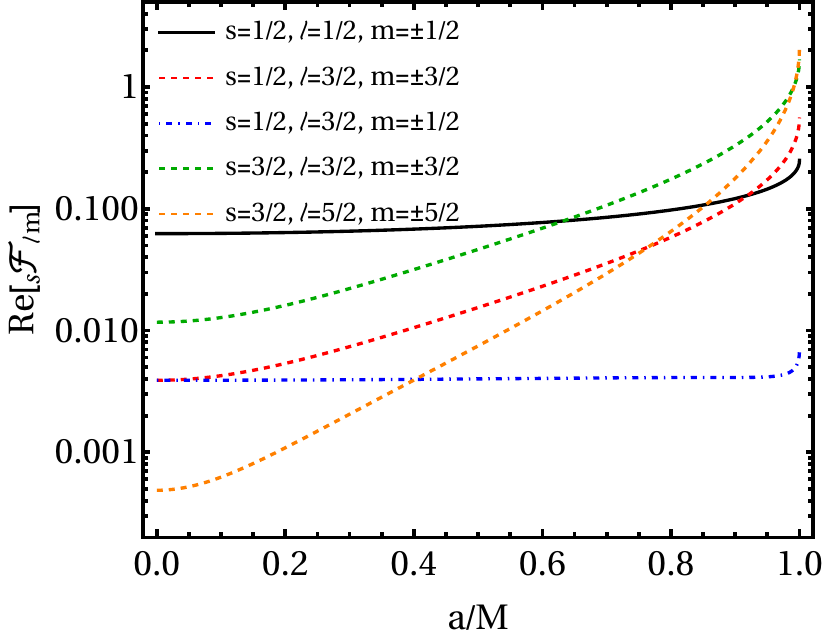}
	\caption{Bosonic (left) and fermionic (right) static response of a Kerr black hole to massless spin-$s$ perturbations as a function of the black-hole dimensionless angular momentum for representative values of $(s,\ell,m)$. Note that ${}_{s}\mathcal{F}_{\ell m}$ is finite in the extremal ($a\to M$) limit and depends on ${\rm sign}(m)$ only in the bosonic sector. }
	\label{fig:response}
\end{figure}

Examples of the response are shown in~\ref{fig:response} for representative values of $(s,\ell,m)$.

%%%%%%%%%%%%%%%%%%%%%%%
%%%%%%%%%%%%%%%%%%%%%%%
%%%%%%%%%%%%%%%%%%%%%%%
%%%%%%%%%%%%%%%%%%%%%%%
%%%%%%%%%%%%%%%%%%%%%%%
%%%%%%%%%%%%%%%%%%%%%%%
%%%%%%%%%%%%%%%%%%%%%%%
%%%%%%%%%%%%%%%%%%%%%%%

%%%%%%%%%%%%%%%%%%%%%%%%%%%%%%%%%%%%%%%%%%%%%%%%%%%%%%%%%%%%%%%%%%%%%%%%%%%%%%
%%%%%%%%%%%%%%%%%%%%%%%%%%%%%%%%%%%%%%%%%%%%%%%%%%%%%%%%%%%%%%%%%%%%%%%%%%%%%%
%%%%%%%%%%%%%%%%%%%%%%%%%%%%%%%%%%%%%%%%%%%%%%%%%%%%%%%%%%%%%%%%%%%%%%%%%%%%%%
\section{Dynamical tidal response of black holes}\label{sec:dynBHs}
In the previous sections we have discussed the \emph{static} tidal response of a BH. However, the assumption of a static tidal field is an idealization.
For example, in a binary system the tidal field has a characteristic frequency given, at the leading Newtonian order, by the Keplerian one, $\Omega=\sqrt{M_{\rm tot}/r^3}$. Therefore, the tidal field is static only if the tidal sources are located at infinity.
Realistic situations will involve the source of the tidal field to be located at a finite distance from the deformed object, as well as having a relative motion with respect to it. The analysis of the tidal response in such a dynamical context will be the main focus of this section. The dynamical tidal deformations of a compact object can be determined in several possible manners, each involving different schemes of approximation. Here we will review the existing methods for the computation of the dynamical LNs and the associated open problems. 
We will discuss only the \emph{bosonic} dynamical response, since the fermionic sector has not been explored yet in dynamical setups.

%%%%%%%%%%%%%%%%%%%%%%%%%%%%%%%%%%%%%%%%%%%%%%%%%%%%%%%%%%%%%%%%
%%%%%%%%%%%%%%%%%%%%%%%%%%%%%%%%%%%%%%%%%%%%%%%%%%%%%%%%%%%%%%%%
%%%%%%%%%%%%%%%%%%%%%%%%%%%%%%%%%%%%%%%%%%%%%%%%%%%%%%%%%%%%%%%%
\subsection{Near-zone approach}\label{nearzoneapp}

We now present the dynamical tidal deformation of a Kerr BH. Gravitational perturbations of the Kerr background in ingoing null Kerr coordinates \citep{Teukolsky:1972my, Teukolsky:1973ha, Press:1973zz, Teukolsky:1974yv} have been discussed in the context of the static tidal case in~\ref{sec:staticBHs}. Here, we focus exclusively on gravitational perturbations; the extension to other types of perturbations proceeds in an analogous manner.

In the dynamical case, the radial function associated with the Weyl scalar $\Psi_{4}$ satisfies the following equation \citep{Teukolsky:1974yv}
%%%%%%%%%%%%%%%%%%%%%%%%%%%%%%%%%%%%%%%%%%%%%%%%%%%%%%%%%%%%%%%%%%%%%%%%%%%%%%
\begin{equation}
\Delta\dfrac{\mathrm{d}^2\,_{-2}R_{\ell m}}{\mathrm{d}r^2}-2\Big[(r-M)+iK\Big]\dfrac{\mathrm{d}\,_{-2}R_{\ell m}}{\mathrm{d}r}+\Bigg[\frac{8i(r-M)K}{\Delta}-10i\omega r-\lambda\Bigg]\,_{-2}R_{\ell m}=0\,,
\label{TEqbis}
\end{equation}
%%%%%%%%%%%%%%%%%%%%%%%%%%%%%%%%%%%%%%%%%%%%%%%%%%%%%%%%%%%%%%%%%%%%%%%%%%%%%%
where $K\equiv (r^2+a^2)\omega -am$, and $\lambda\equiv E_{\ell m}-2am\omega+a^{2}\omega^{2}-2$. The constant $E_{\ell m}$ refers to the eigenvalue of the angular equation and has the following expansion in the small-frequency regime \citep{Press:1973zz, Fackerell:1977shn, Seidel:1988ue, Berti:2005gp},
%%%%%%%%%%%%%%%%%%%%%%%%%%%%%%%%%%%%%%%%%%%%%%%%%%%%%%%%%%%%%%%%%%%%%%%%%%%%%%
\begin{equation}
E_{\ell m}=\ell(\ell+1)-\left(\frac{8m}{\ell(\ell+1)}\right)a\omega+\mathcal{O}\left(a^{2}\omega^{2}\right)\,.
\end{equation}
%%%%%%%%%%%%%%%%%%%%%%%%%%%%%%%%%%%%%%%%%%%%%%%%%%%%%%%%%%%%%%%%%%%%%%%%%%%%%%
Note that the Weyl scalar $\Psi_{4}$ corresponds to Teukolsky perturbations with $s=-2$. Using $\Delta=(r-r_+)(r-r_-)$, we can re-express the radial Teukolsky equation as \citep{Chia:2020yla},
%%%%%%%%%%%%%%%%%%%%%%%%%%%%%%%%%%%%%%%%%%%%%%%%%%%%%%%%%%%%%%%%%%%%%%%%%%%%%%
\begin{multline}\label{Teqrdyn}
\dfrac{\mathrm{d}^2\,_{2}R_{\ell m}}{\mathrm{d}r^2}+\left(\frac{2iP_+-1}{r-r_+}-\frac{2iP_-+1}{r-r_-}-2i\omega\right)\dfrac{\mathrm{d}\,_{-2}R_{\ell m}}{\mathrm{d}r}
\\ 
\quad+\left[-\frac{4iP_+}{(r-r_+)^2}+\frac{4iP_-}{(r-r_-)^2}+\frac{A_-}{(r-r_-)(r_+-r_-)}-\frac{A_+}{(r-r_+)(r_+-r_-)}\right]\,_{-2}R_{\ell m}=0\,,
\end{multline}
%%%%%%%%%%%%%%%%%%%%%%%%%%%%%%%%%%%%%%%%%%%%%%%%%%%%%%%%%%%%%%%%%%%%%%%%%%%%%%
where we have introduced four constants,
%%%%%%%%%%%%%%%%%%%%%%%%%%%%%%%%%%%%%%%%%%%%%%%%%%%%%%%%%%%%%%%%%%%%%%%%%%%%%%
\begin{equation}\label{def_PpmN}
P_{\pm}=\frac{am-2r_{\pm}M\omega}{r_{+}-r_{-}}=-\frac{2Mr_{\pm}\bar{\omega}_{\pm}}{r_{+}-r_{-}}\,,
\qquad 
A_{\pm}=2i\omega r_{\pm}+\lambda\,.
\end{equation}
%%%%%%%%%%%%%%%%%%%%%%%%%%%%%%%%%%%%%%%%%%%%%%%%%%%%%%%%%%%%%%%%%%%%%%%%%%%%%%
Here, $\bar{\omega}_{\pm}\equiv\omega-m(a/2Mr_{\pm})$ is the angular frequency at the event (plus) and Cauchy (minus) horizons in the co-rotating reference frame. We introduce the re-scaled dimensionless radial coordinate $z\equiv (r-r_+)/(r_+-r_-)$, in terms of which the Teukolsky equation becomes (applying the near-zone condition, $M\omega z\ll 1$\footnote{The above near-zone condition is not free from ambiguity. For example, with this approximation, one cannot account for all $\mathcal{O}(M^{2}\omega^{2})$ terms consistently. Since near the horizon $M\omega z \approx M^{2}\omega^{2}$, ignoring $M\omega z$ while keeping terms of $\mathcal{O}(M^{2}\omega^{2})$ is inconsistent. Secondly, to obtain the LNs, we have to expand the near-zone solution to the intermediate zone by assuming $z\gg 1$, which may not be compatible with $M\omega z\ll1$. Thus, for the validity of this approximation, the intermediate zone must satisfy $1\ll z\ll 1/(M\omega)$.} and small frequency approximation $M\omega\ll1$),
%%%%%%%%%%%%%%%%%%%%%%%%%%%%%%%%%%%%%%%%%%%%%%%%%%%%%%%%%%%%%%%%%%%%%%%%%%%%%%
\begin{multline}\label{GMATE}
\frac{\mathrm{d}^{2}\,_{-2}R_{\ell m}}{\mathrm{d}z^{2}}+\left[\frac{2iP_{+}-1}{z}-\frac{2i\left(P_{+}+2\omega r_{+}\right)+1}{1+z}\right]\frac{\mathrm{d}\,_{-2}R_{\ell m}}{\mathrm{d}z}
+\left[-\frac{4iP_{+}}{z^2}+\frac{4i\left\{P_{+}+\omega (M+r_{-})\right\}}{(z+1)^2}\right.
\\
\left.-\frac{\ell(\ell+1)-2}{z(1+z)}+\frac{2am\omega}{z(1+z)}\left\{1+\frac{4}{\ell(\ell+1)}\right\}-\frac{2i\omega r_{+}}{z(1+z)}\right]\,_{-2}R_{\ell m}=0\,.
\end{multline}
%%%%%%%%%%%%%%%%%%%%%%%%%%%%%%%%%%%%%%%%%%%%%%%%%%%%%%%%%%%%%%%%%%%%%%%%%%%%%%
The above radial Teukolsky equation has three regular singular points at $z=-1,\,0,\,\text{and}\,\infty$, respectively, and hence the solution can be written in terms of the Gauss hypergeometric functions as
%%%%%%%%%%%%%%%%%%%%%%%%%%%%%%%%%%%%%%%%%%%%%%%%%%%%%%%%%%%%%%%%%%%%%%%%%%%%%%
\begin{multline}\label{radialfngen}
R(z)=(z+1)^{2-N_{3}}\left[c_{1}z^{2}\,\,_{2}F_{1}\left(3+\ell-N_{2},2-\ell-N_{1};3+2iP_{+};-z\right)\right.
\\
\left.+c_{2}z^{-2iP_{+}}\,\,_{2}F_{1}\left(\ell+1-2iP_{+}-N_{2},-\ell-2iP_{+}-N_{1};-1-2iP_{+};-z\right)\right]\,,
\end{multline}
%%%%%%%%%%%%%%%%%%%%%%%%%%%%%%%%%%%%%%%%%%%%%%%%%%%%%%%%%%%%%%%%%%%%%%%%%%%%%%
where $c_1$ and $c_2$ are integration constants, and only terms up to the linear order in $M\omega$ have been kept, consistently with the approximation. The above solution involves three quantities, $N_{1}$, $N_{2}$, and $N_{3}$, all of which are linear functions of the frequency $\omega$, 
%%%%%%%%%%%%%%%%%%%%%%%%%%%%%%%%%%%%%%%%%%%%%%%%%%%%%%%%%%%%%%%%%%%%%%%%%%%%%%
\begin{align}
N_{1}&=2\omega\Bigg[-\frac{4am}{\ell(\ell+1)(2\ell+1)}+\frac{4i\ell M-am-i(2\ell-1)r_+}{2\ell+1}
\nonumber
\\
&\qquad \qquad +\frac{2\left(r_{+}-M\right)\left(r_{+}-M+iam\right)}{am+2i(r_{+}-M)}+\frac{4i\left(r_{+}-M\right)}{2\ell+1}\Bigg]\,,\label{def_N1}\\
%%%%
N_{2}&=2\omega\Bigg[\frac{4am}{\ell(\ell+1)(2\ell+1)}+\frac{am+4i(\ell+1)M-i(2\ell+3)r_{+}}{2\ell+1}
\nonumber
\\
&\qquad \qquad \qquad +\frac{2\left(r_{+}-M\right)\left(r_{+}-M+iam\right)}{am+2i(r_{+}-M)}-\frac{4i\left(r_{+}-M\right)}{2\ell+1}\Bigg]\,,\label{def_N2}\\
%%%
N_{3}&=\frac{12\omega\left(r_{+}-M\right)^{2}}{am+2i(r_{+}-M)}\label{def_N3}\,.
\end{align}
%%%%%%%%%%%%%%%%%%%%%%%%%%%%%%%%%%%%%%%%%%%%%%%%%%%%%%%%%%%%%%%%%%%%%%%%%%%%%%
%%%%%%%%%%%%%%%%%%%%%%%%%%%%%%%%%%%%%%%%%%%%%%%%%%%%%%%%%%%%%%%%%%%%%%%%%%%%%%
One of the arbitrary constants, $c_{1}$ and $c_{2}$, in~\ref{radialfngen} can be fixed by considering the near-horizon limit. This is obtained by taking $z\to 0$ limit of~\ref{radialfngen}, and using properties of the hypergeometric function (see~\ref{hypzero} in~\ref{app:identities}), such that the radial perturbation variable $\,_{-2}R_{\ell m}(z)$ can be approximately written as
%%%%%%%%%%%%%%%%%%%%%%%%%%%%%%%%%%%%%%%%%%%%%%%%%%%%%%%%%%%%%%%%%%%%%%%%%%%%%%
\begin{equation}
\,_{-2}R^{\rm near}_{\ell m}(z) \sim c_1 z^{2}+c_2 z^{-2 i P_+}\,.
\end{equation}
%%%%%%%%%%%%%%%%%%%%%%%%%%%%%%%%%%%%%%%%%%%%%%%%%%%%%%%%%%%%%%%%%%%%%%%%%%%%%%
The first term, namely $z^{2}$, is equivalent to $\Delta^{2}$, and hence corresponds to a purely ingoing term in these coordinates \citep{Teukolsky:1973ha}. The second term of $\,_{-2}R^{\rm near}_{\ell m}$ can be expressed as $\exp(-2iP_{+}\ln z)$, which represents an outgoing solution. Since there are no outgoing modes at the event horizon of a BH, we need to impose the condition $c_{2}=0$, and~\ref{radialfngen} reduces to
%%%%%%%%%%%%%%%%%%%%%%%%%%%%%%%%%%%%%%%%%%%%%%%%%%%%%%%%%%%%%%%%%%%%%%%%%%%%%%
\begin{equation}
\,_{-2}R_{\ell m}(z)=c_{1}z^{2}(z+1)^{2-N_{3}}\,\,_{2}F_{1}\left(3+\ell-N_{2},2-\ell-N_{1};3+2iP_{+};-z\right)\,.
\end{equation}
%%%%%%%%%%%%%%%%%%%%%%%%%%%%%%%%%%%%%%%%%%%%%%%%%%%%%%%%%%%%%%%%%%%%%%%%%%%%%%
Given the radial Teukolsky function, one determines the dynamical tidal response by computing the radial part of the Newman-Penrose scalar $\Psi_{4}$ in the intermediate regime; denoted by $\Psi_{4}^{\rm (int)}$, obtained by taking the large $r$ (or, equivalently large $z$) limit. This yields
%%%%%%%%%%%%%%%%%%%%%%%%%%%%%%%%%%%%%%%%%%%%%%%%%%%%%%%%%%%%%%%%%%%%%%%%%%%%%%
\begin{align}\label{psi4gen}
\Psi_{4}^{\rm (int)}&\propto z^{\ell-2+N_{1}-N_{3}}\left\{\frac{\Gamma\left(3+2iP_{+}\right)\Gamma \left(1+2\ell+N_{1}-N_{2}\right)}{\Gamma\left(1+\ell+2iP_{+}+N_{1}\right)\Gamma\left(3+\ell-N_{2}\right)}\right\}
\Bigg[1+\left(\frac{r}{r_{+}-r_{-}}\right)^{-2\ell-1+N_{2}-N_{1}}
\nonumber
\\
&\qquad\times\left\{\frac{\Gamma\left(-1-2\ell-N_{1}+N_{2}\right)\Gamma\left(1+\ell+2iP_{+}+N_{1}\right)\Gamma\left(3+\ell-N_{2}\right)}{\Gamma\left(2-\ell-N_{1}\right)\Gamma\left(-\ell+2iP_{+}+N_{2}\right)\Gamma\left(1+2\ell+N_{1}-N_{2}\right)}\right\}\Bigg]\,.
\end{align}
%%%%%%%%%%%%%%%%%%%%%%%%%%%%%%%%%%%%%%%%%%%%%%%%%%%%%%%%%%%%%%%%%%%%%%%%%%%%%%
In order to determine the tidal response function, one must compare the above expansion with the one presented in~\ref{psi_4_intermediate}, and one can explicitly observe the existence of a $r^{-2\ell-1}$ term in~\ref{psi4gen}, albeit with a frequency dependent correction. Note that, up to linear order in $M\omega$, the fall-off behavior of the term inside bracket in~\ref{psi4gen} becomes $r^{-2\ell-1}\{1+(N_{2}-N_{1})\ln r\}$. Thus, the dynamical tidal response function, when compared with~\ref{psi_4_intermediateB_genericspin}, reads
%%%%%%%%%%%%%%%%%%%%%%%%%%%%%%%%%%%%%%%%%%%%%%%%%%%%%%%%%%%%%%%%%%%%%%%%%%%%%%
\begin{align}\label{resp_func_arb_rot}
\,_{-2}\mathcal{F}_{\ell m}(\omega)&=\frac{\Gamma\left(-1-2\ell-N_{1}+N_{2}\right)\Gamma\left(1+\ell+2iP_{+}+N_{1}\right)\Gamma\left(3+\ell-N_{2}\right)}{\Gamma\left(2-\ell-N_{1}\right)\Gamma\left(-\ell+2iP_{+}+N_{2}\right)\Gamma\left(1+2\ell+N_{1}-N_{2}\right)}
\nonumber
\\
&\times\left[1+(N_{2}-N_{1})\ln \left(\frac{r}{r_{+}-r_{-}}\right)\right]\times \left(\frac{r_{+}-r_{-}}{r_{+}}\right)^{2\ell+1}\,.
\end{align}
%%%%%%%%%%%%%%%%%%%%%%%%%%%%%%%%%%%%%%%%%%%%%%%%%%%%%%%%%%%%%%%%%%%%%%%%%%%%%%
The above expression differs from that presented in \citet{Chia:2020yla} by the presence of the terms involving $N_1$ and $N_2$, both of which are linear in $M\omega$. These terms must be included in the above expression for consistency up to $\mathcal{O}(M\omega)$. Interestingly, the LNs also has a logarithmic running, due to the inclusion of these terms, the physical importance of which will be discussed in more detail later. In the zero frequency limit, we have $N_{1}=N_{2}=0$ and $P_{+}=am/(r_{+}-r_{-})$, in which case~\ref{resp_func_arb_rot} reduces to~\ref{resp_func_arb_rotstatic}, recovering the result that the static bosonic LNs of a Kerr BH vanishes. 

In the non-static regime, the above response function can be further simplified, as our analysis is valid only up to linear order in $M\omega$, and hence we should expand the above Gamma functions in powers of $M\omega$ (for details of this calculation, see \citep{Bhatt:2024yyz, Bhatt:2024rpx}). This yields the following non-zero expression for the dynamical LNs of a Kerr BH, 
%%%%%%%%%%%%%%%%%%%%%%%%%%%%%%%%%%%%%%%%%%%%%%%%%%%%%%%%%%%%%%%%%%%%%%%%%%%%%%
\begin{align}\label{dynamic_LNs}
&\medmath{\,_{-2}k^{\rm Kerr\,(dyn)}_{\ell m}=(am)^2\,k^{(0)}_{\ell m}\times \left(\frac{r_{+}-r_{-}}{r_{+}}\right)^{2\ell+1}+ am\omega\,k^{(1)}_{\ell m}\times \left(\frac{r_{+}-r_{-}}{r_{+}}\right)^{2\ell+1}}
\nonumber
\\
&\medmath{+\frac{12am\omega}{2\ell+1}\left(\frac{r_{+}-M}{r_{+}}\right)\ln \left(\frac{r}{r_{+}-r_{-}}\right)\frac{\Gamma(\ell-1)\Gamma(\ell+3)}{\Gamma(2\ell+1)\Gamma(2\ell+2)}\prod_{j=1}^{\ell}\left\{j^{2}\left(1-\frac{r_{-}}{r_{+}}\right)^{2}+\left(\frac{2am}{r_{+}}\right)^{2}\right\}+ \mathcal{O}(M^{2}\omega^{2})~,}
\end{align}
%%%%%%%%%%%%%%%%%%%%%%%%%%%%%%%%%%%%%%%%%%%%%%%%%%%%%%%%%%%%%%%%%%%%%%%%%%%%%%
where the explicit forms of $k^{(0)}_{\ell m}$ and $k^{(1)}_{\ell m}$ are given in \citet{Bhatt:2024yyz}, and are functions of BH parameters as well as of the angular numbers $(\ell,m)$. The first thing to notice from the above expression is that it is non-zero, signaling dynamical deformation for Kerr BHs under external tidal field. Moreover, from the above expression it is clear that the dynamical LNs vanish for --- (i) Schwarzschild BH ($a=0$), (ii) slowly rotating Kerr BH, with $(a/M)\ll 1$ to linear order in $a/M$, and (iii) for axisymmetric tidal perturbation ($m=0$). These results are consistent with the earlier findings \citep{Chia:2020yla, Bhatt:2023zsy}, see also\ref{sec:staticBHs}.

The above LNs remain nonzero for a generic Kerr BH in the limit $\omega \to 0$, which is inconsistent with the results in~\ref{sec:staticBHs}. 
This inconsistency originates from a tacit assumption made in deriving~\ref{dynamic_LNs}, namely that $\ell \to \mathbb{Z}^{+}$ while $\omega \neq 0$. 
As a consequence, the near-zone computation exhibits a noncommutative behavior,
%%%%%%%%%%%%%%%%%%%%%%%%%%%%%%%%%%%%%%%%%%%%%%%%%%%%%%%%%%%%%%%%%%%%%%%%%%%%%%
\begin{align}
\lim_{\omega \to 0}\lim_{\ell\to \mathbb{Z}^{+}}k^{\rm (dyn)}_{\ell m}
\neq 
\lim_{\ell\to \mathbb{Z}^{+}}\lim_{\omega \to 0}k^{\rm (dyn)}_{\ell m}\,.
\end{align}
%%%%%%%%%%%%%%%%%%%%%%%%%%%%%%%%%%%%%%%%%%%%%%%%%%%%%%%%%%%%%%%%%%%%%%%%%%%%%%
A similar singular behavior of the $\omega \to 0$ (static) limit in connection with the LNs has been noted previously in various contexts, such as for magnetic LNs \citep{Pani:2018inf} and for ultra-compact objects \citep{Chakraborty:2023zed}, both of which will be discussed later. 
In the present case, since $\ell$ must be assumed to be an integer, the LNs defined above become ambiguous owing to the possible mixing between the tidal and response contributions at the same order. 
The emergence of a non-smooth static limit can thus be attributed to this ambiguity. 

Moreover, the approximation $M\omega z \ll 1$ is not fully consistent with~\ref{radialfngen}. 
This can be seen by multiplying the differential equation in~\ref{radialfngen} by $z(1+z)$, which modifies both the solutions and the corresponding LNs. 
We therefore conclude that the near-zone method is not ideally suited for computing the LNs. 
These conclusions hold for generic spin perturbations as well, with different values of $k^{(0)}_{\ell m}$, $k^{(1)}_{\ell m}$, and of the logarithmic contribution (see \citealt{Bhatt:2024rpx} for further details).

Finally, we would like to point out the presence of a $\ln r$ term in the expression of dynamical LNs, which, as we will discuss throughout this section, is universal to the dynamical tidal deformation. The logarithmic piece is proportional to the frequency $\omega$, and hence vanishes in the static limit. 

Again, in the extremal limit one has $r_{+}\to r_{-}$, so at first sight the dynamical LNs defined in~\ref{dynamic_LNs} appear to vanish identically. While this is indeed true for the logarithmic contribution, it does not apply to the remaining terms in~\ref{dynamic_LNs}. The reason is that both $k^{(0)}_{\ell m}$ and $k^{(1)}_{\ell m}$ contain a common overall multiplicative factor of the form
$(r_{+}-r_{-})^{-1}\,
\frac{(\ell+s)!(\ell-s)!}{(2\ell+1)!(2\ell)!}
\prod_{j=1}^{\ell}\left[j^{2}+\frac{4a^{2}m^{2}}{(r_{+}-r_{-})^{2}}\right]$,
see \citet{Bhatt:2024yyz}. As a consequence, this factor scales as $(r_{+}-r_{-})^{-1-2\ell}$ in the extremal limit. Therefore, despite the apparent suppression, the response function remains finite as $r_{+}\to r_{-}$, similarly to the static case, yielding a nonvanishing result.

%%%%%%%%%%%%%%%%%%%%%%%%%%%%%%%%%%%%
%%%%%%%%%%%%%%%%%%%%%%%%%%%%%%%%%%%%
%%%%%%%%%%%%%%%%%%%%%%%%%%%%%%%%%%%%
\subsection{Dynamical Love numbers from ladder symmetry}

The vanishing of static LNs for vacuum four-dimensional BHs in GR was first explained using ladder symmetry in \citet{Hui:2021vcv}. This result was later understood at a much broader level, in connection with $SL(2,\mathbb{R})$ symmetries, horizon symmetries, conformal Killing vectors, and related structures, as discussed in \citet{Charalambous:2021kcz, BenAchour:2022uqo, Sharma:2024hlz, Combaluzier-Szteinsznaider:2024sgb, Sharma:2025xii, Rai:2024lho, Sharma:2025xii}. The demonstration of ladder symmetry in the dynamical case was attempted in \citet{Hui:2022vbh}, where the vanishing of LNs for Kerr BHs at leading order in the frequency was suggested. This conclusion, however, is not consistent with more recent results in the literature \citep{Bhatt:2024yyz}. We therefore provide here a fresh perspective on the problem, following \citet{Ghosh:2026vig}. 

The starting point is a general second order ordinary differential operator, which is expressed in the following form:
%%%%%%%%%%%%%%%%%%%%%%%%%%%%%%%%%%%%%
\begin{equation} \label{Hl}
H_\ell \equiv \Delta^2(x)\,\partial_x^2 +\, p_\ell(x)\,\partial_x + q_\ell(x)\,,
\end{equation}
%%%%%%%%%%%%%%%%%%%%%%%%%%%%%%%%%%%%%
where $\Delta$ is a generic function, $x$ is the variable, corresponding to the radial coordinate $r$ in our context. The ladder is in the index $\ell$, which requires one to assume that $\ell$ is a discrete quantity and that arises in any problem with non-trivial boundary conditions. As the equation of motion of the system is given by $H_{\ell}\Psi_{\ell}=0$, the solution $\Psi_{\ell}$ also satisfies $\Omega_{\ell}H_{\ell}\Psi_{\ell}=0$, where $\Omega_{\ell}$ is an arbitrary function of $\ell$. Thus, the Hamiltonian can always be rescaled by an overall multiplicative factor. For this reason, the coefficient of the $\partial_x^{2}$ term carries no $\ell$ dependence: any $\ell$-dependent factor multiplying $\partial_x^{2}$ can be absorbed into an overall rescaling of the Hamiltonian. Since the solutions are invariant under such a global multiplicative redefinition of the Hamiltonian, this choice entails no loss of generality. We can therefore construct the following ladder operators \citep{Ghosh:2026vig}:
%%%%%%%%%%%%%%%%%%%%%%%%%%%%%%%%%%%%%
\begin{equation}\label{ladder_oparetors}
\begin{split}
&D_\ell^+ \equiv \Delta(x)\, \partial_x + W_\ell^+(x)\,, \\
&D_\ell^- \equiv \Delta(x)\, \partial_x + W_\ell^-(x)\,,
\end{split}
\end{equation}
%%%%%%%%%%%%%%%%%%%%%%%%%%%%%%%%%%%%%
where $W_{\ell}^{\pm}(x)$ are two arbitrary functions that need to be determined. To climb up the ladder structure one must impose the following conditions:
%%%%%%%%%%%%%%%%%%%%%%%%%%%%%%%%%%%%%
\begin{equation}\label{ladder_structure}
\begin{split}
&H_{\ell+1}\, D_\ell^+ = D_\ell^+\, H_\ell,\quad H_{\ell-1}\, D_\ell^- = D_\ell^-\, H_\ell\,,\\
&H_\ell = D_{\ell-1}^+\, D_\ell^- + E_\ell(x), \quad H_\ell = D_{\ell+1}^-\,D_\ell^+ + \widetilde{E}_\ell(x)\,.
\end{split}
\end{equation} 
%%%%%%%%%%%%%%%%%%%%%%%%%%%%%%%%%%%%%
The above ladder structure can be considered as a generalization of the operator relations in supersymmetric quantum mechanics \citep{DeLuca:2025zqr}. Note that $E_\ell(x)$ and $\widetilde{E}_\ell(x)$ are two unknowns that need to be determined. Using the Hamiltonian from~\ref{Hl} and the raising and lowering operators from~\ref{ladder_oparetors} in~\ref{ladder_structure}, which provides the necessary and sufficient conditions for having the ladder structure, one gets:
%%%%%%%%%%%%%%%%%%%%%%%%%%%%%%%%%%%%%
\begin{align}
&p_{\ell+1}-p_{\ell}=0\,,
\\
&q_{\ell+1}-q_{\ell}=\Delta\partial_{x}W_{\ell+1}^{-}-(\Delta/f)\partial_{x}W_{\ell}^{+}\,,
\\
&\left(H_{\ell}-2q_{\ell}+q_{\ell-1}\right)W_{\ell}^{-}=(\Delta/f)\partial_{x}q_{\ell}\,,
\label{ladder_check}
\\
&W_{\ell-1}^{+}=(p_{\ell}/\Delta)-W_{\ell}^{-}-\partial_{x}\Delta\,,
\\
&E_{\ell+1}=\widetilde{E}_{\ell}=\textrm{constant}\,,
\\
&E_{\ell}=q_{\ell}-\Delta\partial_{x}W_{\ell}^{-}-W_{\ell-1}^{+}W_{\ell}^{-}\,.
\end{align}
%%%%%%%%%%%%%%%%%%%%%%%%%%%%%%%%%%%%
Thus, the quantities $E_{\ell}$ and $\widetilde{E}_{\ell}$ are constants, $p_{\ell}$ is independent of $\ell$, and~\ref{ladder_check} is the necessary and sufficient condition for having a ladder structure. Thus, given the quantities $\Delta$, $p_{\ell}$ and $q_{\ell}$, one can solve~\ref{ladder_check} to obtain $W_{\ell}^{-}$, then the remaining equations help to determine the other quantities. Inversely, if someone provides the ladder operators and the Hamiltonian, one can check if the ladder structure indeed exists, through~\ref{ladder_check}. In what follows we will show how one can determine the ladder structure for dynamical scalar perturbations of a Kerr BH. 

For this purpose, consider a scalar perturbation of the Kerr BH, which has the decomposition in radial and angular parts as in~\ref{gen_eq_decomp}. Substituting $s=0$ in the corresponding equation and using the near-zone approximation (namely, $M\omega (r-r_{+})\ll 1$ and $M^{2}\omega^{2}\ll 1$) we obtain \citep{Ghosh:2026vig}, 
%%%%%%%%%%%%%%%%%%%%%%%%%%%%%%%%%%%%%
\begin{align}
z(1+z)\frac{\mathrm{d}^2{}_{0}R_{\ell m}}{\mathrm{d}z^2} + \left[2iP_{+}+1+2z\right]\frac{\mathrm{d}{}_{0}R_{\ell m}}{\mathrm{d}z} 
+\left[-2i\omega r_{+} -\ell(\ell+1)+2am\omega\right]{}_{0}R_{\ell m}=0~,
\end{align}
%%%%%%%%%%%%%%%%%%%%%%%%%%%%%%%%%%%%
where again $z\equiv (r-r_{+})/(r_{+}-r_{-})$ and $P_{+}\equiv (am-2M\omega r_{+})/(r_{+}-r_{-})$. Introducing $z=-x$, the above equation reduces to a hypergeometric differential equation: 
%%%%%%%%%%%%%%%%%%%%%%%%%%%%%%%%%%%%%
\begin{align}
x(1-x)\frac{\mathrm{d}^2{}_{0}R_{\ell m}}{\mathrm{d}x^2} + \left[2iP_{+}+1-2x\right]\frac{\mathrm{d}{}_{0}R_{\ell m}}{\mathrm{d}x} 
+(\ell+V)(\ell+1-V){}_{0}R_{\ell m}=0~,
\end{align}
%%%%%%%%%%%%%%%%%%%%%%%%%%%%%%%%%%%%
where, $V=2i\omega r_{+}-2am\omega=\mathcal{O}(M\omega)$. Since we are ignoring $M\omega x$ terms, it follows that the above hypergeometric equation has the following parameters: $c_{\ell}=1+2iP_{+}$, $b_{\ell}=-\ell-V$ and $a_{\ell}=\ell+1-V$. Thus, $c_{\ell}$ is independent of $\ell$, while $a_{\ell}$ and $b_{\ell}$ has the following relations: $a_{\ell+1}-a_{\ell}=1$ and $b_{\ell+1}-b_{\ell}=-1$, since $V$ is independent of $\ell$. Multiplying the above equation by $x(1-x)$, we can compare it with the equation $H_{\ell}{}_{0}R_{\ell m}=0$, where $H_{\ell}$ as compared with~\ref{Hl} yields,
%%%%%%%%%%%%%%%%%%%%%%%%%%%%%%%%%%%%%
\begin{align}
\Delta=x(1-x)\,; 
\quad 
p_{\ell}=x(1-x)\left[2iP_{+}+1-2x\right]\,;
\quad
q_{\ell}=x(1-x)(\ell+V)(\ell+1-V)\,.
\end{align}
%%%%%%%%%%%%%%%%%%%%%%%%%%%%%%%%%%%%
First of all, we have $p_{\ell+1}=p_{\ell}$, and subsequently by solving~\ref{ladder_check} one can obtain both $W_{\ell}^{+}$ and $W_{\ell}^{-}$, with the following expressions:
%%%%%%%%%%%%%%%%%%%%%%%%%%%%%%%%%%%%%
\begin{align}
W_{\ell}^{+}&=-(\ell+1-V)x+\frac{(\ell+1-V)(\ell+1+V+2iP_{+})}{2(\ell+1)}\,,
\\
W_{\ell}^{-}&=(\ell+V)x+\frac{(\ell+V)(2iP_{+}+V-\ell)}{2\ell}\,.
\end{align}
%%%%%%%%%%%%%%%%%%%%%%%%%%%%%%%%%%%%
Thus, we observe that scalar perturbation of a Kerr BH has a ladder structure even in dynamical situations. Let us now try to determine the associated conserved quantity and in the dynamical case, as suggested by \citep{Hui:2022vbh}, it follows that the conserved quantity itself is the LN. Given the above ladder structure, it follows that the conserved quantity for the $\ell=0$ case is given by $P_{0}=-2iP_{+}$. Applying the ladder operators appropriately we can determine the conserved quantity associated with the $\ell$th angular harmonics, yielding
%%%%%%%%%%%%%%%%%%%%%%%%%%%%%%%%%%%%%
\begin{align}
P_{\ell}=-2iP_{+}\frac{\Gamma(1+\ell+V)\Gamma(1+\ell-V) \Gamma(1+\ell+V+2iP_{+})\Gamma(1+\ell-2iP_{+}-V)}{\Gamma(1+V)\Gamma(1-V)\Gamma(1+U+2iP_{+})\Gamma(1-2iP_{+}-V)2^{2\ell}(\ell!)^{2}}\,,
\end{align}
%%%%%%%%%%%%%%%%%%%%%%%%%%%%%%%%%%%%
for which, $\textrm{Re}P_{\ell}\neq 0$. The charges satisfy the relation $\partial_{x} P_{\ell} = 0$. However, the most general solution of this conservation equation is $Q_{\ell} = \alpha_{\ell} P_{\ell} + \beta_{\ell}$. Therefore, ladder symmetry alone cannot determine the response function uniquely; a matching with BH perturbation theory is required to fix $\alpha_{\ell}$ and $\beta_{\ell}$. In this sense, ladder symmetry can give rise to the dynamical tidal LN of a Kerr BH, but its exact value can only be obtained after an appropriate matching with BH perturbation theory.

Thus, the ladder symmetry is not a symmetry in the strict sense; rather, it provides a procedure to generate the solution for the $\ell$th mode starting from the $(\ell-1)$th mode. Moreover, the ladder structure is unrelated to the vanishing of LNs. In particular, the dynamical LNs of Kerr BHs are nonzero, while the ladder structure is nevertheless present. 
In the static case, the ladder transformations give rise to Noether charges that satisfy the Schr\"{o}dinger algebra, namely the semidirect product of the Heisenberg algebra and $\mathrm{SL}(2,\mathbb{R})$. By contrast, the corresponding algebra in the dynamical case is presently unknown \citep{Ghosh:2026vig}. 
Finally, the ladder structure extends beyond the strict static limit and holds up to linear order in $M\omega$. It would therefore be interesting to investigate whether this property persists at generic orders in $M\omega$. 
%%%%%%%%%%%%%%%%%%%%%%%%%%%%%%%%%%%%%%%%%%%%%%%%%%%%%%%%%%%%%%%%
%%%%%%%%%%%%%%%%%%%%%%%%%%%%%%%%%%%%%%%%%%%%%%%%%%%%%%%%%%%%%%%%
%%%%%%%%%%%%%%%%%%%%%%%%%%%%%%%%%%%%%%%%%%%%%%%%%%%%%%%%%%%%%%%%
\subsection{Perturbative approach}\label{LN_Dyn_Pert}

Here we will use a perturbative approach within the context of BH perturbation theory to compute the conservative and dissipative dynamical response of a Schwarzschild BH up to quadratic order in frequency, which has been the focus of recent studies \citep{Poisson:2020vap,Perry:2023wmm, Pitre:2023xsr, HegadeKR:2024agt, Katagiri:2024wbg, HegadeKR:2025qwj, Chakraborty:2025wvs,Combaluzier--Szteinsznaider:2025eoc,Kobayashi:2025vgl,Jarequi:2026cyp}. We will first determine the relevant perturbation equations, order by order in dimensionless frequency $M\omega$ and then we will solve them with appropriate boundary conditions (which in the case of BHs demands regularity at their horizon). After that, we will take the asymptotic expansion of the full solution and hence determine the dynamical response function, leading to the LNs and dissipation numbers.

For this purpose, we start with the Teukolsky formalism and write down the associated radial perturbation equation in ingoing null coordinates, given in~\ref{Teqrdyn} and further simplified in \citet{Chakraborty:2025wvs}, for the Schwarzschild background, as
%%%%%%%%%%%%%%%%%%%%%%%%%%%%%%%%%%%%%
\begin{align}
\label{EOMs-2l2}
\medmath{z(1+z)\dfrac{d^{2}R_{\ell}}{dz^{2}}-\left[(1+2z)+4iM\omega(1+z)^{2}\right]\dfrac{dR_{\ell}}{dz}
-\Big[(\ell-1)(\ell+2)+4iM\omega\frac{(1+z)(z-2)}{z}\Big]R_{\ell}=0\,,}
\end{align}
%%%%%%%%%%%%%%%%%%%%%%%%%%%%%%%%%%%%
where $z=(r/2M)-1$, the standard rescaling of radial coordinate we have performed in this review. Most notably, in the exact equation above there are no terms of $\mathcal{O}(M^{2}\omega^{2})$. This feature is specific to the gravitational perturbations of a Schwarzschild BH, expressed in terms of Teukolsky formalism in advanced null coordinates \citep{Teukolsky:1974yv}. 

In the perturbative approach, we first expand the radial function in powers of $M\omega$, such that,
%%%%%%%%%%%%%%%%%%%%%%%%%%%%%%%%%%%%
\begin{equation}
R_{\ell}=R^{(0)}_{\ell}+M\omega\,R^{(1)}_{\ell}+M^{2}\omega^{2}\,R^{(2)}_{\ell}+\cdots\,.
\end{equation}
%%%%%%%%%%%%%%%%%%%%%%%%%%%%%%%%%%%%
Substituting the above perturbative series in~\ref{EOMs-2l2}, we arrive at the following recurrence relations:
%%%%%%%%%%%%%%%%%%%%%%%%%%%%%%%%%%%%
\begin{align}
\mathcal{D}_{0}R^{(0)}_{\ell}&=0\,,
\label{basic}
\\
\mathcal{D}_{0}R^{(1)}_{\ell}&=4i\mathcal{D}_{1}R^{(0)}_{\ell}\,,
\label{recursion}
\\
\mathcal{D}_{0}R^{(2)}_{\ell}&=4i\mathcal{D}_{1}R^{(1)}_{\ell}\,,
\label{recursion2}
\end{align}
%%%%%%%%%%%%%%%%%%%%%%%%%%%%%%%%%%%%
which continues recursively, such that the radial function associated with the ${\cal O}(M\omega)^{n}$ correction is related to the radial function connected with ${\cal O}(M\omega)^{n-1}$ as: $\mathcal{D}_{0}R^{(n)}_{\ell}=4i\mathcal{D}_{1}R^{(n-1)}_{\ell}$. Thus, the full recurrence relations can be determined in terms of two differential operators $\mathcal{D}_{0}$ and $\mathcal{D}_{1}$, which are given by
%%%%%%%%%%%%%%%%%%%%%%%%%%%%%%%%%%%%
\begin{align}
\mathcal{D}_{0}&=z(1+z)\dfrac{d^{2}}{dz^{2}}-(1+2z)\dfrac{d}{dz}-(\ell-1)(\ell+2)\,, 
\\
\mathcal{D}_{1}&=(1+z)^{2}\dfrac{d}{dz}-\left(\frac{2}{z}+1-z\right)\,.
\end{align}
%%%%%%%%%%%%%%%%%%%%%%%%%%%%%%%%%%%%
This feature makes this approach particularly convenient and there seems to be a possibility that corrections at any order can be computed recursively and might be potentially re-summed to provide the dynamical response at all order in frequency \citep{Chakraborty:2025wvs}. Moreover, the $i$ factor in the source term shows that, for $n=\textrm{even}$, the response function is real, while for $n=\textrm{odd}$, the response function is purely imaginary. Thus, the LNs of a Schwarzschild BH can be non-zero at $\mathcal{O}(M^{2}\omega^{2})$, while they are zero at $\mathcal{O}(M\omega)$, in agreement with the analysis of the previous section.

For static perturbations,~\ref{basic} can be solved exactly to determine the zeroth order radial function, 
%%%%%%%%%%%%%%%%%%%%%%%%%%%%%%%%%%%%
\begin{align}\label{sol_zero}
R^{(0)}_{\ell}=z(1+z)\Big[c_{1}P_{\ell}^{2}(1+2z)+c_{2}Q_{\ell}^{2}(1+2z)\Big]\,,
\end{align}
%%%%%%%%%%%%%%%%%%%%%%%%%%%%%%%%%%%%
which also follows from a combination of~\ref{Hext} with the result $R=z(1+z)H_{0}$. The above expression can be written in terms of hypergeometric functions, which is given by~\ref{staticSchhyp}, clearly identifying the tidal part as well as the associated response. As discussed in detail in~\ref{stat_Love_Kerr}, for a Schwarzschild BH, the static tidal response function identically vanishes, owing to the regularity condition at the horizon (which, in this case sets $c_{2}=0$). Thus, we retrieve that $\,_{-2}k^{(0)}_{\ell}=0=_{-2}\nu_{\ell}^{(0)}$. 

Given the zeroth-order radial function, one can obtain the first-order radial function $R^{(1)}_{\ell}$, by substituting $R^{(0)}_{\ell}$ in~\ref{recursion}. Unfortunately, for generic choices of $\ell$, we are unable to obtain an analytical solution for $R^{(1)}_{\ell}$. However, the latter can be found for any integer value of $\ell$. Focusing on the quadrupolar case, we consider the $\ell=2$ zeroth-order solution, which is regular at the horizon, and substitute it in~\ref{recursion}, yielding the following solution for the first-order radial perturbation:
%%%%%%%%%%%%%%%%%%%%%%%%%%%%%%%%%%%%
\begin{align}\label{sol_oneBH}
R^{(1)}_{2}&=-12 i c_{3}z^{2}(1+z)^{2}+\frac{i c_{4}}{2}(1+2z)[6z(1+z)-1]-6 i c_{4} z^{2}(1+z)^{2}\ln \left(\frac{1+z}{z}\right)
\nonumber
\\
&-ic_{1}\{24z^{2}(1+z)^{2}\ln(1+z)+2z^{3}[28+z(47+20z)]\}\,,
\end{align}
%%%%%%%%%%%%%%%%%%%%%%%%%%%%%%%%%%%%
where the first line depicts the homogeneous solution, while the second line represents the particular solution. Note that the regularity of the zeroth-order solution at the BH horizon demands $c_{2}=0$, and hence the first-order solution consists of three arbitrary constants: $c_{3}$, $c_{4}$ (from homogeneous solution) and $c_{1}$ (from the source term). The $i$ factor in front of $c_1$ can be attributed to the $4i$ term on the right-hand side of~\ref{recursion}.

The regularity of the first-order solution at the horizon requires $c_{4}=0$. The reason is twofold---(a) the second derivative of the first-order solution is irregular at the horizon if $c_{4}\neq 0$, and (b) the radial perturbation behaves as $iM\omega \ln z$, corresponding to an outgoing mode at the horizon, unless $c_{4}=0$. Thus, the first-order solution has only two arbitrary constants, $c_{1}$ and $c_{3}$. The asymptotic behavior of the radial function up to first order in the frequency $M\omega$, which is regular at the horizon, reads: 
%%%%%%%%%%%%%%%%%%%%%%%%%%%%%%%%%%%%
\begin{align}\label{asymplinear}
R^{(0)}_{2}&+M\omega R^{(1)}_{2}\approx-40c_{1}(iM\omega)\left(\frac{r}{2M}\right)^{5}-\frac{1}{3}\Big[36c_{1}+\big\{282c_{1}+36 c_{3}
\nonumber
\\
&\qquad +72c_{1}\ln \left(\frac{r}{2M}\right)\big\}(iM\omega)\Big]\left(\frac{r}{2M}\right)^{4}+\dots-\frac{8c_{1}(iM\omega )}{10}\left(\frac{2M}{r}\right)+\mathcal{O}(r^{-2})\,.
\end{align}
%%%%%%%%%%%%%%%%%%%%%%%%%%%%%%%%%%%%
The dots denote sub-leading terms, which are not relevant for the determination of the tidal response. Therefore, the linear-in-frequency tidal response function can be determined from the ratio between the $r^{-1}$ term and the $r^{4}$ term, and takes the form
%%%%%%%%%%%%%%%%%%%%%%%%%%%%%%%%%%%%
\begin{align}\label{response}
\,_{-2}{\cal F}_{2}(\omega)=\frac{iM\omega}{15}+\mathcal{O}(M^{2}\omega^{2})\,.
\end{align}
%%%%%%%%%%%%%%%%%%%%%%%%%%%%%%%%%%%%
Since the response function is purely imaginary, the linear-in-frequency LN associated with the quadrupolar mode vanishes identically, while the dissipation number, $\,_{-2}\nu_{2}^{(1)}$, is non-zero and equal to $1/15$. This result agrees with the computation based on the scattering approach as well \citep{Saketh:2023bul}. This conclusion can be generalized to arbitrary $\ell$ modes, leading to the following expressions \citep{Chia:2020yla, Charalambous:2021mea}.
%%%%%%%%%%%%%%%%%%%%%%%%%%%%%%%%%%%%
\begin{align}
\,_{-2}k_{\ell}&=0+\mathcal{O}(M^{2}\omega^{2})\,,
\\
\,_{-2}\nu_{\ell}&=\frac{2(\ell+2)!(\ell-2)!(\ell!)^{2}}{(2\ell+1)!(2\ell)!}+\mathcal{O}(M^{2}\omega^{2})\,.
\label{dissi}
\end{align}
%%%%%%%%%%%%%%%%%%%%%%%%%%%%%%%%%%%%

To obtain the second-order radial perturbation, we need to substitute the first-order solution in the source term of~\ref{recursion2}. The resulting solution for the second-order radial perturbation associated with the $\ell=2$ mode can be found in \citet{Chakraborty:2025wvs}. In this case also, the particular solution depends on $(c_{1},c_{3})$, along with the homogeneous solution, depending on two further constants, $c_5$ and $c_6$. Repeating the previous cases, the homogeneous solution of second-order radial perturbation involves a logarithmic piece, which must vanish to ensure regularity at the horizon, leading to $c_6 = 0$. The asymptotic expansion of the Teukolsky radial perturbation, up to second-order in frequency, provides us the following dynamical response function \citep{Chakraborty:2025wvs}, 
%%%%%%%%%%%%%%%%%%%%%%%%%%%%%%%%%%%%
\begin{align}
\label{response-dyn}
\,_{-2}\mathcal{F}_{2}(\omega)=\frac{iM\omega}{15}+\frac{2M^{2}\omega^{2}}{15}\left[\ln \left(\frac{r}{2M}\right)+\frac{137}{21}\right]\,.
\end{align}
%%%%%%%%%%%%%%%%%%%%%%%%%%%%%%%%%%%%
Interestingly, the above dynamical response function does not depend on any of the arbitrary constants. For example, even though the arbitrary constants $c_3$ and $c_5$ appear in the full solution \citep{Chakraborty:2025wvs}, there is no effect on the response function, when restricted to quadratic order in the frequency. This unique feature is a consequence of defining the response function using both the homogeneous and the particular solution (at variance with other approaches \citep{Pitre:2023xsr,Katagiri:2024wbg, HegadeKR:2024agt}) and also relies on the fact that ${}_{-2} k_\ell^{(0)}=0$, so it might not occur for systems with nonzero LNs in the static case. In addition, the use of the advanced null coordinates have significantly simplified the recursive relations, which are particularly convenient to solve the iterative problem in frequency.

In this case as well, we observe the presence of a logarithmic term in the dynamical tidal response, signaling a running behavior. This feature differs from the approach adopted in \citet{Katagiri:2023umb}, while being more closely aligned with the scattering approach \citep{Ivanov:2022hlo}. Such running originates from the identification of the LNs and dissipation numbers as couplings in the EFT action, together with the presence of UV divergences in the corresponding graviton scattering processes derived from the EFT (see, e.g., \citet{Ivanov:2024sds, Saketh:2023bul}). As already emphasized, a comparison with \citet{Saketh:2023bul} confirms the running behavior and yields exact agreement in the coefficients of the corresponding beta functions for the EFT couplings. 

For instance, the matching between scattering amplitudes and BH perturbation theory produces an identical dissipative coefficient $_{-2}\nu_{2}^{(1)}$, as obtained in this section, at leading order in the frequency. The same agreement holds for the coefficient of the logarithmic term in the dynamical LN at $\mathcal{O}(M^{2}\omega^{2})$. This equivalence between BH perturbation theory and scattering amplitudes appears to persist for generic values of $\ell$ \citep{Mano:1996vt, Mano:1996gn, Mano:1996mf, Sasaki:2003xr}. 

Let us now compute the dynamical LNs of a Schwarzschild BH from~\ref{response-dyn}, which leads to the following result for the $\ell=2$ mode:
\begin{equation}
\,_{-2}k_{2}=\frac{M^{2}\omega^{2}}{15}\left[\ln \left(\frac{r}{2M}\right)+\alpha\right]\,.
\label{eq:DLN}
\end{equation}
Within BH perturbation theory one finds $\alpha=137/21$ \citep{Chakraborty:2025wvs}. However, as discussed above, this computation is affected by gauge ambiguities, which can be removed through an appropriate matching to the EFT coefficients. This procedure was recently carried out in \citet{Combaluzier--Szteinsznaider:2025eoc}, yielding
\begin{equation}
\alpha=\frac{787}{2520}\,.
\end{equation}
The logarithmic term in~\ref{eq:DLN}, on the other hand, is universal and encodes the running of the tidal coupling through the non-linearities of Einstein's equations. Its coefficient corresponds to the $\beta$ function governing the renormalization group flow of the tidal coupling \citep{Mandal:2023hqa, Jakobsen:2023pvx}.
In the EFT matching, one typically identifies $r\sim\mu^{-1}$, where $\mu$ is the cutoff scale of the problem, so that the logarithmic contribution becomes $-\ln(2M\mu)$ \citep{Combaluzier--Szteinsznaider:2025eoc}. Recently, the matching between the worldline EFT and BH perturbation theory was carried out explicitly in the magnetic sector for arbitrary $\ell$ using the MST approach \citep{Kobayashi:2025vgl}.

The results presented above apply to gravitational perturbations (with spin-weight $s=-2$) of a Schwarzschild BH, but they can be extended to generic spin-$s$ perturbations as well. Evidence from \citet{Chakraborty:2025wvs} suggests that perturbations with arbitrary spin-$s$ exhibit qualitatively similar behavior. The analysis can also be generalized to higher values of $\ell$ and to different parities \citep{Kobayashi:2025vgl}.

Since our discussion is restricted to a fixed value of $\ell$, mixing between the tidal field and the response function may occur. To avoid such mixing, which typically leads to ambiguities in the determination of the LNs, a matching to gauge- and coordinate-independent quantities is required. This is where worldline EFTs \citep{Goldberger:2004jt,Goldberger:2005cd, Goldberger:2020fot, Porto:2016pyg,Porto:2016zng,Porto:2007qi,Hui:2020xxx, Steinhoff:2016rfi,Goldberger:2020fot} play an important role. In the EFT framework, tidal responses appear as coupling constants of higher-derivative operators encoding dynamical tidal interactions. By computing graviton scattering in a BH background, one can determine the corresponding Wilson coefficients and match them to BH perturbation theory, thereby extracting gauge-invariant LNs.

For a test scalar field on a Schwarzschild background, such a matching between perturbation theory and EFT has been explicitly demonstrated in the dynamical regime, showing that the dynamical LNs of a Schwarzschild BH under scalar perturbations are nonzero \citep{Ivanov:2024sds, Caron-Huot:2025tlq,Akhtar:2025nmt}. This analysis has been extended to the gravitational case, confirming that the conservative response is likewise nonzero at ${\cal O}(\omega^{2})$ and exhibits a universal running term \citep{Combaluzier--Szteinsznaider:2025eoc, Kobayashi:2025vgl}. See also \citet{Parra-Martinez:2025bcu, Ivanov:2025ozg} for further discussion of the logarithmic contributions to the response function.

In light of these results, it is important to assess the impact of nonvanishing dynamical LNs of a Schwarzschild BH on the GW waveform, an issue that we will examine in detail in~\ref{sec:GW}.

%%%%%%%%%%%%%%%%%%%%%%%%%%%%%%%%%%%%%%%%%%%%%%%%%%%%%%%%%%%%%%%%
%%%%%%%%%%%%%%%%%%%%%%%%%%%%%%%%%%%%%%%%%%%%%%%%%%%%%%%%%%%%%%%%
%%%%%%%%%%%%%%%%%%%%%%%%%%%%%%%%%%%%%%%%%%%%%%%%%%%%%%%%%%%%%%%%
\subsection{Scattering amplitude approach}\label{scatampapp}

In this section, we will briefly show the derivation of the dynamical tidal deformations using a scattering approach \citep{Ivanov:2022hlo, Saketh:2023bul, Ivanov:2024sds}. We will use a gravitational tidal field and shall present the results for $\ell=2$ case only, while in the EFT action we will keep both the $\ell=2$ and the $\ell=3$ terms. The starting point is the finite size action for a spinning BH, perturbed by an external tidal field. The quadrupolar tidal field, in the gravitational context is determined by the electric and magnetic parts of the Weyl scalar: $E_{\mu \nu}=W_{\mu \alpha \nu \beta}u^{\alpha}u^{\beta}$ and $B_{\mu \nu}=(1/2)u^{\gamma}\epsilon_{\gamma \langle \mu |\alpha \beta}W^{\alpha \beta}_{\nu\rangle \delta}u^{\delta}$, where $u^{\mu}$ is the four velocity of the point particle (namely, the BH in this case). Owing to the symmetries of the Weyl tensor, it follows that both $E_{\mu \nu}$ and $B_{\mu \nu}$ are spatial tensors in the rest frame of the BH. To the leading order, the finite size action reads 
%%%%%%%%%%%%%%%%%%%%%%%%%%%%%%%%%%%%%%%%%%%%%%%%%%%%%%%%%%%%%%%%%%%%%%%%%%%%%%
\begin{align}
\mathcal{A}_{\rm finite}=\int d\tau \left(M_{ij}E^{ij}+S_{ij}B^{ij}+M_{ijk}E^{ijk}+S_{ijk}B^{ijk}\right)\,,
\end{align}
%%%%%%%%%%%%%%%%%%%%%%%%%%%%%%%%%%%%%%%%%%%%%%%%%%%%%%%%%%%%%%%%%%%%%%%%%%%%%%
where we have kept both the quadrupole and octupole moments of mass and current, associated with the electric and magnetic parts of the external tidal field, respectively. All of these quantities are evaluated on the trajectory of the BH. Note that, under parity transformation, $E^{ij}\to E^{ij}$, $B^{ij}\to -B^{ij}$, $E^{ijk}\to -E^{ijk}$ and $B^{ijk}\to B^{ijk}$. The connection between the mass quadrupole moment with the corresponding tidal moments are given by \citep{Endlich:2015mke} 
%%%%%%%%%%%%%%%%%%%%%%%%%%%%%%%%%%%%%%%%%%%%%%%%%%%%%%%%%%%%%%%%%%%%%%%%%%%%%%
\begin{align}
M_{ij}&=-M\left[(GM)^{4}\Lambda^{\rm E}_{ijkl}E^{kl}-(GM)^{5}(\Lambda^{\rm E}_{\omega})_{ijkl}\dfrac{DE^{kl}}{D\tau}+(GM)^{6}(\Lambda^{\rm E}_{\omega^{2}})_{ijkl}\dfrac{D^{2}E^{kl}}{D\tau^{2}}+\cdots\right]
\nonumber
\\
&-M(GM)^{5}({\Upsilon}^{\rm E})_{ij\langle kl}\hat{s}_{m\rangle}B^{klm}+\cdots\,,
\end{align}
%%%%%%%%%%%%%%%%%%%%%%%%%%%%%%%%%%%%%%%%%%%%%%%%%%%%%%%%%%%%%%%%%%%%%%%%%%%%%%
where $(D/D\tau)=u^{\mu}\nabla_{\mu}$ is the derivative in the co-moving frame of the BH, with $(\Lambda^{E}_{\omega^{n}})_{ijkl}$ being the tidal coefficient tensor associated with the $n$th power of the frequency, while, ${\Upsilon}^{\rm E}_{ijkl}$ are the electric mixing coefficients, connecting terms with opposite parity and different angular momenta through linear coupling to the spin vector. A similar expression holds for the spin quadrupole moments as well, with $E^{kl}\to B^{kl}$, $(\Lambda^{E}_{\omega^{n}})_{ijkl}\to (\Lambda^{B}_{\omega^{n}})_{ijkl}$, $B^{klm}\to E^{klm}$ and $({\Upsilon}^{\rm E})_{ijkl}\to ({\Upsilon}^{\rm B})_{ijkl}$. Similarly, the electric and magnetic octupole moments induced on the compact object will be generated from quadrupolar tidal fields,
%%%%%%%%%%%%%%%%%%%%%%%%%%%%%%%%%%%%%%%%%%%%%%%%%%%%%%%%%%%%%%%%%%%%%%%%%%%%%%
\begin{align}
M_{ijk}&=-M(GM)^{5}\hat{s}_{\langle k}(\varsigma^{\rm E})_{ij\rangle \ell m}B^{\ell m}+\cdots\,,\\
S_{ijk}&=-M(GM)^{5}\hat{s}_{\langle k}(\varsigma^{\rm B})_{ij\rangle \ell m}E^{\ell m}+\cdots\,;
\end{align}
%%%%%%%%%%%%%%%%%%%%%%%%%%%%%%%%%%%%%%%%%%%%%%%%%%%%%%%%%%%%%%%%%%%%%%%%%%%%%%
with $(\varsigma^{\rm E/B})_{ij\ell m}$ being another set of mixing coefficient tensors. Here, we have introduced the unit spin three-vector $\hat{s}_{i}$, which is defined as, $\hat{s}^{i}\equiv (s^{i}/J)$, where $s^{i}=(1/2)\epsilon^{ijk}S_{jk}$ and $J$ is the angular momentum of the BH. It is possible to define a unit spin tensor, as $\hat{S}_{ij}\equiv S_{ij}/J$. One can then express all the tidal coefficient tensors, as well as mixing coefficient tensors, as 
%%%%%%%%%%%%%%%%%%%%%%%%%%%%%%%%%%%%%%%%%%%%%%%%%%%%%%%%%%%%%%%%%%%%%%%%%%%%%%
\begin{align}
\left(\mathcal{T}^{\rm E/B}\right)^{ij}_{kl}=\mathcal{T}^{\rm E/B}\delta^{\langle i}_{\langle k}\delta^{j\rangle}_{l\rangle}
+\mathcal{T}^{\rm E/B}_{\mathbb{S}}\hat{S}^{\langle i}_{\langle k}\delta^{j\rangle}_{l\rangle}
+\mathcal{T}^{\rm E/B}_{\mathbb{S}^{2}}\hat{s}^{\langle i}\hat{s}_{\langle k}\delta^{j\rangle}_{l\rangle}
+\mathcal{T}^{\rm E/B}_{\mathbb{S}^{3}}\hat{s}^{\langle i}\hat{s}_{\langle k}\hat{S}^{j\rangle}_{l\rangle}
+\mathcal{T}^{\rm E/B}_{\mathbb{S}^{4}}\hat{s}^{\langle i}\hat{s}_{\langle k}\hat{s}^{j\rangle}\hat{s}_{l\rangle}\,,
\end{align}
%%%%%%%%%%%%%%%%%%%%%%%%%%%%%%%%%%%%%%%%%%%%%%%%%%%%%%%%%%%%%%%%%%%%%%%%%%%%%%
where $\mathcal{T}^{\rm E/B}$ is a proxy for all the tidal/mixing coefficients, $\Lambda^{\rm E/B}$, $\Upsilon^{\rm E/B}$ and $\varsigma^{\rm E/B}$, respectively. Thus, each of these tidal and mixing coefficient tensors can be expressed in terms of five numbers, these are called the tidal and mixing\footnote{The mixing response coefficients are associated with the so-called \emph{rotational} LNs \citep{Poisson:2014gka,Pani:2015hfa,Pani:2015nua} discussed in~\ref{sec:RTLNs}.} response coefficients. Among these, the tidal response coefficients that does not change sign under time reversal are the LNs, while the ones that change sign are the tidal dissipation numbers. Since, under time reversal, both $\omega \to -\omega$ and $\hat{s}^{i}\to -\hat{s}^{i}$, it follows that we have eight LNs and seven tidal dissipation numbers, both for the electric and the magnetic sectors, which can be collectively expressed as 
%%%%%%%%%%%%%%%%%%%%%%%%%%%%%%%%%%%%%%%%%%%%%%%%%%%%%%%%%%%%%%%%%%%%%%%%%%%%%%
\begin{align}
k^{\rm E/B}_{2m}&=\textrm{combinations\,of}\left\{\underbrace{\Lambda^{\rm E/B},\Lambda^{\rm E/B}_{\mathbb{S}^{2}},\Lambda^{\rm E/B}_{\mathbb{S}^{4}}}_{\textrm{static\,Love\,Numbers}},\underbrace{\Lambda^{\rm E/B}_{\omega,\mathbb{S}},\Lambda^{\rm E/B}_{\omega,\mathbb{S}^{3}},\Lambda^{\rm E/B}_{\omega^{2}},\Lambda^{\rm E/B}_{\omega^{2},\mathbb{S}^{2}},\Lambda^{\rm E/B}_{\omega^{2},\mathbb{S}^{4}}}_{\rm dynamical\,Love\,numbers}\right\}\,,
\\
\nu^{\rm E/B}_{2m}&=\textrm{combinations\,of}\left\{\Lambda^{\rm E/B}_{\mathbb{S}},\Lambda^{\rm E/B}_{\mathbb{S}^{3}},\Lambda^{\rm E/B}_{\omega},\Lambda^{\rm E/B}_{\omega,\mathbb{S}^{2}},\Lambda^{\rm E/B}_{\omega,\mathbb{S}^{4}},\Lambda^{\rm E/B}_{\omega^{2},\mathbb{S}},\Lambda^{\rm E/B}_{\omega^{2},\mathbb{S}^{3}}\right\}\,.
\end{align}
%%%%%%%%%%%%%%%%%%%%%%%%%%%%%%%%%%%%%%%%%%%%%%%%%%%%%%%%%%%%%%%%%%%%%%%%%%%%%%
Here, the term $\Lambda^{\rm E/B}_{\omega^{n},\mathbb{S}^{m}}$ (related to either $k^{\rm E/B}_{2m}$ or $\nu^{\rm (E/B)}_{2m}$ by $(M/R)^{2\ell+1}$, where $M$ is the mass and $R$ is the radius) refers to the coefficient of the $n$th time derivative of the tidal field, and is associated with $m$ spin degrees of freedom, where $m\leq 2\ell$. Thus, for $\ell=2$, the maximum value of $m$ will be $4$. Also note that the LNs $k_{\ell m}$ and dissipation numbers $\nu_{\ell m}$ are related to those defined otherwise in this review, by a factor of two. This is because, we had defined the LN and the dissipation numbers with a relative factor of (1/2), see~\ref{I_def}, which is absent in these definitions. Similarly, the mixing LNs and mixing tidal dissipation numbers are given by 
%%%%%%%%%%%%%%%%%%%%%%%%%%%%%%%%%%%%%%%%%%%%%%%%%%%%%%%%%%%%%%%%%%%%%%%%%%%%%%
\begin{align}
k^{\rm E/B}_{2m_{1}3m_{2}\,\textrm{(mix)}}&=\textrm{combinations\,of}\left\{{\Upsilon}^{\rm E/B},{\Upsilon}^{\rm E/B}_{\mathbb{S}^{2}},{\Upsilon}^{\rm E/B}_{\mathbb{S}^{4}},\varsigma^{\rm E/B},\varsigma^{\rm E/B}_{\mathbb{S}^{2}},\varsigma^{\rm E/B}_{\mathbb{S}^{4}}\right\}\,,
\\
\nu^{\rm E/B}_{2m_{1}3m_{2}\,\textrm{(mix)}}&=\textrm{combinations\,of}\left\{{\Upsilon}^{\rm E/B}_{\mathbb{S}},{\Upsilon}^{\rm E/B}_{\mathbb{S}^{3}},\varsigma^{\rm E/B}_{\mathbb{S}},\varsigma^{\rm E/B}_{\mathbb{S}^{3}}\right\}\,.
\end{align}
%%%%%%%%%%%%%%%%%%%%%%%%%%%%%%%%%%%%%%%%%%%%%%%%%%%%%%%%%%%%%%%%%%%%%%%%%%%%%%
It is worth emphasizing that the mass multipole moments and the electric tidal moments transform identically under parity transformations. Since one typically expects the EFT action to be parity invariant, the mass moments associated with a given $\ell$ mode will, to linear order in the spin, relate to the electric moments of the same $\ell$ mode and to the magnetic moments of the $(\ell\pm1)$ modes, giving rise to the mixing coefficients discussed above. The mixing coefficients $(\Upsilon,\varsigma)$ are normalized with respect to the mass of the compact object, whereas the LNs and tidal dissipation numbers ($k_{\ell m}^{\rm E/B}$ and $\nu_{\ell m}^{\rm E/B}$) have been normalized using the object radius. Consequently, converting between these conventions introduces an overall normalization factor involving appropriate powers of the ratio $(R/M)$. 

Given the EFT action for the compact object placed in an external tidal field, the finite part of the action depicts the interaction between the compact object and the tidal field. These interactions can be presented in a diagrammatic way as described in~\ref{sec:scattering}. The idea is to compute the scattering amplitude following appropriate diagrams with Feynman-like rules and then equating them to the equivalent quantities derived from BH perturbation theory, as in~\ref{scatteringmatch}. We have already described the results, arising from such a procedure for static perturbations in~\ref{sec:scattering}, while we discuss a few results for the dynamical tides in this section.

For example, consider the determination of the dynamical tidal dissipation number for the Schwarzschild BH. As evident from~\ref{diss_static}, the tidal and the dissipation numbers vanish for the Schwarzschild BH under static perturbation and hence we need to study linear-in-frequency response. The corresponding diagram is identical to~\ref{static_scat}, however, with the vertex being represented by $i\omega \Lambda^{ij}_{kl}$. Thus, the amplitude in the spherical basis becomes \citep{Saketh:2023bul, Ivanov:2022hlo, Ivanov:2024sds},
%%%%%%%%%%%%%%%%%%%%%%%%%%%%%%%%%%%%%%%%%%%%%%%%%%%%%%%%%%%%%%%%%%%%%%%%%%%%%%
\begin{align}
i\mathcal{A}(\omega;\ell=2,m,h\to \ell=2,m,h)=\frac{\omega^{6}}{40M_{\rm pl}^{2}\pi}M(GM)^{5}\frac{1}{3}\left(\frac{r_{+}^{5}}{M^{6}}\right)\nu^{\rm E}_{2,\omega}+\textrm{magnetic}\,.
\end{align}
%%%%%%%%%%%%%%%%%%%%%%%%%%%%%%%%%%%%%%%%%%%%%%%%%%%%%%%%%%%%%%%%%%%%%%%%%%%%%%
This can be equated to the corresponding amplitude determined using BH perturbation theory. Since, we are interested in a background Schwarzschild BH, the linear perturbations are independent of $m$ and hence we can equate the above result with the $m=0$ result from BH perturbation theory, which yields, $i\mathcal{A}=(2/225)(2GM\omega)^{6}+\mathcal{O}(M^{7}\omega^{7})$. Using $M_{\rm pl}^{-2}=64\pi G$, we obtain
%%%%%%%%%%%%%%%%%%%%%%%%%%%%%%%%%%%%%%%%%%%%%%%%%%%%%%%%%%%%%%%%%%%%%%%%%%%%%%
\begin{align}
\nu_{2,\omega}^{\rm E/B}=\frac{16}{45}\times \frac{3}{2^{5}}=\frac{1}{30}\,,
\end{align}
%%%%%%%%%%%%%%%%%%%%%%%%%%%%%%%%%%%%%%%%%%%%%%%%%%%%%%%%%%%%%%%%%%%%%%%%%%%%%%
while the LNs vanish at the linear-order-in-frequency as well. This is consistent with our previous results using BH perturbation theory, except for a factor of two, see~\ref{LN_Dyn_Pert}. Moreover, considering the quadratic-in-frequency correction, and applying the same diagram as~\ref{static_scat}, with coupling $-\omega^{2}\Lambda^{ij}_{kl}$, we obtain, 
%%%%%%%%%%%%%%%%%%%%%%%%%%%%%%%%%%%%%%%%%%%%%%%%%%%%%%%%%%%%%%%%%%%%%%%%%%%%%%
\begin{align}
\medmath{i\mathcal{A}(\omega;\ell=2,m,h\to \ell=2,m,h)=i\frac{\omega^{7}}{40M_{\rm pl}^{2}\pi}M(GM)^{6}\left(\frac{2}{3}\frac{r_{+}^{5}}{M^{7}}k_{2,\omega^{2}}^{\rm E}+i\frac{r_{+}^{5}}{3M^{6}}\nu^{\rm E}_{2,\omega^{2}}\right)+\textrm{magnetic}\,.}
\end{align}
%%%%%%%%%%%%%%%%%%%%%%%%%%%%%%%%%%%%%%%%%%%%%%%%%%%%%%%%%%%%%%%%%%%%%%%%%%%%%%
Noticing that the scattering amplitude $\,_{2}r_{2m}$ has no contribution at $\mathcal{O}(M^{7}\omega^{7})$, except for a tail term, only the phase contributes \citep{Saketh:2023bul}. Thus, we obtain, 
%%%%%%%%%%%%%%%%%%%%%%%%%%%%%%%%%%%%%%%%%%%%%%%%%%%%%%%%%%%%%%%%%%%%%%%%%%%%%%
\begin{align}
i\frac{\omega^{7}}{40M_{\rm pl}^{2}\pi}M(GM)^{6}\left(\frac{2}{3}\frac{r_{+}^{5}}{M^{7}}k_{2,\omega^{2}}^{\rm E}+i\frac{r_{+}^{5}}{3M^{6}}\nu^{\rm E}_{2,\omega^{2}}\right)=i\frac{2}{225}\ln(4GM\omega)(2GM\omega)^{7}\,.
\end{align}
%%%%%%%%%%%%%%%%%%%%%%%%%%%%%%%%%%%%%%%%%%%%%%%%%%%%%%%%%%%%%%%%%%%%%%%%%%%%%%
This yields a vanishing dissipation number at $\mathcal{O}(M^{2}\omega^{2})$, and a non-zero dynamical LN for Schwarzschild BH at quadratic order in frequency:
%%%%%%%%%%%%%%%%%%%%%%%%%%%%%%%%%%%%%%%%%%%%%%%%%%%%%%%%%%%%%%%%%%%%%%%%%%%%%%
\begin{align}
\dfrac{\partial k_{2,\omega^{2}}^{\rm E/B}}{\partial \ln(4M\omega)}=\frac{1}{30}\,.
\end{align}
%%%%%%%%%%%%%%%%%%%%%%%%%%%%%%%%%%%%%%%%%%%%%%%%%%%%%%%%%%%%%%%%%%%%%%%%%%%%%%
The above result shows that the coefficient of $\omega^{2}\ln(M\omega)$ in the dynamical response function is the same the linear-in-frequency dissipation number, a result already presented in the context of BH perturbation theory in~\ref{LN_Dyn_Pert}. The log-independent piece of the dynamical tides was recently determined in \citet{Combaluzier--Szteinsznaider:2025eoc,Kobayashi:2025vgl} for gravitational perturbations. 

Besides the determination of dynamical tides, there has been recent interest in the role of tail effects. These tails arise from the scattering of GWs off the long-range Newtonian potential, and they introduce characteristic logarithmic terms in the waveform \citep{Poisson:1994yf}. Interestingly, these logarithmic contributions have been shown to be related to anomalous dimensions in the EFT \citep{Parra-Martinez:2025bcu}.

%%%%%%%%%%%%%%%%%%%%%%%%%%%%%%%%%%%%%%%%%%%%%%%%%%%%%%%%%%%%%%%%%%%%%%%%%%%%%%

%%%%%%%%%%%%%%%%%%%%%%%%%%%%%%%%%%%%%%%%%%%%%%%%%%%%%%%%%%%%%%%%
%%%%%%%%%%%%%%%%%%%%%%%%%%%%%%%%%%%%%%%%%%%%%%%%%%%%%%%%%%%%%%%%
%%%%%%%%%%%%%%%%%%%%%%%%%%%%%%%%%%%%%%%%%%%%%%%%%%%%%%%%%%%%%%%%
\subsection{Worldline effective field theory approach}

Here we will derive the dynamical LNs of a Kerr BH under scalar perturbations using the worldline EFT approach. Generalization to generic spin can be found in \citet{SumantaToAppear2}. For this purpose, we will closely follow \citep{Creci:2021rkz} and determine the dynamical LNs of a Kerr BH under scalar perturbations. The computation of the dynamical LNs follows the following steps: (a) determining the solution of the Klein--Gordon equation in the near-zone ($M\omega \ll 1$ and $\omega(r-r_{+})\ll 1$) satisfying purely ingoing boundary condition; (b) determining the solution of the Klein Gordon equation in the far-zone ($r\gg M$); (c) matching the far-zone solution with the near-zone and matching the far-zone solution with EFT one obtains the response function and hence the dynamical LNs. 

%%%%%%%%%%%%%%%%%%%%%
%%%%%%%%%%%%%%%%%%%%%
%%%%%%%%%%%%%%%%%%%%%
%%%%%%%%%%%%%%%%%%%%%
\begin{figure}
    \centering
    \includegraphics[width=\textwidth]{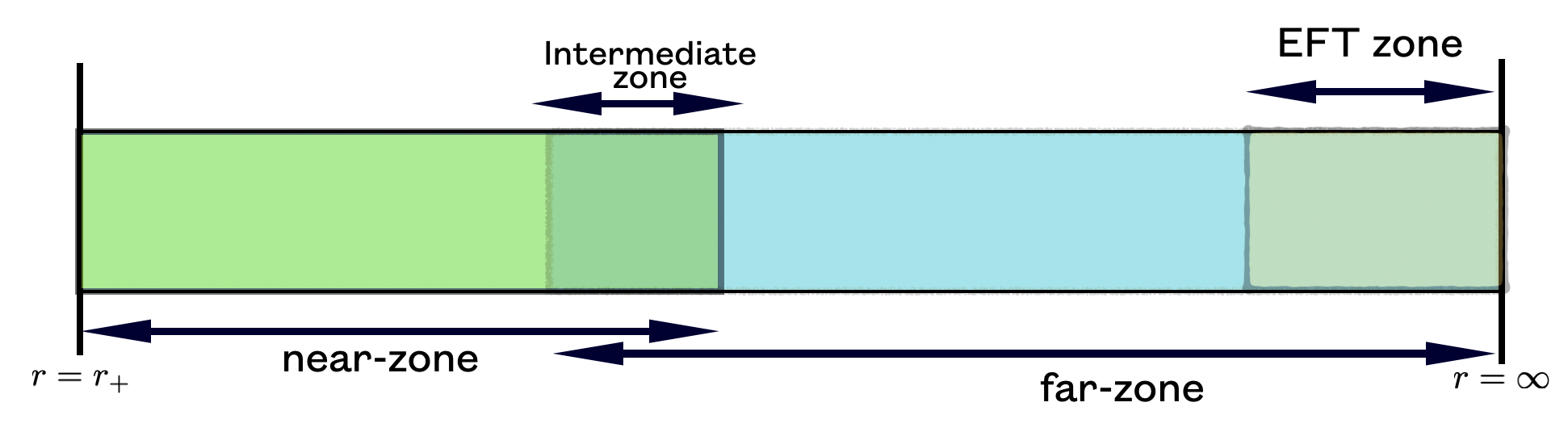}
    \caption{The schematic diagram depicts the near-zone ($\omega r\ll 1$) and the far-zone ($r\gg M$), mixing in the intermediate zone with $r\sim \omega^{-1}$. The EFT is defined in the asymptotic region and has to be matched with the far zone in order to determine the LNs (see also Fig.~1 in~\citet{Creci:2021rkz}).
    }
\label{fig:eftzone}
\end{figure}
%%%%%%%%%%%%%%%%%%%%%
%%%%%%%%%%%%%%%%%%%%%
%%%%%%%%%%%%%%%%%%%%%
%%%%%%%%%%%%%%%%%%%%%

\paragraph{Near-zone solution}~--- As we will be dealing with scalar perturbations, the starting point is the $s=0$ Teukolsky equation in the advanced null coordinates, whose radial part, upon using the coordinate $z\equiv\{(r-r_{+})/(r_{+}-r_{-})\}$, reduces to the following form in the near-zone regime (see~\ref{fig:eftzone}),
%%%%%%%%%%%%%%%%%%%%%%%%%%%%%%%%%%%%%%%%%%%%%%%%%%%%%%%%
\begin{align}
\dfrac{d}{dz}\left[z(1+z)\dfrac{dR^{\rm (near)}_{\ell m}}{dz}\right]+\frac{\left[P_{+}^{2}-\lambda z(1+z)\right]}{z(1+z)}R^{\rm (near)}_{\ell m}=0~,
\end{align}
%%%%%%%%%%%%%%%%%%%%%%%%%%%%%%%%%%%%%%%%%%%%%%%%%%%%%%%%
where we have defined $P_{+}\equiv(2Mr_{+}\omega-am)/(r_{+}-r_{-})=2Mr_{+}\bar{\omega}/(r_{+}-r_{-})$. Here, $\bar{\omega}\equiv \omega-m\Omega_{+}$, with $\Omega_{+}\equiv (a/2Mr_{+})$ being the angular velocity of the event horizon, located at $r=r_{+}$. The solution of the near-zone, that is purely ingoing at the BH horizon, is given by
%%%%%%%%%%%%%%%%%%%%%%%%%%%%%%%%%%%%%%%%%%%%%%%%%%%%%%%%
\begin{align}\label{radial_scalar_near}
R^{\rm (near)}_{\ell m}&=A_{\rm in}^{\rm H}(1+z)^{iP_{+}}z^{-iP_{+}}\,_{2}F_{1}\left(-\hat{\ell},\hat{\ell}+1;1-2iP_{+};-z\right)~,
\end{align}
%%%%%%%%%%%%%%%%%%%%%%%%%%%%%%%%%%%%%%%%%%%%%%%%%%%%%%%%
where, $\hat{\ell}=\ell-\{2am\omega/(2\ell+1)\}+\mathcal{O}(M^{2}\omega^{2})$ and $A_{\rm in}^{\rm H}$ is simply an overall arbitrary constant. This solution, in the limit of large $r$, such that $\omega r\ll 1$, can be expressed as, 
%%%%%%%%%%%%%%%%%%%%%%%%%%%%%%%%%%%%%%%%%%%%%%%%%%%%%%%%
\begin{align}\label{radial_near_far}
\medmath{R_{\ell m\,\textrm{(int)}}^{\rm (near)}=A_{\rm in}^{\rm H}\left[\frac{\Gamma(1-2iP_{+})\Gamma(1+2\hat{\ell})}{\Gamma(1+\hat{\ell})\Gamma(1+\hat{\ell}-2iP_{+})}\frac{r^{\hat{\ell}}}{(r_{+}-r_{-})^{\hat{\ell}}}+\frac{\Gamma(1-2iP_{+})\Gamma(-1-2\hat{\ell})}{\Gamma(-\hat{\ell})\Gamma(-\hat{\ell}-2iP_{+})}\frac{r^{-\hat{\ell}-1}}{(r_{+}-r_{-})^{-\hat{\ell}-1}}\right]\,,}
\end{align}
%%%%%%%%%%%%%%%%%%%%%%%%%%%%%%%%%%%%%%%%%%%%%%%%%%%%%%%%
where, the subscript $\textrm{(int)}$ refers to the solution in the intermediate zone (see~\ref{fig:eftzone}), which ends at $r\sim \omega^{-1}$.
Thus, we have a growing piece $r^{\hat{\ell}}$ and a decaying piece $r^{-\hat{\ell}-1}$, along with several sub-leading pieces, which we have neglected.

\paragraph{Far-zone solution}~--- In the far zone ($r\gg M$), the radial Teukolsky equation in the Kerr background satisfies the following differential equation,
%%%%%%%%%%%%%%%%%%%%%%%%%%%%%%%%%%%%%%%%%%%%%%%%%%%%%%%%
\begin{align}
\dfrac{d^{2}R^{\rm (far)}_{\ell m}}{dr^{2}}+\frac{2}{r}\dfrac{dR^{\rm (far)}_{\ell m}}{dr}&+\left[\omega^{2}-\frac{\lambda+2ma\omega}{r^{2}}\right]R^{\rm (far)}_{\ell m}=\mathcal{O}(M^{2}\omega^{2}r^{-2},r^{-3})~.
\end{align}
%%%%%%%%%%%%%%%%%%%%%%%%%%%%%%%%%%%%%%%%%%%%%%%%%%%%%%%%
Up to linear order in $M\omega$, we have $\lambda+2am\omega=E_{\ell m}+\mathcal{O}(M^{2}\omega^{2})=\ell(\ell+1)+\mathcal{O}(M^{2}\omega^{2})$. Hence, the above differential equation has the following general solution, 
%%%%%%%%%%%%%%%%%%%%%%%%%%%%%%%%%%%%%%%%%%%%%%%%%%%%%%%%
\begin{align}\label{scalar_far}
R_{\ell m}^{\rm (far)}=\frac{A^{\infty}_{\rm reg}}{\sqrt{r}}J_{\ell+\frac{1}{2}}(\omega r)+\frac{A^{\infty}_{\rm irreg}}{\sqrt{r}}Y_{\ell+\frac{1}{2}}(\omega r)~.
\end{align}
%%%%%%%%%%%%%%%%%%%%%%%%%%%%%%%%%%%%%%%%%%%%%%%%%%%%%%%%
The arbitrary constant $A^{\infty}_{\rm reg}$ is connected with the Bessel function $J_{\alpha}(x)$, which is regular as $x\to 0$, while the arbitray constant $A^{\infty}_{\rm irreg}$ is associated with $Y_{\alpha}(x)$, which is irregular as $x\to 0$. The $\omega r\ll 1$ limit of the above solution will bring it to the intermediate zone, where it is matched with near-zone, and in the $r\to \infty$ limit, we will match it with the EFT.  

\paragraph{Matching at the intermediate zone}~--- Here we will match the near zone and the far-zone solutions at the intermediate zone, as both the far and near-zone solutions has a growing part ($\sim r^{\ell}$) and a decaying part ($\sim r^{-\ell-1}$) in appropriate limits. Therefore, by matching the respective coefficients we obtain $A^{\infty}_{\rm reg}$ and $A^{\infty}_{\rm irreg}$ of the far-zone solution, in terms of $A_{\rm H}^{\rm in}$ coming from the near-zone solution. This yields \citep{SumantaToAppear2}, 
%%%%%%%%%%%%%%%%%%%%%%%%%%%%%%%%%%%%%%%%%%%%%%%%%%%%%%%%
\begin{align}\label{matching_scalar}
\frac{A^{\infty}_{\rm irreg}}{A^{\infty}_{\rm reg}}&=-\pi\left(\frac{\omega(r_{+}-r_{-})}{2}\right)^{2\ell+1}\frac{\Gamma(-1-2\hat{\ell})\Gamma(1+\hat{\ell})\Gamma(1+\hat{\ell}-2iP_{+})}{\Gamma(-\hat{\ell})\Gamma(1+2\hat{\ell})\Gamma(\ell+\frac{1}{2})\Gamma(\ell+\frac{3}{2})\Gamma(-\hat{\ell}-2iP_{+})}
\nonumber
\\
&\times \left\{1+2(\hat{\ell}-\ell)\ln \left[\omega(r_{+}-r_{-})\right]\right\}~.
\end{align}
%%%%%%%%%%%%%%%%%%%%%%%%%%%%%%%%%%%%%%%%%%%%%%%%%%%%%%%%
Logarithmic terms, like the one in the above expression, arises in several other contexts as well, see~\ref{eq:DLN} and the discussions therein. Note that here we have replaced $r\to \omega^{-1}$ in the logarithmic term, as $r\sim \omega^{-1}$ is the boundary of the intermediate regime, as well as, this makes the ratio of two constants to be another constant.  

\paragraph{EFT and its matching with BH perturbation theory}~--- In the coarse-grained worldline EFT description, tidal perturbations are modeled by an object moving along a worldline in flat spacetime, while additional degrees of freedom account for finite-size effects. The action encoding these finite-size contributions is given by $A_{\rm finite}$ in~\ref{EFTactiontotal}, which can also be written in the form \citep{Creci:2021rkz}, 
%%%%%%%%%%%%%%%%%%%%%%%%%%%%%%%%%%%%%%%%%%%%%%%%%%%%%%%%
\begin{align}
\mathcal{A}_{\rm finite}=-\int d\tau \sqrt{-u_{\mu}u^{\mu}}\sum_{\ell=0}^{\infty}\frac{1}{\ell!}M^{L}\nabla_{L}\phi~.
\end{align}
%%%%%%%%%%%%%%%%%%%%%%%%%%%%%%%%%%%%%%%%%%%%%%%%%%%%%%%%
Here, the tangent to the worldine is given by $u_{\mu}$, the proper time along the worldline is given by $\tau$, and the mass multipole moments on the worldline are denoted by $M^{L}=M^{i_{1}i_{2}\cdots i_{\ell}}$. The variation of the finite size action with respect to the scalar $\phi$, for asymptotically flat spacetimes, leads to the desired field equation for the scalar field \citep{Creci:2021rkz}, which can be solved as,
%%%%%%%%%%%%%%%%%%%%%%%%%%%%%%%%%%%%%%%%%%%%%%%%%%%%%%%%
\begin{align}
\Phi=\sum_{\ell,m}\sqrt{2\pi \omega}\,\omega^{\ell}Y_{\ell m}(-1)^{\ell}e^{i\omega t}R_{\ell m}^{\rm (flat)}(r)~.
\end{align}
%%%%%%%%%%%%%%%%%%%%%%%%%%%%%%%%%%%%%%%%%%%%%%%%%%%%%%%%
The radial function $R_{\ell m}^{\rm (flat)}(r)$ can be solved identically and it will have two independent solutions, one of which is irregular in the $r\to 0$ limit, and is related to the mass quadrupole moment $M_{L}$, while the other solution is regular in the $r\to 0$ limit and is the tidal field $E_{L}$, with these two being related by: $Q_L=-{\cal F}_{\ell m}(\omega) E_L$, where ${\cal F}_{\ell m}(\omega)$ is the frequency dependent response function. 

To determine the frequency dependent response function, we start with the radial function $R_{\ell m}^{\rm (flat)}(r)$, satisfying the following differential equation , 
%%%%%%%%%%%%%%%%%%%%%%%%%%%%%%%%%%%%%%%%%%%%%%%%%%%%%%%%
\begin{align}
r^{2}\dfrac{d^{2}R^{\rm (flat)}_{\ell m}}{dr^{2}}+2r\dfrac{dR^{\rm (flat)}_{\ell m}}{dr}+\left[\omega^{2}r^{2}-\ell(\ell+1)\right]R^{\rm (flat)}_{\ell m}=0~,
\end{align}
%%%%%%%%%%%%%%%%%%%%%%%%%%%%%%%%%%%%%%%%%%%%%%%%%%%%%%%%
which has the following general solution, 
%%%%%%%%%%%%%%%%%%%%%%%%%%%%%%%%%%%%%%%%%%%%%%%%%%%%%%%%
\begin{align}\label{eftradial}
R_{\ell m}^{\rm (flat)}=\frac{B^{\rm EFT}_{\textrm{reg}}}{\sqrt{r}}J_{\ell+\frac{1}{2}}(\omega r)+\frac{B^{\rm EFT}_{\textrm{irreg}}}{\sqrt{r}}Y_{\ell+\frac{1}{2}}(\omega r)~.
\end{align}
%%%%%%%%%%%%%%%%%%%%%%%%%%%%%%%%%%%%%%%%%%%%%%%%%%%%%%%%
As before, the `reg' and the `irreg' in the subscript of the arbitrary constants refers to the regularity of the Bessel functions in the $r\to 0$ limit.

This radial function in the $r\to \infty$ limit must be matched with the far-zone solution, as in~\ref{scalar_far}, in the same limit. This immediately yields, $A^{\infty}_{\rm reg}=B^{\rm EFT}_{\rm reg}$ and $A^{\infty}_{\rm irreg}=B^{\rm EFT}_{\rm irreg}$, and hence the mass multipole moment must scale as $M_{L}\sim B^{\rm EFT}_{\rm irreg}$, while the tidal field scales as $E_{L}\sim B^{\rm EFT}_{\rm reg}$. Therefore, the response function becomes, 
%%%%%%%%%%%%%%%%%%%%%%%%%%%%%%%%%%%%%%%%%%%%%%%%%%%%%%%%
\begin{align}
{\cal F}_{\ell m}(\omega)=\frac{M_{L}}{E_{L}}=\mathcal{N}_{\ell m}\frac{A^{\infty}_{\rm irreg}}{A^{\infty}_{\rm reg}}\,,
\end{align}
%%%%%%%%%%%%%%%%%%%%%%%%%%%%%%%%%%%%%%%%%%%%%%%%%%%%%%%%
where, the factor $\mathcal{N}_{\ell m}=-(4\sqrt{\pi}/2^{\ell})(2/\omega)^{1+2\ell}\Gamma(\ell+3/2)$ can be determined from the result that $\square \Phi=(1/\sqrt{2\pi})\sum_{\ell=0}^{\infty}(-1)^{\ell}(Q^{L}/\ell!)\partial_{L}\delta (x^{i})$ \citep{Creci:2021rkz}. Hence using the normalization, as well as~\ref{matching_scalar}, we obtain, 
%%%%%%%%%%%%%%%%%%%%%%%%%%%%%%%%%%%%%%%%%%%%%%%%%%%%%%%%
\begin{align}\label{EFT_dyn_tide}
\medmath{{\cal F}_{\ell m}(\omega)=\frac{4\pi^{3/2}}{2^{\ell}}\left(r_{+}-r_{-}\right)^{2\ell+1}\left\{1+2(\hat{\ell}-\ell)\ln \left[\omega(r_{+}-r_{-})\right]\right\}
\frac{\Gamma(-1-2\hat{\ell})\Gamma(1+\hat{\ell})\Gamma(1+\hat{\ell}-2iP_{+})}{\Gamma(-\hat{\ell})\Gamma(1+2\hat{\ell})\Gamma(\ell+\frac{1}{2})\Gamma(-\hat{\ell}-2iP_{+})}\,,}
\end{align}
%%%%%%%%%%%%%%%%%%%%%%%%%%%%%%%%%%%%%%%%%%%%%%%%%%%%%%%%
where terms of $\mathcal{O}(M^{2}\omega^{2})$ have been neglected. 

\paragraph{The dynamical tides}~--- As evident from~\ref{EFT_dyn_tide}, the dynamical tidal response function of a Kerr BH under scalar perturbations is non-zero. To compute the corresponding dynamical LNs, we first rescale the response function as $\widetilde{{\cal F}}_{\ell m}=\{2^{\ell}\Gamma(\ell+1/2)/4\pi^{3/2}(r_{+}-r_{-})^{2\ell+1}\}$, and then re-express the dynamical tidal response function as,
%%%%%%%%%%%%%%%%%%%%%%%%%%%%%%%%%%%%%%%%%%%%%%%%%%%%%%%%
\begin{align}
\label{eq:decompFtildescalar}
&\widetilde{{\cal F}}_{\ell m}(\omega)=\widetilde{{\cal F}}_{\ell m}|_{\rm static}+\widetilde{{\cal F}}_{\ell m}|_{\rm dyn}\times M\omega
+\widetilde{{\cal F}}_{\ell m}|_{\rm log}\times M\omega \ln [\omega(r_{+}-r_{-})]+\mathcal{O}(M^{2}\omega^{2})\,.
\end{align}
%%%%%%%%%%%%%%%%%%%%%%%%%%%%%%%%%%%%%%%%%%%%%%%%%%%%%%%%
The detailed expressions for each of these terms above can be found in \citet{SumantaToAppear2}, and it follows that $\widetilde{{\cal F}}_{\ell m}|_{\rm static}$ and $\widetilde{{\cal F}}_{\ell m}|_{\rm log}$ are purely imaginary and hence do not contribute to the conservative part of the tidal response function, while $\widetilde{{\cal F}}_{\ell m}|_{\rm dyn}$ has both real and imaginary parts. This shows that --- (a) static LNs of a Kerr BH are non-zero; (b) the dynamical LNs of a Kerr BH do not involve any logarithmic term at linear order in $M\omega$, and (c) the dynamical LNs of a Kerr BH are in general non-zero and given by
%%%%%%%%%%%%%%%%%%%%%%%%%%%%%%%%%%%%%%%%%%%%%%%%%%%%%%%%
\begin{align}
\widetilde{k}_{\ell m}&=ma\omega\left(\frac{\Gamma(1+\ell)^{2}}{\Gamma(2+2\ell)\Gamma(1+2\ell)}\right)\prod_{k=1}^{\ell}\left(k^{2}+4\{P_{+}^{a}\}^{2}\right)
\Bigg[\coth(2\pi P_{+}^{a})\left(\frac{\pi P_{+}^{a}}{(2\ell+1)}\right)
\nonumber
\\
&+\left(\frac{2Mr_{+}}{(r_{+}-r_{-})^{2}}\right)\textrm{Re}\Big\{\Psi(1+\ell+2iP^{a}_{+})-\Psi(1+\ell-2iP^{a}_{+})\Big\}\Bigg]\,,
\end{align}
%%%%%%%%%%%%%%%%%%%%%%%%%%%%%%%%%%%%%%%%%%%%%%%%%%%%%%%%
where, $\widetilde{k}_{\ell m}=(1/2)\textrm{Re}\widetilde{{\cal F}}_{\ell m}$. Note that the non-zero LNs scale as $ma\omega$, and has an factor $\sim \{(\ell!)^{2}/(2\ell+1)!(2\ell)!\}$, as it happens for all the other cases, see~\ref{nearzoneapp} and~\ref{scatampapp}, as well as \citep{Charalambous:2021mea, Saketh:2023bul, Bhatt:2024yyz}. However, there are terms inside the square bracket which can differ from one approach to another (compare with the results of~\ref{nearzoneapp}). This is due to the approximation scheme and the imperfect matching procedure in the BH perturbation theory side. This can be overcome by using a more refined matching approach, namely through the MST procedure \citep{Sasaki:2003xr, Ivanov:2022hlo, Saketh:2023bul,Kobayashi:2025vgl}. 

%%%%%%%%%%%%%%%%%%%%%%%%%%%%%%%%%%%%%%%%%%%%%%%%%%%%%%%%%%%%%%%%
%%%%%%%%%%%%%%%%%%%%%%%%%%%%%%%%%%%%%%%%%%%%%%%%%%%%%%%%%%%%%%%%
%%%%%%%%%%%%%%%%%%%%%%%%%%%%%%%%%%%%%%%%%%%%%%%%%%%%%%%%%%%%%%%%
\subsection{Dissipative tides: Tidal heating for black holes}\label{heatingBH}

In the presence of dynamical tidal fields, BHs also experience dissipation through the absorption of GWs emitted in the center-of-mass frame of the binary, resulting in an increase of the mass of the individual BHs. 
There are two complementary approaches to capture this effect: one based on the PN formalism \citep{Hartle:1973zz,Hartle:1974gy,Poisson:2004cw}, which applies to comparable--mass binaries, and the other based on the Teukolsky equation and the Newman--Penrose formalism \citep{Newman:1961qr,Teukolsky:1974yv,Hughes:2001jr,Alvi:2001mx,Chatziioannou:2012gq,Chatziioannou:2016kem}, which requires the secondary to be much lighter than the primary so that its stress--energy tensor can be treated as a perturbation (see also \citealt{Poisson:1994yf} for earlier work combining both PN and point-particle approximations). In what follows we will primarily discuss the Teukolsky formalism, while briefly commenting on the PN treatment at the end of this section for completeness. 

For this purpose, we introduce the following notation: $\Omega^{(i)}$ denotes the angular velocity of the $i$-th BH in the binary, while $\Omega_{\textrm{BH}i}$ denotes the angular velocity of the horizon of the $i$-th BH. In terms of these quantities, three cases can be distinguished: 
(a) the stationary approximation, for which $\Omega^{(i)}=0$ but $\Omega_{\textrm{BH}i}\neq 0$; 
(b) non-rotating binary BHs, for which $\Omega^{(i)}\neq 0$ but $\Omega_{\textrm{BH}i}=0$; 
and (c) the general case, in which both $\Omega^{(i)}\neq 0$ and $\Omega_{\textrm{BH}i}\neq 0$.

In the stationary approximation, the BHs do not move with respect to each other, although there is a relative exchange of angular momentum. Therefore, $\Delta E=0$ while $\Delta J\neq 0$. This implies $(dM_i/dt)=0$ but $(dJ_i/dt)\neq 0$, and consequently $(dA_{\textrm{BH}i}/dt)\neq 0$, where $A_{\textrm{BH}i}$ is the horizon area of the $i$-th BH. This suggests that $(dM_i/dt)\propto \Omega^{(i)}$. 

On the other hand, for non-rotating BHs one must have $(dM_i/dt)\neq 0$ but $(dJ_i/dt)=0$, which suggests $(dJ_i/dt)\propto \Omega_{\textrm{BH}i}$. In the general case, it is possible to have $\Omega^{(i)}=\Omega_{\textrm{BH}i}$. In this situation there is no relative motion between the BHs, implying $(dM_i/dt)=0=(dJ_i/dt)$ and hence $(dA_{\textrm{BH}i}/dt)=0$. 

Combining these considerations with the requirement that the rate of change of the horizon area be non-negative, we are led to the scalings
\[
\frac{dM_i}{dt}\propto (\Omega^{(i)}-\Omega_{\textrm{BH}i})~,\qquad
\frac{dJ_i}{dt}\propto (\Omega^{(i)}-\Omega_{\textrm{BH}i})~,\qquad
\frac{dA_{\textrm{BH}i}}{dt}\propto (\Omega^{(i)}-\Omega_{\textrm{BH}i})^{2}.
\]
These considerations suggest the following expressions for the rates of change of the mass, angular momentum, and area: 
%%%%%%%%%%%%%%%%%%%%%%%%%%%%%%%%%%%%%%%%%%%%%%%%%%%%%%%%%%%%%%%%%%%%%%%%%%%%%%
\begin{align}\label{areadervdynI}
\frac{\kappa_{+i}}{8\pi}\frac{dA_{\textrm{BH}i}}{dt}
&=(\Omega^{(i)}-\Omega_{\textrm{BH}i})^{2} I(\theta_{0})~,
\end{align}
%%%%%%%%%%%%%%%%%%%%%%%%%%%%%%%%%%%%%%%%%%%%%%%%%%%%%%%%%%%%%%%%%%%%%%%%%%%%%%
%%%%%%%%%%%%%%%%%%%%%%%%%%%%%%%%%%%%%%%%%%%%%%%%%%%%%%%%%%%%%%%%%%%%%%%%%%%%%%
\begin{align}\label{massangdyn}
\frac{dM_i}{dt}
&=\Omega^{(i)}(\Omega^{(i)}-\Omega_{\textrm{BH}i}) I(\theta_{0})~,
\qquad
\frac{dJ_i}{dt}
=(\Omega^{(i)}-\Omega_{\textrm{BH}i}) I(\theta_{0})~.
\end{align}
%%%%%%%%%%%%%%%%%%%%%%%%%%%%%%%%%%%%%%%%%%%%%%%%%%%%%%%%%%%%%%%%%%%%%%%%%%%%%%
Here $\kappa_{+i}$ is the surface gravity of the $i$-th BH, and $\theta_{0}$ denotes the angle of the binary plane. As is evident, the rate of change of the horizon area is positive definite, and the above expressions for the rates of change of mass, angular momentum, and area satisfy the laws of BH mechanics. In particular, in the stationary limit ($\Omega^{(i)}\to 0$), the rate of change of the mass of both BH1 and BH2 vanishes identically. These results will first be used in the stationary case to determine $I(\theta_{0})$, after which the general case can be obtained using the above relations.

To compute the mass and angular--momentum fluxes absorbed by one BH (denoted BH1) in a binary system, we begin by considering the simplified case in which the companion BH (denoted BH2) is stationary. We will relax this assumption later. The tidal field exerted by BH2 on BH1 is most conveniently determined in the Local Asymptotic Rest Frame (LARF \citep{PoissonWill}) of BH1. In this coordinate system, the spatial position of the stationary BH2 is specified by $(b,\theta_0,\phi_0)$, with $b$ the binary separation. In this setup, the Newtonian gravitational potential of BH2---effectively the $g_{00}$ component in the LARF---as experienced by any gravitating body in the region $r<b$, reads

%%%%%%%%%%%%%%%%%%%%%%%%%%%%%%%%%%%%%%%%%%%%%%%%%%%%%%%%%%%%%%%%%%%%%%%%%%%%%%
\begin{align}\label{newt_pot}
\Phi(r,\theta,\phi)=-\left(\frac{4\pi M_{2}}{b}\right)\sum_{\ell=0}^{\infty}\sum_{m=-\ell}^{\ell}\left(\frac{1}{2\ell+1}\right)\left(\frac{r}{b}\right)^{\ell}Y^{*}_{\ell m}(\theta_{0},\phi_{0})Y_{\ell m}(\theta,\phi)~.
\end{align}
%%%%%%%%%%%%%%%%%%%%%%%%%%%%%%%%%%%%%%%%%%%%%%%%%%%%%%%%%%%%%%%%%%%%%%%%%%%%%%
The above result simply follows from the Green's function of the Laplacian operator in the spherical polar coordinate system, see e.g., \citep{Jackson:1998nia}.
The above potential, evaluated at the location of the BH1, gives rise to a tidal force that BH1 will experience. The latter is governed by the tidal tensor $\mathcal{E}_{ij} =\partial_{i}\partial_{j}\Phi$. Expanding the tidal field $\mathcal{E}_{ij}$ in a spherical harmonic basis, and keeping in mind that the most relevant angular mode for the gravitational perturbation corresponds to $\ell=2$, we obtain the following behavior of the Weyl scalar $\Psi_{0}$ \citep{Thorne:1972}:
%%%%%%%%%%%%%%%%%%%%%%%%%%%%%%%%%%%%%%%%%%%%%%%%%%%%%%%%%%%%%%%%%%%%%%%%%%%%%%
\begin{align}\label{boundaryintmpsi0}
\Psi_{0}^{\rm (intermediate)}=\frac{8\pi\sqrt{6}M_{2}}{5b^{3}}\sum_{m=-2}^{2}\,_{2}Y_{2m}(\theta,\phi)Y^{*}_{2m}(\theta_{0},\phi_{0})~.
\end{align}
%%%%%%%%%%%%%%%%%%%%%%%%%%%%%%%%%%%%%%%%%%%%%%%%%%%%%%%%%%%%%%%%%%%%%%%%%%%%%%
Here, the superscript denotes that the Weyl scalar is computed in the intermediate zone, which is the region far from the tidal field as well as far from the affected body. Moreover, the existence of the term $(M_{2}/b^{3})$ in the above equation follows from the double derivative of~\ref{newt_pot}, as required for the determination of Weyl scalar from Newtonian potential.
Note that, alike the asymptotic boundary condition for the computation of LNs, here also the boundary condition is imposed in the intermediate regime, i.e., where $M_{1}\ll r\ll b$. In other words, the boundary condition is imposed far from the BH1, but deep within the tidal field. We will assume that the same condition holds for the BH2 as well. The reason for considering $\Psi_{0}$ and not $\Psi_{4}$ for our computation stems from the fact that we will need to determine the horizon flux, i.e., the GW energy propagating through the horizon, and the Weyl scalar $\Psi_{0}$ is best suited for this purpose \citep{Hartle:1973zz}. Given that the background spacetime is stationary and axisymmetric, the Weyl scalar $\Psi_{0}$ can be expressed as, 
%%%%%%%%%%%%%%%%%%%%%%%%%%%%%%%%%%%%%%%%%%%%%%%%%%%%%%%%%%%%%%%%%%%%%%%%%%%%%%
\begin{align}
\Psi_{0}=\int d\omega \sum_{\ell m}e^{-i\omega t}e^{im\phi}\,_{2}R_{\ell m}(r)\,_{2}S_{\ell m}(\theta)~,
\end{align}
%%%%%%%%%%%%%%%%%%%%%%%%%%%%%%%%%%%%%%%%%%%%%%%%%%%%%%%%%%%%%%%%%%%%%%%%%%%%%%
where $(t,r,\theta,\phi)$ are the Boyer-Lindquist coordinates. When substituted in the perturbed Einstein's equations, the angular and the radial parts neatly separate out, with the angular part $\,_{2}S_{\ell m}(\theta)$ satisfying the following equation, 
%%%%%%%%%%%%%%%%%%%%%%%%%%%%%%%%%%%%%%%%%%%%%%%%%%%%%%%%%%%%%%%%%%%%%%%%%%%%%%
\begin{align}
\frac{1}{\sin \theta}\dfrac{d}{d\theta}\left[\sin \theta \dfrac{d\,_{2}S_{\ell m}}{d\theta}\right]
+\Big[(a\omega \cos \theta)^{2}-4a\omega \cos \theta+2+2A_{\ell m}-\frac{(m+2\cos \theta)^{2}}{1-\cos^{2}\theta}\Big]\,_{2}S_{\ell m}=0~,
\end{align}
%%%%%%%%%%%%%%%%%%%%%%%%%%%%%%%%%%%%%%%%%%%%%%%%%%%%%%%%%%%%%%%%%%%%%%%%%%%%%%
while the radial part satisfies one of the  Teukolsky equations, namely
%%%%%%%%%%%%%%%%%%%%%%%%%%%%%%%%%%%%%%%%%%%%%%%%%%%%%%%%%%%%%%%%%%%%%%%%%%%%%%
\begin{align}\label{radteuktime}
\frac{1}{\Delta^{2}}\dfrac{d}{dr}\left[\Delta^{3}\dfrac{d\,_{2}R_{\ell m}}{dr}\right]+\Big[\frac{K^{2}-4i(r-M)K}{\Delta}+4i\dfrac{dK}{dr}-\lambda\Big]\,_{2}R_{\ell m}=0~.
\end{align}
%%%%%%%%%%%%%%%%%%%%%%%%%%%%%%%%%%%%%%%%%%%%%%%%%%%%%%%%%%%%%%%%%%%%%%%%%%%%%%
Note that, the above radial and angular equations differ from the ones presented in the previous section, since the equations in the context of LNs are written in the ingoing null coordinates, while the present discussion uses the Boyer-Lindquist time coordinate. Here, $\Delta\equiv r^{2}+a^{2}-2Mr$, with $M$ being the mass and $J=aM$ is the angular momentum of the BH. 

Since we wish to present analytical results associated with the rate of change of mass of BH1, due to its stationary companion BH2, we will have to resort to small frequency approximation, i.e., alike the previous sections, we will assume $M\omega \ll 1$. The computation is straightforward in the $z$ coordinate, which we have introduced earlier, $z=(r-r_{+})/(r_{+}-r_{-})$, where $r_{\pm}$ are the roots of the equation $\Delta=0$. Also, the quantity $P_{+}=-\{2Mr_{+}/(r_{+}-r_{-})\}\bar{\omega}$, defined in the context of dynamical LNs, will play a major role also in the present discussion (this quantity was denoted by $-i\gamma_{m}$ in \citet{Alvi:2001mx, Chakraborty:2021gdf}). In terms of the new coordinate $z$, and the parameter $P_{+}$, the radial Teukolsky equation in Boyer--Lindquist coordinates, for the $\ell=2$ mode, reduces to
%%%%%%%%%%%%%%%%%%%%%%%%%%%%%%%%%%%%%%%%%%%%%%%%%%%%%%%%%%%%%%%%%%%%%%%%%%%%%%
\begin{align}
\frac{1}{z(1+z)}\partial_{z}\left[\left\{z(1+z)\right\}^{3}\partial_{z}\,_{2}R_{2m}\right]
+\left[P_{+}^{2}+2iP_{+}(1+2z)\right]\,_{2}R_{2m}=0~.
\end{align}
%%%%%%%%%%%%%%%%%%%%%%%%%%%%%%%%%%%%%%%%%%%%%%%%%%%%%%%%%%%%%%%%%%%%%%%%%%%%%%
This equation can be solved exactly in terms of the hypergeometric functions, and the general solution reads 
%%%%%%%%%%%%%%%%%%%%%%%%%%%%%%%%%%%%%%%%%%%%%%%%%%%%%%%%%%%%%%%%%%%%%%%%%%%%%%
\begin{align}\label{radialpsi0}
\,_{2}R_{2m}&=(1+z)^{iP_{+}}\Big[A_{m}z^{-iP_{+}}\,_{2}F_{1}\left(0,5;3-2iP_{+};-z\right)
\nonumber
\\
&\qquad+C_{m}z^{-2+iP_{+}}\,_{2}F_{1}\left(-2+2iP_{+},3+2iP_{+};-1+2iP_{+};-z\right)\Big]\,.
\end{align}
%%%%%%%%%%%%%%%%%%%%%%%%%%%%%%%%%%%%%%%%%%%%%%%%%%%%%%%%%%%%%%%%%%%%%%%%%%%%%%
Using the corollary of~\ref{hypnegativea} and~\ref{hypdiffarg}, the above solution involving hypergeometric functions can be reduced to a simpler form, where the coefficient of $A_{m}$ involves a simple power law solution $z^{-iP_{+}}(1+z)^{iP_{+}}$, while the coefficient of $C_{m}$ involves $z^{-2+iP_{+}}(1+z)^{-2-iP_{+}}$, along with a hypergeometric function $\,_{2}F_{1}(1,-4;-1+2iP_{+};-z)$. The asymptotic form of the above solution can be obtained by considering~\ref{hypasymp}, while the near-horizon behavior is determined from~\ref{hypzero}, yielding,
%%%%%%%%%%%%%%%%%%%%%%%%%%%%%%%%%%%%%%%%%%%%%%%%%%%%%%%%%%%%%%%%%%%%%%%%%%%%%%
\begin{align}
\lim_{z\to 0}\,_{2}R_{2m}&\equiv \,_{2}R_{2m}^{\rm (h)}=A_{m}e^{i\bar{\omega}r_{*}}+\frac{C_{m}}{\Delta^{2}}e^{-i\bar{\omega}r_{*}}
\label{horizonradial}
\\
\lim_{z\to \infty}\,_{2}R_{2m}&\equiv \,_{2}R_{2m}^{\rm (intm)}=A_{m}-\frac{6iC_{m}}{P_{+}(1+iP_{+})(1+2iP_{+})}~.
\label{intermediateradial}
\end{align}
%%%%%%%%%%%%%%%%%%%%%%%%%%%%%%%%%%%%%%%%%%%%%%%%%%%%%%%%%%%%%%%%%%%%%%%%%%%%%%
Since we are dealing with BHs, it follows that the perturbation should be purely ingoing at the horizon. This requires setting $A_{m}=0$, as $A_{m}$ captures the outgoing mode at the horizon, as evident from~\ref{horizonradial}. The coefficient $C_{m}$ is then determined by matching the asymptotic behavior of the Weyl scalar $\Psi_{0}$, obtained through the radial function in~\ref{intermediateradial}, with the one presented in~\ref{boundaryintmpsi0}. This yields, 
%%%%%%%%%%%%%%%%%%%%%%%%%%%%%%%%%%%%%%%%%%%%%%%%%%%%%%%%%%%%%%%%%%%%%%%%%%%%%%
\begin{align}\label{Cm}
C_{m}=i\frac{8\pi M_{2}}{5\sqrt{6}b^{3}}P_{+}(1+iP_{+})(1+2iP_{+})Y^{*}_{2m}(\theta_{0},\phi_{0})~.
\end{align}
%%%%%%%%%%%%%%%%%%%%%%%%%%%%%%%%%%%%%%%%%%%%%%%%%%%%%%%%%%%%%%%%%%%%%%%%%%%%%%
This provides the complete expression for the Weyl scalar $\Psi_{0}$, which is consistent with the purely ingoing boundary condition at the horizon and matches with the tidal field due to the secondary object in the intermediate regime. However, there is still one issue with the above analysis, the radial part of the Weyl scalar $\Psi_{0}$ scales as $z^{-2}$, and hence diverges at the horizon, which is undesirable for the computation of the horizon flux. Therefore, one transforms to the Hartle--Hawking tetrad \citep{Hartle:1972ya}, which introduces an overall factor of $(\Delta^{2}/4)(r^{2}+a^{2})^{-2}$. Since $\Delta=(r_{+}-r_{-})^{2}z(1+z)$, the overall $\Delta^{2}$ factor takes care of the divergent behavior of $\Psi_{0}$ at the event horizon and hence can be used in the horizon flux computation. This rescaled Weyl scalar is denoted as $\Psi_{0}^{\rm HH}$, which is obtained by substituting $A_{m}=0$ and $C_{m}$ from~\ref{Cm} in~\ref{radialpsi0}, and reads
%%%%%%%%%%%%%%%%%%%%%%%%%%%%%%%%%%%%%%%%%%%%%%%%%%%%%%%%%%%%%%%%%%%%%%%%%%%%%%
\begin{align}\label{psi0hh}
\Psi_{0}^{\rm HH}&=i\frac{8\pi M_{2}}{5\sqrt{6}b^{3}}\left(\frac{\Delta^{2}}{4(r^{2}+a^{2})^{2}}\right)\int d\omega \sum_{m=-2}^{2}P_{+}(1+iP_{+})(1+2iP_{+})e^{-i\omega t}e^{im\phi}\,Y^{*}_{2m}(\theta_{0},\phi_{0})
\nonumber
\\
&\quad \times \,_{2}S_{2m}(\theta)\, (1+z)^{iP_{+}}z^{-2+iP_{+}}\,_{2}F_{1}\left(-2+2iP_{+},3+2iP_{+};-1+2iP_{+};-z\right)~.
\end{align}
%%%%%%%%%%%%%%%%%%%%%%%%%%%%%%%%%%%%%%%%%%%%%%%%%%%%%%%%%%%%%%%%%%%%%%%%%%%%%%
The first order gravitational perturbation to BH1, induced by the tidal effects of BH2, will also cause the horizon to be perturbed. Since the horizon is generated by null geodesics, it follows that the convergence and the shear of the null geodesics generating the horizon will be modified. Moreover, the convergence of the null horizon generators is related to the rate of change of area of the horizon, from which, it follows that \citep{Hartle:1972ya},
%%%%%%%%%%%%%%%%%%%%%%%%%%%%%%%%%%%%%%%%%%%%%%%%%%%%%%%%%%%%%%%%%%%%%%%%%%%%%%
\begin{align}\label{areaderv}
\frac{dA_{\rm BH1}}{dt}=\frac{1}{\kappa_{+1}}\int d^{2}x\,\sqrt{h^{(2)}_{+1}}|\sigma^{\rm HH}_{+1}|^{2}~.
\end{align}
%%%%%%%%%%%%%%%%%%%%%%%%%%%%%%%%%%%%%%%%%%%%%%%%%%%%%%%%%%%%%%%%%%%%%%%%%%%%%%
Here $h^{(2)}_{+1}$ is the induced metric and $\sigma^{\rm HH}_{+1}$ is the shear on the horizon of BH1. The shear in the Hartle-Hawking frame is related to the Weyl scalar $\Psi_{0}^{\rm HH}$ through the following relation: $|\sigma^{\rm HH}_{+1}|^{2}=(1/\kappa_{+1}^{2})|\Psi_{0}^{\rm HH}|^{2}$. Given the expression for $\Psi_{0}^{\rm HH}$ in~\ref{psi0hh}, one can obtain the shear in the Hartle-Hawking frame and then averaging over the angular coordinates, we obtain the rate of change of horizon area as, 
%%%%%%%%%%%%%%%%%%%%%%%%%%%%%%%%%%%%%%%%%%%%%%%%%%%%%%%%%%%%%%%%%%%%%%%%%%%%%%
\begin{align}\label{areadervfinal}
\frac{dA_{\rm BH1}}{dt}=\left(\frac{64\pi M_{1}^{5}M_{2}^{2}}{5b^{6}}\right)\,\frac{\chi_{1}^{2}\sin^{2}\theta_{0}}{\sqrt{1-\chi_{1}^{2}}}\left[1-\frac{3}{4}\chi_{1}^{2}+\frac{15}{4}\chi_{1}^{2}\sin^{2}\theta_{0}\right]~.
\end{align}
%%%%%%%%%%%%%%%%%%%%%%%%%%%%%%%%%%%%%%%%%%%%%%%%%%%%%%%%%%%%%%%%%%%%%%%%%%%%%%
Note that in the limit $\chi\to 0$, corresponding to a non-rotating BH, the rate of change of the horizon area vanishes identically. This follows because, in the stationary approximation, the BH mass does not change. Moreover, for a non-rotating BH the rate of change of the area is proportional to the rate of change of the mass, which therefore also vanishes identically.
For rotating BHs, under the stationary approximation, the rate of change of area is directly related to the rate of change of the angular momentum, 
%%%%%%%%%%%%%%%%%%%%%%%%%%%%%%%%%%%%%%%%%%%%%%%%%%%%%%%%%%%%%%%%%%%%%%%%%%%%%%
\begin{align}\label{angmomchange}
\frac{dJ_{1}}{dt}=-\left(\frac{8M_{1}^{5}M_{2}^{2}}{5b^{6}}\right)\,\chi_{1}\sin^{2}\theta_{0}\left[1-\frac{3}{4}\chi_{1}^{2}+\frac{15}{4}\chi_{1}^{2}\sin^{2}\theta_{0}\right]~.
\end{align}
%%%%%%%%%%%%%%%%%%%%%%%%%%%%%%%%%%%%%%%%%%%%%%%%%%%%%%%%%%%%%%%%%%%%%%%%%%%%%%
Thus, the angular momentum decreases, while the BH area increases. Obviously, for non-rotating BHs, the rate of change of angular momentum identically vanishes. In both the expressions for $(dA_{\rm BH1}/dt)$ and $(dJ_{1}/dt)$, $b$ is the separation between the binaries, and $\theta_{0}$ is the angle of the BH2 in the reference frame of the BH1.

We now turn to the dynamical evolution of the binary, thereby going beyond the stationary approximation. We start by considering the binary BHs to be moving around each other with angular velocity $\Omega$, which we approximate to its Newtonian value $\Omega_{\rm N}=(v^{3}/M_{\rm T})$, where $M_{\rm T}=M_{1}+M_{2}$ is the total mass of the binary and $v=\sqrt{M_{\rm T}/b}$ is the relative velocity between the binary components. Therefore, for individual BHs in the binary, the angular velocities are given by $\Omega^{(i)}=(\hat{\bf L}_{\rm orb}\cdot\hat{\bf J}_{i})\Omega_{\rm N}$, where $i=1,2$, with ${\bf L}_{\rm orb}$ being the orbital angular momentum and ${\bf J}_{i}$ is the angular momentum of the $i$-th BH in the binary. In this case, the expressions for the rate of change of area, mass and angular momentum are given by~\ref{areadervdynI} and~\ref{massangdyn}.

%Thus, the rate of change of area for the binary system, evolving dynamically, can be expressed as, 
%%%%%%%%%%%%%%%%%%%%%%%%%%%%%%%%%%%%%%%%%%%%%%%%%%%%%%%%%%%%%%%%%%%%%%%%%%%%%%
%\begin{align}\label{areadervdynI}
%\frac{\kappa_{+1}}{8\pi}\frac{dA_{\rm BH1}}{dt}=(\Omega^{(1)}-\Omega_{\rm BH1})^{2}I(\theta_{0})~,
%\end{align}
%%%%%%%%%%%%%%%%%%%%%%%%%%%%%%%%%%%%%%%%%%%%%%%%%%%%%%%%%%%%%%%%%%%%%%%%%%%%%%
%where $\theta_{0}$ is the plane of the binary and $\Omega_{\rm BH1}$ is the angular velocity of the BH1. As evident, the rate of change of area is a positive definite quantity. Similarly, the rate of change of mass and angular momentum of the BH1 in the dynamical scenario becomes,
%%%%%%%%%%%%%%%%%%%%%%%%%%%%%%%%%%%%%%%%%%%%%%%%%%%%%%%%%%%%%%%%%%%%%%%%%%%%%%
%\begin{align}\label{massangdyn}
%\frac{dM_{1}}{dt}=\Omega^{(1)}(\Omega^{(1)}-\Omega_{\rm BH1})I(\theta_{0})~;
%\qquad
%\frac{dJ_{1}}{dt}=(\Omega^{(1)}-\Omega_{\rm BH1})I(\theta_{0})~.
%\end{align}
%%%%%%%%%%%%%%%%%%%%%%%%%%%%%%%%%%%%%%%%%%%%%%%%%%%%%%%%%%%%%%%%%%%%%%%%%%%%%%

The quantity $I(\theta_{0})$, appearing in~\ref{areadervdynI} and~\ref{massangdyn} is assumed to be independent of the dynamics, i.e., it should have the same value as in the stationary situation. Thus, it can be determined by taking the stationary limit and then matching with~\ref{areadervfinal}, which yields
%%%%%%%%%%%%%%%%%%%%%%%%%%%%%%%%%%%%%%%%%%%%%%%%%%%%%%%%%%%%%%%%%%%%%%%%%%%%%%
\begin{align}\label{Idyn}
I(\theta_{0})=\left(\frac{16M_{1}^{5}M_{2}^{2}}{5b^{6}}\right)r_{+1}\sin^{2}\theta_{0}\left[1-\frac{3}{4}\chi_{1}^{2}+\frac{15}{4}\chi_{1}^{2}\sin^{2}\theta_{0}\right]~.
\end{align}
%%%%%%%%%%%%%%%%%%%%%%%%%%%%%%%%%%%%%%%%%%%%%%%%%%%%%%%%%%%%%%%%%%%%%%%%%%%%%%
Moreover, the difference between the angular velocities of individual BHs in the binary and the angular velocity of the horizon can be determined, which will have the following structure for BH1, $\Omega^{(1)}-\Omega_{\rm BH1}=-(\chi_{1}/2r_{+1})+\mathcal{O}(v^{3})$. Therefore, substitution of the expression for $\Omega^{(1)}-\Omega_{\rm BH1}$ and of $I(\theta_{0})$ from~\ref{Idyn}, in~\ref{massangdyn}, yields the following expressions for the rate of change of mass and angular momentum in the dynamical context,
%%%%%%%%%%%%%%%%%%%%%%%%%%%%%%%%%%%%%%%%%%%%%%%%%%%%%%%%%%%%%%%%%%%%%%%%%%%%%%
\begin{align}
\frac{dM_{1}}{dt}&=\left(\dfrac{dE}{dt}\right)_{\rm N}\left(\frac{M_{1}}{M_{\rm T}}\right)^{3}\frac{v^{5}}{4}\Bigg\{-\chi_{1}\left(\hat{\bf L}_{\rm orb}\cdot\hat{\bf J}_{i}\right)+2\left(\frac{r_{+1}}{M_{\rm T}}\right)\sin^{2}\theta_{0}\,v^{3}\Bigg\}
\nonumber
\\
&\qquad \qquad \times \left[1-\frac{3}{4}\chi_{1}^{2}+\frac{15}{4}\chi_{1}^{2}\sin^{2}\theta_{0}\right]~,
\\
\frac{dJ_{1}}{dt}&=\frac{(dJ/dt)_{\rm N}}{(dE/dt)_{\rm N}}\left(\hat{\bf L}_{\rm orb}\cdot\hat{\bf J}_{i}\right)\frac{dM_{1}}{dt}~.
\end{align}
%%%%%%%%%%%%%%%%%%%%%%%%%%%%%%%%%%%%%%%%%%%%%%%%%%%%%%%%%%%%%%%%%%%%%%%%%%%%%%
The corresponding expression for the rate of change of area can be determined using the above results and the laws of BH mechanics. Further, the quantity $(dJ/dt)_{\rm N}$ is the angular momentum loss by GWs at quadrupolar order and has the following expression: $(dJ/dt)_{\rm N}=(32/5)\varsigma^{2}M_{\rm T}v^{7}$, where $\varsigma\equiv (M_{1}M_{2}/M_{\rm T}^{2})$. Similarly, $(dE/dt)_{\rm N}$ is the quadrupolar energy loss by GWs and has the following expression: $(dE/dt)_{\rm N}=(v^{3}/M_{\rm T})(dJ/dt)_{\rm N}$. The corresponding expressions for the BH2 can be obtained by simply interchanging $1\leftrightarrow 2$ in the above results.

For rotating BHs,  the rate of change of BH mass appears at $\mathcal{O}(v^{5})$ in the leading order, and hence is a 2.5PN correction over and above the quadrupolar contribution. For non-rotating BHs, instead, the leading order contribution arises at $\mathcal{O}(v^{8})$, which corresponds to a relative 4PN term. The leading order tidal heating for rotating and non-rotating BHs on equatorial orbits, in vacuum GR are \citep{Alvi:2001mx}
%%%%%%%%%%%%%%%%%%%%%%%%%%%%%%%%%%%%%%%%%%%%%%%%%%%%%%%%%%%%%%%%%%%%%%%%%%%%%%
\begin{align}
\left(\frac{dM_{1}}{dt}\right)^{\rm 2.5PN}_{\rm rotating}&=\frac{32}{5}\varsigma^{2}v^{10}\left[-\chi_{1}(1+3\chi_{1}^{2})\left(\hat{\bf L}_{\rm orb}\cdot\hat{\bf J}_{i}\right)\left(\frac{M_{1}}{M_{\rm T}}\right)^{3}\frac{v^{5}}{4}\right]~,
\label{heatbranerot}
\\
\left(\frac{dM_{1}}{dt}\right)^{\rm 4PN}_{\rm non-rotating}&=\frac{32}{5}\varsigma^{2}v^{10}
\left[\left(\frac{M_{1}}{M_{\rm T}}\right)^{4}v^{8}\right]~.
\label{heatbranenonrot}
\end{align}
%%%%%%%%%%%%%%%%%%%%%%%%%%%%%%%%%%%%%%%%%%%%%%%%%%%%%%%%%%%%%%%%%%%%%%%%%%%%%%
Note that, for $m=2$, the static dissipation number derived from the scattering approach reads, $\nu^{\rm E/B}_{\ell=2,m=2}=-(8/45)\chi(1+3\chi^{2})$, see~\ref{diss_static}. Interestingly, except for some normalization factor, this exactly matches with the structure of the tidal heating term, for a rotating BH, at 2.5 PN order, see~\ref{heatbranerot}. This connects the dissipation numbers obtained from EFT/scattering amplitude/BH perturbation theory with the tidal heating phenomenon.

The interplay between tidal coupling and superradiance \citep{Brito:2015oca} was investigated in \citet{Cardoso:2012zn}, where it was shown that tidal acceleration of the orbit can occur when the BH rotational energy is extracted through horizon superradiance.

\paragraph{Example: tidal heating for braneworld black holes}~---
Later on we will use these results to differentiate between BHs and ECOs and also to distinguish between BHs in different theories of gravity. Since the tidal heating directly affects the phasing of the GWs, it follows that one can incorporate tight constraints on the departure from the BH paradigm, or on modified theories of gravity, using tidal heating. We will provide here one such example, namely tidal heating for braneworld BHs, in the context of modified theories of gravity. The main roadblock in the study of tidal heating in modified theories of gravity is to determine the equivalent of Teukolsky equation for gravitational perturbations in those theories. There have been recent attempts to write down the Teukolsky equation for a certain class of modified theories of gravity, which may pave way for the study involving tidal heating in those theories. However, the basic features can be found in the example of braneworld BH to be considered here. 

As long as the four-dimensional gravitational perturbation of the braneworld BH is considered, and since the bulk effects appear at a much higher energy scale than the brane, one can safely assume that the gravitational perturbations satisfy the Teukolsky equations, with the metric elements modified according to the BH metric on the brane. Similar to the discussion involving non-rotating braneworld BHs (see~\ref{sec:braneworld}) in the rotating case as well, the metric is equivalent to that of a Kerr-Newman BH with $Q^{2}$ replaced by $-q$, so that the term $\Delta$ in~\ref{radteuktime} becomes $r^{2}-2Mr+a^{2}-M^{2}q$. As a consequence, it follows that the extremality condition becomes, $a^{2}=M^{2}(1+q)$. Thus, for $q>0$, it is possible to have the dimensionless rotation parameter ($\chi\equiv a/M$) to be larger than unity, a distinct feature of these braneworld BHs. Proceeding as outlined above, it follows that the Weyl scalar $\Psi_{0}$ in the Hartle--Hawking tetrad can be expressed exactly as in~\ref{psi0hh}, with $\Delta$ and $P_{+}$ depending implicitly on the charge $q$. Therefore, the rate of change of area will also have a very similar form, as that of~\ref{areadervfinal}, but now with explicit dependence on the charge $q$:
%%%%%%%%%%%%%%%%%%%%%%%%%%%%%%%%%%%%%%%%%%%%%%%%%%%%%%%%%%%%%%%%%%%%%%%%%%%%%%
\begin{align}\label{areadervbrane}
\frac{dA_{\rm BH1}}{dt}=\left(\frac{64\pi M_{1}^{5}M_{2}^{2}}{5b^{6}}\right)\,\frac{\chi_{1}^{2}\sin^{2}\theta_{0}}{\sqrt{1-\chi_{1}^{2}+q}}\left[1-\frac{3}{4}\chi_{1}^{2}+q\left(2-\frac{3}{4}\chi_{1}^{2}+q\right)+\frac{15}{4}\chi_{1}^{2}\left(1+q\right)\sin^{2}\theta_{0}\right]~.
\end{align}
%%%%%%%%%%%%%%%%%%%%%%%%%%%%%%%%%%%%%%%%%%%%%%%%%%%%%%%%%%%%%%%%%%%%%%%%%%%%%%
Clearly, as $q\to 0$ we retrieve~\ref{areadervfinal}. Since the charge $q$ is directly related to the size of the extra dimension, and the energy scale of the bulk is much higher than the energy scale on the brane, we will assume that, under the gravitational perturbation on the brane, the bulk is not affected and hence $q$ does not change. Thus, the above rate of change of the area is directly related to the rate of change of angular momentum (for a stationary companion), or to both the rate of change of angular momentum and mass (in the dynamical context). Since the thermodynamic relations do not change under this assumption, the area law is still valid in this context. The corresponding expressions in the dynamical context can be determined using~\ref{areadervdynI} and~\ref{massangdyn}, with the quantity $I(\theta_{0})$ being determined using the stationary relation, presented in~\ref{areadervbrane}. Therefore, the leading order tidal heating terms for both rotating and non-rotating braneworld BH finally reads: 
%%%%%%%%%%%%%%%%%%%%%%%%%%%%%%%%%%%%%%%%%%%%%%%%%%%%%%%%%%%%%%%%%%%%%%%%%%%%%%
\begin{align}
\left(\frac{dM_{1}}{dt}\right)^{\rm (brane)}_{\rm spin}&=\frac{32}{5}\varsigma^{2}v^{10}\left[-\chi_{1}\left\{1+3\chi_{1}^{2}+q\left(2+3\chi_{1}^{2}+q\right)\right\}\left(\hat{\bf L}_{\rm orb}\cdot\hat{\bf J}_{i}\right)\left(\frac{M_{1}}{M_{\rm T}}\right)^{3}\frac{v^{5}}{4}\right]~,
\\
\left(\frac{dM_{1}}{dt}\right)^{\rm (brane)}_{\rm static}&=\frac{32}{5}\varsigma^{2}v^{10}
\left[\left\{1+\frac{q}{2}+\sqrt{1+q} \right\}\left(\frac{M_{1}}{M_{\rm T}}\right)^{4}v^{8}\right]~.
\end{align}
%%%%%%%%%%%%%%%%%%%%%%%%%%%%%%%%%%%%%%%%%%%%%%%%%%%%%%%%%%%%%%%%%%%%%%%%%%%%%%
This shows how the tidal heating can be modified in theories of gravity beyond GR. The implications of these results and possible constraints on the charge $q$ will be presented in~\ref{sec:GW}.

%%%%%%%%%%%%%%%%%%%%%%%%%%%%%%%%%%%%%%%%%%%%%%%%%%%%%%%%%%%%%
%%%%%%%%%%%%%%%%%%%%%%%%%%%%%%%%%%%%%%%%%%%%%%%%%%%%%%%%%%%%%
%%%%%%%%%%%%%%%%%%%%%%%%%%%%%%%%%%%%%%%%%%%%%%%%%%%%%%%%%%%%%
\subsection{Summary: dynamical tidal response of a black hole}

Here we summarize the dynamical tidal response of BHs.  
For a Schwarzschild BH, the response to a generic spin-$s$ bosonic perturbation in the low-frequency regime schematically reads \citep{Combaluzier--Szteinsznaider:2025eoc}
%%%%%%%%%%%%%%%%%%%%%%%%%%%%%%%%%%%%%%%%%%%%%
\begin{equation}
{}_s{\cal F}^{\rm non\mbox{-}rotating}_{\ell m} 
= {}_s a_{\ell m} \left[i\,\omega M + 2\,\omega^{2} M^{2}
\left({}_s b_{\ell m} - \log (\mu M)\right)\right]
+ {\cal O}(\omega^{3} M^{3})\,,
\end{equation}
%%%%%%%%%%%%%%%%%%%%%%%%%%%%%%%%%%%%%%%%%%%%%
where ${}_s a_{\ell m}$ and ${}_s b_{\ell m}$ are ${\cal O}(1)$ coefficients depending on $(\ell,s)$ (the response is $m$--degenerate for a non-spinning BH), and $\mu$ is an arbitrary renormalization scale associated with the running term.  
The latter is twice the ${\cal O}(\omega M)$ dissipative response for all $(\ell,m,s)$.  
Thus, a Schwarzschild BH exhibits nontrivial dissipation at ${\cal O}(\omega M)$ and nontrivial bosonic LNs at ${\cal O}(\omega^{2}M^{2})$, entering the GW phase at 4PN and 8PN order, respectively (see~\ref{sec:GW}).

For spinning BHs, frame dragging introduces dissipation already in the static limit.  
At ${\cal O}(\omega M)$ both dissipative and conservative contributions appear.  
Consequently, rotating BHs possess non-zero bosonic LNs at ${\cal O}(M\omega)$.    
Schematically, one finds
%%%%%%%%%%%%%%%%%%%%%%%%%%%%%%%%%%%%%%%%%%%%%
\begin{align}
{}_s{\cal F}^{\rm rotating}_{\ell m}
&= i\, {}_{s}\mathbb{A}_{\ell m}
+ i\,M\omega\, {}_s\mathbb{B}_{\ell m}
+ \omega M \left( {}_s \mathbb{C}_{\ell m}
  + 4\, {}_{s}\mathbb{A}_{\ell m} \log (\mu M) \right)\nonumber\\
&+ {}_{s}\mathbb{D}_{\ell m}\, \omega^{2} M^{2} \log (\mu M)
+ {\cal O}(\omega^{2} M^{2})\,.
\end{align}
%%%%%%%%%%%%%%%%%%%%%%%%%%%%%%%%%%%%%%%%%%%%%
Here, both the static dissipation coefficient ${}_{s}\mathbb{A}_{\ell m}$ and the running bosonic LN at ${\cal O}(M\omega)$ are related by an overall factor of $4$ \citep{Saketh:2023bul}.  
All coefficients ${}_{s}\mathbb{A}_{\ell m}$, ${}_s\mathbb{B}_{\ell m}$, ${}_s\mathbb{C}_{\ell m}$, and ${}_{s}\mathbb{D}_{\ell m}$ are real, order unity, and ${}_{s}\mathbb{A}_{\ell m}$ and ${}_{s}\mathbb{C}_{\ell m}$ vanish in the non-rotating case.  
The dissipative coefficient and the non-running part of the conservative bosonic LNs at ${\cal O}(\omega^{2} M^{2})$ have not been computed yet.

%%%%%%%%%%%%%%%%%%%%%%%%%%%%%%%%%%%%%%%%%%%%%%%%%%%%%%%%%%%%%%%%%%%%%%%%%%%%%%
%%%%%%%%%%%%%%%%%%%%%%%%%%%%%%%%%%%%%%%%%%%%%%%%%%%%%%%%%%%%%%%%%%%%%%%%%%%%%%
%%%%%%%%%%%%%%%%%%%%%%%%%%%%%%%%%%%%%%%%%%%%%%%%%%%%%%%%%%%%%%%%%%%%%%%%%%%%%%
\section{Love numbers of neutron stars}\label{sec:NSs}
We now review the main aspects of the tidal LNs of NSs. This topic is of paramount importance, as it originally motivated much of the research on tidal deformability of compact objects in GR. Our discussion will therefore be necessarily brief, since several comprehensive and specialized reviews and textbooks on this subject already exist; see, for example, \citet{Yagi:2016bkt,Chatziioannou:2020pqz,PoissonWill}.

The relativistic theory of tidal deformations of a NS was developed in some seminar papers \citep{Hinderer:2007mb,Binnington:2009bb,Damour:2009vw}, in the latter case building on the formalism in \citet{Damour:1991yw,Damour:1990pi,Damour:1992qi,Damour:1993zn}. We refer to~\ref{conventionLN} for the different conventions used in various papers to defined the tidal deformability parameters and the tidal LNs. 

\subsection{Static Love numbers of non-rotating neutron stars}\label{LNNS_Static}

In the case of a non-rotating NS, the exterior geometry, being static and spherically symmetric, is described by the Schwarzschild metric. The geometry inside the star is also static and spherically symmetric and can be written in the following form:
%%%%%%%%%%%%%%%%%%%%%%%%%%%%%%%%%%%%%%%%%%%%%%%%%%%%%%%%%%%%%%%%%%%%%%%%%%%%%%%%%
\begin{equation}
d\bar{s}^{2}=-e^{\bar{\nu}}dt^{2}+e^{\bar{\lambda}}dr^{2}+r^{2}d\Omega^{2}\,,
\end{equation}
%%%%%%%%%%%%%%%%%%%%%%%%%%%%%%%%%%%%%%%%%%%%%%%%%%%%%%%%%%%%%%%%%%%%%%%%%%%%%%%%%
where $\bar{\nu}$ and $\bar{\lambda}$ are functions of the radial coordinate $r$ alone. These metric functions must smoothly map to the exterior Schwarzschild geometry. Further, the interior of the NS is assumed to be filled up by an isotropic perfect fluid such that the associated stress-energy tensor becomes
%%%%%%%%%%%%%%%%%%%%%%%%%%%%%%%%%%%%%%%%%%%%%%%%%%%%%%%%%%%%%%%%%%%%%%%%%%%%%%%%%
\begin{equation}
\bar{T}_{\mu \nu}^{\rm (ns)}=(\bar{\rho}+\bar{p})\bar{u}_{\mu}\bar{u}_{\nu}+\bar{p}\bar{g}_{\mu \nu}\,,
\end{equation}
%%%%%%%%%%%%%%%%%%%%%%%%%%%%%%%%%%%%%%%%%%%%%%%%%%%%%%%%%%%%%%%%%%%%%%%%%%%%%%%%%
where $\bar{u}^{\mu}$ is the four-velocity of the fluid inside the unperturbed NS along with $\bar{\rho}$ and $\bar{p}$, being the unperturbed energy density and unperturbed pressure, respectively. Expressing, $e^{-\bar{\lambda}}=1-2\bar{m}(r)/r$, where $\bar{m}(r)$ is the mass function of the unperturbed NS, the background Einstein's equations reduce to the well-known Tolman--Oppenheimer--Volkoff (henceforth as TOV) equations,
%%%%%%%%%%%%%%%%%%%%%%%%%%%%%%%%%%%%%%%%%%%%%%%%%%%%%%%%%%%%%%%%%%%%%%%%%%%%%%%%%
\begin{equation}\label{TOV}
\left\{
\begin{array}{l}
\bar{m}'(r)=4\pi r^{2}\bar{\rho}(r)
\\
\bar{p}'(r)=-\left(\bar{p}+\bar{\rho}\right)\left(\frac{\bar{m}(r)+4\pi r^{3}\bar{p}(r)}{r\left[r-2\bar{m}(r)\right]}\right)
\\
\bar{\nu}'(r)=-\frac{2\bar{p}'(r)}{\bar{p}(r)+\bar{\rho}(r)}
\end{array}
\right.\,,
\end{equation}
%%%%%%%%%%%%%%%%%%%%%%%%%%%%%%%%%%%%%%%%%%%%%%%%%%%%%%%%%%%%%%%%%%%%%%%%%%%%%%%%%
where a prime denotes derivative with respect to $r$. 
We have four unknown functions in our system, $(\bar{m},\bar{\nu},\bar{\rho},\bar{p})$, while only three independent equations. Therefore, in order to close the system, the TOV equations must be supplemented by a barotropic EoS of the form $\bar{p}=\bar{p}(\bar{\rho})$. 

For a given EoS, one can solve the TOV equations presented in~\ref{TOV} and determine the mass and radius of a NS for a specified central density (with the corresponding central pressure fixed by the EoS), together with the matching to the Schwarzschild exterior at the stellar surface. By varying the central density, one obtains a mass--radius relation for NSs corresponding to that EoS, where the radius $R$ is defined such that $\bar p(R)=0$ and the mass is $M=\bar m(R)$. 

% We present such mass--radius relations in~\ref{}, which illustrate that the observation of NSs with masses $\gtrsim 2M_{\odot}$ places stringent constraints on the EoS and can rule out several specific models.

\paragraph{Polar sector}~--- The gravitational perturbations of the interior of the NS need to be matched with those in the exterior Schwarzschild geometry.
Since our interest is in static configurations, for polar perturbations we can focus on the perturbation of the $g_{tt}$ component of the metric, which we parametrize as\footnote{We use barred quantities to denote variables in the interior.}
$\delta g_{tt}=-e^{\bar{\nu}}\bar{H}(r)$, together with
$\delta g_{rr}=e^{\bar{\lambda}}\bar{H}(r)$.   
The latter follows from the perturbed Einstein equations, in particular from the equality
$\delta G^{\theta}_{\theta}=\delta G^{\phi}_{\phi}$.
Finally, subtracting the $(t,t)$ and $(r,r)$ components of the perturbed Einstein equations, one obtains \citep{Hinderer:2007mb}
%%%%%%%%%%%%%%%%%%%%%%%%%%%%%%%%%%%%%%%%%%%%%%%%%%%%%%%%%%%%%%%%%%%%%%%%%%%%%%
\begin{align}\label{metric_NS_polar}
\bar{H}''&+\left[\frac{2}{r}+e^{\bar{\lambda}}\left(\frac{2\bar m(r)}{r^{2}}+4\pi r(\bar p-\bar \rho)\right)\right]\bar{H}'
\nonumber\\
&+\left[-\ell(\ell+1)\frac{e^{\bar{\lambda}}}{r^{2}}
+4\pi e^{\bar{\lambda}}\left(5\bar \rho+9\bar p+\frac{\bar \rho+\bar p}{d\bar p/d\bar \rho}\right)
-\bar{\nu}'^{2}\right]\bar{H}=0~.
\end{align}
%%%%%%%%%%%%%%%%%%%%%%%%%%%%%%%%%%%%%%%%%%%%%%%%%%%%%%%%%%%%%%%%%%%%%%%%%%%%%%
Given an EoS, one can solve the background equations~\ref{TOV} and determine the metric functions, the mass function, and the density profile as functions of the radial coordinate $r$. Substituting these background quantities into~\ref{metric_NS_polar} yields a second-order differential equation whose solution determines $\bar{H}(r)$.

Since both the interior equation for $\bar{H}$ and the exterior equation for $H$ are second-order, the full solution involves four integration constants. Regularity of the interior solution at the center of the NS fixes one of them. In particular, near the center the interior perturbation behaves as
%%%%%%%%%%%%%%%%%%%%%%%%%%%%%%%%%%%%%%%%%%%%%%%%%%%%%%%%%%%%%%%%%%%%%%%%%%%%%%
\begin{align}
\bar{H}(r\approx 0)=\bar{A}_{0} r^{\ell}\Bigg[1
-\frac{2\pi}{7}\left(5\bar{\rho}(0)+9\bar{p}(0)
+\frac{\bar{\rho}(0)+\bar{p}(0)}{(d\bar{p}/d\bar{\rho})|_{0}}\right)r^{2}
+\mathcal{O}(r^{3})\Bigg]~,
\end{align}
%%%%%%%%%%%%%%%%%%%%%%%%%%%%%%%%%%%%%%%%%%%%%%%%%%%%%%%%%%%%%%%%%%%%%%%%%%%%%%
where $\bar{A}_{0}$ is an arbitrary integration constant.
Among the remaining constants, two are fixed by imposing continuity\footnote{The matching between the interior and exterior solutions is dictated by Israel's junction conditions \citep{Israel:1966rt}, whose explicit form depends on the behavior of the pressure and energy density at the stellar surface. 
When the pressure and the energy density vanish smoothly at the surface, the junction conditions simply require the continuity of the metric and of its first radial derivative across the stellar boundary. However, if the energy density is discontinuous, as in quark-star models, the implementation of junction conditions is more involved, see, e.g., \citet{Damour:2009vw,Hinderer:2009ca,Pitre:2025qdf}.
} of the perturbation and its derivative at the stellar surface $r=R$:
$\bar{H}(R)=H(R)$ and $\bar{H}'(R)=H'(R)$.
These conditions uniquely determine the LNs as functions of the central density.

Using the known exterior solution for $H$ in terms of associated Legendre functions (see~\ref{Hext}), the $\ell=2$ electric LN of a NS (or, more generally, of any compact object with mass $M$ and radius $R$) is given by \citep{Hinderer:2007mb}
%%%%%%%%%%%%%%%%%%%%%%%%%%%%%%%%%%%%%%%%%%%%%%%%%%%%%%%%%%%%%%%%%%%%%%%%%%%%%%
\begin{align}\label{k2NS}
k_{2}^{\rm E}&=\frac{8}{5}\left(\frac{M}{R}\right)^{5}\left(1-\frac{2M}{R}\right)^{2}
\left[2-y+\frac{2M}{R}(y-1)\right]
\nonumber\\
&\times\Bigg[
\frac{2M}{R}\left(6-3y+\frac{3M}{R}(5y-8)\right)
+4\left(\frac{M}{R}\right)^{3}\left(13-11y+\frac{M}{R}(3y-2)
+2\left(\frac{M}{R}\right)^{2}(1+y)\right)
\nonumber\\
&\qquad
+3\left(1-\frac{2M}{R}\right)^{2}\left(2-y+\frac{2M}{R}(y-1)\right)
\log\!\left(1-\frac{2M}{R}\right)
\Bigg]^{-1},
\end{align}
%%%%%%%%%%%%%%%%%%%%%%%%%%%%%%%%%%%%%%%%%%%%%%%%%%%%%%%%%%%%%%%%%%%%%%%%%%%%%%
where
$y\equiv R H'(R)/H(R)=R\bar{H}'(R)/\bar{H}(R)$
is a dimensionless quantity encoding the information about the NS interior.

The Newtonian LN can be obtained by taking the limit $c\to\infty$, or equivalently
$(GM/c^{2}R)\to 0$, yielding
%%%%%%%%%%%%%%%%%%%%%%%%%%%%%%%%%%%%%%%%%%%%%%%%%%%%%%%%%%%%%%%%%%%%%%%%%%%%%%
\begin{align}
k_{2}^{\rm N}=\frac{1}{2}\left(\frac{2-y}{3+y}\right)~.
\end{align}
%%%%%%%%%%%%%%%%%%%%%%%%%%%%%%%%%%%%%%%%%%%%%%%%%%%%%%%%%%%%%%%%%%%%%%%%%%%%%%

\paragraph{Axial sector}~--- In the axial sector, one can simply use~\ref{eqaxialBHDMup}, along with the fact that we are dealing with isotropic fluid, to obtain a differential equation for the axial metric perturbation $\bar{h}_{0}$ inside the NS,
%%%%%%%%%%%%%%%%%%%%%%%%%%%%%%%%%%%%%%%%%%%%%%%%%%%%%%%%%%%%%%%%%%%%%%%%%%%%%%
\begin{align}\label{static_magnetic_NSh0}
e^{-\bar \lambda}\bar{h}_{0}''-4\pi r\left(\bar p+\bar \rho\right)\bar{h}_{0}'-e^{\bar \lambda}\left[\frac{\ell(\ell+1)}{r^{2}}-\frac{4\bar m}{r^{3}}+8\pi\left(\bar \rho+\bar p\right)\right]\bar{h}_{0}=0\,.
\end{align}
%%%%%%%%%%%%%%%%%%%%%%%%%%%%%%%%%%%%%%%%%%%%%%%%%%%%%%%%%%%%%%%%%%%%%%%%%%%%%%
Note that this assumes $\delta u^{\mu}_{\rm axial}=0$, referred to as the \emph{static} fluid, an assumption that will be further discussed in~\ref{dyntideNS}. Originally, \citet{Damour:2009vw} computed the axial LNs of a NS by defining the variable $\psi\equiv r\bar{h}_{0}'-2\bar{h}_{0}$, where $\bar{h}_{0}$ is the perturbed $g_{t\phi}$ component of the metric, whose equation has been presented above. In the static limit, it follows that $\psi$ satisfies the following differential equation: 
%%%%%%%%%%%%%%%%%%%%%%%%%%%%%%%%%%%%%%%%%%%%%%%%%%%%%%%%%%%%%%%%%%%%%%%%%%%%%%
\begin{align}\label{static_magnetic_NS}
\psi''+\frac{e^{\bar \lambda}}{r^{2}}\left[2\bar m+4\pi r^{3}\left(\bar p-\bar \rho\right)\right]\psi'-e^{\bar \lambda}\left[\frac{\ell(\ell+1)}{r^{2}}-\frac{6\bar m}{r^{3}}+4\pi\left(\bar \rho-\bar p\right)\right]\psi=0
\end{align}
%%%%%%%%%%%%%%%%%%%%%%%%%%%%%%%%%%%%%%%%%%%%%%%%%%%%%%%%%%%%%%%%%%%%%%%%%%%%%%
Again, this equation can be solved for $\psi$ for a given a solution to the TOV equation. Imposing regularity of $\psi$ at the origin and matching with the exterior solution presented in~\ref{h0e}, yields the magnetic LNs. For the $\ell=2$ mode,
%%%%%%%%%%%%%%%%%%%%%%%%%%%%%%%%%%%%%%%%%%%%%%%%%%%%%%%%%%%%%%%%%%%%%%%%%%%%%%
\begin{align}
\frac{1}{k_{2}^{\rm M}}&=\frac{5\left\{12(y+1)\left(\frac{M}{R}\right)^{4}+2(y-3)\left(\frac{M}{R}\right)^{3}+2(y-3)\left(\frac{M}{R}\right)^{2}+3(y-3)\left(\frac{M}{R}\right)-3y+9\right\}}{4\left(\frac{M}{R}\right)^{4}\left(\frac{2M}{R}-1\right)(y-3)}
\nonumber
\\
&\qquad +\frac{15\log\left(1-\frac{2M}{R}\right)}{8\left(\frac{M}{R}\right)^{5}}\,. 
\end{align}
%%%%%%%%%%%%%%%%%%%%%%%%%%%%%%%%%%%%%%%%%%%%%%%%%%%%%%%%%%%%%%%%%%%%%%%%%%%%%%
where now $y\equiv R \bar h_0'(R)/\bar h_0(R)$. In the Newtonian regime the axial LN vanishes, being associated with axial quantities (e.g., currents) that do not produce a gravitational field in Newton's theory.

\subsubsection{Dependence on the equation of state}

Electric and magnetic tidal LNs with different multipolar index $\ell$ were first computed for various polytropic EoS in \citet{Binnington:2009bb,Damour:2009vw} and for incompressible fluids and some tabulated, nuclear-physics based EoS in \citet{Damour:2009vw}.
A more comprehensive analysis --~including more realistic EoS and strange quark matter~-- subsequently appeared in \citet{Hinderer:2010ih} for the most phenomenologically relevant $\ell=2$ electric LN. Such an analysis was extended to $\ell>2$ and to magnetic LNs in \citet{Yagi:2013sva}.

\begin{figure}[htbp]
\begin{center}
\includegraphics[width=0.45\textwidth,clip=true]{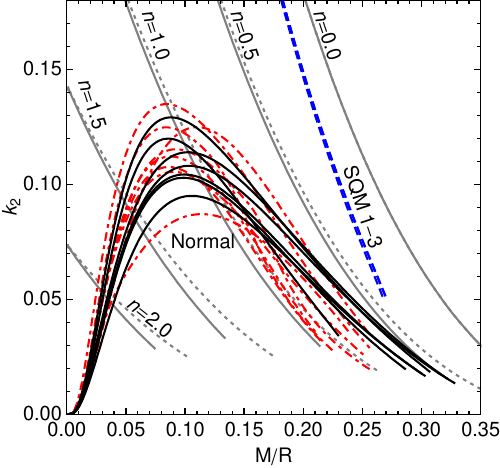}  
\includegraphics[width=0.44\textwidth,clip=true]{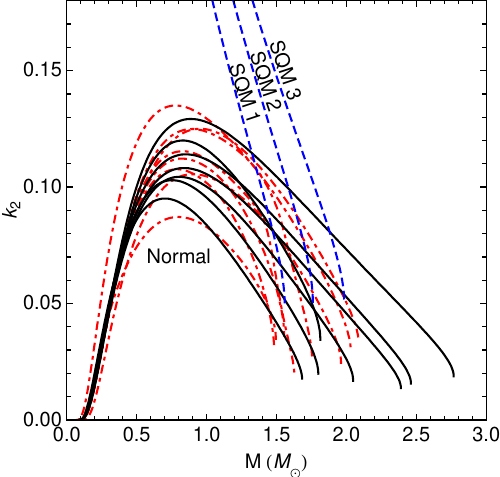}  
\\
\includegraphics[width=0.45\textwidth,clip=true]{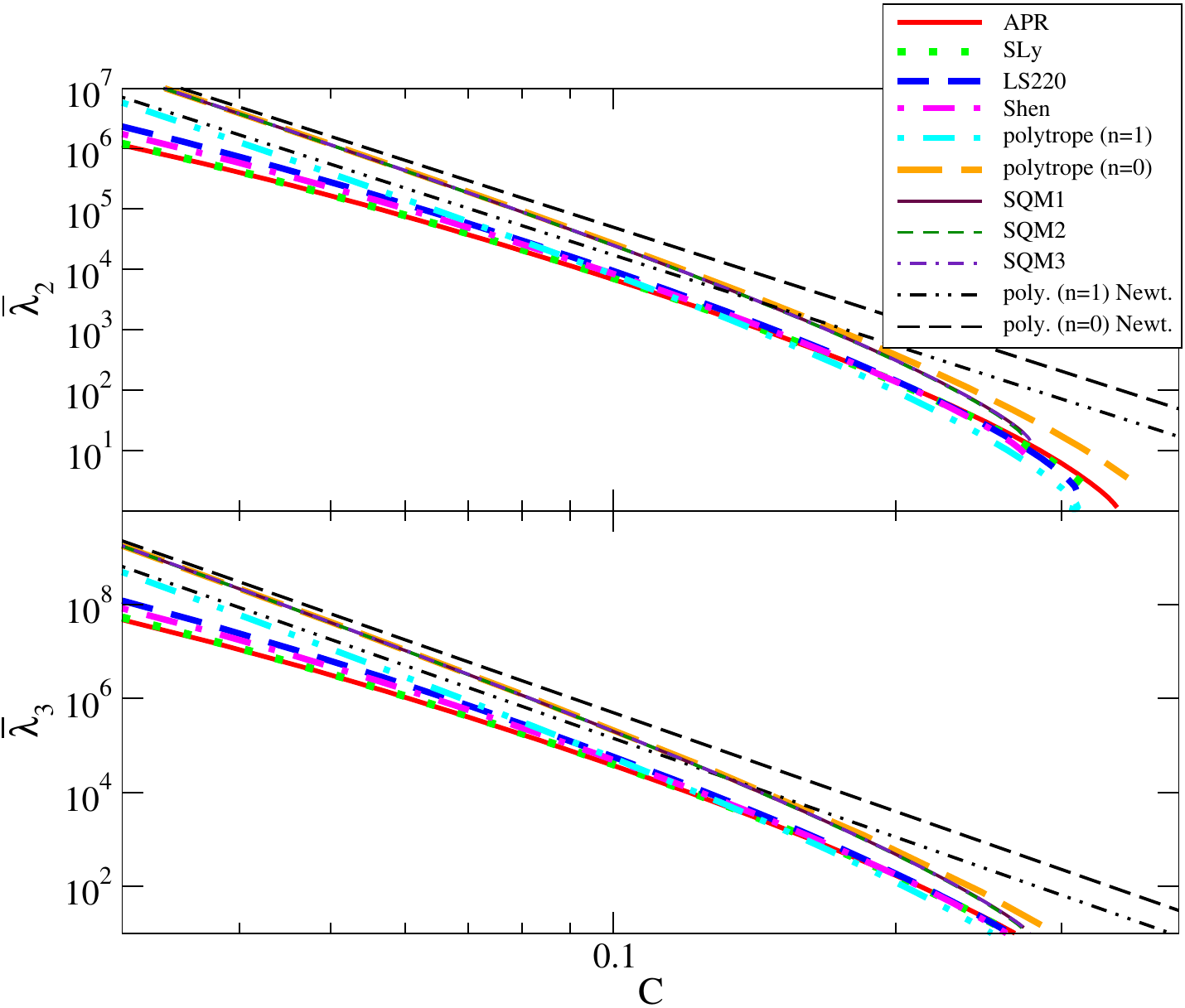}  
\includegraphics[width=0.51\textwidth,clip=true]{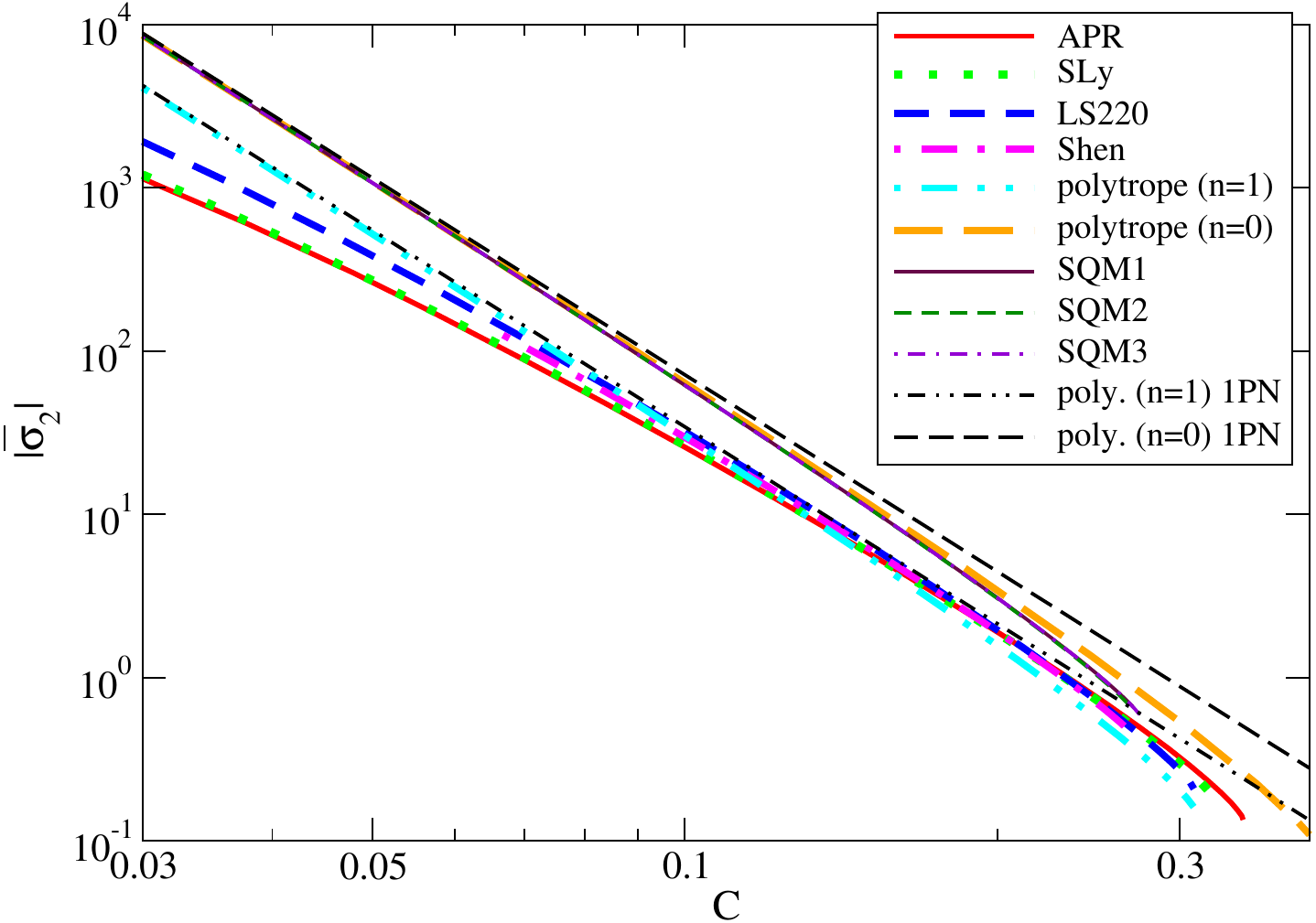}  
\caption{ 
Top panels (from \citet{Hinderer:2010ih}): Quadrupolar electric LN as a function of the
stellar compactness (left) and mass (right). Gray dotted curves correspond to polytropic
EoS ($p=K\rho^{1+1/n}$), while gray solid curves denote rest-mass density polytropes
($p=K\rho_{\rm rm}^{1+1/n}$), with $\rho$ and $\rho_{\rm rm}$ the energy and rest-mass
densities, respectively. The two polytropic descriptions coincide for $n=0$. Solid black
curves refer to realistic EoS with $npe\mu$ matter only, whereas dot-dashed curves include
$\pi$/hyperon/quark matter. EoS with strange quark matter are shown as dashed curves and
appear only in the right panel, differing by an overall scale factor.
Bottom panels (from \citet{Yagi:2013sva}): electric quadrupolar ($\bar{\lambda}_2$) and
octupolar ($\bar{\lambda}_3$) LNs (left) and magnetic quadrupolar ($\bar{\sigma}_2$) LN
(right) as functions of the compactness for various EoS.
See \citet{Hinderer:2010ih,Yagi:2013sva} for details of the adopted EoS and
Sec.~\ref{conventionLN} for the LN conventions.
\label{fig:LNvsC}
}
\end{center}
\end{figure}

In~\ref{fig:LNvsC} we present a selection of these results, based on \citep{Hinderer:2010ih,Yagi:2013sva}. We refer to~\ref{conventionLN} for the different conventions adopted for the LNs.
As a rule of thumb, the dimensionless LNs $\bar\lambda_\ell$ and $\bar\sigma\ell$ are monotonically decreasing functions of the compactness, consistently with the fact that they vanish in the BH-limit, $C\to1/2$, which is however not reached by ordinary NSs.
The LNs $k_\ell$ also decrease at large compactness but, due to the different normalization (see~\ref{conventionLN}) it approaches zero also in the Newtonian limit, $C\to0$.

\subsubsection{Approximate EoS-independent relations} \label{sec:quasiuniversal}

Despite its strong dependence on the NS compactness and EoS (see~\ref{fig:LNvsC}), the LNs (normalized by suitable powers of the NS mass) enjoy a series of approximately EoS-independent relations.
The most known ones are the so-called ``I-Love-Q'' relations among the moment of inertia $I$, the spin-induced quadrupole moment $Q$, and the electric, quadrupolar LN of a NS \citep{Yagi:2013bca,Yagi:2013awa, Katagiri:2025qze}.

\begin{figure}[htb]
\begin{center}
\includegraphics[width=0.48\textwidth,clip=true]{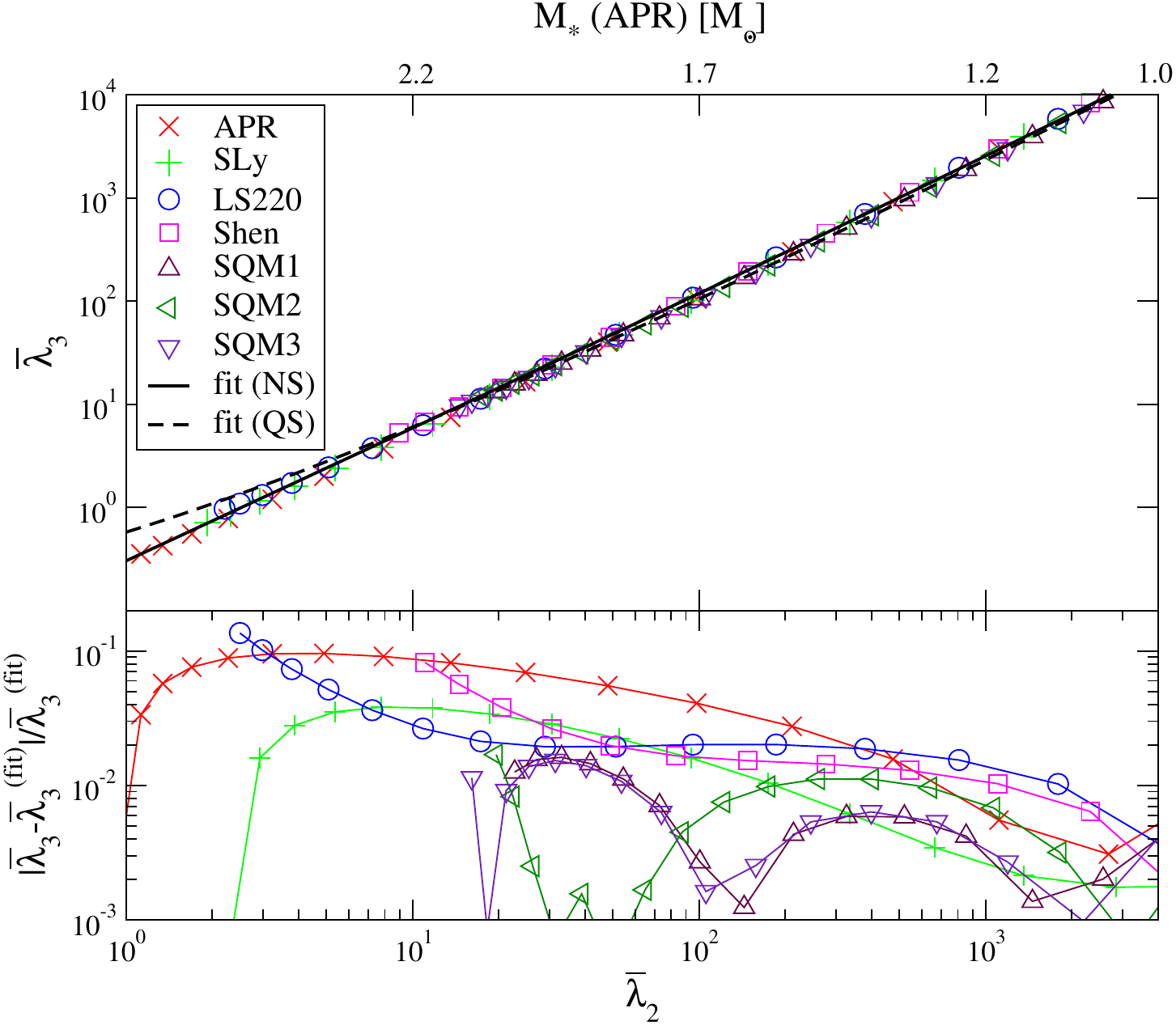} 
\includegraphics[width=0.48\textwidth,clip=true]{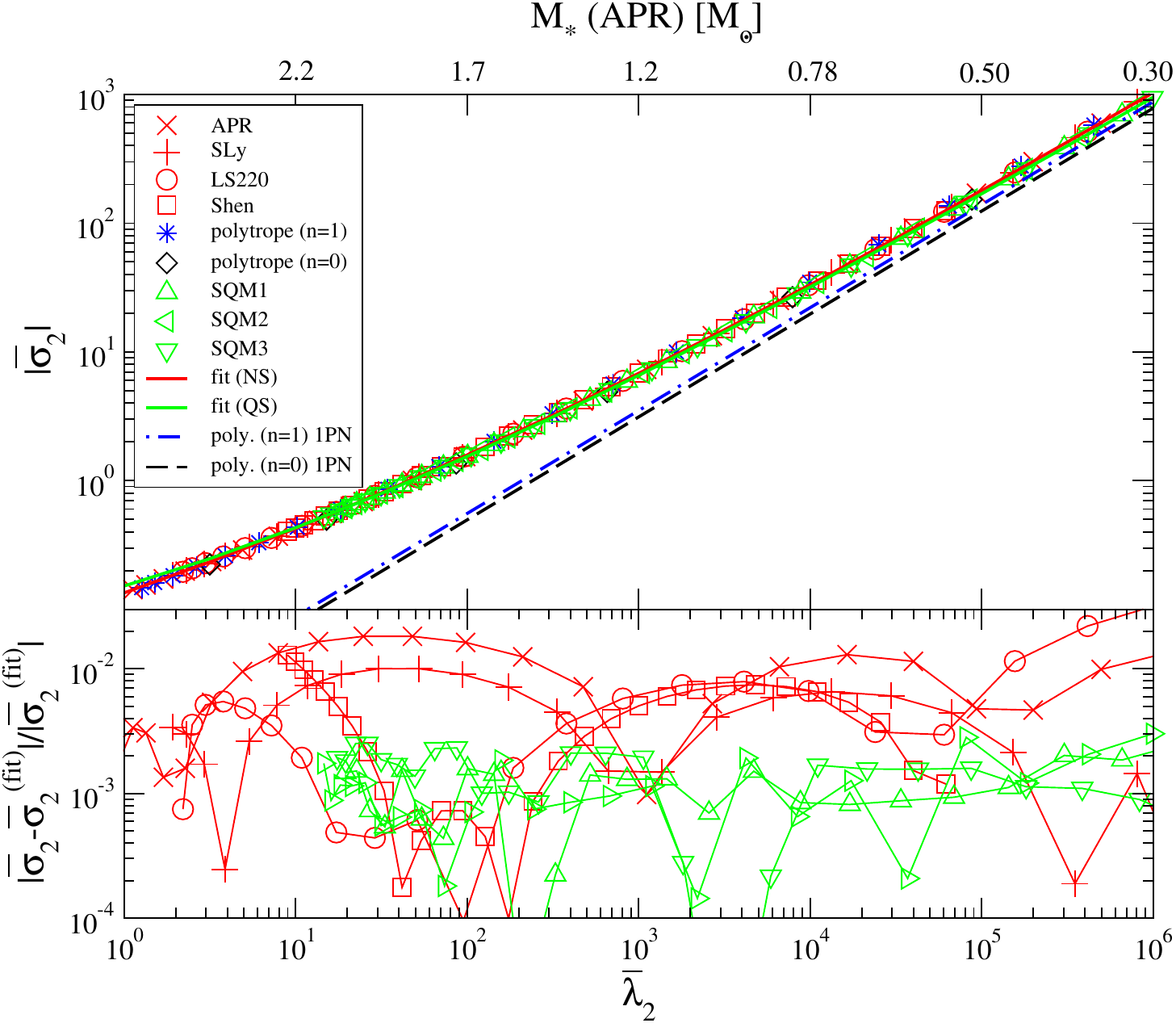} 
\caption{
Approximately EoS-independent relation between the $\ell=2$ and $\ell=3$ dimensionless electric (left panel) and between the $\ell=2$ electric and magnetic (right panel) tidal deformabilities for NSs with various EoS. The single parameter along the curves is the mass or compactness. As a reference, the top axes show the NS mass for the APR EoS.
The bottom panels show the fractional difference of each curve to the fitting formulas, which is polynomial in a log-log scale.
From \citet{Yagi:2013sva}.
\label{fig:universality} 
}
\end{center}
\end{figure}

In addition, there exist approximately EoS-independent relations among the LNs of different parity and different harmonic index, which are typically polynomial in a log-log scale \citep{Yagi:2013sva}.
As example of such relations is given in~\ref{fig:universality}.

For a detailed analysis of various approximately EoS-independent relations in NSs, including those involving the LNs, we refer the reader to the review \citep{Yagi:2016bkt}.

%%%%%%%%%%%%%%%%%%%%%%%%%%%%%%%%%%%%%%%%%%%%%%%%%%%
\subsubsection{Quadratic Love numbers}
%%%%%%%%%%%%%%%%%%%%%%%%%%%%%%%%%%%%%%%%%%%%%%%%%%%
Although originally formulated within linear perturbation theory, recent work has extended the analysis of tidal deformations of relativistic NSs into the \emph{nonlinear} regime, leading to the definition and computation of \emph{quadratic LNs} \citep{Poisson:2020vap,Pitre:2023xsr,Pani:2025qxs}.

This line of investigation was initiated in \citet{Pitre:2023xsr} which, based on the formalism developed in~\citet{Poisson:2020vap}, computed the nonlinear tidal deformation induced by an external quadrupolar tidal field up to quadratic order for NSs described by a polytropic EoS. These results were subsequently extended in \citet{Pani:2025qxs} to realistic EoS and matched to a worldline EFT. The general strategy consists in expanding all perturbations in spherical harmonics and projecting the field equations onto definite multipoles. At quadratic order, the source terms involve integrals over products of three spherical harmonics and their derivatives, which encode the underlying $SO(3)$ selection rules and can be evaluated through integrations by parts (see \citealt{Pani:2025qxs,Iteanu:2024dvx,Pani:2013pma}).

More explicitly, let ${}^{(1)}\!X_1^{\ell_1 m_1}$ and ${}^{(1)}\!X_2^{\ell_2 m_2}$ denote two generic linear perturbations, solutions of the first-order equations with harmonic indices $(\ell_1,m_1)$ and $(\ell_2,m_2)$, respectively. The quadratic corrections to the polar perturbations are governed by an equation of the form~\ref{metric_NS_polar}, supplemented by a source term that schematically reads
\begin{equation}
    S^{\ell m}(r)=\sum_{\ell_1,\ell_2,m_1,m_2}
    I^{m m_1 m_2}_{\ell \ell_1 \ell_2}\,
    S_{\ell \ell_1 \ell_2}
    \!\left[{}^{(1)}\!X_1^{\ell_1 m_1}(r),{}^{(1)}\!X_2^{\ell_2 m_2}(r)\right],
\end{equation}
where
$I^{m m_1 m_2}_{\ell \ell_1 \ell_2}\equiv\int d^2\Omega\,Y^*_{\ell m}Y_{\ell_1 m_1}Y_{\ell_2 m_2}$,
and the explicit form of $S_{\ell \ell_1 \ell_2}$ must be computed for each specific choice of $(\ell,\ell_1,\ell_2)$ \citep{Pani:2025qxs}. The above sum is restricted by angular-momentum selection rules, namely $m=m_1+m_2$, $|\ell_1-\ell_2|\le\ell\le\ell_1+\ell_2$, and $\ell\ge|m|$, with $I^{m m_1 m_2}_{\ell \ell_1 \ell_2}$ vanishing unless $\ell_1+\ell_2+\ell$ is even.

As a consequence of these selection rules, an $\ell=2$ electric tidal field induces quadratic corrections not only to the mass quadrupole ($\ell=2$), but also to the mass monopole ($\ell=0$) and to the hexadecapole ($\ell=4$). Moreover, the degeneracy in the azimuthal number $m$ is broken at quadratic order.

Focusing on the quadrupolar corrections sourced by an $\ell=2$ tidal field, the $tt$ component of the metric perturbation at leading order in $M/R$ reads \citep{Pani:2025qxs}
\begin{align}
	\delta g^{(\ell=2,m)}_{tt}(r)=&
	\mathscr{E}_{2,m}\left(r^2+12\frac{\lambda_2}{r^3}\right)\nonumber\\
	&-\sum_{m_1 m_2}I^{m m_1 m_2}_{2 2 2}\,
	\mathscr{E}_{2,m_1}\mathscr{E}_{2,m_2}
	\left[
	\frac{1}{2}r^4
	+\left(
	12\frac{\lambda_2}{r}
	+\frac{63}{2}\frac{\lambda_{222}}{r^3}
	+72\frac{\lambda_2^2}{r^6}
	\right)
	\right],
\label{eq:deltagttEFT}
\end{align}
where $\mathscr{E}_{2,m}$ denotes the amplitude of the linear tidal field. The tidal deformability coefficients are related to their corresponding dimensionless LNs through
\begin{equation}
k_2^{\rm E}\equiv\frac{6\lambda_2}{R^5},
\qquad
p_2\equiv\frac{9\lambda_{222}M}{R^8},
\label{dimscale}
\end{equation}
with $k_2^{\rm E}$ the standard (linear) electric LN and $p_2$ the dimensionless version \citep{Poisson:2020vap,Pitre:2025qdf} of the quadratic tidal deformability $\lambda_{222}$ introduced in \citet{Pani:2025qxs}. Qualitatively, we have the following relation between the quadrupole moment and the tidal field: $Q^{\rm linear}_{ij}\propto -\lambda_{1}E_{ij}$ and $Q_{ij}^{\rm quad}\propto-\lambda_{222}E_{ik}E^{k}_{j}$.

\begin{figure}[t] 
\begin{center} \includegraphics[width=0.437\textwidth]{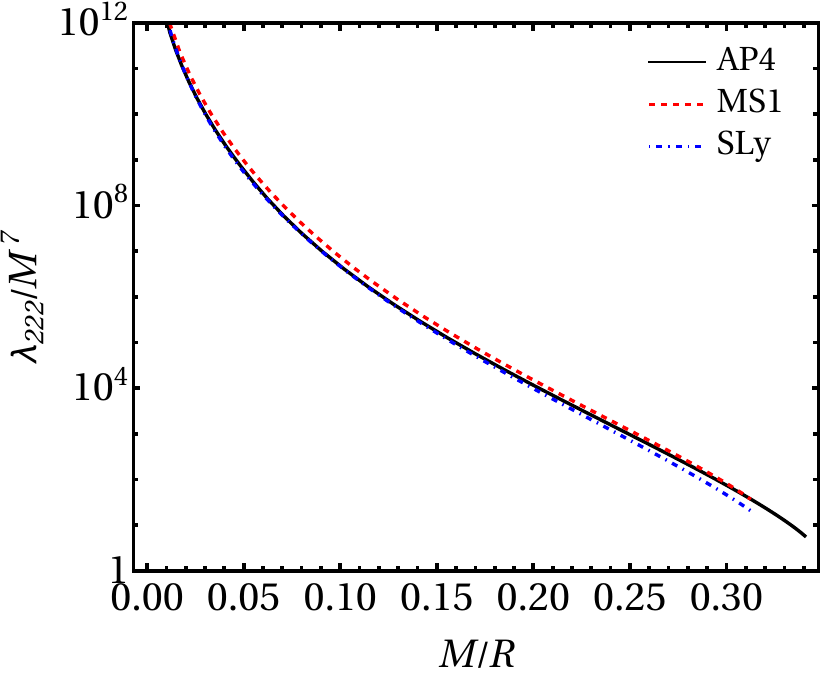} \hspace{1cm} \includegraphics[width=0.4275\textwidth]{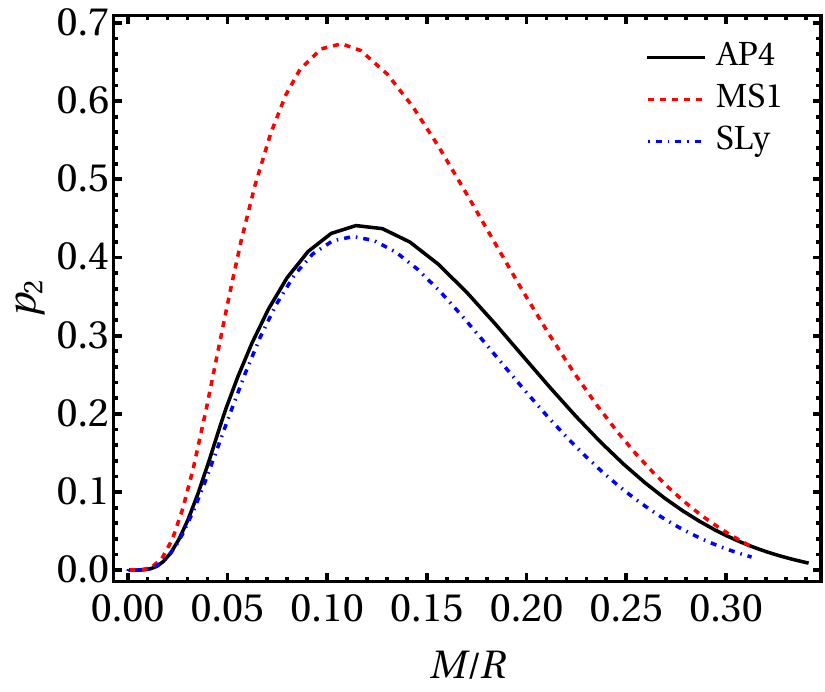} 
\end{center} 
%%%%
\caption{Left: Quadratic LN normalized as $\lambda_{222}/M^7$ as a function of the compactness, for different EoS. 
Right: Dimensionless coefficient $p_2$ capturing the leading quadratic tidal deformation as a function of compactness, for different EoS. 
Adapted from \citet{Pani:2025qxs}. 
} \label{fig:p2-222} 
\end{figure}

The quadratic tidal deformabilities $\lambda_{222}$ and $p_2$ for representative EoS are shown in~\ref{fig:p2-222}.
Note that $p_2$, like the linear LN $k_2^{\rm E}$ is of ${\cal O}(0.1)$.
As discussed in~\ref{sec:pheno_NS}, the GW phase depends on coefficients proportional to the dimensionless combination $\lambda_2/M^5$ (for the usual tidal phase) and $\lambda_{222}/M^7$ (for the quadratic correction) \citep{Pani:2025qxs}. Hence,~\ref{dimscale} implies that the quadratic correction in the waveform scales as $p_2/C^8$, and is therefore more strongly enhanced in the small-compactness limit than the linear tidal term in the GW phase, scaling as $k_2^{\rm E}/C^5$. This enhancement makes quadratic tidal effects potentially relevant despite entering the GW phase at 8PN order, i.e.\ three PN orders beyond the linear tidal contribution (see~\ref{sec:pheno_NS}). Indeed, \citet{Pani:2025qxs} estimated that quadratic LNs can contribute at the level of $\sim10\%$ of the linear tidal effects during the late inspiral.

%%%%%%%%%%%%%%%%%%%%%%%%%%%%%%%%%%%%%%
\subsubsection{Static Love numbers of neutron stars beyond General Relativity}\label{sec:NSBGR}
%%%%%%%%%%%%%%%%%%%%%%%%%%%%%%%%%%%%%%
The LNs of NSs have also been extensively investigated in the context of modified theories of gravity.
In particular, the LNs of NSs were computed in scalar--tensor theories exhibiting spontaneous scalarization \citep{Pani:2014jra}, in which NS solutions can differ significantly from their GR counterparts \citep{Damour:1993hw}.
Remarkably, it was shown that the approximate EoS-independent (or quasi-universal) relations in this class of scalar-tensor theories agree with the GR ones at the level of a few percent, even for spontaneously scalarized configurations with the largest couplings still compatible with current binary-pulsar constraints.

After performing the standard spherical-harmonic decomposition on a spherically-symmetric background, the equations governing tidal perturbations in scalar-tensor theories take the schematic form \citep{Pani:2014jra}
%%%%
\begin{align}
    H_0'' + c_1 H_0' + c_0 H_0 &= c_s \Phi \,, \\
    \Phi'' + d_1 \Phi' + d_0 \Phi &= d_s H_0 \,,
\end{align}
%%%%
where $\Phi$ denotes the radial part of the scalar quadrupolar perturbation, $H_0$ is the radial part of the quadrupolar perturbation of $g_{tt}$, and the coefficients $c_i$ and $d_i$ are radial functions depending on the background metric and scalar field.
In particular, the source terms $c_s$ and $d_s$ vanish when the background scalar field is zero, i.e., for unscalarized NSs.

Due to the coupling between metric and scalar perturbations, an external tidal field generically induces not only a gravitational quadrupole moment, but also a \emph{scalar} quadrupole moment.
Likewise, an external quadrupolar scalar field can induce a mass quadrupole moment on the metric.
This leads to the appearance of novel classes of \emph{scalar LNs}, which have been studied in detail in \citet{Creci:2023cfx,Creci:2024wfu}.

The tidal deformability of NSs has also been explored in theories with quadratic curvature corrections coupled to a scalar field, most notably in Chern--Simons \citep{Gupta:2017vsl,Yagi:2013bca,Yagi:2013awa}, Gauss--Bonnet \citep{Saffer:2021gak} and scalar-tensor \citep{Diedrichs:2025vhv} gravity.
In the former case, the polar LNs remain unaffected by the pseudo-scalar coupling to the Chern-Simons term, whereas in the latter both polar and axial sectors are generically modified.

 \citet{Sham:2013cya} investigated the tidal deformability of NSs in Eddington-inspired Born--Infeld gravity \citep{Banados:2010ix}, exploiting the fact that, when coupled to a perfect fluid, the theory can be mapped to GR with an effective, albeit contrived, EoS \citep{Delsate:2012ky}.

More recently, \citet{Ajith:2022uaw, Vylet:2023pkp} studied the tidal deformability in a theory with Lorentz-symmetry breaking in the gravitational sector, showing that the LN is unaffected in a certain limit of the theory.

Overall, in all these theories one finds the existence of approximately EoS-independent relations for different values of the coupling parameters.
Deviations from the GR quasi-universal relations therefore provide, in principle, a powerful means to discriminate between GR and alternative theories of gravity using tidal observables.

%%%%%%%%%%%%%%%%%%%%%%%%%%%%%%%%%%%%%%
\subsection{Static Love numbers of slowly-spinning neutron stars} \label{sec:RTLNs}
%%%%%%%%%%%%%%%%%%%%%%%%%%%%%%%%%%%%%%

The tidal response of a spinning NS has been investigated within a perturbative expansion in the angular momentum. This is also motivated by the fact that the expansion parameter, namely the dimensionless angular momentum $\chi\equiv J/M^2$, is typically small even for the fastest spinning NSs known to date\footnote{As a reference, for one of the fastest spinning NS observed
so far, namely the most massive companion of the double pulsar system PSR J0737-3039A \citep{Burgay:2003jj}, the spin period is $\approx 23\,{\rm ms}$, which corresponds to $\chi\approx (0.02,0.05)$,
depending on the EoS.}.

The coupling between the external tidal field and the spin of the object gives rise to several novel features: in particular, it induces couplings between electric and magnetic distortions and leads to the emergence of new classes of spin-induced tidal LNs, commonly referred to as \emph{rotational} LNs.

As a result of the external perturbation, the mass and current multipole moments \citep{Geroch:1970cd,Hansen:1974zz,Thorne:1980ru} of the compact object are deformed. 
In linear perturbation theory, if the object is non-rotating and hence spherically symmetric, the azimuthal number $m$ is degenerate and both parity and the angular momentum number $\ell$ are conserved. In this case, an electric (i.e.\ even-parity) tidal field with harmonic index $\ell$ can only induce a mass multipole moment of the same order $\ell$, whereas a magnetic (i.e.\ odd-parity) tidal field with harmonic index $\ell$ can only induce a current multipole moment of order $\ell$.
When the central object is spinning, this degeneracy is lifted and selection rules allow for a more general tidal response, which can be systematically characterized in terms of an extended set of tidal LNs \citep{Poisson:2014gka,Pani:2015nua}. In particular:
\begin{enumerate}
 \item At first order in the spin, a tidal field with multipolar index $\ell$ generically sources a response with multipolar index $\ell\pm1$ and opposite parity. In addition, for non-axisymmetric perturbations ($m\neq0$), a response with the same $\ell$ and the same parity is also allowed.
 \item At second order in the spin, there are spin-induced corrections to the standard (static) LNs for any $m$, as well as genuinely new rotational LNs associated with multipolar indices $\ell\pm2$.
\end{enumerate}

Within this perturbative framework, an analytical study of tidal deformations of a spinning BH was initiated in \citet{Poisson:2014gka}, where the response of a Kerr BH immersed in a generic quadrupolar tidal field was computed to first order in the spin. These results were extended in \citet{Pani:2015hfa} to include second-order spin corrections for an axisymmetric quadrupolar electric tidal field, as well as the response to both electric and magnetic tidal fields with $\ell=3,4$ at first order in the spin.

These analyses provided the first explicit evidence that the rotational tidal LNs of a Kerr BH vanish identically up to quadratic order in the spin \citet{Landry:2015zfa,Pani:2015hfa,Pani:2015nua}, and led to the conjecture that this result holds to \emph{all} orders in the spin. This conjecture was later confirmed by a general proof valid for arbitrary static bosonic perturbations of a Kerr BH \citep{Chia:2020yla,LeTiec:2020bos,LeTiec:2020spy} (see~\ref{stat_Love_Kerr}).

\begin{figure}[t]
\begin{center}
\includegraphics[width=0.65\textwidth]{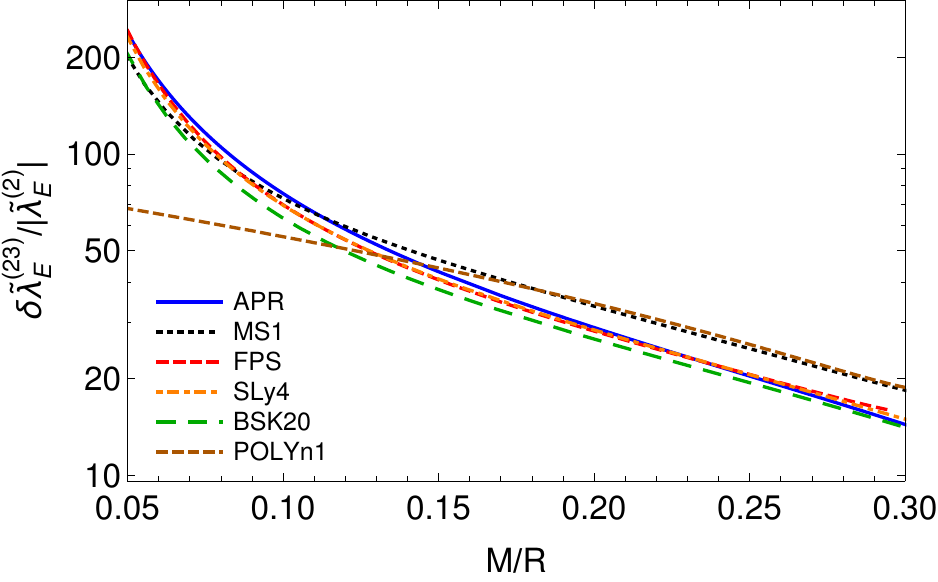}
\caption{Ratio between the (dimensionless) rotational tidal LN $\delta\tilde\lambda_E^{(23)}$ and the standard electric quadrupolar tidal LN $\tilde\lambda_E^{(2)}$ for the various EoS. From \citet{Pani:2015nua}, to which we refer for the definitions.}
\label{fig:RTLNs}
\end{center}
\end{figure}

In the case of NSs, the exterior metric of a spinning material object immersed in a generic quadrupolar (both electric and magnetic) tidal field was computed to first order in the spin in \citet{Landry:2015zfa}. Subsequently, \citet{Pani:2015nua} focused on axisymmetric tidal fields with $\ell=2,3,4$ and computed the corresponding rotational LNs to first order in the spin for a variety of tabulated EoS. In~\ref{fig:RTLNs} we show
the rotational LN associated with a mass quadrupolar deformation induced by an external octupolar magnetic tidal field.

%%%%%%%%
%%%%%%%%%

The implications of these results for GW measurements of the tidal deformability of NSs will be discussed in~\ref{sec:pheno_NS}. Finally, we note that spin-tidal couplings tend to spoil the approximate EoS-independent relations satisfied by NSs in the static limit (see~\ref{sec:quasiuniversal}), although their impact is expected to be modest for the relatively small spins typical of astrophysical NSs.

%%%%%%%%%%%%%%%%%%%%%%%%%%%%%%%%%%%%%%%%%%%%%%%%%%%%%%%%%%%%%%%%
%%%%%%%%%%%%%%%%%%%%%%%%%%%%%%%%%%%%%%%%%%%%%%%%%%%%%%%%%%%%%%%%
%%%%%%%%%%%%%%%%%%%%%%%%%%%%%%%%%%%%%%%%%%%%%%%%%%%%%%%%%%%%%%%%
\subsection{Dynamical tides: Love numbers, dissipation, and resonances}\label{dyntideNS}

Here we review dynamical tidal perturbations of NSs and the derivation of the associated observables, such as dynamical LNs, dissipation coefficients, and the presence of resonances. We begin with an overview of the Newtonian regime, then introduce the relevant perturbation equations in the NS interior in full GR.

%%%%%%%%%%%%%%%%%%%%%%%%%%%%%%%%%%%%%%%
\subsubsection{Newtonian dynamical tides}\label{sec:Newtonian_dynamical_tides}
%%%%%%%%%%%%%%%%%%%%%%%%%%%%%%%%%%%%%%%
Although Newtonian gravity cannot capture strong-field effects quantitatively, many results on the dynamical tidal response of NSs have been obtained in the Newtonian approximation \citep{Ho:1998hq, Andersson:2019ahb, Andersson:2019dwg, Passamonti:2020fur, Passamonti:2022yqp, Pnigouras:2022zpx, Yu:2024uxt}.
These
results provide essential physical insight into the nature of dynamical tides. They clarify
the role of stellar oscillation modes, motivate simple effective descriptions of the
frequency-dependent tidal response, and serve as a conceptual bridge to relativistic
treatments of dynamical LNs discussed in later sections.

When the orbital frequency becomes
comparable to the characteristic oscillation frequencies of the star, the fluid response
lags behind the applied tidal field and genuinely \emph{dynamical} tidal effects arise.
Within Newtonian gravity, this problem admits a particularly transparent formulation~\citep{Lai:1993di,Ho:1998hq,Lai:1997wh},
which has been revisited and systematized in a series of recent works
\citep{Chakrabarti:2013xza,Andersson:2019ahb,Andersson:2019dwg,Passamonti:2020fur,Passamonti:2022yqp,Pnigouras:2022zpx}.

A complementary, physically intuitive route is to model the star as a dynamical ellipsoid (affine approach) and embed the resulting finite-size dynamics in a PN description of the binary, see \citep{Ferrari:2011as}.

In the Newtonian framework, the response of a non-rotating, self-gravitating fluid star
to a time-dependent tidal potential can be expressed as a superposition of the star's
normal modes of oscillation \citep{Andersson:2019ahb}, which form a complete basis~\citep{Kokkotas:1999bd}. The Lagrangian displacement of the fluid is expanded over
a complete set of eigenmodes, each characterized by an eigenfrequency and a tidal overlap
integral that quantifies how efficiently the mode couples to the external quadrupolar field.
This leads to a mode-sum representation of the tidal response, which is written as a frequency-dependent quantity obtained by summing
over all relevant modes,
%%%
\begin{equation}
    {\cal F}_{\ell m}(\omega) \sim \sum_j \frac{Q_j}{\omega_j^2-\omega^2} \label{eq:summodes}
\end{equation}
%%%
where $\omega_j$ are the \emph{normal} modes of a perfect-fluid star \citep{Kokkotas:1999bd}, and $Q_j$ are related to the overlap integral between the
external tidal field and the fluid oscillation modes of the star.
In the low-frequency limit, this construction reduces smoothly
to the familiar static (adiabatic) tidal deformability, while at higher frequencies it
captures resonant and near-resonant excitations of the stellar fluid (see~\ref{sec:resonances} below). We also note that ensuring a finite response at resonance requires including an imaginary part in the mode frequencies. These imaginary parts are generically nonzero due to the dissipative boundary conditions imposed in the underlying eigenvalue problem. See \citet{Flanagan:2006sb} for a treatment of the mode-orbit coupling in NS binaries that consistently incorporates the imaginary part of the $f$-mode frequency.

A robust outcome of these Newtonian analyses is the dominant role played by the
fundamental fluid mode (the $f$-mode) in shaping the dynamical tidal response
\citep{Andersson:2019ahb}. Owing to its global nature and relatively low frequency, the
$f$-mode has the largest tidal coupling and provides the leading correction to the static
LN already well before an actual resonance is reached. Higher-frequency
pressure ($p$-) modes contribute negligibly, while gravity ($g$-) modes, although potentially numerous,
are typically weakly coupled to the quadrupolar tide in barotropic Newtonian stars.
As a result, the frequency dependence of the tidal deformability can be accurately
captured by an effective model in which the static LN is supplemented by a
single dominant $f$-mode contribution.

Building on this physical picture, \citet{Andersson:2019dwg} introduced a simple and
useful phenomenological description of Newtonian dynamical tides. In this approach, the frequency-dependent tidal response is
encoded in an \emph{effective} LN that depends explicitly on the frequency of the tidal source (e.g., the orbital frequency of a binary system) and
the $f$-mode parameters, namely its eigenfrequency and overlap coefficient. 

Newtonian studies have also explored how additional microphysics modifies the dynamical
tide. The inclusion of an elastic crust introduces new classes of oscillation modes,
including interface and shear modes, which can be excited by the tidal field
\citep{Passamonti:2020fur, Pnigouras:2025muo}. While these modes do not typically dominate the global tidal
response, they can lead to significant local stresses in the crust and may, in principle,
trigger crust failure prior to merger during a NS coalescence. Similarly, extensions to two-fluid Newtonian models,
incorporating superfluid neutrons in the core and inner crust, show that entrainment and
superfluid degrees of freedom alter the mode spectrum and tidal couplings
\citep{Passamonti:2022yqp}. Nonetheless, even in these more realistic settings, the overall
dynamical tidal response during the inspiral remains largely governed by the ordinary
$f$-mode.

Rotation introduces further qualitative features already at the Newtonian level.
For spinning stars, the tidal forcing couples to both prograde and retrograde modes, and
the notion of mode orthogonality must be generalized. A consistent treatment of
dynamical tides in rotating Newtonian stars was developed in
\citet{Pnigouras:2022zpx}, where it was shown that rotational effects enter the dynamical
tidal response at linear order in the spin. This is in sharp contrast with static tides, for
which rotational corrections appear only at quadratic order. The resulting frequency
dependence of the effective LN reflects the splitting of the mode spectrum and
the modified resonance conditions in the rotating frame. See \citet{Gupta:2020lnv, Steinhoff:2021dsn} for relativistic generalizations.

%%%%%%%%%%%%%%%%%%%%%%%%%%%%%%%%
%%%%%%%%%%%%%%%%%%%%%%%%%%%%%%%%
%%%%%%%%%%%%%%%%%%%%%%%%%%%%%%%%
\subsubsection{Relativistic dynamical tides} \label{sec:NSdynRel}

While Newtonian models provide valuable physical intuition, a quantitatively accurate
description of tidal interactions in compact binaries ultimately requires a relativistic
treatment. In GR, tidal deformability is encoded in relativistic LN, defined through the asymptotic response of the spacetime metric to an
external tidal field, as in the BH case (see~\ref{sec:dynBHs}). As usual, we consider the axial and polar sectors separately.

\paragraph{Axial sector}~---
As in the exterior Schwarzschild spacetime, axial gravitational perturbations can be described in terms of a single master function $\bpsi_{\rm RW}$, which satisfies a Regge--Wheeler-like equation also inside the star \citep{Kojima:1992ie}:
%%%%%%%%%%%%%%%%%%%%%%%%%%%%%%%%%%%%%%%%%%%%%%%%%%%%%%%%%%%%%%%%%%%%%%%%%%%%%%%%%
\begin{align}
e^{\frac{(\bar{\nu}-\bar{\lambda})}{2}}&\dfrac{d}{dr}\left[e^{\frac{(\bar{\nu}-\bar{\lambda})}{2}}\dfrac{d\bpsi_{\rm RW}}{dr}\right] 
+\Bigg[\omega^2-e^{\bar{\nu}}\Bigg\{\frac{\ell(\ell+1)}{r^2}-\frac{6\bar{m}(r)}{r^3}+4\pi\left(\brho-\bp\right)\Bigg\}\Bigg]\bpsi_{\rm RW}=0 \,,
\label{eqNS}
\end{align}
%%%%%%%%%%%%%%%%%%%%%%%%%%%%%%%%%%%%%%%%%%%%%%%%%%%%%%%%%%%%%%%%%%%%%%%%%%%%%%%%%
where the two nonvanishing axial metric perturbations $\bar{h}_{0}$ and $\bar{h}_{1}$ in the NS interior are related to $\bpsi_{\rm RW}$ through
%%%%%%%%%%%%%%%%%%%%%%%%%%%%%%%%%%%%%%%%%%%%%%%%%%%%%%%%%%%%%%%%%%%%%%%%%%%%%%%%%
\begin{equation}
\bar{h}_{0}=-\frac{e^{\frac{(\bar{\nu}-\bar{\lambda})}{2}}}{i\omega}\dfrac{d}{dr}(r\bpsi_{\rm RW}) \,,
\qquad
\bar{h}_{1}=e^{\frac{(\bar{\lambda}-\bar{\nu})}{2}}(r\bpsi_{\rm RW}) \,.
\end{equation}
%%%%%%%%%%%%%%%%%%%%%%%%%%%%%%%%%%%%%%%%%%%%%%%%%%%%%%%%%%%%%%%%%%%%%%%%%%%%%%%%%
For a given EoS, we first solve the TOV equations~\ref{TOV} to determine the background configuration and then substitute the resulting background quantities into~\ref{eqNS}. Equation~\ref{eqNS} is integrated by imposing regularity of $\bpsi_{\rm RW}$ at the stellar center and continuity with the exterior solution at the stellar surface. The resulting observables depend on the choice of EoS and on the central energy density.

A comparison with the axial master equations derived in \citet{Chakraborty:2024gcr} shows that~\ref{eqNS} corresponds to allowing for nonvanishing axial fluid perturbations, $\delta u^{\mu}_{\rm axial}\neq 0$. 
This assumption plays a crucial role in the static limit.

%%%%%%%%%%%%%%%%%%%%%%%%%%%%%%%%
\paragraph{Static versus zero-frequency limit in the axial sector}~--- 
The axial LNs of a NS were originally computed assuming \emph{strictly static} perturbations. In this case one sets $\omega=0$ directly in the perturbation equations, leading to~\ref{static_magnetic_NS}, in agreement with \citep{Binnington:2009bb}, and assumes $\delta u^{\mu}_{\rm axial}=0$.

Alternatively, one may derive the axial perturbation equations in the fully dynamical case and subsequently take the zero-frequency limit $\omega\to 0$. This procedure corresponds to taking the $\omega\to 0$ limit of~\ref{eqNS} and implicitly assumes $\delta u^{\mu}_{\rm axial}\neq 0$. 
Indeed, in the zero-frequency limit Einstein's equations imply \citep{Pani:2018inf}
\begin{equation}
    u^{\phi}_{\rm axial}\propto \bar h_0
\end{equation}
which does not vanish in the $\omega\to0$ limit.
This condition corresponds to modeling the NS interior as an \emph{irrotational} (rather than static) fluid \citep{Pani:2018inf,Landry:2015cva,Damour:2009vw}. 

Thus, interestingly the zero-frequency limit of the dynamical equations does not coincide with the strictly static case discussed above, as explicitly shown by comparing~\ref{eqNS} in the $\omega\to0$ limit with~\ref{static_magnetic_NSh0}. This mismatch can be traced back to the different assumptions on the axial fluid perturbations: $\delta u^{\mu}_{\rm axial}=0$ (i.e., static fluid) in \citet{Binnington:2009bb} versus $\delta u^{\mu}_{\rm axial}\neq 0$ (i.e., irrotational fluid) in \citet{Damour:2009vw}. 
One can therefore define two classes of axial LNs, the ``static'' and the ``irrotational'' ones.
As previously anticipated, for ordinary NSs they are similar in absolute values but have the opposite sign.

Since the irrotational case is obtained as the zero-frequency limit of the Regge--Wheeler equation, we expect it should describe more accurately relevant astrophysical configurations.

%%%%%%%%%%%%%%%%%%%%%%%%%%%%%%%%
%%%%%%%%%%%%%%%%%%%%%%%%%%%%%%%%
\paragraph{Polar sector}~---
In the polar sector, unlike the axial case, perturbations of a NS cannot be described by a single master function. Instead, one obtains a coupled system of four first-order differential equations, involving two metric perturbations and two fluid perturbations. In this context we introduce the rescaled metric perturbations $\hat{H}$, $\hat{K}$ and $\hat{H}_{1}$ \footnote{The full equations in the dynamical case were first given in \citet{Detweiler:1985zz} in terms of the rescaled variables:
\[
\bar{H}\to r^{\ell}\hat{H}\,,\qquad 
\bar{K}\to r^{\ell}\hat{K}\,,\qquad 
\bar{H}_{1}\to -i\omega r^{\ell+1}\hat{H}_{1}\,,
\]
and with the convention for the frequency opposite to the one adopted in this work, namely$\omega\to -\omega$. These rescalings make sure that these perturbations are well-behaved at the center of the NS, i.e., at $r=0$.}, among which, $\hat{H}_{1}$ and $\hat{K}$ are independent, while $\hat{H}$ can be determined algebraically from $\hat{H}_{1}$ and $\hat{K}$.

Polar fluid perturbations are described in terms of perturbations of the four-velocity, which depend on $\hat{H}$ and on two additional variables $\hat{W}(r)$ and $\hat{V}(r)$, such that
%%%%%%%%%%%%%%%%%%%%%%%%%%%%%%%%%%%%%%%%%%%%%%%%%%%%%%%%%%%%%%%%%%%%%%%%%%%%%%%%%
\begin{align}
\delta u^{r}&=\int d\omega \sum_{\ell m} i\omega r^{\ell-1}e^{-\frac{\nu+\lambda}{2}}\hat{W}(r)Y_{\ell m}e^{i\omega t}\,,\\
\delta u^{\theta}&=-\int d\omega \sum_{\ell m} i\omega r^{\ell-2}e^{-\frac{\nu}{2}}\hat{V}(r)\partial_{\theta}Y_{\ell m}e^{i\omega t}\,,\\
\delta u^{\phi}&=-\int d\omega \sum_{\ell m} i\omega \frac{r^{\ell-2}}{\sin^{2}\theta}e^{-\frac{\nu}{2}}\hat{V}(r)\partial_{\phi}Y_{\ell m}e^{i\omega t}\,.
\end{align}
%%%%%%%%%%%%%%%%%%%%%%%%%%%%%%%%%%%%%%%%%%%%%%%%%%%%%%%%%%%%%%%%%%%%%%%%%%%%%%%%%
Rather than working directly with $\hat{V}$, it is advantageous to introduce the fluid variable $\hat{X}$,
%%%%%%%%%%%%%%%%%%%%%%%%%%%%%%%%%%%%%%%%%%%%%%%%%%%%%%%%%%%%%%%%%%%%%%%%%%%%%%%%%
\begin{align}
\hat{X}=(\brho+\bp)c_{\rm s}^{2}\Bigg[&
\frac{1}{r}e^{\frac{\nu-\lambda}{2}}\hat{W}'+\frac{\ell+1}{r^{2}}e^{\frac{\nu-\lambda}{2}}\hat{W}
+\frac{\ell(\ell+1)}{r^{2}}e^{\frac{\nu}{2}}\hat{V}
-e^{\frac{\nu}{2}}\hat{K}
-\frac{1}{2}e^{\frac{\nu}{2}}\hat{H}
\Bigg]\,,
\end{align}
%%%%%%%%%%%%%%%%%%%%%%%%%%%%%%%%%%%%%%%%%%%%%%%%%%%%%%%%%%%%%%%%%%%%%%%%%%%%%%%%%
which can be inverted to express $\hat{V}$ in terms of $\hat{X}$. The independent perturbation variables are then $\hat{H}_{1}$, $\hat{K}$, $\hat{W}$, and $\hat{X}$ (see, e.g., \citealt{Detweiler:1985zz,Mondal:2023wwo}).

Regularity at the center requires that each perturbation variable admits a Taylor expansion of the form $\hat{Y}(r)=\sum_i \hat{Y}_i r^i$, where $\hat{Y}=(\hat{H}_{1},\hat{K},\hat{W},\hat{X})$. Solving the perturbation equations order by order near the center yields relations among the expansion coefficients, such as
%%%%%%%%%%%%%%%%%%%%%%%%%%%%%%%%%%%%%%%%%%%%%%%%%%%%%%%%%%%%%%%%%%%%%%%%%%%%%%%%%
\begin{align}
\hat H_1(0)&=\frac{2 \ell \hat K(0) +16 \pi (\hat \rho_0 + \hat p_0) \hat W(0)}{\ell (\ell+1)} \,,\\
\hat X(0)&=(\hat \rho_0 + \hat p_0) e^{\nu_0/2}
\Bigg[ \left( \frac{4 \pi}{3} (\hat \rho_0 + 3 \hat p_0) - \frac{\omega^2}{\ell} e^{-\nu_0}\right)\hat W(0)
+ \frac{1}{2} \hat K(0) \Bigg] \,.
\end{align}
%%%%%%%%%%%%%%%%%%%%%%%%%%%%%%%%%%%%%%%%%%%%%%%%%%%%%%%%%%%%%%%%%%%%%%%%%%%%%%%%%
The remaining free parameters are fixed by requiring the vanishing of the Lagrangian perturbation of the pressure at the stellar surface, which translates into the condition $\hat{X}(R)=0$ \citep{Detweiler:1985zz,Mondal:2023wwo}. In practice, this is achieved by performing two independent integrations with different initial conditions and linearly combining the resulting solutions so as to enforce $\hat{X}(R)=0$.

In the polar case, the fluid four-velocity vanishes as $\omega\to0$ so, unlike the axial case, the zero-frequency limit of the dynamical equations corresponds to strictly static fluid. Thus, only a class of polar LNs exists.

Recently, the linear tidal response of a NS in GR was derived from a covariant fluid effective action and matched to the quadrupolar worldline EFT, yielding a mode-sum response function and analytic dynamical tidal deformabilities in terms of mode frequencies and overlaps \citep{Martinez-Rodriguez:2026omk}.

%%%%%%%%%%%%%%%%%%%%%%%%%%%%%%%%%%%%%%%
\subsubsection{Relativistic perturbative scheme}
%%%%%%%%%%%%%%%%%%%%%%%%%%%%%%%%%%%%%%%
The dynamical tidal response of a static NS in the relativistic regime has been investigated in \citet{Pitre:2023xsr,HegadeKR:2024agt,HegadeKR:2025qwj,HegadeKR:2026iou, Counsell:2024pua} using a perturbative approach, and very recently in~\citet{Jarequi:2026cyp} using EFT methods for generic compact objects.
A first systematic relativistic formulation of dynamical tides in non-rotating compact
objects was developed in \citet{Pitre:2023xsr}, based on the formalism developed in~\citet{Poisson:2020vap}.
Like the case of the dynamical tidal response of a BH (see~\ref{LN_Dyn_Pert}), the idea is to expand all the perturbation variables as a series in powers of $M\omega\ll1$ and solve the equations perturbatively, order by order in $M\omega$. 

\begin{figure}[!t] 
\centering
\includegraphics[width=0.7\columnwidth]{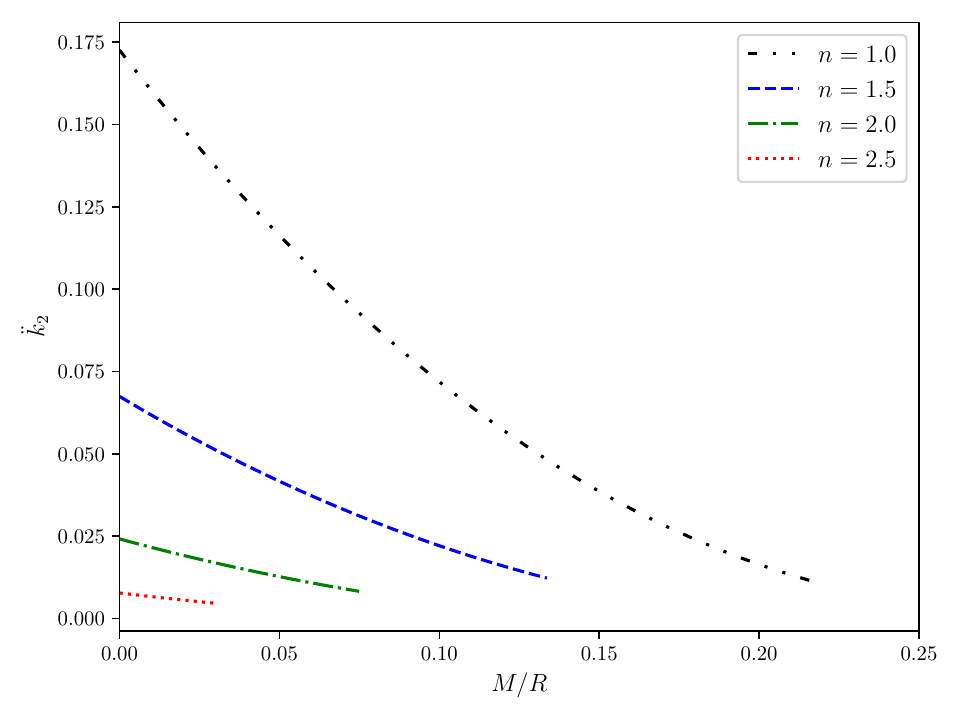}
\caption{
The ${\cal O}(\omega^2 M^2)$ correction to the electric, $\ell=2$ LN, $k_{\ell m}^{(2)}\propto \ddot k_2$, as a function of the NS compactness for some polytropic EoS. From \citet{Pitre:2025qdf,Pitre:2023xsr}.
}
\label{fig:NSdyn}
\end{figure}

The zeroth-order expansion gives the equations for the zero-frequency limit of the LNs (which might differ from the strictly static limit, see the axial case in~\ref{sec:NSdynRel}), while the first-order expansion provides the equations for the dissipation numbers.
In general, the tidal response takes the form in~\ref{def:F} and the tidal LN and dissipation number can be defined by taking its real and imaginary part, respectively, via~\ref{dynamicLNDN}.
For a static and spherically symmetric object, one has 
\begin{align}
    k_{\ell m}(\omega)&=k_{\ell m}^{(0)}+M^{2}\omega^{2}\,k_{\ell m}^{(2)}+\cdots  \label{k2ddot}\\
    \nu_{\ell m}(\omega)&=M\omega\,\nu_{\ell m}^{(1)}+M^{3}\omega^{3}\,\nu_{\ell m}^{(3)}+\cdots
\end{align}
and the dissipation numbers $\nu_{\ell m}$ vanish for perfect-fluid stars, where viscosity is absent. 
Similarly to~\ref{eq:deltagttEFT}, the dynamical quadrupolar LN enters the metric perturbations as
\begin{align}
	\delta g^{(\ell=2,m)}_{tt}(r)=
\mathscr{E}_{2,m}\left(r^2+12\frac{\lambda_2+\omega^2 \ddot\lambda_2}{r^3}\right),
\label{eq:deltagttEFTdyn}
\end{align}
where the coefficient $\ddot\lambda_2$ is related to the quantity $\ddot k_2\propto k_{22}^{(2)}$ defined in \citet{Poisson:2020vap,Pitre:2025qdf,Pitre:2023xsr} by
%%%
\begin{equation}
    \ddot k_2=-6\frac{\ddot\lambda_2 M}{R^8}\,.
\end{equation}
%%%
Like the quadratic LN $p_2$ defined in~\ref{dimscale}, the dynamical tidal LNs enter the GW
phase through a contribution proportional to $\ddot{k}_2/C^8$. Since
$\ddot{k}_2={\cal O}(0.1)$ (see~\ref{fig:NSdyn}), this scaling implies that the impact of
dynamical LNs, much like that of quadratic tidal effects, is parametrically enhanced in
the small-compactness limit when compared to static tidal contributions. This enhancement
partially compensates for their formal PN suppression: dynamical tidal LNs first appear
at 8PN order, that is, three PN orders higher than the leading static tidal LN
contribution (see~\ref{sec:pheno_NS}).

%%%%%%%%%%%%%%%%%%

% \paolo{Possibly add paper by Luca in case. Discuss also running in NS}
Very recently, \citet{Jarequi:2026cyp} extended the EFT framework~\citep{Saketh:2023bul,Saketh:2024juq} to compute dissipation numbers and dynamical LNs for generic compact objects, including NSs.
A key result is that the $\beta$ function governing the running of the dynamical LN splits into two contributions: one proportional to the static LN (which vanishes for BHs), and a universal term independent of the nature of the object, arising from the nonlinear structure of Einstein's equations.
This universal contribution coincides with the coefficient of the logarithmic term in~\ref{eq:DLN}.

An alternative formalism connecting the induced quadrupole moment to the tidal field in the dynamical regime was developed in \citet{Steinhoff:2016rfi}. 
In this approach, the motion of a point particle excites the lowest-lying normal modes of a NS, in particular the fundamental ($f$-) mode. 
As a result, the tidal response exhibits resonant behavior when the orbital frequency approaches the $f$-mode frequency, leading to so-called \emph{orbital-mode} resonances. 
In the frequency domain, the relation between the quadrupole moment and the tidal field takes the form \citep{Steinhoff:2016rfi}
%%%%%%%%%%%%%%%%%%%%%%%%%%%%%%%%%%%%%%%%%%%%%%%%%%%%%%%%%%%
\begin{align}
M^{\mu \nu}
=-\lambda\left(1-\frac{\omega^{2}}{z^{2}\omega_{\rm f}^{2}}\right)^{-1}E^{\mu \nu}\,,
\end{align}
%%%%%%%%%%%%%%%%%%%%%%%%%%%%%%%%%%%%%%%%%%%%%%%%%%%%%%%%%%%
where $\omega_{\rm f}$ is the $f$-mode frequency and $z$ is the redshift factor relating the proper frame of the NS to that of an asymptotic observer. 

The above response function is invariant under $\omega\to -\omega$ and therefore respects time-reversal symmetry. 
Consequently, it encodes both static and dynamical tidal effects in the conservative sector, but does not capture dissipative contributions, which would arise from an imaginary part of the response function. 
A key advantage of this framework is that the tidal degree of freedom can be modeled as a harmonic oscillator with natural frequency $\omega_{\rm f}$, which greatly simplifies the dynamical treatment \citep{Steinhoff:2016rfi}.

Within this approach, the effective tidal LNs of a NS are given by \citep{Hinderer:2016eia}
%%%%%%%%%%%%%%%%%%%%%%%%%%%%%%%%%%%%%%%%%%%%%%%%%%%%%%%%%%%
\begin{align}
k_{\ell}^{\rm eff}
= k_{\ell}\left[
a_{\ell}
+\frac{b_{\ell}}{2}
\left(
\frac{M_{\ell \ell}^{\rm DT}}{M_{\ell \ell}^{\rm AT}}
+\frac{M_{\ell -\ell}^{\rm DT}}{M_{\ell -\ell}^{\rm AT}}
\right)
\right]\,,
\end{align}
%%%%%%%%%%%%%%%%%%%%%%%%%%%%%%%%%%%%%%%%%%%%%%%%%%%%%%%%%%%
where `DT' and `AT' denote dynamical and adiabatic tides, respectively. 
The coefficients $(a_{\ell},b_{\ell})$ are frequency independent and arise from the relative normalization of the tidal field components with $m\neq\ell$ and $m=\ell$ \citep{Maselli:2012zq,Chakrabarti:2013lua}. Also note that \citet{Steinhoff:2014kwa} provided a connection between relativistic tides with \ref{eq:summodes}, with applicability for viscous stars as well. 
Finally, as shown in \citet{Hinderer:2016eia}, $k_{\ell}^{\rm eff}$ reduces to the static LN $k_{\ell}$ in the limit of vanishing orbital frequency $\Omega$, while $k_{2}^{\rm eff}$ increases as $\Omega$ approaches the $f$-mode frequency, reflecting the resonant enhancement of dynamical tides.

\subsubsection{Effects of viscosity and resonances} \label{sec:resonances}
An important refinement of relativistic dynamical tides concerns the role of viscous
dissipation in the stellar interior. In idealized treatments, the tidal response is purely
conservative and the associated LN are real-valued. However, realistic NS
matter is \emph{dissipative}, and viscosity provides a physical mechanism through which tidal
energy can be irreversibly converted into heat. The impact of viscosity on relativistic
dynamical tides has been analyzed in detail in
 \citet{HegadeKR:2024agt,HegadeKR:2025qwj},
developing a relativistic dynamical-response framework where dissipation and
mode damping can be incorporated consistently into the framework of frequency-dependent LN.

In the relativistic perturbative description, viscosity modifies the fluid equations by
introducing dissipative terms that lead to complex mode frequencies, just like the BH quasinormal modes~(QNMs) \citep{Kokkotas:1999bd}. As a result, the
normal modes that dominate the tidal response, most notably the fundamental fluid
mode, acquire a finite damping time. When the tidal forcing frequency approaches a mode frequency, viscosity regulates the resonance, replacing the formally divergent
response of the inviscid theory with a finite-width peak, formally given by~\ref{eq:summodes}, where $\omega_j$ is complex.

Beyond viscous damping and linear perturbative effects, additional sources of systematic uncertainty in the tidal modelling of NSs arise from nonlinear hydrodynamic couplings, background spin effects, and relativistic corrections to mode frequencies. These ingredients can modify the effective tidal response and lead to non-negligible biases in the inferred tidal deformability and EoS constraints if neglected. Recent quantitative studies have highlighted the importance of consistently incorporating such effects in waveform models \citep{Bretz:2026asa}.

Very recently,
\citet{HegadeKR:2026iou} systematically explored how dynamical and dissipative tidal deformations depend on the NS internal structure using a parametrized nuclear EoS and a model for quark strange matter.
They found that while dissipative tides, as predicted by weak-interaction-driven bulk-viscous effects with these EoS, are too small  to be detectable by current or future observations, the conservative dynamical tidal response depends strongly on the slope of the symmetry energy and on other EoS parameters.

% \subsubsection{Resonances} \label{sec:resonances}

% This naturally leads to a
% \emph{frequency-dependent} LN, which reduces to the standard static LN in the
% zero-frequency limit, while capturing resonant behavior associated with relativistic
% quasi-normal modes. In close analogy with the Newtonian case, the fundamental fluid
% mode plays a central role, but relativistic effects shift mode frequencies and modify
% their coupling to the tidal field in a non-trivial way.

%%%%%%%%%%%%%%%%%%%%%%%%%%%%%%%%%%%%%%%%%%%%%%%%%%%%%%%%%%%%%%%%%%%%%%%%%%%%%%
%%%%%%%%%%%%%%%%%%%%%%%%%%%%%%%%%%%%%%%%%%%%%%%%%%%%%%%%%%%%%%%%%%%%%%%%%%%%%%
%%%%%%%%%%%%%%%%%%%%%%%%%%%%%%%%%%%%%%%%%%%%%%%%%%%%%%%%%%%%%%%%%%%%%%%%%%%%%%
\section{Love numbers of exotic compact objects}\label{sec:ECOs}
%%%%%%%%%%%%%%%%%%%%%%%%%%%%%%%%%%%%%%%%%%%%%%%%%%%%%%%%%%%%%%%%%%%%%%%%%%%%%%%%%
Even within GR, any compact object other than a BH is expected to possess a nonvanishing tidal deformability. This is the case for NSs, as well as for generic ECOs and BH mimickers \citep{Cardoso:2017cfl,Cardoso:2019rvt}. 
BH mimickers and horizonless ultra-compact objects have attracted considerable attention in recent years, both as phenomenological tools to probe the nature of dark compact objects and as possible alternatives aiming to address fundamental open issues --~both classical and quantum~-- related to spacetime, horizons, and singularities \citep{Cardoso:2019rvt,Bambi:2025wjx}.
Here we provide an overview of the tidal response of BH mimickers and ECOs, emphasizing their key differences with respect to BHs.

%%%%%%%%%%%%%%%%%%%%%%%%%%%%%%%%%%%%%%%%%%%%%%%%%%%%%%%%%%%%%%%%%%%%%%%%%%%%%%%%%

%%%%%%%%%%%%%%%%%%%%%%%%%%%%%%%%%%%%%%%%%%%%%%%%%%%%%%%%%%%%%
%%%%%%%%%%%%%%%%%%%%%%%%%%%%%%%%%%%%%%%%%%%%%%%%%%%%%%%%%%%%%
%%%%%%%%%%%%%%%%%%%%%%%%%%%%%%%%%%%%%%%%%%%%%%%%%%%%%%%%%%%%%
\subsection{Catalog of static Love numbers of exotic compact objects}\label{staticLNECO}

We start by discussing the static LNs of various ECOs and subsequently generalize the analysis to the dynamical context. 
We have already seen that the most compact objects within GR, namely BHs, possess vanishing static LNs. 
In contrast, the most compact stars, namely NSs, exhibit significantly large LNs (see~\ref{sec:NSs}). 
Since ECOs are less compact than BHs but might still be considerably more compact than NSs, we expect them to have static LNs that are nonzero, though possibly smaller than those of NSs. 
We will examine different classes of ECOs, beginning with those whose compactness overlaps with that of NSs and progressively moving toward objects with compactness approaching that of BHs.

Throughout this section we focus on two simplifying assumptions:  
(a) we restrict ourselves to four-dimensional GR; and  
(b) we consider nonrotating ECOs.  
The first assumption avoids degeneracies, since BHs in modified gravity theories and BHs in higher dimensions can have nonzero static LNs (see~\ref{sec:BHBGR}), which may obscure the comparison with ECOs. 
The second assumption is motivated by simplicity and by the scarcity of spinning ECO geometries. 
As we show below, these assumptions keep the discussion transparent and allow for a clear understanding of the origin and behavior of nonzero static LNs in ECOs.

The results for the LNs of different ECOs will be summarized in~\ref{sec:summaryECOs} below.

%%%%%%%%%%%%%%%%%%%%%%%%%%%%%%%%%%%%%%%
%%%%%%%%%%%%%%%%%%%%%%%%%%%%%%%%%%%%%%%
%%%%%%%%%%%%%%%%%%%%%%%%%%%%%%%%%%%%%%%
\subsubsection{Boson stars} \label{sec:bosonstars}

These hypothetical objects are compact configurations composed of bosonic fields which, in the simplest realization, are described by a complex scalar field $\Phi$ (see \citealt{Liebling:2012fv} for a review). In addition to the standard kinetic term $-g^{\mu\nu}\partial_{\mu}\Phi\,\partial_{\nu}\Phi$, the scalar-field action includes a self-interaction potential $V(|\Phi|)$. Much like NSs, whose macroscopic properties depend on the EoS of nuclear matter, the properties of boson stars are determined by the specific form of the potential $V(|\Phi|)$.

For instance, in the minimal model with $V(|\Phi|)=\mu^{2}|\Phi|^{2}$, the maximum mass of a boson star is \citep{Kaup:1968zz,Ruffini:1969qy}
\begin{equation}
M_{\rm max}\simeq 8\left(\frac{10^{-11}\,\mathrm{eV}}{\mu/\hbar}\right) M_{\odot}\, .
\end{equation}
More general choices of the potential lead to markedly different properties. A commonly studied extension includes a quartic self-interaction term, $(\alpha/4)|\Phi|^{4}$, added to the mass term and is often referred to as the massive model \citep{Colpi:1986ye}. Another example is the solitonic model, in which the mass term is multiplied by $\{1-2|\Phi|^{2}/\sigma_{0}^{2}\}^{2}$ \citep{Friedberg:1986tq}. In each case, the resulting maximum mass and compactness of the boson star differ significantly (see, e.g., \citealt{Cardoso:2017cfl}).

\begin{figure}[!t] 
\centering
\includegraphics[width=0.48\columnwidth]{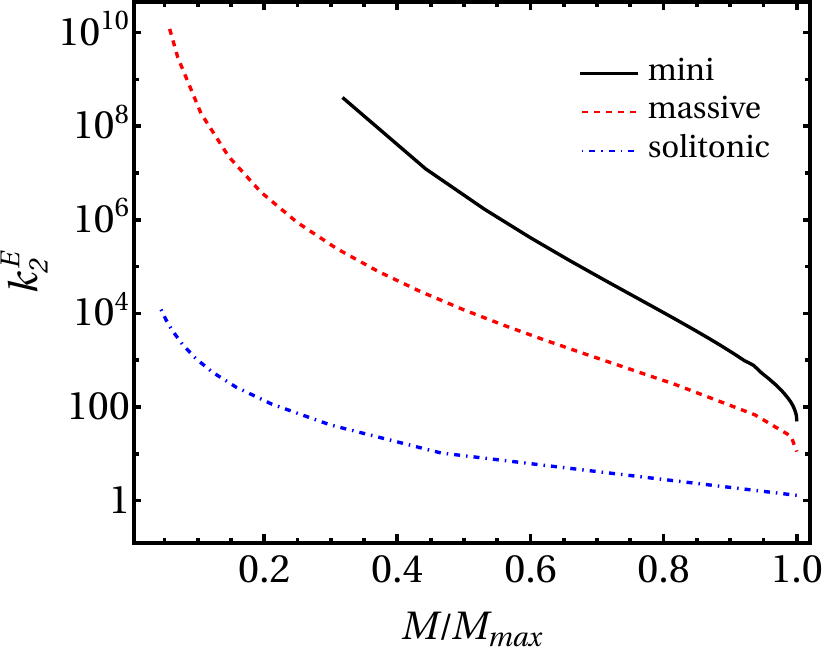}
\includegraphics[width=0.48\columnwidth]{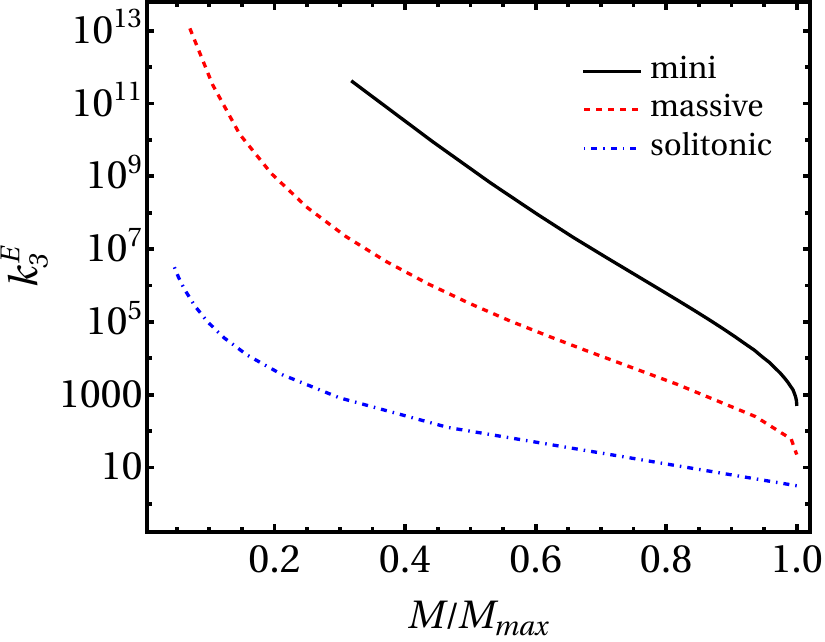}\\
\includegraphics[width=0.48\columnwidth]{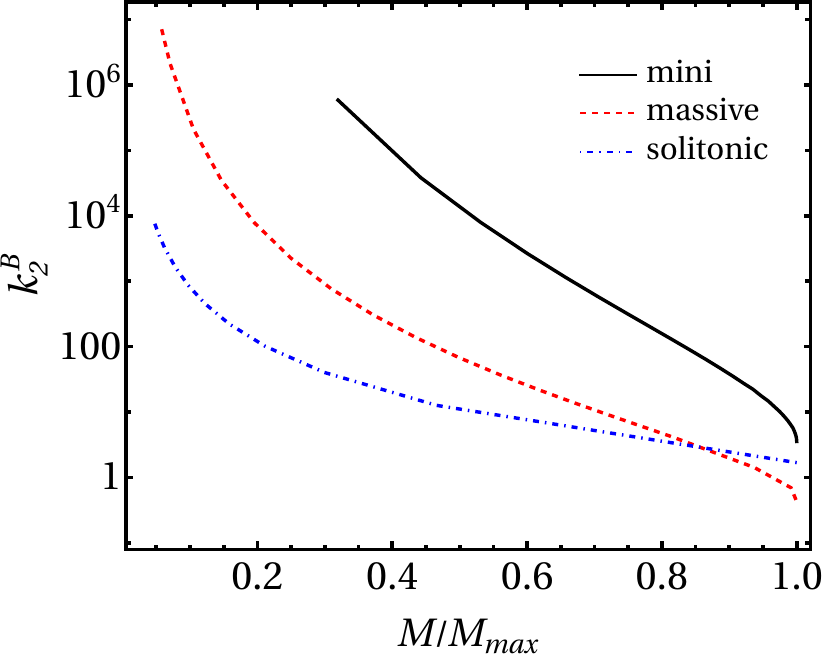}
\includegraphics[width=0.48\columnwidth]{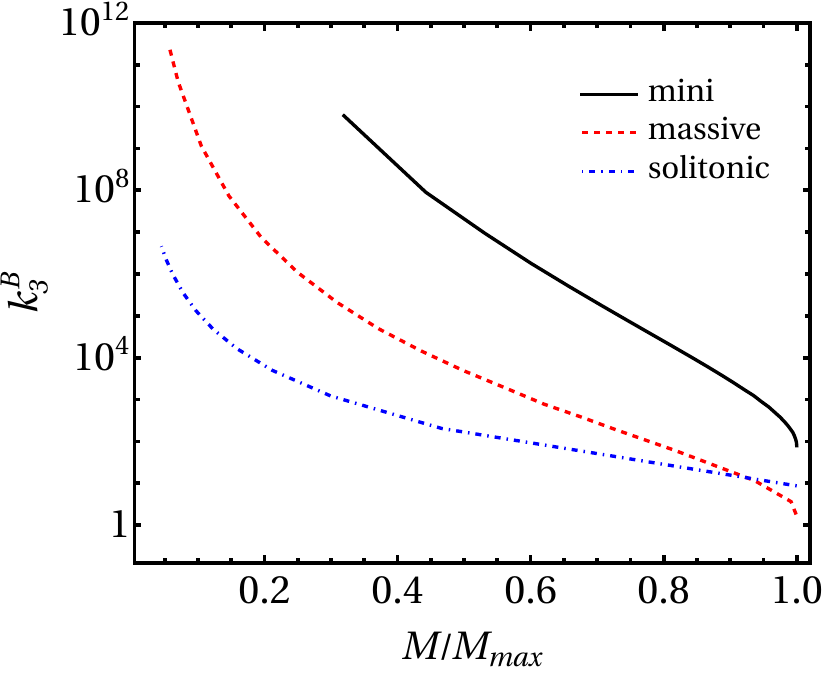}

\caption{
Polar (top panels) and axial (bottom panels) static LNs of various models of boson stars as a function of their mass (normalized by the maximum mass, which is different in each model).
We assume $\alpha=10^4$ and $\sigma_0=0.05$ for the potential of massive and solitonic boson stars, respectively (see \citealt{Sennett:2017etc} for other choices).
Plots adapted from \citet{Cardoso:2017cfl} using our conventions and normalization.}
\label{fig:BS}
\end{figure} 
%%%%%%%%%%%%%%%%%%

Therefore, different contributions to the scalar-field potential give rise to distinct boson-star configurations and compactness values, which can range from those typical of NSs up to values comparable to those of a BH. These configurations are obtained by solving the appropriate TOV equations, with the energy density and the radial and tangential pressures determined by the background scalar field:
%%%%%%%%%%%%%%%%%%%%%%%%%%%%%%%%%%%%%%%%%%%%%%%%%%%%%%%%%%%%%%%%%%%%%%%%%%%%%%%%%
\begin{align}
\bar{\rho}&=g(r)\phi_{0}'^{2}+\frac{\omega^{2}\phi_{0}^{2}}{f}+V(\phi_{0})\,,
\\ 
\bar{p}_{r}&=g(r)\phi_{0}'^{2}+\frac{\omega^{2}\phi_{0}^{2}}{f}-V(\phi_{0})\,,
\\
\bar{p}_{t}&=-g(r)\phi_{0}'^{2}+\frac{\omega^{2}\phi_{0}^{2}}{f}-V(\phi_{0})\,.
\end{align}
%%%%%%%%%%%%%%%%%%%%%%%%%%%%%%%%%%%%%%%%%%%%%%%%%%%%%%%%%%%%%%%%%%%%%%%%%%%%%%%%%
where the background scalar field is taken to be $\Phi^{(0)}=\phi_{0}(r)e^{-i\omega t}$, while the background static and spherically symmetric metric is given by~\ref{met_sph_stat}. One can show that the conservation of the energy-momentum tensor constructed from the corresponding energy density and pressure components is equivalent to the Klein--Gordon equation for the background scalar field. The three unknown functions $(f,g,\phi_{0})$ are determined by the Klein--Gordon equation together with the $(t,t)$ and $(r,r)$ components of Einstein's equations.

Regularity at the center of the star selects a discrete set of allowed values of $\omega$, leading to an oscillatory scalar-field configuration whose stress-energy tensor is nevertheless static. Numerical solutions of the coupled field equations yield both the radial profile of the scalar field and the total mass of the boson star. Typically, boson stars are diffuse objects, with scalar-field configurations extending to spatial infinity, although the bulk of the mass is concentrated within a characteristic radius of order $1/\mu$. The radius $R$ of the boson star is conventionally defined as the radius enclosing $99\%$ of the total mass.

To compute the LNs induced by gravitational perturbations, one must also account for perturbations of the scalar field, $\Phi \to \Phi^{(0)}+\Phi^{(1)}$, where $\Phi^{(1)}$ denotes the scalar perturbation. The latter can be decomposed in a spherical-harmonic basis, see~\ref{scalarpertss}. In the static limit and in the Regge--Wheeler gauge, the polar gravitational perturbations are described by the functions $(H_{0},H_{2},K)$ (see~\ref{pert_grav} for definitions), while the scalar perturbation is encoded in $\delta\phi$ (see~\ref{scalarpertss}). Since $T^{\theta}{}_{\theta}=T^{\phi}{}_{\phi}$, it follows that $H_{0}=H_{2}\equiv H$. Moreover, one can express $K=K(H,\delta\phi)$, so that the problem reduces to two coupled second-order differential equations for the variables $(H,\delta\phi)$ \citep{Cardoso:2017cfl}.

At infinity, the background scalar field is exponentially small so the perturbation equations reduces to a single second-order equation for $H$, whose solutions are given in terms of associated Legendre functions, see~\ref{Hext}. The system is solved by imposing regularity at the origin for the interior solutions and, as in the case of NSs, yields nonvanishing LNs.

An analogous analysis applies to axial perturbations. In this case, the scalar perturbation does not contribute, since scalar fields transform as polar quantities. Both in the interior and in the exterior, one therefore solves a single second-order equation for $h_{0}$, imposing regularity at the center and extracting the LNs at large distance. The resulting polar and axial LNs for $\ell=2,3$ are shown in~\ref{fig:BS}. Useful fitting formulas for certain models can be found in \citet{Sennett:2017etc,Vaglio:2023lrd}.

\subsubsection{Fermion soliton stars}\label{FSS}

\begin{figure}[ht]
	\centering
	\includegraphics[width=0.65\textwidth]{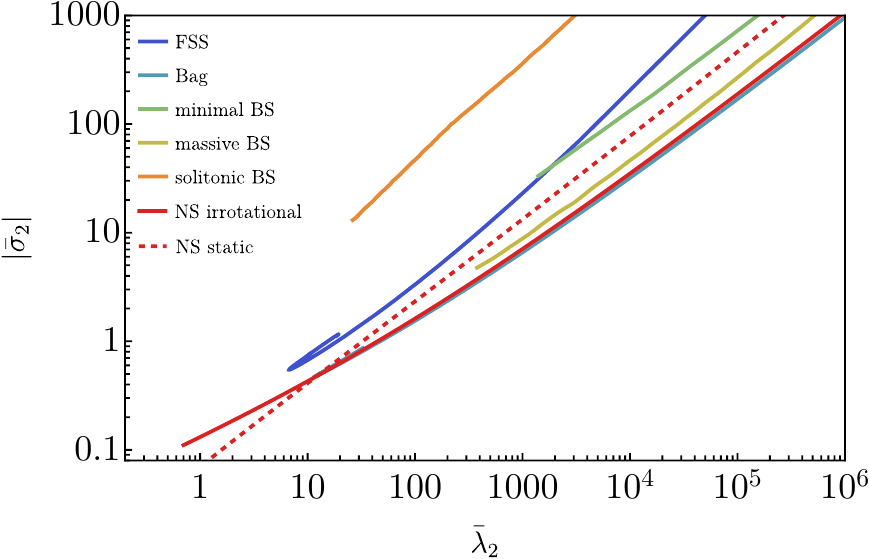}
	\caption{Comparison of the approximately universal relation between axial ($\bar\sigma_2$) and polar ($\bar\lambda_2$) quadrupolar LNs for an example of fermion soliton star (FSS) relative to the universality curves for other compact objects, such as NSs in the irrotational (solid red line) and static configurations (dashed red line), soliton boson stars (orange line), massive boson stars (yellow line), minimal boson stars (green line) and an effective bag model (Bag, water green line).
    from \citet{Berti:2024moe}.
    }
	\label{fig:FSS}
\end{figure}

Recently, \citet{Berti:2024moe} computed the tidal LNs for \emph{fermion soliton stars}, a consistent model of ECOs which involve a nonlinear interaction between a \emph{real} scalar field and fermions through a Yukawa term \citep{DelGrosso:2023trq}. 
As expected, the LNs depend on the model parameters, in particular the fermion-boson coupling and the scalar potential. However, 
 \citet{Berti:2024moe} found some approximately universal relation between axial and polar LNs of a fermion soliton star, which is compared to the analog universal curve of other compact objects, including NSs, in~\ref{fig:FSS}.

%

%%%%%%%%%%%%%%%%%%%%%%%%%%%%%%%%%%%%%%%
%%%%%%%%%%%%%%%%%%%%%%%%%%%%%%%%%%%%%%%
%%%%%%%%%%%%%%%%%%%%%%%%%%%%%%%%%%%%%%%
\subsubsection{Thin-shell gravastars}\label{gravastars}

Gravastars are another class of ECOs, whose interior is described by the static patch of de Sitter, while the exterior is given by the Schwarzschild metric, for a spherically symmetric and non-rotating configuration \citep{Mazur:2001fv, Mazur:2004fk}. For simplicity, we will assume that the two spacetimes are joined across of a shell of vanishing thickness \citep{Visser:2003ge} at $r=R$, however, in general the thickness of the shell need not be small \citep{Uchikata:2016qku}. For a thin-shell gravastar, the metric functions on the two sides of the shell read
%%%%%%%%%%%%%%%%%%%%%%%%%%%%%%%%%%%%%%%%%%%%%%%%%%%%%%%%%%%%%%%%%%%%%%%%%%%%%%%%%
\begin{equation}\label{metric_gravastar}
f=g=
\left\{
\begin{array}{l}
1-\frac{2M}{r} \qquad \quad\qquad  r>R
\\
1-\frac{2M}{R}\left(\frac{r^{2}}{R^{2}}\right) \qquad \,\,r<R
\end{array}
\right.\,,
\end{equation}
%%%%%%%%%%%%%%%%%%%%%%%%%%%%%%%%%%%%%%%%%%%%%%%%%%%%%%%%%%%%%%%%%%%%%%%%%%%%%%%%%
where we have used the metric coefficients of the Schwarzschild metric and fixed the de-Sitter constant $\Lambda=6M/R^{3}$ for continuity of the metric functions across the shell. In this case, the shell has vanishing energy density, but negative surface pressure \citep{Pani:2009ss}. 

The perturbative analysis of a gravastar requires considering perturbations of both the exterior Schwarzschild spacetime and the interior de~Sitter spacetime, together with appropriate matching conditions across the shell. In the static limit, the axial and polar perturbations in the exterior region are governed by the same equations as those of the Schwarzschild spacetime, and are described by the variables $h_{0}$ and $H$, respectively. These satisfy the differential equations~\ref{h0stateq} in the axial sector and~\ref{Hsole} in the polar sector. The corresponding solutions are given by~\ref{h0e} for $h_{0}$ in the $\ell=2$ axial mode, and by~\ref{Hext} for $H$ in the polar sector with generic multipole index $\ell$.

In the interior de~Sitter spacetime, both axial and polar perturbations in the static limit obey second-order differential equations, which can be written as \citep{Pani:2009ss}
%%%%%%%%%%%%%%%%%%%%%%%%%%%%%%%%%%%%%%%%%%%%%%%%%%%%%%%%%%%%%%%%%%%%%%%%%%%%%%%%%
\begin{align}
&\left(1-\frac{2M}{R^{3}}r^{2}\right)\dfrac{d^{2}h_{0}}{dr^{2}}-\left(\frac{\ell(\ell+1)}{r^{2}}-\frac{4M}{R^{3}}\right)h_{0}=0\,,
\\
&\left(1-\frac{2M}{R^{3}}r^{2}\right)^{2}H_{0}''+\frac{2}{r}\left(1-\frac{2M}{R^{3}}r^{2}\right)\left(1-\frac{4M}{R^{3}}r^{2}\right)H_{0}'
\nonumber
\\
&-\left[\frac{\ell(\ell+1)}{r^{2}}\left(1-\frac{2M}{R^{3}}r^{2}\right)+\left(\frac{4M}{R^{3}}\right)^{2}r^{3}+\left(1-\frac{2M}{R^{3}}r^{2}\right)+\frac{12M}{R^{3}}r\right]H_{0}=0\,
\end{align}
%%%%%%%%%%%%%%%%%%%%%%%%%%%%%%%%%%%%%%%%%%%%%%%%%%%%%%%%%%%%%%%%%%%%%%%%%%%%%%%%%
with the following solutions for the axial and polar perturbations, for generic $\ell$ modes, 
%%%%%%%%%%%%%%%%%%%%%%%%%%%%%%%%%%%%%%%%%%%%%%%%%%%%%%%%%%%%%%%%%%%%%%%%%%%%%%%%%
\begin{align}
h_{0}^{\rm int}&=r^{\ell+1}\,_{2}F_{1}\left(\frac{\ell-1}{2},\frac{\ell+2}{2};\ell+\frac{3}{2};\frac{2Mr^{2}}{R^{3}}\right)~,
\label{h0int}
\\
H_{0}^{\rm int}&=\frac{R^{2}r^{\ell}}{R^{2}-(2M/R)r^{2}}\,_{2}F_{1}\left(\frac{\ell-1}{2},\frac{\ell}{2};\ell+\frac{3}{2};\frac{2Mr^{2}}{R^{3}}\right)~,
\label{Hint}
\end{align}
%%%%%%%%%%%%%%%%%%%%%%%%%%%%%%%%%%%%%%%%%%%%%%%%%%%%%%%%%%%%%%%%%%%%%%%%%%%%%%%%%
where we have already neglected the solutions that are irregular at the origin.

Since the energy-momentum tensor on the thin shell does not couple to the axial sector, it follows that both the perturbation and its derivative should be continuous across the shell, i.e., we must have $[[h_{0}]]=0=[[(dh_{0}/dr_{*})]]$, implying $h_{0}^{\rm int}(R)=h_{0}^{\rm ext}(R)$, as well as $(dh^{\rm ext}_{0}/dr_{*})_{R}=(dh_{0}^{\rm int}/dr_{*})_{R}$. Henceforth, we use the symbol $[[\cdots]]$ to denote the discontinuity of a quantity across a surface: $[[A]]\equiv \lim_{y \to 0}A(R+y)-A(R-y)$. These two boundary conditions are sufficient to fix the arbitrary constants of the exterior solution and hence determines the LN. For the polar sector, on the other hand, the discontinuity in the extrinsic curvature vanishes, as the thin-shell has zero surface energy density, and hence, $[[K]]=0$. Similarly, the derivative of the extrinsic curvature will depend on the perturbation of the energy density of the thin-shell \citep{Pani:2009ss}. Assuming that the matter field constituting the thin-shell is stiff in nature, it follows that the energy density of the thin-shell is also not perturbed. Therefore, the relevant boundary conditions read: $[[K]]=0=[[(dK/dr_{*})]]$. These conditions fix the arbitrary constants $\mathcal{P}_{1}$ and $\mathcal{P}_{2}$, arising in the exterior solution for $H_{0}$, see~\ref{Hext} and the arbitrary constants $\mathcal{A}_{1}$ and $\mathcal{A}_{2}$, arising in the exterior solution for $h_{0}$, see~\ref{h0e}. Thus, the LNs can be uniquely obtained once the properties of the thin shell are fixed \citep{Uchikata:2015yma,Uchikata:2016qku}. For still thin shells, the above matching procedure yields~\footnote{Dividing by $2^{2\ell+1}$ to take care of normalization used here when $R\approx 2M$.} \citep{Cardoso:2017cfl},
%%%%%%%%%%%%%%%%%%%%%%%%%%%%%%%%%%%%%%%%%%%%%%%%%%%%%%%%%%%%%%%%%%%%%%%%%%%%%%%%%
\begin{align}
k_{2}^{\rm M}=\frac{1}{5(43-12\ln 2+18\ln \epsilon)}~;
\qquad 
k_{2}^{\rm E}=\frac{1}{10(23-6\ln 2+9\ln \epsilon)}~,
\end{align}
%%%%%%%%%%%%%%%%%%%%%%%%%%%%%%%%%%%%%%%%%%%%%%%%%%%%%%%%%%%%%%%%%%%%%%%%%%%%%%%%%
where we have expressed $R=2M(1+\epsilon)$ and have assumed $\epsilon \ll 1$. Note that the coefficient of $\ln \epsilon$ is identical for both axial and polar LNs, so $k_{2}^{\rm M}=k_{2}^{\rm E}$ in the BH limit ($\epsilon\to0$). This is a general property that will be discussed in~\ref{sec:membrane}.
%%%%%%%%%%%%%%%%%%%%%%%%%%%%%%%%%%%%%%%
%%%%%%%%%%%%%%%%%%%%%%%%%%%%%%%%%%%%%%%
%%%%%%%%%%%%%%%%%%%%%%%%%%%%%%%%%%%%%%%
\subsubsection{Black shell geometries}\label{blackshell}

Similar to gravastars, recent progress has been made in the study of black shell geometries, in which an ultra-compact shell separates an asymptotically flat vacuum exterior from an anti-de~Sitter interior \citep{Danielsson:2017riq}. The motivation for such configurations originates from the nucleation of a bubble via gravitational instantons. During the collapse of a star, a phase transition to a stable anti-de~Sitter vacuum may occur, leading to the formation of a bubble that traps the collapsing matter. The endpoint of this process is a black shell geometry, where a shell of matter separates the exterior vacuum region from the anti-de~Sitter interior.

The resulting spacetime is therefore closely analogous to the gravastar geometry described in~\ref{metric_gravastar}, with the constant scalar curvature of the anti-de~Sitter interior given by $\Lambda = -6M/R^{3}$, which is manifestly negative. The analysis of static perturbations in both the exterior and interior regions then proceeds in close analogy with the gravastar case. In particular, the exterior axial and polar perturbations are again described by the solutions~\ref{h0e} and~\ref{Hext}, respectively, while the interior axial and polar perturbations are given by~\ref{h0int} and~\ref{Hint}, with the argument of the interior solutions being $-(2M/R^{3})r^{2}$.

The remaining step consists of matching the perturbations and their derivatives across the shell, which in general requires including perturbations of the matter fields residing on the shell. In simplified setups, such as the gravastar case, one finds the same characteristic $\ln\epsilon$ behavior for the LNs \citep{Giri:2024cks} (see~\ref{tab:summaryECOs} below).

%%%%%%%%%%%%%%%%%%%%%%%%%%%%%%%%%%%%%%%
%%%%%%%%%%%%%%%%%%%%%%%%%%%%%%%%%%%%%%%
%%%%%%%%%%%%%%%%%%%%%%%%%%%%%%%%%%%%%%%
\subsubsection{Wormholes}\label{wormhole}

Another class of ultra-compact objects is provided by wormhole geometries \citep{VisserBook}. Among the various possibilities, we focus here on the simplest model of Schwarzschild wormhole. The latter is constructed by gluing two Schwarzschild spacetimes across a timelike hypersurface located at a radius $R=2M(1+\epsilon)$, which defines the throat of the wormhole. As a result, the spacetime contains no horizon, and the throat radius can be taken arbitrarily close to the BH horizon, so that the configuration behaves as an ultra-compact object.

The junction of two Schwarzschild geometries at $R$ requires the presence of a thin shell of matter at the throat, characterized by negative energy density and positive pressure. The appearance of negative energy density is a generic feature of wormhole solutions and is usually associated with the presence of exotic matter fields, which are known to be necessary for sustaining wormhole geometries. Both sides of the throat can be conveniently described using the tortoise coordinate $r_{*}$, chosen such that $r_{*}(R)=0$. With this convention, one universe corresponds to $r_{*}>0$, while the other corresponds to $r_{*}<0$.

A key advantage of the Schwarzschild wormhole construction is that the perturbation equations in both the axial and polar sectors on either side of the throat are identical to those of the Schwarzschild spacetime, namely~\ref{h0stateq} and~\ref{Hsole}, respectively. Consequently, the corresponding solutions are again those given in~\ref{h0e} and~\ref{Hext}. The crucial difference is that the universe on the other side of the throat is required to be free of tidal fields. This implies that, for the $\ell=2$ mode, the solutions to the perturbed Einstein equations on that side take the form
%%%%%%%%%%%%%%%%%%%%%%%%%%%%%%%%%%%%%%%%%%%%%%%%%%%%%%%%%%%%%%%%%%%%%%%%%%%%%%%%%
\begin{align}
h_{0\,\mathrm{(int)}}^{(\ell=2)} &=
\mathcal{A}_{\rm int}\frac{12x^{3}-6x^{2}-2x-1}{3x}
+4\mathcal{A}_{\rm int}x^{2}(x-1)\log\!\left(\frac{x-1}{x}\right) \, ,
\label{wormholeintaxial}
\\
H_{\mathrm{(int)}}^{\ell=2} &=
\mathcal{P}_{\rm int}\, Q_{2}^{2}\!\left(2x-1\right) \, ,
\label{wormholeintpolar}
\end{align}
%%%%%%%%%%%%%%%%%%%%%%%%%%%%%%%%%%%%%%%%%%%%%%%%%%%%%%%%%%%%%%%%%%%%%%%%%%%%%%%%%
where $x=r/(2M)$. These solutions correspond to decaying gravitational perturbations as one moves asymptotically away from the throat into the other universe, ensuring the absence of tidal fields there.

On the exterior side, the solutions are identical to those given in~\ref{h0e} and~\ref{Hext} for axial and polar perturbations, respectively, and contain two arbitrary integration constants in each sector. These constants are fixed by imposing appropriate boundary conditions at the throat. Specifically, one requires: (a) continuity of $h_{0}$ and its radial derivative across the throat, and (b) continuity of the extrinsic curvature of the perturbed metric and its derivative across the throat. These conditions relate the exterior constants $(\mathcal{A}_{1},\mathcal{A}_{2})$ in~\ref{h0e} to the interior constant $\mathcal{A}_{\rm int}$ in~\ref{wormholeintaxial}, and similarly $(\mathcal{P}_{1},\mathcal{P}_{2})$ in~\ref{Hext} to $\mathcal{P}_{\rm int}$ in~\ref{wormholeintpolar}. The ratios $(\mathcal{A}_{2}/\mathcal{A}_{1})$ and $(\mathcal{P}_{2}/\mathcal{P}_{1})$ then directly determine the axial and polar LNs.

For the $\ell=2$ mode, and taking $\epsilon\to0$, the resulting LNs are \citep{Cardoso:2017cfl} (after dividing by $2^{2\ell+1}$ to match the normalization adopted here when $R\approx 2M$)
%%%%%%%%%%%%%%%%%%%%%%%%%%%%%%%%%%%%%%%%%%%%%%%%%%%%%%%%%%%%%%%%%%%%%%%%%%%%%%%%%
\begin{align}
k_{2}^{\rm E} \approx \frac{1}{40\,(8+3\ln\epsilon)} \, , 
\qquad
k_{2}^{\rm B} \approx \frac{1}{10\,(31+12\ln\epsilon)} \, .
\end{align}
%%%%%%%%%%%%%%%%%%%%%%%%%%%%%%%%%%%%%%%%%%%%%%%%%%%%%%%%%%%%%%%%%%%%%%%%%%%%%%%%%
Remarkably, the coefficient of the leading $1/\ln\epsilon$ term is the same for both axial and polar perturbations, a feature that also emerges in other classes of ultra-compact objects.

%%%%%%%%
\subsubsection{Black-hole microstates}
%%%%%%%%%%%%
One of the longstanding open problems associated with BHs is the quest for a microscopic understanding of their entropy, which is famously proportional to the horizon area measured in Planck units \citep{Bekenstein:1973ur} and is therefore extraordinarily large for astrophysical BHs. 
A well-motivated explanation is provided by the fuzzball paradigm in string theory, which posits that a classical BH should be viewed as a thermodynamic coarse-grained description of a vast ensemble of regular quantum states \citep{Mathur:2009hf,Bena:2022rna,Bena:2022ldq}. 
In the classical limit, these states correspond to ``microstate geometries'': smooth solitonic solutions that share the same mass and conserved charges as a BH, but feature a radically different interior, where the would-be horizon is replaced by a regular, horizonless cap \citep{Bena:2006kb,Bena:2016ypk,Bena:2017xbt,Bah:2021owp,Bah:2022yji}. 
The absence of an event horizon in this fundamental description naturally resolves Hawking's information loss paradox \citep{Hawking:1976ra,Polchinski:2016hrw}, since information is not destroyed but, in principle, encoded in the nontrivial microstate structure.

In recent years, tractable prototypical examples of regular horizonless microstates have been discovered, most notably topological stars \citep{Bah:2020ogh} and other solitons \citep{Chakraborty:2025ger} in five dimensional supergravity.

The tidal response of topological stars has been studied in \citet{Bianchi:2023sfs}. Interestingly, it has been found that the LNs are zero in the static limit, just like BHs. A similar study for other supergravity solitons is missing, but could be performed extending recent results for their perturbations \citep{Dima:2025tjz}.
%%%%%%%%%%%%%%%%%%%%%%%%%%%%%%%%%%%%%%%%%%%%%%%%%%%%%%%%%%%%%
%%%%%%%%%%%%%%%%%%%%%%%%%%%%%%%%%%%%%%%%%%%%%%%%%%%%%%%%%%%%%
%%%%%%%%%%%%%%%%%%%%%%%%%%%%%%%%%%%%%%%%%%%%%%%%%%%%%%%%%%%%%
\subsection{Dynamical tides of exotic compact objects}

We now provide an overview of the dynamical tides of ECOs, which offer a more general framework to compute their tidal response in genuinely dynamical settings. Unlike static LNs, the approach outlined here relies only on the general properties of the object and is not tied to a specific microscopic model.

Our analysis focuses on Kerr-like rotating compact objects. The corresponding LNs are determined by carefully accounting for the subtleties associated with both the angular momentum and the definition of the reflectivity of the object. The treatment of reflectivity adopted here builds on results developed recently in the context of the QNMs of ECOs \citep{Maggio:2018ivz,Maggio:2020jml}. Although the formalism is valid at arbitrary frequencies, we restrict our attention to the low-frequency regime, where an analytical treatment based on matched asymptotic expansions is possible.

We assume that the exterior geometry of these rotating ECOs is described by the Kerr metric, while the sole deviations from a Kerr BH are encoded in boundary conditions imposed at a finite radius from the would-be horizon. These boundary conditions depend both on the reflectivity of the ECO surface and on its location,
\begin{equation}
    R=r_+(1+\epsilon) \,,\label{surface}
\end{equation}
relative to the would-be BH horizon $r_+$. 

As in the BH case, the starting point of our analysis is the Teukolsky equation for the Weyl scalars, and in particular for $\Psi_{4}$, which describes gravitational perturbations on a Kerr background \citep{Chia:2020yla,Consoli:2022eey,Bhatt:2023zsy}. Solving the resulting master equation, performing the appropriate asymptotic expansions, and imposing suitable boundary conditions at the ECO surface allows us to extract the LNs.

A central ingredient of the analysis is precisely the boundary condition at the ECO surface, which is characterized by two parameters: (a) the reflectivity ${\cal R}$ and (b) the compactness parameter $\epsilon$. Among these, the definition of the reflectivity is particularly subtle, as it requires a description in terms of plane-wave modes near the surface of the object. This requirement is not satisfied by the Weyl scalar $\Psi_{4}$, but it is fulfilled by the associated Detweiler function, which exhibits a wave-like behavior in this region. For this reason, we shall define the reflectivity in terms of the Detweiler function rather than the more commonly used Teukolsky function \citep{Maggio:2020jml}.

%

%%%%%%%%%%%%%%%%%%%%%%%%%%%
%%%%%%%%%%%%%%%%%%%%%%%%%%%
%%%%%%%%%%%%%%%%%%%%%%%%%%%
\subsubsection{Reflectivity of compact objects}\label{sec:reflectivity}

The computation of static LNs for ECOs does not require a detailed discussion of reflectivity, since reflectivity is intrinsically associated with propagating (plane-wave) modes and therefore with genuinely dynamical situations. In the static limit, the reflective character of the object is instead effectively encoded through boundary conditions of Dirichlet or Neumann type imposed at the surface.

In the dynamical case, however, a meaningful definition of reflectivity requires the relevant master function to behave as plane waves near the boundary. This requirement is not satisfied by the radial Teukolsky function associated with the Weyl scalar $\Psi_{4}$. Indeed, for a Kerr BH in the near-horizon limit the radial Teukolsky equation admits two independent solutions \citep{Teukolsky:1974yv}: (a) an ingoing contribution with radial behavior $\Delta^{2}$, and (b) an outgoing contribution behaving as $e^{2 i \bar{\omega} r_*}$, where $r_*$ is the tortoise coordinate and $\bar{\omega}=\omega-m\Omega_{\rm H}$, with $\Omega_{\rm H}$ the angular velocity of the horizon. Accordingly, the near-horizon form of the radial function can be written as
%%%%%%%%%%%%%%%%%%%%%%%%%%%%%%%%%%%%%%%%%%%%%%%%%%%%%%%%%%%%%%%%%
\begin{align}\label{radialnear}
\,_{-2}R_{\ell m}=\mathbb{A}\,\Delta^{2}+\mathbb{B}\,e^{2 i \bar{\omega} r_*}\, .
\end{align}
%%%%%%%%%%%%%%%%%%%%%%%%%%%%%%%%%%%%%%%%%%%%%%%%%%%%%%%%%%%%%%%%%
The coefficient $\mathbb{B}$ is associated with the outgoing mode at the surface. Since a BH absorbs all incoming radiation, regularity at the horizon implies $\mathbb{B}=0$ in the BH case. For non-BH objects, including neutron stars and ECOs \citep{Giudice:2016zpa,Cardoso:2019rvt,Maggio:2021ans}, as well as quantum-corrected BHs \citep{Oshita:2019sat,Chakraborty:2022zlq,Nair:2022xfm}, the presence of a surface generically leads to partial reflection, and hence to outgoing modes with $\mathbb{B}\neq 0$. Thus, BHs have vanishing reflectivity, whereas ECOs---including ordinary neutron stars%
\footnote{The reflectivity of a neutron star is expected to be close to unity due to their viscosity \citep{Maggio:2018ivz}; see however \citep{Ripley:2023qxo} for a recent analysis of out-of-equilibrium tidal dynamics.}
---exhibit a nonzero reflectivity. In general, the reflectivity is characterized by a complex, frequency-dependent function ${\cal R}(\omega)$.

Given the near-horizon behavior in~\ref{radialnear}, it may appear natural to identify the reflectivity with the ratio of amplitudes $(\mathbb{B}/\mathbb{A})$. This identification is, however, incorrect, because neither the ingoing nor the outgoing solutions of the radial Teukolsky equation correspond to plane waves. This is also reflected in the fact that the energy fluxes associated with these modes are not simply proportional to $|\mathbb{A}|^{2}$ and $|\mathbb{B}|^{2}$, but instead involve nontrivial, frequency-dependent prefactors \citep{Teukolsky:1974yv}.

To overcome this difficulty, it is convenient to recast the radial Teukolsky equation into a Schr\"odinger-like form with a real potential that becomes constant in the near-horizon region. In this formulation, the mode solutions are exact plane waves, allowing for an unambiguous definition of reflectivity. This procedure is implemented through the Detweiler function, $\,_{s}X_{\ell m}$, defined as a specific linear combination of the radial Teukolsky function $\,_{s}R_{\ell m}^{(t)}$ (in Boyer-Lindquist coordinates) and its radial derivative \citep{Detweiler:1977gy,Maggio:2018ivz}:
%%%%%%%%%%%%%%%%%%%%%%%%%%%%%%%%%%%%%%%%%%%%%%%%%%%%%%%%%%%%%%%%%
\begin{align}\label{Teu_Det_Trans}
\,_{s}X_{\ell m}
=\frac{\sqrt{r^{2}+a^{2}}}{\Delta}
\left[
\alpha \,_{s}R_{\ell m}^{(t)}
+\frac{\beta}{\Delta}\left(\frac{d}{dr}\,_{s}R_{\ell m}^{(t)}\right)
\right] .
\end{align}
%%%%%%%%%%%%%%%%%%%%%%%%%%%%%%%%%%%%%%%%%%%%%%%%%%%%%%%%%%%%%%%%%
Here $\alpha$ and $\beta$ are functions of $r$, $\omega$, and $(\ell,m,s)$, whose explicit form can be found in \citet{Detweiler:1977gy,Maggio:2018ivz}. With this choice, the Detweiler function satisfies a Schr\"odinger-like equation
%%%%%%%%%%%%%%%%%%%%%%%%%%%%%%%%%%%%%%%%%%%%%%%%%%%%%%%%%%%%%%%%%
\begin{align}
\frac{d^{2}}{dr_{*}^{2}}\,_{s}X_{\ell m}-V(r,\omega)\,_{s}X_{\ell m}=0\, ,
\end{align}
%%%%%%%%%%%%%%%%%%%%%%%%%%%%%%%%%%%%%%%%%%%%%%%%%%%%%%%%%%%%%%%%%
where the effective potential $V(r,\omega)$ is purely real. Importantly, the potential approaches a constant both near the horizon and at spatial infinity, with
$V(r\to r_{+},\omega)\to -\bar{\omega}^{2}$ and $V(r\to\infty,\omega)\to -\omega^{2}$. As a result, the solutions behave as plane waves $\sim e^{\pm i\bar{\omega} r_*}$ in the near-horizon region and as $\sim e^{\pm i\omega r_*}$ at infinity.

Therefore, near the surface of a \emph{ultra-compact} object, located at $r_*=r_*^{0}=r_*(R)$ (such that $|r_*^0|\gg M$, i.e. $R\approx r_+$) the Detweiler function therefore takes the form
%%%%%%%%%%%%%%%%%%%%%%%%%%%%%%%%%%%%%%%%%%%%%%%%%%%%%%%%%%%%%%%%%
\begin{equation}\label{detweiler_nearhorizon}
\,_{s}X_{\ell m}\sim
e^{-i\bar{\omega}(r_{*}-r_*^{0})}
+\mathcal{R}(\omega)\,
e^{i\bar{\omega}(r_{*}-r_*^{0})}\, .
\end{equation}
%%%%%%%%%%%%%%%%%%%%%%%%%%%%%%%%%%%%%%%%%%%%%%%%%%%%%%%%%%%%%%%%%
This plane-wave structure allows for a clean definition of the reflectivity,
%%%%%%%%%%%%%%%%%%%%%%%%%%%%%%%%%%%%%%%%%%%%%%%%%%%%%%%%%%%%%%%%%
\begin{align}\label{def_reflectivity}
\mathcal{R}(\omega)\equiv
\left[
\frac{1-\frac{i}{\bar{\omega}}\left(\frac{1}{X}\frac{dX}{dr_{*}}\right)}
     {1+\frac{i}{\bar{\omega}}\left(\frac{1}{X}\frac{dX}{dr_{*}}\right)}
\right]_{r_{*}=r_{*}^{0}} .
\end{align}
%%%%%%%%%%%%%%%%%%%%%%%%%%%%%%%%%%%%%%%%%%%%%%%%%%%%%%%%%%%%%%%%%
The ratio $(\mathbb{B}/\mathbb{A})$ appearing in~\ref{radialnear} can be related to $\mathcal{R}$ by expressing the radial Teukolsky function in ingoing null coordinates and using the transformation in~\ref{Teu_Det_Trans} between the Teukolsky and Detweiler functions. This relation will be exploited below to determine the LNs of ECOs with generic reflectivity.

As emphasized above and discussed in earlier sections, dynamical effects on the LNs can be treated analytically in the low-frequency regime. Accordingly, throughout this analysis we retain terms only up to linear order in the frequency. The same approximation applies to the reflectivity, which we parametrize as
%%%%%%%%%%%%%%%%%%%%%%%%%%%%%%%%%%%%%%%%%%%%%%%%%%%%%%%%%%%%%%%%%
\begin{equation}\label{reflectivity}
{\cal R}(\omega)
={\cal R}_{0}
+i M \omega\,{\cal R}_{1}
+{\cal O}(M^{2}\omega^{2}) \, ,
\end{equation}
%%%%%%%%%%%%%%%%%%%%%%%%%%%%%%%%%%%%%%%%%%%%%%%%%%%%%%%%%%%%%%%%%
where $M$ is the mass of the compact object. The dimensionless coefficients ${\cal R}_{i}$ encode information about the intrinsic properties of the object. We will show that the LNs in the static limit ($\omega\to 0$) depend on the coefficient ${\cal R}_{1}$ of the ${\cal O}(M\omega)$ term. This implies that, even when computing \emph{static} LNs, a consistent modeling of the tidal response requires knowledge of the frequency dependence of the reflectivity. Therefore, rather than treating ${\cal R}$ as a constant parameter, its frequency dependence plays a central role, with LNs providing a powerful probe of this behavior.

%%%%%%%%%%%%%%%%%%%%%%%%%%%
%%%%%%%%%%%%%%%%%%%%%%%%%%%
%%%%%%%%%%%%%%%%%%%%%%%%%%%
\subsubsection{Dynamical Love numbers of exotic compact objects}\label{dynamical_LN_ECO}

The starting point of this analysis is the radial Teukolsky equation for $\Psi_{4}$, as presented in~\ref{GMATE}, with near-zone ($M\omega z\ll1$) and small-frequency ($M\omega\ll1$) approximations. The solution of the same is in terms of hypergeometric functions, one of which behaves as $\Delta^{s}$ near the horizon, and is purely ingoing, while the other behaves as $e^{2i\bar{\omega}r_{*}}$ near the horizon and hence depicts the outgoing solution. For BHs, we would ignore the outgoing mode and hence derive the response function. On the other hand, for ECOs, there will be contributions from both the ingoing and the outgoing solutions, leading to dependence of the response function on the ratio $(\mathbb{B}/\mathbb{A})$, and hence the reflectivity. Written explicitly, the dynamical LNs of a Kerr-like ECO reads \citep{Chakraborty:2023zed}, %%%%%%%%%%%%%%%%%%%%%%%%%%%%%%%%%%%%%%%%%%%%%%%%%%%%%%%%%%%%%%%%%%%%%%%%%%%%%%%%%%%%%%%%%%%%%%%%%%%%%% 
\begin{align}\label{tidal_love_small} k^{\rm ECO}_{\ell m}=\frac{1}{2}\textrm{Re}\Bigg[\mathcal{F}_{\ell m}^{\rm BH} \left\{\frac{1-\frac{\mathbb{B}}{\mathbb{A}}\Gamma_{1}}{1+\frac{\mathbb{B}}{\mathbb{A}}\Gamma_{1}}\right\}&\Bigg]~; \qquad \Gamma_{1}=\frac{(\ell+2)!}{(\ell-2)!}\frac{\left(3+2iP_{+}\right)_{\ell-2}}{\left(-1-2iP_{+}\right)_{\ell+2}}~, \end{align} 
%%%%%%%%%%%%%%%%%%%%%%%%%%%%%%%%%%%%%%%%%%%%%%%%%%%%%%%%%%%%%%%%%%%%%%%%%%%%%%%%%%%%%%%%%%%%%%%%%%%%%% 
with the $\mathcal{F}_{\ell m}^{\rm BH}$ being the response function of a Kerr BH, and is given by~\ref{resp_func_arb_rot}. Note that the expression for $\Gamma_{1}$ involves the Pochhammer symbol, defined as: $z_{n}\equiv z(z+1)\times \cdots \times (z+n-1)$, and $P_{+}\equiv -2Mr_{+}\bar{\omega}(r_{+}-r_{-})^{-1}$. As evident, setting $\mathbb{B}=0$, i.e., no outgoing wave at the horizon, will reduce the above LNs for Kerr-like ECO to that of Kerr BH. For compact objects other than BHs, the outgoing mode at the horizon exists, and hence $\mathbb{B}$ is non-zero. Thus, the LNs are uniquely determined by the boundary conditions at the surface of the compact object, located at $r=R$, which we now discuss.

\paragraph{Love numbers of non-rotating exotic compact objects}~--- The first step in determining the dynamical LNs for ECOs is to express the quantity $\Gamma_{1}$, defined above, in terms of the frequency $\omega$, which in the non-rotating case reads 
%%%%%%%%%%%%%%%%%%%%%%%%%%%%%%%%%%%%%%%%%%%%%%%%%%%%%%%%%%%%%%%%%%%%%%%%%%%%%%%%%%%%%%%%%%%%%%%%%%%%%% 
\begin{align}\label{Gamma1_freq} \Gamma_{1} 
=\frac{3i(1-2iM\omega)}{M\omega(1+4M^{2}\omega^{2})(1+16M^{2}\omega^{2})}~. 
\end{align} 
%%%%%%%%%%%%%%%%%%%%%%%%%%%%%%%%%%%%%%%%%%%%%%%%%%%%%%%%%%%%%%%%%%%%%%%%%%%%%%%%%%%%%%%%%%%%%%%%%%%%%% 
The second step is to express the ratio $(\mathbb{B}/\mathbb{A})$ in terms of the Detweiler reflectivity $\mathcal{R}(\omega)$. Such a relation is straightforward to obtain in the non-rotating case, 
%%%%%%%%%%%%%%%%%%%%%%%%%%%%%%%%%%%%%%%%%%%%%%%%%%%%%%%%%%%%%%%%%%%%%%%%%%%%%%%%%%%%%%%%%%%%%%%%%%%%%% 
\begin{align}\label{Teu_Det_Rel} 
\frac{\mathbb{B}}{\mathbb{A}}&=\left(\frac{2M\omega}{3}\right)\left(\frac{i+2M\omega+16iM^{2}\omega^{2}+32M^{3}\omega^{3}}{2-iM\omega}\right) \mathcal{R(\omega)}e^{8\pi M\omega-2i\omega\left(r_{*}^{0}-2M\right)}~. 
\end{align} 
%%%%%%%%%%%%%%%%%%%%%%%%%%%%%%%%%%%%%%%%%%%%%%%%%%%%%%%%%%%%%%%%%%%%%%%%%%%%%%%%%%%%%%%%%%%%%%%%%%%%%% 
Given these two inputs, along with the result that in the non-rotating case $\mathcal{F}_{\ell m}^{\rm BH}\sim iM\omega$, we will now determine the static limit of the dynamical LNs for an ECO, and then shall present the results for dynamical LNs. 

Given the expression for $\Gamma_{1}$ in~\ref{Gamma1_freq} and the relation between the Teukolsky and Detweiler reflectivities in~\ref{Teu_Det_Rel}, we can easily obtain the zero-frequency limit of both of these expressions, by keeping the leading order terms in $M\omega$. This yields $\Gamma_{1}=(3i/M\omega)+6+\mathcal{O}(M\omega)$, while the ratio $(\mathbb{B}/\mathbb{A})$ reduces to 
%%%%%%%%%%%%%%%%%%%%%%%%%%%%%%%%%%%%%%%%%%%%%%%%%%%%%%%%%%%%%%%%%%%%%%%%%%%%%%%%%%%%%%%%%%%%%%%%%%%%%% 
\begin{align} 
\frac{\mathbb{B}}{\mathbb{A}}&=\frac{iM\omega\mathcal{R}(\omega)}{3}+\frac{M^2 \omega^2 \mathcal{R}(\omega)}{6}\left(3 + 16 i \pi + 8\epsilon + 8 \ln \epsilon\right)+\mathcal{O}(M^{3}\omega^{3})~. 
\end{align} 
%%%%%%%%%%%%%%%%%%%%%%%%%%%%%%%%%%%%%%%%%%%%%%%%%%%%%%%%%%%%%%%%%%%%%%%%%%%%%%%%%%%%%%%%%%%%%%%%%%%%%% 
Thus, keeping terms up to linear order in $M\omega$, the combination $(\mathbb{B}/\mathbb{A})\Gamma_{1}$, appearing in the expression for the LNs, becomes $(\mathbb{B}/\mathbb{A})\Gamma_{1}=-\mathcal{R}(\omega)+(iM\omega/2)\mathcal{R}(\omega)(7+16\pi i+8\ln \epsilon)$, where we have also ignored terms $\mathcal{O}(\epsilon)$, as we are working with ultra-compact objects with $\epsilon \ll 1$. Hence, the LN for the $\ell=2$ mode in the zero frequency limit becomes
%%%%%%%%%%%%%%%%%%%%%%%%%%%%%%%%%%%%%%%%%%%%%%%%%%%%%%%%%%%%%%%%%%%%%%%%%%%%%%%%%%%%%%%%%%%%%%%%%%%%%% 
\begin{align}\label{tidal_love_nonrot2} k^{\rm ECO}_{2}&=\lim_{\omega \to 0}\textrm{Re}\Bigg\{\left(\frac{iM\omega}{30}\right) \frac{1+\mathcal{R(\omega)}\left[1-\frac{iM\omega}{2}\left(7+16i\pi+8\ln \epsilon\right)\right]}{1-\mathcal{R}(\omega)\left[1-\frac{iM\omega}{2}\left(7+16i\pi+8\ln \epsilon\right)\right]}\Bigg\}~. 
\end{align} %%%%%%%%%%%%%%%%%%%%%%%%%%%%%%%%%%%%%%%%%%%%%%%%%%%%%%%%%%%%%%%%%%%%%%%%%%%%%%%%%%%%%%%%%%%%%%%%%%%%%% 
The above equation shows one of the most striking results: except for a single choice of the zero frequency reflectivity $\mathcal{R}_{0}$, where the above expression has a pole (to be discussed below), the LN identically vanishes in the zero-frequency limit for any other choice of $\mathcal{R}_{0}$, just as in the BH case. This includes all cases of partial reflection, i.e., $|{\cal R}_0|^2<1$. The above result might appear to the reader as a surprise, since even for a perfect-fluid star, which has nonzero LNs and viscosity, effectively corresponds to $|{\cal R(\omega)}|^2<1$. However, the viscosity would affect the reflectivity at ${\cal O}(M\omega)$, leaving ${\cal R}_0$ unaffected. It is the $\mathcal{R}_{0}$ which governs the non-zero value for the LNs. This subtle feature will be further clarified in the context of the membrane paradigm in~\ref{sec:membrane}. 

In the case $\mathcal{R}(\omega)=1+iM\omega \mathcal{R}_{1}$, i.e., $\mathcal{R}_{0}=1$,~\ref{tidal_love_nonrot2} has a pole and the LN for the $\ell=2$ mode, in the zero-frequency limit, becomes 
%%%%%%%%%%%%%%%%%%%%%%%%%%%%%%%%%%%%%%%%%%%%%%%%%%%%%%%%%%%%%%%%%%%%%%%%%%%%%%%%%%%%%%%%%%%%%%%%%%%%%% 
\begin{align}\label{tidal_love_nonrot_f} k^{\rm ECO}_2=\frac{2}{15} 
\textrm{Re}\Bigg[\frac{1}{-2\mathcal{R}_{1}+\left(7 + 16 i \pi + 8 \ln \epsilon\right)}\Bigg]~. 
\end{align} 
%%%%%%%%%%%%%%%%%%%%%%%%%%%%%%%%%%%%%%%%%%%%%%%%%%%%%%%%%%%%%%%%%%%%%%%%%%%%%%%%%%%%%%%%%%%%%%%%%%%%%% 
Therefore, the zero-frequency limit of the frequency-dependent LN differs from the strictly static LN, which we have reported in the previous section, as the branch of the solution to the Teukolsky equation with outgoing behavior close to the horizon becomes ill-defined in this limit. Therefore, we do not expect the LN in the zero frequency limit to match with the results derived in the previous section. Nonetheless, the zero-frequency limit presented here is more natural to consider as the coalescence of binary compact objects is intrinsically frequency-dependent. 

The fact that the above result does not correspond to the strictly static case is also clear from our derivation of~\ref{tidal_love_nonrot_f}, as we have kept terms $\mathcal{O}(M\omega)$ while computing the combination $(\mathbb{B}/\mathbb{A})\Gamma_{1}$, as well as we have taken $\mathcal{R}_1 \neq 0$ in the expression for the reflectivity $\mathcal{R}$, both of which were essential in order to have a non-zero static LN. Furthermore, the magnitude of the reflectivity, in the present context, with $\mathcal{R}_{0}=1$, reads $|\mathcal{R}(\omega)|=1-M\omega\,\textrm{Im}\mathcal{R}_{1}+\mathcal{O}(\omega^{2})$. Thus, if the linear-in-frequency reflectivity $\mathcal{R}_{1}$ is real, it follows that the object is perfectly reflecting. In general, of course $\mathcal{R}_{1}$ can be complex, and with $\textrm{Im}\mathcal{R}_{1}>0$, it follows that the reflectivity is smaller than unity. The other choice, namely $\textrm{Im}\mathcal{R}_{1}<0$ is not physical, since it leads to a reflectivity larger than unity. Thus, in order to have non-zero LNs for a compact objects it is essential that the zero-frequency reflectivity is unity, i.e., $\mathcal{R}_{0}=1$. As we will demonstrate later, this result holds generically for any non-dissipative compact objects, including perfect-fluid stars. 
This is further supported by previous works, involving perfectly reflecting compact objects (see, e.g., \citealt{Cardoso:2017cfl}), for which $\mathcal{R}_{0}=1$ and $\mathcal{R}_{1}=0$, yielding non-zero LNs. Also, it is natural to expect a non-dissipative object to have unit reflectivity for a zero-frequency wave, with no phase shift, leading to $\mathcal{R}_{0}=1$. 

The next important point is that the dynamical LN for $\ell=2$ displays a logarithmic behavior with respect to the compactness parameter $\epsilon$. Whenever $\epsilon$ is very small, so that $|\ln\epsilon|\gg |{\cal R}_1|$, the dynamical LN scales as
%%%%%%%%%%%%%%%%%%%%%%%%%%%%%%%%%%%%%%%%%%%%%%%%%%%%%%%%%%%%%%%%%%%%%%%%%%%%%%%%%%%%%%%%%%%%%%%%%%%%%% 
\begin{equation} 
k^{\rm ECO}_{2}=\frac{1}{60\ln\epsilon}\,. 
\end{equation} 
%%%%%%%%%%%%%%%%%%%%%%%%%%%%%%%%%%%%%%%%%%%%%%%%%%%%%%%%%%%%%%%%%%%%%%%%%%%%%%%%%%%%%%%%%%%%%%%%%%%%%% 
The above logarithmic scaling of the dynamical LN agrees with the discussions in~\ref{staticLNECO} for the strictly static LNs of various classes of ECOs; however, the numerical coefficients differ. 
The above logarithmic scaling of the LNs has also appeared in the effective field theory computations involving tidal effects, see \citep{Saketh:2023bul, Mandal:2023hqa} and also our previous discussions. Though the origin of these terms are entirely different. 

Consider now the properties of the dynamical LN for non-rotating reflective compact objects, associated with the $\ell=2$ mode. Combining the expressions for $\Gamma_1$ and the ratio $(\mathbb{B}/\mathbb{A})$ from~\ref{Gamma1_freq} and~\ref{Teu_Det_Rel}, we obtain, 
%%%%%%%%%%%%%%%%%%%%%%%%%%%%%%%%%%%%%%%%%%%%%%%%%%%%%%%%%%%%%%%%%%%%%%%%%%%%%%%%%%%%%%%%%%%%%%%%%%%%%% 
\begin{align}\label{nonrotfreq} 
k_{2}&=\frac{1}{2}\textrm{Re}\Bigg[\mathcal{F}_{\ell m}^{\rm BH}\left(\frac{1+\mathcal{R}(\omega)G(\omega)}{1-\mathcal{R}(\omega)G(\omega)}\right)\Bigg]~, 
\end{align} 
%%%%%%%%%%%%%%%%%%%%%%%%%%%%%%%%%%%%%%%%%%%%%%%%%%%%%%%%%%%%%%%%%%%%%%%%%%%%%%%%%%%%%%%%%%%%%%%%%%%%%% 
where $\mathcal{F}_{\ell m}^{\rm BH}$ is the response function associated with a BH, and we have introduced a frequency-dependent quantity $G(\omega)$, defined as 
%%%%%%%%%%%%%%%%%%%%%%%%%%%%%%%%%%%%%%%%%%%%%%%%%%%%%%%%%%%%%%%%%%%%%%%%%%%%%%%%%%%%%%%%%%%%%%%%%%%%%% 
\begin{align} G(\omega)&\equiv 2\frac{\exp\left[8\pi M\omega-4iM\omega\left(\epsilon+\ln \epsilon \right)\right]}{2-iM\omega} \left(\frac{1-2iM\omega+16M^{2}\omega^{2}-32iM^{3}\omega^{3}}{1+2iM\omega+16M^{2}\omega^{2}+32iM^{3}\omega^{3}}\right)~, 
\end{align} 
%%%%%%%%%%%%%%%%%%%%%%%%%%%%%%%%%%%%%%%%%%%%%%%%%%%%%%%%%%%%%%%%%%%%%%%%%%%%%%%%%%%%%%%%%%%%%%%%%%%%%% 
which in the zero-frequency limit becomes unity, recovering~\ref{tidal_love_nonrot2}.

We will now depict a connection between the resonances in the dynamical LNs for compact objects with the QNM frequencies. As evident from~\ref{nonrotfreq}, it follows that the dynamical LNs for the $\ell=2$ mode will exhibit resonances whenever $\mathcal{R}(\omega)G(\omega)=1$. On the other hand, following \citet{Maggio:2018ivz}, the real part of the QNMs for a reflective compact object, in the small-$\epsilon$ limit, can be written as
%%%%%%%%%%%%%%%%%%%%%%%%%%%%%%%%%%%%%%%%%%%%%%%%%%%%%%%%%%%%%%%%%%%%%%%%%%%%%%%%%%%%%%%%%%%%%%%%%%%%%%
\begin{equation}\label{QNManalytics} 
\omega_{\rm R}\simeq -\frac{\pi(q+1)}{2|r_*^0| } \,.
\end{equation} 
%%%%%%%%%%%%%%%%%%%%%%%%%%%%%%%%%%%%%%%%%%%%%%%%%%%%%%%%%%%%%%%%%%%%%%%%%%%%%%%%%%%%%%%%%%%%%%%%%%%%%% 
The quantity $q$ appearing in the above expression takes the form $q=2n-1$ (resp., $q=2n$) for perfectly reflecting objects with $\mathcal{R}_0=1$ (resp., $\mathcal{R}_0=-1$) with $n \geq 1$ being the overtone number of the QNMs. 

Intriguingly, one can easily check that the combination $\mathcal{R}(\omega)G(\omega)$ becomes closest to unity at precisely the frequencies in~\ref{QNManalytics}, leading to resonances. For smaller values of the reflectivity, on the other hand, resonances will happen as $G(\omega)$ becomes larger, which in turn implies that these resonances will happen at larger values of $M\omega$. Therefore, as the reflectivity becomes very small, the resonant frequencies will become so large that the small-frequency approximation will break down, and hence the above analysis is invalid in that regime.

\begin{figure}[!t] 
\centering
\includegraphics[width=0.7\columnwidth]{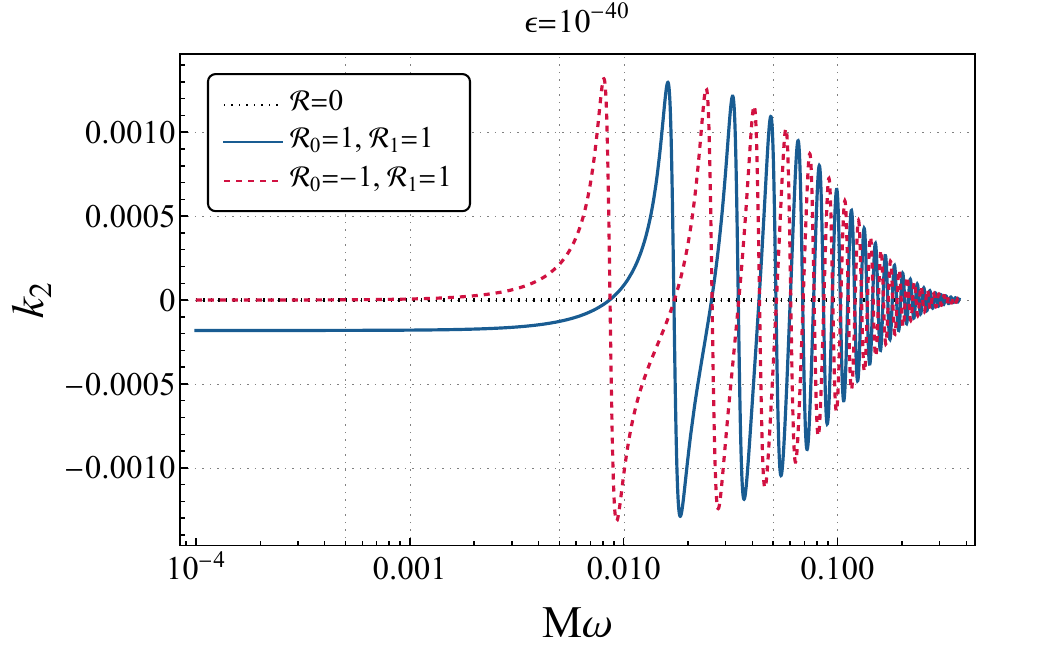}
\caption{
Quadrupolar electric LN as a function of frequency for three different cases: 
(a) a BHs corresponding to $\mathcal{R}=0$; 
(b) a compact object with $\mathcal{R}_{0}=1$; and 
(c) a compact object with $\mathcal{R}_{0}=-1$. 
In the latter two cases we have set $\epsilon=10^{-40}$ and $\mathcal{R}_{1}=1$. 
As evident from the plot, only for $\mathcal{R}_{0}=1$ does the LN approach a nonvanishing value in the zero-frequency limit. 
By contrast, even for $\mathcal{R}_{0}=-1$ --~which still corresponds to a perfectly reflecting object~-- the LN vanishes as the frequency tends to zero. 
In both cases with $\mathcal{R}_{0}\neq 0$, the LN increases with increasing frequency, and in general the frequency-dependent LN does not vanish.
The resonances correspond to the real part of the QNM, see~\ref{QNManalytics}.
From \citet{Chakraborty:2023zed}.
}
\label{fig:LoveResonances}
\end{figure}

\paragraph{Love numbers of rotating exotic compact objects}~--- Having understood the features associated with static as well as dynamical LNs in the case of a non-rotating compact object, we will now briefly touch upon the LNs in the spinning case. The general expression has been provided in~\ref{tidal_love_small}, but for brevity, we will quote here results for the most relevant $\ell=2$ mode. For this case, the tidal response function becomes \citep{Chakraborty:2023zed}, 
%%%%%%%%%%%%%%%%%%%%%%%%%%%%%%%%%%%%%%%%%%%%%%%%%%%%%%%%%%%%%%%%%%%%%%%%%%%%%%%%%%%%%%%%%%%%%%%%%%%%%%
\begin{align}\label{tln_rot_freq} 
k_{2m}&=\textrm{Re}\Bigg[\frac{1}{2}\mathcal{F}_{\ell m}^{\rm BH} \left\{\frac{1+\left(\frac{\mathbb{B}}{\mathbb{A}}\right)\frac{6i\left(1+iP_{+}\right)}{P_{+}\left(1+4P_{+}^{2}\right)\left(1+P_{+}^{2}\right)}}{1-\left(\frac{\mathbb{B}}{\mathbb{A}}\right)\frac{6i\left(1+iP_{+}\right)}{P_{+}\left(1+4P_{+}^{2}\right)\left(1+P_{+}^{2}\right)}}\right\}\Bigg]~, 
\end{align} 
%%%%%%%%%%%%%%%%%%%%%%%%%%%%%%%%%%%%%%%%%%%%%%%%%%%%%%%%%%%%%%%%%%%%%%%%%%%%%%%%%%%%%%%%%%%%%%%%%%%%%% 
where $P_{+}=\{(am-2M\omega r_{+})/(r_{+}-r_{-})\}$ as defined below~\ref{tidal_love_small}.  In the rotating case, the analytical expression for such a relation between the ratio $(\mathbb{B}/\mathbb{A})$ and the reflectivity $\mathcal{R}$ is cumbersome and hence will not be reported here. The LNs for a rotating compact object depends on the azimuthal number $m$ as well, through $P_{+}$. Hence, even in the static limit, $P_{+}$ remains non-zero, leading to non-zero LNs. Of course, the exact zero-frequency limit of the LN will depend on the relation between the Teukolsky and the Detweiler function in that limit. In~\ref{fig:LoveResonances} we present the $\ell=2$ LN of a rotating compact object as a function of the frequency.

%%%%%%%%%%%%%%%%%%%%%%%%%%%
%%%%%%%%%%%%%%%%%%%%%%%%%%%
%%%%%%%%%%%%%%%%%%%%%%%%%%%
\subsubsection{Strictly static Love numbers of non-rotating exotic compact objects}\label{perfrefl}
 
As we have demonstrated, the static limit ($\omega\to0$) of the LNs does not coincide with the LNs obtained from strictly static ($\omega=0$) perturbations of the compact object, even though the frequency-dependent LNs remain continuous throughout the low-frequency regime. This somewhat counterintuitive feature has also been observed in \citet{Pani:2018inf} in the context of the magnetic LNs of perfect-fluid compact objects. In particular, \citet{Pani:2018inf} showed that the static limit of the dynamical LNs (there referred to as irrotational LNs) for neutron stars does not agree with the strictly static LNs (see~\ref{dyntideNS}). This behavior is consistent with earlier results in the literature \citep{Binnington:2009bb,Damour:2009vw} and is related to the existence of two distinct types of magnetic LNs (static and irrotational) for neutron stars.

In the present context, the mismatch between the strictly static LNs and the static limit of the dynamical LNs for ECOs arises from the following observation. The solutions of the Teukolsky equation for a rotating ECO admit both ingoing and outgoing behavior near the horizon. While the ingoing solution possesses a well-defined static limit, the outgoing solution in the $\omega\neq0$ case does not continuously approach the corresponding solution in the strictly static problem. This issue does not affect BHs, since in the BH limit only the ingoing solution is physically relevant, and its zero-frequency limit coincides with the strictly static one. Consequently, for BHs the static limit of the dynamical LNs agrees with the strictly static LNs. The discrepancy between the static limit and the strictly static case therefore originates from the outgoing (or reflected) component of the perturbations, both for ECOs and for neutron stars. In both situations, the static and strictly static LNs do not coincide. This point was also discussed in \citet{LeTiec:2020bos}, where it was emphasized that the two independent solutions of the Teukolsky equation in the dynamical setting cease to be independent in the static limit. This suggests that strictly static LNs must be computed independently and need not agree with the continuous $\omega\to0$ limit of the dynamical LNs.

We also emphasize that, in the $\omega\to0$ limit, only compact objects with unit Detweiler reflectivity, ${\cal R}(\omega\to0)=1$, possess nonvanishing LNs. The static limit considered here therefore corresponds effectively to a perfectly reflecting compact object. We will demonstrate this explicitly through the LN computation below.

The strictly static LNs for a non-rotating compact object must be derived from first principles. To this end, we consider the zero-rotation and zero-frequency limit of the general Teukolsky equation given in~\ref{Teqr}. Introducing the variable $z=(r/2M)-1$, the equation reduces to
\begin{align}\label{teuk_static_n}
\dfrac{d^{2}R_{\ell}}{dz^{2}}
-\left(\frac{1}{z}+\frac{1}{1+z}\right)\dfrac{dR_{\ell}}{dz}
+\left(-\frac{1}{z}+\frac{1}{1+z}\right)\gamma_{\ell}R_{\ell}=0~,
\end{align}
where $\gamma_{\ell}\equiv(\ell+2)(\ell-1)$. Since this equation is independent of $m$, the radial function is likewise independent of $m$, and we denote it simply by $R_{\ell}$. Solving this equation, the far-zone behavior of the radial Teukolsky function is
\begin{align}
R_{\ell}\Big|_{\rm far}
=c_{1}\frac{\Gamma(1+2\ell)}{\Gamma(\ell-1)\Gamma(1+\ell)}
\left(\frac{r}{2M}\right)^{\ell+2}
\left[1+\mathcal{F}_{\ell}\left(\frac{r}{2M}\right)^{-2\ell-1}\right]~.
\end{align}
Here, $\mathcal{F}_{\ell}$ is the (purely real) tidal response function and scales as $\sim(c_{2}/c_{1})$. For BHs, regularity at the horizon requires $c_{2}=0$, implying that the static BH LNs vanish identically. This shows that the ambiguity in defining static versus strictly static LNs does not arise for BH spacetimes.

For non-rotating ECOs, by contrast, one must impose appropriate boundary conditions on the Regge--Wheeler or Zerilli functions. For instance, in the perfectly reflecting case, both the Zerilli and Regge--Wheeler functions vanish at the surface of the compact object. This requires relating the Teukolsky function to the Regge--Wheeler and Zerilli functions in the zero-frequency limit. To do so, we decompose the radial Teukolsky function as $R_{\ell}=R_{\ell}^{\rm axial}+R_{\ell}^{\rm polar}$ and express these components in terms of the Regge--Wheeler and Zerilli functions,
\begin{align}
\frac{R_{\ell}^{\rm axial}}{\sqrt{\gamma_{\ell}(2+\gamma_{\ell})}}
&=\frac{r^{3}}{8}
\Bigg[V^{\rm axial}\Psi^{\rm RW}_{\ell}
+W^{\rm axial}\left(\dfrac{d\Psi_{\ell}^{\rm RW}}{dr_{*}}\right)\Bigg]~,
\label{axial_teu_RW}
\\
\frac{R_{\ell}^{\rm polar}}{\sqrt{\gamma_{\ell}(2+\gamma_{\ell})}}
&=\frac{r^{3}}{8}
\left[V^{\rm polar}\Psi^{\rm Z}_{\ell}
+W^{\rm polar}\left(\dfrac{d\Psi^{\rm Z}_{\ell}}{dr_{*}}\right)\right]~,
\label{polar_Teu_Z}
\end{align}
where $\gamma_{\ell}=(\ell+2)(\ell-1)$. The functions appearing above are
\begin{align}
V^{\rm axial}&=V_{\rm RW}
=\frac{(r-2M)\left[(\gamma_{\ell}+2)r-6M\right]}{r^{4}}~,
\qquad
W^{\rm axial}=\frac{2(r-3M)}{r^{2}}~,
\\
V^{\rm polar}&=V_{\rm Z}
=\frac{(r-2M)}{r^{4}(\gamma_{\ell}r+6M)^{2}}
\Big[\gamma_{\ell}^{2}(\gamma_{\ell}+2)r^{3}
+6M\gamma_{\ell}^{2}r^{2}+36M^{2}\gamma_{\ell}r+72M^{3}\Big]~,
\\
W^{\rm polar}&=\frac{2\gamma_{\ell}r^{2}-6\gamma_{\ell}Mr-12M^{2}}
{r^{2}(\gamma_{\ell}r+6M)}~.
\end{align}
Here, $V_{\rm RW}$ and $V_{\rm Z}$ denote the Regge--Wheeler and Zerilli potentials for axial and polar perturbations, respectively. The computation of the LNs further requires relations connecting the Regge--Wheeler and Zerilli functions to the axial and polar metric perturbations $h_{0}$ and $H_{0}$. In this way, the Teukolsky function can be directly related to the metric perturbations in the zero-frequency limit.

We begin with the polar sector. Using the relations above, one finds \citep{Chakraborty:2023zed}
\begin{align}
R_{\ell}^{\rm polar}
=\frac{r(r-2M)}{4}\sqrt{\gamma_{\ell}(\gamma_{\ell}+2)}\,H_{0}~.
\end{align}
In terms of the variable $z$, this implies $R_{\ell}^{\rm polar}\propto z(1+z)H_{0}$. Using the static solution of the Teukolsky equation in~\ref{sol_zero}, the polar metric perturbation $H_{0}$ can be expressed in terms of associated Legendre functions $P_{\ell}^{2}$ and $Q_{\ell}^{2}$. Imposing the boundary condition $\Psi_{\ell}^{\rm Z}(r_{*}^{0})=0$, as in \citet{Cardoso:2017cfl}, yields the ratio $(c_{2}/c_{1})$ and hence the LN. For $\ell=2$ one finds
\begin{align}
k_{2}^{\rm E}
=\frac{1}{20\left(7+3\ln\epsilon\right)}~,
\end{align}
in agreement with \citep{Cardoso:2017cfl}, and displaying the characteristic logarithmic dependence on the compactness.

For completeness, we also consider the axial sector. In the static limit, the Regge--Wheeler function $\Psi_{\rm RW}$ and the axial metric perturbation $h_{0}$ are related by
\begin{align}
\Psi_{\ell}^{\rm RW}
=\frac{r^{3}}{\gamma_{\ell}}
\dfrac{d}{dr}\left(\frac{h_{0}}{r^{2}}\right)~.
\end{align}
Using this relation together with~\ref{h0stateq}, one finds 
\begin{align}
V^{\rm axial}\Psi^{\rm RW}_{\ell}
+W^{\rm axial}\left(\dfrac{d\Psi_{\ell}^{\rm RW}}{dr_{*}}\right)
=\frac{(r-2M)}{r^{2}}\dfrac{dh_{0}}{dr}
-\frac{2M}{r^{3}}h_{0}~,
\end{align}
so that
\begin{align}\label{axial_staticrad}
R_{\ell}^{\rm axial}
=\sqrt{\gamma_{\ell}(2+\gamma_{\ell})}
\left[\frac{r(r-2M)}{8}\dfrac{dh_{0}}{dr}
-\frac{M}{4}h_{0}\right]~.
\end{align}
For generic $\ell$, the axial metric perturbation involves Meijer-$G$ functions and has a complicated structure. For $\ell=2$, however, the solution simplifies to
\begin{align}
h_{0}
&=c_{1}r^{2}(r-2M)
+\frac{c_{2}}{24M^{5}r}
\Big[4M^{4}+4M^{3}r+6M^{2}r^{2}-6Mr^{3}
\nonumber\\
&\hspace{2cm}
+6Mr^{3}\ln\!\left(1-\frac{2M}{r}\right)
-3r^{4}\ln\!\left(1-\frac{2M}{r}\right)\Big]~.
\end{align}
One can verify that this expression satisfies the static Einstein equations,~\ref{h0stateq}. Imposing the Dirichlet boundary condition $\Psi^{\rm RW}_{2}(r_{*}^{0})=0$ yields
\begin{align}
k_{2}^{\rm B}
=\frac{1}{5\left(25+12\ln\epsilon\right)}~,
\end{align}
which again agrees with \citep{Cardoso:2017cfl} and exhibits the characteristic $\ln\epsilon$ dependence.

In summary, solving the strictly static ($\omega=0$) Teukolsky equation yields LNs with a logarithmic dependence on the compactness, in agreement with earlier results in the literature. However, these strictly static LNs differ from the dynamical LNs evaluated in the static limit ($\omega\to0$), even though the logarithmic dependence on $\epsilon$ persists in both cases. Since binary inspirals are intrinsically dynamical processes, the static limit of the dynamical LNs provides the physically relevant description, whereas the strictly zero-frequency case corresponds to a measure-zero point in the frequency domain.

%%%%%%%%%%%%%%%%%%%%%%%%%%%%%%%%%%%%%%%%%%%%%%%%%%%%%%%%%%%%%
%%%%%%%%%%%%%%%%%%%%%%%%%%%%%%%%%%%%%%%%%%%%%%%%%%%%%%%%%%%%%
%%%%%%%%%%%%%%%%%%%%%%%%%%%%%%%%%%%%%%%%%%%%%%%%%%%%%%%%%%%%%
\subsection{Tidal heating for exotic compact objects}

We have already discussed the effect of tidal heating for BHs, and in particular we have presented the rate of change of the BH mass due to the absorption of GWs through the horizon. In this section, we summarize the corresponding tidal heating for ECOs, or, equivalently, for compact objects endowed with a reflective surface. As discussed above, the LNs of an ECO depend nontrivially on the reflectivity and, in particular, require the static limit of the reflectivity to be unity. As we will show, efficient tidal heating instead requires precisely the opposite behavior, namely that the reflectivity be as different from unity as possible. Thus, LNs and tidal heating are complementary effects. The dominant (static) LNs vanish for BHs but can be large for compact objects with highly reflective surfaces, such as neutron stars and ECOs. Conversely, tidal heating is maximal for BHs and is strongly suppressed for compact objects with highly reflective surfaces.

Our starting point is again the Teukolsky equation in the small-frequency and near-zone regime. Within these approximations, the Teukolsky equation in Boyer--Lindquist coordinates can be solved exactly, yielding the solution reported in~\ref{radialpsi0}. Out of the two integration constants appearing in this solution, one vanishes in the BH case, leading to the tidal heating discussed in~\ref{heatingBH}. In the present context, using the same solution~\ref{radialpsi0} for the $\ell=2$ mode, we recall that $A_{m}$ corresponds to the outgoing wave, while $C_{m}$ describes the ingoing wave at the horizon. By rewriting the solution using appropriate hypergeometric identities, and defining $(C_{m}/C_{m}^{\rm BH})\equiv\mathbb{T}$ as the transmission coefficient and $(A_{m}/C_{m}^{\rm BH})\equiv\mathbb{R}$ as the reflection coefficient, we obtain
%%%%%%%%%%%%%%%%%%%%%%%%%%%%%%%%%%%%%%%%%%%%%%%%%%%%%%%%%%%%%%%%%%%%%%%%%%%%%%
\begin{align}
{}_{2}R^{\rm ECO}_{2m}
&=(1+z)^{iP_{+}}C_{m}^{\rm BH}\Big[
\mathbb{R}z^{-iP_{+}}\,{}_{2}F_{1}\!\left(0,5;3-2iP_{+};-z\right)
\nonumber\\
&\qquad
+\mathbb{T}z^{-2+iP_{+}}(1+z)^{-2-2iP_{+}}\,
{}_{2}F_{1}\!\left(1,-4;-1+2iP_{+};-z\right)
\Big]~.
\end{align}
%%%%%%%%%%%%%%%%%%%%%%%%%%%%%%%%%%%%%%%%%%%%%%%%%%%%%%%%%%%%%%%%%%%%%%%%%%%%%%
By comparing the normalization between the BH and ECO solutions, it follows that $|\mathbb{R}|^{2}+|\mathbb{T}|^{2}=1$. The quantity $C_{m}^{\rm BH}$ is given explicitly in~\ref{heatingBH}. Using the above radial solution, the Weyl scalar in the Hartle-Hawking tetrad can be computed, yielding
%%%%%%%%%%%%%%%%%%%%%%%%%%%%%%%%%%%%%%%%%%%%%%%%%%%%%%%%%%%
\begin{align}
\Psi_{0\,({\rm ECO})}^{\rm HH}
&=\frac{\Delta^{2}}{4(r^{2}+a^{2})^{2}}
\sum_{m=-2}^{2} C_{m}^{\rm BH}\,
{}_{2}Y_{2m}(\theta,\phi)
\nonumber\\
&\qquad\times
\Big[
\mathbb{R}(1+z)^{iP_{+}}z^{-iP_{+}}
+\mathbb{T}z^{iP_{+}-2}(1+z)^{-2-iP_{+}}\,
{}_{2}F_{1}(1,-4;-1+2iP_{+};-z)
\Big]~.
\end{align}
%%%%%%%%%%%%%%%%%%%%%%%%%%%%%%%%%%%%%%%%%%%%%%%%%%%%%%%%%%%
All radial quantities are to be evaluated at the surface location $R=r_{+}(1+\epsilon)$. Expanding the Weyl scalar and the remaining relevant quantities in powers of the compactness parameter $\epsilon$, we obtain, to leading order,
%%%%%%%%%%%%%%%%%%%%%%%%%%%%%%%%%%%%%%%%%%%%%%%%%%%%%%%%%%%
\begin{align}
\left(\frac{dM}{dt}\right)_{\rm ECO}
=\left(1-|\mathbb{R}|^{2}\right)
\left(\frac{dM}{dt}\right)_{\rm BH}~.
\end{align}
%%%%%%%%%%%%%%%%%%%%%%%%%%%%%%%%%%%%%%%%%%%%%%%%%%%%%%%%%%%
Therefore, for $\mathbb{R}=1$ the LNs are non-zero but tidal heating vanishes, whereas in the opposite limit $\mathbb{R}\to0$ tidal heating is non-negligible while the LNs vanish.

%%%%%%%%%%%%%%%%%%%%%%%%%%%%%%%%%%%%%%%%%%%%
\subsection{Membrane paradigm: unifying Love numbers of compact objects}\label{sec:membrane}
%%%%%%%%%%%%%%%%%%%%%%%%%%%%%%%%%%%%%%%%%%%%
In this section we present a complementary and largely model-independent approach to computing the tidal response of compact objects, based on the membrane paradigm. Originally developed for BHs \citep{Damour:1982,Thorne:1986iy,Price:1986yy}, the membrane paradigm has more recently been extended to a variety of horizonless compact objects \citep{Maggio:2020jml,Sherf:2021ppp,Chakraborty:2022zlq}.

Within this framework, a static observer outside the BH horizon replaces the BH interior with a \emph{fictitious} membrane located just outside the horizon, commonly referred to as the stretched horizon. The physical properties of the spacetime inside the horizon are assumed to be encoded on this hypothetical membrane and are related to the exterior spacetime through the Israel-Darmois junction conditions \citep{Darmois1927,Israel:1966rt,VisserBook}. An analogous construction applies to compact objects other than BHs, where the interior of the object is similarly replaced by a membrane located at the surface of the object. Through the junction conditions, the properties of the membrane --~equivalently, those of the underlying compact object~-- are related to observable quantities in the exterior spacetime.

Remarkably, the fictitious membrane can be described as a viscous fluid endowed with an effective density, pressure, shear viscosity, and bulk viscosity. These fluid properties are uniquely fixed by the requirement that the membrane reproduce the observable behavior of the original compact object it replaces \citep{Thorne:1986iy,Jacobson:2011dz}. 

For simplicity, we restrict our discussion to GR and to spherically symmetric background geometries. In this case, the spacetime exterior to the membrane is described by the Schwarzschild metric (see \citealt{Saketh:2024ojw} for an extension to slowly spinning geometries). No specific assumptions are made about the interior geometry of the compact object; instead, the membrane serves as an effective proxy that encodes the interior physics relevant for the exterior tidal response.

%%%%%%%%%%%%%%%%%%%%%%%%%%%%%%%%%%%%%%%%%%%%%
%%%%%%%%%%%%%%%%%%%%%%%%%%%%%%%%%%%%%%%%%%%%%
%%%%%%%%%%%%%%%%%%%%%%%%%%%%%%%%%%%%%%%%%%%%%
\subsubsection{Love numbers of compact objects: General analysis}\label{sec:membrane_bis}
In this section we briefly review the membrane paradigm by replacing a spherically symmetric compact object of mass $M$ with a fictitious membrane, generated by a set of timelike worldlines placed at a fixed radius $R=2M (1+\epsilon)$. The presence of timelike worldlines ensures that the membrane, located at $R>2M$, forms a spacelike hypersurface. The parameter $\epsilon$ is positive definite and quantifies the proximity of the membrane to the would-be BH horizon. For ultra-compact objects, one has $\epsilon\ll1$, while for NSs $\epsilon$ can take arbitrary values\footnote{The parameter $\epsilon$ is also related to the compactness $\mathcal{C}$ of the object via $\mathcal{C}=(1/2)(1+\epsilon)^{-1}$. In the BH limit ($\epsilon\to 0$), one recovers the Schwarzschild BH compactness, $\mathcal{C}=1/2$.}. Accordingly, we will provide results for generic $\epsilon$, relevant for NSs, as well as expansions in $\epsilon\ll1$, applicable to ultra-compact objects.

As emphasized earlier, in the membrane paradigm the properties of the object's interior are projected onto the membrane, and are connected to the exterior geometry through the Israel-Darmois junction conditions \citep{Darmois1927,Israel:1966rt,VisserBook}:
\begin{equation}
[[h_{ab}]]=0\,,\qquad [[K_{ab}-h_{ab}K]]=-8\pi T_{ab}~.
\label{condi}
\end{equation}
The first junction condition guarantees the continuity of the induced metric on the membrane, $h_{ab}=(g_{\mu \nu}-n_{\mu}n_{\nu})e^{\mu}_{a}e^{\nu}_{b}$, while the second condition relates the discontinuity in the extrinsic curvature $K_{ab}$ (and its trace $K=K_{ab}h^{ab}$) to the stress-energy tensor $T_{ab}$ of the membrane fluid. Here, $[[\cdot]]$ denotes the jump across the hypersurface, and $e^{\mu}_{a}$ is the projector from the four-dimensional spacetime onto the membrane.

The energy-momentum tensor of the viscous membrane fluid is
\begin{equation}
T_{ab}= \rho u_a u_b +(p-\zeta \Theta) \gamma_{ab}- 2\eta \sigma_{ab}\,,
\label{T}
\end{equation}
where $\rho$ and $p$ are the fluid density and pressure, and $u_a$ is the 3-velocity, obtained from the 4-velocity $U_\mu$ via $u_a=e^\mu_a U_\mu$. The expansion is $\Theta=D_{a}u^a$, and the shear tensor is defined as 
\[
\sigma_{ab} = \frac{1}{2}\left(D_c u_a \gamma^c{}_b + D_c u_b \gamma^c{}_a - \Theta \gamma_{ab}\right),
\] 
where $\gamma_{ab}=h_{ab}+u_a u_b$ is the induced metric on the 2-surface orthogonal to $u_a$, and $D_a$ is the covariant derivative compatible with $h_{ab}$. The coefficients $\eta$ and $\zeta$ are the shear and bulk viscosities, respectively, which govern the fluid's response to external perturbations. We note that, due to the \emph{fictitious} nature of the membrane, its properties need not resemble those of ordinary materials; for instance, the bulk viscosity associated to a BH within the membrane paradigm is \emph{negative}, reflecting the acausal nature of the membrane \citep{Thorne:1986iy}.

Since spacetime perturbations are generally dynamical, the membrane properties are expected to be frequency-dependent. Accordingly, we assume that, in addition to its location, the viscosity coefficients depend on frequency: $\eta=\eta(\epsilon,\omega)$ and $\zeta=\zeta(\epsilon,\omega)$, a feature that will be crucial in our analysis. Finally, following the standard membrane paradigm, we take the interior extrinsic curvature to vanish, $K^-_{ab}=0$ \citep{Damour:1982,Thorne:1986iy,Price:1986yy}, even for compact objects other than BHs.

%%%%%%%%%%%%%%%%%%%%%%%%%%%
%%%%%%%%%%%%%%%%%%%%%%%%%%%
%%%%%%%%%%%%%%%%%%%%%%%%%%%
\paragraph{Geometry of the background spacetime}~--- 
Since we are interested in static and spherically symmetric spacetimes within vacuum GR, the exterior metric is simply the Schwarzschild solution. In this geometry, the unit normal to the $r=\text{constant}$ hypersurfaces is 
\[
n_{\mu} = \frac{1}{\sqrt{f(r)}} \, \delta^r_\mu~,
\] 
where $f(r)=-g_{00}$.
The components of the induced metric on the membrane, parametrized by $(t,\theta,\phi)$, are then
\[
h_{ab} = \text{diag}\Big(-f({{{R}}}),\, {{{R}}}^2,\, {{{R}}}^2 \sin^2\theta \Big)~,
\] 
where Latin indices $a\in(t,\theta,\phi)$ denote three-dimensional quantities. The three-velocity of the membrane fluid is 
\[
u_a = -\sqrt{f({{{R}}})} \, \delta^t_a~.
\] 
The extrinsic curvature, evaluated from the Schwarzschild exterior, has the form
\begin{equation}
K^+_{ab} = \text{diag}\Big(-\frac{1}{2}\sqrt{f}\, f',\, r\sqrt{f},\, r\sqrt{f} \sin^2\theta \Big)_{{{{R}}}}~.
\end{equation}

Denoting the unperturbed energy density and pressure of the membrane fluid by $\rho_0$ and $p_0$, respectively, the corresponding energy-momentum tensor from~\ref{T} reduces to
\begin{equation}
T_{ab} = \text{diag}\Big( \rho_0 f,\, r^2 p_0,\, r^2 \sin^2\theta \, p_0 \Big)_{r={{{R}}}}~.
\end{equation}
Note that in this static, spherically symmetric background, the dissipative terms involving $\eta$ and $\zeta$ do not contribute; the viscosities are activated only by time-dependent perturbations, which will be discussed below in the context of the LNs.

The junction conditions,~\ref{condi}, yield
\begin{align}
\rho_0 &= -\frac{\sqrt{f({{{R}}})}}{4\pi {{{R}}}}~, 
\qquad
p_0 = \frac{2 f({{{R}}}) + {{{R}}} f'({{{R}}})}{16 \pi {{{R}}} \sqrt{f({{{R}}})}}~.
\label{rho0p0}
\end{align}
It follows that, in the limit ${{{R}}} \to 2M$ (i.e., $\epsilon \to 0$), the energy density vanishes while the pressure diverges. This behavior arises because a static observer cannot exist exactly at the horizon, and thus a static fluid element is impossible at that location.

%%%%%%%%%%%%%%%%%%%%%%
%%%%%%%%%%%%%%%%%%%%%%
%%%%%%%%%%%%%%%%%%%%%%
\paragraph{Perturbing the membrane: Basic equations}~---
The gravitational perturbations of the exterior spacetime have already been discussed in previous sections, and we will not repeat them here. Instead, we focus on the perturbations of the membrane and the junction conditions, which are related to the Schwarzschild metric perturbations and provide the appropriate boundary conditions for the metric components $\delta g_{t\phi}$ and $\delta g_{tt}$.

Perturbations of the Schwarzschild geometry modify the extrinsic curvature of the exterior, and consequently the junction conditions themselves are perturbed. For consistency, one must therefore consider perturbations of the physical quantities associated with the membrane. Unlike the unperturbed case, where the dissipative contributions of the membrane fluid do not affect the energy-momentum tensor, in the perturbed spacetime these dissipative components become relevant. Due to the spacetime perturbations, the pressure $p$, the density $\rho$, and the radial position of the membrane $R_{\textrm{m}}$ are perturbed:
\begin{align}
\rho &= \rho_0 + \delta \rho(t,{{{R}}},\theta)\,, \qquad
p = p_0 + \delta p(t,{{{R}}},\theta)\,, \qquad
R_{\textrm{m}} = R + \delta R(t,{{{R}}},\theta)~.
\end{align}
The perturbed density and pressure are evaluated at the perturbed location $R_{\textrm{m}}$, but the contribution of $\delta R$ enters only at second order, so we evaluate all perturbations at the unperturbed membrane radius ${{{R}}}$. Moreover, $\delta \rho$, $\delta p$, and $\delta R$ are scalars, and thus contribute only to polar perturbations. 

Given the spherical symmetry of the background, these perturbations can be decomposed as
\begin{align}
\delta \rho(t,{{{R}}},\theta) &= \int d\omega \sum_\ell \rho_1({{{R}}}) P_\ell(\cos\theta) e^{-i\omega t}~,\\
\delta p(t,{{{R}}},\theta) &= \int d\omega \sum_\ell p_1({{{R}}}) P_\ell(\cos\theta) e^{-i\omega t}~,\\
\delta R(t,{{{R}}},\theta) &= \int d\omega \sum_\ell R_1({{{R}}}) P_\ell(\cos\theta) e^{-i\omega t}~,
\end{align}
where $\rho_1$, $p_1$, and $R_1$ depend on the unperturbed radius ${{{R}}}$ and the multipole index $\ell$, which is suppressed for brevity. 

Since the membrane location is perturbed, the normal vector $n_\mu$ is also perturbed \citep{Silvestrini:2025lbe}. The three-velocity of the membrane fluid is likewise perturbed, with non-zero components $\delta u^t_{\rm (pol)}$ and $\delta u^\theta_{\rm (pol)}$.

In the polar sector, the perturbed induced metric has non-zero components $\delta h_{tt}$, $\delta h_{\theta\theta}$, and $\delta h_{\phi\phi}$. Correspondingly, the non-zero components of the perturbed extrinsic curvature are $\delta K^+_{tt}$, $\delta K^+_{t\theta}$, $\delta K^+_{\theta\theta}$, and $\delta K^+_{\phi\phi}$, all evaluated at ${{{R}}}$; explicit expressions can be found in \citet{Silvestrini:2025lbe}. The perturbation of the trace, $\delta K^+$, is also non-zero and contributes to the junction conditions. Similarly, the components of the perturbed membrane energy-momentum tensor in the polar sector can be expressed in terms of $\rho_1$, $p_1$, $R_1$, the metric perturbations, and the shear and bulk viscosities, $\eta$ and $\zeta$ \citep{Silvestrini:2025lbe}. 

From~\ref{KHHfd} and~\ref{H1}, it follows that, in the small-frequency limit, the polar metric perturbations $H_1$ and $K$ are determined by $H$ alone. Therefore, in the polar sector, the five variables to solve for are $\rho_1$, $p_1$, $R_1$, $\delta u^\theta_{\rm (polar)}$, and $H$, which are determined from the perturbed junction conditions.

In the axial sector, the membrane location remains fixed, so $\delta n_{\mu\rm (ax)} = 0$. The only non-zero component of the perturbed induced metric is $\delta h_{t\phi} = \delta g_{t\phi}$, and the only non-zero component of the perturbed three-velocity is $\delta u^\phi_{\rm (ax)}$. Consequently, the non-zero components of the perturbed extrinsic curvature are $\delta K^+_{t\phi\,{\rm (ax)}}$ and $\delta K^+_{\theta\phi\,{\rm (ax)}}$, while the trace perturbation $\delta K^+$ vanishes. The corresponding components of the perturbed energy-momentum tensor are determined similarly. 

Thus, in the axial sector there are two variables to solve for: the metric perturbation $h_0$ and the fluid perturbation $\delta u^\phi_{\rm (ax)}$, with $h_1$ determined in terms of $h_0$ via~\ref{h1pfsch}. Finally, both polar and axial perturbations of the membrane fluid depend on the shear viscosity $\eta$, while the bulk viscosity $\zeta$ affects only the polar sector.

%%%%%%%%%%%%%%%%%%%%%%
%%%%%%%%%%%%%%%%%%%%%%
%%%%%%%%%%%%%%%%%%%%%%
\paragraph{Axial Love numbers for the membrane}~---
In the axial sector, the relevant variables are the metric perturbation $h_0$ and the perturbation of the fluid four-velocity $\delta u^\phi_{\rm (ax)}$. As discussed in the previous section, the shear viscosity $\eta$ plays a central role in determining the membrane response. Following the low-frequency expansion of the reflectivity in~\ref{reflectivity}, we expand the shear viscosity as
\begin{equation}
\eta = \eta_0 + i M \omega \eta_1 + \mathcal{O}(M^2 \omega^2)~,
\label{eta}
\end{equation}
where both $\eta_0$ and $\eta_1$ are functions of the compactness parameter $\epsilon$, which we keep arbitrary for generality. For BHs, one has $\eta_0 = 1/(16\pi)$, while $\eta_1$ and all higher-order frequency coefficients vanish \citep{Damour:1982,Thorne:1986iy,Maggio:2020jml}. For compact objects, however, $\eta_1$ plays a crucial role in determining the tidal deformability.

Since $\delta K^+$ vanishes in the axial sector, the perturbed junction condition reduces to
\begin{equation}\label{pertaxial}
\delta K^+_{ab} - K^+ \delta h_{ab} = -8\pi \delta T_{ab}~.
\end{equation}
This condition provides the expression of $\delta u^\phi_{\rm (ax)}$ in terms of $h_0$ and $h_1$, and the boundary condition for the axial metric perturbation $h_0$. The analysis can be divided into two cases: (a) $\eta_0 \neq 0$, and (b) $\eta_0 = 0$. 

For $\eta_0 \neq 0$, the low-frequency boundary condition for $h_0$ becomes
\begin{align}
\left(\frac{h_0'}{h_0}\right)_{{{{R}}}} = \frac{f'}{f}\Big|_{{{{R}}}} = \frac{2M}{{{{R}}}({{{R}}}-2M)} + \mathcal{O}(M\omega)~.
\label{BCaxetagen}
\end{align}
This ratio diverges as ${{{R}}} \to 2M$ and is independent of $\ell$ as well as of $\eta_0$. Consequently, in this case the magnetic LNs are independent of the shear viscosity and scale linearly with the compactness $\epsilon$. 

For $\eta_0 = 0$, the boundary condition becomes
\begin{align}
{{{R}}} \left(\frac{h_0'}{h_0}\right)_{{{{R}}}} = \frac{2 \left({{{R}}}^2 - 3 M {{{R}}} + 8\pi \gamma_\ell \eta_1 M^2 \right)}{{{{R}}}^2 - 3 M {{{R}}} + 8 \pi \gamma_\ell \eta_1 M ({{{R}}} - 2M)}~,
\label{BCaxial}
\end{align}
where terms of $\mathcal{O}(M\omega)$ have been neglected. This boundary condition depends on $\eta_1$ and remains finite as ${{{R}}} \to 2M$. We have discussed a similar behavior in~\ref{dynamical_LN_ECO} for what concerns the low-frequency expansion of the reflectivity \citep{Chakraborty:2023zed}, where only for $\mathcal{R}_0 = 1$ the LNs are non-zero and depend on the reflectivity $\mathcal{R}_1$ (see~\ref{reflectivity}).

The left-hand side of~\ref{BCaxial} can be determined from the exterior solution for $h_0$ presented in~\ref{h0e}. Specializing to $\ell = 2$, one obtains the magnetic LN
\begin{align}
k_2^{\rm M} = \frac{1}{5 (1+\epsilon)^5} \left( \frac{g_2(\epsilon)}{12 g_2(\epsilon) \ln\left(\frac{\epsilon}{1+\epsilon}\right) + \bar{g}_2(\epsilon)} \right)~,
\end{align}
where
\begin{align}
g_2(\epsilon) &\equiv (1+\epsilon)^2 \left( 96 \pi \eta_1 \epsilon^2 + 2\epsilon^3 + 3 \epsilon^2 - 1 \right)~,\\
\bar{g}_2(\epsilon) &\equiv 96 \pi \eta_1 (1 + 2\epsilon) \left[ 6 \epsilon (1+\epsilon) - 1 \right] + 24 \epsilon^3 (3+\epsilon) + 62 \epsilon^2 - 2 \epsilon - 25~.
\end{align}

The procedure generalizes to higher multipoles $\ell$: one first derives the boundary condition at the membrane from the perturbed junction conditions and then matches it with the exterior solution to obtain the LN. The computation for $\ell=3$ is detailed in \citet{Silvestrini:2025lbe}. 

We stress that these expressions for the magnetic LNs are valid for arbitrary $\epsilon$, not necessarily small. Thus, the same formalism applies to BHs, ordinary NSs neutron stars, as well as to ultra-compact objects ($\epsilon\to 0$); in the latter case the magnetic LNs exhibit a logarithmic behavior, which will be discussed in subsequent sections.

%%%%%%%%%%%%%%%%%%%%%%
%%%%%%%%%%%%%%%%%%%%%%
%%%%%%%%%%%%%%%%%%%%%%
\paragraph{Polar Love numbers for the membrane}~---
In the polar sector, there are five unknowns: $\delta u^\theta_{\rm (pol)}$ (perturbation of the membrane fluid three-velocity), $\rho_1$ (perturbation of the energy density), $p_1$ (perturbation of the pressure), $R_1$ (perturbation of the membrane location), and $H$ (metric perturbation). The other even-parity metric perturbations, such as $K$ and $H_1$, are determined in terms of $H$, the perturbation of $g_{tt}$, see~\ref{KHHfd} and~\ref{H1}. In this sector, both $\delta K^+_{ab}$ and $\delta K^+$ contribute to the perturbed junction condition, which reads
\begin{align}\label{pertpolar}
\delta K^+_{ab} - K^+ \delta h_{ab} - h_{ab} \delta K^+ = -8\pi \delta T_{ab}~.
\end{align}

\ref{pertpolar} provides four independent equations, corresponding to the $(t,\theta)$, $(t,t)$, $(\theta,\theta)$, and $(\phi,\phi)$ components. The $(t,\theta)$ component fixes the fluid velocity perturbation $\delta u^\theta_{\rm (pol)}$ \citep{Silvestrini:2025lbe}. The other three components determine $\rho_1$ and $p_1$ in terms of $H$ and its derivative, up to linear order in $M\omega$. The perturbation $R_1$ appears only at $\mathcal{O}(M^2\omega^2)$ and is therefore neglected.

The boundary condition for $H$ at $r={{{R}}}$ follows from relating $p_1$ and $\rho_1$ through an equation of state $p=p(\rho)$:
\begin{align}
p_1 = \left(\frac{\partial p_0}{\partial \rho_0}\right)\rho_1 + \left(\frac{\partial p_0}{\partial R}\right) R_1
= c_s^2 \rho_1 + \left(\frac{\partial p_0}{\partial R}\right) R_1~,
\end{align}
where $c_s = \sqrt{\partial p_0 / \partial \rho_0}$ is the speed of sound in the membrane fluid. For $\eta_0 = 0$ and regular $\zeta$, $R_1$ vanishes in the static limit; in general, $R_1 = \mathcal{O}(M^2 \omega^2)$, leaving only the $c_s^2$ term. Using this relation, the boundary condition for $H$ reads, for any $\eta_0 \neq 0$ and regular $\zeta$,
\begin{align}
\left(\frac{H'}{H}\right)_{{{{R}}}} = \frac{M ({{{R}}}^2 - 6 M {{{R}}} + 6 M^2)}{{{{R}}} ({{{R}}} - 2M) ({{{R}}}^2 - 3 M {{{R}}} + 3 M^2)}~,
\label{BCBH}
\end{align}
while for $\eta_0 = 0$ and arbitrary $\zeta$, keeping linear order in $\omega$, one finds
\begin{align}
\left(\frac{H'}{H}\right)_{{{{R}}}} = 
\frac{-M ({{{R}}}^2 - 6 M {{{R}}} + 6 M^2) + 16 \pi i \omega \zeta {{{R}}}^2 ({{{R}}} - 4 M)({{{R}}} - 2M)}
{{{{R}}} ({{{R}}} - 2M) \left[ -({{{R}}}^2 - 3 M {{{R}}} + 3 M^2) + 16 \pi i \omega \zeta {{{R}}}^2 ({{{R}}} - 2M) \right]}~.
\label{BCH}
\end{align}

In the $\omega \to 0$ limit,~\ref{BCH} reduces to~\ref{BCBH}, which diverges as ${{{R}}} \to 2M$, analogous to the axial sector. Since $\zeta$ enters the boundary condition in the combination $\omega \zeta$, preserving its effect requires assuming a frequency dependence
\begin{equation}
\zeta = i \frac{\zeta_{-1}}{M \omega} + \zeta_0 + \mathcal{O}(M \omega)~,
\label{zeta}
\end{equation}
where $\zeta_{-1}$ and $\zeta_0$ are functions of the compactness $\epsilon$. Although the $1/\omega$ divergence may appear unphysical, the membrane is fictitious and not constrained to represent a conventional fluid. Indeed, for BHs, the bulk viscosity is negative. Remarkably, the $1/\omega$ behavior naturally arises when connecting the membrane paradigm to gravastars and NSs, as we shall discuss.

With~\ref{zeta}, the static boundary condition for $H$ becomes
\begin{align}
\left(\frac{H'}{H}\right)_{{{{R}}}} = 
\frac{M^2 ({{{R}}}^2 - 6 M {{{R}}} + 6 M^2) + 16 \pi \zeta_{-1} {{{R}}}^2 ({{{R}}} - 4 M)({{{R}}} - 2 M)}
{{{{R}}} ({{{R}}} - 2M) \left[ M ({{{R}}}^2 - 3 M {{{R}}} + 3 M^2) + 16 \pi \zeta_{-1} {{{R}}}^2 ({{{R}}} - 2M) \right]}~.
\label{BCHnew}
\end{align}

Matching the exterior solution for $H$~\ref{Hext} to the boundary condition~\ref{BCHnew} gives the electric LN for $\ell = 2$:
\begin{align}\label{k2E}
k_2^{\rm E} = \frac{A(\epsilon)}{5 (1+\epsilon)^4 \left[ B(\epsilon) + C(\epsilon) \log\left( \frac{1}{\epsilon} + 1 \right) \right]}~,
\end{align}
with
\begin{align}
A(\epsilon) &= \epsilon \Big[ 64 \pi \zeta_{-1} \epsilon^4 + 4(64 \pi \zeta_{-1} + 1) \epsilon^3 + (320 \pi \zeta_{-1} + 3) \epsilon^2 + (128 \pi \zeta_{-1} + 3) \epsilon + 1 \Big]~, \nonumber\\
B(\epsilon) &= 768 \pi \zeta_{-1} \epsilon^5 + 48 (72 \pi \zeta_{-1} + 1) \epsilon^4 + (5248 \pi \zeta_{-1} + 60) \epsilon^3 + (3008 \pi \zeta_{-1} + 46) \epsilon^2 + 28 (16 \pi \zeta_{-1} + 1) \epsilon + 3~, \nonumber\\
C(\epsilon) &= -12 \epsilon (\epsilon+1) \Big[ 64 \pi \zeta_{-1} \epsilon^4 + 4(64 \pi \zeta_{-1} + 1) \epsilon^3 + (320 \pi \zeta_{-1} + 3) \epsilon^2 + (128 \pi \zeta_{-1} + 3) \epsilon + 1 \Big]~. \nonumber
\end{align}
In the ultra-compact limit, $A(\epsilon) \approx \epsilon$, $B(\epsilon) \approx 3 + 28 \cdot 16 \pi \zeta_{-1} \epsilon + 28 \epsilon$, and $C(\epsilon) \approx -12 \epsilon$. Unless $\zeta_{-1} \epsilon = -3 / (28 \cdot 16 \pi) + \mathcal{O}(\epsilon)$, the electric LN vanishes. Therefore, to obtain non-zero LNs, the bulk viscosity must scale inversely with both frequency and compactness $\epsilon$. This result holds for generic $\ell$ \citep{Silvestrini:2025lbe}.

%%%%%%%%%%%%%%%%%%%%%%%%%%%%%%%%%%%%%%%%%%%%%%%%%%%%%%%%%%%
%%%%%%%%%%%%%%%%%%%%%%%%%%%%%%%%%%%%%%%%%%%%%%%%%%%%%%%%%%%
%%%%%%%%%%%%%%%%%%%%%%%%%%%%%%%%%%%%%%%%%%%%%%%%%%%%%%%%%%%
\subsubsection{Love numbers of compact objects from the membrane paradigm}\label{sec:LNECOBHmembrane}
We have established the connection between the LNs in both the axial and polar sectors and the shear and bulk viscosity coefficients of the membrane fluid. The frequency dependence of these viscosity coefficients is crucial to obtain non-trivial LNs. As we will show, their dependence on the compactness parameter $\epsilon$ also plays a key role in determining the LNs of ECOs. In the following, we discuss these features in detail, along with the corresponding LNs of BHs.

%%%%%%%%%%%%%%%%%%%%%%
%%%%%%%%%%%%%%%%%%%%%%
%%%%%%%%%%%%%%%%%%%%%%
\paragraph{Membrane paradigm for black holes in vacuum General Relativity}~---
Let us briefly revisit the well-known result that BHs in vacuum GR have vanishing LNs, in the context of the membrane paradigm. First, we emphasize that the membrane paradigm discussed here is restricted to static and spherically symmetric spacetimes, and hence the results apply specifically to Schwarzschild BHs within GR. For such BHs, the viscosity parameters at leading order are $\eta_0 = 1/(16\pi)$ and $\zeta_0 = -1/(16\pi)$.

In the axial sector, the boundary condition in~\ref{BCaxetagen} implies that near the horizon the relevant condition is
\[
\left(\frac{h_0'}{h_0}\right) \sim \frac{1}{2M \epsilon}~,
\] 
which is independent of $\eta_0$. Combining this boundary condition with the exterior solution for $h_0$~\ref{h0e} for $\ell = 2$ yields $k_2^{\rm M} \sim \epsilon^2$. Therefore, in the BH limit $\epsilon \to 0$, the magnetic LN vanishes identically.

Similarly, in the electric sector, the regular zero-frequency behavior of $\zeta$, together with non-zero $\eta_0$, requires the use of~\ref{BCBH} as the boundary condition for $H$. Near the horizon, this boundary condition behaves as
\[
\left(\frac{H'}{H}\right) \to \frac{1}{\epsilon}~,
\] 
which leads to $k_2^{\rm E} \to 0$ in the BH limit. 

Thus, both the electric and magnetic LNs of a Schwarzschild BH vanish within the membrane paradigm, in perfect agreement with previous results, see~\ref{sec:staticBHs}.

%%%%%%%%%%%%%%%%%%%%%%
%%%%%%%%%%%%%%%%%%%%%%
%%%%%%%%%%%%%%%%%%%%%%
\paragraph{Membrane paradigm for reflective exotic compact objects and their Love numbers}~---
In this section we relate the membrane parameters to those of reflective compact objects, as discussed in~\ref{sec:reflectivity}. As noted earlier for BHs, when $\eta_0 \neq 0$, $\zeta_0 \neq 0$, and $\zeta_{-1} = 0$, the LNs vanish. Therefore, a non-zero LN in the non-rotating case requires a one-to-one correspondence between the reflectivity, defined in~\ref{sec:reflectivity} (see also \citealt{Chakraborty:2023zed}), and the membrane viscosity parameters $\eta_1$ and $\zeta_{-1}$.

As explained in~\ref{sec:reflectivity}, the reflectivity $\mathcal{R}$ is defined in terms of the Detweiler function $X(r_*)$, which, in the non-rotating case, decomposes into axial and polar sectors via the Regge--Wheeler and Zerilli functions, respectively. These functions are related to the membrane viscosity parameters through the boundary conditions: (a) in the axial sector \citep{Maggio:2020jml},
\begin{equation}
\left(\frac{1}{\psi_{\rm RW}} \frac{d \psi_{\rm RW}}{dr_*}\right)_{{{{R}}}} = - \frac{i \omega}{16 \pi \eta} - \frac{{{{R}}}^2}{2 ({{{R}}} - 3 M)} V_{\rm RW}({{{R}}})~,
\label{bcax}
\end{equation}
with $V_{\rm RW}$ the usual Regge--Wheeler potential, and (b) in the polar sector \citep{Maggio:2020jml},
\begin{equation}
\frac{1}{\psi_{\rm Z}} \left(\frac{d \psi_{\rm Z}}{dr_*}\right)_{{{{R}}}} = -16 \pi i \eta \omega + \frac{G({{{R}}}, \omega, \eta, \zeta, \ell)}{M}~,
\label{bcpol}
\end{equation}
where $G({{{R}}}, \omega, \eta, \zeta, \ell)$ is a complicated function (see Appendix A of \citealt{Maggio:2020jml}). Expressing the Regge--Wheeler and Zerilli functions in terms of the Detweiler function allows one to relate $(\eta,\zeta)$ to the corresponding reflectivities $\mathcal{R}_{\rm ax}$ and $\mathcal{R}_{\rm pol}al$, which are independent \citep{Silvestrini:2025lbe}.

For an ECO, taking the limit ${{{R}}} \to 2M (1+\epsilon)$ in~\ref{bcax}, and noting that $V_{\rm RW}$ vanishes in this limit, the axial boundary condition becomes independent of $\ell$. For $\ell=2$ and $\epsilon \ll 1$, the shear viscosity is related to the axial reflectivity as \citep{Silvestrini:2025lbe}:
\begin{equation}
\eta = \frac{-4 M \omega (1 + \mathcal{R}_{\rm ax}) + i (\mathcal{R}_{\rm ax} - 1)}{16 \pi \left[ 4 M \omega (\mathcal{R}_{\rm ax} - 1) - i (\mathcal{R}_{\rm ax} + 1) \right]}~.
\label{etaR}
\end{equation}
Expanding in the small-frequency limit $M\omega \ll 1$ yields
\begin{equation}
\eta_0 = \frac{1}{16\pi} \left( \frac{1 - \mathcal{R}^{\rm ax}_0}{1 + \mathcal{R}^{\rm ax}_0} \right)~.
\end{equation}
As expected, for a Schwarzschild BH with perfect absorption ($\mathcal{R}^{\rm ax}_0 = 0$), one has $\eta_0 = 1/(16\pi)$ \citep{Thorne:1986iy}, while for perfect reflection ($\mathcal{R}^{\rm ax}_0 = 1$), $\eta_0 = 0$, the scenario yielding non-zero LNs for ECOs in the zero-frequency limit \citep{Chakraborty:2023zed}. In this case,
\begin{equation}
\eta_1 = - \frac{\mathcal{R}^{\rm ax}_1 + 8}{32 \pi}~,
\label{eta1}
\end{equation}
independent of $\ell$. Thus, a non-zero magnetic LN arises if and only if the membrane shear viscosity scales linearly with $\omega$.

In the polar sector, the bulk viscosity $\zeta(\epsilon, \omega)$ is related to the polar reflectivity $\mathcal{R}_{\rm pol}(\omega)$. For $\ell = 2$ and $\mathcal{R}_0^{\rm pol} = 1$, one finds \citep{Silvestrini:2025lbe}:
\begin{align}
\zeta_{-1}^{\ell=2} = - \frac{3}{448 \pi \epsilon} - \frac{3 (\mathcal{R}_1^{\rm pol} + 18)}{3136 \pi}~,
\label{zetam1l2}
\end{align}
which depends on $\ell$, unlike the axial case. If $\mathcal{R}_0^{\rm pol} \neq 1$, the static electric LNs vanish.

Using these relations, the magnetic and electric LNs can be expressed directly in terms of the reflectivities. For $\ell=2$, one finds:
\begin{align}
k_2^{\rm M} &= \frac{1}{60 \log \epsilon - 15 \mathcal{R}_1^{\rm ax} + 5} + \mathcal{O}(\epsilon)~,
\label{tlnax} \\
k_2^{\rm E} &= \frac{1}{60 \log \epsilon - 15 \mathcal{R}_1^{\rm pol} + 5} + \mathcal{O}(\epsilon)~.
\label{kER}
\end{align}

The leading-order coefficients of $\log \epsilon$ and $\mathcal{R}_1^{\rm ax/pol}$ agree with \citep{Chakraborty:2023zed}, demonstrating that the membrane paradigm yields consistent LNs for ECOs with reflectivity. Interestingly, if $\mathcal{R}_1^{\rm ax} = \mathcal{R}_1^{\rm pol}$, then $k_2^{\rm M} = k_2^{\rm E}$ at leading order in $\epsilon$, in agreement with and generalizing previous results \citep{Cardoso:2017cfl}.

%%%%%%%%%%%%%%%%%%%%%%
%%%%%%%%%%%%%%%%%%%%%%
%%%%%%%%%%%%%%%%%%%%%%
\paragraph{Membrane paradigm for gravastars and their Love numbers}~---
Here we present the mapping between the membrane paradigm and a concrete model of ECOs. Owing to its simplicity and the availability of analytical results, we focus on the aforementioned gravastar model \citep{Chapline:2000en,Mazur:2004fk}. Gravastars are known to be generically stable under radial \citep{Visser:2003ge} as well as non-radial perturbations, in both the axial and polar sectors \citep{Chirenti:2007mk,Pani:2009ss}.

A thin-shell, static and spherically symmetric gravastar \citep{Visser:2003ge} is described by the line element in~\ref{met_sph_stat}, with metric components given in~\ref{metric_gravastar}. The thin shell is located at $r={{{R}}}=2M(1+\epsilon)$ and is characterized by a surface energy density $\Sigma$.
% , corresponding to a surface mass
% \begin{equation}
% M_{\rm s}=4\pi {{{R}}}^{2}\Sigma~.
% \end{equation}
% The interior region has constant energy density $\rho$, yielding a volume mass
% \begin{equation}
% M_{\rm v}=\frac{4\pi}{3}\rho {{{R}}}^{3}~,
% \end{equation}
% so that the ADM mass of the gravastar reads
% \begin{equation}
% M=M_{\rm v}+\frac{M_{\rm s}^{2}}{2{{{R}}}}+M_{\rm s}\sqrt{1-\frac{2M_{\rm v}}{{{{R}}}}}~.
% \end{equation}

Remarkably, for gravastars both axial and polar perturbations in the interior are governed by a single master equation that admits an exact solution. Imposing regularity at $r=0$, this yields \citep{Pani:2009ss}
\begin{align}
\Psi_{\rm int}
&=r^{\ell+1} \left(1-\frac{8\pi\rho r^2}{3}\right)^{i\frac{M\omega}{\sqrt{\hat C}}}
\,_{2}F_{1}\!\left(
\frac{\ell+2+\frac{2iM\omega}{\sqrt{\hat C}}}{2},
\frac{\ell+1+\frac{2iM\omega}{\sqrt{\hat C}}}{2},
\ell+\frac{3}{2},
\frac{\hat C r^{2}}{4M^{2}}
\right)~,
\end{align}
where $\rho=3M/(4\pi R^3)$ is the constant density in the de~Sitter interior, and $\hat C=(2M/{{{R}}})^{3}=(1+\epsilon)^{-3}$.
Using $\Psi_{\rm int}$ together with the axial boundary condition~\ref{bcax}, one finds the shear viscosity associated with a gravastar \citep{Silvestrini:2025lbe}
\begin{equation}
\eta \simeq \frac{\log(\epsilon)}{16\pi}\, i \omega M
+\mathcal{O}(\epsilon,M^{2}\omega^{2})~.
\label{etagravastar}
\end{equation}
Comparing with~\ref{eta}, this shows that the shear viscosity is linear in $\omega$, with $\eta_{0}=0$, which is precisely the condition required for nonvanishing static LNs. Substituting~\ref{etagravastar} into~\ref{tlnax}, we obtain the $\ell=2$ axial LN of a gravastar,
\begin{equation}
k_{2}^{\rm M}
=\frac{1}{5\left(43-12\log 2+18\log\epsilon\right)}
+\mathcal{O}(\epsilon)~,
\label{tlnaxial}
\end{equation}
in agreement with previous results \citep{Cardoso:2017cfl}\footnote{The normalization of the LNs used here differs from that of \citet{Cardoso:2017cfl} by a factor $2^{2\ell+1}$ in the BH limit.}. Likewise, combining~\ref{eta1} and~\ref{etagravastar} yields $\mathcal{R}_{1}^{\rm ax}\simeq -2\log(\epsilon)$, which inserted into~\ref{tlnax} gives $k_{2}^{\rm M}\sim[90\log(\epsilon)]^{-1}$, reproducing the leading behavior of~\ref{tlnaxial}.

Applying the same strategy to the polar sector leads to the following bulk viscosity for a gravastar within the membrane paradigm \citep{Silvestrini:2025lbe}
\begin{equation}
\zeta \simeq \frac{i\zeta_{-1}}{M\omega}+\mathcal{O}(M\omega)~,
\qquad
\zeta_{-1}^{\ell=2}
=-\frac{3i}{448\pi\epsilon}
-\frac{3i\left(3+\log 16-2\log\epsilon\right)}{3136\pi}~.
\label{zetagravastar}
\end{equation}
Thus, the bulk viscosity scales as $(\epsilon\omega)^{-1}$, in agreement with the general membrane result~\ref{zeta}, while the frequency-independent term vanishes, again ensuring non-zero static LNs. Substituting~\ref{zetagravastar} into~\ref{kER}, we obtain the $\ell=2$ polar LN of a gravastar,
\begin{equation}
k_{2}^{\rm E}
=\frac{1}{10\left(23-6\log 2+9\log\epsilon\right)}
+\mathcal{O}(\epsilon)~,
\end{equation}
which coincides with the result of \citet{Cardoso:2017cfl}, up to the different normalization factor $2^{5}$. At leading order one also finds $\mathcal{R}_{1}^{\rm pol}\simeq -2\log(\epsilon)$, which reproduces the same expression when inserted into~\ref{kER}. Finally, the fact that both the shear and bulk viscosities in the membrane description of a gravastar are purely imaginary is consistent with the interpretation of gravastars as non-dissipative objects.

%%%%%%%%%%%%%%%%%%%%%%
%%%%%%%%%%%%%%%%%%%%%%
%%%%%%%%%%%%%%%%%%%%%%
\paragraph{Membrane paradigm for neutron stars, Love numbers, and universality}~---
For a perfect-fluid NS there are no dissipative effects and, consequently, the membrane paradigm can be applied with purely imaginary shear and bulk viscosities. This observation shows that the membrane paradigm is not restricted to BHs ($\epsilon=0$) or ultra-compact objects ($\epsilon\ll 1$), but is also applicable to NSs, for which $\epsilon=\mathcal{O}(1)$. 

To determine the axial and polar LNs, the first step is to solve the TOV equations, as discussed in~\ref{LNNS_Static}, thereby fixing the background spacetime for a given equation of state (EoS). Since nonvanishing static LNs within the membrane paradigm originate from dynamical viscosities, one must then consider dynamical perturbations around this static background. The relevant perturbation equations for both axial and polar sectors are summarized in~\ref{dyntideNS}. Solving these equations with the appropriate boundary conditions, as detailed in~\ref{dyntideNS}, yields the metric perturbations $(H_{1},K)$ at the stellar surface in the polar sector, from which the Zerilli function and its radial derivative at the surface are obtained. Similarly, in the axial sector one computes the interior master function, which can be matched to the exterior Regge--Wheeler function, thereby determining the Regge--Wheeler function and its derivative at the NS surface. These quantities can then be combined with the boundary conditions~\ref{bcpol} and~\ref{bcax} to infer the bulk and shear viscosities of the membrane.

Interestingly, expanding the shear viscosity in powers of $M\omega$ yields $\eta_{0}=0$, with $\eta_{1}$ purely real \citep{Silvestrini:2025lbe}. This provides a reassuring consistency check of the framework, since it naturally leads to nonvanishing LNs for NSs. An analogous low-frequency expansion of the bulk viscosity gives $\zeta_{0}=0$ and a purely real $\zeta_{-1}$. Together, these results imply $|\mathcal{R}|^{2}=1$, with deviations entering only at quadratic order in the dimensionless frequency $M\omega$. The viscosity parameters $\eta_{1}$ and $\zeta_{-1}$ depend on both the angular index $\ell$ and the EoS \citep{Silvestrini:2025lbe}.

\begin{figure}[htbp]
    \centering
    \includegraphics[width=0.45 \textwidth]{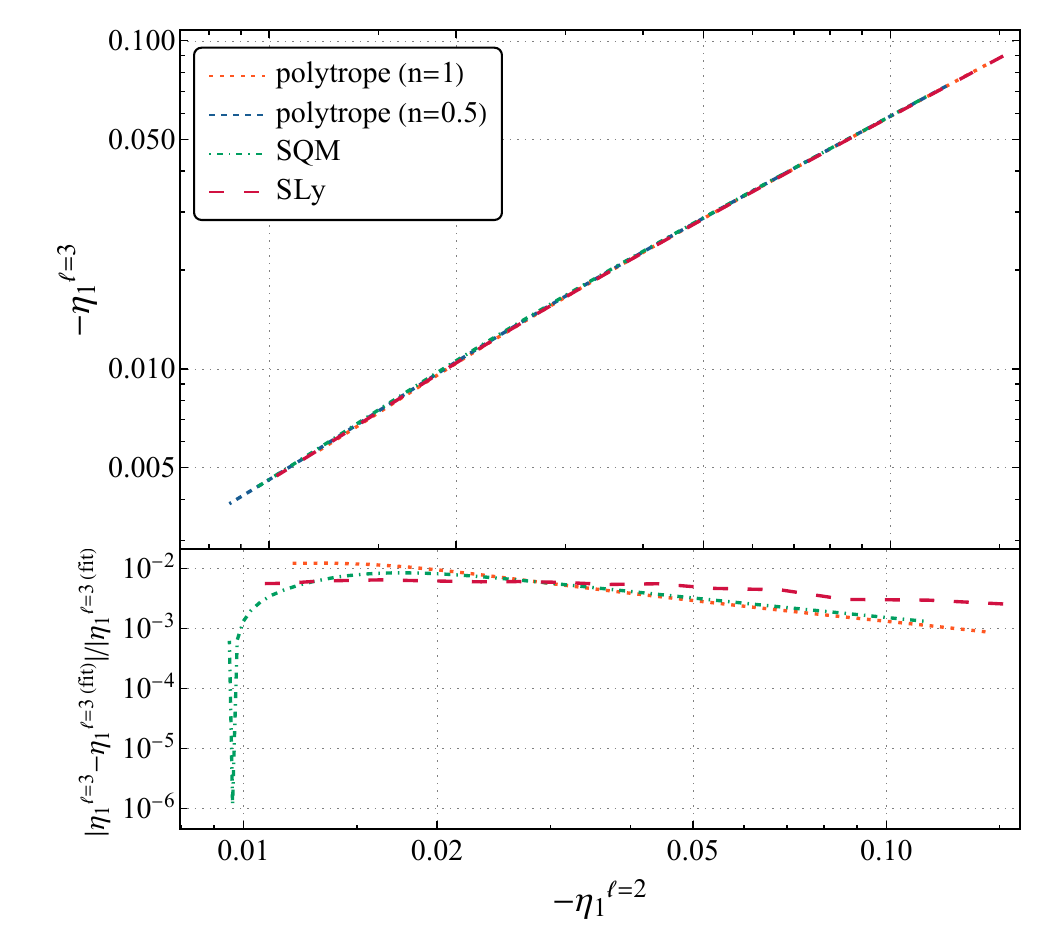}
    \includegraphics[width=0.54 \textwidth]{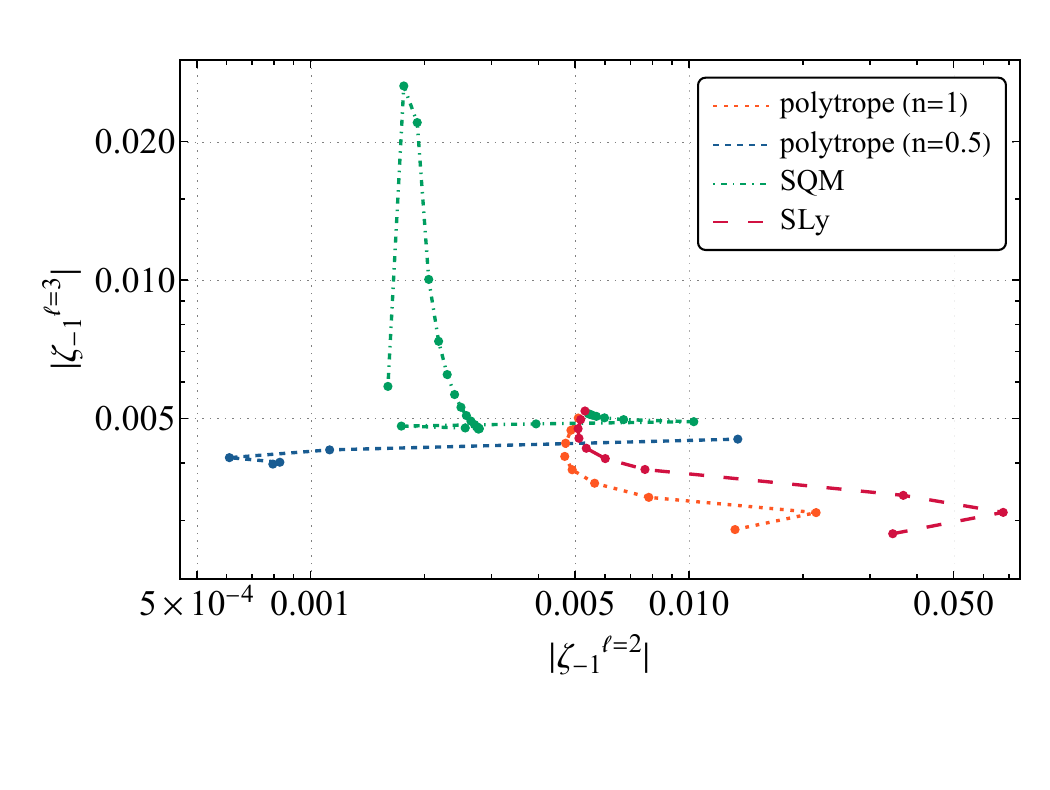}
    \caption{Quasi-universal relation between the $\ell=3$~vs~$\ell=2$ shear (left) and bulk (right) viscosities of a NS within the membrane paradigm, using different EoS. 
    In the axial case, the left bottom panel shows the relative difference with respect to the interpolation of the data for the $n=0.5$ polytropic EoS, taken as a reference for the quasi-universal behavior, which is absent in the polar case.
    Using the membrane paradigm, the viscosity parameters can be mapped to the standard quasi-universal relations among the LNs \citep{Yagi:2013bca,Yagi:2016ejg}.
    From \citet{Silvestrini:2025lbe}.
    }
    \label{fig:universality_viscosity}
\end{figure}

Remarkably, by varying the central density for each EoS, one finds a quasi-universal relation between $\eta_{1}$ for $\ell=3$ and $\eta_{1}$ for $\ell=2$ (see~\ref{fig:universality_viscosity}), with deviations from the universal curve at the level of $\lesssim 1\%$. Although the axial LNs themselves also exhibit universality \citep{Yagi:2013bca,Yagi:2013awa,Yagi:2016ejg}, the relative separation between different EoSs is larger than in the case of the shear-viscosity parameters \citep{Silvestrini:2025lbe}. The polar sector shows the opposite trend: the bulk-viscosity parameter $\zeta_{-1}$ does not display universality and retains a strong dependence on the EoS, see~\ref{fig:universality_viscosity}. This behavior can be traced back to the non-trivial relation between $\zeta_{-1}$ and the polar LNs. Nevertheless, the electric LNs, obtained from relations such as~\ref{k2E}, depend on the central density and the EoS in a manner consistent with previous studies \citep{Yagi:2013bca,Yagi:2016ejg}.

In summary, the membrane paradigm successfully reproduces the axial and polar LNs of NSs and reveals a non-trivial quasi-universal relation for the shear-viscosity parameters in the axial sector. No analogous universality emerges for the bulk-viscosity parameters, although both shear and bulk viscosities ultimately lead to axial and polar LNs that themselves display quasi-universal behavior.

%%%%%%%%%%%%%%%%%%%%%%%%%%%
\subsection{Summary of the Love numbers of exotic compact objects} \label{sec:summaryECOs}
%%%%%%%%%%%%%%%%%%%%%%%%%%%%

In~\ref{tab:summaryECOs}, we summarize the tidal LNs of various classes of ECOs discussed in this section. Note that, for models admitting a BH limit (i.e., with compactness large enough that $(\epsilon\to0$), electric and magnetic LNs with same $\ell$ are the same. This property can be understood within the membrane paradigm discussed in the previous section \citep{Silvestrini:2025lbe}. Furthermore, the LNs of the perfect mirror model are twice those of the wormhole model obtained by gluing together two copies of the Schwarzschild metric. Finally, the perfect mirror case can be obtained from the partly-reflecting case by setting ${\cal R}_1=0$.

\begin{landscape}
\begin{table}[ht!]
\centering
\caption{Tidal LNs of some classes of ECOs, using the normalizations adopted in this work. 
As a comparison, we provide the order of magnitude of the LNs for static NSs with compactness $C\approx0.2$.
For boson stars, the table provides the lowest value of the corresponding LN for a given model, and assuming $\alpha=10^4$ and $\sigma_0=0.05$ for the potential of the massive and solitonic models, respectively (see~\ref{fig:BS}).
For other ECOs, we provide expressions for very compact configurations where the surface $R$
sits at $R\sim 2M(1+\epsilon)$, with $\epsilon\ll1$. See references for the full results. 
}
\label{tab:summaryECOs}
\begin{tabular}{|c|ccc|}
\hline
\hline
                                       &                   $k_2^{\rm E}/C^5$      &   $k_3^{\rm E}/C^7$   & $k_2^{\rm B}/C^5$      %& $k_3^{\rm B}$     
                                       \\
\hline
NSs 	          &  $211$        	 &   $1280$      &   $6.4$       %&       %$70$  
\\
                                       \hline
                Mini boson star \citep{Cardoso:2017cfl}         & $1710$    	 & $7\times10^4$  & $113$  %& $-211.8$ 
                \\
                Massive boson star \citep{Cardoso:2017cfl,Sennett:2017etc}      & $356.1$    	 & $3\times 10^3$  & $13.6$  %& $XXX$ 
                \\
                %%%
                Solitonic boson star \citep{Cardoso:2017cfl,Sennett:2017etc}   & $41.36$    	 & $4\times10^2$  & $53.8$  %& $XXX$ 
                \\
                \hline
                \hline
                &                   $k_2^{\rm E}$      &   $k_3^{\rm E}$   & $k_2^{\rm B}$      %& $k_3^{\rm B}$     
                                       \\
                \hline
                 Gravastar \citep{Cardoso:2017cfl, Pani:2009ss}          &  $\frac{1}{90\log\epsilon}$         	 &  $\frac{1}{630\log\epsilon}$       &  $\frac{1}{90\log\epsilon}$        %&   $\frac{1}{630\log\epsilon}$  
                 \\
                 Black shell \citep{Giri:2024cks}  & $\frac{1}{90\log\epsilon}$ & $\frac{1}{630\log\epsilon}$ & $\frac{1}{90\log\epsilon}$ %& $\frac{1}{630\log\epsilon}$ 
                 \\
                Wormhole \citep{Cardoso:2017cfl}         & $\frac{1}{120\log\epsilon}$         	 & $\frac{1}{840\log\epsilon}$         & $\frac{1}{120\log\epsilon}$          %& $\frac{1}{840\log\epsilon}$         
                \\
                 Reflecting ECOs \citep{Chakraborty:2023zed}     & $\frac{1}{15(-\mathcal{R}_{1}+4\log\epsilon)}$         	 & $\frac{1}{105(-\mathcal{R}_{1}+16\ln \epsilon)}$         & $\frac{1}{15(-\mathcal{R}_{1}+4\log\epsilon)}$          %& $\frac{32}{7 (197+60 \log\epsilon)}$ 
                 \\
                 Perfect mirror \citep{Cardoso:2017cfl, Chakraborty:2023zed}     & $\frac{1}{60\log\epsilon}$         	 & $\frac{1}{1680\log\epsilon}$          & $\frac{1}{60\log\epsilon}$          %& $\frac{1}{420\log\epsilon}$ 
                 \\
                 \hline
                 Membrane paradigm \citep{Silvestrini:2025lbe}  & $\frac{1}{60\ln \epsilon}$  & $\frac{1}{1680\ln \epsilon}$  & $\frac{1}{60(\ln \epsilon+8\pi \eta_{1})}$ %& XX 
                 \\
%%%%
\hline
\hline
\end{tabular}    
\end{table}
\end{landscape}

%%%%%%%%%%%%%%%%%%%%%%%%%%%%%%%%%%%%%%%%%%%%%%%%%%%%%%%%%%%%%%%%%%%%%%%%%%%%%%
\section{Applications of tidal effects to gravitational-wave physics}\label{sec:GW}
%%%%%%%%%%%%%%%%%%%%%%%%%%%%%%%%%%%%%%%%%%%%%%%%%%%%%%%%%%%%%%%%%%%%%%%%%%%%%%
In this section we examine how the different contributions to the tidal response of compact objects discussed above imprint themselves on the GW signal from a binary coalescence, and how these imprints can be exploited to probe nuclear physics and the fundamental nature of the compact objects involved.

\subsection{Modelling tidal effects in gravitational-wave signals} \label{sec:modelling}

We start by discussing the various tidal contributions to the GW signal emitted by a binary system, mostly considering a PN framework (see \citealt{Blanchet:2013haa} for a review on the PN formalism).

%%%%%%%%%%%%%%%%%%%%%%%%%%%%%%%%%%%%%%%%%%%%%%%%%%%%%%%
\subsubsection{Tidal effects in post-Newtonian waveforms} \label{sec:PN}
%%%%%%%%%%%%%%%%%%%%%%%%%%%%%%%%%%%%%%%%%%%%%%%%%%%%%%%
Since the seminal work in \citet{Flanagan:2007ix}, tidal effects in the GW signal have been mostly studied within a PN framework (see also \citealt{1973ApJ...185...83M, Hinderer:2009ca, Damour:2012yf}). While originally developed to include only the leading-order effect of the electric, static, quadrupolar LNs,  the framework has been formalized and extended in various ways, including subleading PN terms \citep{Vines:2010ca,Vines:2011ud,Henry:2020ski,Henry:2020pzq, Pnigouras:2019wmt}, higher multipoles \citep{Flanagan:2007ix,Hinderer:2009ca,Abdelsalhin:2018reg}, magnetic LNs \citep{Yagi:2013sva,Banihashemi:2018xfb,Abdelsalhin:2018reg}, dynamical effects \citep{Chakraborty:2025wvs}, dissipation \citep{Maselli:2017cmm,Ripley:2023lsq,Ripley:2023qxo,Chia:2024bwc,HegadeKR:2024slr}, effects of rotation \citep{Abdelsalhin:2018reg}, and nonlinearities \citep{Pani:2025qxs}.

Here we present a detailed computation of the leading PN contribution of the (electric, quadrupolar) static and dynamical LNs to the GW phase, and provide an overview of the remaining subleading tidal effects. A summary is given in~\ref{tab:PN} below.

\paragraph{Working example: Static and dynamical Love numbers}~---
As an explicit example, we derive the leading-order impact of static and dynamical LNs on the gravitational waveform, focusing on the phenomenologically most relevant contribution, namely the electric quadrupolar LN. Accordingly, we restrict our analysis to the mass quadrupole moment of the compact objects, neglecting higher-order multipoles as well as current-type multipole moments.

We consider a spherically symmetric compact body of mass $M$, which is embedded in a quadrupolar tidal field (assumed to be of electric-type) $E_{\mu\nu}=W_{\mu\alpha\nu\beta}u^{\alpha}u^{\beta}$ (where the four-velocity of the compact object is given by $u^{\alpha}$ and $W_{\mu\alpha\nu\beta}$ is the Weyl tensor). The compact object acquires a non-zero quadrupole moment. For a slowly varying and sufficiently weak tidal field, the induced quadrupole moment can be expressed in terms of the tidal field and its time $(\tau)$ derivatives in a linear fashion as \citep{PoissonWill, Chia:2020yla, Bhatt:2023zsy, Saketh:2023bul, Chakraborty:2025wvs} 
%%%%%%%%%%%%%%%%%%%%%%%%%%%%%%%%%%%%%%%%%%%%%%%%%%%%%%%%%%%%%%%%%%%%%%%%%%%%%%%%%
\begin{align}
&M^{\mu \nu}=-M^{5}\sum_{n=0}^{\infty} (-1)^{n}M^n \lambda_{(n)} \frac{d^{n}E^{\mu\nu}}{d\tau^{n}} \nonumber \\
&=-M^{5}\lambda_{(0)}E^{\mu \nu}+M^{6}\lambda_{(1)}\dot E^{\mu \nu}-M^{7}\lambda_{(2)}\ddot E^{\mu \nu} +\mathcal{O}(\dddot E)\,.
\label{eq:genQE}
\end{align}
%%%%%%%%%%%%%%%%%%%%%%%%%%%%%%%%%%%%%%%%%%%%%%%%%%%%%%%%%%%%%%%%%%%%%%%%%%%%%%%%%
For simplicity, since we only consider quadrupolar deformations, the index $\ell=2$ in the LNs have been left implicit \citep{Hinderer:2007mb, Chia:2020yla, Saketh:2023bul}. One can compare the definitions of $\lambda_{(0)}$ with the electric-type LNs $\bar{\lambda}^{\rm YY}_{2}$ defined in~\ref{eq:conventionElectric}, obtaining $\lambda_{0}=\bar{\lambda}_{2}^{\rm YY}$. Thus, as in~\ref{eq:conventionElectric}, the LN $k_{2}^{\rm E}$ is related to $\lambda_{(0)}$ by the factor of $C^{5}$, where the compactness $C$ is given by the ratio (mass/radius) of a given object.
In particular, the static LN $k_{2}^{\rm E}$ is related to the tidal deformability $\lambda_{(0)}$ by the following explicit formula $\lambda_{(0)}=(2/3)k_{2}^{\rm E}C^{-5}$; the result that the Wilson coefficient $\lambda_{(0)}\propto k_{2}^{\rm E}$ also follows from a matching with the EFT (see \citealt{Hui:2020xxx} for an extensive analysis). The dissipation number $\nu_{2}^{\rm E}$ appears at linear order in $M\omega$ and is related to $\lambda_{(1)}$ by $\lambda_{(1)}=(1/3)\nu_{2}^{\rm E}C^{-5}$. In the case of BHs, $\lambda_{(1)}$ corresponds to the tidal heating coefficient \citep{Hartle:1973zz, Alvi:2001mx, Maselli:2017cmm,Chia:2020yla, saketh2022modeling-cf8}, whereas for NSs it is associated to viscosity \citep{Ripley:2023lsq,Ripley:2023qxo,Chia:2024bwc,HegadeKR:2024slr}. Finally, the quadratic-in-frequency LN $k_{2}^{\rm E, \omega^{2}}$ is related to $\lambda_{(2)}$ by $\lambda_{(2)}=-(2/3)k_{2}^{\rm E, \omega^{2}}C^{-5}$.

Given the structure of the quadratic-in-frequency LN, see \ref{eq:DLN}, one can express the Wilson coefficient $\lambda_{(2)}$ as \citep{Chakraborty:2025wvs} 
%%%%%%%%%%%%%%%%%%%%%%%%%%%%%%%%%%%%%%%%%%%%%%%%%%%%%%%%%%%%%%%%%%%%%%%%%%%%%%%%%
\begin{align}
\lambda_{(2)}={\bar\lambda}_{(2)}+\beta_{(2)}\ln \left(\frac{r}{\bar{r}}\right)\,.
\label{eq:ansatzTLN}
\end{align}
%%%%%%%%%%%%%%%%%%%%%%%%%%%%%%%%%%%%%%%%%%%%%%%%%%%%%%%%%%%%%%%%%%%%%%%%%%%%%%%%%
As previously discussed, the logarithmic term is akin to the running of coupling constants. The explicit determination of the coefficients $(\bar{\lambda}_{(2)}, \beta_{(2)})$ requires matching the EFT with the results from perturbation theory incorporating the UV nature of the compact objects \citep{Mandal:2023hqa, Combaluzier--Szteinsznaider:2025eoc}. In the following, we keep these parameters generic, thus remaining agnostic on the nature of the compact objects involved.

To determine the effect of the static and dynamical LN on the GW waveform, we consider the motion of a non-rotating compact object in a binary system, which can be described by the following worldline action \citep{Goldberger:2004jt,Goldberger:2005cd, Porto:2016pyg,Porto:2016zng} 
%%%%%%%%%%%%%%%%%%%%%%%%%%%%%%%%%%%%%%%%%%%%%%%%%%%%%%%%%%%%%%%%%%%%%%%%%%%%%%%%%
\begin{align}
\mathcal{A}&=\mathcal{A}_\text{PP} + \frac{1}{4} \int {\rm d} \tau \, M^{\mu \nu} E_{\mu \nu} 
= -M\int {\rm d} \tau + \frac{1}{4} \int {\rm d} \tau \,  M^{\mu \nu} E_{\mu \nu}\,,\label{eq:WLAction}
\end{align}
%%%%%%%%%%%%%%%%%%%%%%%%%%%%%%%%%%%%%%%%%%%%%%%%%%%%%%%%%%%%%%%%%%%%%%%%%%%%%%%%%
where $\mathcal{A}_\text{PP}$ is the point-particle action, $\tau$ is the proper time along the worldline, $M^{\mu\nu}$ is the mass quadrupole moment, and $E^{\mu \nu}$ is the quadrupolar tidal field. The connection between $M^{\mu \nu}$ and $E^{\mu \nu}$ is given by~\ref{eq:genQE}. The above action, combined with the orbital energy-balance law of the form \citep{Vines:2010ca, Vines:2011ud, Hinderer:2010ih, Abdelsalhin:2018reg}
%%%%%%%%%%%%%%%%%%%%%%%%%%%%%%%%%%%%%%%%%%%%%%%%%%%%%%%%%%%%%%%%%%%%%%%%%%%%%%%%%
\begin{equation}
\frac{d\mathcal{E}}{dt}=- F\,,
\label{eq:eblaw}
\end{equation}
%%%%%%%%%%%%%%%%%%%%%%%%%%%%%%%%%%%%%%%%%%%%%%%%%%%%%%%%%%%%%%%%%%%%%%%%%%%%%%%%%
enables one to derive the phase of the gravitational waveform due to static and dynamical LNs.
In~\ref{eq:eblaw}, the binding energy $\mathcal{E}$ of the orbit arises from the conservative part of~\ref{eq:WLAction}. On the other hand the dissipative part, associated to GW emission or other possible dissipative effects (see below), comes from the flux $F$. 

Let us consider two non-rotating compact objects with masses $M_1$ and $M_2$, and symmetric mass ratio $\eta \equiv M_1 M_2 / M_{\rm T}^2$, where $M_{\rm T} = M_1 + M_2$ denotes the total mass \citep{Vines:2010ca}. Since the two bodies are widely separated, we may work in standard Cartesian coordinates with Euclidean metric $\delta_{ij}$, and describe their worldlines by $x^i = z_A^i(t)$ ($A=1,2$). The relative separation vector is then $z^i \equiv z_2^i - z_1^i$, with relative velocity and acceleration given by $v^i = \dot z^i$ and $a^i = \ddot z^i$, respectively. 

Exploiting the central nature of the Newtonian interaction, we introduce planar polar coordinates, defining $r = \sqrt{z^i z^i}$, the orbital phase $\varphi$, and the angular velocity $\omega \equiv \dot\varphi$, with unit separation vector $n^i \equiv z^i/r$. As in the Kepler problem, the tidal contribution to the Lagrangian induced by body 2 can then be written, retaining only static and dynamical tidal effects \citep{Blanchet:2013haa,Vines:2010ca, Vines:2011ud, Abdelsalhin:2018reg, Chakraborty:2025wvs}, as 
%%%%%%%%%%%%%%%%%%%%%%%%%%%%%%%%%%%%%%%%%%%%%%%%%%%%%%%%%%%%%%%%%%%%%%%%%%%%%%%%%
\begin{align}
L_{\rm eff}&=\frac{1}{2} \eta M_{\rm T} \dot{r}^2 + \frac{1}{2} \eta M_{\rm T} r^2 \dot{\varphi}^2 + \frac{\eta M_{\rm T}^2}{r} \frac{1}{4}M_{2}^{5} 
\left[ \lambda_{(0)} E^{i j} E^{i j} (r) - M_2^2 \lambda_{(2)} (r) \dot{E}^{i j} \dot{E}^{i j} (r)  \right]\,,\label{eq:effL}
\end{align}
%%%%%%%%%%%%%%%%%%%%%%%%%%%%%%%%%%%%%%%%%%%%%%%%%%%%%%%%%%%%%%%%%%%%%%%%%%%%%%%%%
where the quadrupolar tidal field and its time derivative have the following expressions \citep{Vines:2011ud}
%%%%%%%%%%%%%%%%%%%%%%%%%%%%%%%%%%%%%%%%%%%%%%%%%%%%%%%%%%%%%%%%%%%%%%%%%%%%%%%%%
\begin{align}
E^{i j} &= 
- \frac{3M_1}{r^3} \left( n^i n^j - \frac{1}{3}\delta^{ij} \right)\,, 
\nonumber 
\\ 
\dot{E}^{i j} & 
 = \frac{M_1}{r^4} \left[ \dot r \left(15 n^i n^j - 3\delta^{ij} \right) -  3\left( v^i  n^j+ v^j n^i \right) \right]\,.
\end{align}
%%%%%%%%%%%%%%%%%%%%%%%%%%%%%%%%%%%%%%%%%%%%%%%%%%%%%%%%%%%%%%%%%%%%%%%%%%%%%%%%%
The tidal effect of the first body, and its corresponding contribution to the gravitational waveform, can be obtained straightforwardly by repeating the above analysis with the labels $1$ and $2$ exchanged. We further note that the results derived above apply to circular orbits. Accordingly, imposing $\dot r = \ddot r = 0$ and using the Euler--Lagrange equation $\dot\omega = 0$, we obtain the relation between the orbital separation $r$ and the circular-orbit frequency,
%%%%%%%%%%%%%%%%%%%%%%%%%%%%%%%%%%%%%%%%%%%%%%%%%%%%%%%%%%%%%%%%%%%%%%%%%%%%%%%%%
\begin{align}
\medmath{r(\omega)=\frac{M_{\rm T}^{1/3}}{\omega^{2/3}} \left[1+3\frac{(M_{1}M_{2}^{4})\omega^{10/3}}{M_{\rm T}^{5/3}}\left(\lambda_{(0)2}- 3M_{2}^{2}\lambda_{(2)2}\omega^2+\frac{1}{2}M_{2}^{2}M_{\rm T}^{1/3}\omega^{4/3}\frac{d\lambda_{(2)2}}{dr}\right)+(1\leftrightarrow 2)\right]\,.}
\label{eq:rw}
\end{align}
%%%%%%%%%%%%%%%%%%%%%%%%%%%%%%%%%%%%%%%%%%%%%%%%%%%%%%%%%%%%%%%%%%%%%%%%%%%%%%%%%
The first term is the standard Kepler's second law (namely, orbital time period)$^{2}$=(radial separation)$^{3}$), while the subsequent terms are the leading-order PN corrections due to the tidal LNs considered here. Clearly, the above result neglects subleading point-particle PN corrections as well as other tidal effects, to be discussed below.

One can then compute the Hamiltonian, which is also the binding energy of the system. This can be transformed to a function of the frequency, by substituting the radial coordinate $r$ in~\ref{eq:rw} into the binding energy. The final expression for the binding energy reads 
%%%%%%%%%%%%%%%%%%%%%%%%%%%%%%%%%%%%%%%%%%%%%%%%%%%%%%%%%%%%%%%%%%%%%%%%%%%%%%%%%
\begin{align}
&\medmath{\mathcal{E}(\omega)=-\frac{1}{2}\eta M_{\rm T}^{5/3}\omega^{2/3}\Big\{1-9\frac{(M_{1} M_{2})\omega^{10/3}}{M_{\rm T}^{5/3}}\Big[(M_{1}^{3}\lambda_{(0)1}+M_{2}^{3}\lambda_{(0)2})-5(M_{1}^{5}\bar\lambda_{(2)1}+M_{2}^{5}\bar\lambda_{(2)2})\omega^2 }
\nonumber 
\\
&\medmath{+\frac{2}{3}(M_{1}^{5}\beta_{(2)1}+M_{2}^{5}\beta_{(2)2})\omega^{2}-5\left\{M_{1}^{5}\beta_{(2)1} \ln \left(M_{\rm T}^{1/3}/\omega^{2/3}\bar{r}_1\right)+M_{2}^{5}\beta_{(2)2}\ln\left(M_{\rm T}^{1/3}/\omega^{2/3}\bar{r}_2\right)\right\}\omega^2\Big]\Big\}\,.}\label{eq:ew}
\end{align}
%%%%%%%%%%%%%%%%%%%%%%%%%%%%%%%%%%%%%%%%%%%%%%%%%%%%%%%%%%%%%%%%%%%%%%%%%%%%%%%%%
Here we have replaced $d\lambda_{(2)A}/dr=\beta_{(2)A}\omega^{2/3}/M_{\rm T}^{1/3}$ (with $A=1,2$), obtained by using~\ref{eq:ansatzTLN} and the Newtonian formula connecting time period with the radial separation between the two compact objects. 

Moving to the dissipative corrections, the GW flux emitted by the binary, at leading PN order, is given by the standard quadrupole formula \citep{Blanchet:2013haa}
%%%%%%%%%%%%%%%%%%%%%%%%%%%%%%%%%%%%%%%%%%%%%%%%%%%%%%%%%%%%%%%%%%%%%%%%%%%%%%%%%
\begin{align}
F \equiv\dot E = \frac{1}{5} \frac{d^3 M_{\rm T}^{ij}}{d t^3} \frac{d^3 M_{\rm T}^{ij}}{d t^3}\,,
\end{align}
%%%%%%%%%%%%%%%%%%%%%%%%%%%%%%%%%%%%%%%%%%%%%%%%%%%%%%%%%%%%%%%%%%%%%%%%%%%%%%%%%
where $M_{\rm T}^{ij}$ is the total quadrupole moment, which is the sum of the quadrupole moments of the individual compact objects in the binary and the orbital one due to the reduced mass $\mu=(M_{1}M_{2}/M_{\rm T})$,
%%%%%%%%%%%%%%%%%%%%%%%%%%%%%%%%%%%%%%%%%%%%%%%%%%%%%%%%%%%%%%%%%%%%%%%%%%%%%%%%%
\begin{equation}
M^{ij}_T=M_{1}^{ij}+M_{2}^{ij}+\mu r^{2}\left(n^{i}n^{j}-\frac{1}{3}\delta^{ij}\right)\,.
\end{equation}
%%%%%%%%%%%%%%%%%%%%%%%%%%%%%%%%%%%%%%%%%%%%%%%%%%%%%%%%%%%%%%%%%%%%%%%%%%%%%%%%%
Subsequently computing the third time derivatives of the mass moments, keeping in mind that these moments are computed for circular orbits, we finally obtain \citep{Chakraborty:2025wvs} 
%%%%%%%%%%%%%%%%%%%%%%%%%%%%%%%%%%%%%%%%%%%%%%%%%%%%%%%%%%%%%%%%%%%%%%%%%%%%%%%%%
\begin{align}
&F(\omega)= \frac{32}{5} \mu^2 M_{\rm T}^{4/3} \omega^{10/3} \Bigg\{1 + \left[M_1^4 \left(12 \frac{M_2}{M_{\rm T}}+6\right) \lambda_{(0)1} + M_2^4 \left(12 \frac{M_1}{M_{\rm T}}+6\right) \lambda_{(0)2} \right]\frac{\omega^{10/3}}{M_{\rm T}^{2/3}}  
\nonumber \\
&  + 12 \left[M_1^6 \left(\frac{M_2}{M_{\rm T}}\frac{1}{2} \beta_{(2)1}- \left\{3\frac{M_2}{M_{\rm T}} +2 \right\}\bar{\lambda}_{(2)1}  \right) + M_2^6 \left( \frac{M_1}{M_{\rm T}}\frac{1}{2} \beta_{(2)2} - \left\{3\frac{M_1}{M_{\rm T}}+2\right\}\bar{\lambda}_{(2)2} \right)\right] \frac{\omega^{16/3}}{M_{\rm T}^{2/3}}  \nonumber \\
& - 12 \left[M_1^6 \left(3\frac{M_2}{M_{\rm T}} + 2 \right) \beta_{(2)1} \ln \left(\frac{M_{\rm T}^{1/3}}{\omega^{2/3}\bar{r}_1} \right) + M_2^6 \left(3\frac{M_1}{M_{\rm T}} + 2 \right) \beta_{(2)2} \ln \left(\frac{M_{\rm T}^{1/3}}{\omega^{2/3}\bar{r}_2} \right) \right] \frac{\omega^{16/3}}{M_{\rm T}^{2/3}} \Bigg\}\,.
\end{align}
%%%%%%%%%%%%%%%%%%%%%%%%%%%%%%%%%%%%%%%%%%%%%%%%%%%%%%%%%%%%%%%%%%%%%%%%%%%%%%%%%
Given the above expressions for the binding energy as a function of the frequency, as well as that of the gravitational flux, we can compute the GW phase through the stationary phase approximation \citep{Blanchet:2013haa}, i.e., $(d^{2}\psi/d\omega^{2})=-(2/F)(d\mathcal{E}/d\omega)$, yielding \citep{Vines:2011ud, Chakraborty:2025wvs},
%%%%%%%%%%%%%%%%%%%%%%%%%%%%%%%%%%%%%%%%%%%%%%%%%%%%%%%%%%%%%%%%%%%%%%%%%%%%%%%%%
\begin{align}
\label{psi-gen}
&\psi(x)= \frac{3}{128 \, \eta \, x^{5/2}} \left\{1 -\underbrace{\frac{24}{M_{\rm T}^4}
\left[M_1^3(M_1 +11 \mu)\lambda_{(0)1} + M_2^3(M_2 + 11 \mu)\lambda_{(0)2} \right]}_{(39/2)\Lambda_{(0)}} x^{5} \right. 
\nonumber 
\\
&+\frac{15}{11 M_{\rm T}^6} \Big[M_1^5(8M_1 +147 \mu)\bar \lambda_{(2)1}+M_2^5(8M_2 +147 \mu)\bar \lambda_{(2)\,2}
\nonumber
\\
&+\underbrace{M_1^5(\frac{38}{11}M_1 +\frac{1253}{44}\mu)\beta _{(2)\,1}+M_2^5(\frac{38}{11}M_2 +\frac{1253}{44}\mu)\beta _{(2)2}\Big]}_{(15/11)\Lambda_{(2)}} x^{8} 
\nonumber
\\
&\left. + \frac{15}{11} \left[  \underbrace{\frac{M_1^5(8M_1 + 147 \mu)}{M_{\rm T}^{6}} \beta_{(2)1}}_{B_{(2)1}} \ln \left(\frac{M_{\rm T}}{\bar{r}_1 x}\right)  + \underbrace{\frac{M_2^5(8M_2 + 147 \mu)\beta _{(2)2}}{M_{\rm T}^{6}}}_{B_{(2)2}} \ln \left(\frac{M_{\rm T}}{\bar{r}_2 x}\right)\right] x^{8}  
\right\}\,. 
\end{align}
%%%%%%%%%%%%%%%%%%%%%%%%%%%%%%%%%%%%%%%%%%%%%%%%%%%%%%%%%%%%%%%%%%%%%%%%%%%%%%%%%
The above expression for the phase of the GW waveform is now expressed in terms of the dimensionless frequency parameter $x = (M_{\rm T} \omega)^{2/3}=(v/c)^{2}$, where $v$ is the orbital binary. We have also collected the individual PN pieces using a standard nomenclature \citep{Hinderer:2009ca}. Note that the first line of~\ref{psi-gen} reproduces the standard result for the leading-order effect of the (static, electric, quadrupolar) tidal LN, entering the GW phase at 5PN order \citep{Flanagan:2007ix}.

While $k_2^{\rm E}=0$ for BHs, $k_2^{\rm E}={\cal O}(0.1)$ for ordinary NSs (see~\ref{fig:LNvsC}). Therefore, although the static LNs enter the GW waveform at 5PN order, the tidal term for ordinary neutron stars is enhanced by the fifth power of the inverse compactness, $C^{-5}\approx 10^4$ for $C\approx 0.15$. This makes the tidal term much more relevant than other sub-leading point-particle contributions entering at lower PN order, especially in the last stages of the binary NS coalescence.

As evident from~\ref{psi-gen}, the leading-order correction from the dynamical LN enters at 8PN order, consistent with the fact that it is suppressed by a further term of $\mathcal{O}(M_{\rm T}^{2}\omega^2)\sim \mathcal{O}(v^{6}/c^{6}))$ \citep{Pitre:2023xsr,Pitre:2025qdf,Chakraborty:2025wvs}. 
This might suggest that the dynamical LNs are suppressed in the GW waveform. However, the LN $k_{2}^{\rm E, \omega^{2}}$ has the same compactness falloff as the static one $k_{2}^{E}$ and hence enhanced by a factor $C^{-5}$. Actually, the compactness enhancement is even stronger: the quantity entering the waveform is $\lambda_{(2)}\sim \ddot k_2^E C^{-8}$, where $\ddot k_{2}^{\rm E}=C^{3}k_{2}^{\rm E,\omega^{2}}$ was introduced in \citet{Poisson:2020vap,Pitre:2023xsr,Pitre:2025qdf}.
Since $\ddot k_2^E={\cal O}(0.1)$ for NSs (and nonvanishing for BHs), see~\ref{fig:NSdyn}, the contribution of dynamical LN to the GW waveform is therefore much larger than its nominal PN order, especially for low-compactness objects.

In addition, the 8PN term in~\ref{psi-gen} has an intrinsic ambiguity associated with the term $\Lambda_{(2)}$ in the tidal PN phase. This is because the effective renormalization scales $\bar{r}_A$ in the logarithmic tidal coefficient can be absorbed within $\Lambda_{(2)}$, the 8PN tidal coefficient. Resolving this ambiguity requires a proper matching procedure between the underlying EFT and the perturbative computation incorporating the UV nature of the compact object, which have been achieved very recently only for BHs \citep{Combaluzier--Szteinsznaider:2025eoc,Kobayashi:2025vgl}. 

%%%%%%%%%%%%%%%%%%%%%%%%%%%%%%%%%%%%%%%%%%%%%%%%
\paragraph{Subleading PN corrections}~---
%%%%%%%%%%%%%%%%%%%%%%%%%%%%%%%%%%%%%%%%%%%%%%%%
For the \emph{static} (electric and quadrupolar) LNs, the PN computation sketched above has been generalized to the next-to \citep{Vines:2010ca,Vines:2011ud} and next-to-next-to \citep{Henry:2020ski,Henry:2020pzq} leading PN, entering at 6PN and 7PN, respectively, with ${\cal O}(1)$ coefficients proportional to $\lambda_{(0)A}$ ($A=1,2$). This requires computing the binding energy and fluxes up to the corresponding PN order in the tidal term.

The next-to-leading (6PN) correction is particularly important. Indeed, the 5PN coefficient $\Lambda_{(0)}$ introduced in~\ref{psi-gen} depends on a combination of $\lambda_{(0)A}$ ($A=1,2$). As such, $\Lambda_{(0)}$ alone does not allow measuring $\lambda_{(0)1}$ and $\lambda_{(0)2}$ independently. On the other hand, the 6PN coefficient depends on a \emph{different} combination of $\lambda_{(0)1}$ and $\lambda_{(0)2}$, breaking this degeneracy.
In this context, certain dimensionless (symmetric and antisymmetric) combinations of the tidal deformability have been shown to be only mildly sensitive to the NS EoS \citep{Yagi:2015pkc, Yagi:2016bkt, Yagi:2016qmr}.

%%%%%%%%%%%%%%%%%%%%%%%%%%%%%%%%%%%%%%%%%%%%%%%%
\paragraph{Higher-$\ell$ electric terms}~---
%%%%%%%%%%%%%%%%%%%%%%%%%%%%%%%%%%%%%%%%%%%%%%%%
The GW phase corrections from higher-$\ell$ electric LNs have been computed in \citet{Hinderer:2009ca} (see also \citealt{Abdelsalhin:2018reg}), and only for the static LN. In this case the tidal term in the finite size Lagrangian is proportional to $\hat\lambda_{(0)}E^{ijk}E^{ijk}$ and hence leads to (keeping only the leading order static contribution)
%%%
\begin{align}
&M^{\mu \nu \rho}=-M^{7}\hat \lambda_{(0)}E^{\mu \nu \rho}+...\,.
\label{eq:genQEl3}
\end{align}
%%%
where $E^{\mu \nu \rho}$ and $M^{\mu \nu \rho}$ are the static, octupolar, electric tidal field and the corresponding induced mass octupole, respectively. The dimensionless tidal coefficient $\hat \lambda_{(0)}$ is proportional to the $\ell=3$ static electric LN, $\hat \lambda_{(0)}\propto k_E^3 C^{-7}$ (note that, for ease of notation, we introduced $\hat \lambda_{(0)}\equiv \bar\lambda^{\rm YY}_3$ as defined in~\ref{eq:conventionElectric}).
 The above extra piece $\hat\lambda_{(0)}E^{ijk}E^{ijk}$ in the effective action ultimately leads to the following GW phase under the stationary-phase approximation \citep{Hinderer:2009ca,Abdelsalhin:2018reg} 
%%%%%%%%%%%%%%%%%%%%%%%%%%%%%%%%%%%%%%%%
\begin{equation}
\psi(x) \supset - \frac{3}{128 \, \eta \, x^{5/2}}\left(\frac{4000}{9}-\frac{4000 M_{\rm T}}{11M_1}\right)\left(\frac{M_{1}}{M_{\rm T}}\right)^7\hat\lambda_{(0)1} x^7 +(1\leftrightarrow 2)/,.
\end{equation}
%%%%%%%%%%%%%%%%%%%%%%%%%%%%%%%%%%%%%%%%
The above correction enters at 7PN order and is enhanced by factors $\sim C^{-7}$ for each binary component. This scaling can be generalized: an electric LN of multipolar order $\ell$ enters at $(2\ell+1)$PN order and is enhanced by $\sim C^{-(2\ell+1)}$.

\paragraph{Static magnetic Love numbers}~---
The GW phase corrections from static, quadrupolar magnetic LNs have been computed in \citet{Yagi:2013sva} (see also \citealt{Abdelsalhin:2018reg}). In this case,~\ref{eq:genQE} should be replaced by (keeping only the first term)
%%%
\begin{align}
&S^{\mu \nu}=-M^{5}\sigma_{(0)}B^{\mu \nu}+...\,.
\label{eq:genQBl3}
\end{align}
%%%
where $B^{\mu \nu}$ and $S^{\mu \nu}$ are the static, quadrupolar, \emph{magnetic} tidal field and the corresponding induced \emph{current} quadrupole, respectively. The tidal coefficient $\sigma_{(0)}$ is proportional to the $\ell=2$ static magnetic LN, $\sigma_{(0)}\propto k_B^2 (R/M)^5$ (see~\ref{eq:conventionMagnetic}, where for ease of notation we introduced $\sigma_{(0)}=\bar\sigma_2^{\rm YY}$).

In this case the Lagrangian of the effective action will contain an additional term proportional to $\sigma_{(0)}B^{ij}B^{ij}$. This extra piece affects the GW phase, which in the stationary-phase-approximation reads \citep{Yagi:2013sva,Abdelsalhin:2018reg}
%%%
\begin{equation}
    \psi(x) \supset -\frac{3}{128 \, \eta \, x^{5/2}}\left(\frac{6920}{7}-\frac{20740 M_{\rm T}}{21M_1}\right)\left(\frac{M_{1}}{M_{\rm T}}\right)^{5}\sigma_{(0)1} x^6 +(1\leftrightarrow 2)
\end{equation}
%%%
entering at 6PN order. This can be understood since both $S^{\mu\nu}$ and $B^{\mu\nu}$ are suppressed by a factor $(v/c)$ relative to their electric counterparts $M^{\mu\nu}$ and $E^{\mu\nu}$, yielding an overall $(v^2/c^2)$ suppression relative to the 5PN usual, electric, LN.
Although this term is formally the same PN order as the next-to-leading quadrupolar electric term, note that the magnetic LN is typically smaller than the electric one (see~\ref{fig:LNvsC}).

\paragraph{Rotational/Mixing Love numbers}~---
Spin-tidal effects on the waveforms were computed in \citet{Abdelsalhin:2018reg}. As previously discussed, they mix terms with opposite parity and different angular momentum. The interaction Lagrangian in the effective field theory now contains two new couplings:
\begin{align}
    {\cal L}_{\rm int} \supset \alpha S^i M^{jk}S^{ijk} +\beta  S^i S^{jk}M^{ijk} \label{eq:RTLNs}
\end{align}
%%%
where $S^i$ is the spin vector of the object, introduced in~\ref{scatampapp}. The coupling constants $\alpha$ ($\beta$) regulates the coupling between the mass (current) quadrupole and the current (mass) octupole. These terms enter at 6.5 PN order, together with the terms due to the coupling between the spins and the ordinary LNs \citep{Abdelsalhin:2018reg}.
The PN order can be explained by noticing that the electric quadrupole moments enters at 2PN and is proportional to the magnetic octupolar tidal field, which scales as a 4.5PN term. Likewise, the magnetic quadrupole moments enters at 3PN and is proportional to the electric octupolar tidal field, which scales as a 3.5PN term. In both cases the combination yields a 6.5PN correction.

Overall, the spin-tidal terms arising from $\ell$-pole rotational LNs to linear
order in the spin enter the GW phase at $(2\ell + 1/2 +
2\delta_{\ell2}$)PN order \citep{Abdelsalhin:2018reg}. Therefore, for any $\ell \geq 3$, this contribution
enters half PN order \emph{earlier} than the standard electric
TLNs of order $\ell$ (the latter entering at $(2\ell + 1)$PN order).
Likewise,
the
spin-tidal terms arising from $\ell$-pole standard LNs (both electric
and magnetic) to linear order in the spin enter the GW
phase at $(2\ell + 5/2)$PN order \citep{Abdelsalhin:2018reg}.

Finally, note that~\ref{eq:RTLNs} implies that only \emph{two} coupling constants ($\alpha$ and $\beta$) regulates \emph{four} induced moments, namely the $\ell=2,3$ mass and current moments respectively induced by the $\ell=3,2$ magnetic and electric tidal field at first order in the spin. From the point of view of perturbation theory, this is surprising since it means that perturbations belonging to opposite parity sectors are related.
This ``hidden'' symmetry, suggested on the basis of a Lagrangian description of the tidal interaction in a binary system, was actually shown to hold independently of the EoS of the star \citep{Castro:2021wyc}.

%%%%%%%%%%%%%%%%%%%%%%%%%%%%%%%%%%%
\paragraph{Quadratic Love numbers}~---
%%%%%%%%%%%%%%%%%%%%%%%%%%%%%%%%%%%
Nonlinear tidal effects on the GW signal can be computed by including higher-order tidal terms in the interaction Lagrangian. In particular, quadratic effects are associated with a worldline tidal action containing a cubic term in the tidal field \citep{Pani:2025qxs,Pitre:2025qdf}
\begin{equation}
    S_{\rm tidal}\supset \int d\tau \lambda_{222} {E_\mu}^\rho {E_\rho}^\nu{E_\nu}^\mu
\end{equation}
%%%
where $\tau$ is the proper time along the particle worldline and the quadratic tidal deformability parameter $\lambda_{222}$ is related to its dimensionless parameter through~\ref{dimscale}.

In this case, the contribution to the GW phase in the stationary-phase-approximation reads \citep{Pani:2025qxs} 
\begin{equation}
    \psi(x) \supset \psi_{\rm N}(x)\frac{90 \eta_2}{11}\left(34-\frac{35}{\eta_1}\right) \frac{\lambda_{222}^{(1)}}{M_{\rm T}^7} x^8 +(1\leftrightarrow 2)
\end{equation}
where $\lambda_{222}^{(1)}$ is the quadratic tidal parameter of the first body. As anticipated, this term enters at 8PN order but also scales as $p_2/C^8$, where $p_2={\cal O}(0.1)$ (see~\ref{fig:p2-222}).

\paragraph{Modelling tidal heating: black holes}~--- 
We have so far discussed the conservative part of the tidal response, and here we specialize to the dissipative effects, namely inclusion of tidal heating to the gravitational waveform \citep{Hartle:1973zz, Hughes:2001jr, Maselli:2017cmm, Alvi:2001mx, Chatziioannou:2012gq, Chatziioannou:2016kem, Datta:2020gem, Mukherjee:2022wws, HegadeKR:2024agt, Chia:2024bwc, Mukherjee:2025wxa, saketh2022modeling-cf8}. We start with the leading dissipative effects for both rotating and non-rotating BHs in GR. For this purpose, we use~\ref{heatbranerot} and~\ref{heatbranenonrot}, both of which, at the leading order, can be expressed as,
%%%%%%%%%%%%%%%%%%%%%%%%%%%%%%%%%%%%%%%%%%%%%%%%%%%%%%%%%%%
\begin{align}
\left(\frac{dM_{1}}{dt}\right)_{\rm rot}&=-\left(\dfrac{dE}{dt}\right)_{\rm N}\left(\frac{M_{1}}{M}\right)^{3}\frac{v^{5}}{4}
\chi_{1}(1+3\chi_{1}^{2})\left(\hat{\bf{L}}_{\rm orb}\cdot \hat{\bf{J}}_{1}\right)\,,
\\
\left(\frac{dM_{1}}{dt}\right)_{\rm non-rot}&=\left(\dfrac{dE}{dt}\right)_{\rm N}\left(\frac{M_{1}}{M_{\rm T}}\right)^{4}v^{8}\,,
\end{align}
%%%%%%%%%%%%%%%%%%%%%%%%%%%%%%%%%%%%%%%%%%%%%%%%%%%%%%%%%%%
where $(dE/dt)_{\rm N}=(32/5)\varsigma^{2}v^{10}$ provides us the energy loss due to the quadrupole radiation, with $M_{\rm T}\equiv M_{1}+M_{2}$ is the total mass, $\varsigma\equiv(M_{1}M_{2}/M_{\rm T}^{2})$ is the symmetric mass ratio, and $v=\sqrt{M_{\rm T}/b}$ is the relative velocity of the binary BH system. The corresponding expression for $(dM_{2}/dt)$ can be obtained by interchanging $M_{1}\leftrightarrow M_{2}$. We thus obtain the total GW flux going into the horizon of rotating and non-rotating BHs, 
%%%%%%%%%%%%%%%%%%%%%%%%%%%%%%%%%%%%%%%%%%%%%%%%%%%%%%%%%%%
\begin{align}\label{flux_horizon}
F_{\rm BH}^{\rm rot}&=\sum_{i}\left(\frac{dM_{i}}{dt}\right)
=-\left(\dfrac{dE}{dt}\right)_{\rm N}\underbrace{\left\{\sum_{i}\left(\frac{M_{i}}{M_{\rm T}}\right)^{3}
\chi_{i}(1+3\chi_{i}^{2})\left(\hat{\bf{L}}_{\rm orb}\cdot \hat{\bf{J}}_{i}\right)\right\}}_{\mathcal{G}_{5}}\frac{v^{5}}{4}\,,
\\
F_{\rm BH}^{\rm non-rot}&=\sum_{i}\left(\frac{dM_{i}}{dt}\right)=\left(\dfrac{dE}{dt}\right)_{\rm N}\underbrace{\left\{\sum_{i}2\left(\frac{M_{i}}{M_{\rm T}}\right)^{4}\right\}}_{\mathcal{G}_{8}}\frac{v^{8}}{2}~.
\end{align}
%%%%%%%%%%%%%%%%%%%%%%%%%%%%%%%%%%%%%%%%%%%%%%%%%%%%%%%%%%%
Having derived the flux of the GW through the horizon, leading to tidal heating, we will now compute the corresponding modifications to the GW phase. Following \citet{Tichy:1999pv}, we express the phase of the GW in terms of the relative velocity as
%%%%%%%%%%%%%%%%%%%%%%%%%%%%%%%%%%%%%%%%%%%%%%%%%%%%%%%%%%%
\begin{align}\label{phase_integral}
\psi(v)=-2\int^{v}d\mathtt{v}\left(v^{3}-\mathtt{v}^{3}\right)\frac{(dE_{\rm orb}/d\mathtt{v})}{F_{\infty}(\mathtt{v})+F_{\rm BH}(\mathtt{v})}~.
\end{align}
%%%%%%%%%%%%%%%%%%%%%%%%%%%%%%%%%%%%%%%%%%%%%%%%%%%%%%%%%%%
Here the orbital energy of the binary BH system is given by $E_{\rm orb}$, The GW flux at infinity is described by $F_{\infty}$ and we have already defined $F_{\rm H}$ as the horizon absorption flux. For the orbital energy $E_{\rm orb}$, we obtain the following PN expansion of the orbital energy \citep{Damour:1997ub},
%%%%%%%%%%%%%%%%%%%%%%%%%%%%%%%%%%%%%%%%%%%%%%%%%%%%%%%%%%%
\begin{align}
E_{\rm orb}(\mathtt{v})=-\frac{\varsigma}{2}\mathtt{v}^{2}\left[1-\frac{9+\varsigma}{12}\mathtt{v}^{2}\right]~.
\end{align}
%%%%%%%%%%%%%%%%%%%%%%%%%%%%%%%%%%%%%%%%%%%%%%%%%%%%%%%%%%%
Such that, 
%%%%%%%%%%%%%%%%%%%%%%%%%%%%%%%%%%%%%%%%%%%%%%%%%%%%%%%%%%%
\begin{align}
\dfrac{dE_{\rm orb}}{d\mathtt{v}}=-\varsigma \mathtt{v}\left[1-\frac{9+\varsigma}{6}\mathtt{v}^{2}\right]~.
\end{align}
%%%%%%%%%%%%%%%%%%%%%%%%%%%%%%%%%%%%%%%%%%%%%%%%%%%%%%%%%%%
Similarly, the PN expansion for the GW flux through infinity takes the following expression \citep{Isoyama:2017tbp}, 
%%%%%%%%%%%%%%%%%%%%%%%%%%%%%%%%%%%%%%%%%%%%%%%%%%%%%%%%%%%
\begin{align}
F_{\infty}(\mathtt{v})&=\frac{32}{5}\varsigma^{2}\mathtt{v}^{10}\Big[1-\left(\frac{1247}{336}+\frac{35}{12}\varsigma\right)\mathtt{v}^{2}+(4\pi+F_{\rm SO})\mathtt{v}^{3}\Big]~,
\end{align}
%%%%%%%%%%%%%%%%%%%%%%%%%%%%%%%%%%%%%%%%%%%%%%%%%%%%%%%%%%%
where, $\varsigma$ is the symmetric mass ratio and $F_{\rm SO}$ is the leading order spin-orbit coupling term, which takes the following expression: 
%%%%%%%%%%%%%%%%%%%%%%%%%%%%%%%%%%%%%%%%%%%%%%%%%%%%%%%%%%%
\begin{align}
F_{\rm SO}=-(1+\Delta)^{2}\chi_{1}-(1-\Delta)^{2}\chi_{2}-\frac{5\Delta}{4} \left\{-\frac{1+\Delta}{2}\chi_{1}+\frac{1-\Delta}{2}\chi_{2}\right\}\,.
\end{align}
%%%%%%%%%%%%%%%%%%%%%%%%%%%%%%%%%%%%%%%%%%%%%%%%%%%%%%%%%%%
Here $\Delta=-\sqrt{1-4\varsigma}$. As evident, in the non-rotating case the spin-orbit term vanishes identically. The above provides the flux of GW at infinity, while for computing the phase of the GW we also need the GW flux through the horizon, which for rotating and non-rotating BH, respectively yields, 
%%%%%%%%%%%%%%%%%%%%%%%%%%%%%%%%%%%%%%%%%%%%%%%%%%%%%%%%%%%
\begin{align}
F_{\rm BH}^{\rm rot}(\mathtt{v})=-\frac{32}{5}\varsigma^{2}\mathtt{v}^{10}\left(\frac{\mathtt{v}^{5}}{4}\mathcal{G}_{5}\right)\,;
\quad
F_{\rm BH}^{\rm non-rot}(\mathtt{v})=\frac{32}{5}\varsigma^{2}\mathtt{v}^{10}\left(\frac{\mathtt{v}^{8}}{2}\mathcal{G}_{8}\right)~.
\end{align}
%%%%%%%%%%%%%%%%%%%%%%%%%%%%%%%%%%%%%%%%%%%%%%%%%%%%%%%%%%%
For the rotating case, the contribution of tidal heating from the $(1/\textrm{Flux})$ term in the phase integral in~\ref{phase_integral} yields 
%%%%%%%%%%%%%%%%%%%%%%%%%%%%%%%%%%%%%%%%%%%%%%%%%%%%%%%%%%%
\begin{align}
\frac{1}{F_{\infty}+F_{\rm BH}^{\rm rot}}&=\frac{5}{32\varsigma^{2}}\frac{1}{\mathtt{v}^{10}}\Bigg[\frac{\mathcal{G}_{5}}{4}\mathtt{v}^{5}
+\left(\frac{1247}{336}+\frac{35}{12}\varsigma\right)\frac{\mathcal{G}_{5}}{2}\mathtt{v}^{7}-\left\{\frac{\mathcal{G}_{5}}{2}\left(4\pi+F_{\rm SO}\right)\right\}\mathtt{v}^{8}\Bigg]~.
\end{align}
%%%%%%%%%%%%%%%%%%%%%%%%%%%%%%%%%%%%%%%%%%%%%%%%%%%%%%%%%%%
Therefore the contribution of tidal heating to the phase of the GW becomes 
%%%%%%%%%%%%%%%%%%%%%%%%%%%%%%%%%%%%%%%%%%%%%%%%%%%%%%%%%%%
\begin{align}\label{phase_integral_heating}
\psi^{\rm rot}_{\rm TH}(v)&=\frac{10}{32\varsigma}\Bigg[-\frac{\mathcal{G}_{5}}{12}\left(1+3\ln v\right)-\frac{3\mathcal{G}_{5}}{8}\left(\frac{995}{168}+\frac{952}{168}\varsigma\right)v^{2}-\left\{\frac{\mathcal{G}_{5}}{2}\left(4\pi+F_{\rm SO}\right)\right\}\frac{v^{3}}{3}\left(3\ln v-1\right)\Bigg]~.
\end{align}
%%%%%%%%%%%%%%%%%%%%%%%%%%%%%%%%%%%%%%%%%%%%%%%%%%%%%%%%%%%
For the non-rotating case, the inverse of the total flux to infinity and to the horizon is given by,
%%%%%%%%%%%%%%%%%%%%%%%%%%%%%%%%%%%%%%%%%%%%%%%%%%%%%%%%%%%
\begin{align}
\frac{1}{F_{\infty}+F_{\rm BH}^{\rm non-rot}}&=-\frac{5}{32\varsigma^{2}}\frac{1}{\mathtt{v}^{10}}\left(\frac{\mathcal{G}_{8}}{2}\right)\mathtt{v}^{8}~,
\end{align}
%%%%%%%%%%%%%%%%%%%%%%%%%%%%%%%%%%%%%%%%%%%%%%%%%%%%%%%%%%%
and hence the corresponding contribution to the phase reads, 
%%%%%%%%%%%%%%%%%%%%%%%%%%%%%%%%%%%%%%%%%%%%%%%%%%%%%%%%%%%
\begin{align}\label{phase_integral_heating2}
\psi^{\rm non-rot}_{\rm TH}(v)&=-\frac{10}{32\varsigma}\left(\frac{\mathcal{G}_{8}}{2}\right)
\frac{v^{3}}{3}\left(3\ln v-1\right)~.
\end{align}
%%%%%%%%%%%%%%%%%%%%%%%%%%%%%%%%%%%%%%%%%%%%%%%%%%%%%%%%%%%
Note that the above contribution is degenerate with the tail-of-the-memory term~\citep{Blanchet:2023sbv}. 

All the phasing formulas discussed so far apply to BHs, we now turn to other compact objects, namely horizonless ultra-compact objects and NSs.

\paragraph{Modelling tidal heating for ultra-compact objects}~--- 
For compact objects other than BHs, the above results for tidal heating are modified. 
We first discuss ultra-compact objects and then turn to NSs. 

Ultra-compact objects are typically characterized by a (complex) reflectivity $\mathcal{R}(\omega)$. 
For instance, wormholes, gravastars, and area-quantized BHs exhibit a nontrivial, frequency-dependent reflectivity. 
Since $\omega$ maps to the PN velocity through $\omega\sim v^{3}/M_{\rm T}$, this implies an implicit dependence on the PN parameter. 
The presence of reflectivity redirects part of the incoming radiation back to infinity, thereby reducing the absorption rate and, consequently, $\dot{M}$. 
At leading order,
%%%%%%%%%%%%%%%%%%%%%%%%%%%%%%%%%%%%%%%%%%%%%%%%%%%%%%%%%%%
\begin{align}
\dot{M}\big|_{\rm ECO}
=
\left(1-|\mathcal{R}|^{2}\right)\dot{M}\big|_{\rm BH}\,.
\end{align}
%%%%%%%%%%%%%%%%%%%%%%%%%%%%%%%%%%%%%%%%%%%%%%%%%%%%%%%%%%%
Assuming a binary composed of two ECOs, the flux through their surface follows from the corresponding modifications of~\ref{flux_horizon}. 
We obtain
%%%%%%%%%%%%%%%%%%%%%%%%%%%%%%%%%%%%%%%%%%%%%%%%%%%%%%%%%%%
\begin{align}\label{flux_ECO}
F_{\rm ECO}^{\rm rot}
&=
-\left(\frac{dE}{dt}\right)_{\rm N}
\,\widetilde{\mathcal{G}}_{5}(v)\,
\frac{v^{5}}{4}\,,
\qquad
F_{\rm ECO}^{\rm non\text{-}rot}
=
\left(\frac{dE}{dt}\right)_{\rm N}
\,\widetilde{\mathcal{G}}_{8}(v)\,
\frac{v^{8}}{2}\,,
\\[4pt]
\widetilde{\mathcal{G}}_{5}
&=
\sum_{i}
\left(1-|\mathcal{R}_{i}|^{2}\right)
\left(\frac{M_{i}}{M_{\rm T}}\right)^{3}
\chi_{i}(1+3\chi_{i}^{2})
\left(\hat{\mathbf{L}}_{\rm orb}\cdot \hat{\mathbf{J}}_{i}\right)\,,
\\
\widetilde{\mathcal{G}}_{8}
&=
\sum_{i}
2\left(1-|\mathcal{R}_{i}|^{2}\right)
\left(\frac{M_{i}}{M_{\rm T}}\right)^{4}\,,
\end{align}
%%%%%%%%%%%%%%%%%%%%%%%%%%%%%%%%%%%%%%%%%%%%%%%%%%%%%%%%%%%
which is obviously zero for a zero-absorption object, namely $|{\cal R}|=1$. For viscous objects, the key difference with respect to BHs is that $\widetilde{\mathcal{G}}_{5}$ and $\widetilde{\mathcal{G}}_{8}$ inherit the frequency dependence of $\mathcal{R}(\omega)$. 
As a result, the phasing induced by tidal heating becomes
%%%%%%%%%%%%%%%%%%%%%%%%%%%%%%%%%%%%%%%%%%%%%%%%%%%%%%%%%%%
\begin{align}
\psi_{\rm TH}^{\rm rot}\big|_{\rm ECO}
&=
\frac{10}{32\varsigma}
\int^{v} d\mathtt{v}
\left(
\frac{v^{3}-\mathtt{v}^{3}}{\mathtt{v}^{4}}
\right)
\Bigg[
\frac{\widetilde{\mathcal{G}}_{5}(\mathtt{v})}{4}
+
\frac{\widetilde{\mathcal{G}}_{5}(\mathtt{v})}{4}
\left(
\frac{995}{168}
+
\frac{952}{168}
\right)
\mathtt{v}^{2}
\nonumber\\
&\hspace{4cm}
-
\frac{\widetilde{\mathcal{G}}_{5}(\mathtt{v})}{2}
\left(4\pi+F_{\rm SO}\right)
\mathtt{v}^{3}
\Bigg]\,,
\\
\psi_{\rm TH}^{\rm non\text{-}rot}\big|_{\rm ECO}
&=
-\frac{5}{32\varsigma}
\int^{v} d\mathtt{v}
\left(
\frac{v^{3}-\mathtt{v}^{3}}{\mathtt{v}}
\right)
\widetilde{\mathcal{G}}_{8}(\mathtt{v})\,.
\end{align}
%%%%%%%%%%%%%%%%%%%%%%%%%%%%%%%%%%%%%%%%%%%%%%%%%%%%%%%%%%%

Unlike the BH case, for ECOs both the magnitude and the effective PN order of the phasing correction depend on the detailed frequency dependence of $\mathcal{R}(\omega)$. 
Therefore, unless the functional form of the reflectivity is specified, the phase cannot be expressed in closed analytic form.

For gravastars, wormholes, and solitonic stars, the reflectivity depends on the microscopic details of the model. 
In contrast, for BHs with quantum modifications at the horizon, specific proposals for $\mathcal{R}(\omega)$ have been put forward. 
Three representative classes are:

\begin{itemize}
\item[(a)] \emph{Boltzmann reflectivity} \citep{Oshita:2019sat}:
\begin{align}
\mathcal{R}(\omega)
=
\exp\!\left(
-\frac{\hbar \omega}{k_{\rm B}T_{\rm H}}
\right),
\end{align}
where $k_{\rm B}$ is the Boltzmann constant and $T_{\rm H}$ is the Hawking temperature.

\item[(b)] \emph{Area-quantized BHs} \citep{Bekenstein:1995ju,Cardoso:2019apo,agullo2021potential-5f8,page1976particle-286,Datta:2020rvo}, 
for which the horizon area is quantized as
\begin{align}
A=\alpha N L_{\rm p}^{2}\,,
\end{align}
with $N\in\mathbb{Z}$ and $\alpha={\cal O}(1)$. 
In this case, GW absorption occurs only at discrete frequencies $\omega_{n}$, leading to
\begin{align}
|\mathcal{R}(\omega)|^{2}
=
\sum_{n=0}^{n_{\rm max}}
\beta(\omega-\omega_{n},\Gamma)\,,
\end{align}
where $\beta$ selects narrow frequency windows of width $\Gamma$ centered at $\omega_{n}$. 
The absorption starts at $\omega_{0}=2\Omega_{\rm H}$ and extends up to the frequency corresponding to the ISCO.

\item[(c)] \emph{Minimum-length BHs}, where absorption is allowed only if the horizon area changes by at least \citep{Krishnendu:2025byo}
\begin{align}
A_{\rm min}=4\pi \beta L_{\rm p}^{2}\,.
\end{align}
In this scenario,
\begin{align}
|\mathcal{R}|^{2}
=
\begin{cases}
1 & \text{for }\omega<\omega_{\rm L}\,,\\
0 & \text{for }\omega>\omega_{\rm L}\,,
\end{cases}
\end{align}
with $\omega_{\rm L}=(\kappa \beta^{2}/2)+2\Omega_{\rm H}$ and $\kappa$ is the surface gravity. 
Accordingly, the phase integral acquires a lower limit corresponding to the velocity $v_{\rm L}$ associated with $\omega_{\rm L}$.
\end{itemize}

\paragraph{Modelling tidal heating for neutron stars}~---
For NSs modeled as perfect fluids there is no tidal heating. 
However, realistic NS models include dissipative effects (e.g., shear and bulk viscosity, crust-core interactions, superfluid mutual friction), which lead to nonvanishing energy absorption. 
Within the EFT framework, such finite-size effects can be encoded in a worldline action for a non-rotating NS of the form
%%%%%%%%%%%%%%%%%%%%%%%%%%%%%%%%%%%%%%%%%%%%%%%%%%%%%%%%%%%
\begin{align}
\mathcal{A}_{\rm eff}
=\int d\tau\left(-m+M^{\mu \nu}E_{\mu \nu}+S^{\mu \nu}B_{\mu \nu} \right)\,,
\end{align}
%%%%%%%%%%%%%%%%%%%%%%%%%%%%%%%%%%%%%%%%%%%%%%%%%%%%%%%%%%%
where $E_{\mu\nu}$ and $B_{\mu\nu}$ are the electric and magnetic components of the Weyl tensor evaluated on the worldline, and $M^{\mu\nu}$ and $S^{\mu\nu}$ are the induced mass and current quadrupole moments.

The action alone is not sufficient: one must also specify the constitutive relations linking the induced multipoles to the tidal fields. 
Focusing on the electric sector, the response can be written schematically as
%%%%%%%%%%%%%%%%%%%%%%%%%%%%%%%%%%%%%%%%%%%%%%%%%%%%%%%%%%%
\begin{align}
M^{\mu \nu}
=-\bar{\lambda}_{2}^{\rm YY}M^{5}E^{\mu \nu}
+\sum_{n=1}^{\infty}M^{n}\tau_{2}^{(n)}
\frac{d^{n}}{d\tau^{n}}E^{\mu \nu}\,.
\end{align}
%%%%%%%%%%%%%%%%%%%%%%%%%%%%%%%%%%%%%%%%%%%%%%%%%%%%%%%%%%%
Here $\bar{\lambda}_{2}^{\rm YY}$, defined in~\ref{eq:conventionElectric}, corresponds to the static LN in the conventions of \citet{Yagi:2013awa,Yagi:2013sva}. 
The coefficients $\tau_{2}^{(n)}$ are dimensionless and encode dynamical corrections associated with higher powers of frequency in the response function. 
Since $n$ can be even or odd, the response generically contains both real (conservative) and imaginary (dissipative) parts, the latter describing tidal heating.

In what follows we restrict attention to leading-order effects. 
The conservative response is governed by the static LN $\bar{\lambda}_{2}^{\rm YY}$, while the leading dissipative contribution is parameterized by $\nu_{2}$, defined through $\nu_{2}=-\tau_{2}^{(1)}$. 
This quantity is proportional to the characteristic delay time $\tau_{d}$ of the NS. 
Typically, the orbital time scale is much longer than $\tau_{d}$, justifying the expansion in time derivatives.

NS tidal heating has also been investigated from a scattering-amplitude perspective by computing GW scattering off a NS (gravitational Raman scattering) \citep{Saketh:2024juq}.

Concerning the impact on the GW waveform, the dimensionless tidal deformability 
$\bar{\lambda}_{2}^{\rm YY}$ and the dissipation number $\Upsilon_{2}$ are related 
to the LN and to the time delay of NSs as \citep{Ripley:2023qxo,Ripley:2023lsq,HegadeKR:2024slr}
%%%%%%%%%%%%%%%%%%%%%%%%%%%%%%%%%%%%%%%%%%%%%%%%%%%%%%%%%%%
\begin{align}
\bar{\lambda}_{2,A}^{\rm YY}=\frac{2}{3}\frac{k_{2,A}}{C^{5}_{A}}\,; 
\qquad
\Upsilon_{2,A}=\frac{2}{3}\frac{\nu_{2,A}}{C_{A}^{5}}
=\frac{2}{3}\frac{k_{2,A}\tau_{d,A}}{C_{A}^{6}R_A}\,,
\end{align}
%%%%%%%%%%%%%%%%%%%%%%%%%%%%%%%%%%%%%%%%%%%%%%%%%%%%%%%%%%%
where $A=1,2$ labels the two stars and $C_{A}=M_{A}/R_{A}$ is the compactness of the $A$th NS. 
From these quantities one constructs the symmetric and antisymmetric combinations 
$\bar{\Lambda}$ and $\bar{\Upsilon}$, defined in \citet{Ripley:2023qxo,Ripley:2023lsq,HegadeKR:2024slr}, 
which enter directly in the GW phasing.

Following the same procedure outlined at the beginning of this section, 
the leading-order tidal contribution to the phase reads
%%%%%%%%%%%%%%%%%%%%%%%%%%%%%%%%%%%%%%%%%%%%%%%%%%%%%%%%%%%
\begin{align}
\psi_{\rm dissipative}
=-\frac{3}{128\,\varsigma}
\left[
\frac{75}{32}\,\bar{\Upsilon}\,v^{3}\ln v
+\frac{39}{2}\,\bar{\Lambda}\, v^{5}
\right]\,.
\end{align}
%%%%%%%%%%%%%%%%%%%%%%%%%%%%%%%%%%%%%%%%%%%%%%%%%%%%%%%%%%%
The PN order of the dissipative term matches~\ref{phase_integral_heating2}: 
the corresponding 4PN contribution without the $\ln v$ factor can be absorbed 
into a redefinition of the coalescence time. 
Similarly, the conservative term agrees with~\ref{psi-gen}. 

For NSs, the time delay can be expressed as
\[
\tau_{d}=\frac{q_{2}\,\langle \eta\rangle}{\langle \rho \rangle\, C}\,,
\]
where $q_{2}$ is a dimensionless coefficient whose relation to the shear viscosity is given in \citet{Ripley:2023lsq,HegadeKR:2024slr}, 
$\langle \eta\rangle$ is the average shear viscosity, and $\langle \rho \rangle$ is the average density of nuclear matter inside the NS.

Given the above phasing, the leading dissipative and conservative tidal effects can be incorporated into waveform models by augmenting the point-particle phase (for example in the \texttt{IMRPhenomPv2\_NRTidal} model \citep{Dietrich:2019kaq}). 
Using strain data from the binary NS merger GW170817 \citep{LIGOScientific:2017vwq,LIGOScientific:2019lzm}, 
 \citet{Ripley:2023lsq} obtained the constraint
\[
\bar{\Upsilon}<1200\,,
\]
which in turn can be translated into bounds on the shear viscosity of NS matter \citep{Ripley:2023lsq}. The above constraint was later refined
using next-to-leading order tidal heating effects, allowing to constrain the individual viscosity parameters of the binary components \citep{HegadeKR:2024slr}.

These constraints are particularly relevant for NSs because: 
(i) the dissipative term appears at 4PN$\times\ln v$, i.e.\ at lower order than the conservative tidal correction, and 
(ii) it scales as $C^{-6}$, leading to a strong enhancement for low-compactness objects.

\paragraph{Resonant tidal excitations}~---
As we have discussed in the previous paragraph, there can be instances where the orbital motion excites internal oscillation modes of a compact object (see~\ref{dyntideNS}). A systematic analysis of this regime in a non-relativistic context was performed by \citet{Flanagan:2006sb}. In this framework, the stellar mode amplitude $q_{n\ell m}$ is treated as a driven and damped harmonic oscillator,
%%%%%%%%%%%%%%%%%%%%%%%%%%%%%%%%%%%%%%%%%%%%%%%%%%%%%%%%%%%%%%%%%%%%%%%%%%%%%%%%%
\begin{equation}
\ddot q_{n\ell m}
+ 2\gamma_{n\ell m}\dot q_{n\ell m}
+ \omega_{n\ell m}^2 q_{n\ell m}
=
F_{n\ell m}(t) ,
\label{eq:mode_oscillator}
\end{equation}
%%%%%%%%%%%%%%%%%%%%%%%%%%%%%%%%%%%%%%%%%%%%%%%%%%%%%%%%%%%%%%%%%%%%%%%%%%%%%%%%%
where $\gamma_{n\ell m}$ is the viscous damping rate and $F_{n\ell m}$ is proportional to
the external tidal field through a mode overlap coefficient.
Near resonance, the mode absorbs a finite amount of orbital energy, controlled by the
competition between the inspiral timescale and the damping timescale
$\tau_{\rm damp}\sim\gamma_{n\ell m}^{-1}$.

The impact of resonant dynamical tides on the GW signal manifests primarily as a
localized correction to the inspiral phase.
Using the balance equation between the orbital binding energy $E$ and the GW luminosity
$\dot E$,
%%%%%%%%%%%%%%%%%%%%%%%%%%%%%%%%%%%%%%%%%%%%%%%%%%%%%%%%%%%%%%%%%%%%%%%%%%%%%%%%%
\begin{equation}
\frac{d^2\psi}{df^2}
=
\frac{2\pi}{\dot E}
\frac{dE}{df} ,
\label{eq:phase_balance_res}
\end{equation}
%%%%%%%%%%%%%%%%%%%%%%%%%%%%%%%%%%%%%%%%%%%%%%%%%%%%%%%%%%%%%%%%%%%%%%%%%%%%%%%%%
the dephasing induced by a resonance at frequency $f_{\rm res}$ can be estimated as \citep{Flanagan:2006sb,Ma:2020oni,Gupta:2023oyy} 
%%%%%%%%%%%%%%%%%%%%%%%%%%%%%%%%%%%%%%%%%%%%%%%%%%%%%%%%%%%%%%%%%%%%%%%%%%%%%%%%%
\begin{equation}
\Delta\psi_{\rm res}
\simeq
-2\pi
\frac{\Delta E_{\rm mode}}
{dE/df}
\bigg|_{f_{\rm res}} .
\label{eq:delta_phi_res}
\end{equation}
%%%%%%%%%%%%%%%%%%%%%%%%%%%%%%%%%%%%%%%%%%%%%%%%%%%%%%%%%%%%%%%%%%%%%%%%%%%%%%%%%
Because $dE/df$ decreases at low frequencies, resonances occurring early in the inspiral
can produce large phase shifts, even if the absolute energy transfer is modest.

The corresponding relativistic counterpart, neglecting viscous dissipation, was developed in \citet{Steinhoff:2016rfi,Hinderer:2016eia,Gupta:2023oyy,Pratten:2021pro}. In this approach the tidal response of the NS is modeled by promoting the point particle to a harmonic oscillator with oscillation frequency $\omega_{\rm osc}$ and no dissipation. 
Since tidal interactions contain both electric and magnetic sectors, the electric sector naturally couples to the $f$-mode frequency of the NS. Because the $f$-mode frequency is typically high, this coupling becomes relevant only toward the late stages of the inspiral. 
The magnetic sector, instead, is associated with frame-dragging effects and can excite rotational modes of the NS~\citep{Kokkotas:1999bd}. These modes are proportional to the stellar spin and therefore can lead to resonances that affect the GW phasing already during the inspiral \citep{Gupta:2023oyy}.

Resonances in the electric (parity-even) sector and their impact on the GW waveform were studied in \citet{Steinhoff:2016rfi,Hinderer:2016eia}. In this case a resonance occurs whenever the $f$-mode frequency coincides with the orbital frequency, leading to the so-called \emph{orbital-mode resonances}. The resonance condition reads
%%%%%%%%%%%%%%%%%%%%%%%%%%%%%%%%%%%%%%%%%%%%%%%%%%%%%%%%%%%%%%%%%%%%%%%%%%%%%%%%%
\begin{equation}
f_{n\ell m} \simeq m\,f_{\rm orb}\,,
\label{eq:resonance_condition}
\end{equation}
%%%%%%%%%%%%%%%%%%%%%%%%%%%%%%%%%%%%%%%%%%%%%%%%%%%%%%%%%%%%%%%%%%%%%%%%%%%%%%%%%
where $f_{\rm orb}$ is the orbital frequency and $f_{n\ell m}$ denotes the frequency of the stellar mode. 
As the binary inspirals due to GW emission, this condition can be satisfied only over a finite time interval, during which the energy transfer between the orbit and the stellar mode is maximized.
The resulting phase correction depends on the detuning between the orbital frequency and the mode frequency, thereby providing a link between the inspiral dynamics and the internal oscillation spectrum of the star. In this framework, effective dynamical LNs can be defined \citep{Steinhoff:2016rfi,Hinderer:2016eia}; these depend both on the static LNs and on the $f$-mode frequency of the NS. 
Quadrupolar dynamical tides leave imprints in both NS-BH and NS-NS binaries and encode additional information on the internal structure of the star \citep{Hinderer:2016eia}.

The magnetic sector involves instead \emph{orbital-spin resonances}, which give rise to effective dynamical magnetic LNs. For instance, in the case of aligned-spin binaries the resonances occur at frequencies $f_{n\ell m}=mf_{\rm spin}+f^{\rm B}_{n\ell m}$, where $f^{\rm B}_{n\ell m}$ denotes the frequency of the relevant magnetic mode of the NS in the co-rotating frame (which, in the Newtonian limit, is proportional to the NS spin) and the spin frequency $f_{\rm spin}$ arises from frame dragging.
For generic inclination angles, the GW phase shift associated with these resonances can be written as \citep{Gupta:2023oyy}
%%%%%%%%%%%%%%%%%%%%%%%%%%%%%%%%%%%%%%%%%%%%%%%%%%%%%%%%%%%%%%%%%%%%%%%%%%%%%%%%%
\begin{equation}
\psi^{\rm tidal}_{\rm res}
=-\sum_{A=1,2}\left(1-\frac{f}{f^{\rm res}_{A}}\right)
|\Delta \phi_{A}|\,
\Theta(f-f_{A}^{\rm res})\,,
\label{eq:magnetic_resonance_phase}
\end{equation}
%%%%%%%%%%%%%%%%%%%%%%%%%%%%%%%%%%%%%%%%%%%%%%%%%%%%%%%%%%%%%%%%%%%%%%%%%%%%%%%%%
where $f_{A}^{\rm res}$ is the resonance frequency of the $A$th NS in the binary and $|\Delta \phi_{A}|$ is the corresponding phase shift accumulated as the system crosses the resonance. The resonance frequency depends both on the gravitomagnetic modes and on the orbital dynamics of the binary. 
The analysis of \citet{Gupta:2023oyy} shows that, for equal-mass and aligned-spin systems, rapidly rotating NS binaries can produce significant effects in the GW waveform of events similar to GW170817 when observed with third-generation detectors. Hence, gravitomagnetic tidal resonances may leave detectable imprints in future GW observations. In particular, resonant tides deviate more strongly from the adiabatic approximation as the NS spin increases, since the phase shift induced by the resonance scales as $|\Delta\phi_{\rm res}|\sim \chi^{2/3}$ \citep{Gupta:2023oyy}.
From an observational perspective, resonances imprint narrow-band, frequency-localized features in the GW phase evolution. If not properly modeled, these features could bias parameter estimation. Conversely, their detection would provide direct evidence of internal stellar dynamics beyond the adiabatic regime. Intriguingly, the absence of such resonant signatures would be consistent with the classical BH paradigm, whereas their observation would point to matter effects or to the presence of ECOs.

\subsubsection{Summary: tidal effects in post-Newtonian waveforms}
To summarize the previous section,~\ref{tab:PN} provides an overview of the leading-order PN contribution of each tidal effects discussed above, ranked by PN order.

\begin{landscape}
    
\begin{table}[ht!]
\centering
\caption{
Different tidal contributions to the GW phase of a binary inspiral.
The first column refers to the leading-order PN term in the GW phase of a given effect. The third column refers to the possible enhancement in the small-compactness limit which, for moderately compact objects ($C\approx0.1$) can compensate for the PN suppression. The last column gives the corresponding references, including next-order subleading contributions when available.
}
\label{tab:PN}
\begin{tabular}{||c|c|c|m{8cm}||}
\hline
\hline
%%%%%%%%%
PN  &   Effect  & Enhancement & Reference \\
\hline
2.5$\times\log v$   & dissipation number, spinning ($\nu_2^E$)         & $C^{-6}$    & \citep{Maselli:2017cmm,Chia:2024bwc}\\
\hline
4$\times\log v$     & dissipation number, static ($\nu_2^E$)           & $C^{-6}$    & \citep{Maselli:2017cmm,Ripley:2023lsq,Ripley:2023qxo,Chia:2024bwc}\\
\hline
%%%
5                   & static, electric, $\ell=2$ LN ($k_2^E$)   & $C^{-5}$   & \citep{Flanagan:2007ix,Hinderer:2009ca,Vines:2010ca,Vines:2011ud,Damour:2012yf,Henry:2020ski,Henry:2020pzq}\\
\hline
6                   & static, magnetic, $\ell=2$ LN ($k_2^B$)   & $C^{-5}$   & \citep{Yagi:2013sva,Banihashemi:2018xfb,Abdelsalhin:2018reg}\\
\hline
\multirow{2}{*}{6.5}                 & rotational/mixing LN                     & $C^{-6}$ & \citep{Pani:2015nua,Abdelsalhin:2018reg}\\
                 & linear spin corrections ($k_2^E\times \chi$)                   & $C^{-5}$ & \citep{Abdelsalhin:2018reg} \\
\hline
%%%% 
\multirow{2}{*}{7}  & static, electric, $\ell=3$ LN ($k_3^E$)   & $C^{-7}$   & \citep{Flanagan:2007ix,Hinderer:2009ca,Abdelsalhin:2018reg}\\
                    & quadratic spin correction ($k_2^E\times \chi^2$)   & $C^{-5}$   &none (but see \citealt{Pani:2015hfa,Pani:2015nua})\\
%%%%
\hline
\multirow{3}{*}{8}  & quadratic LN ($p_2$)   & $C^{-8}$   & \citep{Pani:2025qxs}\\
                    & dynamical LN ($\ddot k_2^E$)   & $C^{-8}$   & \citep{Chakraborty:2025wvs}\\
                    & static, magnetic, $\ell=3$ LN ($k_2^B$)   & $C^{-7}$   & \citep{Abdelsalhin:2018reg}\\
%%%%%%%%%
\hline
\hline
\end{tabular}
\end{table}

\end{landscape}

%%%%%%%%%%%%%%%%%%%%%%%%%%%%%%%%%%%%%%%%%%%%%%%%%%%%%%%%%
\subsubsection{Tidal effects in the effective-one-body and phenomenological models} \label{sec:EOB}
%%%%%%%%%%%%%%%%%%%%%%%%%%%%%%%%%%%%%%%%%%%%%%%%%%%%%%%%%
The effective-one-body~(EOB) formalism provides a unified resummed analytical framework for the relativistic two-body problem in GR, mapping the dynamics of a binary system onto an effective particle in a deformed background metric. Originally introduced in \citet{Buonanno:1998gg, Buonanno:2000ef,Damour:2001tu} to describe the signal emitted by BH coalescences, the EOB approach has since become a cornerstone for waveform modelling in GW astronomy due to its ability to bridge PN, self-force, gravitational self-force, and numerical relativity information into accurate inspiral-merger-ringdown templates (see \citealt{Damour:2012mv,Damour:2016bks} for some reviews).

A key aspect of EOB modelling for binaries containing NS or other extended bodies is the inclusion of tidal effects, which imprint information about the internal structure of the bodies into the late inspiral gravitational waveform, modifying both the conservative dynamics and radiation reaction. The first systematic extension of the EOB Hamiltonian to capture adiabatic tidal effects in inspiralling compact binaries was developed in \citet{Damour:2009wj}. Their formulation augments the EOB radial potential with tidal potentials proportional to dimensionless LN of each body. When compared to quasi-equilibrium data and nonresummed PN expansions, tidal EOB models exhibit increased effective tidal polarizability and better capture the late inspiral phasing relative to traditional PN approximants \citep{Damour:2012yf}.

Building on this, \citet{Bini:2012gu} computed next-to-next-leading-order relativistic corrections to the tidal interaction energy within the effective action approach and mapped these results into the EOB framework, showing that second-order PN effects significantly amplify the effective tidal response in the strong field regime \citep{Bini:2012gu, Hinderer:2016eia}. This work also suggested various resummations to improve the robustness of the tidal sector and highlighted the importance of higher-order tidal contributions for extracting EoS information from the waveform~\citep{Schulze:2026ewu}.

Further progress has leveraged gravitational self-force computations to refine the EOB tidal potentials in the small-mass-ratio limit. By computing tidal invariants to high PN order and translating them into the EOB interaction energy, these studies provide analytic insight into the behavior of tidal factors in the strong-field domain, including sign changes and nonlinear contributions that are not captured at low PN orders \citep{Bini:2014zxa,Nagar:2018gnk,Akcay:2018yyh,Lackey:2018zvw,Tissino:2022thn}.

More recent developments have enriched the EOB description of tidal phenomena beyond the adiabatic approximation. Works that include dynamical tidal effects --~such as f-mode resonances~-- demonstrate that finite-frequency responses can significantly affect the inspiral phasing \citep{Steinhoff:2016rfi,Schmidt:2019wrl}, motivating the inclusion of dynamic tidal degrees of freedom in waveform models calibrated against numerical simulations \citep{Hinderer:2016eia,Yu:2025ptm}. Complementary studies have also introduced \emph{tidal spin} effects, whereby the lag in the tidal bulge induces a secular spin evolution of the NS that feeds back into both the conservative and dissipative sectors of the EOB dynamics, yielding notable phase shifts in the waveform \citep{Yu:2025ptm}.

In parallel, the EOB formalism has been extended to cover a wider set of physical effects that interact with tidal physics, such as eccentricity and aligned spins \citep{Nagar:2018zoe,Gamboa:2024hli,Albanesi:2025txj,Haberland:2025luz}, which are particularly relevant for next-generation GW detector analyses and high-precision tests of the nature of compact objects. These extensions introduce higher PN corrections in the radiation reaction and waveform modes, and refine the tidal sector consistently with the underlying Hamiltonian treatment.

Finally, tidal effects have also been incorporated into alternative phenomenological waveform models, which typically combine analytical PN information with phenomenological parametrizations whose coefficients are calibrated to numerical-relativity simulations \citep{Kawaguchi:2018gvj,Dietrich:2019kaq,Williams:2024twp,Abac:2025brd}.
Further systematic improvements and recalibrations against increasingly accurate numerical simulations \citep{Read:2013zra,Bernuzzi:2014owa,Dietrich:2017aum,Gamba:2023mww,Abac:2023ujg,Abac:2025brd} will be crucial to fully exploit tidal imprints in GW observations of NS binaries and to extract robust constraints on the physics of dense matter.

%%%%%%%%%%%%%%%%%%%%%%%%%%%%%%%%%%%%%%%%%%%%%%%%%%%%%%%%%
\subsubsection{Tidal effects in extreme mass-ratio inspirals} \label{sec:EMRImodelling}
%%%%%%%%%%%%%%%%%%%%%%%%%%%%%%%%%%%%%%%%%%%%%%%%%%%%%%%%%
Extreme mass-ratio inspirals~(EMRIs) provide a qualitatively new regime to probe tidal
deformability with GW observations. These systems consist of a compact binary in which
the primary object is much heavier than the secondary. To leading order in the mass ratio
$q \equiv M_2/M_1 \ll 1$, the motion of the secondary is well described by geodesics of the
primary spacetime, with quasi-adiabatic corrections driven by GW emission.

A distinctive feature of EMRIs is that the total number of orbital cycles accumulated
during the inspiral scales as the inverse mass ratio, $N_{\rm orb}\sim 1/q \gg 1$.
This exceptionally large number of cycles makes EMRIs an unparalleled laboratory to
probe the spacetime geometry in the strong-field region of the primary object and to test
the underlying theory of gravity with exquisite precision \citep{Babak:2017tow}.

The primary astrophysical EMRI sources are stellar-mass compact objects
($M_2 \in [1,100]\,M_\odot$) inspiraling into supermassive ones
($M_1 \in [10^5,10^7]\,M_\odot$), corresponding to typical mass ratios
$q \in [10^{-6},10^{-4}]$. These systems constitute one of the main target sources for
future space-based GW detectors such as the Laser Interferometer Space Antenna~(LISA) \citep{LISA:2024hlh}.

Although the SNR of EMRIs in the LISA band is expected to be moderate --~at least when
compared to massive BH coalescences~-- the large number of cycles accumulated in band acts
as a powerful magnifying glass for small physical effects that would otherwise be undetectable.

In this context, tidal effects display a striking behavior. While in comparable-mass
binaries tidal contributions are suppressed by the high PN order of tidal terms in the GW phase, in the EMRI limit the
tidal deformability of the \emph{central} object can affect the GW phase at leading order
in $1/q$ \citep{Pani:2019cyc}. This enhancement opens the possibility of probing extremely
small tidal LNs of massive compact objects through EMRI observations (however, see \citealt{Datta:2021hvm}).

The physical origin of this result can be understood by analyzing the binding energy
and the GW energy flux. At leading PN order, including tidal corrections, these read
%%%%%%%%%%%%%%%%%%%%%%%%%%%%%%%%%%%%%%%%%%%%%%%%%%%%%%%%%%%%%%%%%%%%%%%%%%%%%%%%%
\begin{align}
E(f) &= -\frac{m_1}{2(1+q)}v^2
\left[
1 - 6\epsilon_c q\,
\frac{k_1 + k_2 q^3}{(1+q)^5}\,v^{10}
\right] ,
\label{eq:E_tidal_EMRI}
\\
\dot E(f) &= -\frac{32}{5}\frac{q^2}{(1+q)^4}v^{10}
\left[
1 + 4{\epsilon_d}\,
\frac{(1+3q)k_1 + (3+q)k_2 q^4}{(1+q)^5}\,v^{10}
\right] ,
\label{eq:Edot_tidal_EMRI}
\end{align}
%%%%%%%%%%%%%%%%%%%%%%%%%%%%%%%%%%%%%%%%%%%%%%%%%%%%%%%%%%%%%%%%%%%%%%%%%%%%%%%%%
where $v$ is the orbital velocity, $k_1$ and $k_2$ are the (static, quadrupolar, electric) LN of central and small object, respectively (previously we denoted this quantity as $k_2^E$, here we make the notation lighter), and
$\epsilon_c,\epsilon_d$ are bookkeeping parameters for conservative and dissipative
tidal corrections.

Combining~\ref{eq:E_tidal_EMRI} and~\ref{eq:Edot_tidal_EMRI} through the balance
equation for the GW phase, one finds that the dominant tidal correction to the phase is
driven by the tidal term in the energy flux. In the EMRI limit, $q\ll1$, the GW phase
can be written as
%%%%%%%%%%%%%%%%%%%%%%%%%%%%%%%%%%%%%%%%%%%%%%%%%%%%%%%%%%%%%%%%%%%%%%%%%%%%%%%%%
\begin{equation}
\psi(f) = \psi_N(f)\,
\left( 1 - 16\,\epsilon_d\,k_1\,v^{10} \right) ,
\qquad q \ll 1 ,
\label{eq:phi_tidal_EMRI}
\end{equation}
%%%%%%%%%%%%%%%%%%%%%%%%%%%%%%%%%%%%%%%%%%%%%%%%%%%%%%%%%%%%%%%%%%%%%%%%%%%%%%%%%
where $\psi_N(f)$ is the leading Newtonian phase. Remarkably, the tidal contribution
proportional to $k_1$ is \emph{not} suppressed\footnote{Terms proportional to $\epsilon_c$ or to $k_2$ are instead suppressed by powers of $q$ \citep{Pani:2019cyc}.} by any additional power of $q$ relative
to the leading radiation-reaction term, implying that the tidal deformability of the
central object enters the waveform at leading order in the mass ratio.
This result underlines the
potential of EMRIs detectable by space-based detectors to place unprecedented
constraints on the tidal deformability of massive compact objects, thereby providing
powerful tests of the BH paradigm and of exotic alternatives (see~\ref{sec:application_ECOs}).

The above discussion is restricted to the standard static LN. An extension to dynamical tides in EMRIs is currently missing.
Likewise, it is known that the PN series converges very poorly in the small mass-ratio limit \citep{Fujita:2011zk}, so it would be important to extend the modelling beyond the PN approximation.

Very recently, \citet{Wu:2026epz} considered tidal effects in EMRIs in the context of first-order phase transition within a NS orbiting a supermassive Kerr BH, which suddenly change the LN of the secondary object.

%%%%%%%%%%%%%%%%%%%%%%%%%%%%%%%%%%%%%%%%%%%%%%%%%
\subsection{Measuring tidal effects in gravitational-wave signals}
\label{sec:Fisher}
\noindent
%%%%%%%%%%%%%%%%%%%%%%%%%%%%%%%%%%%%%%%%%%%%%%%%%
In the following sections we review existing and projected constraints on the tidal
deformability of compact objects from GW observations.
In GW astronomy, such constraints are typically obtained by augmenting waveform
templates with tidal effects --~either as PN corrections to the inspiral phase
(see~\ref{tab:PN}) or through effective phenomenological models~-- and performing
a Bayesian parameter estimation. This procedure yields the posterior probability
distribution of the waveform parameters for a given GW event, or for a catalog of
events.

Within the Bayesian framework, the posterior distribution for the set of waveform
parameters $\bm{\theta}$, given the detector data $d$, is obtained from Bayes' theorem,
%%%%%%%%%%%%%%%%%%%%%%%%%%%%%%%%%%%%%%%%%%%%%%%%%%%%%%%%%%%%%%%%%%%%%%%%%%%%%%%%%
\begin{equation}
p(\bm{\theta}\,|\,d) =
\frac{p(d\,|\,\bm{\theta})\,p(\bm{\theta})}{p(d)} \,,
\end{equation}
%%%%%%%%%%%%%%%%%%%%%%%%%%%%%%%%%%%%%%%%%%%%%%%%%%%%%%%%%%%%%%%%%%%%%%%%%%%%%%%%%
where $p(d\,|\,\bm{\theta})$ is the likelihood, $p(\bm{\theta})$ is the prior distribution,
and $p(d)$ is the evidence. For stationary, Gaussian detector noise, the likelihood
takes the standard form
%%%%%%%%%%%%%%%%%%%%%%%%%%%%%%%%%%%%%%%%%%%%%%%%%%%%%%%%%%%%%%%%%%%%%%%%%%%%%%%%%
\begin{equation}
p(d\,|\,\bm{\theta}) \propto
\exp\!\left[
-\frac{1}{2}\,
\langle d-h(\bm{\theta}) \,|\, d-h(\bm{\theta}) \rangle
\right] ,
\end{equation}
%%%%%%%%%%%%%%%%%%%%%%%%%%%%%%%%%%%%%%%%%%%%%%%%%%%%%%%%%%%%%%%%%%%%%%%%%%%%%%%%%
where $h(\bm{\theta})$ is the GW template and the noise-weighted inner product is
defined below. Tidal parameters, such as the tidal LNs and dissipation numbers, enter
the likelihood through their contribution to the waveform phase and amplitude.

For forecasting constraints from high signal-to-noise~(SNR) events, and especially when the waveform model is
known analytically, a computationally cheaper alternative is provided by the Fisher
information matrix formalism \citep{Vallisneri:2007ev}. In the large-SNR limit, the
posterior distribution approaches a multivariate Gaussian centered on the true
(injected) parameter values, and the Fisher matrix provides an estimate of the expected
statistical uncertainties.

Schematically, the frequency-domain GW waveform can be written as \citep{Sathyaprakash:1991mt,Damour:2000gg}
%%%%%%%%%%%%%%%%%%%%%%%%%%%%%%%%%%%%%%%%%%%%%%%%%%%%%%%%%%%%%%%%%%%%%%%%%%%%%%%%%
\begin{equation}\label{eq:htilde}
\tilde h (f;\bm{\theta}) =
C_\Omega \, \mathcal{A}(f;\bm{\theta}) \,
e^{i [\psi_{\text{\tiny PP}}(f;\bm{\theta}) + \psi_{\rm tidal}(f;\bm{\theta})]} \,,
\end{equation}
%%%%%%%%%%%%%%%%%%%%%%%%%%%%%%%%%%%%%%%%%%%%%%%%%%%%%%%%%%%%%%%%%%%%%%%%%%%%%%%%%
where $\bm{\theta}$ denotes the full set of intrinsic (e.g., masses, spin vectors, tidal
LNs and dissipation numbers) and extrinsic (e.g., sky location, orientation, and distance)
parameters of the binary.
The geometric factor $C_\Omega$ encodes the detector response through its antenna
pattern functions and the binary's extrinsic parameters, while $\psi_{\text{\tiny PP}}$
contains the point-particle contribution to the phase, which depends only on the masses
and spins. The tidal effects are entirely captured by $\psi_{\rm tidal}$.
At leading Newtonian order, the waveform amplitude reads
%%%%%%%%%%%%%%%%%%%%%%%%%%%%%%%%%%%%%%%%%%%%%%%%%%%%%%%%%%%%%%%%%%%%%%%%%%%%%%%%%
\begin{equation}
\mathcal{A}(f;\bm{\theta}) =
\sqrt{\frac{5}{24}}
\,\frac{\mathcal{M}^{5/6}}{\pi^{2/3}\,d_L\,f^{7/6}} \,,
\end{equation}
%%%%%%%%%%%%%%%%%%%%%%%%%%%%%%%%%%%%%%%%%%%%%%%%%%%%%%%%%%%%%%%%%%%%%%%%%%%%%%%%%
where $d_L$ is the luminosity distance and $\mathcal{M}$ is the chirp mass.

Within the Fisher-matrix approach, one computes
%%%%%%%%%%%%%%%%%%%%%%%%%%%%%%%%%%%%%%%%%%%%%%%%%%%%%%%%%%%%%%%%%%%%%%%%%%%%%%%%%
\begin{equation}
\Gamma_{ij} =
\left\langle
\frac{\partial h}{\partial \theta_i}
\bigg\vert
\frac{\partial h}{\partial \theta_j}
\right\rangle
\bigg|_{\bm{\theta}=\bm{\hat{\theta}}} \,,
\end{equation}
%%%%%%%%%%%%%%%%%%%%%%%%%%%%%%%%%%%%%%%%%%%%%%%%%%%%%%%%%%%%%%%%%%%%%%%%%%%%%%%%%
where $h$ is the GW signal in the time domain and $\bm{\hat{\theta}}$ denotes the true
(injected) values of the parameters. The covariance matrix is given by the inverse
Fisher matrix, $\Sigma_{ij}=(\Gamma^{-1})_{ij}$, and the expected $1\sigma$ uncertainty
on the parameter $\theta_i$ is $\sigma_i=\sqrt{\Sigma_{ii}}$.

The noise-weighted inner product appearing above is defined as
%%%%%%%%%%%%%%%%%%%%%%%%%%%%%%%%%%%%%%%%%%%%%%%%%%%%%%%%%%%%%%%%%%%%%%%%%%%%%%%%%
\begin{equation}
\langle h_1 | h_2 \rangle =
4\,\Re
\int_{f_{\rm min}}^{f_{\rm max}}
\frac{\tilde h_1^*(f)\,\tilde h_2(f)}{S_n(f)}\,df \,,
\end{equation}
%%%%%%%%%%%%%%%%%%%%%%%%%%%%%%%%%%%%%%%%%%%%%%%%%%%%%%%%%%%%%%%%%%%%%%%%%%%%%%%%%
where $S_n(f)$ is the one-sided noise power spectral density of the detector.
The signal-to-noise ratio is then defined as $\rho \equiv \sqrt{\langle h|h\rangle}$.
The integration limits $f_{\rm min}$ and $f_{\rm max}$ depend on the specific detector
and on the waveform model adopted; typically, $f_{\rm min}$ is set by the low-frequency
sensitivity of the instrument, while $f_{\rm max}$ is determined either by the breakdown
of the inspiral approximation or by the frequency at merger.

For a comprehensive overview of GW parameter estimation with tidal effects, we refer the reader to the review \citep{Chatziioannou:2020pqz}. In the following, we instead focus on the implications of tidal-deformability constraints from GW observations for nuclear physics, cosmology, and fundamental physics.

% For example, in Einstein Telescope (ET) the choices for minimum and maximum frequencies are (25 Hz,4096 Hz), while for Laser Interferometric Space Antenna (LISA) has the following choice for minimum and maximum frequencies ($10^{-5}$ Hz, $0.3$ Hz), assuming the duration of the mission to be 4-year. Moreover, $\sqrt{\langle h | h \rangle}$ denotes the signal to noise ration (SNR) of a GW signal, and is denoted by $\rho$. Finally the quantity $S_{n}$ corresponds to the noise curves of the detector, and for ET and LISA they follows from \citet{ET:2025xjr,Branchesi:2023mws,Babak:2021mhe}.

\subsection{Tests of the nuclear equation of state} \label{sec:pheno_NS}
Traditionally, the main interest on the tidal deformability of compact objects was related to the possibility of constraining the NS EoS through the measurement of the tidal LN in a NS binary coalescence \citep{Flanagan:2007ix}. This section is devoted to this important topic.

As previously discussed, the leading effect of the tidal LN enters at 5PN order in the GW phase, thus being highly suppressed compared to the point-particle terms. As such, it is possibly important only in the late inspiral phase before the merger. However, this effect is enhanced by a factor $C^{-5}\approx4\times 10^3$ for a typical NS with mass $M\approx 1.4 M_\odot$ and radius $R\approx 11{\rm km}$. This makes the tidal correction much larger than the order 5PN point-particle terms, and also more important than other terms at lower PN order.

\subsubsection{Current constraints}
%%%
Until 2017, GW constraints on the EoS were only hypothetical since no NS coalescence was yet detected. The landscape changed dramatically on  August 17, 2017, when the LVK Collaboration detected GW170817, the first GW signal from two merging NSs \citep{LIGOScientific:2017vwq}.

\begin{figure}
    \centering
    \includegraphics[width=0.6\linewidth]{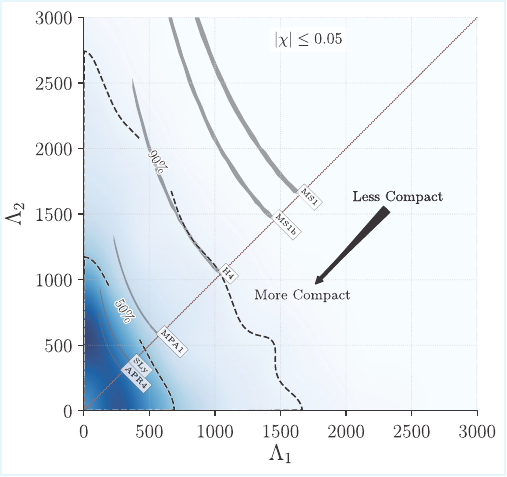}
    \caption{
    Probability density for the tidal deformability parameters of the high and low mass components inferred from GW170817, assuming low spins. For each object $\Lambda=\frac{2}{3}k_2^{\rm E}/C^5$. As a comparison, predictions for tidal deformability given by a set of representative EoS are shown (shaded filled regions). 
    from \citet{LIGOScientific:2017vwq}.    
    }
    \label{fig:GW170817}
\end{figure}

Such a historical detection was exceptional in many aspects, in particular for the relative short distance of the source ($\approx 44\,{\rm Mpc}$), which implied a relatively large SNR event. That allowed a robust upper bound on the 5PN tidal deformability parameter,
%%%%
\begin{equation}
    \Lambda_{(0)}<800
\end{equation}
%%%%
at $90\%$ confidence level. The above bound was obtained assuming the binary components were at most slowly spinning ($|\chi|\leq0.05$); a slightly more stringent bound exists if one assumes high-spin ($|\chi\leq0.89|$) priors.
\ref{fig:GW170817} presents the above constraint in the $\Lambda_1-\Lambda_2$ plane for the tidal deformabilities of the individual components. These upper bounds excluded the stiffest EoS (which provides NSs with larger radii for a given mass) and also allowed for an estimate of the NS radii \citep{Abbott:2018exr,Abbott:2018wiz,De:2018uhw}, which has been further tightened by a join analysis of GW and NICER constraints \citep{Raaijmakers:2021uju}.

To date, GW170817 remains the only confirmed NS binary for which a robust measurement of the tidal deformability has been achieved \citep{LIGOScientific:2025slb}. Owing to the substantially improved sensitivity of the current LVK network relative to the 2017 observing run, a future GW170817-like event is expected to yield significantly tighter constraints on the tidal deformability and, in turn, on the NS EoS.

%%%%%%%%%%%%%%%%%%%%%%%%%%%%%%%%%%%%
\subsubsection{Projected constraints}
%%%%%%%%%%%%%%%%%%%%%%%%%%%%%%%%%%%%
The landscape of NS tidal deformability measurements will dramatically improve with next-generation GW interferometers such as the Einstein Telescope~(ET) \citep{ET:2019dnz,Branchesi:2023mws,ET:2025xjr} and Cosmic Explorer~(CE) \citep{Reitze:2019iox,Evans:2021gyd}.

Thanks to its order-of-magnitude improvement in sensitivity and extended low-frequency coverage, ET is expected to observe binary NS inspirals with very high SNR out to cosmological distances (redshift as large as $z\approx 20$ \citep{ET:2019dnz,Iacovelli:2022bbs}), allowing tidal effects to accumulate over thousands of GW cycles in band and enabling percent-level measurements of the tidal deformability for nearby events \citep{ET:2025xjr,Branchesi:2023mws}. Unlike the current LVK network, ET will routinely constrain individual LNs of the two NSs (rather than only their mass-weighted combination), break degeneracies with masses and spins even for moderately asymmetric systems, and enable precise EoS reconstruction. While $\mathcal{O}(10)$ detections with second-generation detectors at design sensitivity could provide stringent constraints \citep{DelPozzo:2013ala}, even a single GW170817-like event observed by a third-generation detector such as ET is sufficient to rule out several families of nuclear-physics based EoS with very strong statistical significance, and to discriminate among models with similar softness but distinct microphysics \citep{Pacilio:2021jmq}. As a result, ET observations will constrain the properties of nuclear matter to unprecedented levels through both individual high-SNR events and a population of detections \citep{Puecher:2022oiz,Iacovelli:2023nbv,Puecher:2023twf}.
A detailed discussion and a comparison of different ET design in this context can be found in \citet{Branchesi:2023mws,ET:2025xjr}.

The above estimates are based on the leading tidal effect due to the static, electric, quadrupolar LNs. 
The impact of the static (quadrupolar) magnetic LNs (entering at 6PN order) and the resonance associated with low-frequency, gravitomagnetic modes of slowly-rotating NSs was discussed in \citet{Gupta:2023oyy} in the context of third-generation detectors.
The impact of other subleading terms is discussed below.

\paragraph{Love numbers and dark sirens}~---
Future measurements of tidal LNs can also play an important role in GW observations of binary NS systems without an identified electromagnetic counterpart, i.e., so-called \emph{dark sirens} \citep{Mastrogiovanni:2024mqc}. In these events, the absence of a redshift measurement from host-galaxy identification leads to a degeneracy between the source-frame masses and the cosmological redshift, which in turn propagates into uncertainties in the tidal deformability. 
Indeed, in the absence of tidal effects, GW detectors measure the detector-frame masses, related to the source ones by $M_{\rm det}=(1+z)M_{\rm source}$.
Since tidal effects depend on the source-frame masses and on the NS LNs, accurate measurements of tidal contributions to the inspiral phase can partially break this degeneracy, allowing simultaneous constraints on the tidal deformability and cosmological parameters \citep{Messenger:2011gi,DelPozzo:2011vcw,DelPozzo:2015bna}. This requires a precise knowledge of the nuclear EoS and assuming that the detected NSs have the same EoS, in order to invert the $k_2^{\rm E}(M_{\rm source})$ relation and extract the source-frame mass.

\subsubsection{Constraints from putative subsolar binaries}
\begin{figure}[t!]
\centering
\includegraphics[width=0.55\textwidth]{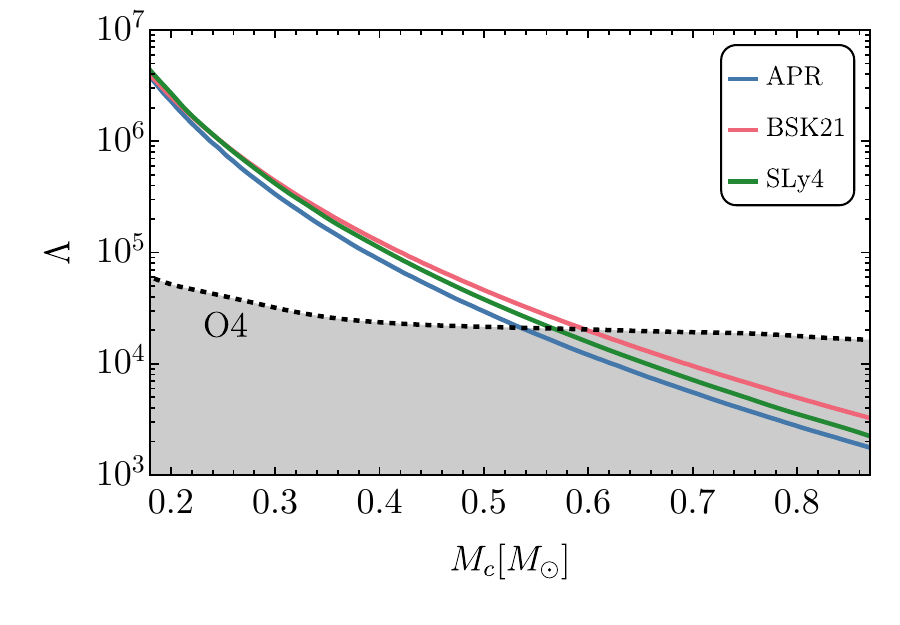}
\includegraphics[width=0.4\textwidth]{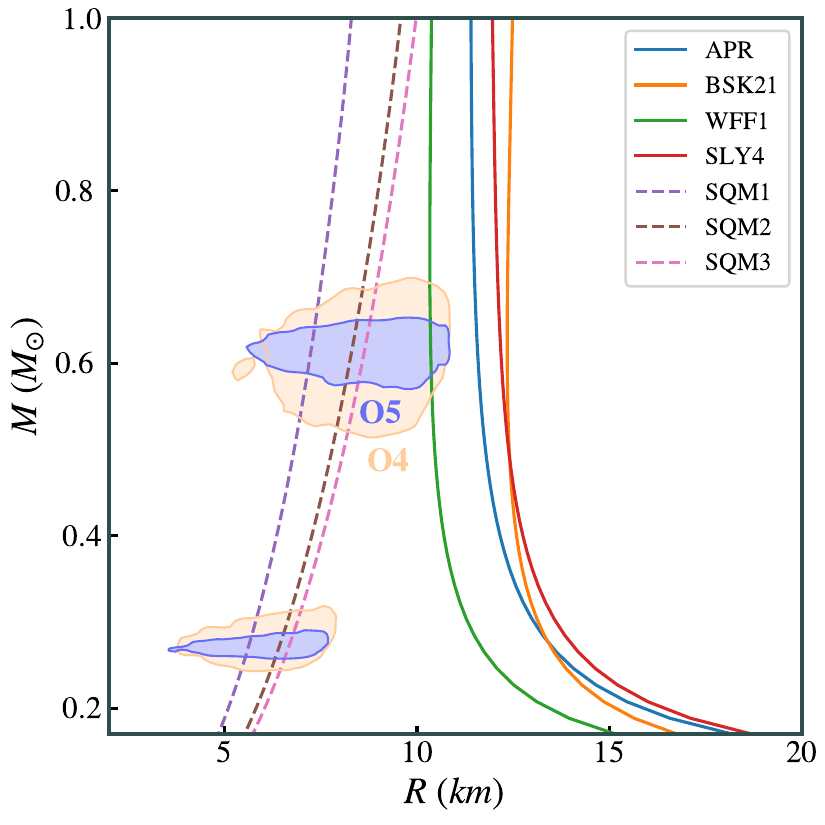}
\caption{ 
Left: Effective deformability parameter for a NS binary with three different EoS as a function of the chirp mass $M_c$, assuming equal mass binaries. Note that $\Lambda\equiv \Lambda_{(0)}$ grows significantly deep in the subsolar range.
The dashed black line indicates the upper bound (at $3 \sigma$) obtained for an equal mass binary with ${\rm SNR} = 12$ in the LVK network at the sensitivity of the fourth observational run. from \citet{Crescimbeni:2024cwh}.
Right:
The projected mass and radius of a subsolar mass NS obtained from the measurement of the tidal deformability. This example assumes $m_1= 0.63 M_\odot$ and $m_2 = 0.27 M_\odot$ with the SQM3 EoS detected with ${\rm SNR} = 25$ (O4) and ${\rm SNR} = 44$ (O5). 
The contours correspond to the corresponding posteriors on the masses and radii of the binary components.
from \citet{Crescimbeni:2024qrq}.
}\label{fig:subsolar}
\end{figure}

Detecting a compact subsolar object would have profound implications in physics, the reach of which depends on the nature of the object. 
 \citet{Crescimbeni:2024cwh,Crescimbeni:2024qrq} have explored
such consequences for a putative subsolar-mass GW event detected by the LVK Collaboration. 
Remarkably, subsolar NSs are also less dense and hence more deformable. Due to their large tidal deformability (see left panel of~\ref{fig:subsolar}),
 \citet{Crescimbeni:2024qrq} showed that the detection of a subsolar NS could rule out or confirm the existence of strange stars made of quarks (see right panel of~\ref{fig:subsolar}).

%%%%%%%%%%%%%%%%%%%%%
\subsubsection{Impact of tidal resonances}
So far, the impact of dynamical tides on the GW measurement of the NS EoS has mostly focused on tidal resonances.
 \citet{Pratten:2021pro} showed that neglecting dynamical tidal effects associated with the fundamental ($f$-) mode leads to large systematic biases in the measured tidal deformability and hence in the inferred NS EOS. In particular, when the orbital frequency approaches the $f$-mode frequency, the adiabatic approximation breaks down and the tidal response is enhanced, producing additional phase shifts that can be partially degenerate with the static tidal deformability (see~\ref{eq:delta_phi_res}), especially for stiff EoS with relatively low $f$-mode frequencies.

In addition to the dominant quadrupolar $f$-mode, higher-order modes and nonlinear couplings have also been investigated, though their impact on parameter estimation appears subdominant for typical binary NS systems \citep{Yu:2022fzw,Yu:2025ptm,Gupta:2023oyy}.

Overall, these studies indicate that tidal resonances introduce systematic waveform corrections that, if not modeled consistently, can bias the inference of the tidal deformability and therefore the NS EoS. This is particularly relevant for next-generation detectors such as ET and CE, whose improved high-frequency sensitivity will probe the regime where dynamical tidal effects become most significant \citep{ET:2025xjr,Gupta:2023oyy}.

%%%%%%%%%%%%
\subsubsection{Impact of dissipative effects}
%%%%%%%%%%%%
Dissipative effects in NSs can arise from several microphysical mechanisms,
including bulk and shear viscosity, crust--core coupling, superfluid mutual
friction, and the excitation and damping of internal modes. In addition,
ECOs may exhibit even stronger absorption due to the presence of effective
horizons or partially absorbing surfaces, leading to tidal heating effects
that can differ qualitatively from those of ordinary NSs.

The impact of such dissipative mechanisms on GW signals has been recently
investigated in \citet{Ripley:2023lsq,Ripley:2023qxo,Chia:2024bwc,Shterenberg:2024tmo}, where tidal
heating and more general absorption effects were consistently incorporated
into inspiral waveform models and studied within a parameter-estimation
framework. \citet{Ripley:2023lsq} derived order-of-magnitude bounds on the
effective, density-averaged bulk ($\zeta$) and shear ($\eta$) viscosities of
the NS in GW170817, and forecast that these constraints could improve by up
to two orders of magnitude with third-generation detectors.

\citet{Ripley:2023qxo} quantified the detectability of tidal heating in
NS binaries. As for the usual LNs, the corresponding tidal dissipation
coefficients are enhanced by large powers of the inverse compactness relative
to their nominal PN order (see~\ref{tab:PN}). However, dissipative effects
enter the GW phase at 2.5PN order for spinning NSs and at 4PN order for
nonspinning NSs, and include an additional $\log v$ dependence, where $v$ is
the orbital velocity (see~\ref{phase_integral_heating}
and~\ref{phase_integral_heating2}). In the absence of this logarithmic term,
the induced phase correction would be exactly degenerate, in the frequency
domain, with shifts in the time and phase of coalescence, $t_c$ and $\psi_c$.
Because the $\log v$ dependence only weakly breaks this degeneracy, a large
fraction of the dissipative contribution can be absorbed by small variations
of these nuisance parameters, substantially degrading the measurability of
tidal heating in realistic data-analysis scenarios.

Overall, the results of \citet{Ripley:2023lsq,Ripley:2023qxo,Chia:2024bwc,Shterenberg:2024tmo}
indicate that, while dissipative tidal effects provide in principle a clean
probe of the internal microphysics of NSs (or of horizon-like absorption in
ECOs), their observability is limited by strong parameter degeneracies and by
their relatively high-PN suppression. Nevertheless, for sufficiently loud
events and with next-generation detectors, these effects could offer
complementary information to that obtained from conservative tidal
deformability measurements (see also~\citealt{HegadeKR:2026iou} for a recent analysis with parametrized EoS).

%%%%%%%%%%%%%%
\subsubsection{Impact of rotational Love numbers}
%%%%%%%%%%%%%%
The impact of rotational tidal LNs on GW signals was first quantified in
 \citet{JimenezForteza:2018rwr} using waveform mismatch estimates, and subsequently
analyzed in greater detail in \citet{Castro:2022mpw} within a Fisher-matrix framework.
These studies demonstrated that, for dimensionless NS spins
$\chi \sim 0.1$ or higher, neglecting tidal-spin couplings can lead to sizable systematic biases
in GW parameter estimation with third-generation detectors such as ET.

In particular, the inclusion of rotational tidal effects was shown to be essential for
accurately recovering intrinsic parameters such as the masses, spins, and tidal LNs of
the binary components. The enhanced sensitivity of ET at high frequencies, where
spin-induced tidal corrections might be relevant during the late inspiral, makes it especially
susceptible to these effects. As a result, rotational LNs may not only need to be
included to avoid biases, but could also become directly measurable with ET, provided
the binary NSs possess sufficiently large spins.

\subsubsection{Impact of quadratic and dynamical Love numbers}
\begin{figure}[t]
\begin{center}
	\includegraphics[width=0.32\textwidth]{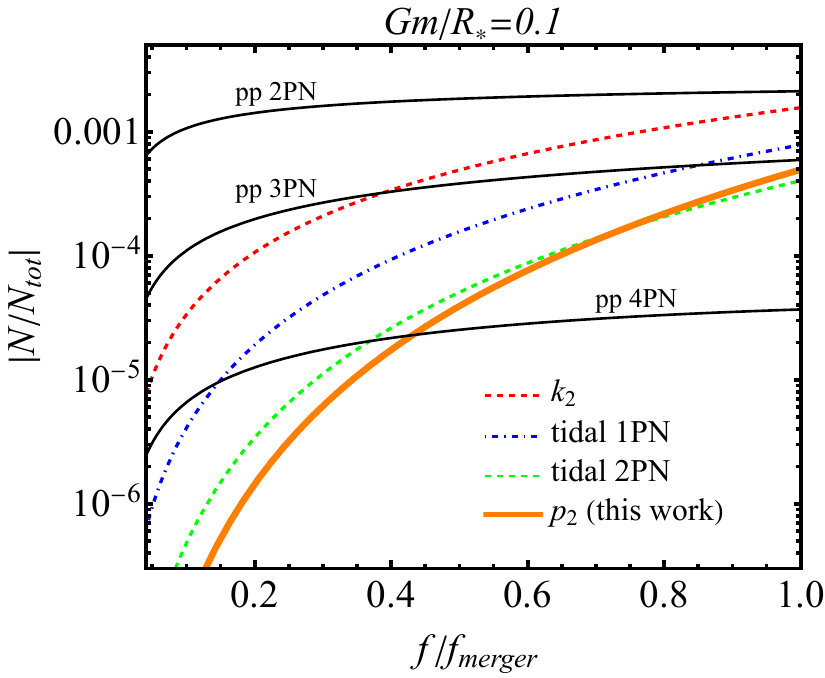}
    \includegraphics[width=0.32\textwidth]{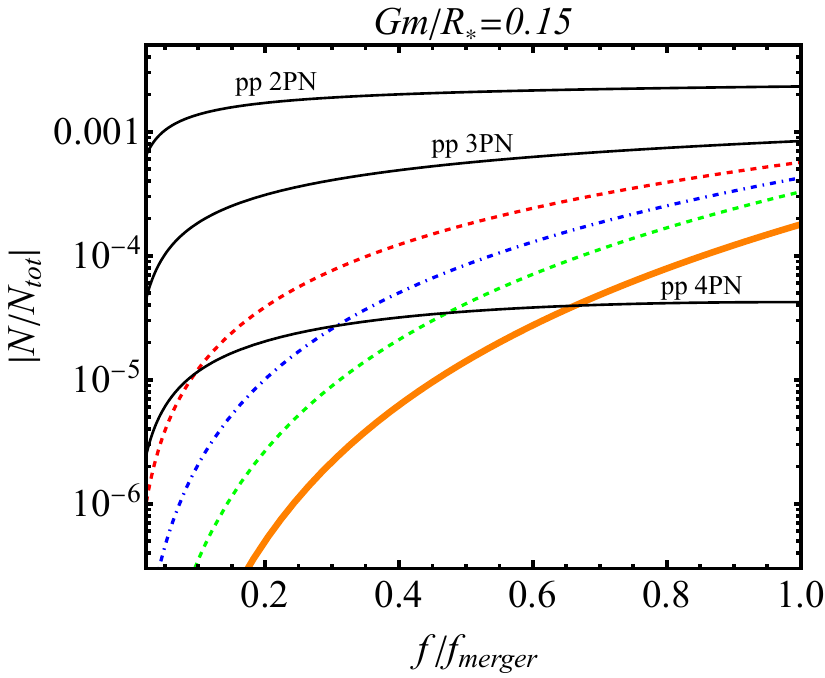}
    \includegraphics[width=0.32\textwidth]{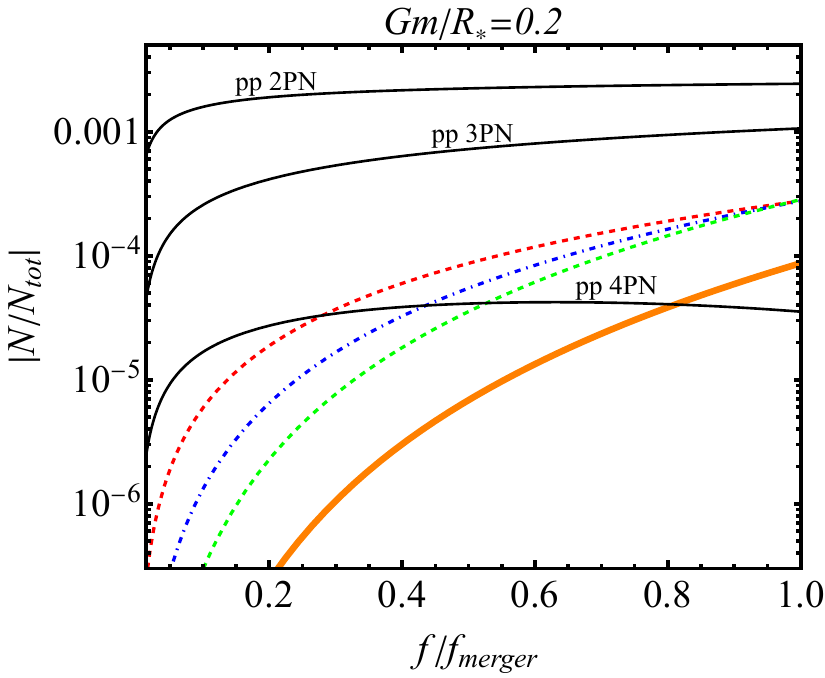}
\end{center}
\caption{Comparison between different tidal and point-particle contributions to the GW cycles $N( f) \equiv [\psi(f) - \psi(f_{\rm entry})]/\pi$, where $f = \omega/\pi$ is the GW frequency and $f_{\rm entry}=30\,{\rm Hz}$ is a reference starting frequency, as function of the GW frequency rescaled by the Newtonian estimate for the merger frequency, normalized by the total number of cycles.
We consider a reference circular binary with $m_1=m_2=1.4M_\odot$, linear LN $k_2=0.1$, and quadratic LN $p_2=0.4$.
Each panel corresponds to a different compactness. As a reference, $C=(0.1,0.15,0.2)$ respectively corresponds to $k_{2}/C^5\approx(10\, 000,1317,313)$, $R_*\approx(20.7,13.8,10.4)\,{\rm km}$, and $f_{\rm merger} \approx (730, 1341, 2065)\,{\rm Hz}$.
The black continuous curves are the 2PN, 3PN and 4PN point-particle contributions, respectively, starting from above.
The red, blue, and green curves correspond to the leading, next-to-leading, and next-to-next-to-leading corrections due to the linear tidal LN $k_2$. The orange thicker curve is the contribution coming from the leading quadratic LN.
Taken from \citet{Pani:2025qxs}.
}
\label{fig:Quadraticcycles}
\end{figure}

As previously discussed, quadratic and dynamical (${\cal O}(\omega^2M^2)$) LNs enter the GW phase at 8PN order. Despite their strong PN suppression,
their contribution to the phase is enhanced by a factor $1/C^8\approx 6\times 10^5$ for a reference NS with $M\approx1.4M_\odot$ and $R\approx 11{\rm km}$.
Because of this, it has been shown that the leading quadratic LN can be as important as the next-to-next-to-leading order linear tidal correction, which enters at 7PN order, and is larger than the subleading point-particle contribution entering at 4PN order \citep{Pani:2025qxs}. In particular, the contribution from quadratic LNs can be as large as $10\%$ of that from linear LNs in the late inspiral phase. 
An example is shown in Fig.~\ref{fig:Quadraticcycles}, which compares the various contributions to the accumulated GW cycles of the binary normalized by the total number of cycles.

A quantitative parameter-estimation analysis for quadratic LNs is still lacking. Moreover, although dynamical LNs have been computed for polytropic NSs \citep{Pitre:2023xsr,Pitre:2025qdf}, their imprint on the GW waveform has not yet been evaluated. As a consequence, the impact of dynamical LNs remains unquantified, although it is expected to be comparable to that of quadratic LNs.
% \paolo{Reminder: Add paper by Luca possibly when it's out. Discuss also running}

%%%%%%%%%%%%%%%%%%%%%%%%%%%%%%%%%%%%%%%%%%%%%%%%%%%%%%%%%%%%%%
\subsection{Tests of the nature of compact objects}\label{testcompact}
%%%%%%%%%%%%%%%%%%%%%%%%%%%%%%%%%%%%%%%%%%%%%%%%%%%%%%%%%%%%%%
Since LNs encode the internal structure of NSs, they likewise characterize the internal properties of any compact object. An important application is therefore the use of LN measurements to probe the nature of coalescing binaries beyond the standard BH and NS scenarios.
In particular, the detection of a nonvanishing LN for an object with mass $M \gtrsim 3M_\odot$ would constitute a smoking-gun signal of physics beyond the standard paradigm: such an object would be too massive to be a NS, while a nonzero LN is incompatible with the BH hypothesis, at least in vacuum.
After excluding observational systematics \citep{Yagi:2013baa} and other potential misidentifications \citep{Gupta:2024gun}, the most plausible interpretations are that the object is either: (i)~neither a BH nor a NS, or (ii)~a BH surrounded by deformable matter.
The former scenario is discussed in this section, while the latter is addressed in~\ref{sec:application_env}.

%%%%%%%%%%%%%%%%%%%%%%%%%%%%%%%%%%%%
\subsubsection{Measuring Love numbers in dark compact objects} \label{sec:application_ECOs}
\begin{figure}[th]
\centering
\includegraphics[width=0.32\textwidth]{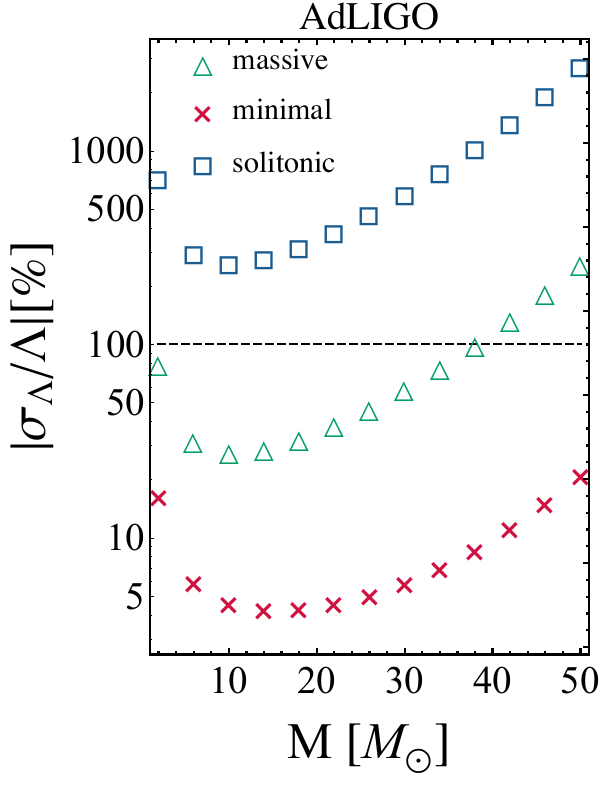}
\includegraphics[width=0.305\textwidth]{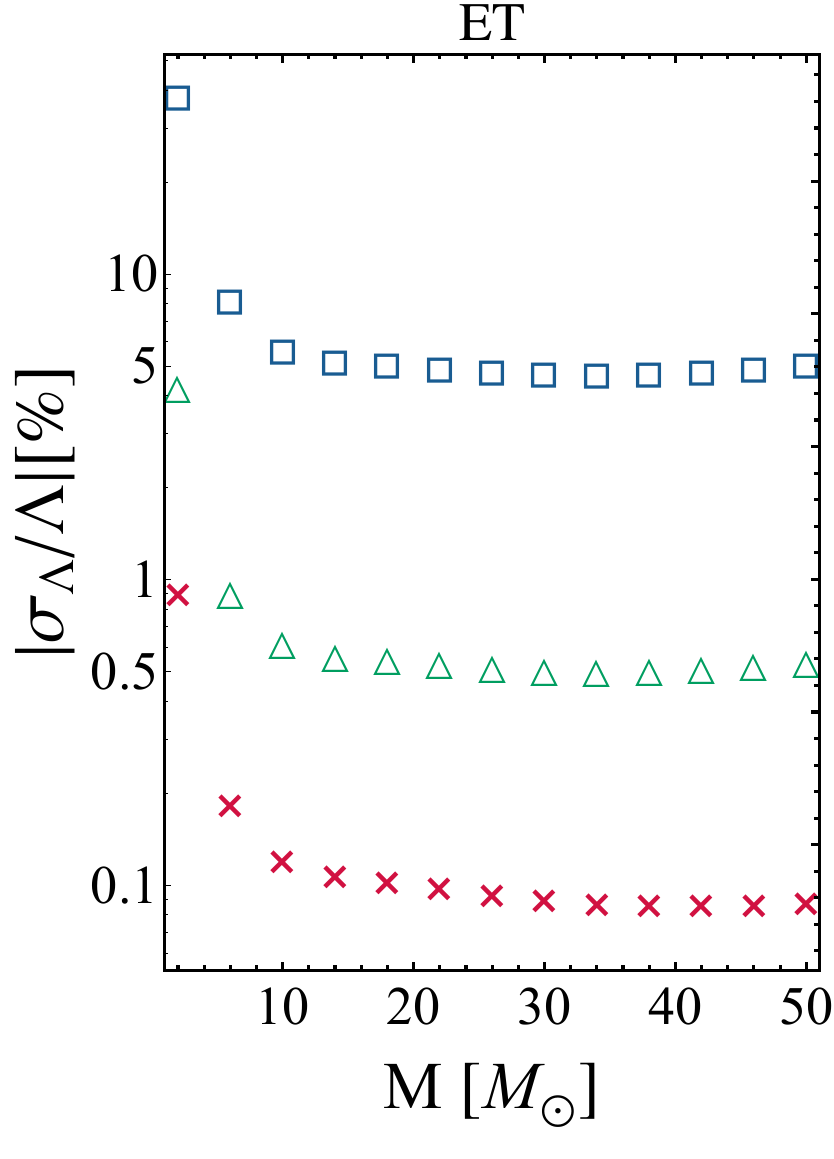}
\includegraphics[width=0.32\textwidth]{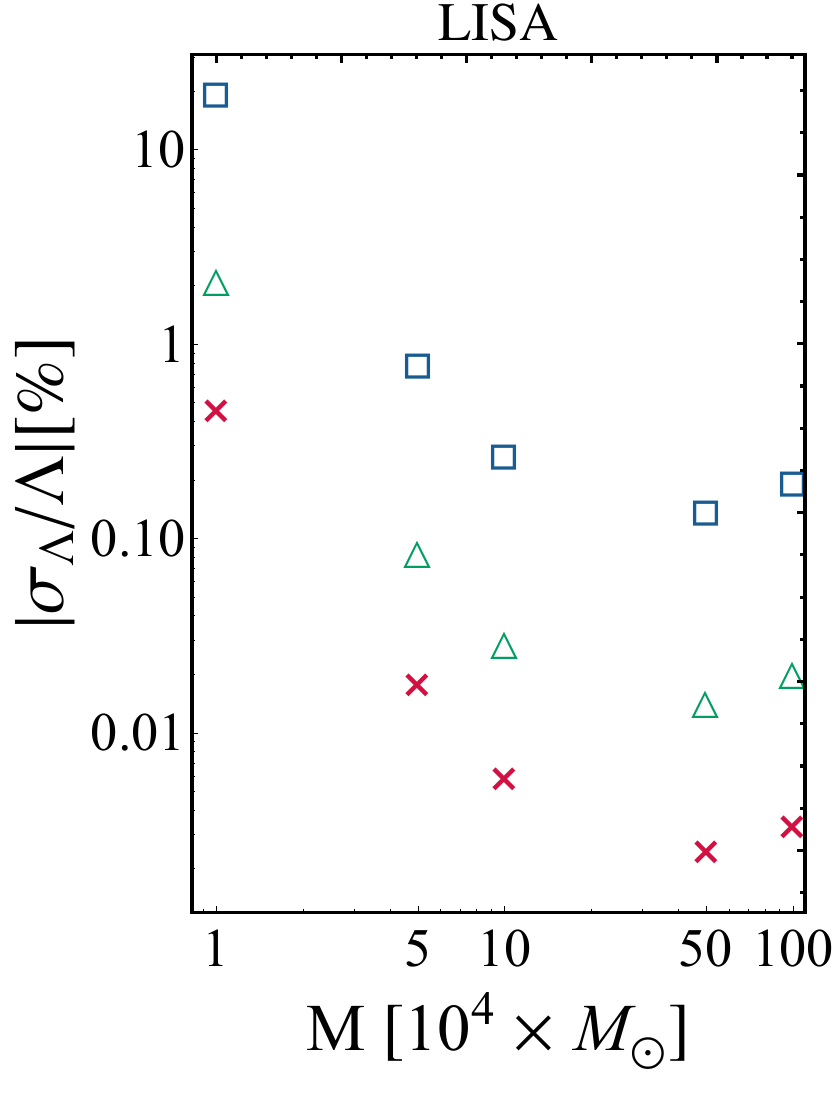}
\caption{Relative percentage errors on the average tidal deformability $\Lambda$ for equal-mass boson-star binaries observed by Advanced LIGO (left), ET (middle), and LISA (right), as a function of the boson-star mass for the models considered. For terrestrial detectors we assume a fiducial source at $d=100\,{\rm Mpc}$, while for LISA the source is located at $d=500\,{\rm Mpc}$. The horizontal dashed line marks the threshold $\sigma_\Lambda/\Lambda=1$. Systems lying below this line would yield a measurement incompatible with zero, allowing the corresponding boson star to be distinguished from BHs. Here $\Lambda\equiv \Lambda_{(0)}$ as defined in~\ref{psi-gen}. from \citet{Cardoso:2017cfl}.}
\label{fig:detectabilityBS}
\end{figure}

%%%%%%%%%%%%%%%%%%%%%%%%%%%%%%%%%%%%
Building on the static LNs computed for various ECO models (see~\ref{sec:ECOs}), \citet{Cardoso:2017cfl} estimated the ability of current and future GW detectors to measure these effects. Similar studies focusing on boson-star binaries were performed in \citet{Wade:2013hoa,Sennett:2017etc} (see also \citealt{Johnson-Mcdaniel:2018cdu}, using polytropic models).
In~\ref{fig:detectabilityBS} we show an illustrative example of the expected constraints on different boson-star models obtained with LIGO, ET, and LISA.

More recently, \citet{Pacilio:2020jza,Vaglio:2023lrd} performed both Fisher-matrix and Bayesian parameter-estimation analyses using coherent inspiral waveforms that consistently include tidal deformability and spin-induced multipole moments for boson-star binaries with large quartic self-interactions, based on the fits of \citet{Sennett:2017etc,Vaglio:2023lrd}. These studies show that coherent waveform models --~rather than phenomenological prescriptions in which tidal terms are added a posteriori to BH waveforms~-- significantly enhance detectability prospects. They also demonstrate that future detectors such as ET and LISA can place strong and complementary constraints on bosonic self-interactions.

A matched-filtering search for binaries with  large tidal deformabilities in the LVK catalog was performed in \citet{Chia:2023tle, Andres-Carcasona:2025bni}, finding no statistically significant candidates for non-NS binaries with nonvanishing tidal LNs. 
Interesting constraints were recently obtained from the large SNR event GW250114, placing a $90\%$ upper limit on the effective tidal deformability of $\Lambda_{(0)}<34.8$.
These bounds are fully consistent with the BH nature of this event and rule out some models of boson stars. 

\paragraph{Probing Planckian corrections at the horizon scale with tidal effects}~---
As discussed in~\ref{sec:ECOs}, the static LN of many ECO models exhibits a logarithmic dependence on the compactness parameter $\epsilon$ in the BH limit. This behavior acts as a magnifying mechanism for the tidal deformability, even in scenarios where $\epsilon$ is motivated by Planck-scale corrections at the horizon \citep{Maselli:2017cmm}. Despite of the fact that in those cases $\epsilon\approx 10^{-40}$, the LNs can be only $\sim4$ orders of magnitude smaller than those of ordinary NSs. Although measuring such small effects is challenging, it may be achievable with LISA, particularly for highly spinning massive binaries \citep{Maselli:2017cmm}. Owing to the logarithmic scaling, the statistical error on $\epsilon$ depends \emph{exponentially} on the LN, implying that resolving Planckian corrections requires extremely accurate measurements of $k_2^{\rm E}$ \citep{Addazi:2018uhd}. Nevertheless, this sensitivity does not preclude meaningful model selection for ECOs, enabling unprecedented tests of putative quantum-gravity effects at the horizon scale \citep{Maselli:2018fay}.

%%%%%%%%%%%%%%%%%%%%%%%%%
\paragraph{ECO tidal constraints with EMRIs}~---
%%%%%%%%%%%%%%%%%%%%%%%%%
The exceptional capability of EMRIs to probe the tidal deformability of supermassive compact objects can be understood by integrating the tidal phase contribution in~\ref{eq:phi_tidal_EMRI} up to the innermost stable circular orbit of the central object. The total accumulated tidal phase is \citep{Pani:2019cyc}
%%%%%%%%%%%%%%%%%%%%%%%%%%%%%%%%%%%%%%%%%%%%%%%%%%%%%%%%%%%%%%%%%%%%%%%%%%%%%%%%%
\begin{equation}
\psi^{\rm tot}_{\rm tidal}
= -\frac{\sqrt{6}}{96}\,k_1\,q^{-1}
\simeq -400\,k_1\,\frac{10^{-5}}{q} ,
\label{eq:phi_tot_tidal_EMRI}
\end{equation}
%%%%%%%%%%%%%%%%%%%%%%%%%%%%%%%%%%%%%%%%%%%%%%%%%%%%%%%%%%%%%%%%%%%%%%%%%%%%%%%%%
which highlights the dramatic enhancement of tidal effects in the $q\ll1$ regime. This estimate suggests that, for typical EMRIs with $q\sim10^{-6}$, even extremely small values of the central object's tidal LN can induce an observable phase shift.

This expectation was confirmed in \citet{Piovano:2022ojl}, which employed accurate semi-analytical EMRI waveforms in the frequency domain together with a detailed Fisher-matrix analysis. That study showed that a LISA detection of an EMRI could constrain the tidal LN of a spinning central object with dimensionless spin $\chi=0.9$ (resp.\ $\chi=0.99$) at a level approximately four (resp.\ six) orders of magnitude stronger than what is achievable with current ground-based detectors observing stellar-mass binaries.

\paragraph{Tidal tests for subsolar-mass gravitational-wave observations}~---
The detection of a subsolar object in a compact binary merger is regarded as one of the smoking gun signatures of a population of primordial BHs. However, these systems could be confused with subsolar NSs, which could also populate the subsolar mass range \citep{Franciolini:2021xbq}. At variance with primordial BHs, the GW signal from stellar binaries is affected by tidal effects, which dramatically grow for moderately compact stars as those expected in the subsolar range (see left panel of~\ref{fig:subsolar}. 
 \citet{Crescimbeni:2024cwh} forecast the capability of constraining tidal effects of putative subsolar neutron star binaries with current and future LVK sensitivities as well as next-generation experiments. They found that, should LVK O4 run observe subsolar NS mergers, it could measure the (large) tidal effects with high significance. In particular, for subsolar NS binaries, O4 and O5 projected sensitivities would allow measuring the effect of tidal disruption on the waveform in a large portion of the parameter space, also constraining the tidal deformability at 
${\cal O}(10\%)$ level, thus excluding a primordial origin of the binary. Vice versa, for subsolar primordial BH binaries, model-agnostic upper bounds on the tidal deformability can rule out NSs or more exotic competitors. O4 projected sensitivity would allow ruling out the presence of NS tidal effects at $\approx 3\sigma$ confidence level, thus strengthening the primordial BH hypothesis of these putative binaries. Future experiments would lead to even stronger ($>5\sigma$) conclusions on potential discoveries of this kind.

More recently, \citet{Crescimbeni:2024qrq} showed that the nature of a subsolar-mass compact binary can be robustly identified using current LVK observations by exploiting the large tidal deformability imprinted in the GW signal (see left panel of~\ref{fig:subsolar}). By performing detailed Bayesian forecasts, it demonstrated that binaries composed of light NSs, primordial BHs, or ECOs occupy well-separated regions of the parameter space, allowing their discrimination with high statistical confidence already at current or near-future sensitivity. These results highlight the key role of tidal effects in breaking degeneracies among different formation scenarios and have important implications for cosmology and dark matter models involving primordial BHs, as well as for placing constraints on the nuclear EoS of low-mass NSs.

\paragraph{Regular BHs}~---
As discussed in~\ref{sec:regularBH}, various models of regular BHs have nonvanishing LNs. The associated dephasing is, schematically,
%%%
\begin{equation}
    \psi_{\rm tidal} \propto\left(\frac{L}{M_{\rm T}}\right)^p\,,
\end{equation}
%%%
where the regularization scale $L$, the exponent $p$ (typically $p\in[2,4]$), and the prefactor depend on the model \citep{Coviello:2025pla}. 
The effect is negligible if $L$ is Planckian, but can be sizable if $L$ saturates the model-dependent, theoretical bound $L={\cal O}(M)$. In the latter case the effect might be detected by third-generation detectors, although it would be degenerate with the ordinary 5PN point-particle terms that are currently unknown.

\paragraph{Measurability of dynamical Love numbers}~---
The above discussion has focused on the standard (static) LNs. 
The measurability of the BH dynamical LNs was recently quantified in \citet{Chakraborty:2025wvs} using a Fisher-matrix analysis. Even under optimistic assumptions (relatively large SNR, absence of spin precession, and circular orbits), these corrections were found to be far too small to be detectable, even with future-generation GW detectors. Moreover, the associated phase contributions are degenerate with unknown point-particle effects entering at the same (eighth) PN order.
Therefore, although theoretically well motivated, the dynamical LNs of BHs are unlikely to have phenomenological relevance in the foreseeable future. 
The situation is expected to be more favorable for NSs, given the enhancement of tidal effects by a factor $C^{-8}$.

%%%%%%%%%%%%%%%%%%
\subsubsection{Tidal heating as a discriminator for horizons}
%%%%%%%%%%%%%%%%%%
%
 \citet{Chia:2024bwc} presented the first constraints on tidal heating for the binary systems using public LVK data. Assuming all events as BH binaries, they obtained a constraint $-13 < \mathcal{H}_0 < 20$ on the dissipation number at the $90\%$ credible interval\footnote{The dissipation number can be expressed, in general, as $\nu^{\rm E/B}_{\ell m} = \nu^{\rm E/B}_{\ell m,\chi} + M\omega \,\nu^{\rm E/B}_{\ell m,\omega}$. Here, $\nu^{\rm E/B}_{\ell m,\chi}$ denotes the static contribution, which is proportional to the spin of the object, whereas $\nu^{\rm E/B}_{\ell m,\omega}$ represents the dynamical contribution. The quantity $\mathcal{H}_{0}$ introduced above is defined as $3\mathcal{H}_{0} = M_{\rm T}^{-4}\left[\frac{r_{+1}^{5}}{M_{1}}\,\nu^{\rm E/B(1)}_{2m,\omega} + \frac{r_{+2}^{5}}{M_{2}}\,\nu^{\rm E/B(2)}_{2m,\omega}\right]$ in terms of the individual dissipation numbers of the binary components.}.
However, this test is not very specific, as the allowed range is two orders of magnitude larger than the value predicted for BHs in GR.
Making no assumption on the nature of the binaries,
the constraints on the dissipation
numbers are further relaxed by an order of magnitude.

\begin{figure*}[th]
\centering
\includegraphics[width=0.6\textwidth]{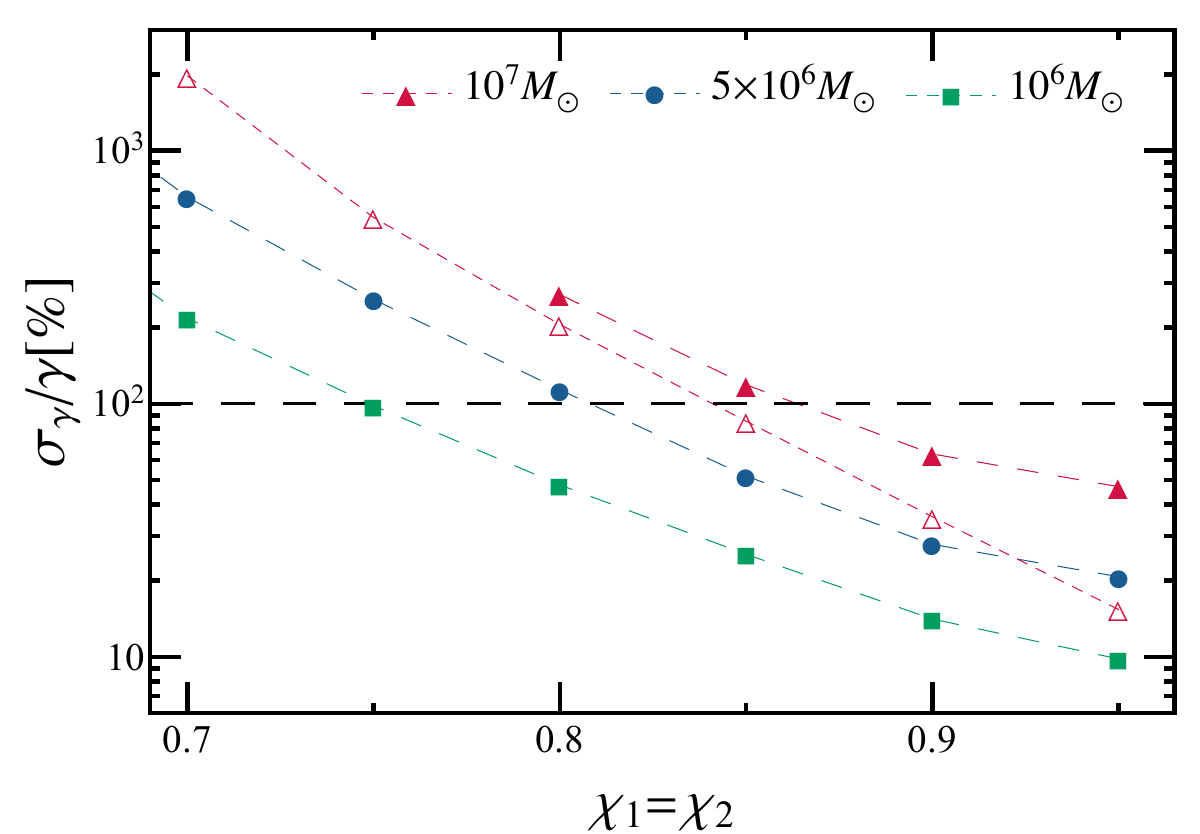}
\caption{Projected percentage errors on the tidal-heating parameter $\gamma$ as a function of the (equal) spin parameter $\chi_1=\chi_2$, for central masses $m_1=(10^6,5\times10^6,10^7)M_\odot$, assuming a future LISA detection. Filled (open) markers correspond to mass ratios $m_1/m_2=1.1$ ($m_1/m_2=2$). Points below the horizontal line indicate measurements that can distinguish a BH from an ECO at better than the $1\sigma$ level. All binaries are placed at a luminosity distance of $2,{\rm Gpc}$; the uncertainty $\sigma_\gamma$ scales inversely with distance, while for $k\ll1$ one has $\sigma_\Lambda\propto 1/\Lambda$ ($k$ is the dimensionless LN and $\Lambda$ is defined in \ref{psi-gen}). The effect is linear in the spin and it would be suppressed by two further PN orders in the nonspinning case. From \citet{Maselli:2017cmm}.
}
\label{fig:heating}
\end{figure*}

Looking ahead, highly-spinning supermassive binaries detectable with a LISA-type interferometer will have a 
large SNR and will place quite stringent constraints on tidal heating \citep{Maselli:2017cmm,Shterenberg:2024tmo}. This is shown in~\ref{fig:heating}, which presents the bounds on parameter $\gamma$, introduced as a book-keeping parameter in front of the tidal-heating phase (see~\ref{phase_integral_heating}): for a BH 
$\gamma=1$, whereas $\gamma=0$ for a perfectly reflecting ECO. 

Constraints on tidal heating with future detectors were also recently explored in~\citet{Shterenberg:2024tmo}.

%

%%%%%%%%%%%%%%%%%%%
\paragraph{Tidal-heating constraints with EMRIs}~---
%%%%%%%%%%%%%%%%%%%
The absence of tidal heating can also leave a potentially detectable imprint in EMRIs \citep{Hughes:2001jr,Datta:2019epe,Maggio:2021uge}. Although horizon absorption is subleading with respect to GW emission to infinity, its putative absence can induce measurable orbital dephasings and, consequently, observable modifications of the GW signal, especially when the central object is rapidly spinning \citep{Hughes:2001jr}.

By studying the orbital dephasing and the GW signal emitted by a point particle on circular, equatorial orbits around a spinning supermassive object, at leading order in the mass ratio, \citet{Datta:2019epe} showed that this effect can be exploited to probe the nature of the central object in a largely model-independent manner. The projected constraints on the reflectivity of ECOs were found to be at the level of ${\cal O}(0.01)\%$. This analysis was later extended to eccentric EMRIs \citep{Datta:2024vll,Xia:2026aty} and to consistent frequency-domain waveforms obtained by solving the Teukolsky equations with suitable (frequency-dependent) boundary conditions \citep{Maggio:2021uge}.

\begin{figure}[ht]
\centering
\includegraphics[width=0.75\textwidth]{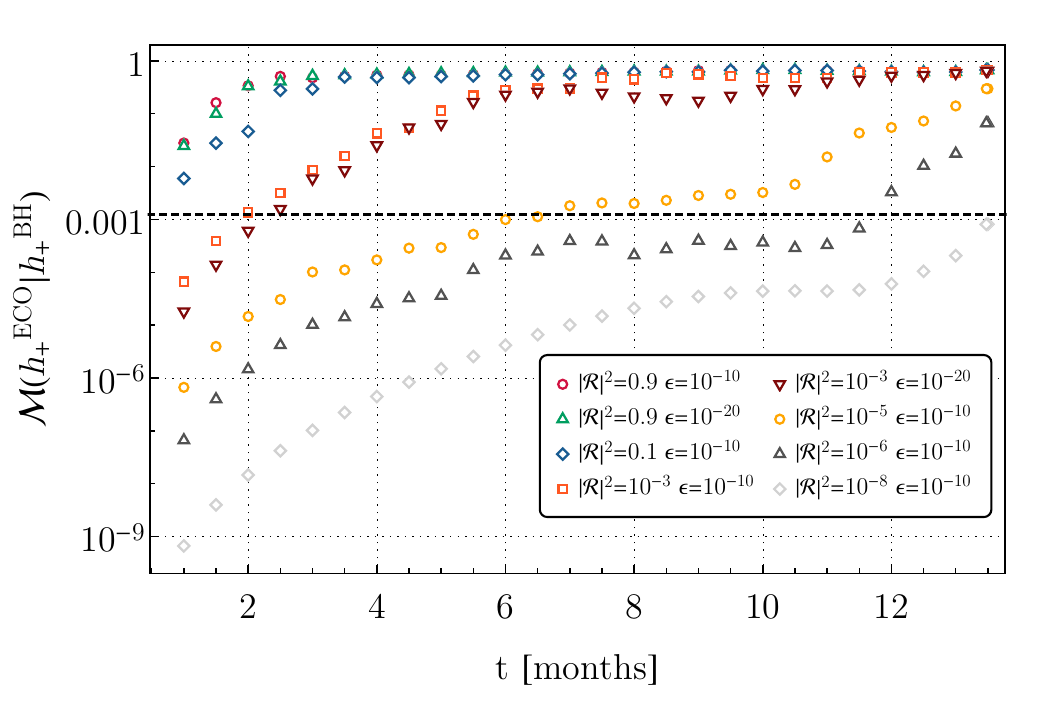}
\caption{Mismatch between the plus polarization of the waveforms with a central ECO and a central BH as a function of time, for primary spin $\chi=0.8$, mass ratio $q=3 \times 10^{-5}$, and several values of the ECO reflectivity ${\cal R}$. from \citet{Maggio:2021uge}.} 
\label{fig:heatingEMRIs}
\end{figure}

This approach reveals that the gravitational fluxes exhibit resonances associated with the low-frequency QNMs of the central object, which can contribute to the accumulated GW phase. A fully consistent approach places even more stringent constraints on the reflectivity, reaching the remarkable level of $10^{-6}\%$ (see~\ref{fig:heatingEMRIs}). These estimates, however, are based on simple mismatch calculations and likely underestimate the impact of waveform-modeling systematics and parameter-estimation uncertainties.
A recent Bayesian analysis using equatorial eccentric EMRIs estimated an upper bound on the reflectivity at the level of $10^{-4}\%$ \citep{Xia:2026aty}.

%

%%%%%%%%%%%%%%%%%%%%%%%%%%%%%
\subsection{Tests of the environment of compact objects}\label{sec:application_env}
%%%%%%%%%%%%%%%%%%%%%%%%%%%%%
If coalescing BHs are surrounded by matter distributions, the latter can be distorted by tidal interactions and contribute a nonvanishing LN.
Furthermore, diluted distributions are expected to be particularly deformable, so the corresponding LN are typically large (see~\ref{sec:BHhalos}).
Thus, quantifying the detectability of this effect is important both to understand the environments in which compact objects live, and to avoid possible misidentification with modified gravity or BH mimickers \citep{Barausse:2014tra,Cannizzaro:2024fpz}.

In this context, \citet{DeLuca:2021ite} computed the TLNs of a BH surrounded by a scalar condensate, considering both scalar and vector tidal perturbations. A pronounced power-law dependence of the TLNs on the scalar-field mass was found (see~\ref{eq:cloud}). Using these results as a proxy for the gravitational tidal response, it was shown that future GW detectors such as ET and LISA can place stringent constraints on the mass of ultralight bosons forming condensates around BHs through accretion or superradiant instabilities. Taken together, LISA and ET are expected to probe tidal signatures of BHs dressed by ultralight bosonic clouds for boson masses in the range $\in(10^{-17},10^{-11})\,{\rm eV}$.

An intriguing feature of tidally deformable environments around BHs is that, due to their large tidal LNs, they are particularly susceptible to tidal disruption. This has motivated effective waveform models in which the tidal deformability interpolates between a finite value (representing a ``dressed'' BH) at relatively low frequencies and a vanishing value (corresponding to a ``naked'' BH) at high frequencies \citep{DeLuca:2021ite,DeLuca:2022xlz}. Applying this framework to the case of a BH surrounded by an ultralight scalar field, \citet{DeLuca:2022xlz} demonstrated that the parameters describing the tidal deformability could be measured with high precision by next-generation detectors, providing a valuable tool to probe dynamical environmental effects during the inspiral of a binary system.  

More recently, \citet{Cannizzaro:2024fpz} studied the tidal deformability of BHs surrounded by thin accretion disks, finding a similarly strong dependence on the compactness (see~\ref{spin0TLNjt=0}). Their results indicate that the disk parameters could be measured by ET and LISA with accuracies of a few to ten percent for sources at gigaparsec distances. This outcome relies on the characteristic evolution of accretion-dressed binaries: due to strong tidal interactions near the Roche radius, also these systems are expected to lose their environment before the merger, continuing their evolution essentially in vacuum. These encouraging prospects suggest that future GW observations will provide unprecedented insights into the environments in which binary BHs evolve.

Finally, in the context of current LVK constraints,
\citet{CanevaSantoro:2023aol} performed the first investigation of environmental effects on various events from the first and second GW transient catalog of the LVK Collaboration, finding no evidence for the presence of environmental effects. The GW170817 event results in the most stringent upper bound on the environmental density ($\approx 21\,{\rm g/cm}^3$). Furthermore, it was found that environmental effects can substantially bias the recovered parameters in a vacuum model, even when these effects are not detectable. 

%%%%%%%%%%%%%%%%%%%%%%%%%%%%%
\subsection{Tests of General Relativity}
%%%%%%%%%%%%%%%%%%%%%%%%%%%%%

\subsubsection{Love numbers as a probe of theories beyond General Relativity}

As discussed in~\ref{sec:BH4DBGR}, the BH LNs in theories beyond GR are generically non-zero.
However, these LNs are proportional to (powers of) the coupling constants of the theory (see~\ref{sec:BHBGR}) and, at variance with their NS counterpart, they enter the waveforms \emph{without} the usual enhancement by large inverse powers of the compactness. As a result, the ordinary (static, electric, quadrupolar) tidal LNs enter the GW phase at 5PN order, with a coefficient that is even \emph{suppressed} relative to the usual GR 5PN point-particle phase by powers of the coupling constant(s).
Therefore, standard beyond-GR tidal effects in BH binaries are degenerate with (and in fact subleading to) \emph{point-particle} 5PN effects in GR, which are currently unknown.
Beside their theoretical interest, the prospects to use them to test GR are weak.

However, in theories in which BHs can have nontrivial hair due to novel degrees of freedom,
the coupling between the latter and the metric perturbations induces novel classes of LNs, and the latter should enter at lower PN order, as it happens for NSs in scalar-tensor theories \citep{Creci:2023cfx,Creci:2024wfu} (see~\ref{sec:NSBGR}). Although the impact of this effect on BH coalescence waveforms and possible tests of gravity has not been explored yet, this is a promising avenue for future work.

A more established avenue concerns tidal tests of gravity with NSs. As discussed in~\ref{sec:NSBGR}, in this case there exist scalar-type (dipolar) LNs entering at lower PN order \citep{Creci:2023cfx,Creci:2024wfu}.
For the theory studied in \citet{Creci:2024wfu}, the various tidal contributions enter the waveform with different signs and scalings with frequency, which generally leads to smaller net tidal GW imprints than for the same binary system in GR. It would be important to quantify if beyond-GR tidal effects can be disentangled from the uncertainty about the NS EoS.

Regarding the latter point, an interesting possibility is offered by the quasi-universal relations among certain NS parameters, including the LNs \citep{Yagi:2013awa,Yagi:2013bca,Yagi:2013sva,Yagi:2016bkt,Yagi:2016qmr}.
These EoS-insensitive relations hold, to certain level of accuracy that depends on the theory, in each given beyond-GR theory, despite the fact that the LNs alone depends on both the theory and the NS EoS (see~\ref{sec:NSBGR}).
This was first proposed for the I-Love-Q relations \citep{Yagi:2013awa,Yagi:2013bca} among certain dimensionless combinations of the moment of inertia $I$, electric quadrupolar LN, and spin-induced quadrupole moments $Q$. Thus, in addition to a measurement of the LN, a test of gravity based on the different I-Love-Q relations beyond GR would require an independent measurement of either $I$ or $Q$. Alternatively, one could use quasi-universal relations among different tidal quantities \citep{Yagi:2013sva,Berti:2024moe}. This requires measuring the quadrupolar magnetic LN or the octupolar electric LN of a given NS, both of which are challenging as they enter at high PN order. Such a measurement should be possible with next-generation instruments such as ET and CE \citep{JimenezForteza:2018rwr,Castro:2022mpw}.
Finally, quasi-universal relations also exist 
between tidal binary parameters and help reduce degeneracies in PN waveforms \citep{Yagi:2016qmr} (see~\ref{sec:PN}).

\subsubsection{Tidal heating in theories beyond General Relativity}
%%%%
Tests of GR based on different tidal heating in modified BH solutions have been so far developed only for braneworld theories. Here we use the results of~\ref{heatingBH} and present possible constraints on the tidal charge parameter $q$. For this purpose, our focus will be on EMRIs and we introduce a parameter $\mathtt{H}$ \citep{Chakraborty:2021gdf}:
%%%%%%%%%%%%%%%%%%%%%%%%%%%%%%%%%%%%%%%%%%
\begin{equation}\label{qconstraint}
1+\mathtt{H}\equiv \frac{\dot{M}_{\rm BH}^{\rm brane}}{\dot{M}_{\rm BH}^{\rm GR}}=\frac{1+3\chi^{2}+q(2+3\chi^{2}+q)}{1+3\chi^{2}}\,,
\end{equation}
%%%%%%%%%%%%%%%%%%%%%%%%%%%%%%%%%%%%%%%%%%
where $\dot{M}_{\rm BH}^{\rm brane}$ and $\dot{M}_{\rm BH}^{\rm GR}$ are respectively the rate of increase of mass for a braneworld BH and for a Kerr BH, due to tidal heating.
The constraints discussed in~\ref{testcompact} can be converted into bounds on the above quantity $\mathtt{H}$, which in turn provide a constraint on the tidal charge $q$. This is achieved by noticing the similarity of $\mathtt{H}$ with the absolute reflectivity of an ECO, on which there exist a strong bound from \citet{Datta:2019epe, Datta:2019euh,Maggio:2021uge}, which tells us that $\mathtt{H}\lesssim 10^{-5}$. This in turn put constraints on the tidal charge $q$, depending on the spin of the primary BH in an EMRI. As an example, considering the primary supermassive BH to have the dimensionless spin parameter $\chi=0.8$ and the SNR of the EMRI as detected by LISA being $\rho_{\rm SNR}=20$, we have a mismatch of $10^{-3}$ for $\mathtt{H}\lesssim 10^{-5}$. In this case,~\ref{qconstraint} leads to $q\lesssim 10^{-6}$, a very stringent constraint on the braneworld scenario \citep{Chakraborty:2021gdf}. 

For alternative theories of gravity, one must first derive the modified Teukolsky equation and compute the energy flux at the horizon to obtain the corresponding tidal heating (see \citealt{Cano:2025zyk, Li:2022pcy, Cano:2023tmv} for an example of a modified Teukolsky equation). By subsequently following the method outlined here, it is then possible to place constraints on the parameter(s) of the modified theory.

%%%%%%%%%%%%%%%%%%%%%%%%%%%%%%%%%%%%%%%%%%%%%%%%%%%%%%%%%%%%%%%%%%%%%%%%%%%%%%
%%%%%%%%%%%%%%%%%%%%%%%%%%%%%%%%%%%%%%%%%%%%%%%%%%%%%%%%%%%%%%%%%%%%%%%%%%%%%%
%%%%%%%%%%%%%%%%%%%%%%%%%%%%%%%%%%%%%%%%%%%%%%%%%%%%%%%%%%%%%%%%%%%%%%%%%%%%%%
\section{Discussion and outlook} \label{sec:conclusion}

Over the past two decades, our understanding of the tidal response of compact objects has undergone a profound transformation. What was once a relatively specialized theoretical subject has evolved into a broad and vibrant research area at the intersection of gravity, nuclear physics, high-energy physics, and GW astronomy. Crucially, theoretical predictions from a wide range of models can now be confronted with observations, as GW measurements provide direct access to tidal effects in the strong-field regime. Looking ahead, the advent of next-generation detectors further motivates the development of increasingly accurate and comprehensive theoretical frameworks.

We hope that this review has conveyed the breadth of this subject, its multifaceted implications across several areas of fundamental physics, and the sense that particularly exciting opportunities lie ahead.

We conclude with a necessarily incomplete and biased list of open problems that we believe are especially promising directions for future research:

\begin{itemize}
    \item \emph{LNs and no-hair theorems}:
    The vanishing of static bosonic LNs for four dimensional classical BHs in vacuum GR has revealed a novel and subtle realization of no-hair theorems, extending them beyond stationary vacuum properties to tidal responses. Understanding the fundamental origin of this result, its robustness under dynamical, quantum, or environmental effects, and its relation to horizon microphysics and effective descriptions remains an open problem. Clarifying whether, and in what sense, deviations from zero LNs provide a genuine violation of BH no-hair theorems is of both conceptual and observational interest.
    \item \emph{Dynamical LNs}:
    While significant progress has been made in recent years, a full computation of dynamical tidal effects in BH binaries is still missing, especially for NS binaries. This would eventually clarify the physical interpretation of the dynamical LNs, their gauge invariance, and relation to EFT descriptions.
    \item \emph{Modeling Love numbers in EMRIs}: Tidal effects in EMRIs have been mostly considered within a PN framework, which is known to converge very poorly in the small mass-ratio limit. A rigorous computation of tidal effects in EMRIs, as well as a consistent incorporation of tidal couplings within gravitational self-force computations, is still missing and will be essential for high-precision EMRI modeling.
    \item \emph{Tidal response in eccentric and generic orbits}: The impact of eccentricity and precession on dynamical tides, mode excitation, and dissipative effects remains marginally explored, especially in strong-field and EMRI regimes.
    \item \emph{Tidal response of rapidly rotating compact objects}: A complete treatment of tidal susceptibilities for rapidly spinning NSs and beyond slow-rotation expansions remains an open challenge, particularly regarding mode couplings and spin-tide interactions.
    \item \emph{Nonlinear tidal response in the strong-field regime}: A fully nonlinear characterization of tidal deformations close to merger, and their extraction from numerical relativity simulations, remains largely unexplored. Clarifying whether meaningful, gauge-invariant tidal observables can be defined beyond perturbation theory is an open problem of both conceptual and phenomenological relevance.
    \item \emph{Fermionic LNs of BHs}: The recent discovery of nonvanishing static fermionic LNs for BHs in GR opens the door to several unexplored directions, including their dynamical response, dissipative properties, and potential observational signatures.
    \item \emph{Convergence at large $\ell$}: 
    The bosonic dissipation numbers (and the fermionic LNs) of a BH grow exponentially in the large $\ell=m$ limit. Presumably, this growth is quenched by the prefactors entering the GW signal in each given term in a PN expansion. However, it would be important to study the convergence of this series in the large-$\ell$ limit and clarify the regime of validity of effective descriptions.
    \item \emph{Full tidal response of a NS}: effects such as mode resonances, dynamical LNs, dissipation, rotational and quadratic LNs have been separately computed, but a full waveform model incorporating all of them is currently missing. Their full impact on GW phasing, possible degeneracies, and their role in high SNR observations remain to be systematically explored.
    \item \emph{Tidal heating and dissipation in exotic compact objects}:
    The role of dissipation and tidal heating in horizonless compact objects is highly model dependent and not yet fully characterized. A better understanding of these effects is needed to assess their discriminating power between BH and BH mimickers.
    \item \emph{Environmental modifications of tidal observables}: The extent to which surrounding realistic matter profiles, dark sectors, or plasma environments modify effective tidal response functions remains insufficiently understood, especially in the context of precision GW tests.
    \item \emph{Quantum corrections to tidal response}: Whether semiclassical or quantum-gravity effects induce parametrically small but universal tidal susceptibilities in BHs, and whether these are observable in principle, remains an important open question.
    \item \emph{LNs as probes of fundamental physics}:
    More broadly, the extent to which tidal observables can be used to test modified theories of gravity, constrain new degrees of freedom, or probe environmental effects remains an active and open area of research. 
    \item \emph{Observational prospects with future detectors}:
    Finally, quantifying the measurability of dynamical, rotational, magnetic, and higher-order tidal effects with future ground- and space-based detectors (such as ET and LISA) is essential to fully exploit their scientific potential.
    Overall, current templates including tidal effects should be significantly improved in order to avoid waveform systematics and extract reliable information from GW signals observed by next generation detectors.
\end{itemize}
%%%%%%%%%%%%%%%%%%%%%%%%%%%%%%%%%%%%%%%%%%%%%%%%%%%%%%%%%%%%%%%%%%%%%%%%%%%%%%
We hope that this review will serve as a useful reference and stimulate further theoretical and observational developments in these and other promising directions.
%%%%%%%%%%%%%%%%%%%%%%%%%%%%%%%%%%%%%%%%%%%%%%%%%%%%%%%%%%%%%%%%%%%%%%%%%%%%%%
%%%%%%%%%%%%%%%%%%%%%%%%%%%%%%%%%%%%%%%%%%%%%%%%%%%%%%%%%%%%%%%%%%%%%%%%%%%%%%
\section*{Acknowledgments}
We are indebted to 
%%%%
% Nils Andersson,
% Emanuele Berti,
% Donato Bini,
Vitor Cardoso,
% Thibault Damour,
Valerio De Luca,
% Valeria Ferrari,
Rossella Gamba,
Leonardo Gualtieri,
% Tanja Hinderer,
% Lam Hui,
Mikhail Ivanov,
Takuya Katagiri,
% Andrea Maselli,
Alessandro Nagar,
% Francesco Pannarale,
% Julio Parra Martinez,
Pantelis Pnigouras,
Eric Poisson,
Rafael Porto,
% Ira Rothstein,
% Massimiliano Maria Riva,
% Luca Santoni,
% Nikola Savic,
Jan Steinhoff,
% Filippo Vernizzi,
Kent Yagi,
Nicolas Yunes,
Zihan Zhou,
Sayak Datta, 
Rajendra Prasad Bhatt, 
Rajes Ghosh, 
Avijit Chowdhury
%%%%
for providing feedback, useful references, for discussions, or for suggesting corrections to an earlier version of this manuscript.
We also thank many participants of the  \href{https://sites.google.com/view/et-ecr-workshop-2026-sapienza}{Einstein Telescope Science Workshop for Early Career Researchers} (Sapienza University of Rome - February 18--20, 2026) for providing relevant references. This research was supported in part by the International Centre for Theoretical Sciences (ICTS) for participating in the program - The Future of Gravitational-Wave Astronomy 2025 (code: ICTS/FGWA2025/10).
SC acknowledges the support by MATRICS (MTR/2023/000049) and Core Research Grants (CRG/2023/000934) from SERB, ANRF, Government of India. SC also thanks the local hospitality at ICTS and IUCAA through the associateship program, where a part of this work was done. 
PP acknowledges support by the MUR FIS2 Advanced Grant ET-NOW (CUP:~B53C25001080001) and by the INFN TEONGRAV initiative. 
Some numerical computations have been performed at the Vera and CHRONOS clusters supported by the Italian Ministry of Research and by Sapienza University of Rome.

%%%%%%%%%%%%%%%%%%%%%%%%%%%%%%%%%%%%%%%%%%%%%%%%%%%%%%%%%%%%%%%%%%%%%%%%%%%%%%
%%%%%%%%%%%%%%%%%%%%%%%%%%%%%%%%%%%%%%%%%%%%%%%%%%%%%%%%%%%%%%%%%%%%%%%%%%%%%%
%%%%%%%%%%%%%%%%%%%%%%%%%%%%%%%%%%%%%%%%%%%%%%%%%%%%%%%%%%%%%%%%%%%%%%%%%%%%%%
\appendix
\labelformat{section}{Appendix~#1} 
%%%%%%%%%%%%%%%%%%%%%%%%%%%%%%%%%%%%%%%%%%%%%%%%%%%%%%%%%%%%%%%%%%%%%%%%%%%%%%
%%%%%%%%%%%%%%%%%%%%%%%%%%%%%%%%%%%%%%%%%%%%%%%%%%%%%%%%%%%%%%%%%%%%%%%%%%%%%%
%%%%%%%%%%%%%%%%%%%%%%%%%%%%%%%%%%%%%%%%%%%%%%%%%%%%%%%%%%%%%%%%%%%%%%%%%%%%%%

\section{Useful mathematical identities} \label{app:identities}
In this appendix the acronym DLMF refers to~\citet{DLMF}.

\begin{itemize}

\item Mirror formula for Gamma function: 
%%%%%%%%%%%%%%%%%%%%%%%%%%%%%%%%%%%%%%%%%%%%%%%%%%%%%%%%%%%%%%%%%%%%%%%%%%%%%%%%%
\begin{align}\label{mirrorgamma}
\Gamma(z)\Gamma(1-z)=\frac{\pi}{\sin(\pi z)}~.
\end{align}
%%%%%%%%%%%%%%%%%%%%%%%%%%%%%%%%%%%%%%%%%%%%%%%%%%%%%%%%%%%%%%%%%%%%%%%%%%%%%%%%%

\item Connection between the associated Legendre polynomials and the hypergeometric functions:
%%%%%%%%%%%%%%%%%%%%%%%%%%%%%%%%%%%%%%%%%%%%%%%%%%%%%%%%%%%%%%%%%%%%%%%%%%%%%%%%%
\begin{align}
P_{\nu}^{\mu}(z)&=\frac{1}{\Gamma(1-\mu)}\left(\frac{z+1}{z-1}\right)^{\mu/2}\,_{2}F_{1}\left(\nu+1,-\nu;1-\mu;\frac{1-z}{2}\right)~,
\label{legPHyp}
\\
P_{\nu}^{m}(z)&=\frac{\Gamma(\nu+m+1)}{2^{m}\Gamma(\nu-m+1)}\frac{\left(z^{2}-1\right)^{m/2}}{\Gamma(m+1)}\,_{2}F_{1}\left(\nu+m+1,m-\nu;m+1;\frac{1-z}{2}\right)~,
\label{legPintHyp}
\\
Q_{\nu}^{\mu}(z)&=\frac{e^{\pi i\mu}2^{\nu}\Gamma(\mu+\nu+1)\Gamma(\nu+1)(z+1)^{\mu/2}}{(z-1)^{1+\nu+(\mu/2)}\Gamma(2+2\nu)}\,_{2}F_{1}\left(\nu+1,\nu+\mu+1;2+2\nu;\frac{2}{1-z} \right)~.
\label{legQHyp}
\end{align}
%%%%%%%%%%%%%%%%%%%%%%%%%%%%%%%%%%%%%%%%%%%%%%%%%%%%%%%%%%%%%%%%%%%%%%%%%%%%%%%%%
The first relation arises from [DLMF 14.3.6], the second one follows from [DLMF 14.3.8], while the third relation arises from [DLMF 14.3.10] and [DLMF 14.3.19]. 

\item Connection between hypergeometric functions with inverse arguments [DLMF 15.8.2]
%%%%%%%%%%%%%%%%%%%%%%%%%%%%%%%%%%%%%%%%%%%%%%%%%%%%%%%%%%%%%%%%%%%%%%%%%%%%%%%%%
\begin{align}\label{hyp_z_1byz}
\,_{2}F_{1}(a,b;c;z)&=\frac{\pi \Gamma(c)}{\sin [\pi(b-a)]}\Big[\frac{(-z)^{-a}}{\Gamma(b)\Gamma(c-a)\Gamma(a-b+1)}\,_{2}F_{1}\left(a,a-c+1;a-b+1;\frac{1}{z}\right)
\nonumber
\\
&-\frac{(-z)^{-b}}{\Gamma(a)\Gamma(c-b)\Gamma(b-a+1)}\,_{2}F_{1}\left(b,b-c+1;b-a+1;\frac{1}{z}\right)\Big]~,
\end{align}
%%%%%%%%%%%%%%%%%%%%%%%%%%%%%%%%%%%%%%%%%%%%%%%%%%%%%%%%%%%%%%%%%%%%%%%%%%%%%%%%%
where $(b-a)\notin \mathbb{Z}$. The above can also be written down as, 
%%%%%%%%%%%%%%%%%%%%%%%%%%%%%%%%%%%%%%%%%%%%%%%%%%%%%%%%%%%%%%%%%%%%%%%%%%%%%%%%%
\begin{align}\label{hyp_z_2byz}
\,_{2}F_{1}(a,b;c;z)&=\Big[\frac{(-z)^{-a}\Gamma(c)\Gamma(b-a)}{\Gamma(b)\Gamma(c-a)}\,_{2}F_{1}\left(a,a-c+1;a-b+1;\frac{1}{z}\right)
\nonumber
\\
&+\frac{(-z)^{-b}\Gamma(c)\Gamma(a-b)}{\Gamma(a)\Gamma(c-b)}\,_{2}F_{1}\left(b,b-c+1;b-a+1;\frac{1}{z}\right)\Big]~,
\end{align}
%%%%%%%%%%%%%%%%%%%%%%%%%%%%%%%%%%%%%%%%%%%%%%%%%%%%%%%%%%%%%%%%%%%%%%%%%%%%%%%%%

\item Limits of hypergeometric functions:
%%%%%%%%%%%%%%%%%%%%%%%%%%%%%%%%%%%%%%%%%%%%%%%%%%%%%%%%%%%%%%%%%%%%%%%%%%%%%%%%%
\begin{align}
\lim_{z\to 0}\,_{2}F_{1}(a,b;c;z)&=\sum_{n=0}^{\infty}\frac{\Gamma(a+n)\Gamma(b+n)\Gamma(c)}{\Gamma(a)\Gamma(b)\Gamma(c+n)}\frac{z^{n}}{\Gamma(n+1)}\approx 1+\mathcal{O}(z)\,,
\label{hypzero}
\\
\lim_{z\to \infty}\,_{2}F_{1}(a,b;c;z)&\propto\frac{(-z)^{-a}}{\Gamma(b)\Gamma(c-a)\Gamma(a-b+1)}-\frac{(-z)^{-b}}{\Gamma(a)\Gamma(c-b)\Gamma(b-a+1)}~,
\label{hypasymp}
\end{align}
%%%%%%%%%%%%%%%%%%%%%%%%%%%%%%%%%%%%%%%%%%%%%%%%%%%%%%%%%%%%%%%%%%%%%%%%%%%%%%%%%
The first identity holds for arbitrary choices of the arguments (a,b,c). While the second identity assumes that $(b-a)\notin \mathbb{Z}$.

\item Hypergeometric function $\,_{2}F_{1}(a,b;c;x)$ with $c=a+b+m$, where $m\in \mathbb{Z}^{+}$, can be expressed in the following manner [DLMF 15.8.10]:
%%%%%%%%%%%%%%%%%%%%%%%%%%%%%%%%%%%%%%%%%%%%%%%%%%%%%%%%%%%%%%%%%%%%%%%%%%%%%%%%%
\begin{align}\label{hyplog}
\,_{2}&F_{1}(a,b;a+b+m;z)=\frac{\Gamma(a+b+m)}{\Gamma(a+m)\Gamma(b+m)}\sum_{k=0}^{m-1}\frac{\Gamma(a+k)\Gamma(b+k)}{\Gamma(a)\Gamma(b)}\frac{(z-1)^{k}}{k!}
\nonumber
\\
&-(z-1)^{m}\frac{\Gamma(a+b+m)}{\Gamma(a)\Gamma(b)}\sum_{k=0}^{\infty}\frac{\Gamma(a+m+k)\Gamma(b+m+k)}{\Gamma(a+m)\Gamma(b+m)}\frac{(1-z)^{k}}{k!(k+m)!}\times \Big[\ln(1-z)
\nonumber
\\
&\qquad \qquad -\psi(1+k)-\psi(k+m+1)+\psi(a+k+m)+\psi(b+k+m)\Big]~.
\end{align}
%%%%%%%%%%%%%%%%%%%%%%%%%%%%%%%%%%%%%%%%%%%%%%%%%%%%%%%%%%%%%%%%%%%%%%%%%%%%%%%%%
A special case of the above identity corresponds to $m=0$, i.e., when $c=a+b$. In this case, the first summation in~\ref{hyplog} vanishes identically, while in the next, at the leading order, only the $\ln(1-z)$ term survives, such that [DLMF 15.4.21]
%%%%%%%%%%%%%%%%%%%%%%%%%%%%%%%%%%%%%%%%%%%%%%%%%%%%%%%%%%%%%%%%%%%%%%%%%%%%%%%%%
\begin{align}\label{hyplog2}
\lim_{z\to 1^{-}}\,_{2}F_{1}(a,b;a+b;z)=-\frac{\Gamma(a+b)}{\Gamma(a)\Gamma(b)}\ln(1-z)~.
\end{align}
%%%%%%%%%%%%%%%%%%%%%%%%%%%%%%%%%%%%%%%%%%%%%%%%%%%%%%%%%%%%%%%%%%%%%%%%%%%%%%%%%

\item Hypergeometric function with argument $(1-z)$, in terms of other hypergeometric functions with argument $z$ [DLMF 15.10.17]:
%%%%%%%%%%%%%%%%%%%%%%%%%%%%%%%%%%%%%%%%%%%%%%%%%%%%%%%%%%%%%%%%%%%%%%%%%%%%%%%%%
\begin{align}\label{hypz1mz}
\,_{2}F_{1}(a,&b;a+b-c+1;1-z)=\frac{\Gamma(1-c)\Gamma(a+b-c+1)}{\Gamma(a-c+1)\Gamma(b-c+1)}\,_{2}F_{1}(a,b;c;z)
\nonumber
\\
&+\frac{\Gamma(c-1)\Gamma(a+b-c+1)}{\Gamma(a)\Gamma(b)}z^{1-c}\,_{2}F_{1}(a-c+1,b-c+1;2-c;z)~.
\end{align}
%%%%%%%%%%%%%%%%%%%%%%%%%%%%%%%%%%%%%%%%%%%%%%%%%%%%%%%%%%%%%%%%%%%%%%%%%%%%%%%%%

\item If in the hypergeometric differential equation, $(a,b)\notin \mathbb{Z}$, but $c\in \mathbb{Z}$, then along with $\,_{2}F_{1}(a,b;n;z)$, the other independent solution will be [DLMF 15.10.8]:
%%%%%%%%%%%%%%%%%%%%%%%%%%%%%%%%%%%%%%%%%%%%%%%%%%%%%%%%%%%%%%%%%%%%%%%%%%%%%%%%%
\begin{align}\label{hypcinteger}
\,_{2}&F_{1}(a,b;a+b-n+1;1-z)=\,_{2}F_{1}(a,b;n;z)\ln z
\nonumber
\\
&-\sum_{k=1}^{n-1}\frac{(n-1)!(k-1)!\Gamma(1-a)\Gamma(1-b)}{(n-k-1)!\Gamma(1-a+k)\Gamma(1-b+k)}(-z)^{-k}
+\sum_{k=0}^{\infty}\frac{\Gamma(a+k)\Gamma(b+k)\Gamma(n)}{\Gamma(a)\Gamma(b)\Gamma(n+k)k!}z^{k}
\nonumber
\\
&\qquad\times \Big[\psi(a+k)+\psi(b+k)-\psi(1+k)-\psi(n+k)\Big]~.
\end{align}
%%%%%%%%%%%%%%%%%%%%%%%%%%%%%%%%%%%%%%%%%%%%%%%%%%%%%%%%%%%%%%%%%%%%%%%%%%%%%%%%%

\item Hypergeometric function with $a=-m$, where $m\in \mathbb{Z}$, but $c$ is not a negative integer [DLMF 15.2.4]: 
%%%%%%%%%%%%%%%%%%%%%%%%%%%%%%%%%%%%%%%%%%%%%%%%%%%%%%%%%%%%%%%%%%%%%%%%%%%%%%%%%
\begin{align}\label{hypnegativea}
\,_{2}F_{1}(-m,b,c,z)=\sum_{n=0}^{m}(-1)^{n}\,^{m}C_{n}\frac{\Gamma(n+b)\Gamma(c)}{\Gamma(b)\Gamma(n+c)}z^{n}~.
\end{align}
%%%%%%%%%%%%%%%%%%%%%%%%%%%%%%%%%%%%%%%%%%%%%%%%%%%%%%%%%%%%%%%%%%%%%%%%%%%%%%%%%
An immediate corollary of the above equation being: $\,_{2}F_{1}(0,b,c,z)=1$. 

\item Hypergeometric function with $z$ as the argument is related to hypergeometric function with $z/(z-1)$ as argument [DLMF 15.8.1]:
%%%%%%%%%%%%%%%%%%%%%%%%%%%%%%%%%%%%%%%%%%%%%%%%%%%%%%%%%%%%%%%%%%%%%%%%%%%%%%%%%
\begin{align}\label{hypztozbyzm1}
\,_{2}F_{1}(a,b,c,z)=(1-z)^{-a}\,_{2}F_{1}\left(a,c-b;c;\frac{z}{z-1}\right)~.
\end{align}
%%%%%%%%%%%%%%%%%%%%%%%%%%%%%%%%%%%%%%%%%%%%%%%%%%%%%%%%%%%%%%%%%%%%%%%%%%%%%%%%%

\item Hypergeometric function with same variable, but different constant arguments: 
%%%%%%%%%%%%%%%%%%%%%%%%%%%%%%%%%%%%%%%%%%%%%%%%%%%%%%%%%%%%%%%%%%%%%%%%%%%%%%%%%
\begin{align}\label{hypdiffarg}
\,_{2}F_{1}(a,b;c;z)=(1-z)^{c-a-b}\,_{2}F_{1}(c-a,c-b;c;z)~.
\end{align}
%%%%%%%%%%%%%%%%%%%%%%%%%%%%%%%%%%%%%%%%%%%%%%%%%%%%%%%%%%%%%%%%%%%%%%%%%%%%%%%%%

\end{itemize}
\vskip 2mm

%%%%%%%%%%%%%%%%%%%%%%%%%%%%%%%%%%%%%%%%%%%%%%%%%%%%%%%%%%%%%%%%%%%%%%%%%%%%%%
%%%%%%%%%%%%%%%%%%%%%%%%%%%%%%%%%%%%%%%%%%%%%%%%%%%%%%%%%%%%%%%%%%%%%%%%%%%%%%
%%%%%%%%%%%%%%%%%%%%%%%%%%%%%%%%%%%%%%%%%%%%%%%%%%%%%%%%%%%%%%%%%%%%%%%%%%%%%%
%\section{Scattering amplitudes using Mano-Suzuki-Takasugi method} \label{App:MST}

%In this appendix we will briefly discuss the MST method for obtaining the scattering amplitude, as discussed in the main text. The starting point of the MST formalism is to write down the purely ingoing wave solution of the radial Teukolsky equation in the Boyer-Lindquist coordinate, which reads, 
%%%%%%%%%%%%%%%%%%%%%%%%%%%%%%%%%%%%%%%%%%%%%%%%%%%%%%%%%%%%%%%%%%%%%%%%%%%%%%%%%
%\begin{align}
%R_{\rm in}&=e^{2iM\sqrt{1-\chi^{2}}\,\omega x}\left(-x\right)^{-s-\frac{2iMr_{+}\bar{\omega}}{r_{+}-r_{-}}}\left(1-x\right)^{i\frac{(2M\omega-\tau)}{2}}
%\nonumber
%\\
%&\qquad \times\sum_{n=-\infty}^{\infty}f_{n}^{\hat{\ell}}\,_{2}F_{1}\left(n+1+\hat{\ell}-i\tau,-n-\hat{\ell}-i\tau;1-s-2iM\omega;x\right)
%\end{align}
%%%%%%%%%%%%%%%%%%%%%%%%%%%%%%%%%%%%%%%%%%%%%%%%%%%%%%%%%%%%%%%%%%%%%%%%%%%%%%%%%

%%%%%%%%%%%%%%%%%%%%%%%%%%%%%%%%%%%%%%%%%%%%%%%%%%%%%%%%%%%%%%%%%%%%%%%%%%%%%%%
%%%%%%%%%%%%%%%%%%%%%%%%%%%%%%%%%%%%%%%%%%%%%%%%%%%%%%%%%%%%%%%%%%%%%%%%%%%%%%%
%%%%%%%%%%%%%%%%%%%%%%%%%%%%%%%%%%%%%%%%%%%%%%%%%%%%%%%%%%%%%%%%%%%%%%%%%%%%%%%
\section{Ratio of out and in scattering amplitudes}\label{App:NearZoneRatioMST}

In this appendix we will determine the ratio of outgoing to the ingoing wave using the MST formalism. The main job is to reduce~\ref{outbyinscatt} to a simple form by considering the low frequency expansion. For this purpose, we introduce a few quantities in the notation of the MST formalism and relate them to our conventional definitions \citep{Mano:1996gn, Mano:1996vt, Sasaki:2003xr}:
%%%%%%%%%%%%%%%%%%%%%%%%%%%%%%%%%%%%%%%%%%
\begin{align}
\epsilon&=2M\omega\,;
\quad
q=\frac{a}{M}=\chi\,;
\quad
\kappa=\sqrt{1-\chi^{2}}=\frac{r_{+}-r_{-}}{2M}\,,
\\
\tau&=\frac{\epsilon-mq}{\kappa}=\frac{2M}{\sqrt{1-\chi^{2}}}\left(\omega-\frac{ma}{2M^{2}}\right)=\frac{2M\bar{\omega}_{\pm}}{\sqrt{1-\chi^{2}}}\mp2Mm\Omega_{\pm}
\\
\bar{\omega}_{\pm}&=\omega-m\Omega_{\pm}\,;
\quad
\Omega_{\pm}=\frac{a}{2Mr_{\pm}}\,;
\quad
\epsilon_{\pm}=\frac{\epsilon\pm \tau}{2}=\pm\frac{2Mr_{\pm}}{r_{+}-r_{-}}\bar{\omega}_{\pm}
\end{align}
%%%%%%%%%%%%%%%%%%%%%%%%%%%%%%%%%%%%%%%%%%
Then, from \citet{Sasaki:2003xr}, we express the quantity $K_{\nu;-s}$ as (we have used $\nu=\hat{\ell}$ in the main text), 
%%%%%%%%%%%%%%%%%%%%%%%%%%%%%%%%%%%%%%%%%%
\begin{align}
K_{\nu;-s}&=\frac{e^{i\epsilon \kappa}(2\epsilon\kappa)^{-s-\nu-r}2^{s}i^{r}\Gamma(1+s-2i\epsilon_{+})\Gamma(r+2\nu+2)}{\Gamma(r+\nu+1+s+i\epsilon)\Gamma(r+\nu+1+i\tau)\Gamma(r+\nu+1-s+i\epsilon)}
\nonumber
\\
&\times \sum_{n=r}^{\infty}(-1)^{n}\frac{\Gamma(n+r+2\nu+1)\Gamma(n+\nu+1-s+i\epsilon)\Gamma(n+\nu+1+i\tau)}{\Gamma(n+\nu+1+s-i\epsilon)\Gamma(n+\nu+1-i\tau)(n-r)!}f_{n;-s}^{\nu}
\nonumber
\\
&\times \left(\sum_{n=-\infty}^{r}\frac{(-1)^{n}(\nu+1-s-i\epsilon)_{n}}{(r-n)!(r+2\nu+2)_{n}(\nu+1+s+i\epsilon)_{n}}f_{n;-s}^{\nu}\right)^{-1}\,,
\end{align}
%%%%%%%%%%%%%%%%%%%%%%%%%%%%%%%%%%%%%%%%%%
and, 
%%%%%%%%%%%%%%%%%%%%%%%%%%%%%%%%%%%%%%%%%%
\begin{align}
K_{-(\nu+1);-s}&=\frac{e^{i\epsilon \kappa}(2\epsilon\kappa)^{-s+\nu+1-r}2^{s}i^{r}\Gamma(1+s-2i\epsilon_{+})\Gamma(r-2\nu)}{\Gamma(r-\nu+s+i\epsilon)\Gamma(r-\nu+i\tau)\Gamma(r-\nu-s+i\epsilon)}
\nonumber
\\
&\times \sum_{n=r}^{\infty}(-1)^{n}\frac{\Gamma(n+r-2\nu-1)\Gamma(n-\nu-s+i\epsilon)\Gamma(n-\nu+i\tau)}{\Gamma(n-\nu+s-i\epsilon)\Gamma(n-\nu-i\tau)(n-r)!}f_{n;-s}^{-\nu-1}
\nonumber
\\
&\times \left(\sum_{n=-\infty}^{r}\frac{(-1)^{n}(-\nu-s-i\epsilon)_{n}}{(r-n)!(r-2\nu)_{n}(-\nu+s+i\epsilon)_{n}}f_{n;-s}^{-\nu-1}\right)^{-1}\,.
\end{align}
%%%%%%%%%%%%%%%%%%%%%%%%%%%%%%%%%%%%%%%%%%
The above quantities depend upon $r$, which fixes the order of the summation. It was argued in \citet{Ivanov:2022qqt, Ivanov:2022hlo} that the final result does not depend on the choice of $r$ and hence we choose $r=0$. Further, since our interest is in the static limit, we consider only a single term in both the sums above, in particular we choose the zeroth term. This yields,
%%%%%%%%%%%%%%%%%%%%%%%%%%%%%%%%%%%%%%%%%%
\begin{align}
K_{-(\nu+1);-s}&\approx\frac{e^{i\epsilon \kappa}(2\epsilon\kappa)^{-s+\nu+1}2^{s}\Gamma(1+s-2i\epsilon_{+})\Gamma(-2\nu)\Gamma(-2\nu-1)}{\Gamma(-\nu+s+i\epsilon)\Gamma(-\nu+s-i\epsilon)\Gamma(-\nu-i\tau)}\,,
\end{align}
%%%%%%%%%%%%%%%%%%%%%%%%%%%%%%%%%%%%%%%%%%
as well as, 
%%%%%%%%%%%%%%%%%%%%%%%%%%%%%%%%%%%%%%%%%%
\begin{align}
K_{\nu;-s}&=\frac{e^{i\epsilon \kappa}(2\epsilon\kappa)^{-s-\nu}2^{s}\Gamma(1+s-2i\epsilon_{+})\Gamma(2\nu+2)\Gamma(2\nu+1)}{\Gamma(\nu+1+s+i\epsilon)\Gamma(\nu+1+s-i\epsilon)\Gamma(\nu+1-i\tau)}\,.
\end{align}
%%%%%%%%%%%%%%%%%%%%%%%%%%%%%%%%%%%%%%%%%%
Given the above simplified versions for $K_{-(\nu+1);-s}$ and $K_{\nu;-s}$, it is straightforward to compute their ratio, which yields,
%%%%%%%%%%%%%%%%%%%%%%%%%%%%%%%%%%%%%%%%%%
\begin{align}
&\frac{K_{-(\nu+1);-s}}{K_{\nu;-s}}
\nonumber
\\
&=(2\epsilon\kappa)^{2\nu+1}\frac{\Gamma(-2\nu-1)\Gamma(-2\nu)}{\Gamma(2\nu+2)\Gamma(2\nu+1)}\times \frac{\Gamma(\nu+1-i\tau)}{\Gamma(-\nu-i\tau)}\times \frac{\Gamma(\nu+1+s-i\epsilon)\Gamma(\nu+1+s+i\epsilon)}{\Gamma(-\nu+s-i\epsilon)\Gamma(-\nu+s+i\epsilon)}
\nonumber
\\
&=\left[2\omega(r_{+}-r_{-})\right]^{2\nu+1}\frac{1}{\Gamma(2\nu+2)^{2}\Gamma(2\nu+1)^{2}}\left(\frac{\pi^{2}}{\sin^{2}(2\pi \nu)}\right)\times \Gamma(\nu+1-i\tau)\Gamma(\nu+1+i\tau)
\nonumber
\\
&\times\frac{\sin[\pi(\nu+i\tau)]}{\pi}\times\Gamma(\nu+1+s-i\epsilon)\Gamma(\nu+1+s+i\epsilon)\Gamma(\nu+1-s+i\epsilon)
\Gamma(\nu+1-s-i\epsilon)
\nonumber
\\
&\times \frac{\sin[\pi(\nu-s+i\epsilon)]\sin[\pi(\nu-s-i\epsilon)]}{\pi^{2}}\,.
\end{align}
%%%%%%%%%%%%%%%%%%%%%%%%%%%%%%%%%%%%%%%%%%
Note that, 
%%%%%%%%%%%%%%%%%%%%%%%%%%%%%%%%%%%%%%%%%%
\begin{align}
\sin[\pi(\nu-s+i\epsilon)]\sin[\pi(\nu-s-i\epsilon)]&=\left\{\sin[\pi(\nu-s)]+i\pi \epsilon\cos[\pi(\nu-s)]\right\}
\nonumber
\\
&\qquad \times\left\{\sin[\pi(\nu-s)]-i\pi \epsilon\cos[\pi(\nu-s)]\right\}
\nonumber
\\
&=\sin^{2}[\pi(\nu-s)]+\mathcal{O}(\epsilon^{2})=\sin^{2}(\pi \nu)+\mathcal{O}(\epsilon^{2})\,.
\end{align}
%%%%%%%%%%%%%%%%%%%%%%%%%%%%%%%%%%%%%%%%%%
Therefore, it follows that,
%%%%%%%%%%%%%%%%%%%%%%%%%%%%%%%%%%%%%%%%%%
\begin{align}
\frac{\sin[\pi(\nu-s+i\epsilon)]\sin[\pi(\nu-s-i\epsilon)]}{\sin^{2}(2\pi \nu)}=\frac{1}{4}\,.
\end{align}
%%%%%%%%%%%%%%%%%%%%%%%%%%%%%%%%%%%%%%%%%%
Note that this result holds for arbitrary $\nu$ and hence is true for any frequencies. Therefore, we obtain the ratio between $K_{-\nu-1}$ and $K_{\nu}$ to yield, 
%%%%%%%%%%%%%%%%%%%%%%%%%%%%%%%%%%%%%%%%%%
\begin{align}
&\frac{K_{-(\nu+1);-s}}{K_{\nu;-s}}=\left[2\omega(r_{+}-r_{-})\right]^{2\nu+1}\frac{1}{\Gamma(2\nu+2)^{2}\Gamma(2\nu+1)^{2}}\times \Gamma(\nu+1-i\tau)\Gamma(\nu+1+i\tau)
\nonumber
\\
&\times\frac{\sin[\pi(\nu+i\tau)]}{4\pi}\times\Gamma(\nu+1+s-i\epsilon)\Gamma(\nu+1+s+i\epsilon)\Gamma(\nu+1-s+i\epsilon)\Gamma(\nu+1-s-i\epsilon)\,.
\end{align}
%%%%%%%%%%%%%%%%%%%%%%%%%%%%%%%%%%%%%%%%%%
Noting that $\nu=\ell+\mathcal{O}(\epsilon^{2})$, we can replace $\nu\to \ell$ in the above expression, as we are interested in the zero frequency limit. Therefore, using the identity: 
%%%%%%%%%%%%%%%%%%%%%%%%%%%%%%%%%%%%%%%%%%
\begin{align}
|\Gamma(1+n+i\gamma)|=\frac{\pi \gamma}{\sinh(\pi \gamma)}\prod_{k=0}^{n}\left(k^{2}+\gamma^{2}\right)\,,
\end{align}
%%%%%%%%%%%%%%%%%%%%%%%%%%%%%%%%%%%%%%%%%%
the above ratio can be expressed as,
%%%%%%%%%%%%%%%%%%%%%%%%%%%%%%%%%%%%%%%%%%
\begin{align}\label{Kratio}
\frac{K_{-(\nu+1);-s}}{K_{\nu;-s}}&=\left[2\omega(r_{+}-r_{-})\right]^{2\ell+1}\frac{1}{\left[\left(2\ell+1\right)!\right]^{2}\left[\left(2\ell\right)!\right]^{2}}\times \frac{\pi \tau}{\sinh(\pi \tau)}\prod_{k=1}^{\ell}\left(k^{2}+\tau^{2}\right)
\nonumber
\\
&\times\frac{(-1)^{\ell}\sin[i\pi \tau]}{4\pi}\times\frac{\pi \epsilon}{\sinh(\pi \epsilon)}\left(\prod_{k=1}^{\ell+s}k^{2}\right)\,\frac{\pi \epsilon}{\sinh(\pi \epsilon)}\left(\prod_{k=1}^{\ell-s}k^{2}\right)\,
\nonumber
\\
&=(-1)^{\ell}\left[2\omega(r_{+}-r_{-})\right]^{2\ell+1}\frac{\left[\left(\ell+s\right)!\right]^{2}\left[\left(\ell-s\right)!\right]^{2}}{\left[\left(2\ell+1\right)!\right]^{2}\left[\left(2\ell\right)!\right]^{2}}\times \frac{i\tau}{4}\prod_{k=1}^{\ell}\left(k^{2}+\tau^{2}\right)
\end{align}
%%%%%%%%%%%%%%%%%%%%%%%%%%%%%%%%%%%%%%%%%%
where we have used the result, $\sin(i\theta)=i\sinh(\theta)$. In this limit, we further obtain, 
%%%%%%%%%%%%%%%%%%%%%%%%%%%%%%%%%%%%%%%%%%
\begin{align}
\frac{\sin[\pi(\nu-s+i\epsilon)]}{\sin[\pi(\nu+s-i\epsilon)]}&=\frac{\sin[\pi(\nu-s)]\left(1+\epsilon^{2}\right)+i\epsilon\cos[\pi(\nu-s)]\left(1+\epsilon^{2}\right)}{\sin[\pi(\nu+s)]\left(1+\epsilon^{2}\right)-i\epsilon\cos[\pi(\nu+s)]\left(1+\epsilon^{2}\right)}
\nonumber
\\
&\approx 1+i\epsilon \left\{\textrm{cot}[\pi(\nu-s)]+\textrm{cot}[\pi(\nu+s)]\right\}\,.
\end{align}
%%%%%%%%%%%%%%%%%%%%%%%%%%%%%%%%%%%%%%%%%%
Thus, in the zero frequency limit, we obtain the above ratio to be unity. This result, along with~\ref{Kratio}, we have used in the main text.  

%%%%%%%%%%%%%%%%%%%%%%%%%%%%%%%%%%%%%%%%%%%%%%%%%%%%%%%%%%%%%%%%%%%%%%%%%%%%%%%
%%%%%%%%%%%%%%%%%%%%%%%%%%%%%%%%%%%%%%%%%%%%%%%%%%%%%%%%%%%%%%%%%%%%%%%%%%%%%%%
%%%%%%%%%%%%%%%%%%%%%%%%%%%%%%%%%%%%%%%%%%%%%%%%%%%%%%%%%%%%%%%%%%%%%%%%%%%%%%%

\section{Surficial Love numbers for stars and black holes} \label{app:surficial}
The tidal LNs described in this work are defined through the asymptotic behavior of the gravitational field and the induced multipole moments of a compact object. 
An alternative and complementary characterization of tidal response is provided by the so-called \emph{surficial (or shape) LNs}, which quantify directly the deformation of the body's surface geometry under an external tidal field. 
This notion, originally rooted in geophysics, was formulated in a relativistic setting in \citet{Damour:2009va,Landry:2014jka, Prasad:2021dfr} (see also \citealt{PoissonWill}).

Consider a non-rotating, self-gravitating body of radius $R$ immersed in an external, static tidal field. 
In the unperturbed configuration, the surface $r=R$ is a round two-sphere with intrinsic metric
\begin{equation}
\Omega_{AB} = R^{2}\,\hat{\Omega}_{AB},
\end{equation}
where $\hat{\Omega}_{AB}$ is the metric on the unit sphere and capital Latin indices refer to angular coordinates. 
Under the action of a tidal perturbation, the intrinsic geometry of the surface is modified,
\begin{equation}
\Omega_{AB} \to \Omega_{AB} + \delta \Omega_{AB}.
\end{equation}
The deformation can be characterized in a gauge-invariant manner by examining the perturbation of the surface's intrinsic scalar curvature $\mathcal{R}$.
Expanding the curvature perturbation in spherical harmonics,
\begin{equation}
\delta \mathcal{R} = \sum_{\ell m} \delta \mathcal{R}_{\ell m} Y_{\ell m}(\theta,\phi),
\end{equation}
one defines the \emph{surficial LNs} $h_{\ell}$ through the linear relation between the curvature perturbation and the external tidal moments $\mathcal{E}_{\ell m}$: 
\begin{equation}
\delta \mathcal{R}_{\ell m} = - \frac{2(\ell+2)}{\ell}\,\frac{R^{\ell+1}}{M} h_{\ell}\, \mathcal{E}_{\ell m}.
\end{equation}
The dimensionless coefficients $h_{\ell}$ therefore quantify the fractional deformation of the surface geometry produced by a given tidal field.

% \paragraph{Relation to standard Love numbers}~---

For a non-rotating body in GR, the surficial LNs are related algebraically to the usual electric-type tidal LNs ${}_2k_{\ell}^{\rm E}$ defined from the asymptotic multipolar structure of the gravitational field \citep{Landry:2014jka}. 
For $\ell\ge2$, one finds schematically
\begin{equation}
h_{\ell} = \Gamma^{(1)}_{\ell}(C)+ 2\Gamma^{(2)}_{\ell}(C) k_{\ell}^{\rm E}\, ,
\end{equation}
where $\Gamma^{(1,2)}_{\ell}(C)$ are compactness-dependent factors determined by the background stellar structure and as usual $C=M/R$ is the compactness of the body. 
In the Newtonian limit ($C\ll1$), $\Gamma^{1,2}_{\ell}(0)\to1$ and we obtain the well-known relation between the surface and potential LNs of classical theory: $h_{\ell}=1+2k_{\ell}$ \citep{PoissonWill, Damour:2009va}.

An important consequence of this relation is that the vanishing of the electric-type LNs implies \emph{nonvanishing} surficial LNs. Therefore, in four-dimensional vacuum GR, static bosonic perturbations of a Schwarzschild BH lead to
\begin{equation}
{}_2k_{\ell}^{\rm E}=0
\qquad \Longrightarrow \qquad
h_{\ell}=\frac{\ell+1}{2(\ell-1)}\frac{(\ell!)^{2}}{(2\ell)!}\,.
\end{equation}
That is, the absence of static tidal polarizability of BHs 
refers to their asymptotic multipole moments, but does not
reflect on the intrinsic geometry of the horizon, which instead is deformed by an external tidal field.

% \paragraph{Black-hole limit}~---

Indeed, in the case of a BH the notion of ``surface'' is replaced by the event horizon. 
One may consider spacelike cross-sections of the horizon and examine their intrinsic curvature under external perturbations. 
In the static limit in four-dimensional vacuum GR, the intrinsic geometry of the horizon does acquire multipolar distortions proportional to the external tidal field. 

Surficial LNs offer a geometrically transparent characterization of the object's tidal response. 
Unlike the standard LNs, which are defined through the asymptotic behavior of the metric, surficial LNs are quasi-local quantities tied directly to the geometry of the object's boundary. 
They therefore provide an alternative diagnostic of tidal deformability that is insensitive to ambiguities in multipole definitions at infinity.

%%%%%%%%%%%%%%%%%%%%%%%%%%%%%%%%%%%%%%%%%%%%%%%%%%%%%%%%%%%%%%%%%%%%%%%%%%%%%%

\phantomsection
\addcontentsline{toc}{section}{References}
%%FS \bibliographystyle{utphys1}
%\bibliographystyle{plainnat}
\bibliography{References}

\begin{thebibliography}{478}
\providecommand{\natexlab}[1]{#1}
\providecommand{\url}[1]{{#1}}
\providecommand{\urlprefix}{URL }
\providecommand{\doi}[1]{\url{https://doi.org/#1}}
\providecommand{\eprint}[2][]{\url{#2}}
 \bibcommenthead

\bibitem[{Abac et~al.(2024)Abac, Dietrich, Buonanno, Steinhoff, and
  Ujevic}]{Abac:2023ujg}
Abac A, Dietrich T, Buonanno A, et~al (2024) {New and robust
  gravitational-waveform model for high-mass-ratio binary neutron star systems
  with dynamical tidal effects}. Phys Rev D 109(2):024062.
  \doi{10.1103/PhysRevD.109.024062},
  {\href{https://arxiv.org/abs/2311.07456}{{arXiv:2311.07456}}} {[gr-qc]}

\bibitem[{Abac et~al.(2025{\natexlab{a}})Abac, Ramis~Vidal, Colleoni, Puecher,
  Gonzalez, and Dietrich}]{Abac:2025brd}
Abac A, Ramis~Vidal FA, Colleoni M, et~al (2025{\natexlab{a}}) {Leveraging
  NRTidalv3 to develop gravitational waveform models with higher-order modes
  for binary neutron star systems}. Phys Rev D 112(10):104026.
  \doi{10.1103/hzn7-39js},
  {\href{https://arxiv.org/abs/2507.15426}{{arXiv:2507.15426}}} {[gr-qc]}

\bibitem[{Abac et~al.(2025{\natexlab{b}})}]{ET:2025xjr}
Abac A, et~al (2025{\natexlab{b}}) {The Science of the Einstein Telescope}.
  arXiv e-prints {\href{https://arxiv.org/abs/2503.12263}{{arXiv:2503.12263}}}
  {[gr-qc]}

\bibitem[{Abac et~al.(2025{\natexlab{c}})}]{LIGOScientific:2025slb}
Abac AG, et~al (2025{\natexlab{c}}) {GWTC-4.0: Updating the Gravitational-Wave
  Transient Catalog with Observations from the First Part of the Fourth
  LIGO-Virgo-KAGRA Observing Run}. arXiv e-prints
  {\href{https://arxiv.org/abs/2508.18082}{{arXiv:2508.18082}}} {[gr-qc]}

\bibitem[{Abbott et~al.(2017)}]{LIGOScientific:2017vwq}
Abbott BP, et~al (2017) {GW170817: Observation of Gravitational Waves from a
  Binary Neutron Star Inspiral}. Phys Rev Lett 119(16):161101.
  \doi{10.1103/PhysRevLett.119.161101},
  {\href{https://arxiv.org/abs/1710.05832}{{arXiv:1710.05832}}} {[gr-qc]}

\bibitem[{Abbott et~al.(2018)}]{Abbott:2018exr}
Abbott BP, et~al (2018) {GW170817: Measurements of neutron star radii and
  equation of state}. Phys Rev Lett 121(16):161101.
  \doi{10.1103/PhysRevLett.121.161101},
  {\href{https://arxiv.org/abs/1805.11581}{{arXiv:1805.11581}}} {[gr-qc]}

\bibitem[{Abbott et~al.(2019)}]{Abbott:2018wiz}
Abbott BP, et~al (2019) {Properties of the binary neutron star merger
  GW170817}. Phys Rev X9(1):011001. \doi{10.1103/PhysRevX.9.011001},
  {\href{https://arxiv.org/abs/1805.11579}{{arXiv:1805.11579}}} {[gr-qc]}

\bibitem[{Abbott et~al.(2021)}]{LIGOScientific:2019lzm}
Abbott R, et~al (2021) {Open data from the first and second observing runs of
  Advanced LIGO and Advanced Virgo}. SoftwareX 13:100658.
  \doi{10.1016/j.softx.2021.100658},
  {\href{https://arxiv.org/abs/1912.11716}{{arXiv:1912.11716}}} {[gr-qc]}

\bibitem[{Abdelsalhin et~al.(2018)Abdelsalhin, Gualtieri, and
  Pani}]{Abdelsalhin:2018reg}
Abdelsalhin T, Gualtieri L, Pani P (2018) {Post-Newtonian spin-tidal couplings
  for compact binaries}. Phys Rev D 98(10):104046.
  \doi{10.1103/PhysRevD.98.104046},
  {\href{https://arxiv.org/abs/1805.01487}{{arXiv:1805.01487}}} {[gr-qc]}

\bibitem[{Addazi et~al.(2019)Addazi, Marciano, and Yunes}]{Addazi:2018uhd}
Addazi A, Marciano A, Yunes N (2019) {Can we probe Planckian corrections at the
  horizon scale with gravitational waves?} Phys Rev Lett 122(8):081301.
  \doi{10.1103/PhysRevLett.122.081301},
  {\href{https://arxiv.org/abs/1810.10417}{{arXiv:1810.10417}}} {[gr-qc]}

\bibitem[{Agullo et~al.(2021)Agullo, Cardoso, Rio, Maggiore, and
  Pullin}]{agullo2021potential-5f8}
Agullo I, Cardoso V, Rio Ad, et~al (2021) Potential gravitational wave
  signatures of quantum gravity. Physical Review Letters 126(4):041302.
  \doi{10.1103/physrevlett.126.041302},
  {\href{https://arxiv.org/abs/2007.13761}{{2007.13761}}}

\bibitem[{Ajith et~al.(2022)Ajith, Yagi, and Yunes}]{Ajith:2022uaw}
Ajith S, Yagi K, Yunes N (2022) {I-Love-Q relations in Ho{\v{r}}ava-Lifshitz
  gravity}. Phys Rev D 106(12):124002. \doi{10.1103/PhysRevD.106.124002},
  {\href{https://arxiv.org/abs/2207.05858}{{arXiv:2207.05858}}} {[gr-qc]}

\bibitem[{Akcay et~al.(2019)Akcay, Bernuzzi, Messina, Nagar, Ortiz, and
  Rettegno}]{Akcay:2018yyh}
Akcay S, Bernuzzi S, Messina F, et~al (2019) {Effective-one-body multipolar
  waveform for tidally interacting binary neutron stars up to merger}. Phys Rev
  D 99(4):044051. \doi{10.1103/PhysRevD.99.044051},
  {\href{https://arxiv.org/abs/1812.02744}{{arXiv:1812.02744}}} {[gr-qc]}

\bibitem[{Akhtar et~al.(2025)Akhtar, Bautista, Iossa, and
  Zhou}]{Akhtar:2025nmt}
Akhtar S, Bautista YF, Iossa C, et~al (2025) {Five-dimensional gravitational
  Raman scattering: Scalar wave perturbations in Schwarzschild-Tangherlini
  spacetime}. Phys Rev D 112(8):085018. \doi{10.1103/rglg-kfxq},
  {\href{https://arxiv.org/abs/2505.21489}{{arXiv:2505.21489}}} {[hep-th]}

\bibitem[{Albanesi et~al.(2025)Albanesi, Gamba, Bernuzzi, Fontbut{\'e},
  Gonzalez, and Nagar}]{Albanesi:2025txj}
Albanesi S, Gamba R, Bernuzzi S, et~al (2025) {Effective-one-body modeling for
  generic compact binaries with arbitrary orbits}. Phys Rev D 112(12):L121503.
  \doi{10.1103/3snf-w1x7},
  {\href{https://arxiv.org/abs/2503.14580}{{arXiv:2503.14580}}} {[gr-qc]}

\bibitem[{Alexander and Yunes(2009)}]{Alexander:2009tp}
Alexander S, Yunes N (2009) {Chern-Simons Modified General Relativity}. Phys
  Rept 480:1--55. \doi{10.1016/j.physrep.2009.07.002},
  {\href{https://arxiv.org/abs/0907.2562}{{arXiv:0907.2562}}} {[hep-th]}

\bibitem[{Alvi(2001)}]{Alvi:2001mx}
Alvi K (2001) {Energy and angular momentum flow into a black hole in a binary}.
  Phys Rev D 64:104020. \doi{10.1103/PhysRevD.64.104020},
  {\href{https://arxiv.org/abs/gr-qc/0107080}{{arXiv:gr-qc/0107080}}} {[gr-qc]}

\bibitem[{Anderson et~al.(2005)Anderson, Balbinot, and
  Fabbri}]{Anderson:2004md}
Anderson PR, Balbinot R, Fabbri A (2005) {Cutoff AdS/CFT duality and the quest
  for braneworld black holes}. Phys Rev Lett 94:061301.
  \doi{10.1103/PhysRevLett.94.061301},
  {\href{https://arxiv.org/abs/hep-th/0410034}{{arXiv:hep-th/0410034}}}

\bibitem[{Andersson and Pnigouras(2020)}]{Andersson:2019ahb}
Andersson N, Pnigouras P (2020) {Exploring the effective tidal deformability of
  neutron stars}. Phys Rev D 101(8):083001. \doi{10.1103/PhysRevD.101.083001},
  {\href{https://arxiv.org/abs/1906.08982}{{arXiv:1906.08982}}} {[astro-ph.HE]}

\bibitem[{Andersson and Pnigouras(2021)}]{Andersson:2019dwg}
Andersson N, Pnigouras P (2021) {The phenomenology of dynamical neutron star
  tides}. Mon Not Roy Astron Soc 503(1):533--539. \doi{10.1093/mnras/stab371},
  {\href{https://arxiv.org/abs/1905.00012}{{arXiv:1905.00012}}} {[gr-qc]}

\bibitem[{Andrade et~al.(2021)Andrade, Pantelidou, Sonner, and
  Withers}]{Andrade:2019rpn}
Andrade T, Pantelidou C, Sonner J, et~al (2021) {Driven black holes: from
  Kolmogorov scaling to turbulent wakes}. JHEP 07:063.
  \doi{10.1007/JHEP07(2021)063},
  {\href{https://arxiv.org/abs/1912.00032}{{arXiv:1912.00032}}} {[hep-th]}

\bibitem[{Andr{\'e}s-Carcasona and
  Caneva~Santoro(2025)}]{Andres-Carcasona:2025bni}
Andr{\'e}s-Carcasona M, Caneva~Santoro G (2025) {No Love for black holes:
  tightest constraints on tidal Love numbers of black holes from GW250114}.
  arXiv e-prints {\href{https://arxiv.org/abs/2512.01918}{{arXiv:2512.01918}}}
  {[gr-qc]}

\bibitem[{Arana et~al.(2025)Arana, Brito, and Castro}]{Arana:2024kaz}
Arana R, Brito R, Castro G (2025) {Tidal Love numbers of gravitational atoms}.
  Phys Rev D 111(4):044013. \doi{10.1103/PhysRevD.111.044013},
  {\href{https://arxiv.org/abs/2410.00968}{{arXiv:2410.00968}}} {[gr-qc]}

\bibitem[{Babak et~al.(2017)Babak, Gair, Sesana, Barausse, Sopuerta, Berry,
  Berti, Amaro-Seoane, Petiteau, and Klein}]{Babak:2017tow}
Babak S, Gair J, Sesana A, et~al (2017) {Science with the space-based
  interferometer LISA. V: Extreme mass-ratio inspirals}. Phys Rev D
  95(10):103012. \doi{10.1103/PhysRevD.95.103012},
  {\href{https://arxiv.org/abs/1703.09722}{{arXiv:1703.09722}}} {[gr-qc]}

\bibitem[{Bah and Heidmann(2021{\natexlab{a}})}]{Bah:2021owp}
Bah I, Heidmann P (2021{\natexlab{a}}) {Smooth bubbling geometries without
  supersymmetry}. JHEP 09:128. \doi{10.1007/JHEP09(2021)128},
  {\href{https://arxiv.org/abs/2106.05118}{{arXiv:2106.05118}}} {[hep-th]}

\bibitem[{Bah and Heidmann(2021{\natexlab{b}})}]{Bah:2020ogh}
Bah I, Heidmann P (2021{\natexlab{b}}) {Topological Stars and Black Holes}.
  Phys Rev Lett 126(15):151101. \doi{10.1103/PhysRevLett.126.151101},
  {\href{https://arxiv.org/abs/2011.08851}{{arXiv:2011.08851}}} {[hep-th]}

\bibitem[{Bah et~al.(2022)Bah, Heidmann, and Weck}]{Bah:2022yji}
Bah I, Heidmann P, Weck P (2022) {Schwarzschild-like topological solitons}.
  JHEP 08:269. \doi{10.1007/JHEP08(2022)269},
  {\href{https://arxiv.org/abs/2203.12625}{{arXiv:2203.12625}}} {[hep-th]}

\bibitem[{Bambi et~al.(2025)}]{Bambi:2025wjx}
Bambi C, et~al (2025) {Black hole mimickers: from theory to observation}. arXiv
  e-prints {\href{https://arxiv.org/abs/2505.09014}{{arXiv:2505.09014}}}
  {[gr-qc]}

\bibitem[{Banados and Ferreira(2010)}]{Banados:2010ix}
Banados M, Ferreira PG (2010) {Eddington's theory of gravity and its progeny}.
  Phys Rev Lett 105:011101. \doi{10.1103/PhysRevLett.105.011101}, [Erratum:
  Phys.Rev.Lett. 113, 119901 (2014)],
  {\href{https://arxiv.org/abs/1006.1769}{{arXiv:1006.1769}}} {[astro-ph.CO]}

\bibitem[{Banihashemi and Vines(2020)}]{Banihashemi:2018xfb}
Banihashemi B, Vines J (2020) {Gravitomagnetic tidal effects in gravitational
  waves from neutron star binaries}. Phys Rev D 101(6):064003.
  \doi{10.1103/PhysRevD.101.064003},
  {\href{https://arxiv.org/abs/1805.07266}{{arXiv:1805.07266}}} {[gr-qc]}

\bibitem[{Barausse et~al.(2014)Barausse, Cardoso, and Pani}]{Barausse:2014tra}
Barausse E, Cardoso V, Pani P (2014) {Can environmental effects spoil precision
  gravitational-wave astrophysics?} Phys Rev D 89(10):104059.
  \doi{10.1103/PhysRevD.89.104059},
  {\href{https://arxiv.org/abs/1404.7149}{{arXiv:1404.7149}}} {[gr-qc]}

\bibitem[{Barbosa et~al.(2025)Barbosa, Brax, Fichet, and
  de~Souza}]{Barbosa:2025uau}
Barbosa S, Brax P, Fichet S, et~al (2025) {Running Love numbers and the
  Effective Field Theory of gravity}. JCAP 07:071.
  \doi{10.1088/1475-7516/2025/07/071},
  {\href{https://arxiv.org/abs/2501.18684}{{arXiv:2501.18684}}} {[hep-th]}

\bibitem[{Barbosa et~al.(2026)Barbosa, Fichet, and de~Souza}]{Barbosa:2026qcv}
Barbosa S, Fichet S, de~Souza L (2026) {Running Love Numbers of Charged Black
  Holes}. arXiv e-prints
  {\href{https://arxiv.org/abs/2602.00349}{{arXiv:2602.00349}}} {[hep-th]}

\bibitem[{Baumann et~al.(2019)Baumann, Chia, and Porto}]{Baumann:2018vus}
Baumann D, Chia HS, Porto RA (2019) {Probing Ultralight Bosons with Binary
  Black Holes}. Phys Rev D 99(4):044001. \doi{10.1103/PhysRevD.99.044001},
  {\href{https://arxiv.org/abs/1804.03208}{{arXiv:1804.03208}}} {[gr-qc]}

\bibitem[{Bekenstein(1972{\natexlab{a}})}]{Bekenstein:1972ky}
Bekenstein JD (1972{\natexlab{a}}) {Nonexistence of baryon number for black
  holes. ii}. Phys Rev D 5:2403--2412. \doi{10.1103/PhysRevD.5.2403}

\bibitem[{Bekenstein(1972{\natexlab{b}})}]{Bekenstein:1971hc}
Bekenstein JD (1972{\natexlab{b}}) {Nonexistence of baryon number for static
  black holes}. Phys Rev D 5:1239--1246. \doi{10.1103/PhysRevD.5.1239}

\bibitem[{Bekenstein(1973)}]{Bekenstein:1973ur}
Bekenstein JD (1973) {Black holes and entropy}. Phys Rev D 7:2333--2346.
  \doi{10.1103/PhysRevD.7.2333}

\bibitem[{Bekenstein and Mukhanov(1995)}]{Bekenstein:1995ju}
Bekenstein JD, Mukhanov VF (1995) {Spectroscopy of the quantum black hole}.
  Phys Lett B 360:7--12. \doi{10.1016/0370-2693(95)01148-J},
  {\href{https://arxiv.org/abs/gr-qc/9505012}{{arXiv:gr-qc/9505012}}} {[gr-qc]}

\bibitem[{Ben~Achour et~al.(2022)Ben~Achour, Livine, Mukohyama, and
  Uzan}]{BenAchour:2022uqo}
Ben~Achour J, Livine ER, Mukohyama S, et~al (2022) {Hidden symmetry of the
  static response of black holes: applications to Love numbers}. JHEP 07:112.
  \doi{10.1007/JHEP07(2022)112},
  {\href{https://arxiv.org/abs/2202.12828}{{arXiv:2202.12828}}} {[gr-qc]}

\bibitem[{Bena et~al.(2006)Bena, Wang, and Warner}]{Bena:2006kb}
Bena I, Wang CW, Warner NP (2006) {Mergers and typical black hole microstates}.
  JHEP 11:042. \doi{10.1088/1126-6708/2006/11/042},
  {\href{https://arxiv.org/abs/hep-th/0608217}{{arXiv:hep-th/0608217}}}

\bibitem[{Bena et~al.(2016)Bena, Giusto, Martinec, Russo, Shigemori, Turton,
  and Warner}]{Bena:2016ypk}
Bena I, Giusto S, Martinec EJ, et~al (2016) {Smooth horizonless geometries deep
  inside the black-hole regime}. Phys Rev Lett 117(20):201601.
  \doi{10.1103/PhysRevLett.117.201601},
  {\href{https://arxiv.org/abs/1607.03908}{{arXiv:1607.03908}}} {[hep-th]}

\bibitem[{Bena et~al.(2018)Bena, Giusto, Martinec, Russo, Shigemori, Turton,
  and Warner}]{Bena:2017xbt}
Bena I, Giusto S, Martinec EJ, et~al (2018) {Asymptotically-flat supergravity
  solutions deep inside the black-hole regime}. JHEP 02:014.
  \doi{10.1007/JHEP02(2018)014},
  {\href{https://arxiv.org/abs/1711.10474}{{arXiv:1711.10474}}} {[hep-th]}

\bibitem[{Bena et~al.(2022{\natexlab{a}})Bena, Martinec, Mathur, and
  Warner}]{Bena:2022rna}
Bena I, Martinec EJ, Mathur SD, et~al (2022{\natexlab{a}}) {Fuzzballs and
  Microstate Geometries: Black-Hole Structure in String Theory}. arXiv e-prints
  {\href{https://arxiv.org/abs/2204.13113}{{arXiv:2204.13113}}} {[hep-th]}

\bibitem[{Bena et~al.(2022{\natexlab{b}})Bena, Martinec, Mathur, and
  Warner}]{Bena:2022ldq}
Bena I, Martinec EJ, Mathur SD, et~al (2022{\natexlab{b}}) {Snowmass White
  Paper: Micro- and Macro-Structure of Black Holes}. arXiv e-prints
  {\href{https://arxiv.org/abs/2203.04981}{{arXiv:2203.04981}}} {[hep-th]}

\bibitem[{Berens et~al.(2023)Berens, Hui, and Sun}]{Berens:2022ebl}
Berens R, Hui L, Sun Z (2023) {Ladder symmetries of black holes and de Sitter
  space: love numbers and quasinormal modes}. JCAP 06:056.
  \doi{10.1088/1475-7516/2023/06/056},
  {\href{https://arxiv.org/abs/2212.09367}{{arXiv:2212.09367}}} {[hep-th]}

\bibitem[{Berens et~al.(2025{\natexlab{a}})Berens, Hui, McLoughlin, Penco, and
  Staunton}]{Berens:2025okm}
Berens R, Hui L, McLoughlin D, et~al (2025{\natexlab{a}}) {Geometric Symmetries
  for the Vanishing of the Black Hole Tidal Love Numbers}. arXiv e-prints
  {\href{https://arxiv.org/abs/2510.18952}{{arXiv:2510.18952}}} {[hep-th]}

\bibitem[{Berens et~al.(2025{\natexlab{b}})Berens, Hui, McLoughlin, Solomon,
  and Staunton}]{Berens:2025jfs}
Berens R, Hui L, McLoughlin D, et~al (2025{\natexlab{b}}) {Ladder Symmetries of
  Higher Dimensional Black Holes}. preprint
  {\href{https://arxiv.org/abs/2510.26748}{{arXiv:2510.26748}}} {[hep-th]}

\bibitem[{Bernuzzi et~al.(2015)Bernuzzi, Nagar, Dietrich, and
  Damour}]{Bernuzzi:2014owa}
Bernuzzi S, Nagar A, Dietrich T, et~al (2015) {Modeling the Dynamics of Tidally
  Interacting Binary Neutron Stars up to the Merger}. Phys Rev Lett
  114(16):161103. \doi{10.1103/PhysRevLett.114.161103},
  {\href{https://arxiv.org/abs/1412.4553}{{arXiv:1412.4553}}} {[gr-qc]}

\bibitem[{Berti et~al.(2006)Berti, Cardoso, and Casals}]{Berti:2005gp}
Berti E, Cardoso V, Casals M (2006) {Eigenvalues and eigenfunctions of
  spin-weighted spheroidal harmonics in four and higher dimensions}. PhysRev D
  73:024013. \doi{10.1103/PhysRevD.73.109902, 10.1103/PhysRevD.73.024013},
  {\href{https://arxiv.org/abs/gr-qc/0511111}{{arXiv:gr-qc/0511111}}} {[gr-qc]}

\bibitem[{Berti et~al.(2024)Berti, De~Luca, Del~Grosso, and
  Pani}]{Berti:2024moe}
Berti E, De~Luca V, Del~Grosso L, et~al (2024) {Tidal Love numbers and
  approximate universal relations for fermion soliton stars}. Phys Rev D
  109(12):124008. \doi{10.1103/PhysRevD.109.124008},
  {\href{https://arxiv.org/abs/2404.06979}{{arXiv:2404.06979}}} {[gr-qc]}

\bibitem[{Berti et~al.(2015)}]{Berti:2015itd}
Berti E, et~al (2015) {Testing General Relativity with Present and Future
  Astrophysical Observations}. Class Quant Grav 32:243001.
  \doi{10.1088/0264-9381/32/24/243001},
  {\href{https://arxiv.org/abs/1501.07274}{{arXiv:1501.07274}}} {[gr-qc]}

\bibitem[{Bertini et~al.(2012)Bertini, Cacciatori, and Klemm}]{Bertini:2011ga}
Bertini S, Cacciatori SL, Klemm D (2012) {Conformal structure of the
  Schwarzschild black hole}. Phys Rev D 85:064018.
  \doi{10.1103/PhysRevD.85.064018},
  {\href{https://arxiv.org/abs/1106.0999}{{arXiv:1106.0999}}} {[hep-th]}

\bibitem[{Bhatt and Singha(2024)}]{Bhatt:2024mvr}
Bhatt RP, Singha C (2024) {Scalar tidal response of a rotating BTZ black hole}.
  JHEP 11:154. \doi{10.1007/JHEP11(2024)154},
  {\href{https://arxiv.org/abs/2407.09470}{{arXiv:2407.09470}}} {[gr-qc]}

\bibitem[{Bhatt et~al.(2023)Bhatt, Chakraborty, and Bose}]{Bhatt:2023zsy}
Bhatt RP, Chakraborty S, Bose S (2023) {Addressing issues in defining the Love
  numbers for black holes}. Phys Rev D 108(8):084013.
  \doi{10.1103/PhysRevD.108.084013},
  {\href{https://arxiv.org/abs/2306.13627}{{arXiv:2306.13627}}} {[gr-qc]}

\bibitem[{Bhatt et~al.(2024)Bhatt, Chakraborty, and Bose}]{Bhatt:2024rpx}
Bhatt RP, Chakraborty S, Bose S (2024) {Response of a Kerr black hole to a
  generic tidal perturbation}. arXiv e-prints
  {\href{https://arxiv.org/abs/2412.15117}{{arXiv:2412.15117}}} {[gr-qc]}

\bibitem[{Bhatt et~al.(2025)Bhatt, Chakraborty, and Bose}]{Bhatt:2024yyz}
Bhatt RP, Chakraborty S, Bose S (2025) {Rotating black holes experience
  dynamical tides}. Phys Rev D 111(4):L041504.
  \doi{10.1103/PhysRevD.111.L041504},
  {\href{https://arxiv.org/abs/2406.09543}{{arXiv:2406.09543}}} {[gr-qc]}

\bibitem[{Bhattacharyya et~al.(2025)Bhattacharyya, Ghosh, Kumar, Kumar, and
  Pal}]{Bhattacharyya:2025slf}
Bhattacharyya A, Ghosh S, Kumar N, et~al (2025) {Love beyond Einstein: metric
  reconstruction and Love number in quadratic gravity using WEFT}. JHEP 11:155.
  \doi{10.1007/JHEP11(2025)155},
  {\href{https://arxiv.org/abs/2508.02785}{{arXiv:2508.02785}}} {[hep-th]}

\bibitem[{Bhattacharyya et~al.(2026)Bhattacharyya, Kumar, and
  Kumar}]{Bhattacharyya:2026itm}
Bhattacharyya A, Kumar N, Kumar S (2026) {Dynamical Tidal Response of Regular
  Black Holes: Perturbative Analysis and Shell EFT Interpretation}. preprint
  {\href{https://arxiv.org/abs/2603.24719}{{arXiv:2603.24719}}} {[hep-th]}

\bibitem[{Bianchi et~al.(2023)Bianchi, Di~Russo, Grillo, Morales, and
  Sudano}]{Bianchi:2023sfs}
Bianchi M, Di~Russo G, Grillo A, et~al (2023) {On the stability and
  deformability of top stars}. JHEP 12:121. \doi{10.1007/JHEP12(2023)121},
  {\href{https://arxiv.org/abs/2305.15105}{{arXiv:2305.15105}}} {[gr-qc]}

\bibitem[{Bini and Damour(2014)}]{Bini:2014zxa}
Bini D, Damour T (2014) {Gravitational self-force corrections to two-body tidal
  interactions and the effective one-body formalism}. Phys Rev D 90(12):124037.
  \doi{10.1103/PhysRevD.90.124037},
  {\href{https://arxiv.org/abs/1409.6933}{{arXiv:1409.6933}}} {[gr-qc]}

\bibitem[{Bini et~al.(2012)Bini, Damour, and Faye}]{Bini:2012gu}
Bini D, Damour T, Faye G (2012) {Effective action approach to higher-order
  relativistic tidal interactions in binary systems and their effective one
  body description}. Phys Rev D 85:124034. \doi{10.1103/PhysRevD.85.124034},
  {\href{https://arxiv.org/abs/1202.3565}{{arXiv:1202.3565}}} {[gr-qc]}

\bibitem[{Binnington and Poisson(2009)}]{Binnington:2009bb}
Binnington T, Poisson E (2009) {Relativistic theory of tidal Love numbers}.
  Phys Rev D 80:084018. \doi{10.1103/PhysRevD.80.084018},
  {\href{https://arxiv.org/abs/0906.1366}{{arXiv:0906.1366}}} {[gr-qc]}

\bibitem[{Blanchet(2014)}]{Blanchet:2013haa}
Blanchet L (2014) {Gravitational Radiation from Post-Newtonian Sources and
  Inspiralling Compact Binaries}. Living Rev Rel 17:2.
  \doi{10.12942/lrr-2014-2},
  {\href{https://arxiv.org/abs/1310.1528}{{arXiv:1310.1528}}} {[gr-qc]}

\bibitem[{Blanchet et~al.(2023)Blanchet, Faye, Henry, Larrouturou, and
  Trestini}]{Blanchet:2023sbv}
Blanchet L, Faye G, Henry Q, et~al (2023) {Gravitational-wave flux and
  quadrupole modes from quasicircular nonspinning compact binaries to the
  fourth post-Newtonian order}. Phys Rev D 108(6):064041.
  \doi{10.1103/PhysRevD.108.064041},
  {\href{https://arxiv.org/abs/2304.11186}{{arXiv:2304.11186}}} {[gr-qc]}

\bibitem[{Blok et~al.(2003)Blok, Bosma, and McGaugh}]{Blok:2002tr}
Blok WJGd, Bosma A, McGaugh SS (2003) {Simulating observations of dark matter
  dominated galaxies: towards the optimal halo profile}. Mon Not Roy Astron Soc
  340:657--678. \doi{10.1046/j.1365-8711.2003.06330.x},
  {\href{https://arxiv.org/abs/astro-ph/0212102}{{arXiv:astro-ph/0212102}}}

\bibitem[{Bonelli et~al.(2022)Bonelli, Iossa, Lichtig, and
  Tanzini}]{Bonelli:2021uvf}
Bonelli G, Iossa C, Lichtig DP, et~al (2022) {Exact solution of Kerr black hole
  perturbations via CFT2 and instanton counting: Greybody factor, quasinormal
  modes, and Love numbers}. Phys Rev D 105(4):044047.
  \doi{10.1103/PhysRevD.105.044047},
  {\href{https://arxiv.org/abs/2105.04483}{{arXiv:2105.04483}}} {[hep-th]}

\bibitem[{Boulware and Deser(1985)}]{Boulware:1985wk}
Boulware DG, Deser S (1985) {String Generated Gravity Models}. Phys Rev Lett
  55:2656. \doi{10.1103/PhysRevLett.55.2656}

\bibitem[{Branchesi et~al.(2023)}]{Branchesi:2023mws}
Branchesi M, et~al (2023) {Science with the Einstein Telescope: a comparison of
  different designs}. JCAP 07:068. \doi{10.1088/1475-7516/2023/07/068},
  {\href{https://arxiv.org/abs/2303.15923}{{arXiv:2303.15923}}} {[gr-qc]}

\bibitem[{Bredberg et~al.(2010)Bredberg, Hartman, Song, and
  Strominger}]{Bredberg:2009pv}
Bredberg I, Hartman T, Song W, et~al (2010) {Black Hole Superradiance From
  Kerr/CFT}. JHEP 04:019. \doi{10.1007/JHEP04(2010)019},
  {\href{https://arxiv.org/abs/0907.3477}{{arXiv:0907.3477}}} {[hep-th]}

\bibitem[{Bretz and Yu(2026)}]{Bretz:2026asa}
Bretz J, Yu H (2026) {Impact on Inferred Neutron Star Equation of State due to
  Nonlinear Hydrodynamics, Background Spin, and Relativity}. arXiv e-prints
  {\href{https://arxiv.org/abs/2602.04951}{{arXiv:2602.04951}}} {[gr-qc]}

\bibitem[{{Breuer} et~al.(1977){Breuer}, {Ryan}, and
  {Waller}}]{1977RSPSA.358...71B}
{Breuer} RA, {Ryan} MPJr., {Waller} S (1977) {Some Properties of Spin-Weighted
  Spheroidal Harmonics}. Proceedings of the Royal Society of London Series A
  358(1692):71--86. \doi{10.1098/rspa.1977.0187}

\bibitem[{Brill and Wheeler(1957)}]{Brill:1957fx}
Brill DR, Wheeler JA (1957) {Interaction of neutrinos and gravitational
  fields}. Rev Mod Phys 29:465--479. \doi{10.1103/RevModPhys.29.465}

\bibitem[{Brito et~al.(2015)Brito, Cardoso, and Pani}]{Brito:2015oca}
Brito R, Cardoso V, Pani P (2015) {Superradiance}. Lect Notes Phys
  906:pp.1--237. \doi{10.1007/978-3-319-19000-6},
  {\href{https://arxiv.org/abs/1501.06570}{{arXiv:1501.06570}}} {[gr-qc]}

\bibitem[{{Brooker} and {Olle}(1955)}]{1955MNRAS.115..101B}
{Brooker} RA, {Olle} TW (1955) {Apsidal-motion constants for polytropic
  models}. \mnras 115:101. \doi{10.1093/mnras/115.1.101}

\bibitem[{Buonanno and Damour(1999)}]{Buonanno:1998gg}
Buonanno A, Damour T (1999) {Effective one-body approach to general
  relativistic two-body dynamics}. Phys Rev D 59:084006.
  \doi{10.1103/PhysRevD.59.084006},
  {\href{https://arxiv.org/abs/gr-qc/9811091}{{arXiv:gr-qc/9811091}}}

\bibitem[{Buonanno and Damour(2000)}]{Buonanno:2000ef}
Buonanno A, Damour T (2000) {Transition from inspiral to plunge in binary black
  hole coalescences}. Phys Rev D 62:064015. \doi{10.1103/PhysRevD.62.064015},
  {\href{https://arxiv.org/abs/gr-qc/0001013}{{arXiv:gr-qc/0001013}}}

\bibitem[{Burgay et~al.(2003)}]{Burgay:2003jj}
Burgay M, et~al (2003) {An Increased estimate of the merger rate of double
  neutron stars from observations of a highly relativistic system}. Nature
  426:531--533. \doi{10.1038/nature02124},
  {\href{https://arxiv.org/abs/astro-ph/0312071}{{arXiv:astro-ph/0312071}}}

\bibitem[{Cai and Wang(2019)}]{Cai:2019npx}
Cai S, Wang KD (2019) {Non-vanishing of tidal Love numbers}. arXiv e-prints
  {\href{https://arxiv.org/abs/1906.06850}{{arXiv:1906.06850}}} {[hep-th]}

\bibitem[{Caneva~Santoro et~al.(2024)Caneva~Santoro, Roy, Vicente, Haney,
  Piccinni, Del~Pozzo, and Martinez}]{CanevaSantoro:2023aol}
Caneva~Santoro G, Roy S, Vicente R, et~al (2024) {First Constraints on Compact
  Binary Environments from LIGO-Virgo Data}. Phys Rev Lett 132(25):251401.
  \doi{10.1103/PhysRevLett.132.251401},
  {\href{https://arxiv.org/abs/2309.05061}{{arXiv:2309.05061}}} {[gr-qc]}

\bibitem[{Cannizzaro et~al.(2024)Cannizzaro, De~Luca, and
  Pani}]{Cannizzaro:2024fpz}
Cannizzaro E, De~Luca V, Pani P (2024) {Tidal deformability of black holes
  surrounded by thin accretion disks}. Phys Rev D 110(12):123004.
  \doi{10.1103/PhysRevD.110.123004},
  {\href{https://arxiv.org/abs/2408.14208}{{arXiv:2408.14208}}} {[astro-ph.HE]}

\bibitem[{Cano(2025)}]{Cano:2025zyk}
Cano PA (2025) {Love numbers beyond GR from the modified Teukolsky equation}.
  JHEP 07:152. \doi{10.1007/JHEP07(2025)152},
  {\href{https://arxiv.org/abs/2502.20185}{{arXiv:2502.20185}}} {[gr-qc]}

\bibitem[{Cano et~al.(2023)Cano, Fransen, Hertog, and Maenaut}]{Cano:2023tmv}
Cano PA, Fransen K, Hertog T, et~al (2023) {Universal Teukolsky equations and
  black hole perturbations in higher-derivative gravity}. Phys Rev D
  108(2):024040. \doi{10.1103/PhysRevD.108.024040},
  {\href{https://arxiv.org/abs/2304.02663}{{arXiv:2304.02663}}} {[gr-qc]}

\bibitem[{Capuano et~al.(2024)Capuano, Santoni, and Barausse}]{Capuano:2024qhv}
Capuano L, Santoni L, Barausse E (2024) {Perturbations of the Vaidya metric in
  the frequency domain: Quasinormal modes and tidal response}. Phys Rev D
  110(8):084081. \doi{10.1103/PhysRevD.110.084081},
  {\href{https://arxiv.org/abs/2407.06009}{{arXiv:2407.06009}}} {[gr-qc]}

\bibitem[{Carballo-Rubio et~al.(2018)Carballo-Rubio, Di~Filippo, Liberati,
  Pacilio, and Visser}]{Carballo-Rubio:2018pmi}
Carballo-Rubio R, Di~Filippo F, Liberati S, et~al (2018) {On the viability of
  regular black holes}. JHEP 07:023. \doi{10.1007/JHEP07(2018)023},
  {\href{https://arxiv.org/abs/1805.02675}{{arXiv:1805.02675}}} {[gr-qc]}

\bibitem[{Cardoso and Duque(2020)}]{Cardoso:2019upw}
Cardoso V, Duque F (2020) {Environmental effects in gravitational-wave physics:
  Tidal deformability of black holes immersed in matter}. Phys Rev D
  101(6):064028. \doi{10.1103/PhysRevD.101.064028},
  {\href{https://arxiv.org/abs/1912.07616}{{arXiv:1912.07616}}} {[gr-qc]}

\bibitem[{Cardoso and Gualtieri(2016)}]{Cardoso:2016ryw}
Cardoso V, Gualtieri L (2016) {Testing the black hole no-hair hypothesis}.
  Class Quant Grav 33(17):174001. \doi{10.1088/0264-9381/33/17/174001},
  {\href{https://arxiv.org/abs/1607.03133}{{arXiv:1607.03133}}} {[gr-qc]}

\bibitem[{Cardoso and Pani(2013)}]{Cardoso:2012zn}
Cardoso V, Pani P (2013) {Tidal acceleration of black holes and superradiance}.
  Class Quant Grav 30:045011. \doi{10.1088/0264-9381/30/4/045011},
  {\href{https://arxiv.org/abs/1205.3184}{{arXiv:1205.3184}}} {[gr-qc]}

\bibitem[{Cardoso and Pani(2019)}]{Cardoso:2019rvt}
Cardoso V, Pani P (2019) {Testing the nature of dark compact objects: a status
  report}. Living Rev Rel 22(1):4. \doi{10.1007/s41114-019-0020-4},
  {\href{https://arxiv.org/abs/1904.05363}{{arXiv:1904.05363}}} {[gr-qc]}

\bibitem[{Cardoso et~al.(2008)Cardoso, Pani, Cadoni, and
  Cavaglia}]{Cardoso:2008kj}
Cardoso V, Pani P, Cadoni M, et~al (2008) {Instability of hyper-compact
  Kerr-like objects}. Class Quant Grav 25:195010.
  \doi{10.1088/0264-9381/25/19/195010},
  {\href{https://arxiv.org/abs/0808.1615}{{arXiv:0808.1615}}} {[gr-qc]}

\bibitem[{Cardoso et~al.(2017{\natexlab{a}})Cardoso, Franzin, Maselli, Pani,
  and Raposo}]{Cardoso:2017cfl}
Cardoso V, Franzin E, Maselli A, et~al (2017{\natexlab{a}}) {Testing
  strong-field gravity with tidal Love numbers}. Phys Rev D 95(8):084014.
  \doi{10.1103/PhysRevD.95.084014},
  {\href{https://arxiv.org/abs/1701.01116}{{arXiv:1701.01116}}} {[gr-qc]}

\bibitem[{Cardoso et~al.(2017{\natexlab{b}})Cardoso, Houri, and
  Kimura}]{Cardoso:2017qmj}
Cardoso V, Houri T, Kimura M (2017{\natexlab{b}}) {Mass Ladder Operators from
  Spacetime Conformal Symmetry}. Phys Rev D 96(2):024044.
  \doi{10.1103/PhysRevD.96.024044},
  {\href{https://arxiv.org/abs/1706.07339}{{arXiv:1706.07339}}} {[hep-th]}

\bibitem[{Cardoso et~al.(2018)Cardoso, Kimura, Maselli, and
  Senatore}]{Cardoso:2018ptl}
Cardoso V, Kimura M, Maselli A, et~al (2018) {Black Holes in an Effective Field
  Theory Extension of General Relativity}. Phys Rev Lett 121(25):251105.
  \doi{10.1103/PhysRevLett.121.251105}, [Erratum: Phys.Rev.Lett. 131, 109903
  (2023)], {\href{https://arxiv.org/abs/1808.08962}{{arXiv:1808.08962}}}
  {[gr-qc]}

\bibitem[{Cardoso et~al.(2019)Cardoso, Foit, and Kleban}]{Cardoso:2019apo}
Cardoso V, Foit VF, Kleban M (2019) {Gravitational wave echoes from black hole
  area quantization}. JCAP 08:006. \doi{10.1088/1475-7516/2019/08/006},
  {\href{https://arxiv.org/abs/1902.10164}{{arXiv:1902.10164}}} {[hep-th]}

\bibitem[{Cardoso et~al.(2022)Cardoso, Destounis, Duque, Macedo, and
  Maselli}]{Cardoso:2021wlq}
Cardoso V, Destounis K, Duque F, et~al (2022) {Black holes in galaxies:
  Environmental impact on gravitational-wave generation and propagation}. Phys
  Rev D 105(6):L061501. \doi{10.1103/PhysRevD.105.L061501},
  {\href{https://arxiv.org/abs/2109.00005}{{arXiv:2109.00005}}} {[gr-qc]}

\bibitem[{Caron-Huot et~al.(2025)Caron-Huot, Correia, Isabella, and
  Solon}]{Caron-Huot:2025tlq}
Caron-Huot S, Correia M, Isabella G, et~al (2025) {Gravitational Wave
  Scattering via the Born Series: Scalar Tidal Matching to $\mathcal{O}(G^7)$
  and Beyond}. arXiv e-prints
  {\href{https://arxiv.org/abs/2503.13593}{{arXiv:2503.13593}}} {[hep-th]}

\bibitem[{Torres~del Castillo and
  Silva-Ortigoza(1990)}]{TorresdelCastillo:1990aw}
Torres~del Castillo GF, Silva-Ortigoza G (1990) {Rarita-Schwinger fields in the
  Kerr geometry}. Phys Rev D 42:4082--4086. \doi{10.1103/PhysRevD.42.4082}

\bibitem[{Castro et~al.(2021)Castro, Gualtieri, and Pani}]{Castro:2021wyc}
Castro G, Gualtieri L, Pani P (2021) {Hidden symmetry between rotational tidal
  Love numbers of spinning neutron stars}. Phys Rev D 104(4):044052.
  \doi{10.1103/PhysRevD.104.044052},
  {\href{https://arxiv.org/abs/2103.16595}{{arXiv:2103.16595}}} {[gr-qc]}

\bibitem[{Castro et~al.(2022)Castro, Gualtieri, Maselli, and
  Pani}]{Castro:2022mpw}
Castro G, Gualtieri L, Maselli A, et~al (2022) {Impact and detectability of
  spin-tidal couplings in neutron star inspirals}. Phys Rev D 106(2):024011.
  \doi{10.1103/PhysRevD.106.024011},
  {\href{https://arxiv.org/abs/2204.12510}{{arXiv:2204.12510}}} {[gr-qc]}

\bibitem[{Chakrabarti et~al.(2013{\natexlab{a}})Chakrabarti, Delsate, and
  Steinhoff}]{Chakrabarti:2013xza}
Chakrabarti S, Delsate T, Steinhoff J (2013{\natexlab{a}}) {Effective action
  and linear response of compact objects in Newtonian gravity}. Phys Rev D
  88:084038. \doi{10.1103/PhysRevD.88.084038},
  {\href{https://arxiv.org/abs/1306.5820}{{arXiv:1306.5820}}} {[gr-qc]}

\bibitem[{Chakrabarti et~al.(2013{\natexlab{b}})Chakrabarti, Delsate, and
  Steinhoff}]{Chakrabarti:2013lua}
Chakrabarti S, Delsate T, Steinhoff J (2013{\natexlab{b}}) {New perspectives on
  neutron star and black hole spectroscopy and dynamic tides}. arXiv e-prints
  {\href{https://arxiv.org/abs/1304.2228}{{arXiv:1304.2228}}} {[gr-qc]}

\bibitem[{Chakraborty and Dadhich(2020)}]{Chakraborty:2020ifg}
Chakraborty S, Dadhich N (2020) {Limits on stellar structures in Lovelock
  theories of gravity}. Phys Dark Univ 30:100658.
  \doi{10.1016/j.dark.2020.100658},
  {\href{https://arxiv.org/abs/2005.07504}{{arXiv:2005.07504}}} {[gr-qc]}

\bibitem[{Chakraborty and Heidmann(2025)}]{Chakraborty:2025ger}
Chakraborty S, Heidmann P (2025) {Microstates of non-extremal black holes: a
  new hope}. JHEP 07:101. \doi{10.1007/JHEP07(2025)101},
  {\href{https://arxiv.org/abs/2503.13589}{{arXiv:2503.13589}}} {[hep-th]}

\bibitem[{Chakraborty and SenGupta(2015)}]{Chakraborty:2015bja}
Chakraborty S, SenGupta S (2015) {Effective gravitational field equations on
  $m$-brane embedded in n-dimensional bulk of Einstein and $f(\mathcal {R})$
  gravity}. Eur Phys J C 75(11):538. \doi{10.1140/epjc/s10052-015-3768-z},
  {\href{https://arxiv.org/abs/1504.07519}{{arXiv:1504.07519}}} {[gr-qc]}

\bibitem[{Chakraborty et~al.(2021)Chakraborty, Datta, and
  Sau}]{Chakraborty:2021gdf}
Chakraborty S, Datta S, Sau S (2021) {Tidal heating of black holes and exotic
  compact objects on the brane}. Phys Rev D 104(10):104001.
  \doi{10.1103/PhysRevD.104.104001},
  {\href{https://arxiv.org/abs/2103.12430}{{arXiv:2103.12430}}} {[gr-qc]}

\bibitem[{Chakraborty et~al.(2022)Chakraborty, Maggio, Mazumdar, and
  Pani}]{Chakraborty:2022zlq}
Chakraborty S, Maggio E, Mazumdar A, et~al (2022) {Implications of the quantum
  nature of the black hole horizon on the gravitational-wave ringdown}. Phys
  Rev D 106(2):024041. \doi{10.1103/PhysRevD.106.024041},
  {\href{https://arxiv.org/abs/2202.09111}{{arXiv:2202.09111}}} {[gr-qc]}

\bibitem[{Chakraborty et~al.(2024)Chakraborty, Maggio, Silvestrini, and
  Pani}]{Chakraborty:2023zed}
Chakraborty S, Maggio E, Silvestrini M, et~al (2024) {Dynamical tidal Love
  numbers of Kerr-like compact objects}. Phys Rev D 110(8):084042.
  \doi{10.1103/PhysRevD.110.084042},
  {\href{https://arxiv.org/abs/2310.06023}{{arXiv:2310.06023}}} {[gr-qc]}

\bibitem[{Chakraborty et~al.(2025{\natexlab{a}})Chakraborty, Comp{\`e}re, and
  Machet}]{Chakraborty:2024gcr}
Chakraborty S, Comp{\`e}re G, Machet L (2025{\natexlab{a}}) {Tidal Love numbers
  and quasinormal modes of the Schwarzschild-Hernquist black hole}. Phys Rev D
  112(2):024015. \doi{10.1103/4p2c-rwdh},
  {\href{https://arxiv.org/abs/2412.14831}{{arXiv:2412.14831}}} {[gr-qc]}

\bibitem[{Chakraborty et~al.(2025{\natexlab{b}})Chakraborty, De~Luca,
  Gualtieri, and Pani}]{Chakraborty:2025wvs}
Chakraborty S, De~Luca V, Gualtieri L, et~al (2025{\natexlab{b}}) {Dynamical
  Love numbers of black holes: theory and gravitational waveforms}. arXiv
  e-prints {\href{https://arxiv.org/abs/2507.22994}{{arXiv:2507.22994}}}
  {[gr-qc]}

\bibitem[{Chakraborty et~al.(2025{\natexlab{c}})Chakraborty, Heidmann, and
  Pani}]{Chakraborty:2025zyb}
Chakraborty S, Heidmann P, Pani P (2025{\natexlab{c}}) {Fermionic Love of Black
  Holes in General Relativity}. arXiv e-prints
  {\href{https://arxiv.org/abs/2508.20155}{{arXiv:2508.20155}}} {[gr-qc]}

\bibitem[{Chakraborty et~al.(To Appear)Chakraborty, Saketh, Hinderer, and
  Steinhoff}]{SumantaToAppear2}
Chakraborty S, Saketh M, Hinderer T, et~al (To Appear) {Dynamical Tidal Love
  numbers Under Generic Perturbations: Connecting Black Hole Perturbation
  Theory to Effective Field Theory}  {[gr-qc]}

\bibitem[{Chakravarti et~al.(2019)Chakravarti, Chakraborty, Bose, and
  SenGupta}]{Chakravarti:2018vlt}
Chakravarti K, Chakraborty S, Bose S, et~al (2019) {Tidal Love numbers of black
  holes and neutron stars in the presence of higher dimensions: Implications of
  GW170817}. Phys Rev D 99(2):024036. \doi{10.1103/PhysRevD.99.024036},
  {\href{https://arxiv.org/abs/1811.11364}{{arXiv:1811.11364}}} {[gr-qc]}

\bibitem[{Chamblin et~al.(2000)Chamblin, Hawking, and Reall}]{Chamblin:1999by}
Chamblin A, Hawking SW, Reall HS (2000) {Brane world black holes}. Phys Rev D
  61:065007. \doi{10.1103/PhysRevD.61.065007},
  {\href{https://arxiv.org/abs/hep-th/9909205}{{arXiv:hep-th/9909205}}}

\bibitem[{{Chandrasekhar}(1933)}]{Chandrasekhar1933}
{Chandrasekhar} S (1933) {The equilibrium of distorted polytropes. II. the
  tidal problem}. \mnras 93:449. \doi{10.1093/mnras/93.6.449}

\bibitem[{Chandrasekhar(1976)}]{Chandrasekhar:1976ap}
Chandrasekhar S (1976) {The Solution of Dirac's Equation in Kerr Geometry}.
  Proc Roy Soc Lond A 349:571--575. \doi{10.1098/rspa.1976.0090}

\bibitem[{Chandrasekhar(1985)}]{ChandraBook}
Chandrasekhar S (1985) {The mathematical theory of black holes}. Oxford
  University Press

\bibitem[{Chapline et~al.(2003)Chapline, Hohlfeld, Laughlin, and
  Santiago}]{Chapline:2000en}
Chapline G, Hohlfeld E, Laughlin RB, et~al (2003) {Quantum phase transitions
  and the breakdown of classical general relativity}. Int J Mod Phys A
  18:3587--3590. \doi{10.1142/S0217751X03016380},
  {\href{https://arxiv.org/abs/gr-qc/0012094}{{arXiv:gr-qc/0012094}}}

\bibitem[{Charalambous and Ivanov(2023)}]{Charalambous:2023jgq}
Charalambous P, Ivanov MM (2023) {Scalar Love numbers and Love symmetries of
  5-dimensional Myers-Perry black holes}. JHEP 07:222.
  \doi{10.1007/JHEP07(2023)222},
  {\href{https://arxiv.org/abs/2303.16036}{{arXiv:2303.16036}}} {[hep-th]}

\bibitem[{Charalambous et~al.(2021{\natexlab{a}})Charalambous, Dubovsky, and
  Ivanov}]{Charalambous:2021kcz}
Charalambous P, Dubovsky S, Ivanov MM (2021{\natexlab{a}}) {Hidden Symmetry of
  Vanishing Love Numbers}. Phys Rev Lett 127(10):101101.
  \doi{10.1103/PhysRevLett.127.101101},
  {\href{https://arxiv.org/abs/2103.01234}{{arXiv:2103.01234}}} {[hep-th]}

\bibitem[{Charalambous et~al.(2021{\natexlab{b}})Charalambous, Dubovsky, and
  Ivanov}]{Charalambous:2021mea}
Charalambous P, Dubovsky S, Ivanov MM (2021{\natexlab{b}}) {On the Vanishing of
  Love Numbers for Kerr Black Holes}. JHEP 05:038.
  \doi{10.1007/JHEP05(2021)038},
  {\href{https://arxiv.org/abs/2102.08917}{{arXiv:2102.08917}}} {[hep-th]}

\bibitem[{Charalambous et~al.(2022)Charalambous, Dubovsky, and
  Ivanov}]{Charalambous:2022rre}
Charalambous P, Dubovsky S, Ivanov MM (2022) {Love symmetry}. JHEP 10:175.
  \doi{10.1007/JHEP10(2022)175},
  {\href{https://arxiv.org/abs/2209.02091}{{arXiv:2209.02091}}} {[hep-th]}

\bibitem[{Chatziioannou(2020)}]{Chatziioannou:2020pqz}
Chatziioannou K (2020) {Neutron star tidal deformability and equation of state
  constraints}. Gen Rel Grav 52(11):109. \doi{10.1007/s10714-020-02754-3},
  {\href{https://arxiv.org/abs/2006.03168}{{arXiv:2006.03168}}} {[gr-qc]}

\bibitem[{Chatziioannou et~al.(2013)Chatziioannou, Poisson, and
  Yunes}]{Chatziioannou:2012gq}
Chatziioannou K, Poisson E, Yunes N (2013) {Tidal heating and torquing of a
  Kerr black hole to next-to-leading order in the tidal coupling}. Phys Rev D
  87(4):044022. \doi{10.1103/PhysRevD.87.044022},
  {\href{https://arxiv.org/abs/1211.1686}{{arXiv:1211.1686}}} {[gr-qc]}

\bibitem[{Chatziioannou et~al.(2016)Chatziioannou, Poisson, and
  Yunes}]{Chatziioannou:2016kem}
Chatziioannou K, Poisson E, Yunes N (2016) {Improved next-to-leading order
  tidal heating and torquing of a Kerr black hole}. Phys Rev D 94(8):084043.
  \doi{10.1103/PhysRevD.94.084043},
  {\href{https://arxiv.org/abs/1608.02899}{{arXiv:1608.02899}}} {[gr-qc]}

\bibitem[{Chen and Kotla{\v{r}}{\'\i}k(2023)}]{Chen:2023akf}
Chen CY, Kotla{\v{r}}{\'\i}k P (2023) {Quasinormal modes of black holes
  encircled by a gravitating thin disk}. Phys Rev D 108(6):064052.
  \doi{10.1103/PhysRevD.108.064052},
  {\href{https://arxiv.org/abs/2307.07360}{{arXiv:2307.07360}}} {[gr-qc]}

\bibitem[{Chia(2021)}]{Chia:2020yla}
Chia HS (2021) {Tidal deformation and dissipation of rotating black holes}.
  Phys Rev D 104(2):024013. \doi{10.1103/PhysRevD.104.024013},
  {\href{https://arxiv.org/abs/2010.07300}{{arXiv:2010.07300}}} {[gr-qc]}

\bibitem[{Chia et~al.(2024)Chia, Edwards, Wadekar, Zimmerman, Olsen, Roulet,
  Venumadhav, Zackay, and Zaldarriaga}]{Chia:2023tle}
Chia HS, Edwards TDP, Wadekar D, et~al (2024) {In pursuit of Love numbers:
  First templated search for compact objects with large tidal deformabilities
  in the LIGO-Virgo data}. Phys Rev D 110(6):063007.
  \doi{10.1103/PhysRevD.110.063007},
  {\href{https://arxiv.org/abs/2306.00050}{{arXiv:2306.00050}}} {[gr-qc]}

\bibitem[{Chia et~al.(2025)Chia, Zhou, and Ivanov}]{Chia:2024bwc}
Chia HS, Zhou Z, Ivanov MM (2025) {Tidal heating constraints for black holes
  and exotic compact objects from the LIGO-Virgo-KAGRA data}. Phys Rev D
  111(6):063002. \doi{10.1103/PhysRevD.111.063002},
  {\href{https://arxiv.org/abs/2404.14641}{{arXiv:2404.14641}}} {[gr-qc]}

\bibitem[{Chirenti and Rezzolla(2007)}]{Chirenti:2007mk}
Chirenti CBMH, Rezzolla L (2007) {How to tell a gravastar from a black hole}.
  Class Quant Grav 24:4191--4206. \doi{10.1088/0264-9381/24/16/013},
  {\href{https://arxiv.org/abs/0706.1513}{{arXiv:0706.1513}}} {[gr-qc]}

\bibitem[{Colpi et~al.(1986)Colpi, Shapiro, and Wasserman}]{Colpi:1986ye}
Colpi M, Shapiro SL, Wasserman I (1986) {Boson Stars: Gravitational Equilibria
  of Selfinteracting Scalar Fields}. Phys Rev Lett 57:2485--2488.
  \doi{10.1103/PhysRevLett.57.2485}

\bibitem[{Colpi et~al.(2024)}]{LISA:2024hlh}
Colpi M, et~al (2024) {LISA Definition Study Report}. arXiv e-prints
  {\href{https://arxiv.org/abs/2402.07571}{{arXiv:2402.07571}}} {[astro-ph.CO]}

\bibitem[{Combaluzier-Szteinsznaider
  et~al.(2025{\natexlab{a}})Combaluzier-Szteinsznaider, Glazer, Joyce,
  Rodriguez, and Santoni}]{Combaluzier--Szteinsznaider:2025eoc}
Combaluzier-Szteinsznaider O, Glazer D, Joyce A, et~al (2025{\natexlab{a}})
  {Dynamical Tidal Response of Schwarzschild Black Holes}. arXiv e-prints
  {\href{https://arxiv.org/abs/2511.02372}{{arXiv:2511.02372}}} {[gr-qc]}

\bibitem[{Combaluzier-Szteinsznaider
  et~al.(2025{\natexlab{b}})Combaluzier-Szteinsznaider, Hui, Santoni, Solomon,
  and Wong}]{Combaluzier-Szteinsznaider:2024sgb}
Combaluzier-Szteinsznaider O, Hui L, Santoni L, et~al (2025{\natexlab{b}})
  {Symmetries of vanishing nonlinear Love numbers of Schwarzschild black
  holes}. JHEP 03:124. \doi{10.1007/JHEP03(2025)124},
  {\href{https://arxiv.org/abs/2410.10952}{{arXiv:2410.10952}}} {[gr-qc]}

\bibitem[{Consoli et~al.(2022)Consoli, Fucito, Morales, and
  Poghossian}]{Consoli:2022eey}
Consoli D, Fucito F, Morales JF, et~al (2022) {CFT description of
  BH\textquoteright{}s and ECO\textquoteright{}s: QNMs, superradiance, echoes
  and tidal responses}. JHEP 12:115. \doi{10.1007/JHEP12(2022)115},
  {\href{https://arxiv.org/abs/2206.09437}{{arXiv:2206.09437}}} {[hep-th]}

\bibitem[{Counsell et~al.(2024)Counsell, Gittins, Andersson, and
  Pnigouras}]{Counsell:2024pua}
Counsell R, Gittins F, Andersson N, et~al (2024) {Neutron star g modes in the
  relativistic Cowling approximation}. Mon Not Roy Astron Soc
  536(2):1967--1979. \doi{10.1093/mnras/stae2721},
  {\href{https://arxiv.org/abs/2409.20178}{{arXiv:2409.20178}}} {[gr-qc]}

\bibitem[{Coviello et~al.(2025)Coviello, Vellucci, and
  Lehner}]{Coviello:2025pla}
Coviello C, Vellucci V, Lehner L (2025) {Tidal response of regular black
  holes}. Phys Rev D 111(10):104073. \doi{10.1103/PhysRevD.111.104073},
  {\href{https://arxiv.org/abs/2503.04287}{{arXiv:2503.04287}}} {[gr-qc]}

\bibitem[{Creci et~al.(2021)Creci, Hinderer, and Steinhoff}]{Creci:2021rkz}
Creci G, Hinderer T, Steinhoff J (2021) {Tidal response from scattering and the
  role of analytic continuation}. Phys Rev D 104(12):124061.
  \doi{10.1103/PhysRevD.104.124061}, [Erratum: Phys. Rev. D 105, 109902
  (2022)], {\href{https://arxiv.org/abs/2108.03385}{{arXiv:2108.03385}}}
  {[gr-qc]}

\bibitem[{Creci et~al.(2023)Creci, Hinderer, and Steinhoff}]{Creci:2023cfx}
Creci G, Hinderer T, Steinhoff J (2023) {Tidal properties of neutron stars in
  scalar-tensor theories of gravity}. Phys Rev D 108(12):124073.
  \doi{10.1103/PhysRevD.108.124073}, [Erratum: Phys. Rev. D 111, 089901
  (2025)], {\href{https://arxiv.org/abs/2308.11323}{{arXiv:2308.11323}}}
  {[gr-qc]}

\bibitem[{Creci et~al.(2025)Creci, van Gemeren, Hinderer, and
  Steinhoff}]{Creci:2024wfu}
Creci G, van Gemeren I, Hinderer T, et~al (2025) {Tidal effects in
  gravitational waves from neutron stars in scalar-tensor theories of gravity}.
  SciPost Phys Core 8:042. \doi{10.21468/SciPostPhysCore.8.2.042},
  {\href{https://arxiv.org/abs/2412.06620}{{arXiv:2412.06620}}} {[gr-qc]}

\bibitem[{Crescimbeni et~al.(2024)Crescimbeni, Franciolini, Pani, and
  Riotto}]{Crescimbeni:2024cwh}
Crescimbeni F, Franciolini G, Pani P, et~al (2024) {Can we identify primordial
  black holes? Tidal tests for subsolar-mass gravitational-wave observations}.
  Phys Rev D 109(12):124063. \doi{10.1103/PhysRevD.109.124063},
  {\href{https://arxiv.org/abs/2402.18656}{{arXiv:2402.18656}}} {[astro-ph.HE]}

\bibitem[{Crescimbeni et~al.(2025)Crescimbeni, Franciolini, Pani, and
  Vaglio}]{Crescimbeni:2024qrq}
Crescimbeni F, Franciolini G, Pani P, et~al (2025) {Cosmology and nuclear
  physics implications of a subsolar gravitational-wave event}. Phys Rev D
  111(8):083538. \doi{10.1103/PhysRevD.111.083538},
  {\href{https://arxiv.org/abs/2408.14287}{{arXiv:2408.14287}}} {[astro-ph.HE]}

\bibitem[{Csaki(2004)}]{Csaki:2004ay}
Csaki C (2004) {TASI lectures on extra dimensions and branes}. In: {Theoretical
  Advanced Study Institute in Elementary Particle Physics (TASI 2002): Particle
  Physics and Cosmology: The Quest for Physics Beyond the Standard Model(s)},
  pp 605--698,
  {\href{https://arxiv.org/abs/hep-ph/0404096}{{arXiv:hep-ph/0404096}}}

\bibitem[{Cveti{\v{c}} et~al.(2026)Cveti{\v{c}}, Liao, and
  Stetsko}]{Cvetic:2026wht}
Cveti{\v{c}} M, Liao MA, Stetsko MM (2026) {Tidal perturbations and Love
  Symmetry for five-dimensional charged rotating black holes}. arXiv e-prints
  {\href{https://arxiv.org/abs/2601.20514}{{arXiv:2601.20514}}} {[hep-th]}

\bibitem[{Dadhich and Pons(2015)}]{Dadhich:2015nua}
Dadhich N, Pons JM (2015) {Static pure Lovelock black hole solutions with
  horizon topology S$^{(n)}\times$ S$^{(n)}$}. JHEP 05:067.
  \doi{10.1007/JHEP05(2015)067},
  {\href{https://arxiv.org/abs/1503.00974}{{arXiv:1503.00974}}} {[gr-qc]}

\bibitem[{Dadhich et~al.(2000)Dadhich, Maartens, Papadopoulos, and
  Rezania}]{Dadhich:2000am}
Dadhich N, Maartens R, Papadopoulos P, et~al (2000) {Black holes on the brane}.
  Phys Lett B 487:1--6. \doi{10.1016/S0370-2693(00)00798-X},
  {\href{https://arxiv.org/abs/hep-th/0003061}{{arXiv:hep-th/0003061}}}

\bibitem[{{Damour}(1982)}]{Damour:1982}
{Damour} T (1982) {Surface Effects in Black-Hole Physics}. In: {Ruffini} R (ed)
  Proceedings of the Second Marcel Grossmann Meeting of General Relativity.
  North Holland, Amsterdam, pp 587--608

\bibitem[{Damour(2001)}]{Damour:2001tu}
Damour T (2001) {Coalescence of two spinning black holes: an effective one-body
  approach}. Phys Rev D 64:124013. \doi{10.1103/PhysRevD.64.124013},
  {\href{https://arxiv.org/abs/gr-qc/0103018}{{arXiv:gr-qc/0103018}}}

\bibitem[{Damour(2014)}]{Damour:2012mv}
Damour T (2014) The general relativistic two body problem and the effective one
  body formalism. In: Bi{\v{c}}{\'a}k J, Ledvinka T (eds) General Relativity,
  Cosmology and Astrophysics: Perspectives 100 years after Einstein's stay in
  Prague, Fundamental Theories of Physics, vol 177. Springer, Cham, p 111--145,
  \doi{10.1007/978-3-319-06349-2_5},
  {\href{https://arxiv.org/abs/1212.3169}{{arXiv:1212.3169}}}

\bibitem[{Damour and Esposito-Farese(1993)}]{Damour:1993hw}
Damour T, Esposito-Farese G (1993) {Nonperturbative strong field effects in
  tensor - scalar theories of gravitation}. Phys Rev Lett 70:2220--2223.
  \doi{10.1103/PhysRevLett.70.2220}

\bibitem[{Damour and Lecian(2009)}]{Damour:2009va}
Damour T, Lecian OM (2009) {On the gravitational polarizability of black
  holes}. Phys Rev D 80:044017. \doi{10.1103/PhysRevD.80.044017},
  {\href{https://arxiv.org/abs/0906.3003}{{arXiv:0906.3003}}} {[gr-qc]}

\bibitem[{Damour and Nagar(2009)}]{Damour:2009vw}
Damour T, Nagar A (2009) {Relativistic tidal properties of neutron stars}. Phys
  Rev D 80:084035. \doi{10.1103/PhysRevD.80.084035},
  {\href{https://arxiv.org/abs/0906.0096}{{arXiv:0906.0096}}} {[gr-qc]}

\bibitem[{Damour and Nagar(2010)}]{Damour:2009wj}
Damour T, Nagar A (2010) {Effective One Body description of tidal effects in
  inspiralling compact binaries}. Phys Rev D 81:084016.
  \doi{10.1103/PhysRevD.81.084016},
  {\href{https://arxiv.org/abs/0911.5041}{{arXiv:0911.5041}}} {[gr-qc]}

\bibitem[{Damour and Nagar(2016)}]{Damour:2016bks}
Damour T, Nagar A (2016) The effective-one-body approach to the general
  relativistic two body problem. In: Haardt F, Gorini V, Moschella U, et~al
  (eds) Astrophysical Black Holes, Lecture Notes in Physics, vol 905. Springer,
  Cham, p 273--312, \doi{10.1007/978-3-319-19416-5_7}

\bibitem[{Damour et~al.(1991)Damour, Soffel, and Xu}]{Damour:1990pi}
Damour T, Soffel M, Xu Cm (1991) {General relativistic celestial mechanics. 1.
  Method and definition of reference systems}. Phys Rev D 43:3273--3307.
  \doi{10.1103/PhysRevD.43.3273}

\bibitem[{Damour et~al.(1992)Damour, Soffel, and Xu}]{Damour:1991yw}
Damour T, Soffel M, Xu Cm (1992) {General relativistic celestial mechanics. 2.
  Translational equations of motion}. Phys Rev D 45:1017--1044.
  \doi{10.1103/PhysRevD.45.1017}

\bibitem[{Damour et~al.(1993)Damour, Soffel, and Xu}]{Damour:1992qi}
Damour T, Soffel M, Xu Cm (1993) {General relativistic celestial mechanics. 3.
  Rotational equations of motion}. Phys Rev D 47:3124--3135.
  \doi{10.1103/PhysRevD.47.3124}

\bibitem[{Damour et~al.(1994)Damour, Soffel, and Xu}]{Damour:1993zn}
Damour T, Soffel M, Xu Cm (1994) {General relativistic celestial mechanics. 4:
  Theory of satellite motion}. Phys Rev D 49:618--635.
  \doi{10.1103/PhysRevD.49.618}

\bibitem[{Damour et~al.(1998)Damour, Iyer, and Sathyaprakash}]{Damour:1997ub}
Damour T, Iyer BR, Sathyaprakash BS (1998) {Improved filters for gravitational
  waves from inspiralling compact binaries}. Phys Rev D 57:885--907.
  \doi{10.1103/PhysRevD.57.885},
  {\href{https://arxiv.org/abs/gr-qc/9708034}{{arXiv:gr-qc/9708034}}}

\bibitem[{Damour et~al.(2000)Damour, Iyer, and Sathyaprakash}]{Damour:2000gg}
Damour T, Iyer BR, Sathyaprakash BS (2000) {Frequency domain P approximant
  filters for time truncated inspiral gravitational wave signals from compact
  binaries}. Phys Rev D 62:084036. \doi{10.1103/PhysRevD.62.084036},
  {\href{https://arxiv.org/abs/gr-qc/0001023}{{arXiv:gr-qc/0001023}}}

\bibitem[{Damour et~al.(2012)Damour, Nagar, and Villain}]{Damour:2012yf}
Damour T, Nagar A, Villain L (2012) {Measurability of the tidal polarizability
  of neutron stars in late-inspiral gravitational-wave signals}. Phys Rev D
  85:123007. \doi{10.1103/PhysRevD.85.123007},
  {\href{https://arxiv.org/abs/1203.4352}{{arXiv:1203.4352}}} {[gr-qc]}

\bibitem[{Danielsson et~al.(2017)Danielsson, Dibitetto, and
  Giri}]{Danielsson:2017riq}
Danielsson UH, Dibitetto G, Giri S (2017) {Black holes as bubbles of AdS}. JHEP
  10:171. \doi{10.1007/JHEP10(2017)171},
  {\href{https://arxiv.org/abs/1705.10172}{{arXiv:1705.10172}}} {[hep-th]}

\bibitem[{Darmois(1927)}]{Darmois1927}
Darmois G (1927) Les équations de la gravitation einsteinienne.
  Gauthier-Villars, \urlprefix\url{http://eudml.org/doc/192556}

\bibitem[{Datta(2020)}]{Datta:2020rvo}
Datta S (2020) {Tidal heating of Quantum Black Holes and their imprints on
  gravitational waves}. Phys Rev D 102(6):064040.
  \doi{10.1103/PhysRevD.102.064040},
  {\href{https://arxiv.org/abs/2002.04480}{{arXiv:2002.04480}}} {[gr-qc]}

\bibitem[{Datta(2022)}]{Datta:2021hvm}
Datta S (2022) {Probing horizon scale quantum effects with Love}. Class Quant
  Grav 39(22):225016. \doi{10.1088/1361-6382/ac9ae4},
  {\href{https://arxiv.org/abs/2107.07258}{{arXiv:2107.07258}}} {[gr-qc]}

\bibitem[{Datta and Bose(2019)}]{Datta:2019euh}
Datta S, Bose S (2019) {Probing the nature of central objects in
  extreme-mass-ratio inspirals with gravitational waves}. arXiv e-prints
  {\href{https://arxiv.org/abs/1902.01723}{{arXiv:1902.01723}}} {[gr-qc]}

\bibitem[{Datta et~al.(2020)Datta, Brito, Bose, Pani, and
  Hughes}]{Datta:2019epe}
Datta S, Brito R, Bose S, et~al (2020) {Tidal heating as a discriminator for
  horizons in extreme mass ratio inspirals}. Phys Rev D 101(4):044004.
  \doi{10.1103/PhysRevD.101.044004},
  {\href{https://arxiv.org/abs/1910.07841}{{arXiv:1910.07841}}} {[gr-qc]}

\bibitem[{Datta et~al.(2021)Datta, Phukon, and Bose}]{Datta:2020gem}
Datta S, Phukon KS, Bose S (2021) {Recognizing black holes in
  gravitational-wave observations: Challenges in telling apart impostors in
  mass-gap binaries}. Phys Rev D 104(8):084006.
  \doi{10.1103/PhysRevD.104.084006},
  {\href{https://arxiv.org/abs/2004.05974}{{arXiv:2004.05974}}} {[gr-qc]}

\bibitem[{Datta et~al.(2024)Datta, Brito, Hughes, Klinger, and
  Pani}]{Datta:2024vll}
Datta S, Brito R, Hughes SA, et~al (2024) {Tidal heating as a discriminator for
  horizons in equatorial eccentric extreme mass ratio inspirals}. Phys Rev D
  110(2):024048. \doi{10.1103/PhysRevD.110.024048},
  {\href{https://arxiv.org/abs/2404.04013}{{arXiv:2404.04013}}} {[gr-qc]}

\bibitem[{De et~al.(2018)De, Finstad, Lattimer, Brown, Berger, and
  Biwer}]{De:2018uhw}
De S, Finstad D, Lattimer JM, et~al (2018) {Tidal Deformabilities and Radii of
  Neutron Stars from the Observation of GW170817}. Phys Rev Lett 121(9):091102.
  \doi{10.1103/PhysRevLett.121.091102}, [Erratum: Phys.Rev.Lett. 121, 259902
  (2018)], {\href{https://arxiv.org/abs/1804.08583}{{arXiv:1804.08583}}}
  {[astro-ph.HE]}

\bibitem[{De~Luca and Pani(2021)}]{DeLuca:2021ite}
De~Luca V, Pani P (2021) {Tidal deformability of dressed black holes and tests
  of ultralight bosons in extended mass ranges}. JCAP 08:032.
  \doi{10.1088/1475-7516/2021/08/032},
  {\href{https://arxiv.org/abs/2106.14428}{{arXiv:2106.14428}}} {[gr-qc]}

\bibitem[{De~Luca et~al.(2023{\natexlab{a}})De~Luca, Khoury, and
  Wong}]{DeLuca:2022tkm}
De~Luca V, Khoury J, Wong SSC (2023{\natexlab{a}}) {Implications of the weak
  gravity conjecture for tidal Love numbers of black holes}. Phys Rev D
  108(4):044066. \doi{10.1103/PhysRevD.108.044066},
  {\href{https://arxiv.org/abs/2211.14325}{{arXiv:2211.14325}}} {[hep-th]}

\bibitem[{De~Luca et~al.(2023{\natexlab{b}})De~Luca, Khoury, and
  Wong}]{DeLuca:2023mio}
De~Luca V, Khoury J, Wong SSC (2023{\natexlab{b}}) {Nonlinearities in the tidal
  Love numbers of black holes}. Phys Rev D 108(2):024048.
  \doi{10.1103/PhysRevD.108.024048},
  {\href{https://arxiv.org/abs/2305.14444}{{arXiv:2305.14444}}} {[gr-qc]}

\bibitem[{De~Luca et~al.(2023{\natexlab{c}})De~Luca, Maselli, and
  Pani}]{DeLuca:2022xlz}
De~Luca V, Maselli A, Pani P (2023{\natexlab{c}}) {Modeling frequency-dependent
  tidal deformability for environmental black hole mergers}. Phys Rev D
  107(4):044058. \doi{10.1103/PhysRevD.107.044058},
  {\href{https://arxiv.org/abs/2212.03343}{{arXiv:2212.03343}}} {[gr-qc]}

\bibitem[{De~Luca et~al.(2024{\natexlab{a}})De~Luca, Franciolini, and
  Riotto}]{DeLuca:2024uju}
De~Luca V, Franciolini G, Riotto A (2024{\natexlab{a}}) {Flea on the elephant:
  Tidal Love numbers in subsolar primordial black hole searches}. Phys Rev D
  110(10):104041. \doi{10.1103/PhysRevD.110.104041},
  {\href{https://arxiv.org/abs/2408.14207}{{arXiv:2408.14207}}} {[gr-qc]}

\bibitem[{De~Luca et~al.(2024{\natexlab{b}})De~Luca, Garoffolo, Khoury, and
  Trodden}]{DeLuca:2024ufn}
De~Luca V, Garoffolo A, Khoury J, et~al (2024{\natexlab{b}}) {Tidal Love
  numbers and Green{\textquoteright}s functions in black hole spacetimes}. Phys
  Rev D 110(6):064081. \doi{10.1103/PhysRevD.110.064081},
  {\href{https://arxiv.org/abs/2407.07156}{{arXiv:2407.07156}}} {[gr-qc]}

\bibitem[{De~Luca et~al.(2025)De~Luca, Khek, Khoury, and
  Trodden}]{DeLuca:2024nih}
De~Luca V, Khek B, Khoury J, et~al (2025) {Tidal Love numbers of analog black
  holes}. Phys Rev D 111(4):044069. \doi{10.1103/PhysRevD.111.044069},
  {\href{https://arxiv.org/abs/2412.08728}{{arXiv:2412.08728}}} {[gr-qc]}

\bibitem[{De~Luca et~al.(2026)De~Luca, Khek, Khoury, and
  Trodden}]{DeLuca:2025zqr}
De~Luca V, Khek B, Khoury J, et~al (2026) {Hidden symmetries for tidal Love
  numbers: Generalities and applications to analog black holes}. Phys Rev D
  113(4):044006. \doi{10.1103/5bdk-xclx},
  {\href{https://arxiv.org/abs/2512.06082}{{arXiv:2512.06082}}} {[gr-qc]}

\bibitem[{Deka et~al.(2025)Deka, Chakraborty, Kapadia, Shaikh, and
  Ajith}]{Deka:2024ecp}
Deka U, Chakraborty S, Kapadia SJ, et~al (2025) {Probing the charge of compact
  objects with gravitational microlensing of gravitational waves}. Phys Rev D
  111(6):064028. \doi{10.1103/PhysRevD.111.064028},
  {\href{https://arxiv.org/abs/2401.06553}{{arXiv:2401.06553}}} {[gr-qc]}

\bibitem[{Del~Grosso et~al.(2023)Del~Grosso, Franciolini, Pani, and
  Urbano}]{DelGrosso:2023trq}
Del~Grosso L, Franciolini G, Pani P, et~al (2023) {Fermion soliton stars}. Phys
  Rev D 108(4):044024. \doi{10.1103/PhysRevD.108.044024},
  {\href{https://arxiv.org/abs/2301.08709}{{arXiv:2301.08709}}} {[gr-qc]}

\bibitem[{Del~Pozzo(2012)}]{DelPozzo:2011vcw}
Del~Pozzo W (2012) {Inference of the cosmological parameters from gravitational
  waves: application to second generation interferometers}. Phys Rev D
  86:043011. \doi{10.1103/PhysRevD.86.043011},
  {\href{https://arxiv.org/abs/1108.1317}{{arXiv:1108.1317}}} {[astro-ph.CO]}

\bibitem[{Del~Pozzo et~al.(2013)Del~Pozzo, Li, Agathos, Van Den~Broeck, and
  Vitale}]{DelPozzo:2013ala}
Del~Pozzo W, Li TGF, Agathos M, et~al (2013) {Demonstrating the feasibility of
  probing the neutron star equation of state with second-generation
  gravitational wave detectors}. Phys Rev Lett 111(7):071101.
  \doi{10.1103/PhysRevLett.111.071101},
  {\href{https://arxiv.org/abs/1307.8338}{{arXiv:1307.8338}}} {[gr-qc]}

\bibitem[{Del~Pozzo et~al.(2017)Del~Pozzo, Li, and
  Messenger}]{DelPozzo:2015bna}
Del~Pozzo W, Li TGF, Messenger C (2017) {Cosmological inference using only
  gravitational wave observations of binary neutron stars}. Phys Rev D
  95(4):043502. \doi{10.1103/PhysRevD.95.043502},
  {\href{https://arxiv.org/abs/1506.06590}{{arXiv:1506.06590}}} {[gr-qc]}

\bibitem[{Delsate and Steinhoff(2012)}]{Delsate:2012ky}
Delsate T, Steinhoff J (2012) {New insights on the matter-gravity coupling
  paradigm}. Phys Rev Lett 109:021101. \doi{10.1103/PhysRevLett.109.021101},
  {\href{https://arxiv.org/abs/1201.4989}{{arXiv:1201.4989}}} {[gr-qc]}

\bibitem[{Deruelle and Piran(1984)}]{Deruelle:1984hq}
Deruelle N, Piran T (eds) (1984) {Gravitational Radiation. Proceedings, Summer
  School, NATO Advanced Study Institute, Les Houches, France, June 2-21, 1982}

\bibitem[{Detweiler(1977)}]{Detweiler:1977gy}
Detweiler SL (1977) {Resonant oscillations of a rapidly rotating black hole}.
  Proc Roy Soc Lond A 352:381--395. \doi{10.1098/rspa.1977.0005}

\bibitem[{Detweiler and Lindblom(1985)}]{Detweiler:1985zz}
Detweiler SL, Lindblom L (1985) {On the nonradial pulsations of general
  relativistic stellar models}. Astrophys J 292:12--15. \doi{10.1086/163127}

\bibitem[{Dey et~al.(2020)Dey, Chakraborty, and Afshordi}]{Dey:2020lhq}
Dey R, Chakraborty S, Afshordi N (2020) {Echoes from braneworld black holes}.
  Phys Rev D 101(10):104014. \doi{10.1103/PhysRevD.101.104014},
  {\href{https://arxiv.org/abs/2001.01301}{{arXiv:2001.01301}}} {[gr-qc]}

\bibitem[{Diedrichs et~al.(2025)Diedrichs, Tsujikawa, and
  Yagi}]{Diedrichs:2025vhv}
Diedrichs RF, Tsujikawa S, Yagi K (2025) {Tidal Love numbers of neutron stars
  in Horndeski theories}. Phys Rev D 112(4):044023. \doi{10.1103/cmb4-chn3},
  {\href{https://arxiv.org/abs/2501.07998}{{arXiv:2501.07998}}} {[gr-qc]}

\bibitem[{Dietrich et~al.(2017)Dietrich, Bernuzzi, and
  Tichy}]{Dietrich:2017aum}
Dietrich T, Bernuzzi S, Tichy W (2017) {Closed-form tidal approximants for
  binary neutron star gravitational waveforms constructed from high-resolution
  numerical relativity simulations}. Phys Rev D 96(12):121501.
  \doi{10.1103/PhysRevD.96.121501},
  {\href{https://arxiv.org/abs/1706.02969}{{arXiv:1706.02969}}} {[gr-qc]}

\bibitem[{Dietrich et~al.(2019)Dietrich, Samajdar, Khan, Johnson-McDaniel,
  Dudi, and Tichy}]{Dietrich:2019kaq}
Dietrich T, Samajdar A, Khan S, et~al (2019) {Improving the NRTidal model for
  binary neutron star systems}. Phys Rev D 100(4):044003.
  \doi{10.1103/PhysRevD.100.044003},
  {\href{https://arxiv.org/abs/1905.06011}{{arXiv:1905.06011}}} {[gr-qc]}

\bibitem[{Dima et~al.(2025)Dima, Heidmann, Melis, Pani, and
  Patashuri}]{Dima:2025tjz}
Dima A, Heidmann P, Melis M, et~al (2025) {W-solitons as prototypical black
  hole microstates}. Phys Rev D 112(12):124056. \doi{10.1103/2wcq-4xny},
  {\href{https://arxiv.org/abs/2509.18245}{{arXiv:2509.18245}}} {[gr-qc]}

\bibitem[{Dolan(2008)}]{dolan2008scattering-43d}
Dolan SR (2008) Scattering and absorption of gravitational plane waves by
  rotating black holes. Class Quantum Grav 25(23):235002.
  \doi{10.1088/0264-9381/25/23/235002},
  {\href{https://arxiv.org/abs/0801.3805}{{0801.3805}}}

\bibitem[{Dyson et~al.(2025)Dyson, Spieksma, Brito, van~de Meent, and
  Dolan}]{Dyson:2025dlj}
Dyson C, Spieksma TFM, Brito R, et~al (2025) {Environmental Effects in
  Extreme-Mass-Ratio Inspirals: Perturbations to the Environment in Kerr
  Spacetimes}. Phys Rev Lett 134(21):211403.
  \doi{10.1103/PhysRevLett.134.211403},
  {\href{https://arxiv.org/abs/2501.09806}{{arXiv:2501.09806}}} {[gr-qc]}

\bibitem[{Emparan et~al.(2000)Emparan, Horowitz, and Myers}]{Emparan:1999wa}
Emparan R, Horowitz GT, Myers RC (2000) {Exact description of black holes on
  branes}. JHEP 01:007. \doi{10.1088/1126-6708/2000/01/007},
  {\href{https://arxiv.org/abs/hep-th/9911043}{{arXiv:hep-th/9911043}}}

\bibitem[{Endlich and Penco(2016)}]{Endlich:2015mke}
Endlich S, Penco R (2016) {Effective field theory approach to tidal dynamics of
  spinning astrophysical systems}. Phys Rev D 93(6):064021.
  \doi{10.1103/PhysRevD.93.064021},
  {\href{https://arxiv.org/abs/1510.08889}{{arXiv:1510.08889}}} {[gr-qc]}

\bibitem[{Ernst(1968{\natexlab{a}})}]{Ernst:1967wx}
Ernst FJ (1968{\natexlab{a}}) {New formulation of the axially symmetric
  gravitational field problem}. Phys Rev 167:1175--1179.
  \doi{10.1103/PhysRev.167.1175}

\bibitem[{Ernst(1968{\natexlab{b}})}]{Ernst:1967by}
Ernst FJ (1968{\natexlab{b}}) {New Formulation of the Axially Symmetric
  Gravitational Field Problem. II}. Phys Rev 168:1415--1417.
  \doi{10.1103/PhysRev.168.1415}

\bibitem[{Evans et~al.(2021)}]{Evans:2021gyd}
Evans M, et~al (2021) {A Horizon Study for Cosmic Explorer: Science,
  Observatories, and Community}. arXiv e-prints
  {\href{https://arxiv.org/abs/2109.09882}{{arXiv:2109.09882}}} {[astro-ph.IM]}

\bibitem[{Fabbri and Procopio(2007)}]{Fabbri:2007kr}
Fabbri A, Procopio GP (2007) {Quantum effects in black holes from the
  Schwarzschild black string?} Class Quant Grav 24:5371--5382.
  \doi{10.1088/0264-9381/24/22/003},
  {\href{https://arxiv.org/abs/0704.3728}{{arXiv:0704.3728}}} {[hep-th]}

\bibitem[{Fackerell and Crossman(1977)}]{Fackerell:1977shn}
Fackerell ED, Crossman RG (1977) {Spin-weighted angular spheroidal functions}.
  J Math Phys 18(9):1849--1854. \doi{10.1063/1.523499}

\bibitem[{Ferrari et~al.(2012)Ferrari, Gualtieri, and Maselli}]{Ferrari:2011as}
Ferrari V, Gualtieri L, Maselli A (2012) {Tidal interaction in compact
  binaries: a post-Newtonian affine framework}. Phys Rev D 85:044045.
  \doi{10.1103/PhysRevD.85.044045},
  {\href{https://arxiv.org/abs/1111.6607}{{arXiv:1111.6607}}} {[gr-qc]}

\bibitem[{Ferrer et~al.(2017)Ferrer, da~Rosa, and Will}]{Ferrer:2017xwm}
Ferrer F, da~Rosa AM, Will CM (2017) {Dark matter spikes in the vicinity of
  Kerr black holes}. Phys Rev D 96(8):083014. \doi{10.1103/PhysRevD.96.083014},
  {\href{https://arxiv.org/abs/1707.06302}{{arXiv:1707.06302}}} {[astro-ph.CO]}

\bibitem[{Figueiredo et~al.(2023)Figueiredo, Maselli, and
  Cardoso}]{Figueiredo:2023gas}
Figueiredo E, Maselli A, Cardoso V (2023) {Black holes surrounded by generic
  dark matter profiles: Appearance and gravitational-wave emission}. Phys Rev D
  107(10):104033. \doi{10.1103/PhysRevD.107.104033},
  {\href{https://arxiv.org/abs/2303.08183}{{arXiv:2303.08183}}} {[gr-qc]}

\bibitem[{Flanagan and Hinderer(2008)}]{Flanagan:2007ix}
Flanagan EE, Hinderer T (2008) {Constraining neutron star tidal Love numbers
  with gravitational wave detectors}. Phys Rev D 77:021502.
  \doi{10.1103/PhysRevD.77.021502},
  {\href{https://arxiv.org/abs/0709.1915}{{arXiv:0709.1915}}} {[astro-ph]}

\bibitem[{Flanagan and Racine(2007)}]{Flanagan:2006sb}
Flanagan EE, Racine E (2007) {Gravitomagnetic resonant excitation of Rossby
  modes in coalescing neutron star binaries}. Phys Rev D 75:044001.
  \doi{10.1103/PhysRevD.75.044001},
  {\href{https://arxiv.org/abs/gr-qc/0601029}{{arXiv:gr-qc/0601029}}}

\bibitem[{Franciolini et~al.(2022)Franciolini, Cotesta, Loutrel, Berti, Pani,
  and Riotto}]{Franciolini:2021xbq}
Franciolini G, Cotesta R, Loutrel N, et~al (2022) {How to assess the primordial
  origin of single gravitational-wave events with mass, spin, eccentricity, and
  deformability measurements}. Phys Rev D 105(6):063510.
  \doi{10.1103/PhysRevD.105.063510},
  {\href{https://arxiv.org/abs/2112.10660}{{arXiv:2112.10660}}} {[astro-ph.CO]}

\bibitem[{Franzin et~al.(2025)Franzin, Frassino, and Rocha}]{Franzin:2024cah}
Franzin E, Frassino AM, Rocha JV (2025) {Tidal Love numbers of static black
  holes in anti-de Sitter}. JHEP 12:224. \doi{10.1007/JHEP12(2024)224},
  {\href{https://arxiv.org/abs/2410.23545}{{arXiv:2410.23545}}} {[hep-th]}

\bibitem[{Friedberg et~al.(1987)Friedberg, Lee, and Pang}]{Friedberg:1986tq}
Friedberg R, Lee TD, Pang Y (1987) {Scalar Soliton Stars and Black Holes}. Phys
  Rev D 35:3658. \doi{10.1103/PhysRevD.35.3658}, [,73(1986)]

\bibitem[{Fujita(2012)}]{Fujita:2011zk}
Fujita R (2012) {Gravitational radiation for extreme mass ratio inspirals to
  the 14th post-Newtonian order}. Prog Theor Phys 127:583--590.
  \doi{10.1143/PTP.127.583},
  {\href{https://arxiv.org/abs/1104.5615}{{arXiv:1104.5615}}} {[gr-qc]}

\bibitem[{Gamba et~al.(2023)}]{Gamba:2023mww}
Gamba R, et~al (2023) {Analytically improved and numerical-relativity informed
  effective-one-body model for coalescing binary neutron stars}. arXiv e-prints
  {\href{https://arxiv.org/abs/2307.15125}{{arXiv:2307.15125}}} {[gr-qc]}

\bibitem[{Gamboa et~al.(2025)}]{Gamboa:2024hli}
Gamboa A, et~al (2025) {Accurate waveforms for eccentric, aligned-spin binary
  black holes: The multipolar effective-one-body model seobnrv5ehm}. Phys Rev D
  112(4):044038. \doi{10.1103/jxrc-z298},
  {\href{https://arxiv.org/abs/2412.12823}{{arXiv:2412.12823}}} {[gr-qc]}

\bibitem[{Gannouji and Dadhich(2014)}]{Gannouji:2013eka}
Gannouji R, Dadhich N (2014) {Stability and existence analysis of static black
  holes in pure Lovelock theories}. Class Quant Grav 31:165016.
  \doi{10.1088/0264-9381/31/16/165016},
  {\href{https://arxiv.org/abs/1311.4543}{{arXiv:1311.4543}}} {[gr-qc]}

\bibitem[{Garraffo and Giribet(2008)}]{Garraffo:2008hu}
Garraffo C, Giribet G (2008) {The Lovelock Black Holes}. Mod Phys Lett A
  23:1801--1818. \doi{10.1142/S0217732308027497},
  {\href{https://arxiv.org/abs/0805.3575}{{arXiv:0805.3575}}} {[gr-qc]}

\bibitem[{Geroch(1970)}]{Geroch:1970cd}
Geroch RP (1970) {Multipole moments. II. Curved space}. JMathPhys
  11:2580--2588. \doi{10.1063/1.1665427}

\bibitem[{Gervalle and Volkov(2024)}]{Gervalle:2024yxj}
Gervalle R, Volkov MS (2024) {Black Holes with Electroweak Hair}. Phys Rev Lett
  133(17):171402. \doi{10.1103/PhysRevLett.133.171402},
  {\href{https://arxiv.org/abs/2406.14357}{{arXiv:2406.14357}}} {[hep-th]}

\bibitem[{Gervalle and Volkov(2025)}]{Gervalle:2025awa}
Gervalle R, Volkov MS (2025) {Black holes with electroweak hair -- the detailed
  derivation}. arXiv e-prints
  {\href{https://arxiv.org/abs/2504.09304}{{arXiv:2504.09304}}} {[hep-th]}

\bibitem[{Ghosh et~al.(2026)Ghosh, Bhatt, Chakraborty, and
  Bose}]{Ghosh:2026vig}
Ghosh R, Bhatt RP, Chakraborty S, et~al (2026) {Universal Ladder Structure
  Across Scales: From Quantum to Black Hole Physics}. preprint
  {\href{https://arxiv.org/abs/2604.06249}{{arXiv:2604.06249}}} {[gr-qc]}

\bibitem[{Giri et~al.(2025)Giri, Danielsson, Lehner, and
  Pretorius}]{Giri:2024cks}
Giri S, Danielsson U, Lehner L, et~al (2025) {Exploring black hole mimickers:
  Electromagnetic and gravitational signatures of AdS black shells}. Phys Rev D
  111(2):024007. \doi{10.1103/PhysRevD.111.024007},
  {\href{https://arxiv.org/abs/2405.08062}{{arXiv:2405.08062}}} {[gr-qc]}

\bibitem[{Giudice et~al.(2016)Giudice, McCullough, and
  Urbano}]{Giudice:2016zpa}
Giudice GF, McCullough M, Urbano A (2016) {Hunting for Dark Particles with
  Gravitational Waves}. JCAP 1610(10):001. \doi{10.1088/1475-7516/2016/10/001},
  {\href{https://arxiv.org/abs/1605.01209}{{arXiv:1605.01209}}} {[hep-ph]}

\bibitem[{Goldberg et~al.(1967)Goldberg, MacFarlane, Newman, Rohrlich, and
  Sudarshan}]{Goldberg:1966uu}
Goldberg JN, MacFarlane AJ, Newman ET, et~al (1967) {Spin-$s$ spherical
  harmonics and $\eth$}. J Math Phys 8:2155. \doi{10.1063/1.1705135}

\bibitem[{Goldberger and Ross(2010)}]{Goldberger:2009qd}
Goldberger WD, Ross A (2010) {Gravitational radiative corrections from
  effective field theory}. Phys Rev D 81:124015.
  \doi{10.1103/PhysRevD.81.124015},
  {\href{https://arxiv.org/abs/0912.4254}{{arXiv:0912.4254}}} {[gr-qc]}

\bibitem[{Goldberger and Rothstein(2006{\natexlab{a}})}]{Goldberger:2004jt}
Goldberger WD, Rothstein IZ (2006{\natexlab{a}}) {An Effective field theory of
  gravity for extended objects}. Phys Rev D 73:104029.
  \doi{10.1103/PhysRevD.73.104029},
  {\href{https://arxiv.org/abs/hep-th/0409156}{{arXiv:hep-th/0409156}}}

\bibitem[{Goldberger and Rothstein(2006{\natexlab{b}})}]{Goldberger:2005cd}
Goldberger WD, Rothstein IZ (2006{\natexlab{b}}) {Dissipative effects in the
  worldline approach to black hole dynamics}. Phys Rev D 73:104030.
  \doi{10.1103/PhysRevD.73.104030},
  {\href{https://arxiv.org/abs/hep-th/0511133}{{arXiv:hep-th/0511133}}}

\bibitem[{Goldberger et~al.(2021)Goldberger, Li, and
  Rothstein}]{Goldberger:2020fot}
Goldberger WD, Li J, Rothstein IZ (2021) {Non-conservative effects on spinning
  black holes from world-line effective field theory}. JHEP 06:053.
  \doi{10.1007/JHEP06(2021)053},
  {\href{https://arxiv.org/abs/2012.14869}{{arXiv:2012.14869}}} {[hep-th]}

\bibitem[{Gondolo and Silk(1999)}]{Gondolo:1999ef}
Gondolo P, Silk J (1999) {Dark matter annihilation at the galactic center}.
  Phys Rev Lett 83:1719--1722. \doi{10.1103/PhysRevLett.83.1719},
  {\href{https://arxiv.org/abs/astro-ph/9906391}{{arXiv:astro-ph/9906391}}}

\bibitem[{Gounis et~al.(2025)Gounis, Kehagias, and Riotto}]{Gounis:2024hcm}
Gounis LR, Kehagias A, Riotto A (2025) {The vanishing of the non-linear static
  love number of Kerr black holes and the role of symmetries}. JCAP 03:002.
  \doi{10.1088/1475-7516/2025/03/002},
  {\href{https://arxiv.org/abs/2412.08249}{{arXiv:2412.08249}}} {[gr-qc]}

\bibitem[{Graham et~al.(2006)Graham, Merritt, Moore, Diemand, and
  Terzic}]{Graham:2005xx}
Graham AW, Merritt D, Moore B, et~al (2006) {Empirical models for Dark Matter
  Halos. I. Nonparametric Construction of Density Profiles and Comparison with
  Parametric Models}. Astron J 132:2685--2700. \doi{10.1086/508988},
  {\href{https://arxiv.org/abs/astro-ph/0509417}{{arXiv:astro-ph/0509417}}}

\bibitem[{Gralla(2018)}]{Gralla:2017djj}
Gralla SE (2018) {On the Ambiguity in Relativistic Tidal Deformability}. Class
  Quant Grav 35(8):085002. \doi{10.1088/1361-6382/aab186},
  {\href{https://arxiv.org/abs/1710.11096}{{arXiv:1710.11096}}} {[gr-qc]}

\bibitem[{Gueven(1980)}]{Gueven:1980be}
Gueven R (1980) {Black holes have no superhair}. Phys Rev D 22:2327.
  \doi{10.1103/PhysRevD.22.2327}

\bibitem[{Guica et~al.(2009)Guica, Hartman, Song, and
  Strominger}]{Guica:2008mu}
Guica M, Hartman T, Song W, et~al (2009) {The Kerr/CFT Correspondence}. Phys
  Rev D 80:124008. \doi{10.1103/PhysRevD.80.124008},
  {\href{https://arxiv.org/abs/0809.4266}{{arXiv:0809.4266}}} {[hep-th]}

\bibitem[{Gupta et~al.(2024)}]{Gupta:2024gun}
Gupta A, et~al (2024) {Possible causes of false general relativity violations
  in gravitational wave observations}. SciPost Phys Comm Rep
  \doi{10.21468/SciPostPhysCommRep.5},
  {\href{https://arxiv.org/abs/2405.02197}{{arXiv:2405.02197}}} {[gr-qc]}

\bibitem[{Gupta et~al.(2021)Gupta, Steinhoff, and Hinderer}]{Gupta:2020lnv}
Gupta PK, Steinhoff J, Hinderer T (2021) {Relativistic effective action of
  dynamical gravitomagnetic tides for slowly rotating neutron stars}. Phys Rev
  Res 3(1):013147. \doi{10.1103/PhysRevResearch.3.013147},
  {\href{https://arxiv.org/abs/2011.03508}{{arXiv:2011.03508}}} {[gr-qc]}

\bibitem[{Gupta et~al.(2023)Gupta, Steinhoff, and Hinderer}]{Gupta:2023oyy}
Gupta PK, Steinhoff J, Hinderer T (2023) {Effect of dynamical gravitomagnetic
  tides on measurability of tidal parameters for binary neutron stars using
  gravitational waves}. Phys Rev D 108(12):124040.
  \doi{10.1103/PhysRevD.108.124040},
  {\href{https://arxiv.org/abs/2302.11274}{{arXiv:2302.11274}}} {[gr-qc]}

\bibitem[{Gupta et~al.(2018)Gupta, Majumder, Yagi, and Yunes}]{Gupta:2017vsl}
Gupta T, Majumder B, Yagi K, et~al (2018) {I-Love-Q Relations for Neutron Stars
  in dynamical Chern Simons Gravity}. Class Quant Grav 35(2):025009.
  \doi{10.1088/1361-6382/aa9c68},
  {\href{https://arxiv.org/abs/1710.07862}{{arXiv:1710.07862}}} {[gr-qc]}

\bibitem[{Haberland et~al.(2025)Haberland, Buonanno, and
  Steinhoff}]{Haberland:2025luz}
Haberland M, Buonanno A, Steinhoff J (2025) {Modeling matter in seobnrv5thm:
  Generating fast and accurate effective-one-body waveforms for spin-aligned
  binary neutron stars}. Phys Rev D 112(8):084024. \doi{10.1103/d3ns-h77x},
  {\href{https://arxiv.org/abs/2503.18934}{{arXiv:2503.18934}}} {[gr-qc]}

\bibitem[{Hansen(1974)}]{Hansen:1974zz}
Hansen R (1974) {Multipole moments of stationary space-times}. J Math Phys
  15:46--52. \doi{10.1063/1.1666501}

\bibitem[{Hartle(1973)}]{Hartle:1973zz}
Hartle JB (1973) {Tidal Friction in Slowly Rotating Black Holes}. Phys Rev D
  8:1010--1024. \doi{10.1103/PhysRevD.8.1010}

\bibitem[{Hartle(1974)}]{Hartle:1974gy}
Hartle JB (1974) {Tidal shapes and shifts on rotating black holes}. Phys Rev D
  9:2749--2759. \doi{10.1103/PhysRevD.9.2749}

\bibitem[{Hartle and Hawking(1972)}]{Hartle:1972ya}
Hartle JB, Hawking SW (1972) {Solutions of the Einstein-Maxwell equations with
  many black holes}. Commun Math Phys 26:87--101. \doi{10.1007/BF01645696}

\bibitem[{Haud and Einasto(1989)}]{Haud:1986yj}
Haud U, Einasto J (1989) {Galactic models with massive corona I. Method}.
  Astron Astrophys 223:89--94

\bibitem[{Hawking(1976)}]{Hawking:1976ra}
Hawking S (1976) {Breakdown of Predictability in Gravitational Collapse}. Phys
  Rev D 14:2460--2473. \doi{10.1103/PhysRevD.14.2460}

\bibitem[{Hegade K.~R. et~al.(2024{\natexlab{a}})Hegade K.~R., Ripley, and
  Yunes}]{HegadeKR:2024slr}
Hegade K.~R. A, Ripley JL, Yunes N (2024{\natexlab{a}}) {Dissipative tidal
  effects to next-to-leading order and constraints on the dissipative tidal
  deformability using gravitational wave data}. Phys Rev D 110(4):044041.
  \doi{10.1103/PhysRevD.110.044041},
  {\href{https://arxiv.org/abs/2407.02584}{{arXiv:2407.02584}}} {[gr-qc]}

\bibitem[{Hegade K.~R. et~al.(2024{\natexlab{b}})Hegade K.~R., Ripley, and
  Yunes}]{HegadeKR:2024agt}
Hegade K.~R. A, Ripley JL, Yunes N (2024{\natexlab{b}}) {Dynamical tidal
  response of nonrotating relativistic stars}. Phys Rev D 109(10):104064.
  \doi{10.1103/PhysRevD.109.104064},
  {\href{https://arxiv.org/abs/2403.03254}{{arXiv:2403.03254}}} {[gr-qc]}

\bibitem[{Hegade K.~R. et~al.(2025)Hegade K.~R., Kwon, Venumadhav, Yu, and
  Yunes}]{HegadeKR:2025qwj}
Hegade K.~R. A, Kwon KJ, Venumadhav T, et~al (2025) {Relativistic and Dynamical
  Love}. arXiv e-prints
  {\href{https://arxiv.org/abs/2507.10693}{{arXiv:2507.10693}}} {[gr-qc]}

\bibitem[{Hegade K.~R. et~al.(2026)Hegade K.~R., Yang, Hippert,
  Noronha-Hostler, Noronha, and Yunes}]{HegadeKR:2026iou}
Hegade K.~R. A, Yang Y, Hippert M, et~al (2026) {Dynamical tidal response of
  neutron stars as a probe of dense-matter properties}. preprint
  {\href{https://arxiv.org/abs/2603.26886}{{arXiv:2603.26886}}} {[gr-qc]}

\bibitem[{Henry et~al.(2020{\natexlab{a}})Henry, Faye, and
  Blanchet}]{Henry:2020pzq}
Henry Q, Faye G, Blanchet L (2020{\natexlab{a}}) {Hamiltonian for tidal
  interactions in compact binary systems to next-to-next-to-leading
  post-Newtonian order}. Phys Rev D 102(12):124074.
  \doi{10.1103/PhysRevD.102.124074},
  {\href{https://arxiv.org/abs/2009.12332}{{arXiv:2009.12332}}} {[gr-qc]}

\bibitem[{Henry et~al.(2020{\natexlab{b}})Henry, Faye, and
  Blanchet}]{Henry:2020ski}
Henry Q, Faye G, Blanchet L (2020{\natexlab{b}}) {Tidal effects in the
  gravitational-wave phase evolution of compact binary systems to
  next-to-next-to-leading post-Newtonian order}. Phys Rev D 102(4):044033.
  \doi{10.1103/PhysRevD.102.044033}, [Erratum: Phys. Rev. D 108, 089901 (2023),
  Erratum: Phys. Rev. D 111, 029901 (2025)],
  {\href{https://arxiv.org/abs/2005.13367}{{arXiv:2005.13367}}} {[gr-qc]}

\bibitem[{Hernquist(1990)}]{Hernquist:1990be}
Hernquist L (1990) {An Analytical Model for Spherical Galaxies and Bulges}.
  Astrophys J 356:359. \doi{10.1086/168845}

\bibitem[{Hinderer(2008)}]{Hinderer:2007mb}
Hinderer T (2008) {Tidal Love numbers of neutron stars}. Astrophys J
  677:1216--1220. \doi{10.1086/533487}, {Erratum: {\it ibid.}
  \href{https://dx.doi.org/10.1088/0004-637X/697/1/964}{{\bf 697}, 964
  (2009)}}, {\href{https://arxiv.org/abs/0711.2420}{{arXiv:0711.2420}}}
  {[astro-ph]}

\bibitem[{Hinderer et~al.(2010{\natexlab{a}})Hinderer, Lackey, Lang, and
  Read}]{Hinderer:2009ca}
Hinderer T, Lackey BD, Lang RN, et~al (2010{\natexlab{a}}) {Tidal deformability
  of neutron stars with realistic equations of state and their gravitational
  wave signatures in binary inspiral}. Phys Rev D 81:123016.
  \doi{10.1103/PhysRevD.81.123016},
  {\href{https://arxiv.org/abs/0911.3535}{{arXiv:0911.3535}}} {[astro-ph.HE]}

\bibitem[{Hinderer et~al.(2010{\natexlab{b}})Hinderer, Lackey, Lang, and
  Read}]{Hinderer:2010ih}
Hinderer T, Lackey BD, Lang RN, et~al (2010{\natexlab{b}}) {Tidal deformability
  of neutron stars with realistic equations of state and their gravitational
  wave signatures in binary inspiral}. Phys Rev D 81(12):101--12

\bibitem[{Hinderer et~al.(2016)}]{Hinderer:2016eia}
Hinderer T, et~al (2016) {Effects of neutron-star dynamic tides on
  gravitational waveforms within the effective-one-body approach}. Phys Rev
  Lett 116(18):181101. \doi{10.1103/PhysRevLett.116.181101},
  {\href{https://arxiv.org/abs/1602.00599}{{arXiv:1602.00599}}} {[gr-qc]}

\bibitem[{Ho and Lai(1999)}]{Ho:1998hq}
Ho WCG, Lai D (1999) {Resonant tidal excitations of rotating neutron stars in
  coalescing binaries}. Mon Not Roy Astron Soc 308:153.
  \doi{10.1046/j.1365-8711.1999.02703.x},
  {\href{https://arxiv.org/abs/astro-ph/9812116}{{arXiv:astro-ph/9812116}}}

\bibitem[{Hughes(2001)}]{Hughes:2001jr}
Hughes SA (2001) {Evolution of circular, nonequatorial orbits of Kerr black
  holes due to gravitational wave emission. II. Inspiral trajectories and
  gravitational wave forms}. Phys Rev D 64:064004.
  \doi{10.1103/PhysRevD.64.064004, 10.1103/PhysRevD.88.109902}, [Erratum: Phys.
  Rev. D 88 (10), 109902 (2013)],
  {\href{https://arxiv.org/abs/gr-qc/0104041}{{arXiv:gr-qc/0104041}}} {[gr-qc]}

\bibitem[{Hui and Nicolis(2013)}]{Hui:2012qt}
Hui L, Nicolis A (2013) {No-Hair Theorem for the Galileon}. Phys Rev Lett
  110:241104. \doi{10.1103/PhysRevLett.110.241104},
  {\href{https://arxiv.org/abs/1202.1296}{{arXiv:1202.1296}}} {[hep-th]}

\bibitem[{Hui et~al.(2021)Hui, Joyce, Penco, Santoni, and
  Solomon}]{Hui:2020xxx}
Hui L, Joyce A, Penco R, et~al (2021) {Static response and Love numbers of
  Schwarzschild black holes}. JCAP 04:052. \doi{10.1088/1475-7516/2021/04/052},
  {\href{https://arxiv.org/abs/2010.00593}{{arXiv:2010.00593}}} {[hep-th]}

\bibitem[{Hui et~al.(2022{\natexlab{a}})Hui, Joyce, Penco, Santoni, and
  Solomon}]{Hui:2021vcv}
Hui L, Joyce A, Penco R, et~al (2022{\natexlab{a}}) {Ladder symmetries of black
  holes. Implications for love numbers and no-hair theorems}. JCAP 01(01):032.
  \doi{10.1088/1475-7516/2022/01/032},
  {\href{https://arxiv.org/abs/2105.01069}{{arXiv:2105.01069}}} {[hep-th]}

\bibitem[{Hui et~al.(2022{\natexlab{b}})Hui, Joyce, Penco, Santoni, and
  Solomon}]{Hui:2022vbh}
Hui L, Joyce A, Penco R, et~al (2022{\natexlab{b}}) {Near-zone symmetries of
  Kerr black holes}. JHEP 09:049. \doi{10.1007/JHEP09(2022)049},
  {\href{https://arxiv.org/abs/2203.08832}{{arXiv:2203.08832}}} {[hep-th]}

\bibitem[{Iacovelli et~al.(2022)Iacovelli, Mancarella, Foffa, and
  Maggiore}]{Iacovelli:2022bbs}
Iacovelli F, Mancarella M, Foffa S, et~al (2022) {Forecasting the Detection
  Capabilities of Third-generation Gravitational-wave Detectors Using GWFAST}.
  Astrophys J 941(2):208. \doi{10.3847/1538-4357/ac9cd4},
  {\href{https://arxiv.org/abs/2207.02771}{{arXiv:2207.02771}}} {[gr-qc]}

\bibitem[{Iacovelli et~al.(2023)Iacovelli, Mancarella, Mondal, Puecher,
  Dietrich, Gulminelli, Maggiore, and Oertel}]{Iacovelli:2023nbv}
Iacovelli F, Mancarella M, Mondal C, et~al (2023) {Nuclear physics constraints
  from binary neutron star mergers in the Einstein Telescope era}. Phys Rev D
  108(12):122006. \doi{10.1103/PhysRevD.108.122006},
  {\href{https://arxiv.org/abs/2308.12378}{{arXiv:2308.12378}}} {[gr-qc]}

\bibitem[{Ishibashi and Kodama(2003)}]{Ishibashi:2003ap}
Ishibashi A, Kodama H (2003) {Stability of higher dimensional Schwarzschild
  black holes}. Prog Theor Phys 110:901--919. \doi{10.1143/PTP.110.901},
  {\href{https://arxiv.org/abs/hep-th/0305185}{{arXiv:hep-th/0305185}}}

\bibitem[{Isoyama and Nakano(2018)}]{Isoyama:2017tbp}
Isoyama S, Nakano H (2018) {Post-Newtonian templates for binary black-hole
  inspirals: the effect of the horizon fluxes and the secular change in the
  black-hole masses and spins}. Class Quant Grav 35(2):024001.
  \doi{10.1088/1361-6382/aa96c5},
  {\href{https://arxiv.org/abs/1705.03869}{{arXiv:1705.03869}}} {[gr-qc]}

\bibitem[{Israel(1966)}]{Israel:1966rt}
Israel W (1966) {Singular hypersurfaces and thin shells in general relativity}.
  Nuovo Cim B44S10:1. \doi{10.1007/BF02710419, 10.1007/BF02712210}, [Nuovo
  Cim.B44,1(1966)]

\bibitem[{Iteanu et~al.(2025)Iteanu, Riva, Santoni, Savi{\'c}, and
  Vernizzi}]{Iteanu:2024dvx}
Iteanu S, Riva MM, Santoni L, et~al (2025) {Vanishing of quadratic Love numbers
  of Schwarzschild black holes}. JHEP 02:174. \doi{10.1007/JHEP02(2025)174},
  {\href{https://arxiv.org/abs/2410.03542}{{arXiv:2410.03542}}} {[gr-qc]}

\bibitem[{Ivanov and Zhou(2023{\natexlab{a}})}]{Ivanov:2022hlo}
Ivanov MM, Zhou Z (2023{\natexlab{a}}) {Revisiting the matching of black hole
  tidal responses: A systematic study of relativistic and logarithmic
  corrections}. Phys Rev D 107(8):084030. \doi{10.1103/PhysRevD.107.084030},
  {\href{https://arxiv.org/abs/2208.08459}{{arXiv:2208.08459}}} {[hep-th]}

\bibitem[{Ivanov and Zhou(2023{\natexlab{b}})}]{Ivanov:2022qqt}
Ivanov MM, Zhou Z (2023{\natexlab{b}}) {Vanishing of Black Hole Tidal Love
  Numbers from Scattering Amplitudes}. Phys Rev Lett 130(9):091403.
  \doi{10.1103/PhysRevLett.130.091403},
  {\href{https://arxiv.org/abs/2209.14324}{{arXiv:2209.14324}}} {[hep-th]}

\bibitem[{Ivanov et~al.(2024)Ivanov, Li, Parra-Martinez, and
  Zhou}]{Ivanov:2024sds}
Ivanov MM, Li YZ, Parra-Martinez J, et~al (2024) {Gravitational Raman
  Scattering in Effective Field Theory: A Scalar Tidal Matching at O(G3)}. Phys
  Rev Lett 132(13):131401. \doi{10.1103/PhysRevLett.132.131401}, [Erratum:
  Phys.Rev.Lett. 134, 159901 (2025)],
  {\href{https://arxiv.org/abs/2401.08752}{{arXiv:2401.08752}}} {[hep-th]}

\bibitem[{Ivanov et~al.(2025)Ivanov, Li, Parra-Martinez, and
  Zhou}]{Ivanov:2025ozg}
Ivanov MM, Li YZ, Parra-Martinez J, et~al (2025) {Resummation of Universal
  Tails in Gravitational Waveforms}. Phys Rev Lett 135(14):141401.
  \doi{10.1103/jzd1-qzkt},
  {\href{https://arxiv.org/abs/2504.07862}{{arXiv:2504.07862}}} {[hep-th]}

\bibitem[{Ivanov et~al.(2026)Ivanov, Li, Parra-Martinez, and
  Zhou}]{Ivanov:2026icp}
Ivanov MM, Li YZ, Parra-Martinez J, et~al (2026) {Gravitational Raman
  Scattering: a Systematic Toolkit for Tidal Effects in General Relativity}.
  arXiv e-prints {\href{https://arxiv.org/abs/2602.06951}{{arXiv:2602.06951}}}
  {[hep-th]}

\bibitem[{Jackson(1998)}]{Jackson:1998nia}
Jackson JD (1998) {Classical Electrodynamics}. Wiley

\bibitem[{Jacobson et~al.(2017)Jacobson, Mohd, and Sarkar}]{Jacobson:2011dz}
Jacobson T, Mohd A, Sarkar S (2017) {Membrane paradigm for
  Einstein-Gauss-Bonnet gravity}. Phys Rev D 95(6):064036.
  \doi{10.1103/PhysRevD.95.064036},
  {\href{https://arxiv.org/abs/1107.1260}{{arXiv:1107.1260}}} {[gr-qc]}

\bibitem[{Jakobsen et~al.(2024)Jakobsen, Mogull, Plefka, and
  Sauer}]{Jakobsen:2023pvx}
Jakobsen GU, Mogull G, Plefka J, et~al (2024) {Tidal effects and
  renormalization at fourth post-Minkowskian order}. Phys Rev D 109(4):L041504.
  \doi{10.1103/PhysRevD.109.L041504},
  {\href{https://arxiv.org/abs/2312.00719}{{arXiv:2312.00719}}} {[hep-th]}

\bibitem[{Jarequi et~al.(2026)Jarequi, Mitra, and Vaidya}]{Jarequi:2026cyp}
Jarequi G, Mitra S, Vaidya V (2026) {Dynamical Tidal response of compact stars
  -- An EFT approach}. preprint
  {\href{https://arxiv.org/abs/2603.12331}{{arXiv:2603.12331}}} {[gr-qc]}

\bibitem[{Jim{\'e}nez~Forteza et~al.(2018)Jim{\'e}nez~Forteza, Abdelsalhin,
  Pani, and Gualtieri}]{JimenezForteza:2018rwr}
Jim{\'e}nez~Forteza X, Abdelsalhin T, Pani P, et~al (2018) {Impact of
  high-order tidal terms on binary neutron-star waveforms}. Phys Rev D
  98(12):124014. \doi{10.1103/PhysRevD.98.124014},
  {\href{https://arxiv.org/abs/1807.08016}{{arXiv:1807.08016}}} {[gr-qc]}

\bibitem[{Johnson-Mcdaniel et~al.(2020)Johnson-Mcdaniel, Mukherjee, Kashyap,
  Ajith, Del~Pozzo, and Vitale}]{Johnson-Mcdaniel:2018cdu}
Johnson-Mcdaniel NK, Mukherjee A, Kashyap R, et~al (2020) {Constraining black
  hole mimickers with gravitational wave observations}. Phys Rev D 102:123010.
  \doi{10.1103/PhysRevD.102.123010},
  {\href{https://arxiv.org/abs/1804.08026}{{arXiv:1804.08026}}} {[gr-qc]}

\bibitem[{Kanti et~al.(1996)Kanti, Mavromatos, Rizos, Tamvakis, and
  Winstanley}]{Kanti:1995vq}
Kanti P, Mavromatos NE, Rizos J, et~al (1996) {Dilatonic black holes in higher
  curvature string gravity}. Phys Rev D 54:5049--5058.
  \doi{10.1103/PhysRevD.54.5049},
  {\href{https://arxiv.org/abs/hep-th/9511071}{{arXiv:hep-th/9511071}}}

\bibitem[{Katagiri et~al.(2023{\natexlab{a}})Katagiri, Kimura, Nakano, and
  Omukai}]{Katagiri:2022vyz}
Katagiri T, Kimura M, Nakano H, et~al (2023{\natexlab{a}}) {Vanishing Love
  numbers of black holes in general relativity: From spacetime conformal
  symmetry of a two-dimensional reduced geometry}. Phys Rev D 107(12):124030.
  \doi{10.1103/PhysRevD.107.124030},
  {\href{https://arxiv.org/abs/2209.10469}{{arXiv:2209.10469}}} {[gr-qc]}

\bibitem[{Katagiri et~al.(2023{\natexlab{b}})Katagiri, Nakano, and
  Omukai}]{Katagiri:2023yzm}
Katagiri T, Nakano H, Omukai K (2023{\natexlab{b}}) {Stability of relativistic
  tidal response against small potential modification}. Phys Rev D
  108(8):084049. \doi{10.1103/PhysRevD.108.084049},
  {\href{https://arxiv.org/abs/2304.04551}{{arXiv:2304.04551}}} {[gr-qc]}

\bibitem[{Katagiri et~al.(2024)Katagiri, Ikeda, and Cardoso}]{Katagiri:2023umb}
Katagiri T, Ikeda T, Cardoso V (2024) {Parametrized Love numbers of nonrotating
  black holes}. Phys Rev D 109(4):044067. \doi{10.1103/PhysRevD.109.044067},
  {\href{https://arxiv.org/abs/2310.19705}{{arXiv:2310.19705}}} {[gr-qc]}

\bibitem[{Katagiri et~al.(2025{\natexlab{a}})Katagiri, Cardoso, Ikeda, and
  Yagi}]{Katagiri:2024fpn}
Katagiri T, Cardoso V, Ikeda T, et~al (2025{\natexlab{a}}) {Tidal response
  beyond vacuum general relativity with a canonical definition}. Phys Rev D
  111(8):084081. \doi{10.1103/PhysRevD.111.084081},
  {\href{https://arxiv.org/abs/2410.02531}{{arXiv:2410.02531}}} {[gr-qc]}

\bibitem[{Katagiri et~al.(2025{\natexlab{b}})Katagiri, Mukkamala, and
  Yagi}]{Katagiri:2025qze}
Katagiri T, Mukkamala GR, Yagi K (2025{\natexlab{b}}) {Theoretical modeling of
  approximate universality of tidally deformed neutron stars}. Phys Rev D
  112(2):023030. \doi{10.1103/8yfc-v76l},
  {\href{https://arxiv.org/abs/2505.05429}{{arXiv:2505.05429}}} {[gr-qc]}

\bibitem[{Katagiri et~al.(2025{\natexlab{c}})Katagiri, Yagi, and
  Cardoso}]{Katagiri:2024wbg}
Katagiri T, Yagi K, Cardoso V (2025{\natexlab{c}}) {Relativistic dynamical
  tides: Subtleties and calibration}. Phys Rev D 111(8):084080.
  \doi{10.1103/PhysRevD.111.084080},
  {\href{https://arxiv.org/abs/2409.18034}{{arXiv:2409.18034}}} {[gr-qc]}

\bibitem[{Kaup(1968)}]{Kaup:1968zz}
Kaup DJ (1968) {Klein-Gordon Geon}. Phys Rev 172:1331--1342.
  \doi{10.1103/PhysRev.172.1331}

\bibitem[{Kawaguchi et~al.(2018)Kawaguchi, Kiuchi, Kyutoku, Sekiguchi, Shibata,
  and Taniguchi}]{Kawaguchi:2018gvj}
Kawaguchi K, Kiuchi K, Kyutoku K, et~al (2018) {Frequency-domain gravitational
  waveform models for inspiraling binary neutron stars}. Phys Rev D
  97(4):044044. \doi{10.1103/PhysRevD.97.044044},
  {\href{https://arxiv.org/abs/1802.06518}{{arXiv:1802.06518}}} {[gr-qc]}

\bibitem[{Kehagias and Riotto(2025)}]{Kehagias:2024rtz}
Kehagias A, Riotto A (2025) {Black holes in a gravitational field: the
  non-linear static love number of Schwarzschild black holes vanishes}. JCAP
  05:039. \doi{10.1088/1475-7516/2025/05/039},
  {\href{https://arxiv.org/abs/2410.11014}{{arXiv:2410.11014}}} {[gr-qc]}

\bibitem[{Kobayashi et~al.(2025)Kobayashi, Mukohyama, Oshita, Takahashi, and
  Yingcharoenrat}]{Kobayashi:2025vgl}
Kobayashi H, Mukohyama S, Oshita N, et~al (2025) {Dynamical Tidal Response of
  Non-rotating Black Holes: Connecting the MST Formalism and Worldline EFT}.
  arXiv e-prints {\href{https://arxiv.org/abs/2511.12580}{{arXiv:2511.12580}}}
  {[gr-qc]}

\bibitem[{Kojima(1992)}]{Kojima:1992ie}
Kojima Y (1992) {Equations governing the nonradial oscillations of a slowly
  rotating relativistic star}. Phys Rev D 46:4289--4303.
  \doi{10.1103/PhysRevD.46.4289}

\bibitem[{Kokkotas and Schmidt(1999)}]{Kokkotas:1999bd}
Kokkotas KD, Schmidt BG (1999) {Quasinormal modes of stars and black holes}.
  Living Rev Rel 2:2. \doi{10.12942/lrr-1999-2},
  {\href{https://arxiv.org/abs/gr-qc/9909058}{{arXiv:gr-qc/9909058}}} {[gr-qc]}

\bibitem[{Kol and Smolkin(2012)}]{Kol:2011vg}
Kol B, Smolkin M (2012) {Black hole stereotyping: Induced gravito-static
  polarization}. JHEP 02:010. \doi{10.1007/JHEP02(2012)010},
  {\href{https://arxiv.org/abs/1110.3764}{{arXiv:1110.3764}}} {[hep-th]}

\bibitem[{Konoplya et~al.(2016)Konoplya, Rezzolla, and
  Zhidenko}]{Konoplya:2016jvv}
Konoplya R, Rezzolla L, Zhidenko A (2016) {General parametrization of
  axisymmetric black holes in metric theories of gravity}. Phys Rev D
  93(6):064015. \doi{10.1103/PhysRevD.93.064015},
  {\href{https://arxiv.org/abs/1602.02378}{{arXiv:1602.02378}}} {[gr-qc]}

\bibitem[{Konoplya et~al.(2018)Konoplya, Stuchl{\'\i}k, and
  Zhidenko}]{Konoplya:2018arm}
Konoplya RA, Stuchl{\'\i}k Z, Zhidenko A (2018) {Axisymmetric black holes
  allowing for separation of variables in the Klein-Gordon and Hamilton-Jacobi
  equations}. Phys Rev D 97(8):084044. \doi{10.1103/PhysRevD.97.084044},
  {\href{https://arxiv.org/abs/1801.07195}{{arXiv:1801.07195}}} {[gr-qc]}

\bibitem[{Kotla\v{r}\'\i{}k and Kofro\v{n}(2022)}]{Kotlarik:2022spo}
Kotla\v{r}\'\i{}k P, Kofro\v{n} D (2022) {Black Hole Encircled by a Thin Disk:
  Fully Relativistic Solution*}. Astrophys J 941(1):25.
  \doi{10.3847/1538-4357/ac9620},
  {\href{https://arxiv.org/abs/2211.04823}{{arXiv:2211.04823}}} {[gr-qc]}

\bibitem[{Kotla\v{r}\'\i{}k et~al.(2018)Kotla\v{r}\'\i{}k, Semer\'ak, and
  \v{C}\'\i{}\v{z}ek}]{Kotlarik:2018nbd}
Kotla\v{r}\'\i{}k P, Semer\'ak O, \v{C}\'\i{}\v{z}ek P (2018) {Schwarzschild
  black hole encircled by a rotating thin disc: Properties of perturbative
  solution}. Phys Rev D 97(8):084006. \doi{10.1103/PhysRevD.97.084006},
  {\href{https://arxiv.org/abs/1804.02010}{{arXiv:1804.02010}}} {[gr-qc]}

\bibitem[{Krishnendu and Chakraborty(2024)}]{Krishnendu:2024jkj}
Krishnendu NV, Chakraborty S (2024) {Probing black hole charge from the binary
  black hole inspiral}. Phys Rev D 109(12):124047.
  \doi{10.1103/PhysRevD.109.124047},
  {\href{https://arxiv.org/abs/2402.15336}{{arXiv:2402.15336}}} {[gr-qc]}

\bibitem[{Krishnendu et~al.(2025)Krishnendu, Perri, Chakraborty, and
  Pesci}]{Krishnendu:2025byo}
Krishnendu NV, Perri A, Chakraborty S, et~al (2025) {Probing the existence of a
  minimal length through compact binary inspiral}. arXiv e-prints
  {\href{https://arxiv.org/abs/2505.22877}{{arXiv:2505.22877}}} {[gr-qc]}

\bibitem[{Lackey et~al.(2019)Lackey, P{\"u}rrer, Taracchini, and
  Marsat}]{Lackey:2018zvw}
Lackey BD, P{\"u}rrer M, Taracchini A, et~al (2019) {Surrogate model for an
  aligned-spin effective one body waveform model of binary neutron star
  inspirals using Gaussian process regression}. Phys Rev D 100(2):024002.
  \doi{10.1103/PhysRevD.100.024002},
  {\href{https://arxiv.org/abs/1812.08643}{{arXiv:1812.08643}}} {[gr-qc]}

\bibitem[{Lai(1994)}]{Lai:1993di}
Lai D (1994) {Resonant oscillations and tidal heating in coalescing binary
  neutron stars}. Mon Not Roy Astron Soc 270:611.
  \doi{10.1093/mnras/270.3.611},
  {\href{https://arxiv.org/abs/astro-ph/9404062}{{arXiv:astro-ph/9404062}}}

\bibitem[{Lai(1997)}]{Lai:1997wh}
Lai D (1997) {Dynamical tides in rotating binary stars}. Astrophys J 490:847.
  \doi{10.1086/304899},
  {\href{https://arxiv.org/abs/astro-ph/9704132}{{arXiv:astro-ph/9704132}}}

\bibitem[{Lan et~al.(2023)Lan, Yang, Guo, and Miao}]{Lan:2023cvz}
Lan C, Yang H, Guo Y, et~al (2023) {Regular Black Holes: A Short Topic Review}.
  Int J Theor Phys 62(9):202. \doi{10.1007/s10773-023-05454-1},
  {\href{https://arxiv.org/abs/2303.11696}{{arXiv:2303.11696}}} {[gr-qc]}

\bibitem[{Landry and Poisson(2014)}]{Landry:2014jka}
Landry P, Poisson E (2014) {Relativistic theory of surficial Love numbers}.
  Phys Rev D 89(12):124011. \doi{10.1103/PhysRevD.89.124011},
  {\href{https://arxiv.org/abs/1404.6798}{{arXiv:1404.6798}}} {[gr-qc]}

\bibitem[{Landry and Poisson(2015{\natexlab{a}})}]{Landry:2015cva}
Landry P, Poisson E (2015{\natexlab{a}}) {Gravitomagnetic response of an
  irrotational body to an applied tidal field}. Phys Rev D 91(10):104026.
  \doi{10.1103/PhysRevD.91.104026},
  {\href{https://arxiv.org/abs/1504.06606}{{arXiv:1504.06606}}} {[gr-qc]}

\bibitem[{Landry and Poisson(2015{\natexlab{b}})}]{Landry:2015zfa}
Landry P, Poisson E (2015{\natexlab{b}}) {Tidal deformation of a slowly
  rotating material body. External metric}. Phys Rev D 91:104018.
  \doi{10.1103/PhysRevD.91.104018},
  {\href{https://arxiv.org/abs/1503.07366}{{arXiv:1503.07366}}} {[gr-qc]}

\bibitem[{Le~Tiec and Casals(2021)}]{LeTiec:2020spy}
Le~Tiec A, Casals M (2021) {Spinning Black Holes Fall in Love}. Phys Rev Lett
  126(13):131102. \doi{10.1103/PhysRevLett.126.131102},
  {\href{https://arxiv.org/abs/2007.00214}{{arXiv:2007.00214}}} {[gr-qc]}

\bibitem[{Le~Tiec et~al.(2021)Le~Tiec, Casals, and Franzin}]{LeTiec:2020bos}
Le~Tiec A, Casals M, Franzin E (2021) {Tidal Love Numbers of Kerr Black Holes}.
  Phys Rev D 103(8):084021. \doi{10.1103/PhysRevD.103.084021},
  {\href{https://arxiv.org/abs/2010.15795}{{arXiv:2010.15795}}} {[gr-qc]}

\bibitem[{Lee(1977)}]{Lee:1977gk}
Lee CH (1977) {Massive Spin 1/2 Wave Around a Kerr-Newman Black Hole}. Phys
  Lett B 68:152--156. \doi{10.1016/0370-2693(77)90189-7}

\bibitem[{Levi(2020)}]{Levi:2018nxp}
Levi M (2020) {Effective Field Theories of Post-Newtonian Gravity: A
  comprehensive review}. Rept Prog Phys 83(7):075901.
  \doi{10.1088/1361-6633/ab12bc},
  {\href{https://arxiv.org/abs/1807.01699}{{arXiv:1807.01699}}} {[hep-th]}

\bibitem[{Li et~al.(2023)Li, Wagle, Chen, and Yunes}]{Li:2022pcy}
Li D, Wagle P, Chen Y, et~al (2023) {Perturbations of Spinning Black Holes
  beyond General Relativity: Modified Teukolsky Equation}. Phys Rev X
  13(2):021029. \doi{10.1103/PhysRevX.13.021029},
  {\href{https://arxiv.org/abs/2206.10652}{{arXiv:2206.10652}}} {[gr-qc]}

\bibitem[{Liebling and Palenzuela(2023)}]{Liebling:2012fv}
Liebling SL, Palenzuela C (2023) {Dynamical boson stars}. Living Rev Rel
  26(1):1. \doi{10.1007/s41114-023-00043-4},
  {\href{https://arxiv.org/abs/1202.5809}{{arXiv:1202.5809}}} {[gr-qc]}

\bibitem[{Liu and Zhang(2025)}]{Liu:2025iby}
Liu Y, Zhang X (2025) {Quasinormal modes and tidal Love numbers of covariant
  effective quantum black holes with cosmological constant}. arXiv e-prints
  {\href{https://arxiv.org/abs/2509.12013}{{arXiv:2509.12013}}} {[gr-qc]}

\bibitem[{Love(1909)}]{Love1909}
Love AEH (1909) The yielding of the earth to disturbing forces. Proceedings of
  the Royal Society of London Series A, Containing Papers of a Mathematical and
  Physical Character 82(551):73--88. \doi{10.1098/rspa.1909.0008}

\bibitem[{Lovelock(1971)}]{Lovelock:1971yv}
Lovelock D (1971) {The Einstein tensor and its generalizations}. J Math Phys
  12:498--501. \doi{10.1063/1.1665613}

\bibitem[{Lupsasca(2025)}]{Lupsasca:2025pnt}
Lupsasca A (2025) {Why there is no Love in black holes}. arXiv e-prints
  {\href{https://arxiv.org/abs/2506.05298}{{arXiv:2506.05298}}} {[gr-qc]}

\bibitem[{Ma et~al.(2025)Ma, Wu, Pang, and Lu}]{Ma:2024few}
Ma L, Wu ZH, Pang Y, et~al (2025) {Charging the Love numbers: Charged scalar
  response coefficients of Kerr-Newman black holes}. Phys Rev D 111(4):044003.
  \doi{10.1103/PhysRevD.111.044003},
  {\href{https://arxiv.org/abs/2408.10352}{{arXiv:2408.10352}}} {[gr-qc]}

\bibitem[{Ma et~al.(2021)Ma, Yu, and Chen}]{Ma:2020oni}
Ma S, Yu H, Chen Y (2021) {Detecting resonant tidal excitations of Rossby modes
  in coalescing neutron-star binaries with third-generation gravitational-wave
  detectors}. Phys Rev D 103(6):063020. \doi{10.1103/PhysRevD.103.063020},
  {\href{https://arxiv.org/abs/2010.03066}{{arXiv:2010.03066}}} {[gr-qc]}

\bibitem[{Maartens(2004)}]{Maartens:2003tw}
Maartens R (2004) {Brane world gravity}. Living Rev Rel 7:7.
  \doi{10.12942/lrr-2004-7},
  {\href{https://arxiv.org/abs/gr-qc/0312059}{{arXiv:gr-qc/0312059}}}

\bibitem[{Maggio et~al.(2019)Maggio, Cardoso, Dolan, and Pani}]{Maggio:2018ivz}
Maggio E, Cardoso V, Dolan SR, et~al (2019) {Ergoregion instability of exotic
  compact objects: electromagnetic and gravitational perturbations and the role
  of absorption}. Phys Rev D 99(6):064007. \doi{10.1103/PhysRevD.99.064007},
  {\href{https://arxiv.org/abs/1807.08840}{{arXiv:1807.08840}}} {[gr-qc]}

\bibitem[{Maggio et~al.(2020)Maggio, Buoninfante, Mazumdar, and
  Pani}]{Maggio:2020jml}
Maggio E, Buoninfante L, Mazumdar A, et~al (2020) {How does a dark compact
  object ringdown?} Phys Rev D 102(6):064053.
  \doi{10.1103/PhysRevD.102.064053},
  {\href{https://arxiv.org/abs/2006.14628}{{arXiv:2006.14628}}} {[gr-qc]}

\bibitem[{Maggio et~al.(2021{\natexlab{a}})Maggio, van~de Meent, and
  Pani}]{Maggio:2021uge}
Maggio E, van~de Meent M, Pani P (2021{\natexlab{a}}) {Extreme mass-ratio
  inspirals around a spinning horizonless compact object}. Phys Rev D
  104(10):104026. \doi{10.1103/PhysRevD.104.104026},
  {\href{https://arxiv.org/abs/2106.07195}{{arXiv:2106.07195}}} {[gr-qc]}

\bibitem[{Maggio et~al.(2021{\natexlab{b}})Maggio, Pani, and
  Raposo}]{Maggio:2021ans}
Maggio E, Pani P, Raposo G (2021{\natexlab{b}}) {Testing the nature of dark
  compact objects with gravitational waves}. arXiv e-prints
  {\href{https://arxiv.org/abs/2105.06410}{{arXiv:2105.06410}}} {[gr-qc]}

\bibitem[{Maggiore et~al.(2020)}]{ET:2019dnz}
Maggiore M, et~al (2020) {Science Case for the Einstein Telescope}. JCAP
  03:050. \doi{10.1088/1475-7516/2020/03/050},
  {\href{https://arxiv.org/abs/1912.02622}{{arXiv:1912.02622}}} {[astro-ph.CO]}

\bibitem[{Mandal et~al.(2024)Mandal, Mastrolia, Silva, Patil, and
  Steinhoff}]{Mandal:2023hqa}
Mandal MK, Mastrolia P, Silva HO, et~al (2024) {Renormalizing Love: tidal
  effects at the third post-Newtonian order}. JHEP 02:188.
  \doi{10.1007/JHEP02(2024)188},
  {\href{https://arxiv.org/abs/2308.01865}{{arXiv:2308.01865}}} {[hep-th]}

\bibitem[{Mano and Takasugi(1997)}]{Mano:1996gn}
Mano S, Takasugi E (1997) {Analytic solutions of the Teukolsky equation and
  their properties}. Prog Theor Phys 97:213--232. \doi{10.1143/PTP.97.213},
  {\href{https://arxiv.org/abs/gr-qc/9611014}{{arXiv:gr-qc/9611014}}}

\bibitem[{Mano et~al.(1996{\natexlab{a}})Mano, Suzuki, and
  Takasugi}]{Mano:1996mf}
Mano S, Suzuki H, Takasugi E (1996{\natexlab{a}}) {Analytic solutions of the
  Regge-Wheeler equation and the postMinkowskian expansion}. Prog Theor Phys
  96:549--566. \doi{10.1143/PTP.96.549},
  {\href{https://arxiv.org/abs/gr-qc/9605057}{{arXiv:gr-qc/9605057}}}

\bibitem[{Mano et~al.(1996{\natexlab{b}})Mano, Suzuki, and
  Takasugi}]{Mano:1996vt}
Mano S, Suzuki H, Takasugi E (1996{\natexlab{b}}) {Analytic solutions of the
  Teukolsky equation and their low frequency expansions}. Prog Theor Phys
  95:1079--1096. \doi{10.1143/PTP.95.1079},
  {\href{https://arxiv.org/abs/gr-qc/9603020}{{arXiv:gr-qc/9603020}}}

\bibitem[{Mart{\'\i}nez-Rodr{\'\i}guez(2026)}]{Martinez-Rodriguez:2026omk}
Mart{\'\i}nez-Rodr{\'\i}guez I (2026) {Neutron Stars as Perfect Fluids:
  Extracting the Linearized Response Function}. arXiv e-prints
  {\href{https://arxiv.org/abs/2602.07115}{{arXiv:2602.07115}}} {[gr-qc]}

\bibitem[{Maselli et~al.(2012)Maselli, Gualtieri, Pannarale, and
  Ferrari}]{Maselli:2012zq}
Maselli A, Gualtieri L, Pannarale F, et~al (2012) {On the validity of the
  adiabatic approximation in compact binary inspirals}. Phys Rev D 86:044032.
  \doi{10.1103/PhysRevD.86.044032},
  {\href{https://arxiv.org/abs/1205.7006}{{arXiv:1205.7006}}} {[gr-qc]}

\bibitem[{Maselli et~al.(2018)Maselli, Pani, Cardoso, Abdelsalhin, Gualtieri,
  and Ferrari}]{Maselli:2017cmm}
Maselli A, Pani P, Cardoso V, et~al (2018) {Probing Planckian corrections at
  the horizon scale with LISA binaries}. Phys Rev Lett 120(8):081101.
  \doi{10.1103/PhysRevLett.120.081101},
  {\href{https://arxiv.org/abs/1703.10612}{{arXiv:1703.10612}}} {[gr-qc]}

\bibitem[{Maselli et~al.(2019)Maselli, Pani, Cardoso, Abdelsalhin, Gualtieri,
  and Ferrari}]{Maselli:2018fay}
Maselli A, Pani P, Cardoso V, et~al (2019) {From micro to macro and back:
  probing near-horizon quantum structures with gravitational waves}. Class
  Quant Grav 36(16):167001. \doi{10.1088/1361-6382/ab30ff},
  {\href{https://arxiv.org/abs/1811.03689}{{arXiv:1811.03689}}} {[gr-qc]}

\bibitem[{{Mashhoon}(1973)}]{1973ApJ...185...83M}
{Mashhoon} B (1973) {Tidal Gravitational Radiation}. \apj 185:83--86.
  \doi{10.1086/152397}

\bibitem[{Mastrogiovanni et~al.(2024)Mastrogiovanni, Karathanasis, Gair,
  Ashton, Rinaldi, Huang, and D{\'a}lya}]{Mastrogiovanni:2024mqc}
Mastrogiovanni S, Karathanasis C, Gair J, et~al (2024) {Cosmology with
  Gravitational Waves: A Review}. Ann Phys 536(2):2200180.
  \doi{10.1002/andp.202200180}

\bibitem[{Mathur(2009)}]{Mathur:2009hf}
Mathur SD (2009) {The Information paradox: A Pedagogical introduction}. Class
  Quant Grav 26:224001. \doi{10.1088/0264-9381/26/22/224001},
  {\href{https://arxiv.org/abs/0909.1038}{{arXiv:0909.1038}}} {[hep-th]}

\bibitem[{Mazur and Mottola(2004)}]{Mazur:2004fk}
Mazur PO, Mottola E (2004) {Gravitational vacuum condensate stars}. Proc Nat
  Acad Sci 101:9545--9550. \doi{10.1073/pnas.0402717101},
  {\href{https://arxiv.org/abs/gr-qc/0407075}{{arXiv:gr-qc/0407075}}} {[gr-qc]}

\bibitem[{Mazur and Mottola(2023)}]{Mazur:2001fv}
Mazur PO, Mottola E (2023) {Gravitational Condensate Stars: An Alternative to
  Black Holes}. Universe 9(2):88. \doi{10.3390/universe9020088},
  {\href{https://arxiv.org/abs/gr-qc/0109035}{{arXiv:gr-qc/0109035}}}

\bibitem[{Messenger and Read(2012)}]{Messenger:2011gi}
Messenger C, Read J (2012) {Measuring a cosmological distance-redshift
  relationship using only gravitational wave observations of binary neutron
  star coalescences}. Phys Rev Lett 108:091101.
  \doi{10.1103/PhysRevLett.108.091101},
  {\href{https://arxiv.org/abs/1107.5725}{{arXiv:1107.5725}}} {[gr-qc]}

\bibitem[{Mitra et~al.(2025)Mitra, Speeney, Chakraborty, and
  Berti}]{Mitra:2025tag}
Mitra S, Speeney N, Chakraborty S, et~al (2025) {Extreme mass ratio inspirals
  in rotating dark matter spikes}. Phys Rev D 112(4):044030.
  \doi{10.1103/ycl1-kx7d},
  {\href{https://arxiv.org/abs/2505.04697}{{arXiv:2505.04697}}} {[gr-qc]}

\bibitem[{Mondal and Bagchi(2024)}]{Mondal:2023wwo}
Mondal S, Bagchi M (2024) {f-mode oscillations of anisotropic neutron stars in
  full general relativity}. Phys Rev D 110(12):123011.
  \doi{10.1103/PhysRevD.110.123011},
  {\href{https://arxiv.org/abs/2309.00439}{{arXiv:2309.00439}}} {[gr-qc]}

\bibitem[{Motaharfar and Singh(2025{\natexlab{a}})}]{Motaharfar:2025typ}
Motaharfar M, Singh P (2025{\natexlab{a}}) {Loop quantum gravitational
  signatures via Love numbers}. Phys Rev D 111(10):106018.
  \doi{10.1103/PhysRevD.111.106018},
  {\href{https://arxiv.org/abs/2501.09151}{{arXiv:2501.09151}}} {[gr-qc]}

\bibitem[{Motaharfar and Singh(2025{\natexlab{b}})}]{Motaharfar:2025ihv}
Motaharfar M, Singh P (2025{\natexlab{b}}) {Love numbers of covariant loop
  quantum black holes}. Phys Rev D 112(6):066008. \doi{10.1103/13lp-hydg},
  {\href{https://arxiv.org/abs/2505.14784}{{arXiv:2505.14784}}} {[gr-qc]}

\bibitem[{Mougiakakos and Vanhove(2024)}]{Mougiakakos:2024nku}
Mougiakakos S, Vanhove P (2024) {Schwarzschild Metric from Scattering
  Amplitudes to All Orders in GN}. Phys Rev Lett 133(11):111601.
  \doi{10.1103/PhysRevLett.133.111601},
  {\href{https://arxiv.org/abs/2405.14421}{{arXiv:2405.14421}}} {[hep-th]}

\bibitem[{Mougiakakos et~al.(2022)Mougiakakos, Riva, and
  Vernizzi}]{Mougiakakos:2022sic}
Mougiakakos S, Riva MM, Vernizzi F (2022) {Gravitational Bremsstrahlung with
  Tidal Effects in the Post-Minkowskian Expansion}. Phys Rev Lett
  129(12):121101. \doi{10.1103/PhysRevLett.129.121101},
  {\href{https://arxiv.org/abs/2204.06556}{{arXiv:2204.06556}}} {[hep-th]}

\bibitem[{Mukherjee et~al.(2022)Mukherjee, Datta, Tiwari, Phukon, and
  Bose}]{Mukherjee:2022wws}
Mukherjee S, Datta S, Tiwari S, et~al (2022) {Toward establishing the presence
  or absence of horizons in coalescing binaries of compact objects by using
  their gravitational wave signals}. Phys Rev D 106(10):104032.
  \doi{10.1103/PhysRevD.106.104032},
  {\href{https://arxiv.org/abs/2202.08661}{{arXiv:2202.08661}}} {[gr-qc]}

\bibitem[{Mukherjee et~al.(2025)Mukherjee, Datta, Bose, and
  Phukon}]{Mukherjee:2025wxa}
Mukherjee S, Datta S, Bose S, et~al (2025) {Binary black holes in the heat of
  merger} {\href{https://arxiv.org/abs/2506.22363}{{arXiv:2506.22363}}}
  {[gr-qc]}

\bibitem[{Myers and Perry(1986)}]{Myers:1986un}
Myers RC, Perry MJ (1986) {Black Holes in Higher Dimensional Space-Times}.
  Annals Phys 172:304. \doi{10.1016/0003-4916(86)90186-7}

\bibitem[{Nagar and Rettegno(2019)}]{Nagar:2018gnk}
Nagar A, Rettegno P (2019) {Efficient effective one body time-domain
  gravitational waveforms}. Phys Rev D 99(2):021501.
  \doi{10.1103/PhysRevD.99.021501},
  {\href{https://arxiv.org/abs/1805.03891}{{arXiv:1805.03891}}} {[gr-qc]}

\bibitem[{Nagar et~al.(2018)}]{Nagar:2018zoe}
Nagar A, et~al (2018) {Time-domain effective-one-body gravitational waveforms
  for coalescing compact binaries with nonprecessing spins, tides and self-spin
  effects}. Phys Rev D 98(10):104052. \doi{10.1103/PhysRevD.98.104052},
  {\href{https://arxiv.org/abs/1806.01772}{{arXiv:1806.01772}}} {[gr-qc]}

\bibitem[{Nair et~al.(2023)Nair, Chakraborty, and Sarkar}]{Nair:2022xfm}
Nair S, Chakraborty S, Sarkar S (2023) {Dynamical Love numbers for area
  quantized black holes}. Phys Rev D 107(12):124041.
  \doi{10.1103/PhysRevD.107.124041},
  {\href{https://arxiv.org/abs/2208.06235}{{arXiv:2208.06235}}} {[gr-qc]}

\bibitem[{Nair et~al.(2024)Nair, Chakraborty, and
  Sarkar}]{nair2024asymptotically-199}
Nair S, Chakraborty S, Sarkar S (2024) Asymptotically de sitter black holes
  have nonzero tidal love numbers. Phys Rev D 109(6):064025.
  \doi{10.1103/physrevd.109.064025},
  {\href{https://arxiv.org/abs/2401.06467}{{2401.06467}}}

\bibitem[{{Navarro} et~al.(1996){Navarro}, {Frenk}, and {White}}]{NFW}
{Navarro} JF, {Frenk} CS, {White} SDM (1996) {The Structure of Cold Dark Matter
  Halos}. \apj 462:563

\bibitem[{Newman and Penrose(1962)}]{Newman:1961qr}
Newman E, Penrose R (1962) {An Approach to gravitational radiation by a method
  of spin coefficients}. J Math Phys 3:566--578. \doi{10.1063/1.1724257}

\bibitem[{Olver et~al.(2010)Olver, Lozier, Boisvert, and Clark}]{DLMF}
Olver FWJ, Lozier DW, Boisvert RF, et~al (2010) {NIST Handbook of Mathematical
  Functions}

\bibitem[{Ori(1991)}]{Ori:1991zz}
Ori A (1991) {Inner structure of a charged black hole: An exact mass-inflation
  solution}. Phys Rev Lett 67:789--792. \doi{10.1103/PhysRevLett.67.789}

\bibitem[{Oshita et~al.(2020)Oshita, Wang, and Afshordi}]{Oshita:2019sat}
Oshita N, Wang Q, Afshordi N (2020) {On Reflectivity of Quantum Black Hole
  Horizons}. JCAP 04:016. \doi{10.1088/1475-7516/2020/04/016},
  {\href{https://arxiv.org/abs/1905.00464}{{arXiv:1905.00464}}} {[hep-th]}

\bibitem[{Pacilio et~al.(2020)Pacilio, Vaglio, Maselli, and
  Pani}]{Pacilio:2020jza}
Pacilio C, Vaglio M, Maselli A, et~al (2020) {Gravitational-wave detectors as
  particle-physics laboratories: Constraining scalar interactions with a
  coherent inspiral model of boson-star binaries}. Phys Rev D 102(8):083002.
  \doi{10.1103/PhysRevD.102.083002},
  {\href{https://arxiv.org/abs/2007.05264}{{arXiv:2007.05264}}} {[gr-qc]}

\bibitem[{Pacilio et~al.(2022)Pacilio, Maselli, Fasano, and
  Pani}]{Pacilio:2021jmq}
Pacilio C, Maselli A, Fasano M, et~al (2022) {Ranking Love Numbers for the
  Neutron Star Equation of State: The Need for Third-Generation Detectors}.
  Phys Rev Lett 128(10):101101. \doi{10.1103/PhysRevLett.128.101101},
  {\href{https://arxiv.org/abs/2104.10035}{{arXiv:2104.10035}}} {[gr-qc]}

\bibitem[{Padmanabhan and Kothawala(2013)}]{Padmanabhan:2013xyr}
Padmanabhan T, Kothawala D (2013) {Lanczos-Lovelock models of gravity}. Phys
  Rept 531:115--171. \doi{10.1016/j.physrep.2013.05.007},
  {\href{https://arxiv.org/abs/1302.2151}{{arXiv:1302.2151}}} {[gr-qc]}

\bibitem[{Page(1976{\natexlab{a}})}]{page1976particle-286}
Page DN (1976{\natexlab{a}}) Particle emission rates from a black hole. {II}.
  massless particles from a rotating hole. Phys Rev D 14(12):3260--3273.
  \doi{10.1103/physrevd.14.3260}

\bibitem[{Page(1976{\natexlab{b}})}]{page1976particle-25c}
Page DN (1976{\natexlab{b}}) Particle emission rates from a black hole:
  Massless particles from an uncharged, nonrotating hole. Physical Review D
  13(2):198--206. \doi{10.1103/physrevd.13.198}

\bibitem[{Pang et~al.(2025)Pang, Tian, Zhang, and Jiang}]{Pang:2025myy}
Pang X, Tian Y, Zhang H, et~al (2025) {Fermionic Love number of
  Reissner-Nordstr{\"o}m black holes}. arXiv e-prints
  {\href{https://arxiv.org/abs/2510.10036}{{arXiv:2510.10036}}} {[gr-qc]}

\bibitem[{Pani(2013)}]{Pani:2013pma}
Pani P (2013) {Advanced Methods in Black-Hole Perturbation Theory}. Int J Mod
  Phys A28:1340018. \doi{10.1142/S0217751X13400186},
  {\href{https://arxiv.org/abs/1305.6759}{{arXiv:1305.6759}}} {[gr-qc]}

\bibitem[{Pani(2015)}]{Pani:2015tga}
Pani P (2015) {I-Love-Q relations for gravastars and the approach to the
  black-hole limit}. Phys Rev D 92(12):124030.
  \doi{10.1103/PhysRevD.92.124030},
  {\href{https://arxiv.org/abs/1506.06050}{{arXiv:1506.06050}}} {[gr-qc]}

\bibitem[{Pani and Berti(2014)}]{Pani:2014jra}
Pani P, Berti E (2014) {Slowly rotating neutron stars in scalar-tensor
  theories}. Phys Rev D 90(2):024025. \doi{10.1103/PhysRevD.90.024025},
  {\href{https://arxiv.org/abs/1405.4547}{{arXiv:1405.4547}}} {[gr-qc]}

\bibitem[{Pani and Maselli(2019)}]{Pani:2019cyc}
Pani P, Maselli A (2019) {Love in Extrema Ratio}. Int J Mod Phys D
  28(14):1944001. \doi{10.1142/S0218271819440012},
  {\href{https://arxiv.org/abs/1905.03947}{{arXiv:1905.03947}}} {[gr-qc]}

\bibitem[{Pani et~al.(2009)Pani, Berti, Cardoso, Chen, and Norte}]{Pani:2009ss}
Pani P, Berti E, Cardoso V, et~al (2009) {Gravitational wave signatures of the
  absence of an event horizon. I. Nonradial oscillations of a thin-shell
  gravastar}. Phys Rev D 80:124047. \doi{10.1103/PhysRevD.80.124047},
  {\href{https://arxiv.org/abs/0909.0287}{{arXiv:0909.0287}}} {[gr-qc]}

\bibitem[{Pani et~al.(2015{\natexlab{a}})Pani, Gualtieri, and
  Ferrari}]{Pani:2015nua}
Pani P, Gualtieri L, Ferrari V (2015{\natexlab{a}}) {Tidal Love numbers of a
  slowly spinning neutron star}. Phys Rev D 92:124003.
  \doi{10.1103/PhysRevD.92.124003},
  {\href{https://arxiv.org/abs/1509.02171}{{arXiv:1509.02171}}} {[gr-qc]}

\bibitem[{Pani et~al.(2015{\natexlab{b}})Pani, Gualtieri, Maselli, and
  Ferrari}]{Pani:2015hfa}
Pani P, Gualtieri L, Maselli A, et~al (2015{\natexlab{b}}) {Tidal deformations
  of a spinning compact object}. Phys Rev D 92(2):024010.
  \doi{10.1103/PhysRevD.92.024010},
  {\href{https://arxiv.org/abs/1503.07365}{{arXiv:1503.07365}}} {[gr-qc]}

\bibitem[{Pani et~al.(2018)Pani, Gualtieri, Abdelsalhin, and
  Jim\'enez-Forteza}]{Pani:2018inf}
Pani P, Gualtieri L, Abdelsalhin T, et~al (2018) {Magnetic tidal Love numbers
  clarified}. Phys Rev D 98(12):124023. \doi{10.1103/PhysRevD.98.124023},
  {\href{https://arxiv.org/abs/1810.01094}{{arXiv:1810.01094}}} {[gr-qc]}

\bibitem[{Pani et~al.(2025)Pani, Riva, Santoni, Savi{\'c}, and
  Vernizzi}]{Pani:2025qxs}
Pani P, Riva MM, Santoni L, et~al (2025) {Nonlinear Relativistic Tidal Response
  of Neutron Stars}. arXiv e-prints
  {\href{https://arxiv.org/abs/2512.14663}{{arXiv:2512.14663}}} {[gr-qc]}

\bibitem[{Parra-Martinez and Podo(2025)}]{Parra-Martinez:2025bcu}
Parra-Martinez J, Podo A (2025) {Naturalness of vanishing black-hole tides}.
  arXiv e-prints {\href{https://arxiv.org/abs/2510.20694}{{arXiv:2510.20694}}}
  {[hep-th]}

\bibitem[{Passamonti et~al.(2021)Passamonti, Andersson, and
  Pnigouras}]{Passamonti:2020fur}
Passamonti A, Andersson N, Pnigouras P (2021) {Dynamical tides in neutron
  stars: The impact of the crust}. Mon Not Roy Astron Soc 504(1):1273--1293.
  \doi{10.1093/mnras/stab870},
  {\href{https://arxiv.org/abs/2012.09637}{{arXiv:2012.09637}}} {[astro-ph.HE]}

\bibitem[{Passamonti et~al.(2022)Passamonti, Andersson, and
  Pnigouras}]{Passamonti:2022yqp}
Passamonti A, Andersson N, Pnigouras P (2022) {Dynamical tides in superfluid
  neutron stars}. Mon Not Roy Astron Soc 514(1):1494--1510.
  \doi{10.1093/mnras/stac1380},
  {\href{https://arxiv.org/abs/2202.05161}{{arXiv:2202.05161}}} {[astro-ph.HE]}

\bibitem[{Pere{\~n}iguez and Cardoso(2022)}]{Pereniguez:2021xcj}
Pere{\~n}iguez D, Cardoso V (2022) {Love numbers and magnetic susceptibility of
  charged black holes}. Phys Rev D 105(4):044026.
  \doi{10.1103/PhysRevD.105.044026},
  {\href{https://arxiv.org/abs/2112.08400}{{arXiv:2112.08400}}} {[gr-qc]}

\bibitem[{Pere{\~n}iguez and Karnickis(2025)}]{Pereniguez:2025jxq}
Pere{\~n}iguez D, Karnickis E (2025) {On the non-zero Love numbers of magnetic
  black holes}. arXiv e-prints
  {\href{https://arxiv.org/abs/2509.12418}{{arXiv:2509.12418}}} {[gr-qc]}

\bibitem[{Pere{\~n}iguez et~al.(2024)Pere{\~n}iguez, de~Amicis, Brito, and
  Panosso~Macedo}]{Pereniguez:2024fkn}
Pere{\~n}iguez D, de~Amicis M, Brito R, et~al (2024) {Superradiant Instability
  of Magnetic Black Holes}. Phys Rev D 110:104001.
  \doi{10.1103/PhysRevD.110.104001},
  {\href{https://arxiv.org/abs/2402.05178}{{arXiv:2402.05178}}} {[gr-qc]}

\bibitem[{Perry and Rodriguez(2023)}]{Perry:2023wmm}
Perry M, Rodriguez MJ (2023) {Dynamical Love Numbers for Kerr Black Holes}.
  arXiv e-prints {\href{https://arxiv.org/abs/2310.03660}{{arXiv:2310.03660}}}
  {[gr-qc]}

\bibitem[{Piovano et~al.(2023)Piovano, Maselli, and Pani}]{Piovano:2022ojl}
Piovano GA, Maselli A, Pani P (2023) {Constraining the tidal deformability of
  supermassive objects with extreme mass ratio inspirals and semianalytical
  frequency-domain waveforms}. Phys Rev D 107(2):024021.
  \doi{10.1103/PhysRevD.107.024021},
  {\href{https://arxiv.org/abs/2207.07452}{{arXiv:2207.07452}}} {[gr-qc]}

\bibitem[{Pitre and Poisson(2024)}]{Pitre:2023xsr}
Pitre T, Poisson E (2024) {General relativistic dynamical tides in binary
  inspirals without modes}. Phys Rev D 109(6):064004.
  \doi{10.1103/PhysRevD.109.064004},
  {\href{https://arxiv.org/abs/2311.04075}{{arXiv:2311.04075}}} {[gr-qc]}

\bibitem[{Pitre and Poisson(2025)}]{Pitre:2025qdf}
Pitre T, Poisson E (2025) {Impact of nonlinearities on relativistic dynamical
  tides in compact binary inspirals}. arXiv e-prints
  {\href{https://arxiv.org/abs/2506.08722}{{arXiv:2506.08722}}} {[gr-qc]}

\bibitem[{Pnigouras(2019)}]{Pnigouras:2019wmt}
Pnigouras P (2019) {Gravitational-wave-driven tidal secular instability in
  neutron star binaries}. Phys Rev D 100(6):063016.
  \doi{10.1103/PhysRevD.100.063016},
  {\href{https://arxiv.org/abs/1909.04490}{{arXiv:1909.04490}}} {[astro-ph.HE]}

\bibitem[{Pnigouras et~al.(2024)Pnigouras, Gittins, Nanda, Andersson, and
  Jones}]{Pnigouras:2022zpx}
Pnigouras P, Gittins F, Nanda A, et~al (2024) {Rotating Love: The dynamical
  tides of spinning Newtonian stars}. Mon Not Roy Astron Soc 527:8409--8428.
  \doi{10.1093/mnras/stad3593},
  {\href{https://arxiv.org/abs/2205.07577}{{arXiv:2205.07577}}} {[gr-qc]}

\bibitem[{Pnigouras et~al.(2025)Pnigouras, Andersson, Gittins, and
  Counsell}]{Pnigouras:2025muo}
Pnigouras P, Andersson N, Gittins F, et~al (2025) {Dynamical neutron star
  tides: the signature of a mode resonance}. Mon Not Roy Astron Soc
  542(2):1375--1387. \doi{10.1093/mnras/staf1285},
  {\href{https://arxiv.org/abs/2508.06416}{{arXiv:2508.06416}}} {[gr-qc]}

\bibitem[{Poisson(2004)}]{Poisson:2004cw}
Poisson E (2004) {Absorption of mass and angular momentum by a black hole:
  Time-domain formalisms for gravitational perturbations, and the small-hole /
  slow-motion approximation}. Phys Rev D 70:084044.
  \doi{10.1103/PhysRevD.70.084044},
  {\href{https://arxiv.org/abs/gr-qc/0407050}{{arXiv:gr-qc/0407050}}} {[gr-qc]}

\bibitem[{Poisson(2005)}]{Poisson:2005pi}
Poisson E (2005) {Metric of a tidally distorted, nonrotating black hole}. Phys
  Rev Lett 94:161103. \doi{10.1103/PhysRevLett.94.161103},
  {\href{https://arxiv.org/abs/gr-qc/0501032}{{arXiv:gr-qc/0501032}}}

\bibitem[{Poisson(2015)}]{Poisson:2014gka}
Poisson E (2015) {Tidal deformation of a slowly rotating black hole}. Phys Rev
  D 91(4):044004. \doi{10.1103/PhysRevD.91.044004},
  {\href{https://arxiv.org/abs/1411.4711}{{arXiv:1411.4711}}} {[gr-qc]}

\bibitem[{Poisson(2021)}]{Poisson:2020vap}
Poisson E (2021) {Compact body in a tidal environment: New types of
  relativistic Love numbers, and a post-Newtonian operational definition for
  tidally induced multipole moments}. Phys Rev D 103(6):064023.
  \doi{10.1103/PhysRevD.103.064023},
  {\href{https://arxiv.org/abs/2012.10184}{{arXiv:2012.10184}}} {[gr-qc]}

\bibitem[{Poisson and Israel(1990)}]{Poisson:1990eh}
Poisson E, Israel W (1990) {Internal structure of black holes}. Phys Rev D
  41:1796--1809. \doi{10.1103/PhysRevD.41.1796}

\bibitem[{Poisson and Sasaki(1995)}]{Poisson:1994yf}
Poisson E, Sasaki M (1995) {Gravitational radiation from a particle in circular
  orbit around a black hole. 5: Black hole absorption and tail corrections}.
  Phys Rev D 51:5753--5767. \doi{10.1103/PhysRevD.51.5753},
  {\href{https://arxiv.org/abs/gr-qc/9412027}{{arXiv:gr-qc/9412027}}}

\bibitem[{Poisson and Will(2014)}]{PoissonWill}
Poisson E, Will C (2014) {Gravity: Newtonian, Post-Newtonian, Relativistic}.
  Cambridge University Press, Cambridge, UK

\bibitem[{Polchinski(2017)}]{Polchinski:2016hrw}
Polchinski J (2017) {The Black Hole Information Problem}. In: {Theoretical
  Advanced Study Institute in Elementary Particle Physics}: {New Frontiers in
  Fields and Strings}, pp 353--397, \doi{10.1142/9789813149441_0006},
  {\href{https://arxiv.org/abs/1609.04036}{{arXiv:1609.04036}}}

\bibitem[{Porto(2006)}]{Porto:2005ac}
Porto RA (2006) {Post-Newtonian corrections to the motion of spinning bodies in
  NRGR}. Phys Rev D 73:104031. \doi{10.1103/PhysRevD.73.104031},
  {\href{https://arxiv.org/abs/gr-qc/0511061}{{arXiv:gr-qc/0511061}}}

\bibitem[{Porto(2008)}]{Porto:2007qi}
Porto RA (2008) {Absorption effects due to spin in the worldline approach to
  black hole dynamics}. Phys Rev D 77:064026. \doi{10.1103/PhysRevD.77.064026},
  {\href{https://arxiv.org/abs/0710.5150}{{arXiv:0710.5150}}} {[hep-th]}

\bibitem[{Porto(2016{\natexlab{a}})}]{Porto:2016pyg}
Porto RA (2016{\natexlab{a}}) {The effective field theorist{\textquoteright}s
  approach to gravitational dynamics}. Phys Rept 633:1--104.
  \doi{10.1016/j.physrep.2016.04.003},
  {\href{https://arxiv.org/abs/1601.04914}{{arXiv:1601.04914}}} {[hep-th]}

\bibitem[{Porto(2016{\natexlab{b}})}]{Porto:2016zng}
Porto RA (2016{\natexlab{b}}) {The Tune of Love and the Nature(ness) of
  Spacetime}. Fortsch Phys 64:723--729. \doi{10.1002/prop.201600064},
  {\href{https://arxiv.org/abs/1606.08895}{{arXiv:1606.08895}}} {[gr-qc]}

\bibitem[{Prada et~al.(2006)Prada, Klypin, Simonneau, Betancort-Rijo, Patiri,
  Gottlober, and Sanchez-Conde}]{Prada:2005mx}
Prada F, Klypin AA, Simonneau E, et~al (2006) {How far do they go? The Outer
  structure of dark matter halos}. Astrophys J 645:1001--1011.
  \doi{10.1086/504456},
  {\href{https://arxiv.org/abs/astro-ph/0506432}{{arXiv:astro-ph/0506432}}}

\bibitem[{Prasad et~al.(2022)Prasad, Gupta, Bose, and
  Krishnan}]{Prasad:2021dfr}
Prasad V, Gupta A, Bose S, et~al (2022) {Tidal deformation of dynamical
  horizons in binary black hole mergers}. Phys Rev D 105(4):044019.
  \doi{10.1103/PhysRevD.105.044019},
  {\href{https://arxiv.org/abs/2106.02595}{{arXiv:2106.02595}}} {[gr-qc]}

\bibitem[{Pratten et~al.(2022)Pratten, Schmidt, and Williams}]{Pratten:2021pro}
Pratten G, Schmidt P, Williams N (2022) {Impact of Dynamical Tides on the
  Reconstruction of the Neutron Star Equation of State}. Phys Rev Lett
  129(8):081102. \doi{10.1103/PhysRevLett.129.081102},
  {\href{https://arxiv.org/abs/2109.07566}{{arXiv:2109.07566}}} {[astro-ph.HE]}

\bibitem[{Press and Teukolsky(1973)}]{Press:1973zz}
Press WH, Teukolsky SA (1973) {Perturbations of a Rotating Black Hole. II.
  Dynamical Stability of the Kerr Metric}. Astrophys J 185:649--674.
  \doi{10.1086/152445}

\bibitem[{Price and Thorne(1986)}]{Price:1986yy}
Price R, Thorne K (1986) {Membrane Viewpoint on Black Holes: Properties and
  Evolution of the Stretched Horizon}. Phys Rev D 33:915--941.
  \doi{10.1103/PhysRevD.33.915}

\bibitem[{Puecher et~al.(2023{\natexlab{a}})Puecher, Dietrich, Tsang,
  Kalaghatgi, Roy, Setyawati, and Van Den~Broeck}]{Puecher:2022oiz}
Puecher A, Dietrich T, Tsang KW, et~al (2023{\natexlab{a}}) {Unraveling
  information about supranuclear-dense matter from the complete binary neutron
  star coalescence process using future gravitational-wave detector networks}.
  Phys Rev D 107(12):124009. \doi{10.1103/PhysRevD.107.124009},
  {\href{https://arxiv.org/abs/2210.09259}{{arXiv:2210.09259}}} {[gr-qc]}

\bibitem[{Puecher et~al.(2023{\natexlab{b}})Puecher, Samajdar, and
  Dietrich}]{Puecher:2023twf}
Puecher A, Samajdar A, Dietrich T (2023{\natexlab{b}}) {Measuring tidal effects
  with the Einstein Telescope: A design study}. Phys Rev D 108(2):023018.
  \doi{10.1103/PhysRevD.108.023018},
  {\href{https://arxiv.org/abs/2304.05349}{{arXiv:2304.05349}}} {[astro-ph.IM]}

\bibitem[{Raaijmakers et~al.(2021)Raaijmakers, Greif, Hebeler, Hinderer,
  Nissanke, Schwenk, Riley, Watts, Lattimer, and Ho}]{Raaijmakers:2021uju}
Raaijmakers G, Greif SK, Hebeler K, et~al (2021) {Constraints on the Dense
  Matter Equation of State and Neutron Star Properties from
  NICER{\textquoteright}s Mass{\textendash}Radius Estimate of PSR J0740+6620
  and Multimessenger Observations}. Astrophys J Lett 918(2):L29.
  \doi{10.3847/2041-8213/ac089a},
  {\href{https://arxiv.org/abs/2105.06981}{{arXiv:2105.06981}}} {[astro-ph.HE]}

\bibitem[{Rai and Santoni(2024)}]{Rai:2024lho}
Rai M, Santoni L (2024) {Ladder symmetries and Love numbers of
  Reissner-Nordstr{\"o}m black holes}. JHEP 07:098.
  \doi{10.1007/JHEP07(2024)098},
  {\href{https://arxiv.org/abs/2404.06544}{{arXiv:2404.06544}}} {[gr-qc]}

\bibitem[{Randall and Sundrum(1999)}]{Randall:1999ee}
Randall L, Sundrum R (1999) {A Large mass hierarchy from a small extra
  dimension}. Phys Rev Lett 83:3370--3373. \doi{10.1103/PhysRevLett.83.3370},
  {\href{https://arxiv.org/abs/hep-ph/9905221}{{arXiv:hep-ph/9905221}}}

\bibitem[{Read et~al.(2013)Read, Baiotti, Creighton, Friedman, Giacomazzo,
  Kyutoku, Markakis, Rezzolla, Shibata, and Taniguchi}]{Read:2013zra}
Read JS, Baiotti L, Creighton JDE, et~al (2013) {Matter effects on binary
  neutron star waveforms}. Phys Rev D 88:044042.
  \doi{10.1103/PhysRevD.88.044042},
  {\href{https://arxiv.org/abs/1306.4065}{{arXiv:1306.4065}}} {[gr-qc]}

\bibitem[{Regge and Wheeler(1957)}]{Regge:1957td}
Regge T, Wheeler JA (1957) {Stability of a Schwarzschild singularity}. Phys Rev
  108:1063--1069. \doi{10.1103/PhysRev.108.1063}

\bibitem[{Reitze et~al.(2019)}]{Reitze:2019iox}
Reitze D, et~al (2019) {Cosmic Explorer: The U.S. Contribution to
  Gravitational-Wave Astronomy beyond LIGO}. Bull Am Astron Soc 51(7):035.
  {\href{https://arxiv.org/abs/1907.04833}{{arXiv:1907.04833}}} {[astro-ph.IM]}

\bibitem[{Ripley et~al.(2023)Ripley, Hegade K.~R., and Yunes}]{Ripley:2023qxo}
Ripley JL, Hegade K.~R. A, Yunes N (2023) {Probing internal dissipative
  processes of neutron stars with gravitational waves during the inspiral of
  neutron star binaries}. Phys Rev D 108(10):103037.
  \doi{10.1103/PhysRevD.108.103037},
  {\href{https://arxiv.org/abs/2306.15633}{{arXiv:2306.15633}}} {[gr-qc]}

\bibitem[{Ripley et~al.(2024)Ripley, Hegade K.~R., Chandramouli, and
  Yunes}]{Ripley:2023lsq}
Ripley JL, Hegade K.~R. A, Chandramouli RS, et~al (2024) {A constraint on the
  dissipative tidal deformability of neutron stars}. Nature Astron
  8(10):1277--1283. \doi{10.1038/s41550-024-02323-7},
  {\href{https://arxiv.org/abs/2312.11659}{{arXiv:2312.11659}}} {[gr-qc]}

\bibitem[{Riva et~al.(2024)Riva, Santoni, Savi\'c, and Vernizzi}]{Riva:2023rcm}
Riva MM, Santoni L, Savi\'c N, et~al (2024) {Vanishing of nonlinear tidal Love
  numbers of Schwarzschild black holes}. Phys Lett B 854:138710.
  \doi{10.1016/j.physletb.2024.138710},
  {\href{https://arxiv.org/abs/2312.05065}{{arXiv:2312.05065}}} {[gr-qc]}

\bibitem[{Rodr\'iguez et~al.(2026)Rodr\'iguez, Santoni, and
  Solomon}]{OtherReview}
Rodr\'iguez M, Santoni L, Solomon AR (2026) Love numbers of black holes and
  compact objects. preprint

\bibitem[{Rosquist(1994)}]{Rosquist:1994yd}
Rosquist K (1994) {A Tensorial lax pair equation and integrable systems in
  relativity and classical mechanics}. In: {7th Marcel Grossmann Meeting on
  General Relativity (MG 7)}, pp 379--385,
  {\href{https://arxiv.org/abs/gr-qc/9410011}{{arXiv:gr-qc/9410011}}}

\bibitem[{Rosquist and Goliath(1998)}]{Rosquist:1997fn}
Rosquist K, Goliath M (1998) {Lax pair tensors and integrable space-times}. Gen
  Rel Grav 30:1521--1534. \doi{10.1023/A:1018817209424},
  {\href{https://arxiv.org/abs/gr-qc/9707003}{{arXiv:gr-qc/9707003}}}

\bibitem[{Ruffini and Bonazzola(1969)}]{Ruffini:1969qy}
Ruffini R, Bonazzola S (1969) {Systems of selfgravitating particles in general
  relativity and the concept of an equation of state}. Phys Rev 187:1767--1783.
  \doi{10.1103/PhysRev.187.1767}

\bibitem[{Sadeghian et~al.(2013)Sadeghian, Ferrer, and
  Will}]{Sadeghian:2013laa}
Sadeghian L, Ferrer F, Will CM (2013) {Dark matter distributions around massive
  black holes: A general relativistic analysis}. Phys Rev D 88(6):063522.
  \doi{10.1103/PhysRevD.88.063522},
  {\href{https://arxiv.org/abs/1305.2619}{{arXiv:1305.2619}}} {[astro-ph.GA]}

\bibitem[{Saffer and Yagi(2021)}]{Saffer:2021gak}
Saffer A, Yagi K (2021) {Tidal deformabilities of neutron stars in
  scalar-Gauss-Bonnet gravity and their applications to multimessenger tests of
  gravity}. Phys Rev D 104(12):124052. \doi{10.1103/PhysRevD.104.124052},
  {\href{https://arxiv.org/abs/2110.02997}{{arXiv:2110.02997}}} {[gr-qc]}

\bibitem[{Saketh and Maggio(2024)}]{Saketh:2024ojw}
Saketh MVS, Maggio E (2024) {Quasinormal modes of slowly-spinning horizonless
  compact objects}. Phys Rev D 110(8):084038.
  \doi{10.1103/PhysRevD.110.084038},
  {\href{https://arxiv.org/abs/2406.10070}{{arXiv:2406.10070}}} {[gr-qc]}

\bibitem[{Saketh et~al.(2022)Saketh, Steinhoff, Vines, and
  Buonanno}]{saketh2022modeling-cf8}
Saketh MVS, Steinhoff J, Vines J, et~al (2022) Modeling horizon absorption in
  spinning binary black holes using effective worldline theory. {arXiv}
  \doi{10.48550/arxiv.2212.13095},
  {\href{https://arxiv.org/abs/2212.13095}{{2212.13095}}}

\bibitem[{Saketh et~al.(2024{\natexlab{a}})Saketh, Zhou, Ghosh, Steinhoff, and
  Chatterjee}]{Saketh:2024juq}
Saketh MVS, Zhou Z, Ghosh S, et~al (2024{\natexlab{a}}) {Investigating tidal
  heating in neutron stars via gravitational Raman scattering}. Phys Rev D
  110:103001. \doi{10.1103/PhysRevD.110.103001},
  {\href{https://arxiv.org/abs/2407.08327}{{arXiv:2407.08327}}} {[gr-qc]}

\bibitem[{Saketh et~al.(2024{\natexlab{b}})Saketh, Zhou, and
  Ivanov}]{Saketh:2023bul}
Saketh MVS, Zhou Z, Ivanov MM (2024{\natexlab{b}}) {Dynamical tidal response of
  Kerr black holes from scattering amplitudes}. Phys Rev D 109(6):064058.
  \doi{10.1103/PhysRevD.109.064058},
  {\href{https://arxiv.org/abs/2307.10391}{{arXiv:2307.10391}}} {[hep-th]}

\bibitem[{Sasaki and Tagoshi(2003)}]{Sasaki:2003xr}
Sasaki M, Tagoshi H (2003) {Analytic black hole perturbation approach to
  gravitational radiation}. Living Rev Rel 6:6. \doi{10.12942/lrr-2003-6},
  {\href{https://arxiv.org/abs/gr-qc/0306120}{{arXiv:gr-qc/0306120}}}

\bibitem[{Sathyaprakash and Dhurandhar(1991)}]{Sathyaprakash:1991mt}
Sathyaprakash BS, Dhurandhar SV (1991) {Choice of filters for the detection of
  gravitational waves from coalescing binaries}. Phys Rev D 44:3819--3834.
  \doi{10.1103/PhysRevD.44.3819}

\bibitem[{Schmidt and Hinderer(2019)}]{Schmidt:2019wrl}
Schmidt P, Hinderer T (2019) {Frequency domain model of $f$-mode dynamic tides
  in gravitational waveforms from compact binary inspirals}. Phys Rev D
  100(2):021501. \doi{10.1103/PhysRevD.100.021501},
  {\href{https://arxiv.org/abs/1905.00818}{{arXiv:1905.00818}}} {[gr-qc]}

\bibitem[{Schulze et~al.(2026)Schulze, Bernuzzi, Rettegno, Fontbut{\'e},
  Placidi, and Damour}]{Schulze:2026ewu}
Schulze M, Bernuzzi S, Rettegno P, et~al (2026) {High-order effective-one-body
  tidal interactions and gravitational scattering}. preprint
  {\href{https://arxiv.org/abs/2603.22467}{{arXiv:2603.22467}}} {[gr-qc]}

\bibitem[{Seahra et~al.(2005)Seahra, Clarkson, and Maartens}]{Seahra:2004fg}
Seahra SS, Clarkson C, Maartens R (2005) {Detecting extra dimensions with
  gravity wave spectroscopy: the black string brane-world}. Phys Rev Lett
  94:121302. \doi{10.1103/PhysRevLett.94.121302},
  {\href{https://arxiv.org/abs/gr-qc/0408032}{{arXiv:gr-qc/0408032}}}

\bibitem[{Seidel(1989)}]{Seidel:1988ue}
Seidel E (1989) {A Comment on the Eigenvalues of Spin Weighted Spheroidal
  Functions}. Class Quant Grav 6:1057. \doi{10.1088/0264-9381/6/7/012}

\bibitem[{Sennett et~al.(2017)Sennett, Hinderer, Steinhoff, Buonanno, and
  Ossokine}]{Sennett:2017etc}
Sennett N, Hinderer T, Steinhoff J, et~al (2017) {Distinguishing Boson Stars
  from Black Holes and Neutron Stars from Tidal Interactions in Inspiraling
  Binary Systems}. Phys Rev D 96(2):024002. \doi{10.1103/PhysRevD.96.024002},
  {\href{https://arxiv.org/abs/1704.08651}{{arXiv:1704.08651}}} {[gr-qc]}

\bibitem[{Sham et~al.(2014)Sham, Lin, and Leung}]{Sham:2013cya}
Sham YH, Lin LM, Leung PT (2014) {Testing universal relations of neutron stars
  with a nonlinear matter-gravity coupling theory}. Astrophys J 781:66.
  \doi{10.1088/0004-637X/781/2/66},
  {\href{https://arxiv.org/abs/1312.1011}{{arXiv:1312.1011}}} {[gr-qc]}

\bibitem[{Sharma et~al.(2024)Sharma, Ghosh, and Sarkar}]{Sharma:2024hlz}
Sharma C, Ghosh R, Sarkar S (2024) {Exploring ladder symmetry and Love numbers
  for static and rotating black holes}. Phys Rev D 109(4):L041505.
  \doi{10.1103/PhysRevD.109.L041505},
  {\href{https://arxiv.org/abs/2401.00703}{{arXiv:2401.00703}}} {[gr-qc]}

\bibitem[{Sharma et~al.(2026)Sharma, Roy, and Sarkar}]{Sharma:2025xii}
Sharma C, Roy S, Sarkar S (2026) {Ladder symmetry: The necessary and sufficient
  condition for vanishing Love numbers}. Phys Rev D 113(2):024066.
  \doi{10.1103/44dg-smt2},
  {\href{https://arxiv.org/abs/2511.09670}{{arXiv:2511.09670}}} {[gr-qc]}

\bibitem[{Sherf(2021)}]{Sherf:2021ppp}
Sherf Y (2021) {Tidal-heating and viscous dissipation correspondence in black
  holes and viscous compact objects}. Phys Rev D 103(10):104003.
  \doi{10.1103/PhysRevD.103.104003},
  {\href{https://arxiv.org/abs/2104.03766}{{arXiv:2104.03766}}} {[gr-qc]}

\bibitem[{Shiromizu et~al.(2000)Shiromizu, Maeda, and
  Sasaki}]{Shiromizu:1999wj}
Shiromizu T, Maeda Ki, Sasaki M (2000) {The Einstein equation on the 3-brane
  world}. Phys Rev D 62:024012. \doi{10.1103/PhysRevD.62.024012},
  {\href{https://arxiv.org/abs/gr-qc/9910076}{{arXiv:gr-qc/9910076}}}

\bibitem[{Shterenberg and Zhou(2025)}]{Shterenberg:2024tmo}
Shterenberg J, Zhou Z (2025) {Fisher forecast of finite-size effects with
  future gravitational wave detectors}. Phys Rev D 111(8):084068.
  \doi{10.1103/PhysRevD.111.084068},
  {\href{https://arxiv.org/abs/2410.00294}{{arXiv:2410.00294}}} {[gr-qc]}

\bibitem[{Silvestrini et~al.(2025)Silvestrini, Maggio, Chakraborty, and
  Pani}]{Silvestrini:2025lbe}
Silvestrini M, Maggio E, Chakraborty S, et~al (2025) {Tidal deformations of
  compact objects from the membrane paradigm}. Phys Rev D 112(12):124021.
  \doi{10.1103/pky4-23jb},
  {\href{https://arxiv.org/abs/2506.16516}{{arXiv:2506.16516}}} {[gr-qc]}

\bibitem[{Singha and Chakraborty(2025)}]{Singha:2025xah}
Singha C, Chakraborty S (2025) {Tidal deformation of black holes in Lovelock
  gravity}. arXiv e-prints
  {\href{https://arxiv.org/abs/2508.14944}{{arXiv:2508.14944}}} {[gr-qc]}

\bibitem[{Speeney et~al.(2022)Speeney, Antonelli, Baibhav, and
  Berti}]{Speeney:2022ryg}
Speeney N, Antonelli A, Baibhav V, et~al (2022) {Impact of relativistic
  corrections on the detectability of dark-matter spikes with gravitational
  waves}. Phys Rev D 106(4):044027. \doi{10.1103/PhysRevD.106.044027},
  {\href{https://arxiv.org/abs/2204.12508}{{arXiv:2204.12508}}} {[gr-qc]}

\bibitem[{{Starobinskij} and {Churilov}(1973)}]{Starobinskij2}
{Starobinskij} AA, {Churilov} SM (1973) {Amplification of electromagnetic and
  gravitational waves scattered by a rotating black hole.} Zhurnal
  Eksperimentalnoi i Teoreticheskoi Fiziki 65:3--11

\bibitem[{Steinhoff(2015)}]{Steinhoff:2014kwa}
Steinhoff J (2015) {Spin and quadrupole contributions to the motion of
  astrophysical binaries}. Fund Theor Phys 179:615--649.
  \doi{10.1007/978-3-319-18335-0_19},
  {\href{https://arxiv.org/abs/1412.3251}{{arXiv:1412.3251}}} {[gr-qc]}

\bibitem[{Steinhoff et~al.(2016)Steinhoff, Hinderer, Buonanno, and
  Taracchini}]{Steinhoff:2016rfi}
Steinhoff J, Hinderer T, Buonanno A, et~al (2016) {Dynamical Tides in General
  Relativity: Effective Action and Effective-One-Body Hamiltonian}. Phys Rev D
  94(10):104028. \doi{10.1103/PhysRevD.94.104028},
  {\href{https://arxiv.org/abs/1608.01907}{{arXiv:1608.01907}}} {[gr-qc]}

\bibitem[{Steinhoff et~al.(2021)Steinhoff, Hinderer, Dietrich, and
  Foucart}]{Steinhoff:2021dsn}
Steinhoff J, Hinderer T, Dietrich T, et~al (2021) {Spin effects on neutron star
  fundamental-mode dynamical tides: Phenomenology and comparison to numerical
  simulations}. Phys Rev Res 3(3):033129.
  \doi{10.1103/PhysRevResearch.3.033129},
  {\href{https://arxiv.org/abs/2103.06100}{{arXiv:2103.06100}}} {[gr-qc]}

\bibitem[{Tan(2020)}]{Tan:2020hog}
Tan HS (2020) {Tidal Love Numbers of Braneworld Black Holes and Wormholes}.
  Phys Rev D 102(4):044061. \doi{10.1103/PhysRevD.102.044061},
  {\href{https://arxiv.org/abs/2001.00403}{{arXiv:2001.00403}}} {[gr-qc]}

\bibitem[{Taylor and Poisson(2008)}]{Taylor:2008xy}
Taylor S, Poisson E (2008) {Nonrotating black hole in a post-Newtonian tidal
  environment}. Phys Rev D 78:084016. \doi{10.1103/PhysRevD.78.084016},
  {\href{https://arxiv.org/abs/0806.3052}{{arXiv:0806.3052}}} {[gr-qc]}

\bibitem[{Teukolsky(1972)}]{Teukolsky:1972my}
Teukolsky SA (1972) {Rotating black holes - separable wave equations for
  gravitational and electromagnetic perturbations}. Phys Rev Lett
  29:1114--1118. \doi{10.1103/PhysRevLett.29.1114}

\bibitem[{Teukolsky(1973)}]{Teukolsky:1973ha}
Teukolsky SA (1973) {Perturbations of a rotating black hole. 1. Fundamental
  equations for gravitational electromagnetic and neutrino field
  perturbations}. Astrophys J 185:635--647. \doi{10.1086/152444}

\bibitem[{Teukolsky and Press(1974)}]{Teukolsky:1974yv}
Teukolsky SA, Press WH (1974) {Perturbations of a rotating black hole. III -
  Interaction of the hole with gravitational and electromagnet ic radiation}.
  Astrophys J 193:443--461. \doi{10.1086/153180}

\bibitem[{Thorne(1972)}]{Thorne:1972}
Thorne K (1972) Nonspherical gravitational collapse: A short review. In:
  Klauder J (ed) Magic Without Magic: John Archibald Wheeler. A Collection of
  Essays in Honor of his Sixtieth Birthday. W.H. Freeman, San Francisco, p
  231--258

\bibitem[{Thorne(1980)}]{Thorne:1980ru}
Thorne KS (1980) {Multipole Expansions of Gravitational Radiation}. Rev Mod
  Phys 52:299--339. \doi{10.1103/RevModPhys.52.299}

\bibitem[{Thorne et~al.(1986)Thorne, Price, and Macdonald}]{Thorne:1986iy}
Thorne KS, Price R, Macdonald D (eds)  (1986) {Black holes: the membrane
  paradigm}. Yale University Press, New Haven

\bibitem[{Tichy et~al.(2000)Tichy, Flanagan, and Poisson}]{Tichy:1999pv}
Tichy W, Flanagan EE, Poisson E (2000) {Can the postNewtonian gravitational
  wave form of an inspiraling binary be improved by solving the energy balance
  equation numerically?} Phys Rev D 61:104015.
  \doi{10.1103/PhysRevD.61.104015},
  {\href{https://arxiv.org/abs/gr-qc/9912075}{{arXiv:gr-qc/9912075}}}

\bibitem[{Tissino et~al.(2023)Tissino, Carullo, Breschi, Gamba, Schmidt, and
  Bernuzzi}]{Tissino:2022thn}
Tissino J, Carullo G, Breschi M, et~al (2023) {Combining effective-one-body
  accuracy and reduced-order-quadrature speed for binary neutron star merger
  parameter estimation with machine learning}. Phys Rev D 107(8):084037.
  \doi{10.1103/PhysRevD.107.084037},
  {\href{https://arxiv.org/abs/2210.15684}{{arXiv:2210.15684}}} {[gr-qc]}

\bibitem[{Uchikata and Yoshida(2016)}]{Uchikata:2015yma}
Uchikata N, Yoshida S (2016) {Slowly rotating thin shell gravastars}. Class
  Quant Grav 33(2):025005. \doi{10.1088/0264-9381/33/2/025005},
  {\href{https://arxiv.org/abs/1506.06485}{{arXiv:1506.06485}}} {[gr-qc]}

\bibitem[{Uchikata et~al.(2016)Uchikata, Yoshida, and Pani}]{Uchikata:2016qku}
Uchikata N, Yoshida S, Pani P (2016) {Tidal deformability and I-Love-Q
  relations for gravastars with polytropic thin shells}. Phys Rev D
  94(6):064015. \doi{10.1103/PhysRevD.94.064015},
  {\href{https://arxiv.org/abs/1607.03593}{{arXiv:1607.03593}}} {[gr-qc]}

\bibitem[{Unruh(1973)}]{Unruh:1973bda}
Unruh W (1973) {Separability of the Neutrino Equations in a Kerr Background}.
  Phys Rev Lett 31(20):1265--1267. \doi{10.1103/PhysRevLett.31.1265}

\bibitem[{Vaglio et~al.(2023)Vaglio, Pacilio, Maselli, and
  Pani}]{Vaglio:2023lrd}
Vaglio M, Pacilio C, Maselli A, et~al (2023) {Bayesian parameter estimation on
  boson-star binary signals with a coherent inspiral template and
  spin-dependent quadrupolar corrections}. Phys Rev D 108(2):023021.
  \doi{10.1103/PhysRevD.108.023021},
  {\href{https://arxiv.org/abs/2302.13954}{{arXiv:2302.13954}}} {[gr-qc]}

\bibitem[{Vallisneri(2008)}]{Vallisneri:2007ev}
Vallisneri M (2008) {Use and abuse of the Fisher information matrix in the
  assessment of gravitational-wave parameter-estimation prospects}. Phys Rev D
  77:042001. \doi{10.1103/PhysRevD.77.042001},
  {\href{https://arxiv.org/abs/gr-qc/0703086}{{arXiv:gr-qc/0703086}}}

\bibitem[{Vines et~al.(2011)Vines, Flanagan, and Hinderer}]{Vines:2011ud}
Vines J, Flanagan EE, Hinderer T (2011) {Post-1-Newtonian tidal effects in the
  gravitational waveform from binary inspirals}. Phys Rev D 83:084051.
  \doi{10.1103/PhysRevD.83.084051},
  {\href{https://arxiv.org/abs/1101.1673}{{arXiv:1101.1673}}} {[gr-qc]}

\bibitem[{Vines and Flanagan(2013)}]{Vines:2010ca}
Vines JE, Flanagan EE (2013) {Post-1-Newtonian quadrupole tidal interactions in
  binary systems}. Phys Rev D 88:024046. \doi{10.1103/PhysRevD.88.024046},
  {\href{https://arxiv.org/abs/1009.4919}{{arXiv:1009.4919}}} {[gr-qc]}

\bibitem[{Visser(1996)}]{VisserBook}
Visser M (1996) {Lorentzian wormholes: From Einstein to Hawking}. AIP, Woodbury

\bibitem[{Visser and Wiltshire(2004)}]{Visser:2003ge}
Visser M, Wiltshire DL (2004) {Stable gravastars: An Alternative to black
  holes?} Class Quant Grav 21:1135--1152. \doi{10.1088/0264-9381/21/4/027},
  {\href{https://arxiv.org/abs/gr-qc/0310107}{{arXiv:gr-qc/0310107}}} {[gr-qc]}

\bibitem[{Vylet et~al.(2024)Vylet, Ajith, Yagi, and Yunes}]{Vylet:2023pkp}
Vylet K, Ajith S, Yagi K, et~al (2024) {I-Love-Q relations in Einstein-aether
  theory}. Phys Rev D 109(2):024054. \doi{10.1103/PhysRevD.109.024054},
  {\href{https://arxiv.org/abs/2306.11930}{{arXiv:2306.11930}}} {[gr-qc]}

\bibitem[{Wade et~al.(2013)Wade, Creighton, Ochsner, and
  Nielsen}]{Wade:2013hoa}
Wade M, Creighton JDE, Ochsner E, et~al (2013) {Advanced LIGO's ability to
  detect apparent violations of the cosmic censorship conjecture and the
  no-hair theorem through compact binary coalescence detections}. Phys Rev D
  88(8):083002. \doi{10.1103/PhysRevD.88.083002},
  {\href{https://arxiv.org/abs/1306.3901}{{arXiv:1306.3901}}} {[gr-qc]}

\bibitem[{Wang et~al.(2026)Wang, Lehner, Micol, and Sturani}]{Wang:2026qst}
Wang L, Lehner L, Micol M, et~al (2026) {Matching Tidal Deformability (Wilson)
  Coefficients to Black Hole Love Numbers in Higher-Curvature Gravity}.
  preprint {\href{https://arxiv.org/abs/2604.04259}{{arXiv:2604.04259}}}
  {[gr-qc]}

\bibitem[{Wang et~al.(2025)Wang, Shi, Xiong, and Li}]{Wang:2025oek}
Wang R, Shi QL, Xiong W, et~al (2025) {Tidal Love numbers for regular black
  holes}. arXiv e-prints
  {\href{https://arxiv.org/abs/2512.05767}{{arXiv:2512.05767}}} {[gr-qc]}

\bibitem[{Weinberg and Witten(1980)}]{Weinberg:1980kq}
Weinberg S, Witten E (1980) {Limits on Massless Particles}. Phys Lett B
  96:59--62. \doi{10.1016/0370-2693(80)90212-9}

\bibitem[{Williams et~al.(2024)Williams, Schmidt, and
  Pratten}]{Williams:2024twp}
Williams N, Schmidt P, Pratten G (2024) {Phenomenological model of
  gravitational self-force enhanced tides in inspiraling binary neutron stars}.
  Phys Rev D 110(10):104013. \doi{10.1103/PhysRevD.110.104013},
  {\href{https://arxiv.org/abs/2407.08538}{{arXiv:2407.08538}}} {[gr-qc]}

\bibitem[{Wu et~al.(2026)Wu, Luo, and Shi}]{Wu:2026epz}
Wu J, Luo L, Shi J (2026) {Analytical derivation of long-term dephasing caused
  by phase transitions in the context of Kerr black holes}. preprint
  {\href{https://arxiv.org/abs/2603.13637}{{arXiv:2603.13637}}} {[gr-qc]}

\bibitem[{Xia et~al.(2025)Xia, Ma, Pang, and Lu}]{Xia:2025zfp}
Xia M, Ma L, Pang Y, et~al (2025) {Full spectrum of Love numbers of
  Reissner-Nordstrom black hole in D-dimensions}. arXiv e-prints
  {\href{https://arxiv.org/abs/2511.09642}{{arXiv:2511.09642}}} {[hep-th]}

\bibitem[{Xia et~al.(2026)Xia, Long, Pan, Jing, and Qian}]{Xia:2026aty}
Xia ZW, Long S, Pan Q, et~al (2026) {Bayesian inference for tidal heating with
  extreme mass ratio inspirals}. arXiv e-prints
  {\href{https://arxiv.org/abs/2602.11039}{{arXiv:2602.11039}}} {[gr-qc]}

\bibitem[{Yagi(2014)}]{Yagi:2013sva}
Yagi K (2014) {Multipole Love Relations}. Phys Rev D 89(4):043011.
  \doi{10.1103/PhysRevD.89.043011}, [Erratum: Phys. Rev. D 96, 129904 (2017),
  Erratum: Phys. Rev. D 97, 129901 (2018)],
  {\href{https://arxiv.org/abs/1311.0872}{{arXiv:1311.0872}}} {[gr-qc]}

\bibitem[{Yagi and Yunes(2013{\natexlab{a}})}]{Yagi:2013bca}
Yagi K, Yunes N (2013{\natexlab{a}}) {I-Love-Q}. Science 341:365--368.
  \doi{10.1126/science.1236462},
  {\href{https://arxiv.org/abs/1302.4499}{{arXiv:1302.4499}}} {[gr-qc]}

\bibitem[{Yagi and Yunes(2013{\natexlab{b}})}]{Yagi:2013awa}
Yagi K, Yunes N (2013{\natexlab{b}}) {I-Love-Q Relations in Neutron Stars and
  their Applications to Astrophysics, Gravitational Waves and Fundamental
  Physics}. Phys Rev D 88(2):023009. \doi{10.1103/PhysRevD.88.023009},
  {\href{https://arxiv.org/abs/1303.1528}{{arXiv:1303.1528}}} {[gr-qc]}

\bibitem[{Yagi and Yunes(2014)}]{Yagi:2013baa}
Yagi K, Yunes N (2014) {Love can be Tough to Measure}. Phys Rev D 89(2):021303.
  \doi{10.1103/PhysRevD.89.021303},
  {\href{https://arxiv.org/abs/1310.8358}{{arXiv:1310.8358}}} {[gr-qc]}

\bibitem[{Yagi and Yunes(2016{\natexlab{a}})}]{Yagi:2015pkc}
Yagi K, Yunes N (2016{\natexlab{a}}) {Binary Love Relations}. Class Quant Grav
  33(13):13LT01. \doi{10.1088/0264-9381/33/13/13LT01},
  {\href{https://arxiv.org/abs/1512.02639}{{arXiv:1512.02639}}} {[gr-qc]}

\bibitem[{Yagi and Yunes(2016{\natexlab{b}})}]{Yagi:2016ejg}
Yagi K, Yunes N (2016{\natexlab{b}}) {I-Love-Q Relations: From Compact Stars to
  Black Holes}. Class Quant Grav 33(9):095005.
  \doi{10.1088/0264-9381/33/9/095005},
  {\href{https://arxiv.org/abs/1601.02171}{{arXiv:1601.02171}}} {[gr-qc]}

\bibitem[{Yagi and Yunes(2017{\natexlab{a}})}]{Yagi:2016qmr}
Yagi K, Yunes N (2017{\natexlab{a}}) {Approximate Universal Relations among
  Tidal Parameters for Neutron Star Binaries}. Class Quant Grav 34(1):015006.
  \doi{10.1088/1361-6382/34/1/015006},
  {\href{https://arxiv.org/abs/1608.06187}{{arXiv:1608.06187}}} {[gr-qc]}

\bibitem[{Yagi and Yunes(2017{\natexlab{b}})}]{Yagi:2016bkt}
Yagi K, Yunes N (2017{\natexlab{b}}) {Approximate Universal Relations for
  Neutron Stars and Quark Stars}. Phys Rept 681:1--72.
  \doi{10.1016/j.physrep.2017.03.002},
  {\href{https://arxiv.org/abs/1608.02582}{{arXiv:1608.02582}}} {[gr-qc]}

\bibitem[{Yu and Lau(2025)}]{Yu:2025ptm}
Yu H, Lau SY (2025) {Effective-one-body model for coalescing binary neutron
  stars: Incorporating tidal spin and enhanced radiation from dynamical tides}.
  Phys Rev D 111(8):084029. \doi{10.1103/PhysRevD.111.084029},
  {\href{https://arxiv.org/abs/2501.13064}{{arXiv:2501.13064}}} {[gr-qc]}

\bibitem[{Yu et~al.(2023)Yu, Weinberg, Arras, Kwon, and
  Venumadhav}]{Yu:2022fzw}
Yu H, Weinberg NN, Arras P, et~al (2023) {Beyond the linear tide: impact of the
  non-linear tidal response of neutron stars on gravitational waveforms from
  binary inspirals}. Mon Not Roy Astron Soc 519(3):4325--4343.
  \doi{10.1093/mnras/stac3614},
  {\href{https://arxiv.org/abs/2211.07002}{{arXiv:2211.07002}}} {[gr-qc]}

\bibitem[{Yu et~al.(2024)Yu, Arras, and Weinberg}]{Yu:2024uxt}
Yu H, Arras P, Weinberg NN (2024) {Dynamical tides during the inspiral of
  rapidly spinning neutron stars: Solutions beyond mode resonance}. Phys Rev D
  110(2):024039. \doi{10.1103/PhysRevD.110.024039},
  {\href{https://arxiv.org/abs/2404.00147}{{arXiv:2404.00147}}} {[gr-qc]}

\bibitem[{Yusmantoro et~al.(2025)Yusmantoro, Zen, and
  Prihadi}]{Yusmantoro:2025ylw}
Yusmantoro, Zen FP, Prihadi HL (2025) {Scalar tidal response of static and
  rotating black holes in anti-de Sitter spacetime}. arXiv e-prints
  {\href{https://arxiv.org/abs/2507.22759}{{arXiv:2507.22759}}} {[gr-qc]}

\bibitem[{{Zahn}(1970)}]{1970A&A.....4..452Z}
{Zahn} JP (1970) {Forced Oscillations in Close Binaries. The Adiabatic
  Approximation}. \aap 4:452

\end{thebibliography}

\end{document}